\documentclass[preprint,3p,times]{elsarticle}
\usepackage{graphicx}
\usepackage{amssymb}
\usepackage{amsthm}
\usepackage{amsmath}
\usepackage{colordvi}

\newcommand{\be}{\begin{equation}}
\newcommand{\ee}{\end{equation}}
\newcommand{\bea}{\begin{eqnarray}}
\newcommand{\eea}{\end{eqnarray}}
\newcommand{\bnum}{\begin{enumerate}}
\newcommand{\enum}{\end{enumerate}}
\newcommand{\ii}{{\rm i}}
\newcommand{\spin}{{\mbox{\boldmath $\sigma$\unboldmath}}}
\newcommand{\gradient}{{\mbox{\boldmath$\nabla$\unboldmath}}}
\newcommand{\bgreek}[1]{\mbox{\boldmath$#1$\unboldmath}}
\newcommand{\up}{\uparrow}
\newcommand{\down}{\downarrow}
\newcommand{\bsigma}{\bgreek{\sigma}}
\newcommand{\bomega}{\bgreek{\omega}}
\newcommand{\bOmega}{\bgreek{\Omega}}
\newcommand{\bnabla}{\bgreek{\nabla}}

\biboptions{sort&compress}
\journal{Physics Reports}
\begin{document}

\begin{frontmatter}
\title{Spin dynamics in semiconductors}
\author{M. W. Wu\corref{cor1}}
\cortext[cor1]{Corresponding author.}
\ead{mwwu@ustc.edu.cn}
\author{J. H. Jiang}
\author{M. Q. Weng}

\address{Hefei National Laboratory for Physical Sciences at
  Microscale, University of Science and Technology of China, Hefei,
  Anhui, 230026, China\\ Department of Physics,
  University of Science and Technology of China, Hefei,
  Anhui, 230026, China}


\begin{abstract}
This article reviews the current status of 
spin dynamics in semiconductors which has achieved a lot of progress in the past
years due to the fast growing field of semiconductor spintronics. 
The primary focus is  the theoretical and
experimental developments of spin relaxation and dephasing in
both spin precession in time domain and spin diffusion and transport
in spacial domain. A fully microscopic many-body investigation
on spin dynamics based on the kinetic spin Bloch equation approach 
is reviewed comprehensively. 
\end{abstract}

\begin{keyword} spin dynamics; spin-orbit coupling; spin relaxation;
 spin dephasing; spin diffusion; spin transport; electron-electron
Coulomb  scattering; magnetic semiconductors; magnetization dynamics;
  ultrafast phenomena; quantum well; quantum wire; quantum dot;
  many-body effect; high field effect.

\end{keyword}

\end{frontmatter}

\section{Introduction}

Since the pioneering works by Lampel \cite{PhysRevLett.20.491} and
Parsons \cite{PhysRevLett.23.1152} and the
following extensive experimental and theoretical works at Ioffe
Institute in St. Petersburg and Ecole Polytechnique in Paris in 1970s
and early 1980s, a lot of understanding on spin dynamics in
semiconductors has been achieved. 
 Some basic spin relaxation and
dephasing mechanisms have also been proposed at that time. A nice
review on these findings can be found in the book {\em ``Optical Orientation''}
\cite{opt-or}.

Starting from the late 1990s, there is a big revival of
research interest in the spin dynamics of semiconductors, jump-started
by some nice experimental works by Awschalom and co-workers
\cite{PhysRevLett.80.4313,kikkawa_99}. An extensive number of 
experimental and theoretical investigations 
 have been carried out on all aspects of spin properties.
Properties of spin relaxation and dephasing in both the
time domain (in the spacial uniform system) and the
spacial domain (in spin diffusion and transport) have been fully explored in
different materials at various conditions (such as  temperature,
external field, doping density/material and strain) and 
dimensions (including bulk materials, 
quantum wells, quantum wires and quantum dots). These materials include
III-V and II-VI zinc-blende and wurtzite semiconductors and silicon/germanium,
as well as diluted magnetic semiconductors. The spin relaxation and dephasing 
mechanisms have been rechecked and  reinvestigated at different conditions. 
The spin-related material properties have also achieved much progress along 
with these investigations which facilitate the better understanding of
the observed phenomena as well as the manipulation of the spin coherence.
Many novel spin-related properties, such as the spin Hall effect
\cite{dyakonov:jetplett.13.467,Dyakonov1971459,PhysRevLett.83.1834,PhysRevLett.85.393,Kato12102004,PhysRevLett.94.047204,Valenzuela2006},
spin Coulomb drag effect \cite{nature.437.1330,amico_00,amico_01}, spin
photogalvanic effect \cite{ganichev_nature417153,ganichev:3146} and persistent 
spin helix effect \cite{bernevig:236601,Koralek2009,shen_09}, have
been discovered. For the sake of realizing the spintronic devices such
as spin transistors, there have been extensive investigations on spin
injection and detection. Much progress has been achieved on the spin
injection from ferromagnetic materials into semiconductors.
Nevertheless, a satisfactory realization of spin transistor which is
crucial to the application of semiconductor spintronics is yet to come.

Despite decades of studies, theoretical understanding of the
spin relaxation and dephasing which is one of the prerequisites of the
realization of spintronics, in both
spacial uniform and nonuniform systems is still not fully
achieved. Many new experimental findings go far beyond the previous
theoretical understandings, thanks to the fast development of
experimental techniques including the sample preparations and
ultrafast optical techniques. Some of the physics requires a many-body
and/or non-equilibrium (even far away from the equilibrium)
theory. However, the previous theories \cite{opt-or} are all based on
the single-particle approach and for systems near the equilibrium which
thus have their strict limits in applying to the nowadays experiments. Some 
even give totally incorrect qualitative predictions.
Wu et al. developed a fully microscopic many-body kinetic spin Bloch
equation approach to study the spin dynamics in semiconductors and
their nanostructures. Unlike other approaches widely used in the
literature which treat scattering using the relaxation time
approximation, the kinetic spin Bloch equation approach treats all
scattering explicitly and self-consistently. Especially,
the carrier-carrier Coulomb scattering is explicitly included in the
theory. This allowed them to study spin dynamics not only near but
also far away from the equilibrium, for example, the spin dynamics in the
presence of high 
electric field (hot-electron condition) and/or with large initial spin
polarization.  They applied this approach to study spin
relaxation/dephasing and spin diffusion/transport in various kinds of
semiconductors and their nanostructures under diversified conditions
and have offered a complete and systematic understanding of spin
dynamics in semiconductors. Many novel effects were predicted and some
of them have been verified experimentally. 

In this review, we try to provide a full coverage of the latest
developments of spin dynamics in semiconductors. Both experimental
and theoretical developments are summarized. In theory, we first
review the studies based on the single-particle approach. Then we
provide a comprehensive review of the results based on the kinetic
spin Bloch equation approach.
We organize the paper as follows: We first review the spin interactions in
semiconductors in Section~2. Then we briefly discuss the physics of spin
dynamics in semiconductors in Section~3.  Here we review the previous
understanding on spin relaxation and dephasing mechanisms.
Differing from the existing nice reviews on these
topics \cite{zuticrmp,Fabianbook,dyakonovbook_08},  we
try to provide a full and up-to-date picture of spin dynamics 
of itinerant carriers in semiconductors. We then
review the latest experimental developments on spin relaxation and
dephasing in time domain as well as the latest 
theoretical investigations based on
the single-particle approach in Section~4. The theoretical investigations
of the spin relaxation and dephasing in the spacial
uniform systems based on the kinetic spin Bloch equation approach 
together with the
 experimental verifications are reviewed in 
Section~5. In Section~6 we review the latest experimental and theoretical
(single-particle theory) results on spin diffusion and transport. Then
in Section~7 we review the investigation on spin diffusion and transport
using the kinetic spin Bloch equation approach and the
related experimental results. We summarize in Section~8.

\section{Spin interactions in semiconductors}

\subsection{A short introduction of spin interactions}

Before introducing the spin interactions, it should be mentioned that
the ``spins'' in semiconductors are not {\it pure} spins, or in other
words, they are angular momentums which consist of both spin and
orbital angular momentums. A direct consequence is that the
$g$-factor is no longer $g_0=2.0023$ of pure electron spins.

There are various spin interactions in semiconductors.
Besides the Zeeman interaction,
there are spin-orbit coupling due to the space inversion asymmetry or
strain, $s(p)$-$d$ exchange interaction with magnetic impurities, hyperfine
interaction with nuclear spins, spin-phonon interaction, electron-hole
and electron-electron exchange interactions. Among these interactions,
spin-orbit coupling usually plays the most important role in spin dynamics.
For electrons in conduction band, spin-orbit coupling consists of the
Dresselhaus, Rashba and strain-induced terms. For valence band
electrons, besides these terms, there is an important spin-orbit
coupling from the spin-dependent terms in the Luttinger
Hamiltonian in cubic semiconductors. These spin-orbit couplings make the motion of
carrier spin couple with the orbital motion and give rise
to many novel effects in coherent (ballistic) and dissipative
(diffusive) electron spin/charge dynamics. A direct consequence is
that the spin polarization of a wave packet oscillates during its
propagation
\cite{PhysRevB.69.125310,weng:410,PhysRevB.67.052407,jiang:113702,cheng:073702,cheng:205328,hruska:075306,PhysRevLett.94.236601}.
Recent studies indicate that as the precession during the electron
propagation correlates the spin polarization with real-space trajectory, the spin lifetime is enhanced at certain spacial
inhomogeneous spin polarization states, such as, some spin grating states
\cite{bernevig:236601,weng:063714,weber:076604,stanescu:125307,PhysRevB.70.155308,PhysRevB.71.155317,Koralek:nature458.610,shen_09}.
Another important spin interaction is the exchange interaction between
carriers due to the permutative antisymmetry of the many-body fermion
wavefunctions. It has been demonstrated to be important for spin
dynamics of electrons in quantum dots \cite{J.R.Petta09302005} and 
localized electrons bound to impurities \cite{PhysRevB.66.245204,0268-1242-23-11-114009}.
The effect of the exchange interactions between {\em free} electrons on spin
dynamics was only observed in experiments
very recently and has attracted interest
\cite{stich:176401,stich:205301,saarikoski:097204}, though it has been
predicted much earlier \cite{PhysRevB.68.075312}.

Spin interactions together with the induced spin structures are
the physical foundations of the vital topic of coherent spin
manipulation which is crucial for spin-based quantum information
processing
\cite{DavidP.DiVincenzo10131995,PhysRevA.57.120,Taylor:nphys174,Trauzettel:nphys544,JPSJ.77.031012,Engel:QIP115,hanson:1217,Hanson:453-1043}
and spintronic device operation \cite{PhysRevLett.97.240501,Fabianbook,datta:665,HyunCheolKoo09182009,bandyopadhyay:1814}.
For example, spin-orbit couplings enable coherent manipulation of spin
by electric field \cite{rashbabook,rashba:apl5295,PhysRevLett.91.126405,Y.Kato:nat50,efros:prb165325,mduckheim:195,duckheim:prb201305,lmeier:650,meier:prb035305,studer:045302,studer:027201,wilamowski:174423,jiang:prb125309,rashba:195302,PhysRevB.74.165319,PhysRevLett.97.240501,bulaev:097202,PhysRevB.76.195307,PhysRevB.77.045310,K.C.Nowack11302007,PhysRevLett.94.226803}.
As electric field is more easily accessible and controllable in
genuine electronic devices at nanoscale, the  electrical
 manipulation of spins is important for semiconductor
  spintronics. The idea has been extended to a large variety of
schemes of electrical or optical manipulation of
spin coherence: coherent control via electrical modulation of
$g$-tensor in nanostructures with spacial-dependent $g$-tensor
\cite{Y.Kato:sci1201,gsalis:619,andlauer:prb045307,de:prl017603,doty:prl197202};
and similar scheme based on spatially varying magnetic field
\cite{tokura:047202,Pioro-Ladriere:776}, or hyperfine field
\cite{laird:246601,rashba:195302,J.R.Petta02052010}; coherent manipulation based on
tuning the exchange interaction between electrons in a double quantum dot by
gate-voltage \cite{J.R.Petta09302005,Petta:prb161301}; coherent spin
manipulation via detuned optical pulses which act as effective
magnetic pulses as the Stark shift is spin-dependent due to
spin-dependent virtual transition of the circularly polarized light 
\cite{J.A.Gupta06292001,J.Berezovsky04182008,carter:201308,pryor:233108,davidpress:218,Xu:nphys1054,Greilich:nphys1226,2009arXiv0905.0297K,2009arXiv0912.2589Z}.
The $p$-$d$ or $s$-$d$ exchange interactions and the induced spin
structures further enable the above schemes to control Mn ion spins
in paramagnetic phase \cite{tang:106803,leger:107401,rcmyers:203} or magnetization
in ferromagnetic phase \cite{chovan:085321,chovan:057402} in magnetic
semiconductors via manipulating carrier spins.

One of the key obstacles for spintronic device operation and spin-based
quantum information processing is the spin relaxation and spin
dephasing (i.e., the decay of the longitudinal
and transverse spin components, respectively).
Spin relaxation and spin
dephasing, in principle, also originate from the spin interactions:
the fluctuation or inhomogeneity in spin interactions leads to spin
relaxation and spin dephasing.\footnote{A practical operating scheme
should both control the spin interactions and avoid their fluctuation
and inhomogeneity carefully. Optimization of the ability of coherent
control and spin lifetime is often needed.}
For example, the spin-orbit coupling  is  the origin of one of
the most efficient spin relaxation mechanisms--the D'yakonov-Perel'
mechanism \cite{dp,DP2}. Spin mixing due to conduction-valence band mixing
together with momentum scattering give rise to the Elliott-Yafet spin relaxation mechanisms
\cite{yafetbook,PhysRev.96.266}. Other spin relaxation mechanisms are
the $s(p)$-$d$ exchange mechanism, the Bir-Aronov-Pikus mechanism
\cite{BAP,aronov} due to the electron-hole exchange interaction
\cite{jetp33.108}, the hyperfine interaction mechanism and the
anisotropic exchange interactions between localized electrons
\cite{PhysRevB.66.245204,0268-1242-23-11-114009}. Spin relaxation and
dephasing in genuine semiconductor structures are the consequence of
the coaction of these mechanisms. Hence it is important to determine
the dominant spin relaxation mechanism under various conditions
\cite{opt-or,dyakonovbook_08,PhysRevB.66.035207,zhou:075318,jiang:125206,jiang:155201,PhysRevB.66.245204,zhou0905.2790}.

Finally, as the focus of this review is spin dynamics in
semiconductors, we only provide an overview for various spin
interactions and explain the related physics using intuitive and
qualitative pictures. For details, we suggest the salient works in the
existing literature: for spin-orbit coupling in semiconductors, the
readers could refer to the book by Winkler \cite{winklerbook}; for 
knowledge of the $s(p)$-$d$ exchange interactions, we suggest the review
article by Jungwirth et al. \cite{jungwirth:809} and that by Furdyna
\cite{furdyna:R29} and references therein; the hyperfine interaction
in semiconductors has been reviewed recently by Fischer et
al. \cite{fischer0903.0527}; details of the spin-phonon interaction
can be found in Ref.~\cite{Titkovbook}; the electron-hole exchange
interaction has been discussed in detail in
Refs.~\cite{jetp33.108,PhysRevB.47.15776,PhysRevB.37.6429}.

\subsection{Spin-orbit interactions in semiconductors and their
  nanostructures}

A consequence of the relativistic effect in atomic and condensed
matter physics is that an electron moving in the atomic potential
$V(\bf r)$ feels an effective magnetic field acting on its spin 
\begin{equation}
  H_{\rm SO} = \frac{1}{4m_0^2c^2} {\bf p}\cdot 
  \left[ \spin \times \left( \gradient V({\bf r}) \right) \right],
\label{soi}
\end{equation}
where $m_0$ and $c$ represent the free electron mass and the velocity
of light in vacuum, respectively, $\spin$
is the Pauli matrices, and ${\bf p}$ stands for the canonical
momentum. $V({\bf r})$ is the total potential generated by
other charges. In semiconductors,
$V({\bf r})$ includes the periodic potentials generated by the
ion-core, the deviation of the periodic potential due to defects and
phonons, and the external potential. This is the origin of various
spin-orbit interactions in semiconductors.

\subsubsection{Electron spin-orbit  coupling term 
  in semiconductors due to space inversion asymmetry}

In semiconductors with space inversion symmetry, the electron spectrum
satisfies the following relation, $\varepsilon_{{\bf k}\uparrow} =
\varepsilon_{{\bf k}\downarrow}$. This is the consequence of the
coaction of the time-reversal symmetry and the space-inversion
symmetry. In the absence of space inversion symmetry,
$\varepsilon_{{\bf k}\uparrow}\ne \varepsilon_{{\bf k}\downarrow}$. In
this case, there should be some term which breaks the spin degeneracy
of the same ${\bf k}$ states. The term should be an odd function of
both ${\bf k}$ and ${\bf \spin}$ as it breaks the space
inversion symmetry whereas keeps the time reversal
  symmetry. This term is the so-called the spin-orbit coupling term.
For electron with
spin $S=1/2$, as $\sigma_i\sigma_j=\delta_{i,j}+\ii
\epsilon_{ijk}\sigma_k$ ($i,j,k=x,y,z$ and $\epsilon_{ijk}$ is
Levi-Civita tensor), only the linear form of $\sigma_i$ can appear in
the spin-orbit coupling term. Hence, the only possible form of the spin-orbit
  coupling term is
\be
H_{\rm SO} = \frac{1}{2}{\bf \Omega}({\bf k})\cdot \spin,
\label{eso}
\ee
where ${\bf \Omega}({\bf k})$ is an odd function of ${\bf k}$, which
acts as an effective magnetic field. In a reversed logic, any electron
spin-orbit coupling which can be written in the form $H_{\rm SO}=
\sum_{n,m,i,j}C_{nm}^{ij}\sigma_i^nk_j^m$, can be reduced to the form
of Eq.~(\ref{eso}). Hence the breaking of the space inversion symmetry is
a prerequisite\footnote{Note that, it was found that there is
  intersubband spin-orbit coupling even in symmetric quantum wells
  \cite{bernardes:076603,calsaverini:155313}, which seems to
  contradict the above conclusion. However, even in symmetric quantum
  wells, the space inversion symmetry is {\em broken} at the boundary of the
  quantum well \cite{bernardes:076603,calsaverini:155313}.} for the appearance of electron spin-orbit
coupling term in  semiconductors.\footnote{Differently, hole has spin 
  $J=3/2$, where $J_i^2\ne 1$. Hence, terms which are quadratic in 
  ${\bf J}$ can appear in the
  spin-orbit coupling. These terms should be even in ${\bf k}$ in
  order to keep the time-reversal symmetry. Hence, hole can have
  spin-orbit coupling terms which do not break space inversion
  symmetry. Actually, the Luttinger Hamiltonian
  \cite{PhysRev.102.1030} only contains spin-dependent terms quadratic
  in ${\bf k}$.}

In semiconductors, there are several kinds of space inversion
asymmetry: (i) the bulk inversion asymmetry due to structures lacking inversion
center \cite{PhysRev.100.580}, such as zinc-blende III-V or II-VI semiconductors; (ii) the structure inversion asymmetry
resulting from the inversion asymmetry of the potentials in
nanostructures including external gate-voltage and/or built-in electric
field \cite{0022-3719-17-33-015,rashba:jetplett.39.78}; and (iii) interface
inversion asymmetry associated with the chemical bonding within
interfaces \cite{PhysRevLett.77.1829,PhysRevB.54.5852}.
Besides, strain can also induce inversion asymmetry and leads to
spin-orbit coupling. The explicit form of the spin-orbit couplings due
to these space inversion asymmetries can be  obtained, e.g., by
the ${\bf k}\cdot{\bf p}$ theory \cite{winklerbook}.

\subsection{Explicit form of the spin-orbit coupling terms in semiconductors}

The explicit form of the spin-orbit coupling in semiconductors can be
derived from the ${\bf k}\cdot {\bf p}$ theory. For common III-V and
II-VI semiconductors, such as GaAs, InAs, CdTe and ZnSe, the minimum
${\bf k}\cdot{\bf p}$ theory to describe carrier and spin dynamics is
the eight-band Kane model \cite{Kane1957249,winklerbook}. In the 
following, we introduce the spin-orbit coupling in the framework
of such model.

\subsubsection{Kane Hamiltonian and block-diagonalization}

In zinc-blende structures, the Kane model is of the following form
\cite{Titkovbook} 
\be
  H_{\rm Kane} = \left[ \begin{array}{ccc} 
     H_{c} & H_{c,v} & H_{c,sv} \\
     H_{c,v}^{\dagger} & H_v & H_{v,sv} \\
     H_{c,sv}^{\dagger} &  H_{v,sv}^{\dagger} & H_{sv}
  \end{array}\right].
\ee
The subscript $c$, $v$, $sv$ denote the conduction, valence and the
split-off valence bands respectively.
As $H_{c,v}$, $H_{v,sv}$ and $H_{c,sv}$ are nonzero, the electron
motions in these bands are coupled together, which are obviously too
complex. Fortunately, the couplings between these
bands are much smaller than their separations. By a perturbative
block-diagonalization, one can decouple the motions in these
bands. The block-diagonalization is achieved by the L\"owdin
partition method \cite{lowdin:1396} which is actually a unitary
transformation. For example, the eigen-energy $E_n({\bf k})$ and
eigen-state $\Psi_n({\bf k})$ of the original Kane Hamiltonian satisfy
the following eigen-equation 
\be
H_{\rm Kane} \Psi_n({\bf k}) = E_n({\bf k}) \Psi_n({\bf k}).
\ee
After a unitary transformation,
\be
 \tilde{H}_{\rm Kane}  \tilde{\Psi}_n({\bf k}) = E_n({\bf k})
 \tilde{\Psi}_n({\bf k}), 
\ee
where $\tilde{H}_{\rm Kane}  = U H_{\rm Kane} U^{\dagger}$ is
block-diagonal. As the couplings between bands $H_{c,v}$, $H_{v,sv}$
and $H_{c,sv}$ are small near the center of the concerned valley (For
most of the III-V and II-VI semiconductors, the concerned valley is
$\Gamma$ valley, where ${\bf k}=0$ is the valley center.), 
the results can be given in a perturbative expansion series. 
Below, we give the results with only the lowest order nontrivial
terms kept, which are enough for the discussion of spin dynamics in
most cases. Besides the transformation of the Hamiltonian, the
wavefunction also changes slightly, $\tilde{\Psi}_n({\bf k}) = U
\Psi_n({\bf k})$. This modification of the wavefunction always
leads to the mixing of different spin states.
 In the presence of spin-mixing, {\em any}
momentum scattering can lead to spin flip and spin relaxation, which
gives the Elliott-Yafet spin relaxation mechanism
\cite{yafetbook,PhysRev.96.266}. 

In the following, we present the conduction (electron) and 
valence (hole) band  parts  
of the block-diagonal Hamiltonian
$\tilde{H}_{\rm Kane}$.

\subsubsection{Electron Hamiltonian and spin-orbit coupling in bulk system}

By choosing $x$, $y$, $z$ axes along the crystal axes of
$[100]$, $[010]$ and $[001]$ respectively, the electron Hamiltonian
can be written as 
\be
  H_e = \frac{ k^2}{2m_e} + \frac{1}{2}{\bf \Omega}_{D}\cdot
    \spin + \frac{1}{2} {\bf \Omega}_{\rm S}\cdot \spin,
\ee
where $m_e$ is the electron (conduction band) effective mass. The second and
third terms in the right hand side of the above equation describe the
Dresselhaus spin-orbit coupling \cite{PhysRev.100.580} and
strain-induced spin-orbit coupling \cite{Titkovbook} respectively.
\be
\Omega_{Dx} = 2\gamma_{D} k_x(k_y^2-k_z^2), \quad\quad
\Omega_{Sx} = 2C_3 (\epsilon_{xy} k_y - \epsilon_{xz} k_z) + 2 D
k_x(\epsilon_{yy}-\epsilon_{zz}), 
\label{strain_soc}
\ee
with other components obtained by cyclic permutation of indices. Here
$\gamma_{D}$ is the Dresselhaus coefficient, which can be expressed formally as $\gamma_{D} =
2\eta / [3m_{cv}\sqrt{2m_eE_g(1-\eta/3)}]$ \cite{Titkovbook},
where $\eta=\Delta_{\rm SO}/(E_g+\Delta_{\rm SO})$ with $E_g$ and
$\Delta_{\rm SO}$ being the bandgap and the spin-orbit
splitting of the valence band respectively. $m_{cv}$ is a
parameter with mass dimension and is related to the
interaction between the conduction and valence bands \cite{Titkovbook}.
The coefficients $\gamma_{D}$, $C_3$ and $D$ are crucial for spin
dynamics. There have been a lot of studies on these quantities,
especially $\gamma_{D}$, in common III-V semiconductors (such as
GaAs, InAs, GaSb and InSb) both theoretically and experimentally
\cite{Titkovbook,PhysRevB.38.1806,PhysRevLett.68.106,PhysRevB.47.16028,PhysRevB.51.4707,PhysRevB.51.5121,PhysRevB.53.3912,PhysRevB.53.12813,PhysRevLett.90.076807,PhysRevB.72.193201,chantis:086405,krich:226802,luo:056405}.
Interestingly, these studies give quite different values of 
$\gamma_{D}$ in GaAs from 7.6 to 36~eV$\cdot$\AA$^3$. Detailed lists of
$\gamma_{D}$ in GaAs in the literature are presented in
Refs.~\cite{chantis:086405,krich:226802,luo:056405}. Latest
theoretical advancements include the first principle calculations
\cite{chantis:086405,luo:056405} and full-Brillion zone investigations
\cite{Fu20082890,luo:056405}. Recently, from fitting the
magnetotransport properties in chaotic GaAs quantum dots
\cite{krich:226802} and from fitting the spin relaxation in bulk GaAs
via the fully microscopic kinetic spin Bloch equation approach \cite{jiang:125206}, $\gamma_{D}$
in GaAs was found to be $9$ and $8.2$~eV$\cdot$\AA$^3$ respectively,
which are close to the value $\gamma_{D}=8.5$~eV$\cdot$\AA$^3$
from {\it ab initio} calculations with GW approximation
\cite{chantis:086405}.\footnote{There are other recent papers \cite{luo:056405,PhysRevLett.104.066405} with
  $\gamma_{D}$ close to $8.5$~eV$\cdot$\AA$^3$.} The coefficient for strain-induced spin-orbit
coupling is $C_3 = 2 c_2 \eta /[3\sqrt{2m_eE_g(1-\eta/3)}]$, where $c_2$ is the interband
deformation-potential constant. $D$ comes from higher
order corrections which is usually smaller than $C_3$. The strain-induced
spin-orbit coupling was studied experimentally in
Refs.~\cite{Y.Kato:sci1201,knotz:241918}, indicating that the strain
induced spin-orbit coupling can be more important than the
Dresselhaus spin-orbit coupling and that both the $C_3$ and $D$ terms
can be dominant. Recently, the coefficients $C_3$ and $D$ from {\it ab
  initio} calculations showed good agreement with the experimental
results \cite{chantis-2008-78}, where $C_3=6.8$~eV$\cdot$\AA\ 
and $D=2.1$~eV$\cdot$\AA.

\subsubsection{Electron spin-orbit coupling in nanostructures}

Electric field can also break the space inversion symmetry and lead to
additional spin-orbit coupling
\cite{0022-3719-17-33-015,rashba:jetplett.39.78}. In III-V and II-VI
heterostructures the induced spin-orbit coupling, named
the Rashba spin-orbit coupling, can be as important as (or even more
important than) the Dresselhaus spin-orbit coupling due to bulk
inversion asymmetry \cite{PhysRevB.72.161311,PhysRevB.70.115328}. The
effect is caused by the total electric field, including the external
electric field due to gate-voltage, the electric field due to built-in
electrostatic potential and the electric field due to interfaces
\cite{winklerbook}. The leading effect 
of interfaces is to produce a spacial variation of band edge, which
thus leads to an effective electric field. Via L\"owdin partition
method, one obtains
\be
H_R = \alpha_0 \spin \cdot ({\bf k}\times{\bf \cal{E}}_v({\bf r})). 
\ee
Here $\alpha_0 = e\eta(2-\eta)P^2 /(3m_0^2E_g^2)$ with $P$
being the interband momentum matrix \cite{winklerbook},
${\bf \cal{E}}_v ({\bf r}) = \gradient V_v({\bf r})/|e|$ where
$V_v({\bf r})$ is the potential felt by valence electron. Averaging
${\bf \cal{E}}_v ({\bf r})$ over, e.g., the subband wavefunction of a
quantum well, one has
\be
H_R = \alpha_{R} \spin \cdot ({\bf k} \times \hat{\bf n}),
\ee
where $\alpha_{R} = \alpha_0\langle {\cal{E}}_v({\bf r}) \rangle$ is the
Rashba parameter and $\hat{\bf n}$ is a unit vector along the growth
direction of the quantum well.

Historically, there is a paradox about the Rashba parameter. According
to the Ehrenfest Theorem, the average of the force acting on a bound
state is zero. Therefore, the Rashba parameter should be very
small. This paradox is resolved by Lassnig \cite{PhysRevB.31.8076},
who pointed out that for the Rashba spin-orbit coupling in conduction
band, $\alpha_{R}$ is related with the average of the electric field in
valence band over the conduction band (subband) wavefunction. As
pointed out above, the electric potential on the valence or conduction
band comprises three parts,
\be
V_v({\bf r}) = V_{\rm ext}({\bf r}) + V_{\rm built}({\bf r}) +
E_v({\bf r}), \quad\quad
V_c({\bf r}) = V_{\rm ext}({\bf r}) + V_{\rm built}({\bf r}) +
E_c({\bf r}),
\ee
where $E_c({\bf r})$ and $E_v({\bf r})$ are the position dependent
conduction and valence band edges. $V_{\rm built}({\bf r})$ is the
built-in electrostatic potential and $V_{\rm ext}({\bf r})$ is the
external electrostatic potential. Let $V_{\rm Coul}({\bf r}) =
V_{\rm ext}({\bf r}) + V_{\rm built}({\bf r})$ [total electrostatic
(Coulomb) potential]. According to the Ehrenfest theorem, one has
$\langle \gradient V_{\rm Coul}({\bf r})\rangle = - \langle \gradient
E_c({\bf r}) \rangle$, as the average over the (conduction)
  electron wavefunction (denoted as $\langle ...\rangle$) of the net
 force felt by electron is zero. If the ratio of the valence band
offset to the conduction band one is $r_{vc}$, then
\be
\langle \gradient V_{v}({\bf r}) \rangle = \langle \gradient (V_{\rm
Coul}({\bf r}) + r_{vc} E_c({\bf r})) \rangle =  (1-r_{vc}) \langle \gradient V_{\rm Coul}({\bf r}) \rangle.
\ee
For example, in GaAs/Al$_x$Ga$_{1-x}$As quantum wells $r_{vc}\simeq -0.5$,
hence $1-r_{vc}\simeq 1.5$. Therefore, the Rashba spin-orbit coupling is
proportional to the average of the sum of the external and built-in 
electric field. The above discussion is just a simple illustration on
the existence of the Rashba spin-orbit coupling, where a few
factors have been ignored. Nevertheless, the indication that the
Rashba spin-orbit coupling can be tuned electrically inspired the
community. Many proposals of spintronic devices based on
electrical manipulation of spin-orbit coupling, such as, the spin field-effect
transistors \cite{datta:665}, were proposed.\footnote{For recent
  advancement in experiment on the spin field-effect transistors, see,
  e.g.,
  Refs.~\cite{appelbaum:262501,Appelbaum:447.295,HyunCheolKoo09182009}.} In this background, the Rashba
spin-orbit couplings in various heterostructures have been studied
extensively. Theoretical investigations on the Rashba spin-orbit coupling
in quantum wells were performed in
Refs.~\cite{PhysRevLett.60.728,PhysRevB.55.16293,PhysRevB.59.15902,PhysRevB.67.165329,yang:113303,yang:193314,li:152107}.
Specifically, it was found that even in symmetric quantum wells, the
built-in electric field contributes an intersubband Rashba spin-orbit
coupling, though it does not contribute to the intrasubband spin-orbit
coupling \cite{bernardes:076603,calsaverini:155313}.

Experimental methods to determine the spin-orbit coupling coefficients consist of magnetotransport (including both the
Shubnikov-de Haas oscillation
\cite{PhysRevB.55.R1958,PhysRevB.57.11911,PhysRevB.60.7736,PhysRevB.61.15588,schapers:4324,sato:8017,PhysRevLett.78.1335,PhysRevLett.84.6074,guzenko:165301,kwon:112505,simmonds:124506}
and weak (anti-)localization
\cite{PhysRevLett.89.046801,pssb.233.436,PhysRevB.70.233311,PhysRevB.71.045328,yu:035304,pssc.5.322,0268-1242-13-4-005,PhysRevB.68.035317,PhysRevLett.90.076807}),
optically probed spin dynamics (spin relaxation
\cite{2007arXiv0707.4493E,2008arXiv0807.4845E,eldridge:125344} and spin precession \cite{Y.Kato:nat50}), electron spin resonance
\cite{PhysRevB.70.155322} and spin-flip Raman scattering
\cite{PhysRevLett.69.848,PhysRevB.51.4707}, etc. Recently, it was
proposed that the radiation-induced oscillatory magnetoresistance can
be used as a sensitive probe of the zero-field spin-splitting
\cite{PhysRevB.69.193304}. The experimental results indicated that the
largest Rashba parameter $\alpha_{R}$ can be 30$\times
10^{-12}$~eV$\cdot$m \cite{sato:8017,PhysRevB.61.15588} or even
40$\times 10^{-12}$~eV$\cdot$m \cite{PhysRevLett.84.6074} for InAs and
14$\times 10^{-12}$~eV$\cdot$m for GaSb \cite{PhysRevB.70.155322}
quantum wells, while the smallest Rashba parameter can be negligible
\cite{2008arXiv0807.4845E,bel'kov:176806}.\footnote{The channel width
  dependence \cite{kwon:112505}, density dependence \cite{schapers:4324,PhysRevB.61.15588,PhysRevB.57.11911,PhysRevB.60.7736,guzenko:165301,PhysRevLett.84.6074,PhysRevB.71.045328,PhysRevLett.89.046801,PhysRevB.70.233311,yu:035304,pssb.233.436},
gate-voltage dependence
\cite{sato:8017,PhysRevB.55.R1958,PhysRevB.57.11911,PhysRevLett.78.1335,yu:035304,2008arXiv0807.4845E},
temperature dependence \cite{2007arXiv0707.4493E} and interface effect
\cite{schapers:4324,PhysRevLett.84.6074,PhysRevB.71.045328,pssc.5.322,2009arXiv0912.2143N}
were investigated in various materials from InAs, InGaAs and GaAs to
GaSb. Rashba spin-orbit coupling in quantum wire was studied in Refs.~\cite{zhang:075304,PhysRevB.67.165318,0022-3727-40-2-030,zhang:205331,guzenko:032102,Guzenko,zhang:155316}.}

In asymmetric quantum wells, both the Rashba and Dresselhaus
spin-orbit couplings exist. After averaging over the lowest subband
wavefunction in (001) quantum well, the Dresselhaus spin-orbit
coupling becomes,
\be 
H_{D} = \beta_{D}(-k_x\sigma_x + k_y\sigma_y) + \gamma_{D} ( k_x k_y^2 \sigma_x - k_y k_x^2 \sigma_y), 
\ee
where $\beta_{D} = \gamma_{D} \langle \hat{k}_z^2 \rangle=\int
dz \phi_1(z)^{\ast}(-\partial_z^2)\phi_1(z)$ with $\phi_1(z)$ being the lowest
subband wavefunction. In narrow
quantum wells, the linear-${\bf k}$ term dominates. The ratio of the 
strength of the Dresselhaus spin-orbit coupling to the Rashba one
$\beta_{D}/\alpha_{R}$ can be tuned by the well width or the electric
field along the growth direction \cite{Koralek:nature458.610,winklerbook}. The
electric field dependence of the Dresselhaus and Rashba spin-orbit
couplings in GaAs/Al$_x$Ga$_{1-x}$As 
quantum wells for varies well widths was studied in the work by Lau
and Flatt\'e \cite{PhysRevB.72.161311}, indicating that under high bias the Rashba
spin-orbit coupling can exceed the Dresselhaus one.\footnote{The effects of structure inversion asymmetry,
heterointerface and gate-voltage on the Dresselhaus and Rashba spin-orbit
couplings in quantum wells were studied experimentally in
Refs.~\cite{wilde:125330,0268-1242-13-4-005,schapers:4324,PhysRevLett.90.076807,eldridge:125344,2008arXiv0807.4845E,bel'kov:176806,Koralek:nature458.610,PhysRevLett.91.246601,HyunCheolKoo09182009}.}
The interference of the two spin-orbit couplings leads to anisotropic spin splitting,
which hence results in anisotropic spin precession
\cite{lmeier:650,meier:prb035305,studer:045302,studer:027201},
spin relaxation
\cite{PhysRevB.60.15582,0957-4484-11-4-304,0953-8984-14-12-202,PhysRevB.68.075322,averkiev:033305,stich:073309},
spin diffusion
\cite{weber:076604,cheng:205328,stanescu:125307,weng:063714,Koralek:nature458.610}
and spin photocurrent
\cite{PhysRevLett.92.256601,giglberger:035327,0268-1242-23-11-114003}.
Reversely, the ratio of $\beta_{D}/\alpha_{\rm
  R}$ can be inferred from the anisotropy of spin precession
\cite{lmeier:650,PhysRevB.69.045317}, spin relaxation
\cite{averkiev:033305,stich:073309}, spin diffusion
\cite{weber:076604,cheng:205328} and spin photocurrent
\cite{PhysRevLett.92.256601,giglberger:035327}.

\subsubsection{Hole Hamiltonian and hole spin-orbit coupling in bulk system}

The hole Hamiltonian can be written as \cite{winklerbook}
\be
H_{h} = H_{\rm L} + H^h_{\rm SO} + H_{\rm hsp},
\label{hso}
\ee
where $H_{\rm L}$ is the Luttinger Hamiltonian, $H^h_{\rm SO}$ is the
spin-orbit coupling due to the space inversion asymmetry and $H_{\rm hsp}$
describes the hole-strain and hole-phonon interactions which can be
found in Ref.~\cite{Titkovbook}. The Luttinger Hamiltonian $H_{\rm L}$ is
given by \cite{PhysRev.102.1030} (in the order of hole spin state $J_z
= -\frac{3}{2}, - \frac{1}{2}, \frac{1}{2}, \frac{3}{2}$ with $z$ axis
along [001] direction)
\be
  H_{\rm L} = \left[ \begin{array}{cccc} 
     P+Q & L & M & 0 \\
     L^{\ast} & P-Q & 0 & M \\
     M^{\ast} & 0 & P-Q & -L \\
     0 & M^{\ast} & -L^{\ast} & P+Q 
   \end{array}\right], 
\ee
where 
\be
P=\frac{\gamma_1}{2m_0}k^2, \quad Q=\frac{\gamma_2}{2m_0}(k_x^2+k_y^2-2k_z^2),\quad
L=-\frac{2\sqrt{3}\gamma_3}{2m_0} (k_x - \ii k_y) k_z, \quad
M = - \frac{\sqrt{3}}{2m_0} [\gamma_2(k_x^2-k_y^2)-2\ii \gamma_3 k_x k_y],
\ee
with $\gamma_1$, $\gamma_2$ and $\gamma_3$ being the Luttinger parameters.
\be
H^h_{\rm SO} = \frac{1}{\eta}({\bf \Omega}_{D} + {\bf \Omega}_S) \cdot {\bf J}
\ee
which is similar to the spin-orbit coupling in the conduction band
except differed by a factor of $1/\eta$. Here ${\bf J} =
(J_x, J_y, J_z)$ is the hole spin operator with $J=3/2$.

Under spherical approximation, the Luttinger Hamiltonian can be
written in a compact form,
\be
H_L^{\rm sp} = \frac{\gamma_1}{2m_0} k^2 + \frac{\bar{\gamma}_2}{2m_0}
[\frac{5}{2}k^2 - 2 ({\bf k}\cdot{\bf J})^2],
\label{sphole}
\ee
where $\bar{\gamma}_2 = (2\gamma_2+3\gamma_3)/5$. Under this
approximation, the spectrum of hole becomes parabolic,
\be
E_{\lambda_1/\lambda_2}({\bf k}) = \frac{k^2}{2m_0}(\gamma_1\pm
2\bar{\gamma}_2) .
\ee
The indices $\lambda_1=\pm \frac{1}{2},\lambda_2=\pm \frac{3}{2}$
denote the hole spin. The four branches with different effective mass
$m_0/(\gamma_1\pm 2\bar{\gamma}_2)$ are the light hole and
heavy hole branches respectively. The two fold degeneracy of the 
spectrum is a consequence of the coexistence of the time-reversal
symmetry and the space inversion symmetry of the Luttinger Hamiltonian.

In zinc-blende III-V or II-VI semiconductors, $\gamma_1$ and
  $\bar{\gamma}_2$ are usually of the same order of magnitude 
  (e.g., in GaAs, $\gamma_1=6.85$ and $\bar{\gamma}_2=2.5$ \cite{madelung}).
Therefore the spin-dependent term in Eq.~(\ref{sphole}) (the
spin-orbit coupling term) is comparable with the spin-independent ones. This indicates that the spin-orbit coupling is
very strong in bulk hole system, which is one of the main differences
between bulk electron and hole systems. The main consequences are
that: change in orbit motion (states) can influence
strongly on spin
dynamics \cite{culcer:106601,0268-1242-23-11-114017,winklerbook,dyakonovbook_08};
the spin relaxation/dephasing is usually much faster in hole
system ($\sim 100$~fs) than in electron system
\cite{PhysRevLett.89.146601,culcer:195204,PhysRevB.71.245312,krauss:256601}.

\subsubsection{Hole spin-orbit coupling in nanostructures}

As mentioned above the hole spin-orbit coupling is important in
determining hole spectrum and its spin state. In nanostructures, it is
always necessary to diagonalize the hole Hamiltonian Eq.~(\ref{hso})
including the confining potential \cite{winklerbook,PhysRevLett.104.066405}. 
The dispersion and spin states can vary markedly with the confinement
geometry \cite{winklerbook}. It was demonstrated that in $p$-GaAs
quantum wells the in-plane and out-of-plane effective masses and
$g$-factors of the lowest heavy-hole subband can be quite different for
different growth directions \cite{winklerbook}.

Similar to the electron case, the electric field can also induce the
Rashba spin-orbit coupling in hole system, which is written as \cite{winklerbook}
\be 
 H^h_R = \alpha_0^h {\bf J}\cdot ({\bf k}\times {\bf \cal{E}}_c) .
\ee
In [001] grown quantum wells, it is reduced to
\be
H^h_R = \alpha_0^h \langle {\cal{E}}_c\rangle ( k_y J_x - k_x J_y ),
\label{falsehrso}
\ee
where $\alpha_0^h$ is a material dependent parameter
\cite{winklerbook} and $\langle {\cal{E}}_c\rangle$ is the average of
the electric field (along the growth direction) acting on conduction
band over the hole subband envelope function. In practice, to
obtain the Rashba spin-orbit coupling in, e.g., the lowest heavy-hole
subband, one must further block-diagonalize the subband Hamiltonian
via the L\"owdin partition method \cite{winklerbook}. Quite different from
the electron case, the subband block-diagonalization gives another
Rashba spin-orbit coupling which originates from the electric field on
other hole subband rather than the conduction band. Practical
calculation indicated that such kind of Rashba spin-orbit coupling is 
more important than that due to the valence-conduction band coupling
[Eq.~(\ref{falsehrso})] \cite{winklerbook}. The expression of such
kind of Rashba spin-orbit coupling in the lowest heavy-hole subband is
given by \cite{winklerbook},
\be
H^{\rm HH}_R = \alpha^{hh} \langle {\cal{E}}_z \rangle \ii
(k_{+}^3\sigma_{-} - k_{-}^3\sigma_{+}),
\ee
where $k_{\pm}=k_x\pm \ii k_y$, $\sigma_{\pm}=\sigma_x\pm \ii
\sigma_y$ with $\spin$ being the Pauli spin matrices in the manifold of
hole spin $J_z=-3/2,3/2$. $\langle {\cal{E}}_z \rangle$ is the
electric field acting on the valence band averaged over the lowest
hole subband envelope function. For an infinite depth rectangular
quantum well \cite{winklerbook},
\be
\alpha^{hh} = \frac{e}{m_0^2} \frac{64}{9\pi^2}
\gamma_3(\gamma_2+\gamma_3)\left[
  \frac{1}{\Delta_{11}^{hl}}(\frac{1}{\Delta_{12}^{hl}}-\frac{1}{\Delta_{12}^{hh}})+\frac{1}{\Delta_{12}^{hl}\Delta_{12}^{hh}} \right],
\label{alpha_hole}
\ee
where $\Delta_{ij}^{hl}$ ($i,j=1,2...$) is the splitting between the $i$-th
heavy-hole subband and the $j$-th light-hole subband. Similar to the
case in electron system, the hole Rashba spin-orbit coupling can also
be tuned by gate-voltage and structure asymmetry but now $\alpha^{hh}$
is also electric field dependent
\cite{PhysRevB.62.4245,winklerbook,habib:3151,PhysRevLett.81.1282,S.J.Papadakis03261999}.

\subsection{Spin-orbit coupling in nanostructures due to interface
inversion asymmetry}

The interface inversion asymmetry is a kind of inversion asymmetry
caused by the inversion asymmetric bonding of atoms at the interfaces
of nanostructures.\footnote{The interface
  inversion asymmetry induced spin-orbit coupling is also reviewed in
  Ref.~\cite{winklerbook}.} It was first pointed out by Ale\u{i}ner and
Ivchenko \cite{Aleiner:55.692} that the symmetry at the interface may
be different from the symmetry away from the interface, and hence may
introduce a new spin-orbit coupling
\cite{PhysRevB.54.5852,PhysRevB.56.R12744,0268-1242-14-3-004}.
The interface-induced spin-orbit coupling was usually studied in
quantum wells. The underlying physics is similar for other
nanostructures.

Let us start by examining the case of GaAs/AlAs quantum wells. Without
the interface effect, the quantum wells possess a $D_{2d}$
symmetry. It is easy to find that due to the difference of Al and Ga
atoms, the symmetry at the interface is reduced to $C_{2v}$
\cite{winklerbook}. For quantum wells which possess a mirror symmetry,
the asymmetry induced by the two interfaces cancels out
\cite{winklerbook}. However, in the situation that the materials in
the well and barrier do not share a common atom, the interface effect
at the two interfaces can not be canceled out as they
are different interfaces \cite{PhysRevLett.77.1829}. For example, in
InAs/GaSb quantum wells, the two interfaces are Sb-In-As and In-As-Ga.
The symmetry of the quantum well is then reduced to $C_{2v}$ and
results in new spin-orbit coupling terms for holes and electrons. It
induces a term $\sim (J_xJ_y+J_yJ_x)$ and independent of ${\bf k}$ in
the hole Hamiltonian \cite{winklerbook}, which mixes the heavy-hole
and light-hole spins effectively in narrow quantum wells
\cite{PhysRevB.56.R12744,0268-1242-14-3-004} and largely suppresses
the hole spin lifetime in these structures
\cite{PhysRevB.56.R12744}. The induced electron spin-orbit
coupling is of the linear-Dresselhaus-type $\sim (k_x\sigma_x -
k_y\sigma_y$) \cite{Rossler2002313}.

The experimental test of the importance of the interface-induced
electron spin-orbit coupling was performed by comparing the spin dynamics
in InGaAs-AlInAs quantum wells (which share one common atom) with that
in InGaAs-InP quantum wells (which have no common atom) with similar
parameters and overall quality \cite{PhysRevB.58.R10179}. The observed
huge difference in spin relaxation times between these two
samples indicates that the interface inversion asymmetry induced
spin-orbit coupling is important in narrow quantum wells without
common atom in well and barrier.

It should be pointed out that in [110] grown quantum wells, even in
the case without common atom in well and barrier, the two interfaces
can be symmetric.\footnote{The study on the quantum well growth 
direction dependence of the interface induced spin-orbit 
coupling is presented in Ref.~\cite{Chen:pssb231.423}.} Hence the
interface inversion asymmetry does not contribute. By comparing the
spin relaxation in InAs/GaSb quantum wells with different growth
directions, with the help of theoretical calculation, it was found
that the interface inversion asymmetry plays an important role in
narrow InAs/GaSb quantum wells
\cite{PhysRevB.64.201301,PhysRevB.68.115311}.

In genuine nanostructures, as interfaces can not be perfect, the
interface inversion asymmetry  exists in most samples. A 
quantitative analysis of such effect  requires the knowledge of
specific interfaces, which is usually very difficult to acquire. 
In general, since it
is an interface effect, the interface inversion asymmetry should
be important to spin-orbit coupling in narrow quantum wells 
with no common atom in the
barrier and well, whereas unimportant in other cases.

\subsection{Zeeman interaction with external magnetic field}

The Zeeman interaction for electron is written
  as 
\be
H_{\rm Z} = \mu_{\rm B}{\bf S}\cdot\hat{g}\cdot{\bf B},
\ee
where $\hat{g}$ is the $g$-tensor and $\mu_{\rm B}$ is the Bohr
magneton. The $g$-tensor determines the spin precession (both
the frequency and the direction) in the presence of an external
magnetic field. It is hence an important quantity for spin
dynamics. For typical III-V semiconductors, such as GaAs, the
conduction band $g$-tensor is isotropic $\hat{g}_e=g_e\hat{1}$. In
GaAs, at conduction band edge, $g_e=-0.44$, whereas for higher energy
$g$ factor increases
\cite{0370-1328-89-2-326,Golubev:61.1214,PhysRevB.47.6807}. An
empirical relation for the energy $\varepsilon$ dependence of the $g$ factor in
GaAs was found from experiments as
$g_e(\varepsilon)=-0.44+6.3\varepsilon$ ($\varepsilon$ in unit of eV)
\cite{PhysRevB.47.6807,PhysRevB.44.9048,0268-1242-2-9-002}. Experiments
indicated that the $g$ factor varies from $-0.45$ to $-0.3$ in the
temperature range of $4\sim 300$~K 
\cite{PhysRevLett.74.2315,lai:192106,2006cond.mat..8534H,litvinenko:033204,zawadzki:245203}.
Recent calculations \cite{zawadzki:245203} indicated that the
temperature dependence of $g$-factor is mainly due to the combined
effects of (i) the energy-dependent $g$ factor and (ii) the thermal
dilatation of lattice.

In quantum wells the $g$ factors of the barrier material and well
material are usually different or even of opposite signs. The $g$
factor can then be tuned by well width. For example, in GaAs/AlAs
quantum wells: at large well width the electronic property is
GaAs-like and $g_e<0$; at small well width the subband wavefunction is
largely penetrated into the AlAs barrier and the electronic property
is AlAs-like, hence $g_e>0$. This phenomenon has been observed in
experiment \cite{PhysRevB.72.235313}. Moreover, due to the confinement
in quantum well, which reduces the symmetry of the system, the
electron $g$-factor usually becomes anisotropic: the in-plane
$g$-factor $g_{\|}$ is different from the out-of-plane one $g_{\perp}$
\cite{pfeffer:233303,PhysRevB.62.2051,ivchenko:8.827,winklerbook}.

In parabolic GaAs/AlGaAs quantum wells, as the Al concentration is
tuned almost continuously with position in the growth direction, the
$g$-tensor becomes position dependent $\hat{g}_e(z)$. The effective
$g$-tensor is an average over position $\hat{g}^{\ast} = \int dz
\hat{g}_e(z)|\psi_e(z)|^2$ where $\psi_e(z)$ is the electron subband
wavefunction. This fact enables efficient tuning of electron
$g$-tensor by modulating the subband wavefunction via, e.g.,
gate-voltage \cite{gsalis:619,PhysRevB.64.041307,studer:027201}. A
time-dependent gate-voltage can induce a modification to the
Zeeman interaction $\delta H_Z = \mu_{\rm B} {\bf
  S}\cdot\delta\hat{g}\cdot{\bf B}$ ($\delta \hat{g}$ is the change of
$g$-tensor due to gate-voltage). In proper geometry, the
time-dependent gate-voltage induces a spin precession perpendicular to
the unperturbed spin precession. In such a way, electron spin
resonance is achieved with time-dependent gate voltage. Such a spin
resonance, which is called $g$-tensor modulation resonance, was first
proposed and realized by Kato et al. \cite{Y.Kato:sci1201}.

The hole Zeeman interaction is complicated. In bulk system
there are two terms which contribute to the hole Zeeman interaction,
\be
 H_{\rm Z}^h = 2\kappa \mu_{\rm B} {\bf B}\cdot {\bf J} + 2 q \mu_{\rm
   B} {\bf B}\cdot {\bf {\cal{T}}}.
\label{Hzv}
\ee
Here $\kappa$ is the isotropic valence band $g$ factor, whereas $q$ is
the anisotropic one. ${\bf {\cal{T}}}=(J_x^3,J_y^3,J_z^3)$. In GaAs
$\kappa=1.2$, whereas $q=0.01$ is much smaller and can always be
neglected. It is noted that there is no direct coupling between
spin-up and -down heavy holes ($J_z = \frac{3}{2}, -\frac{3}{2}$) in the
Zeeman interaction unless in the much smaller anisotropic term [the
second term in Eq.~(\ref{Hzv})].

In (001) quantum wells, the lowest subband is the heavy-hole subband
if no strain is applied. According to the above analysis, for an
in-plane magnetic field, the coupling of spin-up and -down states is
rather small. Within the subband quantization, L\"owdin partition
method gives the dominant terms of the Zeeman interaction in the
lowest heavy-hole subband as
\be
H_{\rm Z}^{\rm HH} = \frac{3\kappa\mu_{\rm B}}{m_0\Delta^{hl}_{11}}
\left\{\sigma_x\left[ B_x\gamma_2\left(k_x^2-k_y^2\right) -
2B_y\gamma_3k_xk_y\right] + \sigma_y\left[ 2B_x\gamma_3k_xk_y +
B_y\gamma_2\left(k_x^2-k_y^2\right) \right]\right\},
\label{hzv001qw}
\ee
where $\Delta_{11}^{hl}$ denotes the splitting between the lowest
heavy-hole subband and the lowest light-hole subband. Note that the
in-plane $g$ factor of the lowest heavy-hole subband is
${\bf k}$-dependent and the average of the $g$ factor is zero.
The in-plane $g$-factor of the lowest heavy-hole
subband varies significantly with the growth direction of the quantum
well as pointed out by Winkler et al. \cite{PhysRevLett.85.4574,winklerbook}.

Zeeman interaction of holes in quantum wires has been investigated
only recently. Experimentally, Danneau et al. found that the $g$-factor of
holes in quantum wires along [$\bar{2}33$] direction is largely
anisotropic \cite{danneau:026403}. Theoretically, Csontos and
Z\"ulicke reported that the hole $g$-factor in quantum wire can vary
largely with subband \cite{csontos:073313} and can be tuned
effectively by confinement \cite{csontos:023108}. These predictions
were partly confirmed recently in the experiment by Chen et
al. \cite{2009arXiv0909.5295C}. Similar results were found in hole
quantum dots \cite{haendel:086403,roddaro:186802}.
The large anisotropy and variation of the hole $g$-factor in these quantum
confined systems are not surprising, as (i) the symmetry of these
structures is reduced and (ii) the hole spin-orbit coupling is very strong.

\subsection{Spin-orbit coupling in Wurtzite semiconductors and other
  materials}

In previous discussions, the spin-orbit coupling and $g$-factor in
bulk and nanostructures of the zinc-blende III-V and II-VI compounds are
reviewed. Besides these two important kinds of materials, there are
several other kinds of materials with different structures which have also
attracted much attention of the semiconductor spintronic community.

One of these materials is the wurtzite structure semiconductors, such
as GaN, AlN and ZnO.\footnote{GaN also has zinc-blende structure phase. The
  spin-orbit coupling in zinc-blende GaN was studied in
  Ref.~\cite{fu:093712}} In contrast to the zinc-blende
semiconductors such as GaAs, the existence of hexagonal $c$-axis in
wurtzite semiconductors leads to an intrinsic wurtzite structure
inversion asymmetry in addition to the bulk inversion asymmetry
\cite{strainbook,weber:262106}. Therefore, the electron spin
splittings include both the Dresselhaus effect
\cite{wang:082110,fu:093712} (cubic in $k$) and Rashba effect (linear
in $k$)
\cite{0022-3719-17-33-015,rashba:jetplett.39.78,voon:10703,Majewski:icps27,kurdak:113308,thillosen:022111,schmult:033302,belyaev:035311}.
In a recent work by Fu and Wu \cite{fu:093712}, a Kane-type
Hamiltonian was constructed and the spin-orbit coupling for electron
and hole bands were investigated in the full Brillion zone in bulk ZnO
and GaN.

For the states near the $\Gamma$ point (${\bf k}=0$), by choosing the $z$
axis along the $c$ axis of the crystal, the electron spin-orbit coupling
in the conduction band  reads \cite{fu:093712}
\be
H^e_{\rm SO} = [\mathbf{\Omega}_e^R(\mathbf{k})+
\mathbf{\Omega}_e^D(\mathbf{k})] \cdot \spin \big/ 2
\ee
with
\be
\mathbf{\Omega}_e^R(\mathbf{k})=2\alpha_e(k_y,-k_x,0), \quad\quad  
\mathbf{\Omega}_e^D(\mathbf{k})=2\gamma_e(bk_z^2-k_{\|}^2)(k_y,-k_x,0).
\label{socwur_e}
\ee
Here $\mathbf{\Omega}_e^R(\mathbf{k})$ and
$\mathbf{\Omega}_e^D(\mathbf{k})$ are the Rashba-type and Dresselhaus
terms, respectively. $\alpha_e$ and $\gamma_e$ are the corresponding
spin-orbit coupling coefficients. $b$ is a coefficient originating from
the anisotropy induced by the lattice structure. It was found that in
ZnO and GaN, $b=3.855$ and $3.959$ respectively \cite{fu:093712}. The
value of $\alpha_e$ and $\gamma_e$ are listed in
Table~\ref{wurtzite_ga} \cite{fu:093712}.

In wurtzite semiconductors, the valence bands consist of three
separated bands: $\Gamma_{9v}$ (heavy-hole, spin $\pm3/2$), $\Gamma_{7v}$
(light hole, spin $\pm 1/2$) and $\Gamma_{7^{\prime}v}$ (split-off
hole, spin $\pm 1/2$). The spin-orbit couplings in these bands at
small $k$ can also be written in the general form of
$\frac{1}{2}[\mathbf{\Omega}_i^R(\mathbf{k})+
\mathbf{\Omega}_i^D(\mathbf{k})]\cdot\spin$, which includes
both the linear Rashba-type term $\frac{1}{2}\mathbf{\Omega}_i^R(\mathbf{k})$,
and the cubic Dresselhaus term $\frac{1}{2}\mathbf{\Omega}_i^D(\mathbf{k})$,
with $i=e,9,7,7^\prime$ being the band index. For heavy-hole band
$\Gamma_{9v}$, the spin-orbit coupling fields are written as \cite{fu:093712}
\be
\mathbf{\Omega}_9^R(\mathbf{k}) = 0, \quad\quad
\mathbf{\Omega}_9^D(\mathbf{k}) = 2\gamma_9 (k_y(k_y^2-3k_x^2),k_x(k_x^2-3k_y^2),0).
\ee
For valence bands $\Gamma_{7v}$ (light-hole) and $\Gamma_{7^\prime v}$
(split-off hole), the spin-orbit coupling fields are the same as that
given in Eq.~(\ref{socwur_e}) with only the coefficients $\alpha_e$
and $\gamma_e$ being replaced by $\alpha_i$ and $\gamma_i$
($i=7,7^\prime$) (the coefficients for $\Gamma_{7v}$ and
$\Gamma_{7^\prime v}$ bands) respectively \cite{fu:093712}. The coefficients $\alpha_i$
and $\gamma_i$ ($i=e,9,7,7^{\prime}$) are listed in
Table~\ref{wurtzite_ga} \cite{fu:093712}.\footnote{Using these
  coefficients, Bu\ss~et al. found reasonable agreement between
  calculated spin relaxation times and the experimentally measured
  ones recently \cite{buss2009}.}

\begin{center}
\begin{table}[htbp]
 \caption{Rashba-type ($\alpha_i$) (in meV$\cdot$\AA) and Dresselhaus ($\gamma_i$) (in
  eV$\cdot$\AA$^3$) spin-orbit coupling coefficients in wurtzite ZnO and GaN at small momentum
with $i=e,9,7,7^\prime$. From Fu and Wu \cite{fu:093712}.\
($^{\mbox{a}}$ from Ref.~\cite{voon:10703}; $^{\mbox{b}}$ from
Ref.~\cite{Majewski:icps27})}
\vskip 0.2cm
\begin{tabular}{lllllllllllllllllll}\hline\hline
&\mbox{} &$\alpha_e$ & \mbox{}& $\alpha_9$ & \mbox{}&$\alpha_7$ & \mbox{}
& $\alpha_{7^\prime}$ & $\gamma_e$ &\mbox{}& $\gamma_9$\mbox{}& &$\gamma_7$\mbox{} &$\gamma_{7^\prime}$ \\ \hline
ZnO: &\mbox{} &1.1$^{\mbox{a}}$&\mbox{}& $0$ &\mbox{}& 35$^{\mbox{a}}$ (21$^{\mbox{b}}$) &\mbox{} &51$^{\mbox{a}}$ &0.33
 &\mbox{} &0.09&\mbox{}& 6.3 & 6.1 & \\
GaN: &\mbox{} &9.0$^{\mbox{b}}$&\mbox{}&$0$ &\mbox{}& 45$^{\mbox{b}}$ &\mbox{} & 32$^{\mbox{b}}$ &0.32
 &\mbox{} &0.07&\mbox{}& 15.3 & 15.0 & \\
\hline\hline\\
\end{tabular}
\label{wurtzite_ga}
\end{table}
\end{center}

Interestingly, in two-dimensional electron system (quantum wells or
heterojunctions) where the growth direction is along the $c$-axis, the
linear-${\bf k}$ spin-orbit coupling is of the same form for both
the Dresselhaus and Rashba-type terms due to the crystal field.
Additional contribution to the spin-orbit coupling comes from the electric
field across the two-dimensional electron system, which is just the
Rashba spin-orbit coupling
\cite{0022-3719-17-33-015,rashba:jetplett.39.78}.
Therefore, all the linear in ${\bf k}$ terms are of the Rashba-type,\footnote{The symmetry of the cubic term is also the same as the
  linear one.}
which makes them indistinguishable in experiments. Indeed, experiments
revealed that the spin-orbit coupling is dominated by the Rashba-type
spin-orbit coupling, where no effect that signals spin-orbit coupling
of the form of $(k_x\sigma_x-k_y\sigma_y)$ was observed
\cite{tang:071920,cho:085327,he:071912,cho:041909,he:147402,weber:262106}.
The measured Rashba parameter lies in the range of 0.6 to 8$\times
10^{-12}$~eV$\cdot$m for various conditions
\cite{tsubaki:3126,tang:073304,tang:172112,cho:222102,zhou:053703,belyaev:035311,0268-1242-22-8-007,zhou:262104,pssc.3.4247,schmult:033302,thillosen:022111,kurdak:113308}.

Another kind of material of special interest is the so-called gapless
semiconductor, such as HgTe, partly because that in HgTe/CdTe
quantum wells the quantum spin Hall effect was observed
\cite{MarkusKonig11022007}. HgTe, which is also a zinc-blende
II-VI semiconductor, has the same symmetry as CdTe and GaAs. However,
the $\Gamma_8$ band, which consists of heavy and light holes in GaAs,
is higher than the conduction band $\Gamma_6$. Moreover, the
heavy-hole band is inverted, i.e., the effective mass of electron in
the heavy-hole band is positive. On the other hand, the conduction band
$\Gamma_6$, becomes hole-like. In HgTe/CdTe quantum wells, the
relative position of the $\Gamma_6$ and $\Gamma_8$ bands can be tuned
by well width. At the crossing band point a massless Dirac
spectrum is realized \cite{JPSJ.77.031007}. The spin-orbit coupling in
these bands was investigated both
theoretically \cite{0295-5075-81-3-38003} and experimentally
\cite{Gui2002416,0268-1242-11-8-009,Radantsev95.491,Becker:pssb.229.775,0268-1242-21-4-015,PhysRevB.63.245305,Radantsev2001989,PhysRevB.70.115328},
where the Rashba spin-orbit coupling was found to be very large ($\simeq
4\times 10^{-11}$~eV$\cdot$m) in two-dimensional electron system.

Spin dynamics in silicon has attracted renewed interest recently due
to the experimental advancement in silicon spintronic devices \cite{Appelbaum:447.295,appelbaum:262501,grenet:032502,huang:072501,erve:212109,Jonker:nphys673}
and also due to recent development in spin qubits based on silicon quantum
dots \cite{liu:prb.77.073310,Shajinphs988,hu:nnano.2.622} or donor bound electrons
\cite{xiao:nature.430.435,PhysRevLett.91.246603,0953-8984-18-21-S06,PhysRevB.68.115322}.
Silicon has a diamond  structure with space inversion symmetry,
hence bulk silicon does not have the Dresselhaus spin-orbit
coupling. In two-dimensional or other nanostructures, the Rashba
spin-orbit coupling due to structure inversion asymmetry emerges
\cite{0022-3719-17-33-015,rashba:jetplett.39.78,PhysRevB.71.075315}.
Moreover, the interface inversion asymmetry also contributes to new
spin-orbit coupling, which can be either the Rashba-type or the
linear-Dresselhaus-type or their combinations
depending on the symmetry of the interface
\cite{PhysRevB.69.115333,nestoklon:235334}.
The effect of the electric field on electron spin-orbit coupling in
Si/SiGe quantum well was discussed in Ref.~\cite{nestoklon:155328}.
In practice, the doping inhomogeneity can induce a random
(in-plane position dependent) electric field along the growth direction even in
nominally symmetrically doped quantum wells. This random electric
field can induce a random Rashba spin-orbit coupling \cite{sherman:209}.
Similarly, the surface roughness can also induces a random spin-orbit
coupling \cite{PhysRevB.69.115333}. Experimentally,
the electron spin-orbit coupling in silicon quantum wells was
determined via the anisotropy of the electron spin resonance
frequency and line-width 
\cite{PhysRevB.66.195315,Wilamowski2002439,Wilamowski2003111}.
Hole spin-orbit coupling in
silicon quantum wells was studied in
Refs.~\cite{PhysRevB.71.035321,zhang:155311}.

\subsection{Hyperfine interaction}

The origin of the hyperfine interaction is that the nuclear spins
feel the magnetic field generated by carriers due to their spins and
orbital angular momentums.  For conduction band electron, which has $s$-wave
symmetry, the magnetic field generated by electron orbital motion is
negligible. The electron hyperfine interaction is of the 
form \cite{opt-or,0953-8984-15-50-R01}
\be 
H_{\rm hf} = \sum_{i,\nu}
\frac{2\mu_0}{3}\beta_{\nu}\gamma_e\gamma_{\nu}
\delta({\bf r} - {\bf R}_{i,\nu}) {\bf S}\cdot {\bf I}_{i,\nu}.
\label{isohf}
\ee
Here $\mu_0$ is vacuum permeability, $\gamma_e=2\mu_{\rm B}$ and $i$ is
the index of the Wigner-Seitz cell. For III-V and II-VI compounds
(such as GaAs), there are two atoms (such as Ga and As) in each
Wigner-Seitz cell. And the nuclei of certain atoms have their
isotopes (such as $^{69}$Ga and $^{71}$Ga). The index $\nu$ labels
both the atomic site in the Wigner-Seitz cell and the
isotopes. $\beta_{\nu}$ is the abundance and $\sum_{\nu}
\beta_{\nu}=N_{\rm atm}$ with $N_{\rm atm}$ being
the number of atoms within a Wigner-Seitz cell. $N_{\rm atm}=2$ for
zinc-blende structures. $\gamma_{\nu} = g_{\nu}
\mu_N$ with $g_{\nu}$ and $\mu_N$ representing the $g$ factor of the
$\nu$ nucleus and the nuclear magneton respectively. ${\bf R}_{i,\nu}$
and ${\bf I}_{i,\nu}$ represent the position and spin of the $\nu$
nucleus in $i$-th Wigner-Seitz cell. In practical calculation it is
assumed that the isotopes are distributed uniformly. Within the
envelope function approximation, the electron hyperfine interaction is
written as
\be
H_{\rm hf} 
= \sum_{i,\nu} A_{\nu} v_0 |\psi({\bf R}_i)|^2 {\bf S}\cdot{\bf I}_{i,\nu}
= {\bf h} \cdot {\bf S},
\ee
where $A_{\nu} = \frac{2\mu_0}{3} \gamma_e\beta_{\nu}\gamma_{\nu}
|u_c({\bf R}_{\nu})|^2 N_0$ and ${\bf h} = \sum_{i,\nu} A_{\nu} v_0
|\psi({\bf R}_i)|^2 {\bf I}_{i,\nu}$. Here $v_0$ is the volume of the
Wigner-Seitz cell and $N_0=1/v_0$, $u_c$ is Bloch wave
amplitude. $A=\sum_{\nu} A_{\nu}$ measures the strength 
of the hyperfine interaction. In GaAs, possible isotopes are
$^{69}$Ga, $^{71}$Ga and $^{75}$As, which all have spin
$I=3/2$. Their natural abundances are $\beta_{^{69}{\rm Ga}}=0.6$,
$\beta_{^{71}{\rm Ga}}=0.4$ and $\beta_{^{71}{\rm As}}=1$. The average strength of the 
hyperfine interaction is $A\simeq 90$~$\mu$eV. In silicon, the
possible isotopes are $^{28}$Si and $^{29}$Si with natural abundances
$\beta_{^{28}{\rm Si}}=0.9533$ and $\beta_{^{29}{\rm
    Si}}=0.0467$. In silicon, the main isotope $^{28}$Si has no
nuclear spin and only the minor isotope $^{29}$Si has spin
$I=1/2$. Hence the average hyperfine interaction in silicon $A\sim
0.2$~$\mu$eV is much smaller than that in GaAs.

For holes, as the valence band Bloch wavefunction is very small at the
nuclei position (which is a property of $p$-wave symmetry), the
contact hyperfine interaction is negligible. However, the long range
dipole-dipole interaction and the interaction between electron orbital
angular momentum and nuclear spin do not vanish for
holes.\footnote{In contrast, these interactions are zero for
  (conduction band) electrons
  due to $s$-wave symmetry.} Within the envelope function
approximation, Fischer et al. showed that the hyperfine interaction
for heavy-holes in quantum well is of the Ising form \cite{fischer:155329}
\be
H^h_{\rm hf} = \sum_{i,\nu} A^h_{\nu} v_0 |\psi({\bf R}_i)|^2 S^z
I^z_{i,\nu},
\ee
where $S^z$ is the $z$-component of the heavy-hole spin and $A^h_{\nu}$ is
the hole hyperfine coupling constant. 
Fischer et al. estimated that the hole hyperfine coupling constant $A^h_{\nu}$ is about
$-10$~$\mu$eV in GaAs \cite{fischer:155329},
about one order of magnitude smaller than the electron one.

Besides inducing spin relaxation, the hyperfine interaction can also
transfer  the nonequilibrium carrier  spin polarization to 
nuclear spin system, which is called the dynamic nuclear polarization
\cite{jetp.15.179,PhysRevB.15.5780,opt-or,JPSJ.77.031011,baugh:096804,petta:067601,2008CRPhy...9..874K,PhysRevB.72.233204,PhysRevLett.91.036602,eble:081306}.
Via the hyperfine interaction, periodic optical pumping of electron
spin polarization acts as a periodic magnetic field on nuclear spin
and induces a nuclear spin resonance \cite{Kikkawa01212000}.
As nuclear spin relaxes much slower than electron spin, nuclear
spin was proposed as long-lived ``quantum memory'' for spin qubits
\cite{PhysRevLett.90.206803,witzel:045218}. Such proposal was realized in
experiment in $^{31}$P donors in isotopically pure $^{28}$Si crystal \cite{nature:455.1085}.

\subsection{Exchange interaction with magnetic impurities} 

The exchange interaction between carrier and magnetic impurities, such
as Mn, is very important in diluted magnetic semiconductors
\cite{0268-1242-16-4-201}, as it is the physical origin of various
magnetic orders in these materials. For example, the $d$ orbital
electrons of Mn impurities mix with $s$ (conduction band) or $p$
(valence band) electrons in the semiconductor matrix, which 
leads to the exchange interactions between carriers and Mn
impurities \cite{PhysRev.124.41,0268-1242-16-4-201}. These
interactions, known as the $s$-$d$ and $p$-$d$ exchange interactions,
can be written as
\be
H_{\rm sd} = - \sum_{i} J_{\rm sd} {\bf S} \cdot {\bf
S}_d^{(i)}\delta({\bf r}-{\bf R}_i),\quad\quad
H_{\rm pd} = - \sum_{i} J_{\rm pd} {\bf J} \cdot {\bf
S}_d^{(i)}\delta({\bf r}-{\bf R}_i).
\ee
Here ${\bf S}_d^{(i)}$ is the $i$th Mn spin at ${\bf R}_i$.
${\bf S}$ and ${\bf J}$ represent the electron and hole spins
respectively. $J_{\rm sd}$ and $J_{\rm pd}$ stand for the $s$-$d$ and
$p$-$d$ exchange constants respectively. 
The exchange constants,
which are basic parameters of magnetic semiconductors, have been
extensively studied both theoretically and experimentally [for review
in (III,Mn)V, see Ref.~\cite{jungwirth:809} and references therein; for
(II,Mn)VI, see Ref.~\cite{furdyna:R29}]. To illustrate the strength of
the $s(p)$-$d$ exchange interaction, we give two examples: in GaMnAs,
$N_0J_{\rm sd}\approx -0.1$~eV \cite{PhysRevLett.95.017204,PhysRevB.72.235313}, $N_0J_{\rm pd}\approx -1$~eV \cite{jungwirth:809}; in
CdMnTe, $N_0J_{\rm sd}\approx 0.22$~eV, $N_0J_{\rm pd}\approx
-0.88$~eV \cite{furdyna:R29} where $N_0=1/v_0$ with $v_0$ being the volume of the unit cell.

\subsection{Exchange interaction between carriers}

The exchange interaction between carriers originates from the
combination of the carrier-carrier Coulomb interaction and the
permutative antisymmetry of the wavefunction of the fermionic carrier
system. The exchange interaction usually increases with the overlap
between the wavefunction of the carriers. Therefore, it is usually strong
in localized carrier system, such as carriers in quantum dots or those
bound to ionized impurities. Experiments have shown that the exchange
interaction between electrons in quantum dots can be tuned via the
magnetic field\footnote{This was
  first predicted by Burkard et al. \cite{PhysRevB.59.2070}.}
\cite{PhysRevLett.91.196802,PhysRevLett.93.256801} or by the gate voltage
\cite{Petta:prb161301,J.R.Petta09302005,PhysRevLett.91.196802},
varying from several meV to very small. The exchange interaction
between electrons bound to adjacent donors is believed to be
responsible for the spin relaxation of the donor bound electrons
\cite{PhysRevB.66.245204,Kavokin:1521221,0268-1242-23-11-114009,PhysRevB.69.075302,PhysRevB.64.075305,PhysRevB.70.113201,PhysRevB.70.113201,PhysRevB.67.033203}.
The underlying physics is that due to the spin-orbit
coupling, the exchange interaction becomes
anisotropic which does not conserve the total spin of the
electron
system any more and hence leads to spin relaxation.

Besides, the exchange interaction between extended carriers also shows
up in spin dynamics. For example, the exchange interaction between
electrons and holes\footnote{The electron-hole exchange interaction is
in fact from the  conduction-valence band mixing
  \cite{jetp33.108}.} leads to electron spin relaxation, which is called
the Bir-Aronov-Pikus mechanism. Specifically, this exchange
interaction consists of two parts: the short-range part and the
long-range part. The short-range part is written as
\cite{PhysRevB.47.15776}
\be
H_{\rm SR} = - \frac{1}{V}\frac{1}{2} \frac{\Delta E_{\rm SR}}{|\phi_{\rm 3D}(0)|^2}
  \hat{\bf J}\cdot \hat{\bf S} \delta_{{\bf K},{\bf K}^{\prime}},
\label{bapisr}
\ee
where $\Delta E_{\rm SR}$ is the exchange splitting of the exciton ground
state and $|\phi_{\rm 3D}(0)|^2=1/(\pi a_0^3)$ with $a_0$ being the
exciton Bohr radius. ${\bf K}={\bf k}_{e}+{\bf k}_{h}$ is the
sum of the electron ${\bf k}_{e}$ and hole ${\bf k}_{h}$ wave-vectors of
the interacting electron-hole pair. $V$ is the volume of the
sample. The long-range part reads \cite{PhysRevB.47.15776}
\be
H_{\rm LR} = \frac{1}{V}\frac{3}{8} \delta_{{\bf K},{\bf K}^{\prime}}
\frac{\Delta E_{\rm LT}}{|\phi_{\rm 3D}(0)|^2} 
  ( \hat{M}_z \hat{S}_z + \frac{1}{2}\hat{M}_{-} \hat{S}_{+} +
  \frac{1}{2}\hat{M}_{+} \hat{S}_{-} ),
\label{bapilr}
\ee 
where $\Delta E_{\rm LT}$ is the longitudinal-transverse splitting;
$\hat{M}_z$, $\hat{M}_{-}$ and $\hat{M}_{+}(=\hat{M}_{-}^{\dagger})$ 
are operators in hole spin space. $\hat{S}_{\pm} =
\hat{S}_x \pm i \hat{S}_y$ are the electron spin ladder operators. The
expressions for $\hat{M}_z$ and $\hat{M}_{-}$ are given below (in the
order of hole spin $J_z=\frac{3}{2}, \frac{1}{2}, -\frac{1}{2}, -\frac{3}{2}$),
\begin{eqnarray}
  \hat{M}_{z} = \frac{1}{K^2}
  \left[ \begin{array}{cccc} 
    - K_{\parallel}^2 & \frac{2}{\sqrt{3}} K_z K_{-} &
      \frac{1}{\sqrt{3}} K_{-}^2 & 0 \\
      \frac{2}{\sqrt{3}}K_z K_{+} &
      (\frac{1}{3}K_{\parallel}^2-\frac{4}{3} K_z^2) & -\frac{4}{3} 
      K_z K_{-} & -\frac{1}{\sqrt{3}} K_{-}^2 \\
      \frac{1}{\sqrt{3}} K_{+}^2 & -\frac{4}{3} K_z K_{+} &
  (\frac{4}{3}K_z^2-\frac{1}{3}K_{\parallel}^2) &  \frac{2}{\sqrt{3}}K_zK_{-} \\
    0 & -\frac{1}{\sqrt{3}}K_{+}^2 & \frac{2}{\sqrt{3}}K_zK_{+} & K_{\|}^2
  \end{array} \right],
\end{eqnarray}
\begin{eqnarray}
  \hat{M}_{-} = \frac{1}{K^2}
  \left[ \begin{array}{cccc} 
    0 & 0 & 0 & 0 \\
    - \frac{2}{\sqrt{3}} K_{\parallel}^2 & \frac{4}{3} K_z K_{-} &
      \frac{2}{3} K_{-}^2 & 0 \\
      \frac{4}{\sqrt{3}}K_z K_{+} & -\frac{8}{3} K_z^2 & -\frac{4}{3}
      K_z K_{-} & 0 \\
      2 K_{+}^2 & -\frac{4}{\sqrt{3}} K_z K_{+} & -\frac{2}{\sqrt{3}}
      K_{\parallel}^2 & 0 
  \end{array}\right].
\end{eqnarray}
Here $K_{\pm} = K_x \pm i K_y$ and $K_{\parallel}^2 = K_x^2 +
K_y^2$. The short-range and long-range parts of the electron-hole exchange
interaction can be written in a compact form ($i={z,+,-}$),
\be
H_{\rm ex} = \frac{1}{V} \delta_{{\bf K},{\bf K}^{\prime}} 
(\hat{{\cal{J}}}_z({\bf K}) \hat{S}_z +
\frac{1}{2}\hat{{\cal{J}}}_{-}({\bf K}) \hat{S}_{+} + 
  \frac{1}{2}\hat{{\cal{J}}}_{+}({\bf K}) \hat{S}_{-} )
\quad {\rm with}\quad
{\cal{J}}_i({\bf K}) = \frac{1}{|\phi_{3D}(0)|^2}\left[ \frac{3}{8} \Delta
    E_{\rm LT} \hat{M}_{i}({\bf K}) - \frac{1}{2} \Delta
    E_{\rm SR} \hat{J}_{i}
    \right].
\label{bapitot}
\ee
In semiconductor nanostructures, e.g., in quantum
wells, the above electron-hole exchange interaction should be written
in the subband bases. Such expressions have been derived in
Ref.~\cite{PhysRevB.47.15776}. Besides, it is shown that the quantum
confinement can enhance the electron-hole exchange interaction
in quantum wells \cite{PhysRevB.37.6429}.

Unlike the electron-hole exchange interaction, the exchange
interaction between free electrons has almost been ignored by the
community until several years ago. It was first pointed out by
Weng and Wu that the exchange interaction between free electrons (the
Coulomb Hartree-Fock term), which acts as an effective magnetic field
along the spin polarization direction, plays important role in spin
dynamics \cite{PhysRevB.68.075312}. For example, if the spin
polarization is along the $z$ axis, the effective magnetic field from
the Hartree-Fock term reads \cite{PhysRevB.68.075312,jiang:125206},
\begin{equation}
  B_{\rm HF} ({\bf k}) = \sum_{{\bf q}} V_{{\bf q}} \left(f_{{\bf
        k}-{\bf q}\uparrow}-f_{{\bf k}-{\bf q}\downarrow}\right) \Big/g_e\mu_{\rm B},
\label{HFB}
\end{equation}
where $f_{{\bf k}-{\bf q}\uparrow}$ and $f_{{\bf k}-{\bf q}\downarrow}$ are
the distributions on the spin-up and spin-down bands respectively. $g_e$
is electron $g$-factor and 
$V_{\bf q}=e^2/[\epsilon_0\kappa_0q^2\epsilon({\bf q})]$ in bulk with
$\epsilon({\bf q})$ being the dielectric function. At large spin 
polarization, $B_{\rm HF}$ can be as large as a few tens of Tesla for
an electron density of $4\times 10^{11}$~cm$^{-2}$ at 120~K in GaAs 
quantum well \cite{PhysRevB.68.075312}. Recently, the effect of 
the exchange interaction between electrons was observed in 
experiments in high mobility two-dimensional electron system
\cite{stich:176401,stich:205301,korn}. The
Hartree-Fock effective magnetic field has also been observed in other
experiments recently \cite{0295-5075-83-4-47006}. Detailed
review on the Hartree-Fock effective magnetic field on spin
dynamics is given in Sec.~5.4.3.

\section{Spin relaxation and spin dephasing in semiconductors}

After briefly introducing various spin interactions, we now move
  to one of the central issues in this review: spin relaxation and 
  dephasing, i.e., the dissipative part of the spin dynamics. As stated
  before, {\em any} fluctuation or inhomogeneity in spin interactions 
  can lead to spin relaxation and spin dephasing. From the spin
  interactions, one can identify the possible spin
  relaxation/dephasing mechanisms.
 In what follows, we first 
introduce the concepts of spin relaxation and  dephasing in time
domain as well as in spacial domain.
 We then introduce the relevant spin relaxation mechanisms
in semiconductors and their nanostructures.

\subsection{Spin relaxation and spin dephasing}

Simply speaking, spin relaxation is
related to the nonequilibrium population decay of the spin-resolved
energy eigenstates, whereas spin dephasing is related to the
destruction of phase coherence of these eigenstates. For single spin
system with isotropic $g$-factor, spin relaxation is related to the
decay of spin polarization parallel to the external magnetic field,
whereas spin dephasing is related to the decay of spin polarization
transverse to the magnetic field. Spin relaxation time (denoted as
$T_1$) and spin dephasing time ($T_2$) are quantities to characterize
the time scale of spin relaxation and spin dephasing.\footnote{Spin decay
  does not have to be exponential. However, in most cases, a
  characteristic decay time can be identified.} Microscopically, spin
relaxation and spin dephasing are usually related to different
processes. Although spin relaxation also inevitably leads to spin
dephasing, there are processes which only contribute to spin
dephasing, called pure dephasing processes \cite{quandissip}. Spin
relaxation and spin dephasing in single spin system 
are caused by the unavoidable coupling with the fluctuating
environment, such as the phonon system, nuclear spin system and other
electrons nearby. These processes are always irreversible.

Often there are many electron spins forming a spin ensemble. For an
ensemble of spins, spin polarization can decay without coupling to any
environment: when spin precession frequencies or directions vary from
part to part within the ensemble, the total spin polarization
gets a free-induction-decay due to destructive interference. This
variation is called the {\em inhomogeneous broadening}
\cite{mrbook,2levelbook}. For a finite ensemble with independent
spins, the decay of total spin polarization can be reversible: after
some time, spins in all parts of the ensemble rotate to the same
direction.\footnote{This phenomenon was recently observed in
  experiments in spin dephasing in an ensemble of self-assemble
  quantum dots \cite{A.Greilich07212006}.} However, quite often, the
ensemble is very large, e.g., there are $\sim 10^{16}$~cm$^{-3}$ electrons in
$n$-doped GaAs. In this regime, the spin decay due to inhomogeneous broadening can not be reversed automatically. However, the inhomogeneous broadening
induced spin decay can still be removed by the technique of
spin echo. Standard spin echo is achieved by a $\pi$-pulse magnetic
field which is perpendicular to the inhomogeneous broadened
(effective) magnetic field, to invert the sign of the phase gained
through spin precession so that the gained phase can be gradually
canceled out in the subsequent evolution. The ensemble spin relaxation and spin
dephasing times including the inhomogeneous broadening are usually denoted as
$T_1^{\ast}$ and $T_2^{\ast}$ respectively, whereas the irreversible (echoed)
spin relaxation and spin dephasing times are denoted as $T_1$ and $T_2$
respectively. In the following, when we need not to discriminate the two,
we simply denote the spin relaxation time and dephasing time as
$\tau_s$ and called it spin lifetime. There are a lot of factors which
contribute to inhomogeneous broadening, such as, the 
${\bf k}$-dependent spin-orbit coupling \cite{JPSJ.70.2195} or the
energy(or ${\bf k}$)-dependent $g$-factor
\cite{margulis:918,wu:epjb.18.373}. In quantum 
dots, due to the spin-orbit coupling, electron or hole $g$-factor is size
and geometry dependent \cite{PhysRevB.69.125302}. The variations of the
size and geometry of quantum dots in a quantum dot ensemble also lead
to the inhomogeneous broadening \cite{A.Greilich07212006}. Finally,
the time and/or space variations of the nuclear hyperfine field also
lead to the inhomogeneous broadening.

Spin diffusion and transport also suffers spin relaxation. In the absence of
the inhomogeneous broadening, spin diffusion length $L_s$ is related to
spin lifetime $\tau_s$ as $L_s=\sqrt{D_s\tau_s}$, where $D_s$
is the spin diffusion constant. However, in semiconductors it has been
shown that the inhomogeneous broadening usually dominates spin
diffusion \cite{PhysRevB.66.235109,weng:410,cheng:073702}, which
breaks down the above relation \cite{PhysRevB.70.155308,weng:063714}. The
inhomogeneous broadening in spin diffusion and transport is different from that in
spin relaxation in time domain: electron at different ${\bf k}$ state has different
velocity and/or spin precession frequency ${\bf \Omega}({\bf k})$
(e.g., due to the spin-orbit coupling), making the spin propagation
frequency along the spin diffusion direction (${\hat{\bf n}}$) 
${\bf k}$-dependent $\sim{\bf \Omega}({\bf k})/({\bf k}\cdot{\hat{\bf n}})$
\cite{PhysRevB.66.235109,weng:410,cheng:073702,cheng:205328,kleinert:045317}.
This inhomogeneous broadening can not be removed by traditional spin-echo.
However, it can be tuned via controlling spin-orbit coupling
\cite{cheng:205328,bernevig:236601,stanescu:125307,weng:063714,bernevig:245123,kleinert:045317}.
For example, when both the Rashba and Dresselhaus spin-orbit couplings exist,
by tuning the two to be comparable, spin diffusion length can be largely increased
\cite{PhysRevLett.90.146801,cartoixa:1462,cheng:205328,bernevig:236601,stanescu:125307,weng:063714,bernevig:245123,kleinert:045317}.
Such enhancement of spin diffusion length has been achieved in a
recent experiment, where the control over both the doping asymmetry
and well width was shown to be effective to manipulate the
relative strength of the Rashba and Dresselhaus spin-orbit couplings and
hence the spin diffusion length \cite{Koralek:nature458.610}. Studies
on spin diffusion in the literature are reviewed in Section~6 and 7.
Below we concentrate on spin relaxation and spin dephasing in time
domain, where the basic physics and the mechanisms responsible for
spin relaxation and spin dephasing are reviewed.

\subsubsection{Single spin relaxation and spin dephasing due to fluctuating
  magnetic field}

As stated above, fluctuations in spin interaction due to the coupling
with the environment are the origin of spin relaxation and spin
dephasing in single spin system. In this subsection, we discuss
generally the spin relaxation and spin dephasing in single spin system,
where the fluctuations in spin interaction are simply characterized as
fluctuating magnetic fields. These fluctuating magnetic fields are
assumed to have a correlation time $\tau_c$ which is much shorter than
the spin relaxation/dephasing time, so that the Markovian
approximation is justified \cite{quanNoise}. We also assume that the
fluctuations are weak so that they can be treated within the Born
approximation. The spin dynamics is then solved by the standard
Born-Markov method \cite{quanNoise}.\footnote{It is noted that
  discussions in this subsection are similar to those in
  Ref.~\cite{Fabianbook} in Sec.~IV~B2.}

The Hamiltonian is simply given by
\be
H = \frac{1}{2}[\omega_0\hat{{\bf n}}_z + \mbox{\boldmath $\omega$\unboldmath}(t)]\cdot\spin,
\ee
where $\hat{{\bf n}}_z$ is a unit vector along $z$
direction. $\omega_0$ and $\mbox{\boldmath $\omega$\unboldmath}(t)$
are spin precession frequencies due to the static and fluctuating
magnetic fields respectively. The correlation of the fluctuating
magnetic fields is assumed to be (for $i,j=x,y,z$)
$\overline{\omega_{i}(t)\omega_{j}(t^{\prime})} = 
\delta_{i,j} \overline{\omega_i^2}e^{-|t-t^{\prime}|/\tau_c}$,
with the overline denoting the ensemble average.
The requirement of the shortness of the correlation time
$\tau_c$ and the weakness of the fluctuating magnetic fields is equivalent to 
$\sqrt{\sum_i \overline{\omega_i^2}}~\tau_c\ll 1$.
Spin relaxation
time and spin dephasing time are obtained by solving the Born-Markov
equation of motion. After standard derivation one obtains that
\cite{Fabianbook},
\be
\frac{1}{T_1} = \frac{(\overline{\omega_x^2} +
\overline{\omega_y^2})\tau_c}{1+\omega_0^2\tau_c^2},\quad\quad
\frac{1}{T_2} = \overline{\omega_z^2}\tau_c +
\frac{(\overline{\omega_x^2}+\overline{\omega_y^2})\tau_c}{2(1+\omega_0^2\tau_c^2)}. 
\label{srt-single}
\ee
From these results, one may find several features of spin
relaxation/dephasing. First, a trivial result, only the noise
magnetic fields perpendicular to the spin polarization lead to spin
decay as only these fields can rotate spin. Second, in the absence of
static magnetic field, spin relaxation/dephasing rate is 
proportional to the noise correlation time $\tau_c$. That is to say that
the more fluctuating the noise (the shorter correlation time) is, the
more ineffective it is. This counter-intuitive result is known as
motional narrowing \cite{mrbook}. Third, a static magnetic field suppresses spin
relaxation/dephasing caused by the fluctuating magnetic field perpendicular to
it. The underlying physics is better understood in the fashion of what
was presented in the review article by Fabian et al. \cite{Fabianbook}:
In the presence of magnetic field, the spin-flip scattering terms
acquire a dynamic phase. However, the scattering terms which conserve
the spin along the magnetic field do not acquire such dynamic
phase. The spin-flip term is then proportional to
\be
\int_0^{t\gg\tau_c} dt^\prime\overline{\omega_i(0)\omega_i(t^\prime)}\exp(-\ii\omega_0
t^\prime) + {\rm H.c.} \propto \overline{\omega_i^2}\frac{\tau_c}{\omega_0^2\tau_c^2+1}.
\ee
Therefore the effect of fluctuating magnetic field perpendicular to
the static magnetic field is suppressed.

In the following, we illustrate several limits which are frequently
encountered in spin relaxation in semiconductors. Eq.~(\ref{srt-single})
can be written as
\be
1/T_2= 1/T_2^\prime + 1/(2T_1),
\ee
where $1/T_2^\prime= \overline{\omega_z^2}\tau_c$. Therefore,
in general case, $T_2\le 2T_1$. In the isotropic noise case,
$\overline{\omega_x^2}=\overline{\omega_y^2}=\overline{\omega_z^2}$,
at very low static magnetic field $\omega_0\tau_c\ll 1$, spin
relaxation time is equal to spin dephasing time, $T_1 = T_2$
\cite{PhysRev.100.1014}. At very large static magnetic field, spin
relaxation vanishes $1/T_1\to 0$, whereas the spin dephasing
rate is finite, $1/T_2=1/T_2^\prime$. This is a pure dephasing case.
The noise can be anisotropic. For example, when there are only
transverse noises, $\omega_x\ne 0$, $\omega_y\ne 0$ and
$\omega_z=0$, $T_2= 2T_1$.\footnote{An example is the spin relaxation
  in GaAs quantum dots due to the spin-orbit coupling and phonon
  scattering as demonstrated by Golovach et
  al. \cite{PhysRevLett.93.016601}. This conclusion was later
 generalized to spin relaxation due to arbitrary phonon scattering at
 low temperature when only the lowest two Zeeman levels are relevant by
  Jiang et al. \cite{jiang:035323}.}
Another example, there are only
longitudinal noises, $\omega_x=\omega_y=0$ and $\omega_z\ne 0$. In
this case, there is no spin relaxation $1/T_1=0$ but spin dephasing
rate is still finite, i.e., $1/T_2=\overline{\omega_z^2}\tau_c$. This is
another pure dephasing case. If
  $\overline{\omega_x^2}+\overline{\omega_y^2} \ll
  \overline{\omega_z^2}$, then $T_2\ll T_1$.

\subsubsection{Ensemble spin relaxation and spin dephasing: a simple model}

To illustrate the role of the inhomogeneous broadening in the ensemble spin relaxation and spin dephasing, here we
introduce a simple model which actually has been well studied in the
context of semiconductor optics. Denote the conduction band as ``spin
up'', the valence band as ``spin down'' and their distributions as
$f^{\uparrow}_{\bf k}$ and $f^{\downarrow}_{\bf k}$. The interband
coherence, like spin coherence, is a complex variable 
$\rho_{\bf k} = \rho_{cv}({\bf k})e^{-\ii \omega t}$ 
($\omega$ is the laser light frequency). Hence the model system is equivalent to a spin
ensemble. The optical relaxation and dephasing is
 described by the Bloch equation  \cite{shahbook,haugkoch}
\be
\partial_t {\bf S}_{\bf k} = \omega_{\bf k} \times {\bf S}_{\bf k} -
\left[ \begin{array}{ccc}\frac{1}{T_2}&0&0 \\[8pt] 0& \frac{1}{T_2}&0\\
  0&0&\frac{1}{T_1} 
\end{array}\right]
({\bf S}_{\bf k} - {\bf S}_{\bf k}^0).
\ee
In semiconductor optics, 
$\omega_{\bf k}=(-d_{cv}{\cal{E}}, 0, \delta_{\bf k})$ where $d_{cv}$ is the
interband optical dipole, $\cal{E}$ is the laser electric field and
$\delta_{\bf
  k}=\varepsilon_{c,k}-\varepsilon_{v,k}-\omega=k^2/2m_R+E_g-\omega$
is the detuning with the frequency of light. For simplicity, take
$\omega=E_g$ and hence $\delta_{\bf k}=k^2/(2m_R)$. $m_R$ is the reduced effective mass of
electron-hole pair. ${\bf S}_{\bf k}=({\rm Re}\rho_{\bf k}, -{\rm
  Im}\rho_{\bf k}, [f^{\uparrow}_{\bf  k}-f^{\downarrow}_{\bf k}]/2)$
and the equilibrium polarization is ${\bf S}_{\bf
  k}^0=(0,0,-1/2)$. $T_1$ and $T_2$ represent the irreversible 
decay due to fluctuations. In the absence of laser field
${\cal{E}}=0$, the solution of the equation gives the transverse
polarization as
\be
\left[ \begin{array}{c}
S_{\bf k}^x (t) \\ S_{\bf k}^y(t) 
\end{array} \right]
= \left[ \begin{array}{cc}
\cos(\delta_{\bf k}t) & -\sin(\delta_{\bf k}t) \\
\sin(\delta_{\bf k}t) & \cos(\delta_{\bf k}t) 
\end{array}\right]
\left[ \begin{array}{c}
S_{\bf k}^x (0) \\ S_{\bf k}^y(0) 
\end{array}\right]
e^{-t/T_2}.
\ee
As the oscillation frequency $\delta_{\bf k}$ varies with ${\bf k}$,
the total transverse polarizations $S_x(t)=\sum_{\bf k} S_{\bf k}^x(t)$
and $S_y(t)=\sum_{\bf k}S_{\bf k}^y(t)$ get an additional
decay due to inhomogeneous broadening besides $e^{-t/T_2}$. In two-dimensional system,
assuming a simplified initial condition: $S_{\bf k}^x(0) = 0$,
$S_{\bf k}^y(0)=-1$ for $k\le k_m$ and $S_{\bf k}^y(0)=0$ for $k>k_m$,
one obtains
\be
S_x(t) = e^{-t/T_2}\sum_{\bf k}\sin(\delta_{\bf k}t)=e^{-t/T_2}
\frac{k_m^2}{4\pi}\frac{1-\cos(\delta_{k_m}t)}{t/T_2^{\prime\prime}}. 
\ee
The inhomogeneous broadening induces a power law decay with a
characteristic time $T_2^{\prime\prime} = 2m_R /k_m^2$. Now
the optical dephasing rate is 
\be
1/T_2^{\ast} = 1/T_2 + 1/T_2^{\prime\prime}.
\ee
The inhomogeneous broadening induced dephasing
$1/T_2^{\prime\prime}$ can be removed by photon echo. Exerting a
$\pi$-pulse along the $y$-axis at time $T$, the transverse polarization
after the pulse is
\be
\left[ \begin{array}{c} 
S_{\bf k}^x (T) \\ S_{\bf k}^y(T) 
\end{array}\right]
= \left[ \begin{array}{cc} 
\cos(\delta_{\bf k}T) & \sin(\delta_{\bf k}T) \\
-\sin(\delta_{\bf k}T) & \cos(\delta_{\bf k}T) 
\end{array} \right]
\left[ \begin{array}{c}
S_{\bf k}^x (0) \\ S_{\bf k}^y(0) 
\end{array}\right]
e^{-T/T_2}.
\ee
The transverse polarization at time $t=2T$ is given by
\be
\left[ \begin{array}{c} 
S_{\bf k}^x (2T) \\ S_{\bf k}^y(2T) 
\end{array}\right]
= \left[ \begin{array}{cc} 
\cos(\delta_{\bf k}T) & -\sin(\delta_{\bf k}T) \\
\sin(\delta_{\bf k}T) & \cos(\delta_{\bf k}T) 
\end{array}\right]
\left[ \begin{array}{c}
S_{\bf k}^x (T) \\ S_{\bf k}^y(T) 
\end{array}\right]
e^{-T/T_2}=\left[ \begin{array}{c}
S_{\bf k}^x (0) \\ S_{\bf k}^y(0) 
\end{array}\right]
e^{-2T/T_2}.
\ee
Now the inhomogeneous broadening induced dephasing is removed and
only the irreversible dephasing $1/T_2$ remains. Theoretically,
the decay rate of the incoherently summed optical coherence
$P(t)=\sum_{\bf k}|\rho_{\bf k}(t)|=\sum_{\bf k}|S^x_{\bf k}(t)+iS^y_{\bf
  k}(t)|$, can be used to obtain the irreversible optical dephasing rate $1/T_2$
\cite{PhysRevLett.69.977}, as the phase of $\rho_{\bf k}(t)$ is removed.

Similar to the case of optical dephasing, the inhomogeneous broadening
of the spin precession induces spin dephasing in spin ensemble
\cite{wu:epjb.18.373}. This inhomogeneous broadening can be removed by
spin echo \cite{pershin:165320}. As pointed out by Wu and co-workers, 
the irreversible spin dephasing can be obtained from the decay of the
incoherently summed spin coherence
\cite{wu:pssb.222.523,wu:epjb.18.373,PhysRevB.66.235109,PhysRevB.61.2945},
similar to that in optical dephasing.

In system with isotropic $g$-tensor, the spin polarization parallel to
the magnetic field does not precess. However, in general, the $g$-tensor
can be anisotropic and spin polarization parallel to the magnetic field
can also precess. In this case, spin polarization along the magnetic
field may also suffer the inhomogeneous broadening induced decay. We then
have both $T_1^{\ast}$ and $T_2^{\ast}$ in such situation.\footnote{A
  simple example demonstrating the same physics is presented in
  Ref.~\cite{PhysRevB.72.121303}.} An example of such ensemble is
the spins in self-assembled quantum dots
\cite{PhysRevLett.91.257901,masumoto:205332,PhysRevB.72.161303}.

\subsection{Spin relaxation mechanisms}

In this subsection, we introduce spin relaxation mechanisms in
  semiconductors.
Generally speaking, any fluctuation or inhomogeneity of spin
interaction can induce spin relaxation and dephasing. However, there are several
mechanisms which are more efficient than others. For the materials widely
used in spintronics, such as III-V and II-VI semiconductors, there are
only a few relevant spin relaxation/dephasing mechanisms, such as the Elliott-Yafet mechanism
\cite{yafetbook,PhysRev.96.266}, the D'yakonov-Perel' mechanism
\cite{dp,DP2}, the Bir-Aronov-Pikus mechanism \cite{BAP,aronov} and
the $g$-tensor inhomogeneity mechanism
\cite{margulis:918,wu:epjb.18.373}. These 
mechanisms dominate spin relaxation and dephasing in metallic regime. In the
insulating regime, other mechanisms such as the anisotropic exchange
interaction \cite{PhysRevB.69.075302,0268-1242-23-11-114009} and the
hyperfine interaction \cite{opt-or} prevail. In semiconductor
quantum dots, spin relaxation/dephasing is dominated by the
electron-phonon
scattering and the hyperfine interaction as well as
their combinations. In magnetically doped semiconductors, the exchange
interaction with magnetic impurities can also play an important role
in spin relaxation and dephasing. Below we give a brief
introduction to these mechanisms.

\subsubsection{Elliott-Yafet mechanism}

The Elliott-Yafet mechanism was first proposed by Elliott
\cite{PhysRev.96.266} and Yafet \cite{yafetbook} during their study on
spin relaxation in silicon and alkali metals. It was pointed out by
Elliott that due to the spin-orbit interaction the electronic eigenstates
(Bloch states) mix spin-up and spin-down states. For example \cite{zuticrmp,PhysRev.96.266},
\bea
\Psi_{{\bf k}n\uparrow}({\bf r}) &=& [a_{{\bf k}n}({\bf r})|\uparrow\rangle +
b_{{\bf k}n}({\bf r})|\downarrow\rangle] e^{\ii {\bf k}\cdot{\bf r}},\\
\Psi_{{\bf k}n\downarrow}({\bf r}) &=& [a_{-{\bf k}n}^{\ast}({\bf
  r})|\downarrow\rangle - b_{-{\bf k}n}^{\ast}({\bf
  r})|\uparrow\rangle] e^{\ii {\bf k}\cdot{\bf r}},
\eea
where $a_{{\bf k}n}({\bf r})$ and $b_{{\bf k}n}({\bf r})$ possess the
lattice periodicity. These two Bloch states are connected by time
reversal and space inversion operators \cite{zuticrmp,PhysRev.96.266}. Usually
the spin mixing is very small, i.e., $|b|\ll 1$ and $|a|\approx
1$. When the spin-orbit interaction is much smaller than the band
splitting, an estimation of $|b|$ from perturbation theory gives \cite{zuticrmp,simon:177003}
$|b|\sim {\rm max}\{L_{\rm SO}/\Delta E\}$, where $L_{\rm SO}$ is the
spin-orbit interactions with other bands and $\Delta E$ is the band
distances. In the presence of such spin mixing,
any spin-independent scattering can cause spin flip and hence spin
relaxation. It should be emphasized that without scattering the
spin-mixing alone can not lead to any spin relaxation. Besides, there is
another process leading to spin relaxation: the phonon modulation of
the spin-orbit interaction. As the spin-orbit interaction is induced
by the periodic lattice ions, the lattice vibration can then directly
couple to spin and lead to spin-flip. Such kind of process was first
considered by Overhauser within jullium model for metal
\cite{PhysRev.89.689} and then by Yafet with specific band structure
\cite{yafetbook}. This process is called the Yafet process, whereas
the previous one due to the spin mixing is called the Elliott process.

Elliott observed a relation, the ``Elliott relation'', between spin
lifetime $\tau_s$ and the deviation of the electron $g$-factor
from that of the free electron $g_0=2.0023$ \cite{PhysRev.96.266}:
$1/\tau_s \approx (\Delta g)^2/\tau_p$ where $\Delta g=g-g_0$ and
$\tau_p$ is the average momentum scattering time. The relation is
based on the observation that $\Delta g\sim|b|$ by perturbation
theory. The Elliott relation was verified experimentally by Monod and
Beuneu, who observed an empirical factor of 10, i.e., $1/\tau_s \simeq
10 (\Delta g)^2/\tau_p$ \cite{PhysRevB.18.2422}. Of course these
relations are rough, the specific ratio of $\tau_s /\tau_p$ depends on
specific scattering and band structure. Systematic theoretical study
was given by Yafet \cite{yafetbook}, where the microscopic band
structure and electron-phonon scattering were considered to obtain the
spin lifetime. Yafet gave a relation between the spin
lifetime $\tau_s$ and the resistivity $\rho$: $1/\tau_s\sim
\langle b^2\rangle\rho$ \cite{yafetbook}. The Yafet relation was tested experimentally
by Monod and Beuneu \cite{PhysRevB.19.911}. It was found
that in many materials the Yafet relation agrees well with
experiments \cite{PhysRevB.19.911,zuticrmp}. However, it is not
consistent with the experimental results in MgB$_2$ where the spin
relaxation rate is not proportional to the resistivity above
150~K. The problem was solved by Simon et al. \cite{simon:177003}
recently via generalizing the Elliott-Yafet theory to the regime where
the scattering-induced spectral broadening of the quasi-particle
$1/\tau_p$ is comparable with the band gap. This condition is
satisfied in MgB$_2$ because one of the band across the Fermi surface
is very close to the nearest band ($\simeq 0.2$~eV) and the
electron-phonon interaction is strong. In this regime, the spin
lifetime is given by,
\be 
 \tau_s^{-1} = \frac{L_{\rm
     eff}^2 \tau_p}{1+\Delta\omega_{\rm eff}^2\tau_p^2},
\ee
where $\Delta\omega_{\rm eff}$ is the average band gap and $L_{\rm
  eff}$ is the interband spin-orbit interaction. In the weak scattering
or large band-gap regime, $\Delta\omega_{\rm eff}\tau_p\gg 1$, the
above equation returns to the Yafet relation, $1/\tau_s =
(L_{\rm eff}/\Delta\omega_{\rm eff})^2/\tau_p$. Spin relaxation due to
the Elliott-Yafet mechanism in polyvalent metals (such as aluminum) was
studied by Fabian and Das Sarma
\cite{PhysRevLett.81.5624,PhysRevLett.83.1211}. Through realistic
calculation, they found that the electron spin relaxation is
significantly enhanced at the Brillouin zone boundaries, special
symmetry points and lines of accidental degeneracy. The total spin
relaxation rate is then determined by the spin relaxation rates at
these ``spin-hot-spots'' \cite{PhysRevLett.81.5624,PhysRevLett.83.1211}.

In
III-V or II-VI semiconductors zinc-blende structure, for realistic
calculation of the Elliott-Yafet spin relaxation, one should start from
the Kane Hamiltonian. By L\"owdin partitioning (block-diagonalization),
the conduction band wavefunction is transformed to
$\tilde{\Psi}_c({\bf k}) = U \Psi_n({\bf k})$ where $U$ is the unitary
matrix for L\"owdin partitioning. This transformation mixes spin-up and
-down states a little, which enables spin flip by spin-independent
fluctuation. After the transformation $U$, the conduction band matrix
element of a spin-independent fluctuation of the lattice potential
(including the electron-electron interaction) $V_{c{\bf k},c{\bf
    k}^{\prime}} =\langle c{\bf k}|V|c{\bf  k}^{\prime}\rangle$,
changes into $\tilde{V}_{c{\bf k},c{\bf k}^{\prime}} = \langle c{\bf k}|U^{\dagger}VU|c{\bf
    k}^{\prime}\rangle$,
which may contain spin-flip part. The spin-flip interaction consists of
long-range and short-range parts. The long-range part comes from
the first order perturbation of the wavefunction (both in $U$ and
$U^{\dagger}$) and the intraband spin-independent fluctuation. The
short-range part comes from the combination of $U$ (or $U^{\dagger}$)
and the interband electron-phonon interactions. For example, for
$U=e^{S_{\bf k}}$, the lowest order long-range part is given by $\langle c{\bf
  k}|\frac{1}{2}(S_{\bf k}^{\dagger}S_{\bf k}^{\dagger}V + 2S_{\bf k}^{\dagger}V
S_{{\bf k}^{\prime}} + V S_{{\bf k}^{\prime}} S_{{\bf k}^{\prime}})
|c{\bf k}^{\prime}\rangle$, whereas the lowest order short-range 
part is given by $\langle c{\bf k}|S_{\bf k}^{\dagger}V+VS_{{\bf
    k}^{\prime}}|c{\bf k}^{\prime}\rangle$. The long-range
interaction is the Elliott process, whereas the short-range
interaction is the Yafet process.

After the transformation, the leading term of the long-range
part is given by \cite{Titkovbook}
\be
\tilde{V}_{c{\bf k},c{\bf k}^{\prime}} = 
V_{c{\bf k},c{\bf k}^{\prime}}[1-\ii \lambda_c({\bf k}\times{\bf
  k}^{\prime})\cdot\spin], 
\label{EYV}
\ee
where $\lambda_c=\eta(1-\eta/2) /[3m_eE_g(1-\eta/3)]$ with
$\eta=\Delta_{\rm SO}/(\Delta_{\rm SO} + E_g)$. The spin lifetime
due to long-range part of the Elliott-Yafet mechanism is
\be
  \tau_s^{-1} = A \frac{\langle \varepsilon_{\bf
        k}\rangle^2}{E_g^2} \eta^2
  \left(\frac{1-\eta/2}{1-\eta/3}\right)^2 \tau_p^{-1}, 
\label{eyapp}
\ee
where the numerical factor $A\sim 1$ depending on the specific
scattering. The short-range interaction due to
the interband electron--optical-phonon interaction is given by
\be 
\tilde{V}^{\rm OP}_{{\bf k},{\bf k}^{\prime}} = - \frac{1}{3} \frac{\eta
  d_2 [{\bf U}\times ({\bf k}+{\bf k}^{\prime})]\cdot \spin}{2m_eE_g(1-\eta/3)]^{1/2}}.
\ee
Here ${\bf U}$ is the relative displacement of the two atoms in a
unit cell which can be expressed as combinations of phonon creation
and annihilation operators \cite{Titkovbook}. Assuming that the longitudinal and
transversal optical phonons have the same frequency $\omega_{\rm
  LO}=\omega_{\rm TO}$, the spin relaxation due to the short-range
interaction reads \cite{Titkovbook}
\be
\tau_s^{-1} = A_{\rm sr} \frac{\langle \varepsilon_{\bf
    k}\rangle^2}{E_g E_0} \frac{\eta^2}{1-\eta/3} \tau_p^{-1}.
\label{srey}
\ee
Here $E_0=C^2/(4d_2m_e)$ with $C=e\omega_{\rm LO}\sqrt{4\pi
D(\kappa_{\infty}^{-1}-\kappa_{0}^{-1})}$ ($D$ is the
volume density, $\kappa_{0}$ and $\kappa_{\infty}$ are the static and
high frequency dielectric constants) and $A_{\rm sr}\sim 1$. The
short-range interaction due to the interband electron--acoustic-phonon
interaction can be derived similarly. The explicit form is given in
Ref.~\cite{Titkovbook}. Usually the long-range interaction is more
important than the short-range one in III-V semiconductors
\cite{Titkovbook,zuticrmp}.

The Elliott-Yafet spin relaxation in in bulk InSb at low
temperature was first studied by Chazalviel \cite{PhysRevB.11.1555},
where only the electron-impurity scattering is considered. 
Very recently, Jiang
and Wu studied the Elliott-Yafet spin relaxation in bulk III-V
semiconductors with all relevant scatterings (i.e., electron-impurity,
electron-phonon, electron-electron Coulomb and electron-hole Coulomb
scatterings) included in a fully microscopic fashion
\cite{jiang:125206}. The Elliott-Yafet spin relaxation in bulk silicon
was studied systematically by Cheng et al. \cite{2009arXiv0906.4054C},
where the Elliott and the Yafet processes
interfere destructively and the spin relaxation is largely
suppressed \cite{2009arXiv0906.4054C}.

Similarly, both the admixture of different spin states due to the L\"owdin
partitioning and the phonon modulation of spin-orbit interaction lead to 
the Elliott-Yafet spin relaxation for holes and split-off
holes.

\subsubsection{D'yakonov-Perel' mechanism}

In III-V and II-VI semiconductors, due to the bulk inversion asymmetry
in lattice structure, spin-orbit coupling emerges in conduction
band. In semiconductor nanostructures, the structure and interface
inversion asymmetry further contributes additional spin-orbit
coupling. Spin-orbit coupling is equivalent to a ${\bf k}$-dependent
effective magnetic field $H_{\rm SO} = \frac{1}{2}{\bf \Omega}({\bf k})\cdot\spin$,
where ${\bf \Omega}({\bf k})$ is the spin precession frequency. In the
presence of momentum scattering, electron changes its momentum
${\bf k}$ randomly, hence spin precesses randomly between
adjacent scattering events. This random-walk-like evolution of spin
phase leads to spin relaxation. This spin relaxation mechanism is
called the D'yakonov-Perel' mechanism \cite{dp,DP2}. There are
two regimes for the D'yakonov-Perel' spin relaxation: (i) strong
scattering regime where $\langle \Omega\rangle \tau_p\ll 1$ 
($\langle...\rangle$ stands for the average over the electron ensemble),
and (ii) weak scattering regime where $\langle \Omega\rangle
\tau_p\gtrsim 1$. Here $\tau_p$ is the momentum scattering time. The
spin lifetime in regime (i) can be estimated as
$1/\tau_s=\langle \Omega^2\rangle \tau_p$ according to the
random walk theory. This spin relaxation has the salient feature of
motional narrowing, i.e., stronger momentum scattering leads to longer
spin lifetime. In regime (ii) the momentum scattering no longer
impedes spin relaxation. In contrast, via the spin-orbit coupling,
momentum scattering provides a spin relaxation channel
\cite{wu:epjb.18.373,lue:125314}. In this regime, stronger momentum
scattering leads to shorter spin lifetime \cite{lue:125314}.
Analytic results with only the electron-impurity scattering give that the
irreversible spin lifetime is $\tau_s=2\tau_p$
\cite{lue:125314,Gridnev:jetplett.74.380,PhysRevB.72.075307}. As
$\langle \Omega\rangle \tau_p\gtrsim 1$, the spin
precession due to the spin-orbit coupling is not inhibited by the
scattering. As a consequence, the ensemble spin polarization
oscillates at zero magnetic field
\cite{lue:125314,Gridnev:jetplett.74.380}. Besides, such prominent
spin precession leads to the free induction decay. The ensemble spin
lifetime is then limited by both the irreversible decay and the free
induction decay $\tau_s^{-1}\simeq \sqrt{\langle\Omega^2\rangle}$
\cite{lue:125314,zuticrmp,PhysRevB.72.075307}. Both regime (i) and
regime (ii) have been realized experimentally in GaAs quantum wells
\cite{harleybook,leyland:195305}. Especially, the crossover from
regime (ii) to regime (i) has been observed by Brand et
al. \cite{PhysRevLett.89.236601}.

The characteristic of the D'yakonov-Perel' spin relaxation is that during
adjacent momentum scatterings spins precess coherently. Hence a
spin-echo at a time scale comparable to or smaller than the momentum
scattering time $\tau_p$ can efficiently
suppress the D'yakonov-Perel' spin relaxation.
Such a proposal was given in a recent work by Pershin
\cite{pershin:165320}.

In most cases the system is in the strong scattering regime. In this regime,
some analytical results about spin lifetime can be obtained for
isotropic band, if one assumes that (1) the carrier-carrier 
scattering can be neglected,\footnote{It was first pointed out by Wu
  and Ning \cite{wu:epjb.18.373} that although normal (not umklapp)
  carrier-carrier scattering does not contribute to the
  mobility, it contributes to the D'yakonov-Perel' spin relaxation as it
  randomizes the momentum.} (2) the system is near
equilibrium\footnote{That is, electron distribution is close to the
  equilibrium distribution and the spin polarization is very
  small.} and (3) the electron-phonon scattering can be treated in the
elastic scattering approximation. Within these assumptions and
approximations, for three-dimensional case, following Pikus and
Titkov, the evolution of spin polarization along $z$ direction is
\cite{Titkovbook}
\be
\partial_t S_z = - \tilde{\tau}_l [S_z \langle \Omega^{l}_x\Omega^{l}_x
+\Omega^{l}_y\Omega^{l}_y\rangle -S_x \langle \Omega^{l}_x\Omega^{l}_z\rangle -
S_y\langle\Omega^{l}_y\Omega^{l}_z\rangle],
\label{szt}
\ee
where $\langle...\rangle$ represents the average over the direction of 
${\bf k}$. The equations for the evolution of $S_x$ and $S_y$ can be
obtained by index permutation. The momentum scattering time
$\tilde{\tau}_l$ is given by
\be
\tilde{\tau}_l^{-1} = \int_{-1}^{1} \
W(\theta)[1-P_l(\cos\theta)] d \cos\theta.
\label{tau_l}
\ee
The integer $l$ is determined by the angular dependence of the
spin-orbit field. The above results assume that the spin-orbit
coupling Hamiltonian satisfies
$
\hat{H}_{\rm SO} = \frac{1}{2}{\bf \Omega}^{l}\cdot\spin= \sum_{m}
\hat{C}_{lm} Y_{m}^{l}(\theta_{\bf k},\phi_{\bf k})$,
where $\theta_{\bf k}$ and $\phi_{\bf k}$ are the angular coordinates of ${\bf k}$
in spherical coordinates, $Y_{m}^l$ stands for the spherical hamonics
and the coefficients $\hat{C}_{lm}$ are 2$\times$2 matrices.
Here the summation is taken over $m$, whereas $l$ is fixed. For
example, $l=1$ for linear spin-orbit coupling due to strain and $l=3$
for Dresselhaus cubic spin-orbit coupling.
If there are both linear and cubic spin-orbit couplings, the
summation over $l$ should also be added into the above equations. According to
Eq.~(\ref{szt}), with the summation over $l$, the spin relaxation
tensor is given by \cite{opt-or,zuticrmp,Fabianbook}
\be
(\tau_{s}^{-1})_{ij} = \sum_l \frac{\tilde{\tau}_p}{\gamma_l}
[\langle|{\bf \Omega}^{l}|^2\rangle\delta_{ij}-\langle \Omega^{l}_i\Omega^{l}_j\rangle],
\label{dpsrt-tensor}
\ee
where $\gamma_l=\tilde{\tau}_p/\tilde{\tau}_l$ and
$\tilde{\tau}_p=\tilde{\tau}_1$. For two-dimensional case, expanding
the spin-orbit coupling as
$\hat{H}_{\rm SO} = \sum_{l} \frac{1}{2}{\bf \Omega}^{l} e^{\ii
  l\theta_{\bf k}}\cdot\spin$, 
one obtains a similar result \cite{d'yakonov:110,0268-1242-23-11-114002}
\be
(\tau_{s}^{-1})_{ij} = \sum_l \frac{\tilde{\tau}_p}{\gamma_l}
[|{\bf \Omega}^{l}|^2 \delta_{ij} - \Omega^{l}_i\Omega^{-l}_j],
\ee
with $\tilde{\tau}_l^{-1} = \int_0^{2\pi} W(\theta) (1-\cos l\theta)d\theta$.
To give an example, consider electron spin relaxation due to the
D'yakonov-Perel' mechanism in bulk III-V semiconductors. In the
absence of strain, the electron spin-orbit coupling comes solely from
the Dresselhaus term,
\be
{\bf \Omega}({\bf k})=2\gamma_{D}[k_x(k_y^2-k_z^2),
k_y(k_z^2-k_x^2), k_z(k_x^2-k_y^2)].
\ee
In this case spin relaxation is isotropic
\be
(\tau_s^{-1})_{ij} = \delta_{ij} \frac{\tilde{\tau}_p}{\gamma_3}
\frac{8}{105} (2\gamma_{D})^2k^6.
\ee
Here $\gamma_3$ depends on relevant scattering: for ionized-impurity
scattering $\gamma_3\simeq 6$; for acoustic-phonon scattering
$\gamma_3\simeq 1$; for optical-phonon scattering $\gamma_3\simeq
41/6$. After the average over electron distribution, the spin
relaxation rate is given by \cite{zuticrmp},
\be
\tau_s^{-1} \simeq Q^{\prime} \tau_m \alpha^2 \langle
  \varepsilon_{\bf k}^3\rangle /E_g.
\label{dpapp}
\ee
$\alpha = 2\gamma_{D}\sqrt{2m_e^3E_g}$ is a
dimensionless parameter. $\tau_m=
\langle\tilde{\tau}_p(\varepsilon_{\bf k})\varepsilon_{\bf
  k}\rangle/\langle\varepsilon_{\bf k}\rangle$.\footnote{In the
  remaining part of the paper, we denote all these
  quantities ($\tilde{\tau}_l$, $\tilde{\tau}_p$ and
  $\tau_m$) as $\tau_p$ in qualitative discussions, if we can.}
$Q^{\prime}\sim 0.1$  
is a numerical constant depending on the relevant momentum
scattering \cite{zuticrmp} $Q^{\prime} =
\frac{128}{3675}\gamma_3^{-1}(\nu+\frac{7}{2})(\nu+\frac{5}{2})$,
where the power law $\tilde{\tau}_{p}\sim \varepsilon_{\bf k}^{\nu}$ is
assumed for each kind of momentum scattering. For nondegenerate
electron system, one obtains ($Q=\frac{105}{8}Q^{\prime}$) 
\be
\tau^{-1}_s \simeq Q \tau_m \alpha^2 (k_{\rm
    B}T)^3 /E_g,
\label{dpapp-nondeg}
\ee
where $Q\sim 1$ depending on relevant scattering: $Q\simeq 1.5$ for
ionized impurity scattering, $Q\simeq 3$ for longitudinal optical
phonon scattering \cite{PhysRevLett.93.216402}, $Q\simeq 0.8$ for piezoelectric acoustic phonon
scattering, and $Q\simeq 2.7$ for acoustic phonon scattering due to
deformation potential \cite{opt-or,zuticrmp}.

Finally, it should be noted that {\em caution} must be taken on the
assumptions and approximations used in the above results. For example,
in intrinsic bulk GaAs at temperature below $100$~K, the
electron-electron and electron-hole scatterings dominate the momentum
scattering. The assumption that the carrier-carrier scattering can be
neglected does not hold. Therefore, the above results fail
\cite{jiang:125206}. It has been shown that the electron-electron
scattering is important for spin relaxation in high mobility
two-dimensional system where the above results also fail
\cite{zhou:045305,harleybook}.

The influence of the magnetic field on the D'yakonov-Perel' spin
relaxation comes from two factors: the Zeeman interaction and the
orbital effect.\footnote{Effect of magnetic field on the
  D'yakonov-Perel' spin relaxation was studied in the semiclassical limit in
  Refs.~\cite{Titkovbook,ivchenko.spss.15.1048,margulis:918,PhysRevB.66.233206,PhysRevB.67.155309,PhysRevB.69.245312,PhysRevB.70.195314,PhysRevB.71.235318,PhysRevB.71.195329,Glazov2007531,0295-5075-76-1-102} as well as in the quantum limit in Refs.~\cite{PhysRevB.43.14228,PhysRevB.46.4253,PhysRevLett.82.3324,dickman:jetp.lett.78.452,dikman:128,sherman:205335}.}
The Zeeman interaction leads to the Larmor spin precession and 
induces a slowdown of the relaxation for spin component parallel
to the Larmor spin precession direction \cite{Titkovbook},
\be
\tau_s(B) \simeq \tau_s(0) [1 + (\omega_{\rm L}\tau_c)^2],
\label{dp-B}
\ee
where $\omega_{\rm L}$ is the Larmor frequency. $\tau_c=\gamma_l^{-1}
\tilde{\tau}_p$ is the correlation time of the random spin precession
due to spin-orbit coupling. This result is similar to
Eq.~(\ref{srt-single}). The Larmor spin precession also mixes the
relaxation rates of the spin components perpendicular to the Larmor
spin precession direction. This may lead to considerable effect on the
spin relaxation in semiconductor nanostructures where the spin relaxation tensor
is usually anisotropic
\cite{PhysRevB.70.195314,PhysRevLett.93.147405,stich:073309}.
In non-quantizing magnetic field, the orbital effect induces the
cyclotron motion where the electron velocity (hence its ${\bf k}$)
rotates under the Lorentz force. As the spin-orbit effective magnetic
field ${\bf \Omega}({\bf k})$ is ${\bf k}$ dependent, the rotation in
${\bf k}$ leads to the rotation in the spin-orbit field. This leads to
the reduction of the D'yakonov-Perel' spin relaxation which is related
to the rotating component of ${\bf \Omega}({\bf k})$
\cite{Titkovbook,PhysRevB.69.245312}
\be
\tau_s(B) \simeq \tau_s(0) [1 + (\omega_{c}\tau_c)^2],
\label{dp-B2}
\ee
where $\omega_c=eB/m_e$ is the cyclotron frequency. Note that the
ratio of the Larmor frequency and the cyclotron frequency is 
$\omega_{\rm L}/\omega_c= |g_e| m_e / (2 m_0)$. In semiconductors, as 
$m_e$ is usually smaller than $m_0$ and often $|g_e|<2$, the
ratio is always much smaller than 1 (in GaAs, it is $\simeq
0.015$). Hence the cyclotron effect is stronger than the Zeeman
interaction \cite{Titkovbook}. The reduction of spin relaxation rate is
saturated at high magnetic field, where Landau level quantization
dominates. At such high magnetic field, electron spin
relaxes similar to that in localized states or that in quantum dot
\cite{PhysRevB.43.14228,PhysRevB.46.4253,PhysRevLett.82.3324,dickman:jetp.lett.78.452,dikman:128,sherman:205335}.

The hole D'yakonov-Perel' spin relaxation is different from the
electron one mainly in two aspects: (1) the hole spin-orbit coupling
is usually much stronger, and (2) hole has spin $J=3/2$. Due to the
strong spin-orbit coupling, the D'yakonov-Perel' mechanism is very
efficient in hole system and the spin lifetime is usually very
short. For example, hole spin lifetime in intrinsic bulk GaAs is
observed to be $\simeq 0.1$~ps \cite{PhysRevLett.89.146601}. As the
spin-orbit coupling is strong, the hole system is usually in the weak
scattering regime of the D'yakonov-Perel' spin relaxation
\cite{lue:125314,0268-1242-23-11-114017,culcer:195204}. L\"u et
al. found that in two-dimensional hole system the spin polarization
oscillates without an exponential decay in the weak scattering
regime \cite{lue:125314}. In this case, the spin lifetime is not
easily characterized. However, the incoherently summed spin coherence
$P(t)=\sum_{\bf k}|\rho_{\bf k}(t)|$ ($\rho_{\bf k}$ denotes the spin
coherence) shows good exponential decay and from its decay rate the
irreversible spin dephasing rate is obtained \cite{lue:125314}.
Another salient feature of hole spin system is that
hole has spin $J=3/2$. As a consequence the spin density matrix of
hole system is $4\times 4$ which contains spin coherence that does not
correspond to any spin polarization. Also, the hole spin-orbit
coupling may not be equivalent to an effective magnetic field as it
can be high powers of the hole spin operator ${\bf J}$ (such as
$({\bf k}\cdot{\bf J})^2$ in the Luttinger Hamiltonian
[Eq.~(\ref{sphole})]). In the work of Winkler
\cite{PhysRevB.70.125301}, the hole spin density matrix is decomposed
by the spin multipole series to provide a more systematic understanding
on the spin-dependent phenomena in hole system. Winkler found that the
hole spin-orbit coupling can induce transfer between different spin
multipoles, while the magnetic field can only induce spin precession
of the spin dipole (i.e., the spin polarization) in bulk hole system
\cite{PhysRevB.70.125301}. The transfer of the hole spin multipoles
under the hole spin-orbit coupling was found to influence the hole
spin {\it polarization} relaxation due to the D'yakonov-Perel' mechanism
\cite{culcer:195204}.

\subsubsection{Bir-Aronov-Pikus mechanism}

It was proposed by Bir, Aronov and Pikus that the electron-hole
exchange scattering can lead to efficient electron spin relaxation in
$p$-type semiconductors \cite{BAP,aronov}. According to the electron-hole exchange
interaction given in Eq.~(\ref{bapitot}), within elastic scattering
approximation, spin lifetime limited by the Bir-Aronov-Pikus mechanism
is given by the Fermi Golden rule,
\be
\frac{1}{\tau_{s}({\bf k})} = 4\pi\hspace{-3pt} \sum_{{\bf
    q},{\bf k}^{\prime}\atop m,m^{\prime}}\hspace{-3pt}
\delta(\varepsilon_{{\bf k}} +\varepsilon^{h}_{{\bf k}^{\prime}
  m^{\prime}}-\varepsilon_{{\bf k}-{\bf q}} - \varepsilon^{h}_{{\bf
    k}^{\prime}+{\bf q}m}) |{\cal{J}}^{(+)\ {\bf k}^{\prime}
  m^{\prime}}_{{\bf k}^{\prime}+{\bf q}m}|^2 f^{h}_{{\bf k}^{\prime}
  m^{\prime}} (1 - f^{h}_{{\bf k}^{\prime}+{\bf q}m}).
\label{BAPes}
\ee
Here $\varepsilon^{h}_{{\bf k}m}$ is hole energy with spin index $m$,
${\cal{J}}$ is the electron-hole exchange interaction matrix element
[see Eq.~(\ref{bapitot}) in Sec.~2.9] and $f_h$ is hole distribution
function. As hole spin and momentum relax very fast, $f_h$ is taken as
hole equilibrium distribution. In bulk semiconductors, there are several
facts which could help to reduce the above complicated equation: (i)
the heavy-hole density of states is much larger than the light-hole
one (thanks to the much larger heavy-hole effective mass), which makes the
contribution to the spin relaxation rate mainly from the
heavy hole; (ii) the heavy-hole effective mass is much larger than the
electron one, which enables the elastic scattering approximation. Using
these approximations and including only the short-range exchange interaction,
one obtains a simple result for nondegenerate hole system \cite{Titkovbook}
\be
\tau_s^{-1} = \frac{2}{\tau_0} n_h a_{\rm B}^3 \frac{\langle v_{\bf
  k}\rangle}{v_{\rm B}}, 
\label{bapapp1}
\ee
where $a_{\rm B}$ is the exciton Bohr radius, $1/\tau_0=(3\pi/64)\Delta
E_{\rm SR}^2 / ( E_{\rm B})$ with $E_{\rm B}$ being the exciton
Bohr energy, $n_h$ is the hole density, $\langle v_{\bf k}\rangle =
\langle  k/m_e\rangle$ is the average electron velocity, and
$v_{\rm B}=1/(m_R a_{\rm B})$ with $m_R\approx m_e$ being the
reduced mass of the interacting electron-hole pair. In the presence of
localized holes, the equation is improved to be \cite{Titkovbook}
\be
\tau_s^{-1} = \frac{2}{\tau_0} N_A a_{\rm B}^3 \frac{\langle v_{\bf
  k}\rangle}{v_{\rm B}} \left(\frac{n_h}{N_A} + \frac{5}{3}\frac{N_A-n_h}{N_A}\right),
\ee
where $N_A$ is the acceptor density. For degenerate hole system \cite{Titkovbook},
\be
\tau_s^{-1} \simeq \frac{3}{\tau_0} n_h a_{\rm B}^3 \frac{\langle v_{\bf
  k}\rangle}{v_{\rm B}} \frac{k_{\rm B} T}{E_{\rm F}^{h}}, 
\label{bapapp2}
\ee
with $E_{\rm F}^{h}$ denoting the hole Fermi energy. In an interacting
electron-hole plasma, electron and hole attract each other and there
is an enhancement of the electron-hole exchange interaction due to this
attraction. This enhancement is described by the Sommerfeld
factor. For unscreened Coulomb potential, the Sommerfeld factor is
$|\psi(0)|^2=\frac{2\pi}{\varepsilon_{\bf k}/E_{\rm
    B}}\big/[1-\exp(-\frac{2\pi}{\varepsilon_{\bf k}/E_{\rm B}})]$.
With this factor, spin relaxation is enhanced,
$1/\tau_s^{\prime} = |\psi(0)|^4 /\tau_s$. However,
for a completely screened Coulomb potential, there
is no enhancement, $|\psi(0)|^2=1$. The effect of the Sommerfeld
factor was discussed in Refs.~\cite{PhysRevB.54.1967,Titkovbook}.
It should be noted that the long-range electron-hole exchange
interaction can not be neglected, often (such as in GaAs) it is more
important than the short-range one \cite{jiang:125206}. Hence the
above analytical formulae is quite {\em limited}, unless $E_{\rm SR}$ is
substituted by some proper average of the whole (both the short-range
and the long-range) electron-hole exchange interaction. The analytical
expressions for the Bir-Aronov-Pikus spin relaxation in bulk materials
and in quantum wells with both short-range and long-range interactions
are given in Ref.~\cite{PhysRevB.54.1967}.

Spin relaxation rate due to the Bir-Aronov-Pikus mechanism is
usually calculated via Eq.~(\ref{BAPes}), which actually implies the
elastic scattering approximation \cite{PhysRevB.54.1967,PhysRevB.47.15776,PhysRevB.54.1967,PhysRevB.55.13771}. Recently
Zhou and Wu \cite{zhou:075318} reinvestigated the Bir-Aronov-Pikus
spin relaxation without such approximation from
the fully microscopic kinetic spin
Bloch equation approach 
\cite{PhysRevB.61.2945,wu:epjb.18.373,PhysRevB.68.075312}. They found
that spin relaxation rate was largely overestimated
at low temperature in the previous theories. The underlying physics is that at low temperature
the Pauli blocking impedes spin-flip in fully occupied states
\cite{zhou:075318}. The spin-flip is only allowed around the chemical
potential, which suppresses the Bir-Aronov-Pikus
spin relaxation at low
temperature. Amo et al. found similar arguments and results \cite{amo:085202}.

Finally, holes also suffer the Bir-Aronov-Pikus
spin relaxation. However, as hole spin-orbit coupling is strong, the
D'yakonov-Perel' mechanism is very efficient and the Bir-Aronov-Pikus
mechanism rarely shows up. The Bir-Aronov-Pikus mechanism may
become important for holes in heavily $n$-doped quantum wells, where the
D'yakonov-Perel' mechanism is suppressed by the hole-electron and
hole-impurity scatterings and the spin lifetime can be rather long
($\gtrsim 500$~ps) \cite{Baylac199557,PhysRevB.46.7292,0268-1242-13-8-004}.

\subsubsection{$g$-tensor inhomogeneity}

Under a given magnetic field, spin precession (both direction and
frequency) is determined by the $g$-tensor. Margulis and Margulis
first proposed a spin relaxation mechanism in the presence of
magnetic field due to the momentum-dependent $g$-factor
\cite{margulis:918}. Later Wu and Ning also pointed out that
energy-dependent $g$-factor gives rise to an inhomogeneous
broadening \cite{wu:epjb.18.373}
in spin precession and hence results in spin relaxation. Without 
scattering the inhomogeneous broadening only leads to a reversible spin
relaxation. Any scattering, including the electron-electron scattering, 
which randomizes spin precession, results in irreversible spin relaxation
\cite{wu:epjb.18.373}. In the irreversible regime, under a magnetic field 
${\bf B}$ along $j$ direction, the relaxation rate of the spin
component along the $i$ direction [$i,j=x,y,z$] can be obtained in
analogy with Eq.~(\ref{srt-single}), 
\be
\frac{1}{\tau_{s,{ii}}} = (\mu_{\rm B}B)^2 \sum_{l\ne i} 
\frac{(\overline{{g}_{lj}^2}-\overline{{g}_{lj}}^2)\tau_c}{1+
  (\mu_{\rm B}B)^2\sum_{n\ne
    l}\overline{{g}_{nj}}^2\tau_c^2 }.
\ee
Here $l,n=x,y,z$, $g_{ij}$ is the $g$-tensor and $\tau_c$ is
the correlation time of spin precession limited by
scattering. $\overline{{g}_{lj}}$ denotes the ensemble average of
${g}_{lj}$. The above equation holds only in the motional narrowing
regime, i.e.,\\ $(\mu_{\rm B}B) \sqrt{\sum_{l\ne i} (\overline{{g}_{lj}^2}-\overline{{g}_{lj}}^2) }~\tau_c\ll 1$ for any $i$. In III-V or II-VI semiconductors and their
nanostructures, the electron $g$-tensor is ${\bf k}$-dependent
\cite{0370-1328-89-2-326,Golubev:61.1214}, hence $\tau_c$ is limited
by momentum scattering ($\tau_c\simeq \tau_p$)
\cite{margulis:918,PhysRevB.66.233206}.\footnote{If the $g$-tensor is
  only energy dependent, then $\tau_c$ is only limited
  by inelastic scattering.} In the case of isotropic $g$-tensor, the
$g$-tensor inhomogeneity only leads to spin relaxation transverse
to the magnetic field. For example, in III-V or II-VI bulk
semiconductors, the ${\bf k}$-dependent $g$-factor limited electron
spin dephasing time is
\be
T^{-1}_2 \simeq (\mu_{\rm B}B)^2
(\overline{g^2}-\overline{g}^2) \tau_p.
\ee
It is noted that the induced spin dephasing rate increases with
increasing magnetic field. The $g$-tensor inhomogeneity mechanism
  usually dominates at high magnetic field \cite{PhysRevB.66.233206}.

Another limit  is that there
is no scattering or the scattering is very weak. In this case, the
spin lifetime is limited by free induction decay.\footnote{Out of the strong scattering and no scattering limits, the
  spin lifetime is limited by both the free induction decay time and
  the irreversible spin relaxation time.} In this regime, if the
inhomogeneity of the $g$-tensor can be characterized by a Gaussian
distribution, the coherent spin precession leads to a Gaussian decay
$\sim \exp(-t^2/\tau_s^2)$ with the spin lifetime being
\be
\frac{1}{\tau_{s,ii}} \simeq \frac{\mu_{\rm B}B}{\sqrt{2}} 
\sqrt{\sum_{l\ne i}(\overline{{g}_{lj}^2}-\overline{{g}_{lj}}^2)}.
\ee
In self-assembled quantum dot ensemble, the $g$-tensor
inhomogeneity mechanism leads to
efficient spin relaxation, masking the intrinsic irreversible one
\cite{A.Greilich07212006,PhysRevB.66.125307,PhysRevB.59.R10421}.
However, such spin relaxation can be removed by spin echo or
mode-locking techniques in experiments \cite{A.Greilich07212006}. The
$g$-tensor inhomogeneity mechanism is also important for the spin relaxation of localized
electrons in $n$-type quantum wells \cite{chen:nphys537} and localized
holes in $p$-type quantum wells
\cite{PhysRevB.66.113302,syperek:187401}.\footnote{The ${\bf
    k}$-dependent hole $g$-factor in [001] quantum wells is given by
  Eq.~(\ref{hzv001qw}).}

\subsubsection{Hyperfine interaction}

Recalling that the hyperfine interaction between electron spin and
nuclear spins is
\be
H_{\rm hf} = \sum_{i,\nu} A_{\nu} v_0 |\psi({\bf R}_i)|^2 {\bf
  S}\cdot{\bf I}_{i,\nu} = {\bf h}\cdot {\bf S}.
\ee
Here $\nu$ labels both the atomic site in the Wigner-Seitz cell and
the isotopes, while $i$ is the index of the cell. $A_{\nu}$ is the
coupling constant of the hyperfine interaction. $v_0$ is the volume of the
Wigner-Seitz cell and $N_0=1/v_0$. $\psi({\bf r})$ is the envelope
function of the electron wavefunction. ${\bf h}$ is the effective
hyperfine field acting on electron spin, which is also called
the Overhauser field. The hyperfine interaction exchanges electron spin
with nuclear spin and hence leads to electron spin relaxation.

There are different scenarios of the hyperfine interaction induced
electron spin relaxation acting in several regimes. There are four
regimes: (i) the non-interacting spin ensemble without spin echo
\cite{J.R.Petta09302005}; (ii) the non-interacting spin ensemble with
spin echo \cite{J.R.Petta09302005}; (iii) the hopping regime where
electrons at different local sites are weakly connected; and (iv) the
metallic regime where most electrons are in extended states.

In regime (i), spins are separated in space or time and do not
interact with each other, i.e., they evolve independently. Examples are
spins in a number of singly charged self-assembled quantum dots
\cite{PhysRevLett.94.116601} and single
spin in a quantum dot (or two spins in a double quantum dot) measured
at different times \cite{koppens:236802,J.R.Petta09302005}. In this
regime, the spin relaxation is determined by the free induction decay
due to the random local Overhauser field. For the quantum dots ensemble,
each electron spin in a quantum dot interacts with about $10^3\sim 10^6$
nuclear spins. According to central limit theorem, the distribution of
the Overhauser field is Gaussian
\be
P({\bf h}) = \frac{1}{(2\pi \sigma_h^2)^{3/2}} \exp(-{\bf h}^2/2\sigma_h^2).
\ee
The variance $\sigma_h$ is given by
\be
\sigma_h = 
\sqrt{ \frac{1}{3} \sum_{\nu} A^2_{\nu} I_{\nu}(I_{\nu}+1) v_0 \int
 d^3{\bf r} |\psi({\bf r})|^4 } = h_1/ \sqrt{N_L},
\label{sigma_h}
\ee
where $h_1=\sqrt{\frac{2}{3}\sum_{\nu}A^2_{\nu}I_{\nu}(I_{\nu}+1)}$ 
and $N_L=2/[v_0\int d{\bf r}|\psi({\bf r})|^4]$ is the effective
number of the nuclei. Assume that the Overhauser field is
quasi-static, i.e., the Overhauser field fluctuates at the time scale
much longer than the electron spin lifetime
\cite{PhysRevB.65.205309}. Under such an approximation, the electron
spin polarization decays as $\exp(-t^2/\tau_s^2)$, where the
spin lifetime is
\be
\tau_s^{-1} = \sqrt{2}\sigma_h.
\label{hypefine1}
\ee
For $N_{L}\sim 10^5$, calculation gives a short spin lifetime on the
time scale of $10$~ns ($1$~$\mu$s) in GaAs (silicon) quantum
dots.\footnote{This timescale is much shorter than both the time scale
  of the nuclear spin precession under the hyperfine field generated
  by electron $\sim 1$~$\mu$s ($\sim 100$~$\mu$s) and
  nuclear spin-flip time due to the nuclear spin dipole-dipole
  interaction $\sim 100$~$\mu$s ($\sim 6000$~$\mu$s) in GaAs (silicon)
  \cite{PhysRevB.15.5780,hayashi:153201}, which justifies the
  quasi-static treatment of the Overhauser field \cite{PhysRevB.65.205309}.}
Electron spin relaxation under external (static or time-dependent)
magnetic field can be different from the zero field case, but is still
well described by averaging over spin precession under the coaction of
the external and Overhauser field with the distribution $P({\bf h})$
\cite{PhysRevB.65.205309,koppens:106803,fischer:155329}.\footnote{The
  magnetic field may also change the electronic wavefunction 
  and hence the spin relaxation as $N_L$ depends on the wavefunction
  \cite{fischer:155329}.}
It was proposed that high nuclear spin polarization
\cite{PhysRevLett.91.017402} or state-narrowing of
the nuclear distribution \cite{stepanenko:136401,klauser:205302} can
markedly narrow the distribution of the Overhauser field and
suppress the spin relaxation. Recent experiment in double quantum dot
demonstrated the enhancement of the electron spin lifetime by a factor
of $\sim$70 through polarizing the nuclear spins
\cite{D.J.Reilly08082008}.

In regime (ii), the inhomogeneous broadening of the Overhauser field
is removed by spin-echo. The spin relaxation is then caused by the
temporal fluctuation of the Overhauser field. The fluctuation is
caused by the nuclear spin dynamics due to the hyperfine interaction
and nuclear spin dipole-dipole interaction. As nuclear spin relaxes
much slower than electron spin, the electron spin dynamics due to the
hyperfine interaction is non-Markovian. Furthermore, the fluctuation
in nuclear spin system is a many-body problem, where the collective
excitation is not obvious. These factors make it difficult to
calculate the spin lifetime limited by the hyperfine
interaction. Nevertheless, recent developments attacked the problem
via the equation of motion approach
\cite{PhysRevB.67.195329,PhysRevB.70.195340,jiang:035323,coish:125329,deng:245301},
Green function 
approach \cite{PhysRevLett.88.186802,PhysRevB.67.195329,deng:241303} and the quantum cluster expansion approach
\cite{PhysRevB.68.115322,PhysRevB.72.161306,witzel:035322,yao:195301,saikin:125314,yang:085302,yang:085315,cywinski:057601}.
It was predicted theoretically that via designed magnetic pulses, the
electron spin coherence lost through the hyperfine interaction can be
restored
\cite{1367-2630-9-7-226,yang-2008-101,witzel:241303,PhysRevB.72.045330,lee:160505,witzel:077601,yao:077602,zhang:201302,PhysRevB.79.245314}.
Recent experiment confirmed the theoretical prediction \cite{du.nature.461.1265}.
Other experimental investigations in regime (ii) involve: the study
of the decay of spin echo of a single spin in a quantum dot
\cite{koppens:236802}, two spins in a double quantum dot
\cite{J.R.Petta09302005,reilly:236803} and spins of donor bound
electrons in silicon \cite{0953-8984-18-21-S06}; the development of
an optical mode-locking method to study the irreversible electron spin
dephasing in self-assembled quantum dots \cite{A.Greilich07212006,hernandez:041303}.

Regime (iii) is the hopping regime where electrons at different local
sites are weakly connected. The hopping between the local sites
induces the exchange interaction between electrons. The exchange
interaction interrupts the spin precession induced by the random
local Overhauser field. If the exchange interaction is weak, the
electron spin ensemble suffers both the free induction decay and the
irreversible spin decay due to the randomization caused by the
exchange interaction. If the exchange interaction is strong, the spin
precession around the random local Overhauser field is motional
narrowed
\cite{dp:jetp65.362,PhysRevB.66.245204,0268-1242-23-11-114009,PhysRevB.64.075305}.
An example of regime (iii) is the localized electron spin ensemble in the
insulating phase of a $n$-doped semiconductor.

The exchange interaction between electrons limits the
correlation time $\tau_c$ of the spin precession due to the random
Overhauser field. In the following, we focus on the motional narrowing
limit, which is also the focus in the literature. In this limit,
$\langle \omega_{\rm hf}\tau_c\rangle \ll 1$ ($\langle ...\rangle$
denoting the electron ensemble average) with $\omega_{\rm hf}$ being the spin
precession frequency due to the hyperfine interaction. In this regime,
the spin lifetime is given by \cite{dp:jetp65.362,opt-or} 
\be
\tau_s^{-1} = \frac{2}{3}\langle \omega_{\rm hf}^2 \rangle \tau_c.
\label{hyperfine2}
\ee
Here the prefactor $\frac{2}{3}$ is due to the facts that only
transverse fluctuation can lead to spin relaxation and that the
angular distribution of the fluctuation field is uniform. Roughly
$\tau_c\simeq 1/\langle J_{ij}\rangle$, where $\langle
J_{ij}\rangle$ is the averaged exchange coupling strength with
$i$ and $j$ being the site indices
\cite{PhysRevB.66.245204,0268-1242-23-11-114009}.

In regime (iv), the metallic regime, most of the electrons are
extended and ${\bf k}$ is a good quantum number. In this regime, the
spin relaxation due to the hyperfine interaction is described as a
spin-flip scattering. For example, in bulk semiconductors, the spin
lifetime due to the hyperfine interaction is given by \cite{PhysRevB.16.820}, 
\be
\tau_s^{-1}=\langle\frac{1}{\tau_s(\varepsilon_{\bf k})}\rangle =
\frac{m_e\langle k\rangle}{3\pi}\nu_0
\sum_{\nu} \beta_{\nu} A_{\nu}^2 I_{\nu}(I_{\nu}+1).
\label{hyperfine3}
\ee
Here $\langle k\rangle = \sum_{\bf k} k f_{\bf k}(1-f_{\bf
  k})/\sum_{\bf k} f_{\bf k}(1-f_{\bf k})$ with $f_{\bf k}$ the
Fermi distribution. Calculation by Fishman and Lampel
\cite{PhysRevB.16.820} in bulk GaAs shows that the spin lifetime
is on the order of $10^{3}\sim 10^{4}$~ns in the metallic regime,
whereas the measured spin lifetime is less than 300~ns
\cite{PhysRevB.66.245204,zuticrmp}. Therefore the hyperfine
interaction mechanism is irrelevant to the spin relaxation in the
metallic regime in bulk GaAs.

Hole spin relaxation due to the hyperfine interaction attracted
a lot of interest recently
\cite{fischer:155329,fischer0903.0527,eble:146601,2009arXiv0903.3874T,2009arXiv0905.1586K,2009arXiv0905.1743C}.
For a long time, it was believed that holes interact weakly with
nuclear spin and the hyperfine interaction is negligible to hole spin
relaxation \cite{opt-or}. However, recent calculations indicated that
the hole-nucleus hyperfine interaction is only one order of magnitude
smaller than the electron one \cite{fischer:155329}, which still
provides considerable effects on hole spin relaxation
\cite{fischer:155329}. In principle, the above discussions on electron
spin relaxation due to the hyperfine interaction are also applicable
to hole system. Differently, the heavy-hole hyperfine interaction in
quantum wells is of the Ising form $H^h_{\rm hf} = \sum_{i,\nu}
A^h_{\nu} v_0 |\psi({\bf R}_i)|^2 S^z I^z_{i,\nu}$ ($z$
axis is along the growth direction). Consequently, the hole spin
relaxation due to the hyperfine interaction is different from the
electron one: it is anisotropic \cite{fischer:155329}. Further studies
on the hole spin relaxation due to the hyperfine interaction,
especially in quantum dots where hole spin is a promising candidate
for qubits \cite{heiss:241306,nature:451.441}, are required.

\subsubsection{Anisotropic exchange interaction}

The anisotropic exchange interaction is an efficient spin
relaxation mechanism in the insulating phase of doped
semiconductors
\cite{PhysRevB.69.075302,0268-1242-23-11-114009,PhysRevB.64.075305,PhysRevB.70.113201,PhysRevB.70.113201,PhysRevB.66.245204,tamborenea:085209,PhysRevB.67.033203,2009arXiv0908.2961}.
The interaction comes from a correction to the Heisenberg exchange
interaction between carriers bound to adjacent dopants due to the
spin-orbit coupling. It can be written as
\cite{0268-1242-23-11-114009,PhysRevB.69.075302}, 
\be
H_{\rm anex} \approx - J_{ij}\Big[\sin(\gamma_{ij})\frac{\vec{\gamma}_{ij}}{\gamma_{ij}}\cdot({\bf S}_i\times{\bf
  S}_j) +
(1-\cos\gamma_{ij})\big(\frac{\vec{\gamma}_{ij}}{\gamma_{ij}}\cdot{\bf
  S}_i\big)\big(\frac{\vec{\gamma}_{ij}}{\gamma_{ij}}\cdot{\bf S}_j\big)\Big],
\label{aniso-ex}
\ee
in which $J_{ij}$ is the isotropic exchange constant.
$\vec{\gamma}_{ij}$ is a vector due to
spin-orbit coupling. It is estimated as $\vec{\gamma}_{ij} \approx m_e{\bf \Omega}_{\rm SO} 
(\sqrt{2m_eE_B}\frac{{\bf r}_{ij}}{r_{ij}})r_{ij}/\sqrt{2m_eE_B}$, 
where ${\bf \Omega}_{\rm SO}({\bf k})$ is the spin precession frequency,
$E_B>0$ is the electron binding energy and ${\bf r}_{ij}$
is the vector connecting the positions of site $i$ and $j$. The first
term is the Dzhyaloshinskii-Moriya interaction and the second one is the
scalar exchange interaction. The scalar exchange interaction does not
conserve the spin perpendicular to $\vec{\gamma}_{ij}$,
whereas the Dzhyaloshinskii-Moriya interaction does not conserve spin
along any direction. Consequently the anisotropic
exchange interaction leads to decay of total carrier spin
polarization \cite{PhysRevB.69.075302}. As
$\gamma_{ij}$ is usually very small in insulating phase, $\sin
(\gamma_{ij}) \approx \gamma_{ij}$ and $1-\cos(\gamma_{ij}) \approx
\frac{1}{2} \gamma_{ij}^2$. One should note that $\gamma_{ij}$ is
proportional to the strength of the spin-orbit coupling.

In insulating phase, electron spin precesses around the anisotropic exchange
field produced by a randomly oriented adjacent electron spin, which
results in a random-walk-like spin precession. The correlation time of
spin precession, is limited by hopping time as well as the isotropic
exchange interaction with adjacent electrons. The latter perturbs the
spin precession of individual electron but conserves the total spin
polarization. Studies indicate that the latter is much more efficient
than the former in GaAs \cite{0268-1242-23-11-114009}. Hence an
estimation of the correlation time is $\tau_c \approx 1/\langle
J_{ij} \rangle$, and the spin precession frequency is $\Omega \approx
\langle J_{ij} \gamma_{ij}\rangle$. Therefore spin relaxation
time reads \cite{0268-1242-23-11-114009}
\be
\tau_s^{-1} \approx \frac{2}{3} \langle J_{ij}\rangle
  \langle\gamma_{ij}\rangle^2.
\label{tau_aniso_ex}
\ee
Here the prefactor $\frac{2}{3}$ is due to the facts that only transverse
fluctuation can lead to spin relaxation and that the angular
distribution of the random exchange field is uniform (as the adjacent
electron spins orient randomly).  It should be noted that the spin
relaxation rate, which $\propto \langle\gamma_{ij}\rangle^2$, is
proportional to the square of the strength of the spin-orbit coupling.

The anisotropic exchange interaction mechanism has also been reviewed by
Kavokin in Ref.~\cite{0268-1242-23-11-114009}, where many aspects of
spin relaxation of localized electrons were discussed. However, we note
that there was no theoretical study on the anisotropic exchange
interaction for holes in insulating phase up till now. As the
spin-orbit coupling is stronger, one would expect that the anisotropic
exchange interaction is more efficient in hole system. Recently there
are a few experimental studies
\cite{syperek:187401,yugova:167402,2009arXiv0909.3711K} on hole
spin relaxation in the insulating phase in two-dimensional hole
system. Theoretical investigation on the role of the anisotropic
exchange interaction in hole spin
relaxation under experimental conditions is desired.

\subsubsection{Exchange interaction with magnetic impurities}

In magnetic semiconductors, the exchange interaction with magnetic
impurities can be an important source of the spin relaxation. In
(II,Mn)VI magnetic semiconductors, the $s$-$d$ exchange interaction
has been recognized to be responsible for electron spin relaxation
\cite{PhysRevLett.75.505,PhysRevB.41.7899,PhysRevB.64.085331,ronnburg:117203}.
Recently, it was found that in $p$-type paramagnetic GaMnAs quantum
wells, the $s$-$d$ exchange interaction dominates electron spin
relaxation at low temperature \cite{jiang:155201}.
In general, the spin relaxation due to the $s$-$d$ ($p$-$d$)
exchange interaction is similar to that due to the hyperfine
interaction. There can be several regimes, where different
scenarios take place. However, there are only two regimes which
are commonly encountered: the insulating regime where most carriers
are localized and the metallic regime where most carriers are
extended. In the insulating regime, the exchange interaction with
randomly oriented magnetic impurity spins induces random spin
precessions. There are several processes interrupt such random spin
precessions: (i) exchange interaction between carriers; (ii) carrier
hopping; (iii) fluctuation or diffusion of the magnetic impurity
spins. These processes limit the correlation time $\tau_c$ of the
random spin precession. If $\tau_c$ is small, so that 
$\langle{\bf \Omega}_{s(p)\mbox{-}d}\rangle\tau_c\ll 1$ 
(${\bf \Omega}_{s(p)\mbox{-}d}$ is the spin precession frequency due to
the $s(p)$-$d$ exchange interaction and $\langle...\rangle$ represents
the ensemble average). In this regime, the spin relaxation
rate reads
\be
\tau_s^{-1} = \frac{2}{3} \langle\Omega_{s(p)\mbox{-}d}^2\rangle\tau_c.
\ee
In metallic regime, spin relaxation rate is given by the Fermi Golden
rule. For example, in bulk system, the electron spin relaxation rate
is \cite{J.Kossut.pssb,PhysRevB.39.10918}
\be
\tau_s^{-1} = \frac{m_e\langle k\rangle}{3\pi} J_{\rm sd}^2
N_M S_M(S_M+1),
\ee
where $J_{\rm sd}$ is the $s$-$d$ exchange constant, $N_M$ is the
density of magnetic ions with spin $S_M$, $\langle k\rangle =
\sum_{\bf k} k f_{\bf k}(1-f_{\bf k})/\sum_{\bf k} f_{\bf
  k}(1-f_{\bf k})$ with $f_{\bf k}$ being the Fermi distribution. Hole
spin relaxation rate due to the $p$-$d$ exchange interaction can also
be calculated via the Fermi Golden rule, only that the spectrum and
the eigen-spinor are more complicated. In principle, the same scheme
is applicable to electrons (holes) in nanostructures
\cite{PhysRevB.41.7899}.

\subsubsection{Other spin relaxation mechanisms}

The main remaining spin relaxation mechanisms are various spin-phonon
interactions (except those which have been grouped into the Elliott-Yafet
mechanism). Generally the origin of the spin-phonon interactions is
that when spin interactions depend on the positions of atomic ions
(both host and dopant), lattice vibration couples directly with
spin. There are several such kinds of spin interactions, such as
the strain-induced spin-orbit coupling and the hyperfine interaction. The
strain also modifies $g$-tensor \cite{PhysRev.118.1534}. All such
lattice dependent spin interactions can be the origin of the spin-phonon
interactions
\cite{PhysRev.118.1534,PhysRevB.64.125316,PhysRevB.66.155327,PhysRevB.70.085305,romano:033301,PhysRevLett.95.166603}.
In nanostructure, the gate-voltage and structure induced modification
of spin-orbit coupling \cite{winklerbook} and $g$-tensor
\cite{winklerbook,doty:prl197202} further induce more spin-phonon
interactions \cite{jiang:035323,romano:033301}. Besides, spin mixing
associated with spin-independent phonon scattering also leads to
the spin-flip phonon scattering.
 For example, for localized electrons
which are bound to impurities or confined in quantum dots, spin-orbit
coupling and hyperfine interaction can induce spin mixing and enables
spin-flip electron-phonon scattering
\cite{PhysRevB.61.12639,jiang:035323}.

In existing literature, there are a lot of studies on the spin-phonon
interaction induced spin relaxation. Most of them focus on spin
relaxation of donor-bound electrons
\cite{PhysRevB.70.113201,PhysRev.118.1534,PhysRevB.66.035314,PhysRev.155.816,PhysRevB.65.245213,Guseinov200765}
and electrons/holes in quantum dots
\cite{jiang:035323,PhysRevB.66.161318,romano:033301,PhysRevLett.95.166603,PhysRevB.69.125330,meza-montes:205307,0953-8984-19-44-445007,Roszak:0903.0783,weng:155309,PhysRevLett.93.016601,PhysRevLett.92.026601,PhysRevB.70.075313,PhysRevLett.95.076603,PhysRevB.70.085305,PhysRevB.69.115318,PhysRevB.72.115326,PhysRevB.64.125316,PhysRevB.66.155327,PhysRevB.68.155330,PhysRevB.70.195320,PhysRevB.71.205324,PhysRevB.61.12639,stano:045320,Khaetskii2000470,stano:186602,wang:165312,jiang:035307,shen:235313,wang:125323,PhysRevB.66.035314,PhysRevB.68.045308,semenov:195342,trif:045434,golovach:045328,borhani:155311,lehmann:045328,PhysRevB.71.075308,PhysRevLett.95.076805}.
A comprehensive study of electron spin relaxation/dephasing due to
various spin-phonon interactions in GaAs quantum dots is presented in
Ref.~\cite{jiang:035323}. The spin-phonon scatterings are usually
limited by the electron-phonon scattering rate and the effect of the
spin-orbit interaction. It was believed that they are unimportant in
metallic regime. The relative importance of different spin-phonon
interactions are compared for various conditions in GaAs quantum dots
\cite{jiang:035323} and in Si/Ge quantum dots
\cite{semenov:195342}. Among those spin-phonon interactions, the
phonon induced $g$-tensor fluctuation was found to be {\em irrelevant} in
GaAs quantum dots \cite{jiang:035323}, whereas it is {\em important} in
Si/Ge quantum dots \cite{semenov:195342}. Finally, the gate noise can also
lead to spin relaxation in gated quantum dots in the presence of the
hyperfine interaction, spin-orbit interaction or electron-electron
exchange interaction \cite{PhysRevB.71.165325,hu:100501}.

\subsubsection{Relative efficiency of spin relaxation mechanisms in
  bulk semiconductors}

In this subsection, we briefly discuss and review the relative
efficiency of various spin relaxation mechanisms. We give some
rough discussions based on the analytical formulae which are based on
some approximations.\footnote{Besides competition, there are
  also some cooperation between different spin relaxation
  mechanisms. For example, spin-flip momentum scatterings due to the
  Elliott-Yafet, the Bir-Aronov-Pikus, the $s(p)$-$d$ exchange interaction and
  other mechanisms also lead to the randomization of momentum,
  they thus contribute to the D'yakonov-Perel' spin
  relaxation as well. However, as these spin-flip momentum scatterings are
  usually much weaker than the spin-conserving ones, they only lead to
  a small modification of the D'yakonov-Perel' spin relaxation
  \cite{zhou:075318,jiang:155201}.} For simplicity, we restrict the
discussion here on the bulk III-V semiconductors. The situation in
III-V semiconductor nanostructures or II-VI semiconductors and their
nanostructures may be different. Nevertheless, some insight can be
gained from these discussions. We also comment on hole spin
relaxation and carrier spin relaxation in centrosymmetric
semiconductors such as silicon and germanium.

For spin relaxation in bulk III-V semiconductors in the insulating
regime, as stated before, the anisotropic exchange interaction
dominates at high doping density, whereas the hyperfine interaction
dominates at low doping density. In metallic regime, the spin
relaxation mechanisms are the D'yakonov-Perel', the Elliott-Yafet and
the Bir-Aronov-Pikus mechanisms. In $n$-doped semiconductors the
Bir-Aronov-Pikus mechanism is ineffective due to lack of holes. Hence
the relevant mechanisms are the Elliott-Yafet and the D'yakonov-Perel'
mechanisms. From Eqs.~(\ref{eyapp}) and (\ref{dpapp}), the ratio of
the spin lifetimes due to the Elliott-Yafet and the
D'yakonov-Perel' mechanisms in bulk III-V semiconductors is given by
\cite{jiang:125206,Titkovbook},
\be
\frac{\tau_{\rm EY}}{\tau_{\rm DP}} \approx A_{1}
(\frac{m_e}{m_{cv}})^2 \langle \varepsilon_{\bf k}\rangle
  \tau_p^2 E_g \frac{(1-\eta/3)}{(1-\eta/2)^2}, 
\ee
where the prefactor $A_{1}\sim 1$. It is inferred from the above
expression that the Elliott-Yafet spin relaxation is more important in
semiconductors with smaller $E_g$, such as InAs and InSb. The factor
$\langle \varepsilon_{\bf k}\rangle \tau_p^2$ indicates that the
Elliott-Yafet mechanism is more important at lower temperature with
higher impurity density.

In $p$-type and intrinsic semiconductors the relative efficiencies of
the Bir-Aronov-Pikus and the D'yakonov-Perel' mechanisms are always
compared. The Bir-Aronov-Pikus mechanism is believed to be important
for high hole density at low temperature \cite{Titkovbook,zuticrmp}. In
such case the hole system is degenerate, thus it is meaningful to
compare the D'yakonov-Perel' and the Bir-Aronov-Pikus mechanisms for
degenerate holes. From Eqs.~(\ref{dpapp}) and (\ref{bapapp2}) one
finds
\be
\frac{\tau_{\rm BAP}}{\tau_{\rm DP}} \approx A_{2}
\tau_p\tau_0\frac{\langle \varepsilon_{\bf k}\rangle^{5/2}E_{\rm B}^{1/2}}{E_g}(\frac{m_e}{m_{cv}})^2\frac{\eta^2}{1-\eta/3}\frac{1}{n_ha_{\rm B}^3}\frac{E_{\rm F}^h}{k_{\rm B}T}, 
\ee
where the prefactor $A_{2}\sim 1$ and $E_{\rm F}^h$ is hole Fermi
energy. It is seen from the above formula that the Bir-Aronov-Pikus mechanism is
more important in semiconductors with larger band gap $E_g$, smaller
spin-orbit interaction (i.e., smaller $\eta$) and stronger
electron-hole exchange interaction (i.e., smaller $\tau_0$). From the
above equation, one finds four controlling factors: $\tau_p$,
$\langle \varepsilon_{\bf k}\rangle$, $n_h$ and $T$ (Note that
$E_{\rm F}^h\sim n_h^{2/3}$). The Bir-Aronov-Pikus mechanism is more important
at higher hole/impurity density and lower electron density. For nondegenerate 
electron, as $\langle \varepsilon_{\bf k}\rangle\sim k_{\rm B}T$, the
Bir-Aronov-Pikus mechanism is more important at low temperature. However, for
degenerate electron, the Bir-Aronov-Pikus mechanism is more important at high
temperature, as $\langle \varepsilon_{\bf k}\rangle\sim E_{\rm F}$.
Such qualitative different behavior for nondegenerate and
degenerate electron was illustrated in a recent study
\cite{jiang:125206} (see also Sec.~\ref{jiang-wu-bulk}).

The Elliott-Yafet mechanism can be comparable to the Bir-Aronov-Pikus mechanism in $p$-type
semiconductors. The ratio of the two spin lifetimes for
degenerate hole is 
\be
\frac{\tau_{\rm EY}}{\tau_{\rm BAP}} \approx A_{3}
\frac{\tau_p}{\tau_0}n_ha_{\rm B}^3\frac{E_g^2}{E_{\rm B}^{1/2}\langle
\varepsilon_{\bf k}\rangle^{3/2}} \frac{1}{\eta^2}
(\frac{1-\eta/3}{1-\eta/2})^2\frac{k_{\rm B}T}{E_{\rm F}^h},
\ee
where $A_{3}\sim 1$. According to the above equation, the Elliott-Yafet mechanism is
more important in semiconductors with larger $\tau_0$ (i.e., weaker
electron-hole exchange interaction), smaller band-gap $E_g$ and larger
$\eta$ (i.e., larger spin-orbit interaction). Hence the Elliott-Yafet mechanism
may exceed the Bir-Aronov-Pikus mechanism in semiconductors with small band-gap and
large spin-orbit interaction, such as InSb and InAs. For a given
material, the Elliott-Yafet mechanism is more important at stronger momentum
scattering (i.e., smaller $\tau_p$) and lower hole density. For
degenerate electrons, the Elliott-Yafet mechanism is more important at larger
electron density and lower temperature, as $\langle \varepsilon_{\bf
  k}\rangle\sim E_{\rm F}$. However, for nondegenerate electrons, the
Elliott-Yafet mechanism is more important at higher temperature, as $\langle
\varepsilon_{\bf k}\rangle\sim k_{\rm B}T$.

In brief, some understanding of the topic can be obtained from
the above
discussions. Nevertheless, as there are a lot of approximations
in these formulae, they can {\em not} give a picture of whole
parameter range. In some case the many-body effects which are absent
in the above analysis plays an important role. Moreover, the relative
efficiency relies strongly on the genuine material parameters. Only a realistic
calculation can fully resolve the problem. A systematic calculation
from the fully microscopic kinetic spin Bloch equation approach by
Jiang and Wu \cite{jiang:125206} indicated that the Elliott-Yafet mechanism is
{\em less} important than the D'yakonov-Perel' mechanism in {\em all}
parameter regime in $n$-type bulk III-V semiconductors in metallic
regime. In the same work, the Bir-Aronov-Pikus mechanism was found to
be unimportant in intrinsic samples, whereas it dominates spin
relaxation in $p$-type samples at low temperature and high hole
density when the electron 
density is low. However, the Bir-Aronov-Pikus mechanism is ineffective when the
electron density is high \cite{jiang:125206}. It was also found via
the same approach that
in intrinsic or $p$-type GaAs quantum wells when electron density is
comparable with hole density, the Bir-Aronov-Pikus mechanism is unimportant
\cite{zhou:075318}. In a recent work \cite{zhou0905.2790},
the relative importance of the Bir-Aronov-Pikus mechanism and the
D'yakonov-Perel' mechanism in $p$-type GaAs quantum wells was compared
comprehensively. Systematic comparisons of various spin relaxation
mechanisms in paramagnetic GaMnAs quantum wells was performed in
Ref.~\cite{jiang:155201} also via the kinetic spin Bloch equation approach.

Finally, in centrosymmetric semiconductors, such as silicon and germanium, the
situation is totally different. In bulk system, as the spin-orbit
coupling term in conduction band vanishes, the D'yakonov-Perel' mechanism
is absent in electron spin relaxation. Hence in $n$-type bulk silicon and
germanium, the Elliott-Yafet mechanism dominates electron spin relaxation in
metallic regime \cite{zuticrmp,Fabianbook,2009arXiv0906.4054C}. In
insulating regime, electron spin relaxation is dominated by other
mechanisms such as the hyperfine interaction and the spin-phonon interaction
\cite{PhysRev.118.1534,PhysRev.155.816,PhysRevB.68.115322,Guseinov200765}.
In nanostructures, which breaks the centro-inversion symmetry,
spin-orbit coupling term shows up
\cite{PhysRevB.66.195315,sherman:209,PhysRevB.69.115333}. Experiment
indicated that in silicon quantum wells the D'yakonov-Perel' mechanism is
dominant in high mobility regime, whereas the Elliott-Yafet mechanism is more
important in low mobility regime \cite{Wilamowski2003111}. Due to the
nature of indirect band, optical spin orientation is not easily accesible. Hence
the electron spin dynamics was studied only in $n$-type system,
whereas the hole spin dynamics in $p$-type system. In $p$-type system,
like the situation in III-V semiconductors, hole spin-orbit coupling
in the Luttinger Hamiltonian is large and the D'yakonov-Perel'
mechanism is believed to be dominant. The D'yakonov-Perel' mechanism
also dominates hole spin relaxation in nanostructures of silicon and germanium
\cite{zhang:155311}.

\section{Spin relaxation: experimental studies and single-particle theories}

The study of spin lifetime and spin diffusion length is one of
the central focuses in spintronics. A large quantity of papers can
be found in the past decades. Recent
development of spin grating measurement gives new insight into spin
diffusion \cite{PhysRevLett.76.4793,nature.437.1330,weber:076604},
while spin noise spectroscopy enables measurement of the intrinsic
carrier spin lifetime with negligible disturbance on the carrier
system \cite{nature.431.49,PhysRevLett.95.216603,romer:103903,PhysRevLett.104.036601}. Also
the latest technique of tomographic Kerr rotation achieves spin state
tomography \cite{nature.457.702}. We expect that these new
experimental techniques will inspire and enable more important
findings.

In this section
we review experimental studies and single-particle theories on spin
relaxation and dephasing times. Here, the term ``single-particle theory'' refers to the
theory where the carrier-carrier scattering is not
considered. It will be shown in next section that the carrier-carrier
 scattering plays an important role in spin relaxation. We
focus on spin relaxation in III-V and II-VI semiconductors where the
bulk inversion asymmetry induced spin-orbit coupling plays an important
role. Spin relaxation in centrosymmetric semiconductors, such as
silicon and germanium, is also reviewed. The discussions in
 the previous section will be used frequently.

\subsection{Carrier spin relaxation at low temperature in insulating regime}

Carriers in semiconductors are mostly ionized from dopants. At low
temperature, dopants can trap those carriers. When doping density is
low, carriers are bound to
isolated dopants and the system behaves as an insulator. At elevated
doping density, carrier system becomes metallic after the
metal-insulator transition at some critical density $n_c$. In metallic
regime carrier transport is band-like, whereas in insulating regime it
is dominated by carrier hopping. It is then not difficult to
understand that the relevant carrier spin relaxation mechanisms in
insulating regime are different from those in the metallic regime. In
this subsection, we review spin relaxation at low temperature in
insulating regime. Remarkably, the longest spin lifetimes in GaAs was
reported as $\tau_s\simeq 300$~ns \cite{PhysRevLett.88.256801,jetp.lett.74.182} at
zero magnetic field. At high magnetic field in lightly doped GaAs, the
spin lifetime can reach $\tau_s\simeq 19$~$\mu$s
\cite{PhysRevB.69.121307}. In high purity GaAs, the spin relaxation
time of donor-bound electrons can be as long as several milliseconds
\cite{fu:121304}.\footnote{The ensemble spin dephasing time ($T_2^{\ast}\simeq
  5$~ns \cite{PhysRevLett.95.187405}) and irreversible spin dephasing
  time ($T_2=7$~$\mu$s \cite{clark:247601}) are much shorter under the same
  condition.}

It has been identified that
spin relaxation in insulating regime is dominated by anisotropic
exchange interaction mechanism at high doping density whereas by
the hyperfine interaction mechanism at low doping density, at zero
magnetic field
\cite{PhysRevB.64.075305,PhysRevB.66.245204,0268-1242-23-11-114009}.
If $\sum_f \sqrt{\langle\omega_f^2\rangle} \tau_c\ll 1$, spin
relaxation rate is given by
\cite{opt-or,dp:jetp65.362} 
\be
\tau_s^{-1} \simeq \sum_f \frac{2}{3} \langle \omega_f^2 \rangle \tau_c,
\ee
where $\omega_f$ is the electron spin precession frequency due to
the random effective magnetic field. The index $f$ denotes the mechanism
leading to the random fields: it can be the hyperfine interaction
with the nuclei and the anisotropic exchange interaction with adjacent
electrons. $\langle ... \rangle$ represents ensemble average. $\tau_c$
is the correlation time of the random spin precessions. $\tau_c$ can be
limited by disturbance of electron spin state or the random field
$\omega_f$. Realistic calculation indicated that in GaAs $\tau_c$ is
mainly limited by the former due to the isotropic exchange interaction
\cite{PhysRevB.66.245204,0268-1242-23-11-114009}.\footnote{In fact such
  disturbance of electron spin state can be viewed as spin diffusion
  in a network coupled by the isotropic exchange interaction
  \cite{0268-1242-23-11-114009}. From the random-walk theory, $\tau_c$ is
  related to the spin diffusion constant $D_s$ as
  $\tau_c\simeq (3n_D^{2/3}D_s)^{-1}$
  \cite{0268-1242-23-11-114009}.} Therefore $\tau_c 
\simeq 1 /\langle J_{ij}\rangle$ where $J_{ij}$ is the isotropic
exchange coupling constant between $i$ and $j$ donor-bound
electrons. The average of $J_{ij}$ is taken over the donor
configuration. Spin relaxation rate due to the anisotropic exchange
interaction is then given by [see also Eq.~(\ref{tau_aniso_ex})] 
\be
\tau_s^{-1} \approx \frac{2}{3} \langle J_{ij}\rangle
  \langle\gamma_{ij}\rangle^2,
\label{tau_aniso_ex2}
\ee
where $\gamma_{ij}$ is the spin precession angle during the tunnel
hoping from $i$ to $j$ donor site (see Sec.~3.2.6). The hyperfine
interaction-limited spin lifetime is
\be
\tau_s^{-1} \simeq \frac{2}{3} \sigma_h^2 \frac{1}{\langle
  J_{ij}\rangle}, 
\label{tau_hf_ex2}
\ee
where $\sigma_h$ [given by Eq.~(\ref{sigma_h})] is the variance of the
Overhauser field. It is noted that the spin relaxation rates due to the
two mechanisms have different dependences on $\langle
J_{ij}\rangle$. Naively, it is expected that $\langle J_{ij}\rangle$
increases with decreasing inter-donor distance (i.e., increasing doping
density). It is then easy to understand that the anisotropic exchange
interaction dominates spin relaxation at high doping density, whereas
the hyperfine interaction dominates at low doping density
\cite{PhysRevB.64.075305,PhysRevB.66.245204,0268-1242-23-11-114009}.
For hydrogen-like centers, the exchange coupling constant is given by
\cite{PhysRevB.67.033203}
\be
J_{ij} \simeq 0.82 \frac{e^2}{4\pi\epsilon_0\kappa_0 a}
(\frac{r_{ij}}{a})^{5/2} \exp(-\frac{2r_{ij}}{a}),
\ee
where $a$ is the Bohr radius of the hydrogen-like wavefunction. It is
noted that the dependence on inter-donor distance $r_{ij}$ is
exponential.\footnote{As stated in the previous footnote, $\tau_c$ is
  actually limited by spin diffusion due to exchange coupling in the
  network of randomly distributed donors. In some clusters where the
  inter-donor distance is shorter than the average one, spin diffusion
  becomes much faster due to the exponential dependence. Therefore
  this kind of cluster serves as easy passages of spin diffusion. At
  low doping density spin diffusion is dominated by such kind of small
  cluster. In this case the spin diffusion should be treated with
  percolation theory \cite{0268-1242-23-11-114009}. Such kind of studies
  were performed in Refs.~\cite{jetplet.85.55,0268-1242-23-11-114009}.}

Usually both localized and free (extended) electrons coexist and they
are also coupled by exchange interactions. It was found that the
exchange coupling between the localized and free electrons is so
efficient that the two share a common spin lifetime
\cite{0268-1242-23-11-114009,PhysRevB.70.113201,mahan:075205}. The
spin relaxation rate of the whole electron system is then given by
\cite{PhysRevB.70.113201},
\be
\tau_s^{-1} = \tau_{ls}^{-1} n_l/(n_l+n_f) + \tau_{fs}^{-1}n_f/(n_l+n_f),
\ee
where $n_l$ and $n_f$ are the densities of localized and free
electrons, respectively. $\tau_{ls}$ and $\tau_{fs}$ are the spin
lifetimes of localized and free electrons, separately. In many
cases spin relaxation is dominated by localized electrons as they have
much faster spin relaxation rate
\cite{0268-1242-23-11-114009,PhysRevB.70.113201}. Besides, the exchange
interaction between free and localized electrons can also limit
$\tau_c$ and elongate the spin lifetime
\cite{PhysRevLett.88.256801,jetp.lett.74.182}. The effect of electron
localization on spin relaxation in $n$-doped quantum wells, especially in
symmetric (110) quantum wells, was discussed in a recent theoretical work \cite{2009arXiv0911.2452H}.

In the presence of magnetic field, the situation becomes much more
complicated \cite{jetp.let.88.814,0268-1242-23-11-114009}. First,
magnetic field reduces the isotropic exchange coupling (as the
wavefunction becomes more localized) and hence increases $\tau_c$
\cite{0268-1242-23-11-114009}. The anisotropic exchange interaction is
also reduced. Second, spin relaxation due to the random spin precession is
slowed down by the magnetic field parallel to spin polarization 
[see Eq.~(\ref{srt-single})]. Third, spin relaxation due to $g$-tensor
inhomogeneity increases with the magnetic field. Fourth, at high magnetic
field, spin-flip electron-phonon 
scattering may come into play \cite{PhysRevB.46.4253}. All these
factors complicate the magnetic field dependence of the spin
relaxation. Up till now, there is little discussion on the problem
\cite{0268-1242-23-11-114009,jetp.let.88.814}. Experimental results on
magnetic field dependence of spin lifetime are available in
Ref.~\cite{colton:205201}. Yet the observed magnetic field dependence
is to be explained by theory \cite{colton:205201}. Before turning to
experimental studies, it should be mentioned that until now there are
only a few theoretical works on spin relaxation in insulating regime
\cite{PhysRevB.69.075302,0268-1242-23-11-114009,PhysRevB.64.075305,PhysRevB.70.113201,PhysRevB.70.113201,PhysRevB.66.245204,tamborenea:085209,PhysRevB.67.033203,jetplet.85.55,shklovskii:193201,jetp.let.88.814,PhysRevB.72.155312},
more studies are needed. Especially hole spin relaxation in 
insulating regime was discussed only in Ref.~\cite{PhysRevB.66.113302}
where only the $g$-factor fluctuation mechanism was studied.

We now review the experimental studies on carrier spin relaxation at
low temperature in insulating regime. Many efforts have been devoted to
the spin relaxation in $n$-GaAs at low temperature
\cite{PhysRevLett.80.4313,0268-1242-17-4-302,jetp.lett.13.177,jetp.lett.74.182,PhysRevB.67.165315,PhysRevB.66.245204,PhysRevLett.95.216603,knotz:241918,Hohage:4346,crooker:035208,PhysRevLett.88.256801,colton:205201,PhysRevB.69.121307,Kavokin:1521221,2007arXiv0706.1884S,2009arXiv0911.4084R,Kikkawa2001194}.
In one of the seminal work by Kikkawa and Awschalom
  \cite{PhysRevLett.80.4313}, a surprisingly long spin lifetime
  (130~ns) was observed in $n$-GaAs, which sheds light on the
  possibility of the semiconductor spintronics. This, together with
  the observation of a very long spin diffusion length (100~$\mu$m) in
  GaAs \cite{kikkawa_99} as well as the progress on spin injection into
  semiconductors \cite{fiederling_99,ohno_99}, excited the society and
  led to the rapid rising of the field of semiconductor 
 spintronics. Interestingly,
  most of these works were done in bulk $n$-GaAs with a doping density
  of $10^{16}$~cm$^{-3}$.

\begin{figure}[bth]
  \centering
  \includegraphics[height=7cm]{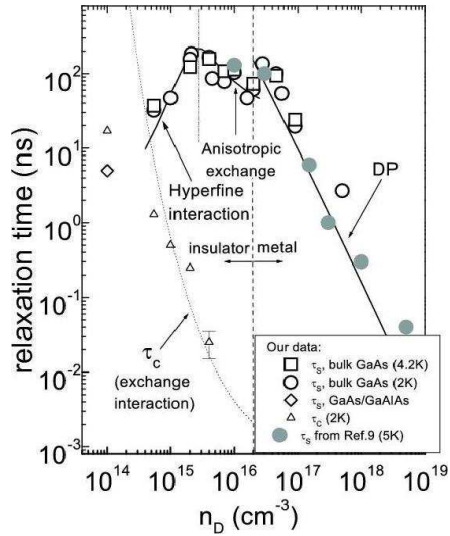}
  \caption{Spin lifetime $\tau_s$ ($\square$, $\bigcirc$, $\Diamond$ and
    $\medbullet$) and spin correlation time $\tau_c$ ($\triangle$) as
    function of donor concentration $n_{\rm D}$ in $n$-GaAs. Solid
    curves: theoretical calculation of $\tau_s$. Dotted curve:
    theoretical calculation of $\tau_c$ due to exchange interaction
    between adjacent electrons. The vertical dashed lines indicate
    the positions of the peaks of the spin lifetime. Especially, the
    dashed line at $n_{\rm D}=2\times 10^{16}$~cm$^{-3}$ also denotes
    the metal-insulator transition. The hyperfine interaction, the
    anisotropic exchange interaction and the D'yakonov-Perel'
    mechanisms dominate the spin relaxation in the low, medium, high
    doping density regimes respectively, as indicated in the
    figure. From Dzhioev et al. \cite{PhysRevB.66.245204}.}
  \label{bulk-lowTne}
\end{figure}

Dzhioev et al. \cite{PhysRevB.66.245204}  measured,
in a wide range ($10^{14}\sim10^{19}$~cm$^{-3}$), the doping density
dependence of spin lifetime at low temperature $T\le
5$~K.\footnote{Similar investigation in CdTe was performed in a recent work \cite{2010arXiv1001.0869S}.} The results
are presented in Fig.~\ref{bulk-lowTne}. Spin lifetime first increases
before reaching its first maximum at doping density $n_D \simeq 3\times
10^{15}$~cm$^{-3}$; it then decreases with doping density; around
the metal-insulator transition point at doping density $n_D=2\times
10^{16}$~cm$^{-3}$, spin lifetime increases rapidly and reaches its
second maximum; after that (i.e., in metallic regime) spin lifetime
decreases monotonically with doping density. This intricate behavior is
due to the different spin relaxation mechanisms in metallic and
insulating regimes. In metallic regime, spin relaxation is dominated
by the D'yakonov-Perel' mechanism where $\tau_s^{-1} \sim \langle
\varepsilon_{\bf k}^3\rangle \tau_p$. At such low temperature, electron
system is degenerate. Hence $\tau_p\sim n_D/k_F^3\sim n_D^0$
(according to Brooks-Herring formula) and $\langle
\varepsilon_{\bf k}^3\rangle\sim n_D^2$. Therefore $\tau_s$ decreases with
doping density $n_D$ as $\tau_s\sim n_D^{-2}$. In insulating regime,
the D'yakonov-Perel'
mechanism is irrelevant. The relevant spin relaxation mechanisms  are
the anisotropic exchange interaction (at high doping density) and the
hyperfine interaction (at low doping density). As indicated by
Eqs.~(\ref{tau_aniso_ex2}) and (\ref{tau_hf_ex2}), spin relaxation due
to these two mechanisms have different doping density dependences (as
$\langle J_{ij}\rangle$ increases with increasing doping
density). Therefore, their competition leads to the nonmonotonic
doping density dependence: spin lifetime first increases then
decreases with increasing doping density. It is noted in the figure
that the measured $\tau_c$ decreases with increasing doping density,
which agrees with the theory as $\tau_c\simeq 1/\langle J_{ij}\rangle$.

It was demonstrated by Dzhioev et al. that different spin relaxation scenarios
can be invoked by optical pumping
\cite{PhysRevLett.88.256801,jetp.lett.74.182}. At low concentration
spin relaxation is limited by the hyperfine interaction mechanism,
where the spin lifetime is $\tau_s=5$~ns. The authors showed
that by injecting additional electrons, the exchange interaction
between those electrons and the donor-bound electrons motionally
narrows the random spin precession caused by the hyperfine interaction. The
spin lifetime is then elongated to $\tau_s=300$~ns.\footnote{As the
  photo-excited electrons may affect the spin relaxation significantly
  at low temperature in insulating regime 
  \cite{PhysRevLett.88.256801,jetp.lett.74.182,0268-1242-23-11-114009},
  it is important to reexamine the spin relaxation times obtained by
  Hanle, or Faraday/Kerr measurements via the spin noise spectroscopy
  method. Such study has been done by R\"omer et
  al. in Ref.~\cite{2009arXiv0911.4084R}, where the spin relaxation
  times measured by different methods are compared in very dilute doped
  ($n_D=10^{14}$~cm$^{-3}$), low doped
  ($n_D=2.7\times10^{15}$~cm$^{-3}$), and doping close to
  metal-insulator transition ($n_D=1.8\times10^{16}$~cm$^{-3}$) samples.}

Recently, Schreiber et al. studied energy resolved electron spin
relaxation in $n$-GaAs near the metal-insulator transition
\cite{2007arXiv0706.1884S}. They found that there are three components
in the time-resolved Kerr rotation under magnetic field parallel to
the surface by examining the spin precession frequency as well as spin
lifetime. The different precession frequencies indicate that
those components should be attributed to electrons at different energy
states. The sequential emergence of the three components confirms such
hypothesis. The authors attributed these three states as: delocalized
states in donor band, localized states in the tail of donor band and
free electron states in the conduction band. The authors also found
that absolute value of the $g$-factor in localized states in the donor
band tails is smaller than those in the other two states.

\begin{figure}[bth]
  \begin{minipage}[h]{0.5\linewidth}
    \centering
    \includegraphics[height=8cm]{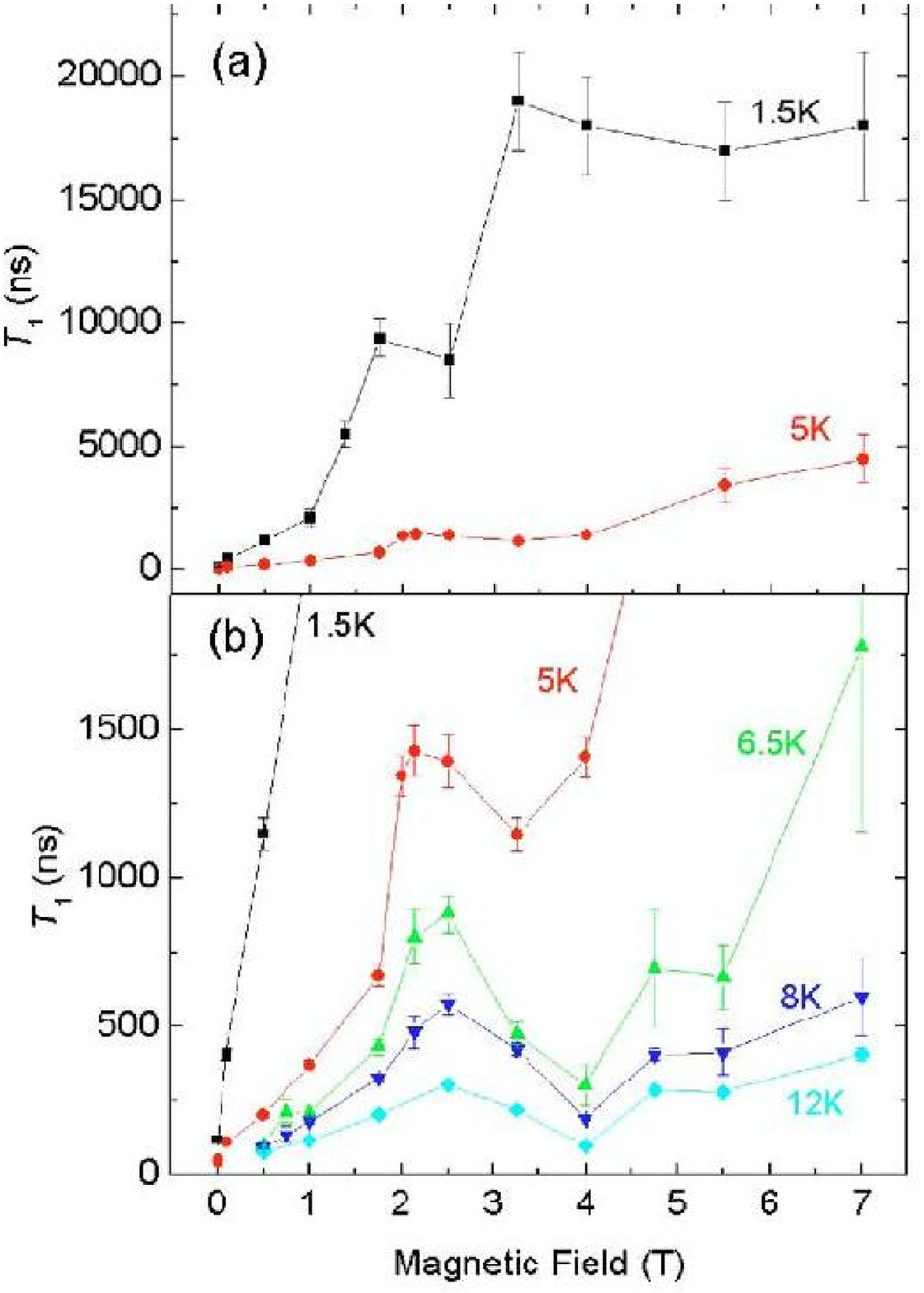}
  \end{minipage}\hfill
  \begin{minipage}[h]{0.5\linewidth}
    \centering
    \includegraphics[height=5.4cm]{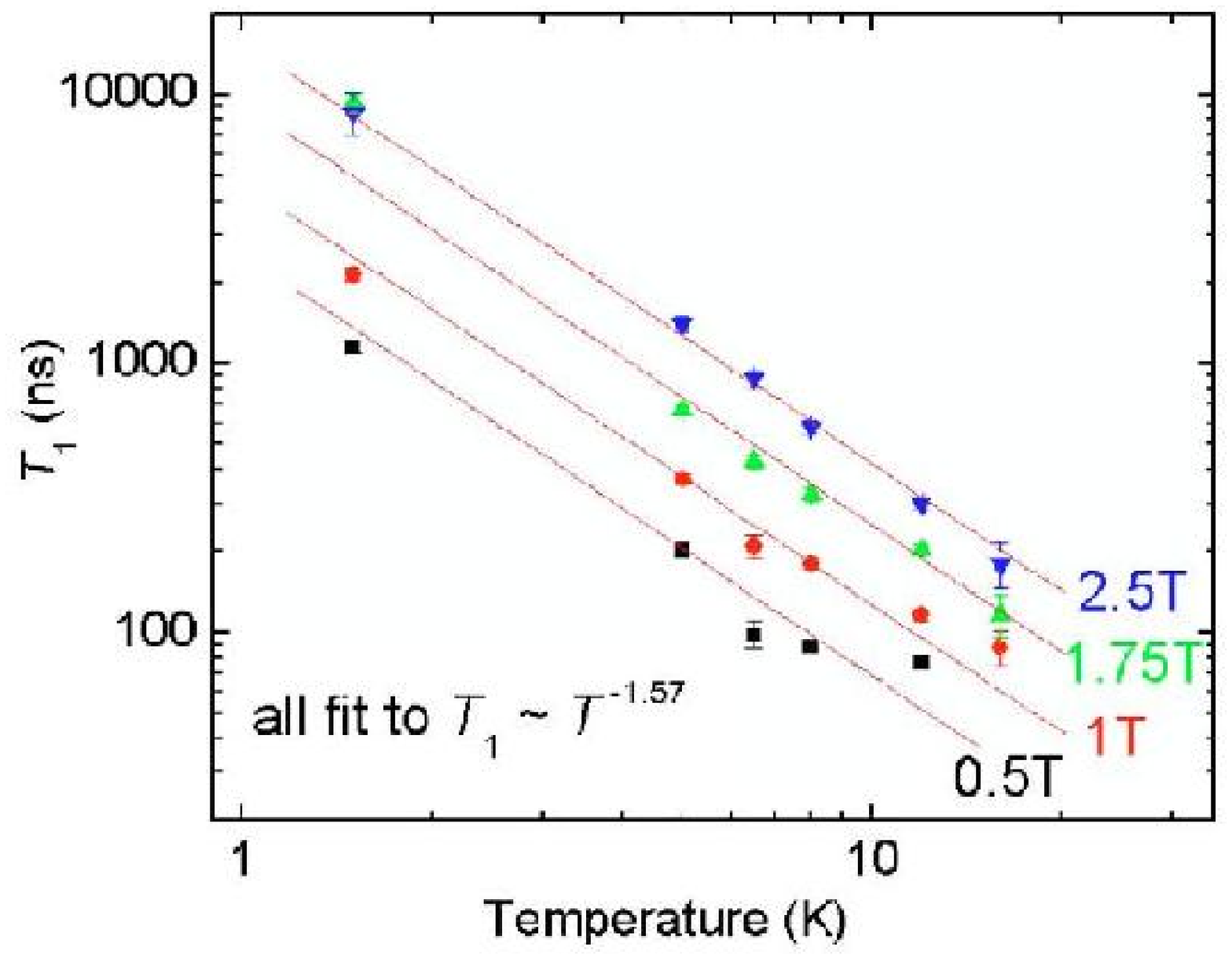}
  \end{minipage}\hfill
  \caption{Left: Spin lifetime $T_1$ as function of magnetic field at
    various temperatures (a) $T=1.5$ and 5~K; (b) $T=1.5$, 5, 6.5, 8
    and 12~K. Right: Spin lifetime $T_1$ as function of temperature for
    various magnetic field. All temperature dependences can be fitted to
    $T_1\sim T^{-1.57}$, as indicated by the thin solid lines. Doping
    density is $n_D=10^{15}$~cm$^{-3}$. From Colton et al. \cite{colton:205201}.}
  \label{colton}
\end{figure}

The magnetic field and temperature dependences of spin relaxation
time in lightly doped $n$-GaAs were measured systematically by Colton
et al. in Refs.~\cite{colton:205201,PhysRevB.69.121307}. The main
results are shown in Fig.~\ref{colton}. Surprisingly a complicated
nonmonotonic magnetic field dependence was observed: spin lifetime
first increases ($B<2$-2.5~T) then decreases (2-$2.5<B<3$-4~T) and
further increases ($B>3$-4~T) with increasing magnetic field
\cite{colton:205201}.
Moreover, spin lifetime decreases rapidly with increasing temperature. For
$1<T<20$~K and $0.5\le B\le 2.5$~T, it exhibits a power law
$\tau_s\sim T^{-1.57}$ \cite{colton:205201}. The underlying physics is
still unclear 
\cite{colton:205201,0268-1242-23-11-114009}. Recently, the temperature
dependence of spin lifetime in quantum wells was studied carefully in
Refs.~\cite{gerlovin:115330,yugova:167402,2010arXiv1001.3228F}. In
these works, {\em differently}, it was found that the temperature
dependence of electron (hole) spin lifetime is quite weak for $T<5$~K
($T<2$~K). At higher temperature, the spin lifetime decreases
dramatically with increasing temperature. It was proposed that the
excited localized states from thermal activation, which have much
shorter spin lifetimes than the ground states, is responsible for the
observed phenomena \cite{2010arXiv1001.3228F}. The temperature
dependence of spin lifetime in insulating regime was also studied
theoretically in Refs.~\cite{0268-1242-23-11-114009,jetplet.85.55}.

The electric field dependence of the spin lifetime in insulating
regime also reflects characteristics of electron localization. Furis
et al. measured such dependence \cite{furis:102102}. Their results
showed that the spin lifetime varies slowly with electric field in
the linear transport regime. However, after the threshold field of
impact ionization, spin lifetime drops rapidly with electric field.
At such threshold, the free electrons get enough energy from the
electric field to ionize the localized electrons
\cite{furis:102102}. After the localized electrons become free and the
electron system acquires enough energy to become a hot electron
system, the D'yakonov-Perel' spin relaxation is accelerated. The
spin lifetime hence drops rapidly.

\begin{figure}[bth]
  \centering
  \includegraphics[height=6.2cm]{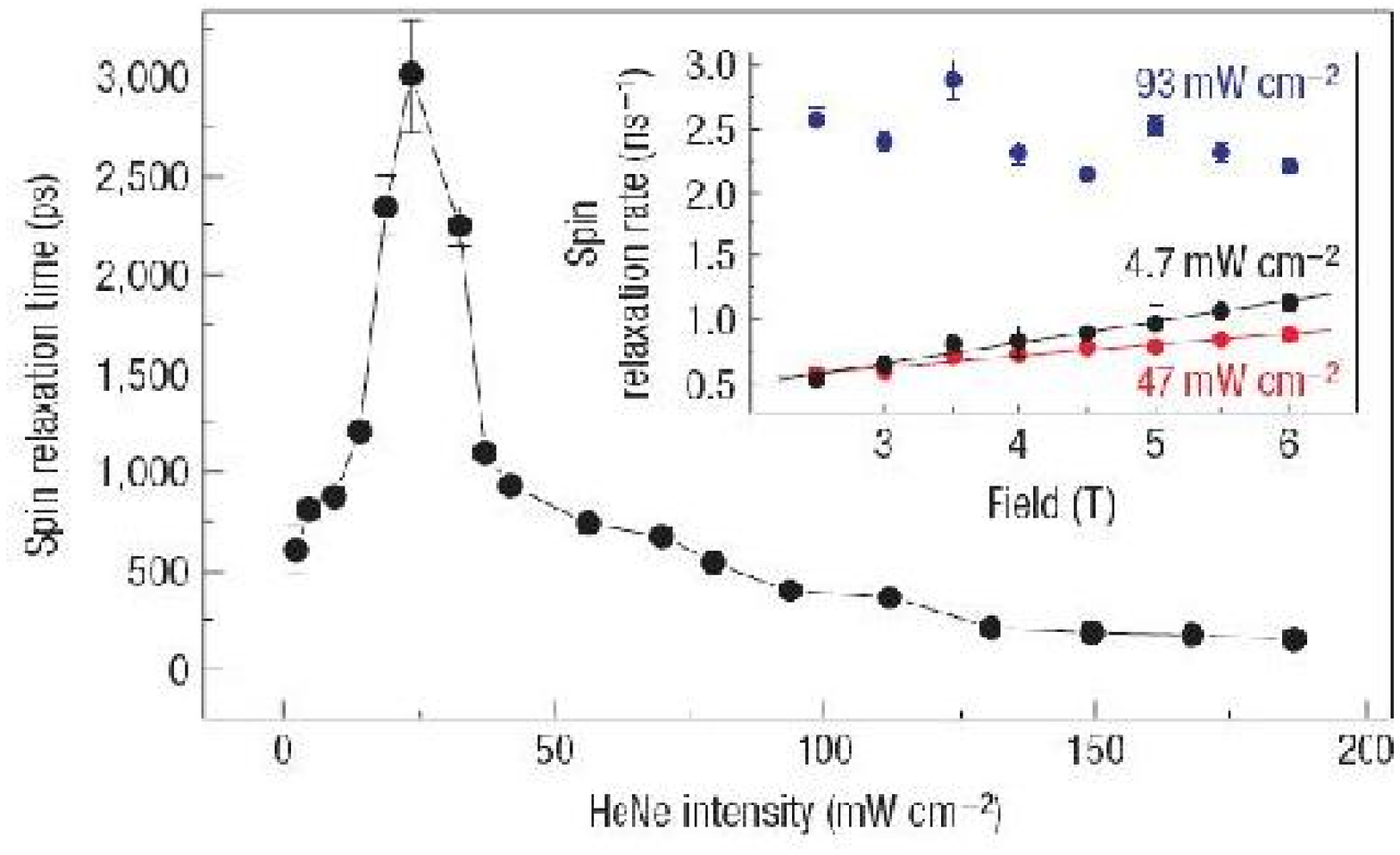}
  \includegraphics[height=7cm]{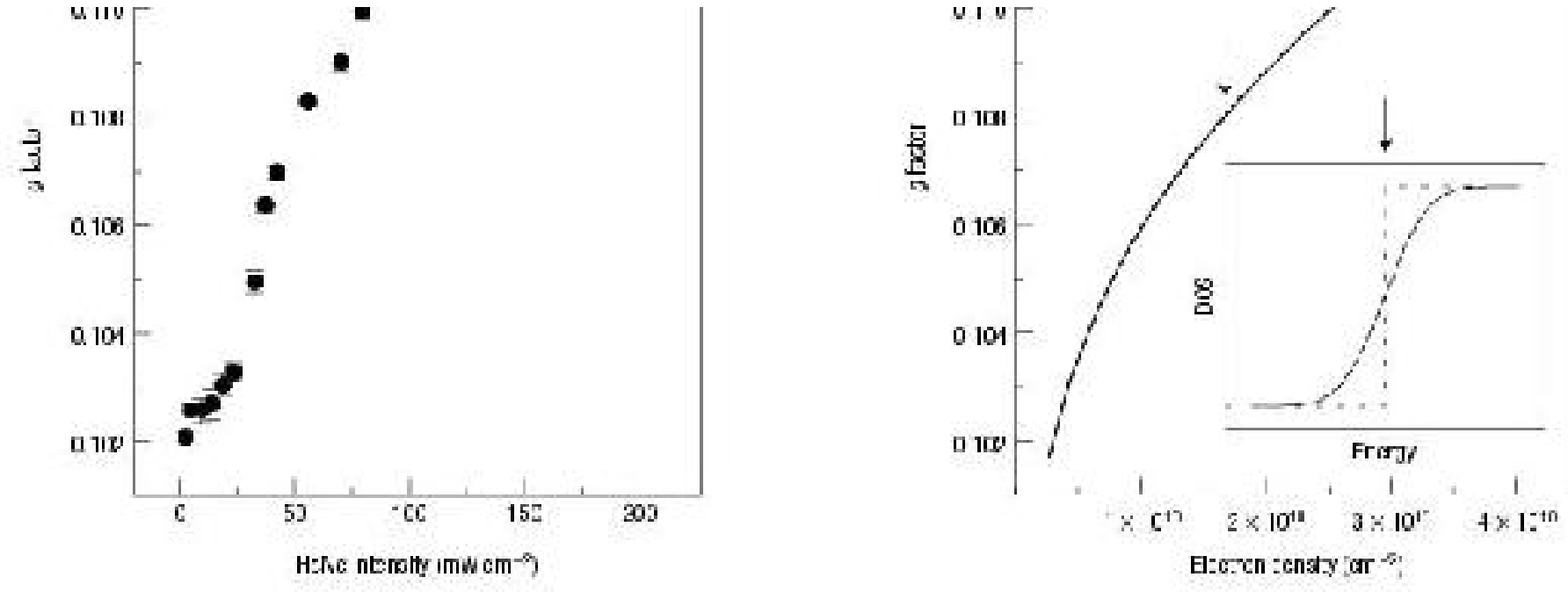}
  \caption{Up: Spin relaxation time as function of HeNe laser excitation
    intensity (HeNe intensity). Inset indicates magnetic field
    dependence of spin relaxation rate for different HeNe laser
    intensities [solid lines denote the fittings according to
    Eq.~(\ref{chen-g})]. Down: The (a) measured and (b) calculated
    $g$-factor as function of HeNe laser intensity (electron
    density). Inset of (b) shows the calculated density of states
    (DOS) of the two-dimensional electron gas with a localization tail
    (solid curve) and that of the unperturbed two-dimensional electron
    gas (dashed curve). The temperature is 4~K.
    From Chen et al. \cite{chen:nphys537}.} 
  \label{chen_nphs}
\end{figure}

Density dependence of spin lifetime in two-dimensional
electron system was studied by Sandhu et al. \cite{PhysRevLett.86.2150}
and Chen et al. \cite{chen:nphys537} (see Fig.~\ref{chen_nphs}), where metal-insulator transition
is involved. In the work of Sandhu et al., metal-insulator transition
was demarcated by conductivity \cite{PhysRevLett.86.2150}. Differently
Chen et al. revealed metal-insulator transition by determining the
density of states at band edge via $g$-factor measurement
\cite{chen:nphys537}. In GaAs $g$-factor has an energy dependence
$g(\varepsilon)=g_c+\beta \varepsilon$. Therefore at zero (low)
temperature the density of states $D(\varepsilon)$ is written as
\be
D(\varepsilon) = \beta (\frac{d g^{\ast}}{d n})^{-1},
\ee
where $g^{\ast}$ is the measured (ensemble averaged) $g$-factor and $n$
is electron density. By measuring the density dependence of $g^{\ast}$,
Chen et al. obtained the density of states near band edge (see Fig.~\ref{chen_nphs}) where the
tail reveals the existence of localized states. Electron localization
also manifests itself in the magnetic field dependence of spin
lifetime: spin dephasing rate in insulating regime was found to
follow 
\be
\frac{1}{T_2^{\ast}(B)}=\frac{1}{T_2^{\ast}(0)}+\frac{\Delta g\mu_{\rm
    B} B}{\sqrt{2}},
\label{chen-g}
\ee
due to the $g$-factor inhomogeneity and localization.\footnote{In metallic
  regime spin relaxation rate either decreases with increasing magnetic field or
  varies as $[1/T_2^{\ast}(B)-1/T_2^{\ast}(0)]\sim B^2$
  due to the $g$-factor inhomogeneity.} Similar dependence was also found
in other two-dimensional structures at low temperature
\cite{tribollet:205304,pssb243.878,zhukov:155318},
two-dimensional hole system \cite{syperek:187401,zhukov:155318}
and in ZnSe epilayers close to GaAs/ZnSe interface
\cite{PhysRevLett.84.1015}. In localized regime, the electron (hole)
spin lifetime in two-dimensional system is very
long. Usually electron spin lifetime is of the order of nanoseconds to
microseconds
\cite{pssb243.858,hoffmann:073407,yugova:167402,zhukov:155318}, whereas
hole spin lifetime can reach nanoseconds
\cite{yugova:167402,2009arXiv0909.3711K}. As temperature increases,
spin lifetime decreases due to delocalization and enhancement of the
D'yakonov-Perel' spin relaxation which increases with electron kinetic
energy
\cite{gerlovin:115330,syperek:187401,yugova:167402,zhukov:205310,kugler-2009}.
Pump density dependence of spin lifetime was studied in
Refs.~\cite{zhukov:205310,hoffmann:073407} where spin lifetime was
found to decrease with pump density at $T<2$~K, similar to that in
bulk materials in insulating regime
\cite{PhysRevLett.80.4313,PhysRevB.66.245204,crooker:035208}.
The underlying physics is similar to the temperature dependence:
the increasing pump density leads to the increase of electron
density as well as the thermalization of the electron system, and
enhances spin relaxation.

Spin relaxation in type-II quantum wells was studied in both GaAs and
ZnSe quantum wells at very low temperature
\cite{PhysRevB.39.8552,mino:153101}. As electrons and holes are
spatially separated, the electron-hole exchange interaction is
weakened and spin relaxation time is elongated. Especially, in GaAs
type-II quantum well, electron spin lifetime can be as long as
7~$\mu$s at 1.7 K \cite{PhysRevB.39.8552}.

Recently, Korn et al. observed extremely long hole spin lifetime,
$\simeq 70$~ns, in two-dimensional hole system in (001) quantum wells
at $0.4$~K \cite{2009arXiv0909.3711K}. They also observed the decrease
of hole spin lifetime with increasing magnetic field roughly following
Eq.~(\ref{chen-g}). The low-field ($\sim 0.2$~T) spin lifetime was
observed to decrease by about one order of magnitude when temperature
increases from 0.4~K to 4.5~K. Interestingly, spin relaxation
of localized holes due to the hyperfine interaction can be inhibited by a
very small in-plane magnetic field ($\sim 10$~mT) due to the Ising
form of the hole hyperfine interaction \cite{fischer:155329}. Also, the
$g$-tensor inhomogeneity mechanism is not dominant at small (but still
much larger than $10$~mT) magnetic field
\cite{2009arXiv0909.3711K}. Hence the most likely spin relaxation
mechanism at such magnetic field is the anisotropic exchange
interaction mechanism.

Finally, it should be mentioned that dynamic nuclear polarization
is more efficient in insulating regime than in metallic regime because the
hyperfine interaction plays a more important role in electron spin
relaxation in insulating regime. The dynamic nuclear spin polarization,
in turn, affects the electron spin dynamics. Experimental studies
involve continuous \cite{opt-or,jetp.15.179} and time-resolved
\cite{Kikkawa01212000,Malinowski2000419,malinowski2001nuclear,testelin:235306}
measurements of electron spin dynamics to reveal the nuclear spin
dynamics.

\subsection{Carrier spin relaxation in metallic regime}

At high temperature and/or high density, carrier system is
in metallic regime where most of the carriers are in
extended states in the conduction/valence band with ${\bf k}$ being a
good quantum number. The relevant spin relaxation mechanisms in the
metallic regime are the Elliott-Yafet, the D'yakonov-Perel' and the
Bir-Aronov-Pikus mechanisms. In the presence of magnetic field, the
$g$-tensor inhomogeneity mechanism can also be important
\cite{PhysRevB.66.233206}. Their relative efficiencies differ in
various materials and structures. The different temperature, mobility and
carrier density dependences of these mechanisms together with their
competition result in various behaviors observed in experiments. In
this subsection, we review electron/hole spin relaxation in metallic
regime in both bulk semiconductors and semiconductor
nanostructures.\footnote{The experiments and the corresponding theory
  of electron spin relaxation in $p$-type bulk semiconductors are
  summarized in the seminal book {\em ``Optical Orientation''}
  \cite{opt-or}. We will not review this topic here.} We focus on 
the experiments and single-particle theories, whereas the many-body
theory is also mentioned partly when it is necessary to understand the
experimental results. Review of the many-body theory and its salient
predictions and results are presented in the next section.

\subsubsection{Electron spin relaxation in $n$-type bulk III-V and
  II-VI semiconductors}

As there are few holes, the Bir-Aronov-Pikus mechanism is irrelevant in
$n$-type III-V and II-VI semiconductors. Relevant mechanisms are the
D'yakonov-Perel' and Elliott-Yafet mechanisms. As has been shown in
recent work \cite{jiang:125206}, the Elliott-Yafet mechanism is less
important than the D'yakonov-Perel' mechanism in $n$-type III-V
semiconductors, even in the narrow band-gap semiconductors such as
InAs and InSb. We neglect the Elliott-Yafet mechanism in the
subsequent discussions. A simple theory of the D'yakonov-Perel' spin
relaxation in bulk system has been presented in Sec.~3.2.2
\cite{Titkovbook,zuticrmp}. The main result is
\be
\tau^{-1}_{\rm DP} \simeq Q \tau_p \alpha^2 \langle
  \varepsilon_{\bf k}^3\rangle/E_g.
\label{dpappne}
\ee
$Q\sim 1$. In the presence of magnetic field, the $g$-tensor inhomogeneity
mechanism also works \cite{PhysRevB.66.233206},
\be
\tau^{-1}_{s} \simeq (\mu_{\rm B}B)^2
(\overline{g^2}-\overline{g}^2) \tau_p.
\ee
The $g$-tensor is assumed to be isotropic. The above equations should
be taken only qualitatively, as they are based on a series of
approximations and assumptions which are to be justified. From these
equations the qualitative behavior of spin relaxation can be
understood. To achieve such understanding, the
knowledge of momentum scattering time $\tau_p$ is needed. In $n$-type
III-V and II-VI semiconductors, momentum scattering mainly consists of
the impurity, phonon, electron-electron scatterings. In earlier
theories based on single particle approach, the electron-electron
scattering is considered irrelevant to spin relaxation.\footnote{In
  fact, electron-electron scattering plays important role in the
  D'yakonov-Perel' spin relaxation in $n$-type bulk III-V
  semiconductors, as was shown by Jiang and Wu in Ref.~\cite{jiang:125206}.}
Moreover, the acoustic phonon and transverse-optical phonon
scatterings are negligible. Therefore, the important scattering
mechanisms are the ionized-impurity scattering and the
longitudinal-optical-phonon scattering. According to the
Brooks-Herring formula, the temperature and density dependence of
ionized-impurity scattering time is \cite{zuticrmp} $\tau_p^{\rm ei}
\sim T^{3/2} n_D^{-1}$ ($n_D$ is doping density) for $T\gg T_F$
(nondegenerate regime) and $\tau_p^{\rm ei} \sim T^{0} n_D^0$ for
$T\ll T_F$ (degenerate regime). Based on these dependences, for the
D'yakonov-Perel' spin relaxation due to the ionized-impurity scattering,
$\tau_s\sim T^{-9/2}n_D$ in nondegenerate regime and $\tau_s\sim
T^{0}n_D^{-2}$ in degenerate regime. For spin relaxation due to the
$g$-tensor inhomogeneity mechanism associated with the ionized-impurity
scattering, roughly, $\tau_s\sim B^{-2}T^{-7/2}n_D$ in nondegenerate
regime and $\tau_s\sim B^{-2}T^{0}n_D^{-4/3}$ in degenerate
regime. The longitudinal-optical-phonon scattering is only important
at high temperature $k_BT\gtrsim \omega_{\rm LO}/4$ ($\omega_{\rm LO}$
is longitudinal-optical-phonon frequency). For example, in GaAs
the longitudinal-optical-phonon scattering is important only at $T\gtrsim
100$~K. The temperature and density dependences of
longitudinal-optical-phonon scattering are complex. Roughly,
the longitudinal-optical-phonon scattering rate increases with
temperature but varies weakly with density.\footnote{Dyson and Ridley
  calculated the longitudinal-optical-phonon scattering time for
  nondegenerate electron system. They found that in nondegenerate
  regime the longitudinal-optical-phonon scattering rate increases
  with electron energy and temperature \cite{PhysRevB.69.125211}.}

\begin{figure}[bth]
\begin{minipage}[h]{0.5\linewidth}
  \centering
  \vskip0.1cm
  \includegraphics[height=4.9cm]{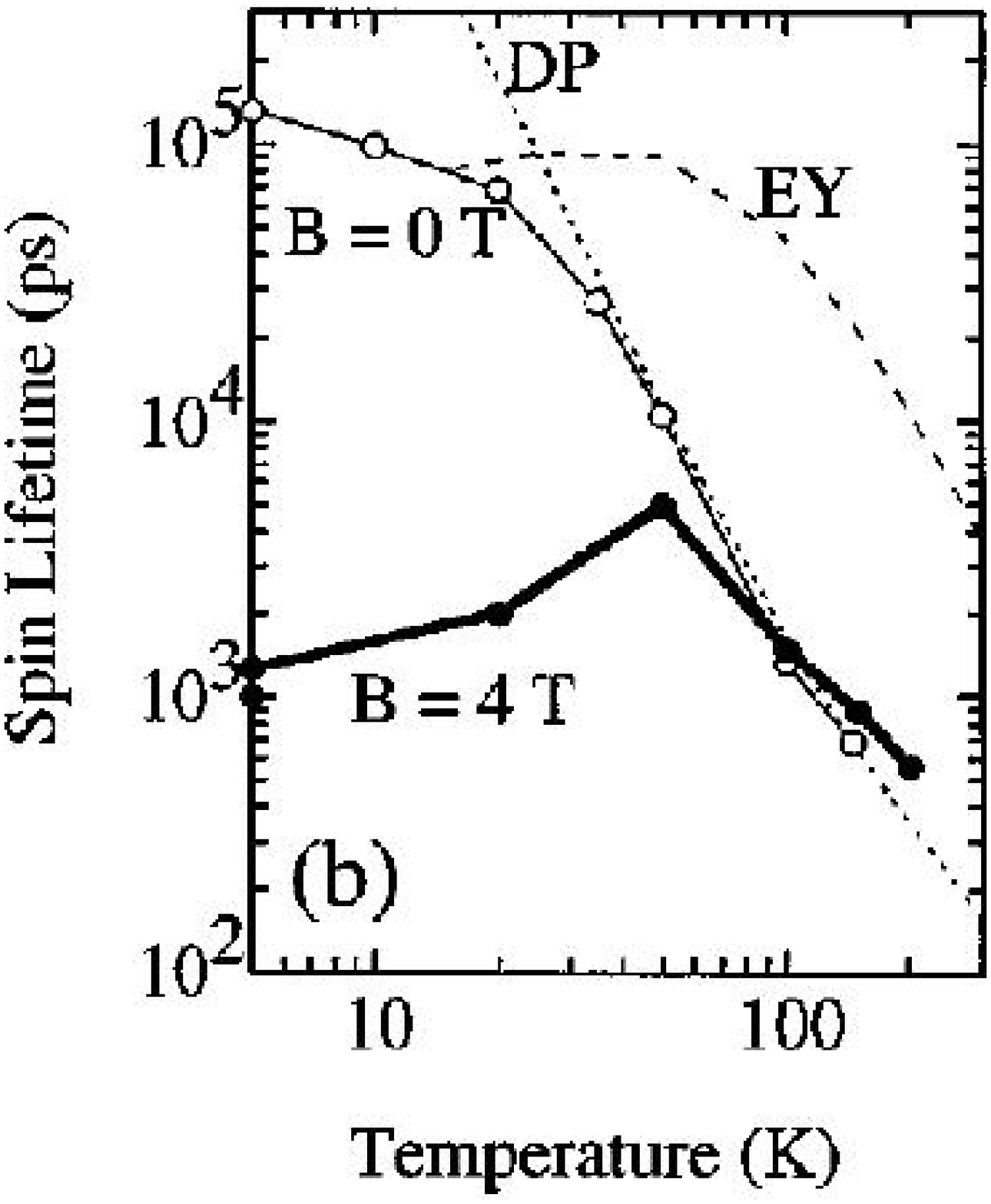}
  \caption{Temperature dependence of spin lifetime for bulk
    $n$-GaAs at $B=0$ and $B=4$~T with doping density
    $n_{\rm D}=10^{16}$~cm$^{-3}$. The dotted and dashed curves
    represent the calculated D'yakonov-Perel' (DP) and Elliott-Yafet
    (EY) spin lifetimes respectively.
    From Kikkawa and Awschalom 
    \cite{PhysRevLett.80.4313}.} 
  \label{Awsc_bulk_T}
  \vskip 0.64cm
  \centering
  \includegraphics[height=4.8cm]{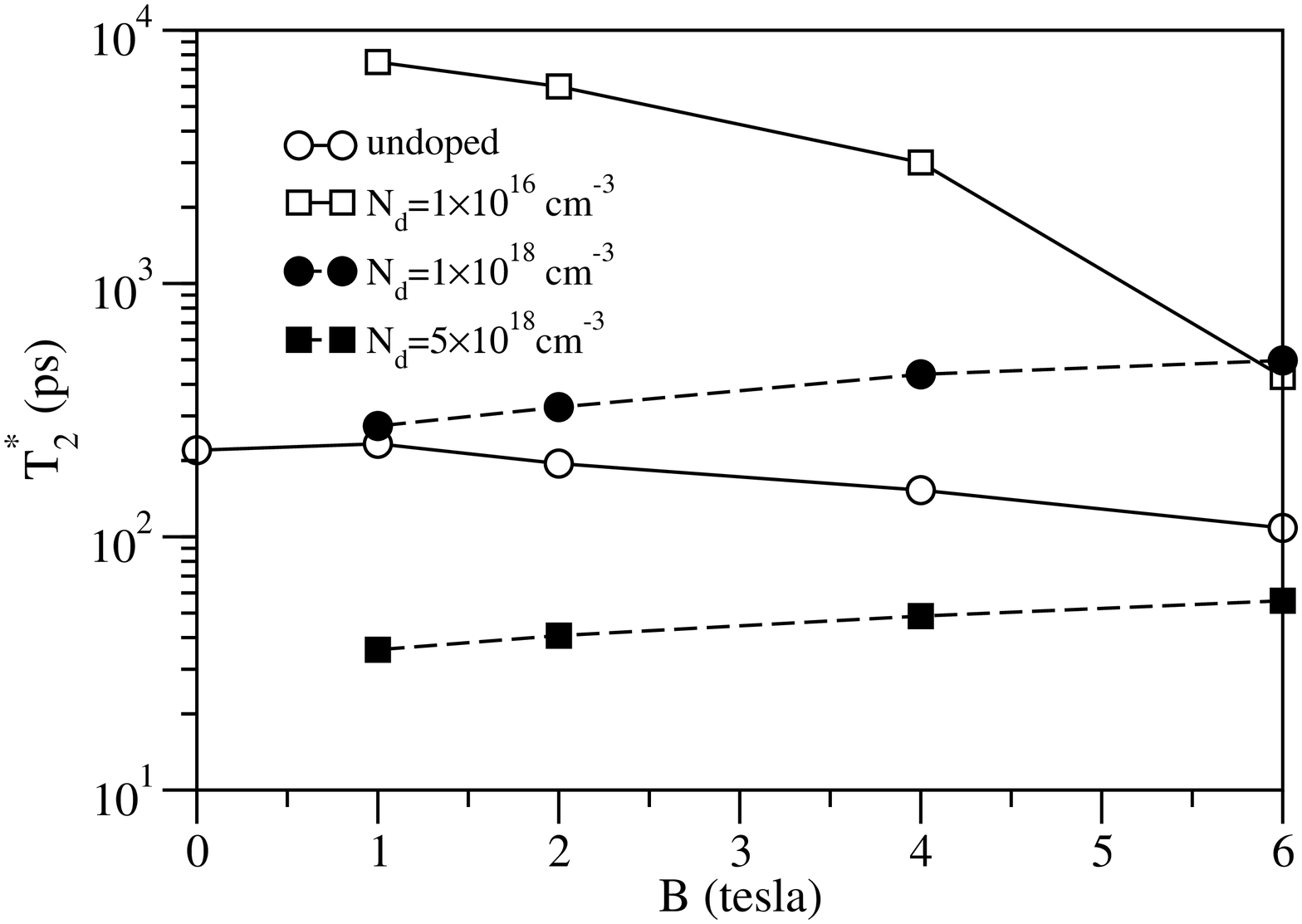}
  \caption{Magnetic field $B$ dependence of ensemble spin dephasing time
    $T_2^\ast$ for bulk GaAs at various doping density at temperature
    $T=5$~K. Experimental data from Kikkawa and Awschalom
    \cite{PhysRevLett.80.4313}(re-plotted by \v{Z}uti\'c
    et al. \cite{zuticrmp}).}
  \label{awsch-bulk2}
\end{minipage}\hfill
  \begin{minipage}[h]{0.4\linewidth}
    \centering
    \includegraphics[height=8.5cm]{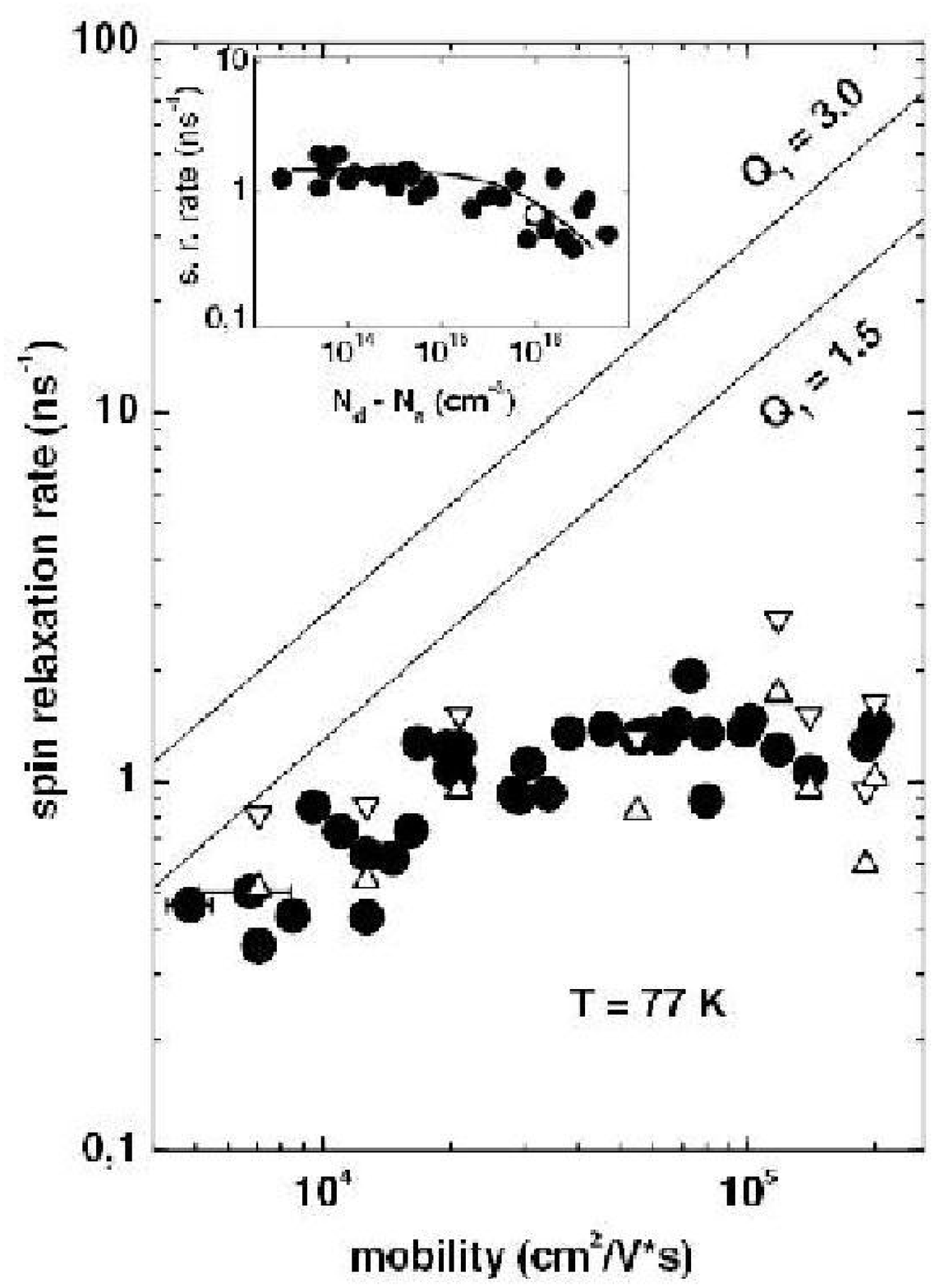}
  \caption{Mobility dependence of spin relaxation rate in bulk
    $n$-GaAs at temperature $T=77$~K. Experimental results are
    represented by dots. Solid lines: calculation from
    $1/\tau_{s} = Q_1 \tau_p \alpha^2 (k_BT)^3 /E_g$ with $Q_1=3$
    (scattering by phonons, upper straight line) and $Q_1=1.5$
    (scattering by ionized impurities, lower straight
    line). Triangles: calculation with $1/\tau_{s} = Q_1 \tau_p
    \alpha^2 (k_BT)^3 /E_g$ using $\tau_p$ measured from
    spin-relaxation suppression in Faraday geometry, 
    assuming scattering by phonons ($\triangledown$), and ionized
    impurities ($\triangle$). Inset: Density dependence
    of the spin relaxation rate (s. r. rate). $N_d$ and $N_a$ denote
    the donor and acceptor densities respectively. Data point
    ($\circ$) is taken from Ref.~\cite{PhysRevLett.80.4313} at
    $T=100$~K. Solid curve is a guide
    to the eyes. From Dzhioev et al. \cite{PhysRevLett.93.216402}.} 
  \label{Dizhov_mob_dep}
\end{minipage}\hfill
\end{figure}

We first review experimental works. Let us start with temperature
dependence of spin lifetime. Spin lifetime was measured as
function of temperature in GaAs
\cite{PhysRevLett.80.4313,PhysRevLett.93.216402,hohage:231101,2009arXiv0911.4084R},
GaSb \cite{0268-1242-24-2-025018}, GaN
\cite{PhysRevB.63.121202,PhysRevB.48.15144,PhysRevB.48.17878,koehl:072110}, 
InAs 
\cite{PhysRevB.72.085346,litvinenko:075331,Litvinenko:461,kini:064318},
InSb
\cite{Litvinenko:461,murdin:096603}, InP
\cite{li:222114}, ZnSe \cite{malajovich:5073}, ZnO
\cite{ghosh:232507}, CdTe \cite{2010arXiv1001.0869S} and HgCdTe
\cite{Murzyn2004220,PhysRevB.67.235202}. It was found
to decrease with increasing temperature at high temperature
(nondegenerate regime) in all these works [e.g., see 
Fig.~\ref{Awsc_bulk_T}].\footnote{At low temperature the system may
  be insulating where spin relaxation is governed by localized
  electrons \cite{PhysRevB.70.113201,harmon:115204}.} This is
consistent with the discussion in above paragraph: $\tau_s\sim
T^{-9/2}$ for ionized-impurity scattering. For longitudinal-optical-phonon scattering,
as $\tau_p^{-1}$ varies slower than $1/\langle \varepsilon_{\bf k}^3 \rangle
\sim T^{-3}$, $\tau_s$ still decreases with temperature. It was observed
that at high density in degenerate regime the spin lifetime varies slowly
with temperature
\cite{0268-1242-24-2-025018,kallaher:075322,malajovich:5073}. This is
understood as that in degenerate regime the ionized-impurity scattering
and longitudinal-optical-phonon scattering times as well as 
$\langle\varepsilon_{\bf k}\rangle$ vary slowly with temperature [see
Eq.~(\ref{dpappne})].

The magnetic field dependence of spin lifetime was studied in
GaAs \cite{PhysRevLett.80.4313,Hohage:4346,2009arXiv0909.3406M} and GaN
\cite{PhysRevB.63.121202,buss2009}. As indicated in Sec.~3.2.2, spin
lifetime due to the D'yakonov-Perel' mechanism increases with increasing magnetic
field. On the other hand, spin lifetime due to the $g$-tensor
inhomogeneity mechanism decreases with increasing magnetic field. These two
mechanisms compete with each other, hence spin lifetime
first increases then decreases with increasing magnetic field
\cite{PhysRevB.66.233206}. The peak magnetic 
field $B_p$ roughly satisfies $\alpha^2 \langle
  \varepsilon_{\bf k}^3\rangle/E_g \simeq (\mu_{\rm
    B}B_p)^2
(\overline{g^2}-\overline{g}^2)$. In III-V and II-VI bulk
semiconductors the electron $g$-factor is energy-dependent $g \simeq
g_c + \beta\varepsilon_{\bf k}$, hence
$(\overline{g^2}-\overline{g}^2) \sim \langle \varepsilon_{\bf k}^2
\rangle$. Therefore, the peak magnetic field $B_p\sim \langle
\varepsilon_{\bf k} \rangle$ which increases with the average
energy. Typical experimental results are shown in
Fig.~\ref{awsch-bulk2}. It is seen that for the undoped sample the spin
lifetime first increases and then decreases with increasing magnetic field. For
$B>1$~T, at low density the spin lifetime decreases with magnetic field
whereas at high density it increases with magnetic
field. This is easily understood since $\langle \varepsilon_{\bf k}
\rangle$ is larger at high density, the peak $B_p$ moves to higher
magnetic field \cite{PhysRevB.66.233206}.

The variation of spin lifetime with mobility has been
investigated in GaAs \cite{PhysRevLett.93.216402} and InAs
\cite{litvinenko:075331} at zero magnetic field.\footnote{The magnetic
field is always zero in this subsection unless otherwise
specified.} In GaAs, the spin relaxation rate was found to increase with
mobility, which is consistent with the D'yakonov-Perel'
mechanism. However, the simple formula, e.g., Eq.~(\ref{dpapp-nondeg})
in Sec.~3.2.2, can not explain the results qualitatively
\cite{PhysRevLett.93.216402} (see Fig.~\ref{Dizhov_mob_dep}). Momentum
scattering that does not contribute to mobility was suggested to be
responsible for the smaller spin relaxation rate at high mobility
regime \cite{PhysRevLett.93.216402}.

The photo-excitation density dependence of spin lifetime was
investigated in $n$-type GaAs
\cite{PhysRevLett.80.4313,pssc.0.1506,krauss0902.0270}, GaSb
\cite{0268-1242-24-2-025018} and InAs \cite{kini:064318}. In all these
measurements, the spin lifetime was found to decrease with
excitation density in $n$-doped samples. The decrease was unexplained
in some works \cite{PhysRevLett.80.4313,pssc.0.1506} or was assigned to the
Elliott-Yafet mechanism \cite{kini:064318}. Other works explained the
decrease as a result of thermalization of the electron system via
photo-excitation \cite{0268-1242-24-2-025018}. The decrease of spin
lifetime may be due to the increase of $\langle
\varepsilon_{\bf k}^3\rangle$ as the electron density increases. However,
the thermalization effect may also play an important role.\footnote{The
  thermalization effect was shown to be crucial in a recent fully
  microscopic calculation \cite{shenpeak}. This is also supported by
  the experimental results that the spin lifetime was observed to
  decrease with photon energy of the pump laser
  \cite{lai:062110,hohage:231101,ma:241112}.}

The doping density dependence was studied in GaAs
\cite{PhysRevLett.80.4313,PhysRevLett.93.216402,PhysRevB.66.245204}
and InAs \cite{PhysRevB.72.085346}. In all these works spin relaxation
time was found to decrease with doping density in degenerate
regime. There is few data in the nondegenerate regime up till now
\cite{PhysRevLett.80.4313,PhysRevB.72.085346}.
The decrease of spin lifetime in degenerate regime is
understood as following: in degenerate regime the ionized-impurity
scattering time and longitudinal-optical-phonon scattering time vary slowly with
density, whereas $\langle\varepsilon_{\bf k}^3\rangle\sim n_D^2$
increases with density. Therefore the spin lifetime $\tau_s\sim
1/(\langle\varepsilon_{\bf k}^3 \rangle\tau_p)$ decreases with density.

Electric field dependence was studied in GaAs
\cite{furis:102102,beck_06}, ZnO \cite{ghosh:162109} and GaN
\cite{koehl:072110}. Usually, spin lifetime decreases with
increasing electric field in metallic regime in GaAs
\cite{furis:102102,beck_06}. This is because the
electric field induces electron drift and the hot-electron effect
which increase the average electron energy and then enhance the
D'yakonov-Perel' spin relaxation \cite{jiang:125206}. However,
the hot-electron effect also gives rise to the enhancement of momentum
scattering \cite{PhysRevB.69.245320} which leads to the suppression of
the D'yakonov-Perel' spin relaxation. These two effects compete
with each other \cite{PhysRevB.69.245320}. It was found that in the
case of cubic-${\bf k}$ spin-orbit coupling, the former effect
dominates because the spin lifetime varies rapidly with the electron
energy \cite{jiang:125206,PhysRevB.72.033311}. The spin lifetime
hence decreases with increasing electric field
\cite{jiang:125206,PhysRevB.72.033311}. This is the case for bulk
GaAs. The situation is more complicated in the case of linear-${\bf
  k}$ spin-orbit coupling: the spin lifetime first increases due to
the enhancement of momentum scattering then decreases due to the
increases of electron energy \cite{PhysRevB.69.245320}. This is the
case for bulk wurzite ZnO where Ghosh et al. observed that the spin lifetime
first increases then decreases with increasing electric field
\cite{ghosh:162109}. Usually the effect of electric field on spin
relaxation is more pronounced at low temperature
\cite{furis:102102,zhou:045305,jiang:125206,ghosh:162109,koehl:072110}.
The spin lifetime can also be reduced by applying strain which
induces additional spin-orbit coupling
\cite{knotz:241918,sih:241316,chantis-2008-78,d'yakonov:655}. It was
found that $1/\tau_s\sim {\cal{\epsilon}}^2$, where ${\cal{\epsilon}}$
is the applied stress
\cite{d'yakonov:655} when the strain-induced spin-orbit coupling
dominates. Under strain field the temperature and density dependence
of spin relaxation becomes weaker as the strain-induced spin-orbit
coupling is linear in ${\bf k}$ [see Eq.~(\ref{strain_soc})]. It is
also found that the spin relaxation can be anisotropic under strain
\cite{d'yakonov:655}.

In wurzite GaN and ZnO, the spin-orbit coupling is different
from that in zinc-blende semiconductors. The spin-orbit coupling in
conduction band reads 
\be
H^e_{\rm SO} = [\mathbf{\Omega}_e^R(\mathbf{k})+
\mathbf{\Omega}_e^D(\mathbf{k})] \cdot \spin \big/2
\ee
with
\be
\mathbf{\Omega}_e^R(\mathbf{k})=2\alpha_e(k_y,-k_x,0), \quad\quad  
\mathbf{\Omega}_e^D(\mathbf{k})=2\gamma_e(bk_z^2-k_{\|}^2)(k_y,-k_x,0).
\ee
It is seen that the symmetry of the spin-orbit coupling is quite
different from that of the Dresselhaus one. Unlike in bulk GaAs, electron
spin relaxation in bulk GaN is anisotropic. Lifetime of spins along
[0001] direction is smaller than those along other directions. The anisotropy of
spin relaxation was demonstrated in recent experiment \cite{buss2009}.

Spin lifetime $\tau_s$ at room temperature is also of
concern due to interest in possible device application. It was
measured in GaAs \cite{hohage:231101,PhysRevB.63.235201} with $\tau_s$
in the range of $15$-$110$~ps, in InAs
\cite{boggess:1333,murzyn:5220,kini:064318} with $\tau_s$ in the range
of $2$-$24$~ps and in InSb
\cite{PhysRevB.67.235202,murdin:096603,Litvinenko:461} with $\tau_s$
in the range of $2$-$300$~ps depending on the doping density.

We now review the single-particle theory on electron spin relaxation
in $n$-type bulk III-V and II-VI semiconductors. Besides the early
works in 1970s and early 1980s, which have been summerized in the book
{\em ``Optical Orientation''} \cite{opt-or}, there are also a lot of
theoretical studies in the past decade after the rising of the semiconductor spintronics
\cite{S.A.Wolf11162001}. These studies try to go beyond the widely
used formulae for the D'yakonov-Perel' [Eqs.~(\ref{dpapp}) and
(\ref{dpapp-nondeg})] and Elliott-Yafet [Eq.~(\ref{eyapp})] spin
relaxation \cite{Titkovbook}. These developments go in the same
paradigm: calculate spin-orbit coupling and scattering via
microscopic theory and then average the spin relaxation rate at
different ${\bf k}$ as
\be
\tau_s^{-1} = \frac{\int d{\bf k}~(f_{\bf k}^{\uparrow}-f_{\bf
    k}^{\downarrow}) \tau_s^{-1}({\bf k})}{\int d{\bf
    k}~(f_{\bf k}^{\uparrow}-f_{\bf k}^{\downarrow})}.
\label{srt-ave}
\ee
The spin relaxation rate at different $k$, $\tau_s^{-1}(k)$, due to the
D'yakonov-Perel' mechanism is given in Eq.~(\ref{dpsrt-tensor}), while
that due to the Elliott-Yafet mechanism is given by
\be
\tau_s^{-1}({\bf k}) = \sum_{{\bf k}^{\prime}}
  [\Gamma({\bf k}\uparrow,{\bf k}^{\prime}\downarrow) +
  \Gamma({\bf k}\downarrow,{\bf k}^{\prime}\uparrow) ],
\ee
with $\Gamma({\bf k}\uparrow,{\bf k}^{\prime}\downarrow)$ and
$\Gamma({\bf k}\downarrow,{\bf k}^{\prime}\uparrow)$ denoting the
spin-flip scattering rates.
Such a paradigm was first applied to
$n$-type bulk GaAs, InAs and GaSb by Lau et al. where the spin-orbit
coupling was calculated via the fourteen-band ${\bf k}\cdot{\bf p}$
theory \cite{PhysRevB.64.161301}. The momentum scattering rates due
to the electron-impurity and electron-phonon scattering were
calculated via standard methods in transport theory. Similar schemes were then applied to bulk GaAs and GaN
by Krishnamurthy et al. \cite{krishnamurthy:1761}. Later the scheme was
developed to the case of arbitrary band structure by Yu et
al. \cite{PhysRevB.71.245312}.\footnote{In this work, the approach
  that the authors used to calculate hole spin relaxation rate seems
  to have some problem, as they concluded that the hole spin
  relaxation is dominated by the Elliott-Yafet mechanism which is not
  true.} The role of scattering by dislocations
in both the D'yakonov-Perel' and Elliott-Yafet spin relaxation was
studied by Jena within such scheme \cite{PhysRevB.70.245203}. A systematic
study on spin relaxation in III-V semiconductors based on such
paradigm was given by Song and Kim \cite{PhysRevB.66.035207}. However,
in their work the Boltzmann statistics was assumed which makes the
discussion at low temperature and/or high density meaningless. The
Elliott-Yafet spin relaxation due to the electron-electron scattering was
studied in Refs.~\cite{Boguslawski1980389,PhysRevB.68.245205} within
such paradigm. Recently spin relaxation in ZnO due to the
D'yakonov-Perel' and Elliott-Yafet mechanisms was also studied in
such paradigm \cite{harmon:115204}. An improvement of the paradigm was
given in Ref.~\cite{oertel:132112}. In that work, it was argued that
the strong scattering condition $\langle\Omega({\bf
  k})\rangle\tau_p\ll 1$ may not be satisfied for large $k$ as both
$\langle\Omega({\bf k})\rangle$ and $\tau_p$ increase with $k$. Hence
the spin relaxation rate should be modified to include the free induction
decay as \cite{oertel:132112} 
\be
\frac{1}{\tau_s^{\prime}(k)} = \left(\tau_s(\varepsilon_{k})
+ \frac{\sqrt{128E_g}}{\alpha\varepsilon_{k}^{3/2}}\right)^{-1}.
\ee
Spin relaxation rate of the electron ensemble is then obtained by
averaging over ${\bf k}$ via Eq.~(\ref{srt-ave}). A closer
examination on the D'yakonov-Perel' spin relaxation due to the
electron-longitudinal-optical-phonon scattering from the equation of
motion was given in Ref.~\cite{PhysRevB.69.125211}, where it was shown
that the elastic scattering approximation can have problem in treating
the longitudinal-optical-phonon scattering to spin relaxation. The effect
of the electron-longitudinal-optical-phonon scattering on the
D'yakonov-Perel' spin relaxation may not be characterized by a single
momentum scattering time used in the 
paradigm. Spin relaxation due to the D'yakonov-Perel' mechanism under
electric field was studied via the Monte Carlo method in
Ref.~\cite{barry:3686}, where spin lifetime was found to
decrease with the electric field.

\subsubsection{Electron spin relaxation in intrinsic bulk III-V and
  II-VI semiconductors}

In intrinsic semiconductors electrons are generated together with
equall number of holes by photo excitation. Due to the large number of
holes, the Bir-Aronov-Pikus mechanism is now relevant to the spin
relaxation. As have been pointed out in Ref.~\cite{jiang:125206} that
the Elliott-Yafet mechanism is unimportant in intrinsic III-V
semiconductors. Similar conclusion should also hold for II-VI
semiconductors. Hence, the relevant spin relaxation mechanisms are the
Bir-Aronov-Pikus and the D'yakonov-Perel' mechanisms. It was debated about
the relative importance of these two mechanisms in intrinsic III-V and
II-VI semiconductors
\cite{zuticrmp,amo:085202,J.M.Kikkawa08291997,PhysRevB.54.1967,0022-3727-42-13-135111,PhysRevB.24.3623}.
Single-particle theory based on the elastic scattering approximation
comes to the conclusion that spin relaxation due to the Bir-Aronov-Pikus
mechanism can be more important at low temperature and high hole
density \cite{PhysRevB.54.1967}. However, as pointed out by Zhou and
Wu \cite{zhou:075318}, the Pauli blocking of electrons at low temperature and/or high hole
density largely reduces the Bir-Aronov-Pikus spin relaxation from a
fully microscopic kinetic spin Bloch equation approach
\cite{PhysRevB.61.2945,wu:epjb.18.373,PhysRevB.68.075312}. After that
Jiang and Wu showed that in intrinsic bulk III-V semiconductors the
Bir-Aronov-Pikus mechanism is 
unimportant via the same approach \cite{jiang:125206}, where similar
conclusion should also hold for II-VI semiconductors. Therefore, the
relevant spin relaxation mechanism is the D'yakonov-Perel' mechanism.
Spin relaxation in intrinsic bulk III-V and II-VI semiconductors can
then be understood easily.

Experimental studies on spin relaxation in intrinsic semiconductors
involve temperature dependence
\cite{PhysRevB.69.165214,oertel:132112,litvinenko:083105,lai:192106} and
density dependence
\cite{oertel:132112,0022-3727-42-13-135111,ma:241112}. The temperature
dependence coincides with what predicted by theory: spin
lifetime varies slowly in degenerate regime and then decreases
rapidly in nondegenerate regime with increasing temperature. The
density dependence, however, indicates an interesting behavior: spin
lifetime increases with electron density
\cite{oertel:132112}. This is totally different from what observed in
$n$-type semiconductors where spin lifetime decreases with
electron density
\cite{PhysRevLett.80.4313,PhysRevLett.93.216402,PhysRevB.66.245204}.
However, in intrinsic semiconductors, the main scattering mechanisms
are the carrier-carrier scattering and the longitudinal-optical-phonon scattering, which
makes spin relaxation difficult to understand via the single-particle theory.
From the many-body kinetic spin Bloch equation approach \cite{PhysRevB.61.2945,wu:epjb.18.373,PhysRevB.68.075312}, Jiang and Wu
found a nonmonotonic density dependence of spin lifetime due
to the nonmonotonic density dependence of the carrier-carrier
scattering rate:
in nondegenerate regime the carrier-carrier scattering rate increases with
increasing density whereas in degenerate regime it
decreases with increasing density \cite{jiang:125206}. After their prediction
Teng et al. observed in intrinsic bulk GaAs that at high excitation
density ($n_e>10^{17}$~cm$^{-3}$) spin lifetime decreases with
density \cite{0022-3727-42-13-135111}. However, Teng et al. assigned
the decrease of spin lifetime to the Bir-Aronov-Pikus mechanism
which is wrong (see comment \cite{jiang-2009}). Later Ma et al. observed
such a nonmonotonic density dependence of spin lifetime in
bulk CdTe \cite{ma:241112}.\footnote{They, however, assigned the
  decrease of the spin lifetime with increasing density to the
  Elliott-Yafet mechanism, which is incorrect (see comment
  \cite{2009arXiv0910.1506J}).}

\subsubsection{Electron spin relaxation in $n$-type and intrinsic III-V and II-VI
  semiconductor two-dimensional structures}

As was found in $n$-type and intrinsic bulk III-V and II-VI
semiconductors that spin relaxation is dominated by the
D'yakonov-Perel' mechanism, it is reasonable to believe that the same
conclusion holds for the two-dimensional case. Indeed, calculations
showed that the Elliott-Yafet mechanism is unimportant in GaAs and
InGaAs quantum wells
\cite{leyland:165309,JJAP.38.4680,jiang:155201}.\footnote{In narrow
  bandgap semiconductor InSb quantum wells, experiment gives some
  evidences that the Elliott-Yafet spin relaxation can be more important
  than the D'yakonov-Perel' one at low mobility samples as revealed by
  Litvinenko et al. in Ref.~\cite{1367-2630-8-4-049}. Nevertheless,
  the two mechanisms are still comparable in their experiment.} Fully
microscopic many-body kinetic spin Bloch equation calculation
indicated that the Bir-Aronov-Pikus mechanism is ineffective in
intrinsic GaAs quantum wells 
\cite{zhou:075318}.\footnote{However, in some wide bandgap
  semiconductors which have weak 
  spin-orbit coupling and strong electron-hole exchange interaction,
  the Bir-Aronov-Pikus mechanism can be important in intrinsic
  samples, especially when excitons are formed. Such as the situation
  in ZnSe \cite{Hagele1999338}. Also in some special cases where the
  D'yakonov-Perel' mechanism is suppressed, the Bir-Aronov-Pikus or
  the Elliott-Yafet mechanisms may be important, such as in (110)
  quantum wells \cite{PhysRevLett.83.4196}. We will return to these
  special cases after the discussion in common cases.} These
evidences enable us to focus on spin relaxation due to the
D'yakonov-Perel' mechanism.

{\bf Anisotropic spin relaxation}\\
\indent One of the salient features of the D'yakonov-Perel' spin relaxation in
two-dimensional structures (quantum wells and heterostructures) is
that it varies significantly with the growth direction and the
structure. The former originates from tailoring the symmetry of the
Dresselhaus spin-orbit coupling by the growth direction. The latter
partly comes from the fact that the strength of the linear Dresselhaus
term can be tuned by quantum confinement. Moreover, the Rashba
spin-orbit coupling also appears in asymmetric two-dimensional
structures. The joint effect of the Dresselhaus and the Rashba
spin-orbit couplings leads to {\em anisotropic} spin relaxation. The
strength of the Rashba spin-orbit coupling can be controlled by the
gate voltage. The tunability of spin relaxation opens route to a
variety of new phenomena and functionalities \cite{Fabianbook,hall:162503}.

Let us start with the spin-orbit coupling which is crucial to the
D'yakonov-Perel' spin relaxation. Assuming that the confinement is
strong enough so that only the lowest subband is involved. Consider a
simple case: (001) quantum well, where the Dresselhaus spin-orbit
coupling is 
\be 
H_{D} = \beta_{D} (-k_x\sigma_x + k_y\sigma_y) + 
\gamma_{D} ( k_x k_y^2 \sigma_x - k_y k_x^2 \sigma_y ).
\ee
$\beta_D=\gamma_{D} \langle\psi_1|\hat{k}_z^2|\psi_1\rangle$ where
$\psi_1$ is the envelope function of the lowest subband and $\hat{k}_z
= -i\partial_z$. $\beta_D\sim \gamma_{D} (\pi/a)^2$ with
$a$ being the well width. The Rashba spin-orbit coupling reads
\be
H_{R} = \alpha_{R} (k_y\sigma_x - k_x\sigma_y). 
\ee
For narrow quantum well when ${\bf k}^2\ll (\pi/a)^2$ 
[${\bf k}=(k_x,k_y)$ is the in-plane wavevector], the linear term dominates 
\be
H_{\rm SOC} = (\alpha_R k_y-\beta_D k_x)\sigma_x + (-\alpha_R k_x +
\beta_D k_y) \sigma_y.
\ee
It was first pointed out by Averkiev and Golub
\cite{PhysRevB.60.15582} that the spin relaxation is anisotropic when the
Rashba and Dresselhaus spin-orbit couplings compete with each other. The spin
relaxation tensor is then
\be
\tau_{zz}^{-1} = 2\tau_{xx}^{-1} = 2\tau_{yy}^{-1} =
8m\langle\tau_p\varepsilon_{\bf k}\rangle
(\alpha_R^2+\beta_D^2),\quad\quad \tau_{xy}^{-1} = -
8m\langle\tau_p\varepsilon_{\bf k}\rangle \alpha_R\beta_D.
\ee
Diagonalizing the spin relaxation tensor, one obtains
\be
\tau_{zz}^{-1} = 8m\langle\tau_p\varepsilon_{\bf k}\rangle
(\alpha_R^2+\beta_D^2), \quad\quad 
\tau_{\pm\pm}^{-1} = 4m\langle\tau_p\varepsilon_{\bf k}\rangle
(\alpha_R\mp\beta_D)^2,
\ee
where the principal axes are 
${\bf n}_{\pm}=\frac{1}{\sqrt{2}}(1,\pm1,0)$ and $z$-axis.
It is meaningful to rewrite the spin-orbit coupling in these principal
axis,
\be
H_{\rm SOC} = (\alpha_R-\beta_D)k_{+}\sigma_{-} - (\alpha_R+\beta_D)
k_{-}\sigma_{+}.
\ee
Recall that $\tau_{ij}^{-1} \sim [\overline{\Omega^2}\delta_{ij}-
\overline{\Omega_i\Omega_j}]$. As averaging over angle
$\overline{k_{+}k_{-}}=0$ and $\overline{k_{+}^2}=\overline{k_{-}^2}$,
one readily finds that $1/\tau_{++} \sim (\alpha_R-\beta_D)^2$,
$1/\tau_{--}\sim (\alpha_R+\beta_D)^2$, $1/\tau_{+-} = 0$ and
$1/\tau_{zz}\sim 2(\alpha_R^2+\beta_D^2)$. If $\alpha_R=\beta_D$
($\alpha_R=-\beta_D$), $1/\tau_{++}=0$ ($1/\tau_{--}=0$), i.e., spin
lifetime is infinite along ${\bf n}_{+}$ (${\bf n}_{-}$). However,
taking into account of the cubic Dresselhaus term, spin lifetime along
this particular direction is finite but still much larger than that along other directions \cite{cheng:083704}. Therefore spin
relaxation is highly anisotropy when $\alpha_R\simeq \pm\beta_D$
\cite{0953-8984-14-12-202,0268-1242-23-11-114002}.

\begin{figure}[bth]
  \centering
  \includegraphics[height=8cm]{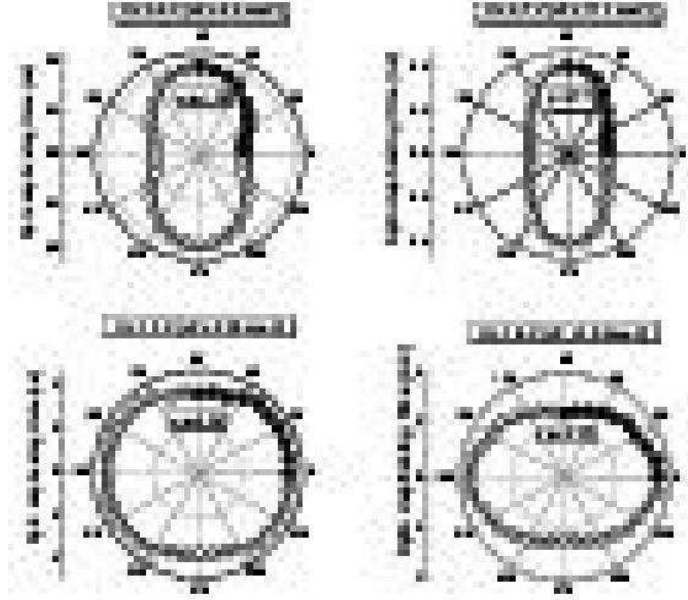}
  \caption{Polar plot of the spin dephasing time measured as a
    function of the angle (in degree) between the magnetic field and the [110]
    axis at four applied biases $U$ in coupled double quantum
    wells. The spin dephasing time at each
    angle is represented by the distance to the center of the polar
    coordinates. Experimental data are shown by points; 
    theoretical values are presented by solid curves. The ratio of the
    Rashba spin-orbit coupling parameter $\alpha$ to the Dresselhaus
    one $\beta$, $\alpha/\beta$, extracted from the anisotropy of spin
    dephasing time is also presented in each figure. The values
    $\Delta E$ represent the exciton Stark shifts for the given
    biases $U$. Temperature is $T=2$~K. From Larionov and Golub
    \cite{larionov:033302}.} 
  \label{Larionov_aniso}
\end{figure}

Experimentally as only the spin polarization along the quantum well
growth direction can be measured,\footnote{Note that a recently
  developed technique of tomographic Kerr rotation can measure spin
  polarization in other directions \cite{nature.457.702}.} spin
relaxation anisotropy is revealed indirectly by comparing spin
relaxation rates along $z$ direction in the two cases with magnetic
field along [110] and [$1\bar{1}0$] directions.
When a moderate magnetic field is along [110] direction, spin
polarization along the growth direction evolves
\be
S_z(t) = S_z(0) \exp[-\frac{1}{2}(\tau^{-1}_{zz}+\tau^{-1}_{--})t]\cos(\omega_0 t+\phi_0).
\ee
The observed envelope decay rate is
$\frac{1}{2}(\tau^{-1}_{zz}+\tau^{-1}_{--})$. Applying
magnetic field along [$1\bar{1}0$] direction, one obtains a decay rate
of $\frac{1}{2}(\tau_{zz}^{-1}+\tau_{++}^{-1})$. Comparing
these two, one can obtain spin lifetimes $\tau_{++}$ and $\tau_{--}$
and the ratio $|\alpha_R/\beta_D|$ can be
extracted.\footnote{Actually, via such method, one can determine the
  ratio but not sure whether it is $|\alpha_R/\beta_D|$ or $|\beta_D/\alpha_R|$
  \cite{averkiev:033305}.} The spin relaxation
anisotropy in (001) quantum wells was first observed by Averkiev et
al. \cite{averkiev:033305} via the Hanle measurements, and then by several
groups via time-resolved measurements
\cite{liu:112111,stich:073309,larionov:033302,studer:045302}.
Among these works, the spin relaxation anisotropy was demonstrated to
be controlled/affected by density \cite{liu:112111}, magnetic field
\cite{stich:073309}, gate-voltage (see Fig.~\ref{Larionov_aniso})
\cite{larionov:033302,sih:081710} and temperature
\cite{liu:112111,studer:045302}. The symmetry of the spin-orbit coupling
in other two-dimensional structures with different growth directions
also has marked effect on spin relaxation. In some particular cases,
such as (111) and (110) quantum wells, spin relaxation can be largely
suppressed. As there are also a lot of papers on those topics, we will
discuss them in detail later while below we only focus on (001) quantum wells.

{\bf Spin relaxation in (001) quantum wells}\\
\indent Widely concerned topics are the temperature, density, mobility,
excitation density, magnetic field, gate-voltage and quantum
subband quantization energy dependences of spin relaxation. These dependences
reveal intriguing underlying physics and provide important knowledges for
spintronic device design. Below we first review experimental studies
and then the single-particle theories. To benefite the understanding,
we first present some simple analytical results which assume only
elastic scattering in the strong scattering regime
\cite{0953-8984-14-12-202}
\be
\tau_{z}^{-1} =
4\left\langle\left[\tilde{\tau}_1(\alpha_R^2+\tilde{\beta}_D^2)k^2
    + \tilde{\tau}_3\gamma_D^2 k^6/16\right]\right\rangle, \quad\quad
\tau_{\pm}^{-1} =
2\left\langle\left[\tilde{\tau}_1(\alpha_R \mp\tilde{\beta}_D)^2 k^2
    + \tilde{\tau}_3\gamma_D^2 k^6/16\right]\right\rangle,
\label{srt-single-2d} 
\ee
Here $\tau_{i}$ ($i=z,+,-$) denotes the relaxation time of the spin
component along $i$ direction and $\langle...\rangle$ represents the
average defined by
Eq.~(\ref{srt-ave}). $\tilde{\beta}_D=\beta_D-\frac{1}{4}\gamma_Dk^2$. It
is noted that when the linear spin-orbit coupling dominates, the spin
lifetime varies as $1/\langle\tau_p\varepsilon_{\bf k}\rangle$ which changes
with temperature and density slowly. However, when the cubic term
becomes important, the spin lifetime can vary as
$1/\langle\tau_p\varepsilon_{\bf k}^2\rangle$ or
$1/\langle\tau_p\varepsilon_{\bf k}^3\rangle$. In the latter case, spin relaxation
varies rapidly with density and temperature.

\begin{figure}[bth]
\begin{minipage}[h]{0.45\linewidth}
  \centering
  \includegraphics[height=6cm]{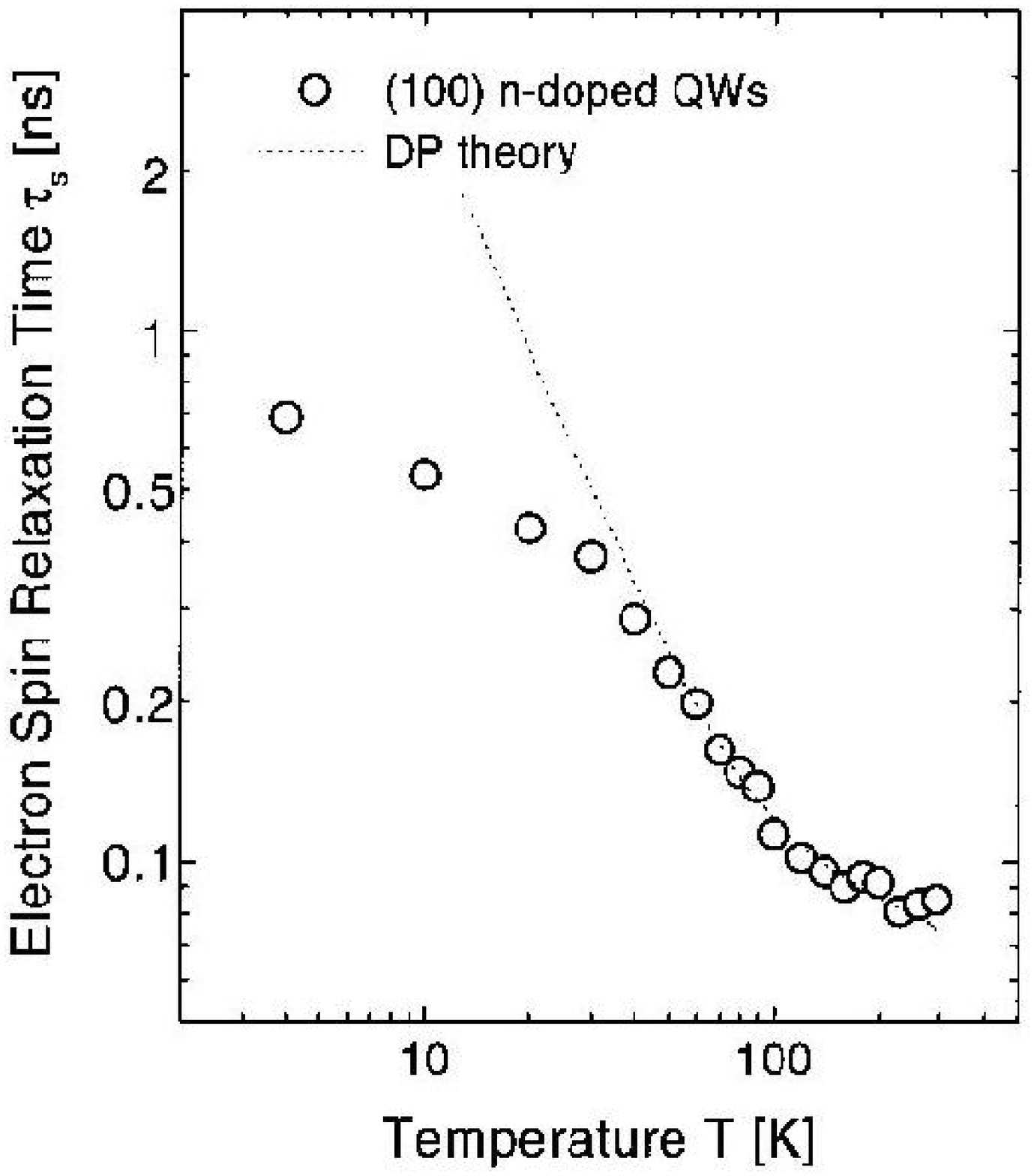}
  \caption{Temperature $T$ dependence of electron spin relaxation time
    $\tau_s$ for $n$-doped (001) GaAs/AlGaAs quantum wells with well
    width 7.5~nm. Electron density is $4\times 10^{10}$~cm$^{-2}$
    in each quantum well. Dotted curve is the  
    calculated result of spin lifetime based on the
    D'yakonov-Perel' theory. From Ohno et al. \cite{Ohno2000817}.}
  \label{Ohno_T_001}
\end{minipage}\hfill
\begin{minipage}[h]{0.45\linewidth}
  \centering
  \includegraphics[height=6cm]{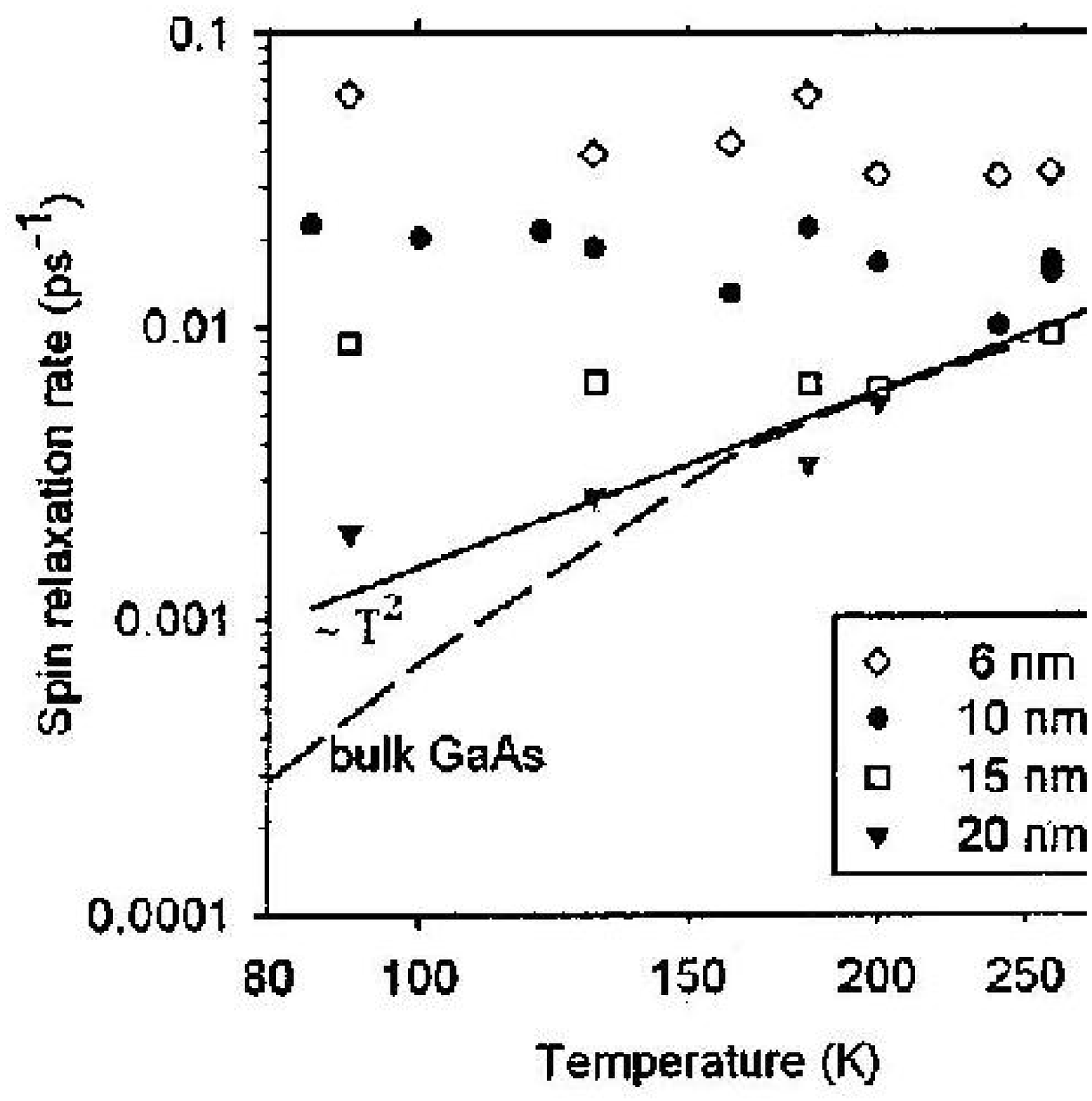}
  \caption{Temperature dependence of spin relaxation rate
    for undoped (001) GaAs/AlGaAs quantum wells with various well
    widths. $\Diamond$: well width 6~nm; $\medbullet$: well width
    10~nm; $\square$: well width 15~nm; $\blacktriangledown$: well
    width 20~nm. Solid curve is a fit with $1/\tau_s\sim T^2$
    dependence. Dashed curve represents the data in 
    intrinsic bulk GaAs by Maruschak et al. (as reproduced in
    Ref.~\cite{opt-or}). From Malinowski et
    al. \cite{PhysRevB.62.13034}.}
  \label{Harley_T_undop}
\end{minipage}\hfill
\end{figure}

{\bf $\bullet$ Temperature dependence.}
In experiments, the temperature dependence was measured in metallic
regime in
Refs.~\cite{PhysRevB.22.863,Ohno2000817,PhysRevB.62.13034,leyland:165309,leyland:195305,ruan:193307,studer:045302,1367-2630-8-4-049}.
In heavily-doped (low mobility) quantum wells, spin lifetime decreases
with increasing temperature.
A good example is the experiment of Ohno et al. \cite{Ohno2000817}
[Fig.~\ref{Ohno_T_001}]. In heavily-doped quantum wells, the momentum
scattering is dominated by the electron-impurity scattering
except that at high temperature the
electron-longitudinal-optical-phonon scattering may become more important.
When the electron-impurity scattering dominates, as both $\tau_p$ and
$\langle\varepsilon_{\bf k}\rangle$ increase with temperature, spin lifetime
decreases with increasing temperature
rapidly. At high temperature the rise of
electron-longitudinal-optical-phonon scattering slows down the
decrease or even leads to an increase of the spin lifetime with
increasing temperature.

In undoped quantum wells, Malinowski et al. \cite{PhysRevB.62.13034}
presented a systematic study on electron spin relaxation. The main
results are shown in Fig.~\ref{Harley_T_undop}. It is seen
that the temperature dependence of spin relaxation rate varies largely
with quantum well width. At small well width, spin relaxation rate
changes slowly with temperature, whereas at large well width, it increases rapidly with temperature. The temperature
dependence can be fitted roughly as $\sim T^0$ for narrow quantum well
and $\sim T^2$ for wide quantum well. The authors explained that as 
the longitudinal-optical-phonon scattering dominates at high temperature for undoped
quantum wells, roughly $\tau_p\sim T^{-1}$
\cite{2008arXiv0807.4845E}. For narrow quantum wells, spin relaxation
rate $\tau_s^{-1}\sim \tau_p\langle \varepsilon_{\bf k}\rangle\sim T^0$,
whereas for wide quantum wells $\tau_s^{-1}\sim \tau_p\langle
\varepsilon_{\bf k}^3\rangle\sim T^2$. It is noted that in the wide quantum
well, the increase of spin relaxation rate is still slower than that
in bulk samples at low temperature, which indicates the crossover of
the leading spin-orbit coupling from linear term to cubic term with
increasing temperature.

\begin{figure}[bth]
  \centering
  \includegraphics[height=5.5cm]{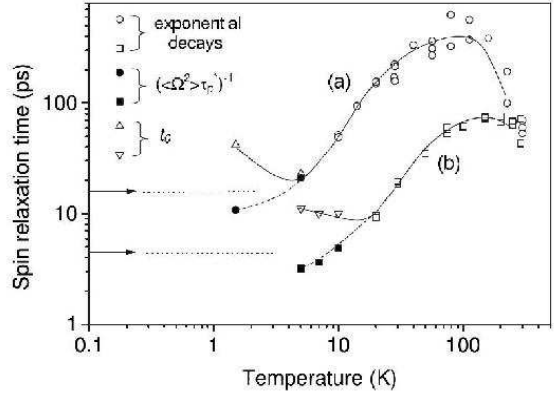}
  \caption{Temperature dependence of spin lifetime for
    sample (a) (well width $20$~nm, electron density $1.8\times
    10^{11}$~cm$^{-2}$) and (b) (well width 10~nm, electron density
    $3.1\times 10^{11}$~cm$^{-2}$). 
    The curves are guides for the eyes. Open circles and squares are
    spin-lifetimes measured in the high temperature regime where
    spin evolution is exponential. Solid symbols are values of
    $(|\Omega(k_F)|^2\tau_p^{\ast})^{-1}$ (denoted in the figure as
    $(\langle \Omega^2\rangle\tau_p^{\ast})^{-1}$) ($\Omega(k_F)$ is
    obtained from analysis of the spin evolution in the
    low-temperature oscillatory regime. $\tau_p^{\ast}$ is the 
    momentum scattering time including the electron-electron
    scattering.). Open triangles are the decay time constant of the
    oscillatory spin evolution, $t_0$, obtained by fitting the
    experimental data with oscillatory exponential decay
    $S_z(t)=S_0\exp(-t/t_0)\cos(\omega t+\phi)$. Arrows
    indicate the value of $|\Omega(k_F)|^{-1}$ for each sample,
    corresponding to the condition $|\Omega(k_F)|\tau_p\simeq
    1$. From Leyland et al. \cite{leyland:195305}.}
  \label{Harley_osT_dep}
\end{figure}

In high mobility two-dimensional electron system at sufficiently low
temperature the momentum scattering can be very weak. In this
situation, the D'yakonov-Perel' spin relaxation is in the weak
scattering regime where spin polarization shows precessional 
decay \cite{leyland:195305,PhysRevLett.89.236601}. Brand et
al. observed the transition from the precessional decay regime to the
motional-narrowing regime \cite{PhysRevLett.89.236601}. As spin
lifetime changes from $\tau_s \sim \tau_p^{-1}$ in motional-narrowing
regime to $\tau_s \sim \tau_p$ in precessional decay regime, the temperature
dependence of spin lifetime has a turning point around the
transition.\footnote{Such behavior also exists in hole spin relaxation
  in Si/Ge quantum wells as reported by Zhang and Wu in Ref.~\cite{zhang:155311}.} This was observed in experiment by Leyland et
al. (see Fig.~\ref{Harley_osT_dep}) \cite{leyland:195305}. Several
works on temperature dependence of spin relaxation in high mobility
two-dimensional electron system revealed the importance of
the electron-electron scattering to spin relaxation
\cite{leyland:165309,ruan:193307,leyland:195305,leyland:205321}.
This important issue will be reviewed in the framework of the
many-body theory in Section 5.

\begin{figure}[bth]
  \centering
  \includegraphics[height=5.2cm]{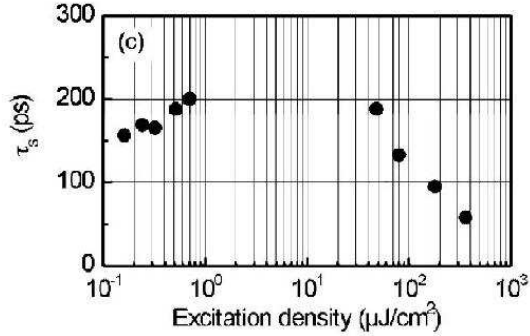}
  \caption{Excitation density dependence of spin lifetime $\tau_s$ in
    undoped (001) InGaAs multiple quantum wells at room
    temperature. From Aleksiejunas et al. \cite{aleksiejunas:102105}.}
  \label{Nex_qw}
\end{figure}

{\bf $\bullet$ Excitation density dependence.}
Excitation density dependence of spin relaxation was investigated in
Refs.~\cite{Snelling1990208,0268-1242-17-4-302,aleksiejunas:102105,studer:045302,0953-8984-19-29-295206,0295-5075-84-2-27006,PhysRevB.47.10907,Lu.5.326,zhao:115321}.
Interestingly, it was discovered that the spin lifetime decreases with
excitation density at low temperature in $n$-doped quantum wells
\cite{studer:045302,0953-8984-19-29-295206,PhysRevB.47.10907}. At room
temperature in undoped quantum wells, Aleksiejunas et al. found that
at low density spin lifetime increases with excitation density
whereas at high density it decreases (see Fig.~\ref{Nex_qw})
\cite{aleksiejunas:102105}. Moreover, Teng et al. observed a peak in
density dependence at room temperature in undoped quantum well
\cite{0295-5075-84-2-27006}. As in intrinsic quantum wells the
dominant scatterings are the many-body electron-electron and
electron-hole Coulomb scatterings, the results can not be explained in
the framework of single-particle approach. Nevertheless, 
many-body theories have revealed the underlying physics: the
electron-electron and electron-hole Coulomb scatterings increase with
density in nondegenerate (low density) regime, whereas decrease in
degenerate (high density) regime \cite{jiang:125206,jiang:155201,zhang:155311,0268-1242-24-11-115010}. The nonmonotonic density dependence
of these many-body carrier-carrier Coulomb scattering results in the
nonmonotonic density dependence of spin lifetime
\cite{jiang:125206,jiang:155201,zhang:155311,0268-1242-24-11-115010}.

\begin{figure}[bth]
\begin{minipage}[h]{0.45\linewidth}
  \centering
  \includegraphics[height=4.2cm]{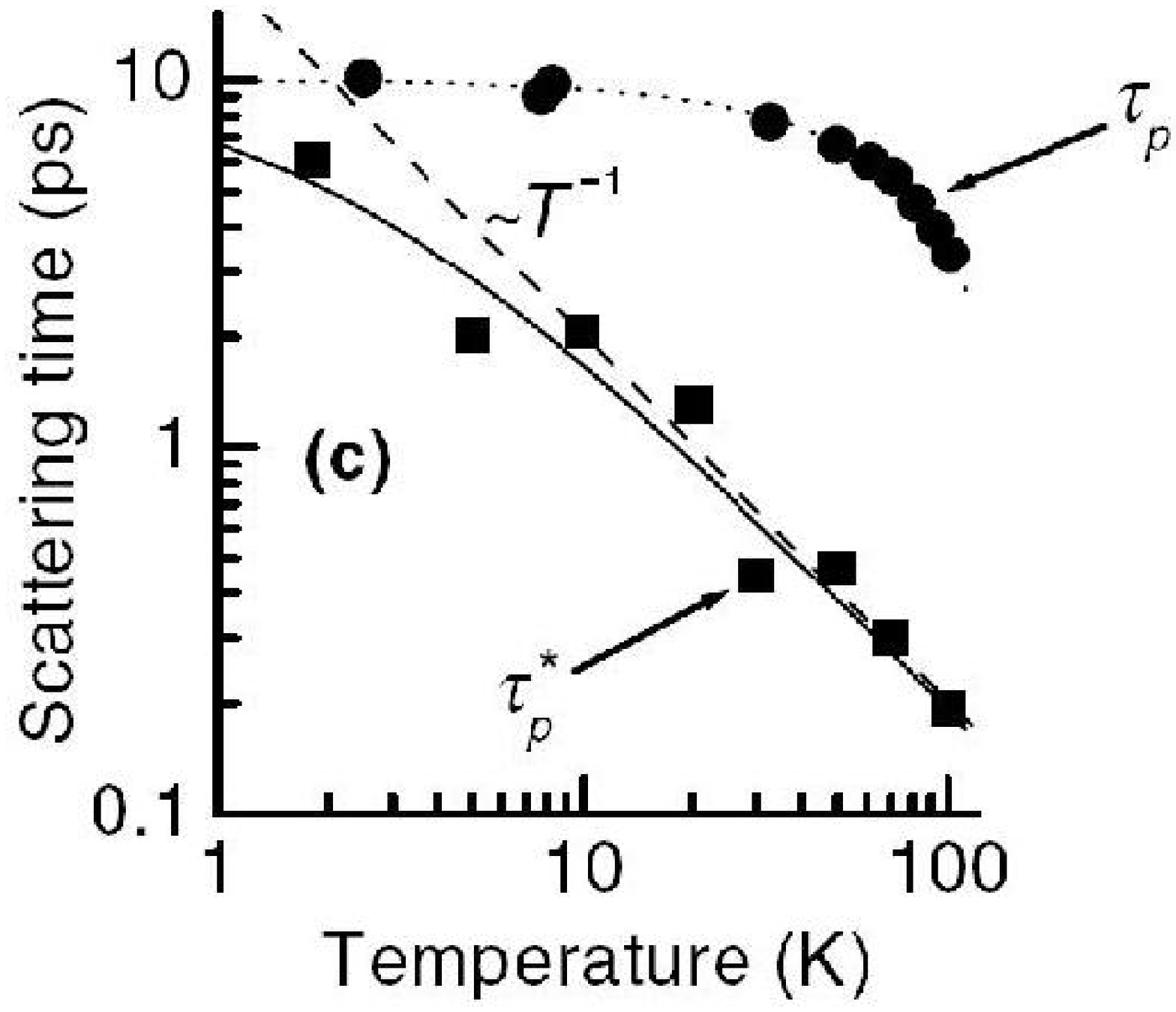}
  \caption{Momentum scattering time $\tau_p^{\ast}$ (the total
    momentum scattering including the electron-electron scattering)
    and $\tau_p$ (the momentum scattering time extracted from
    mobility) as function of temperature. The electron density is
  $1.9\times 10^{11}$~cm$^{-2}$. Dotted curve is a guide to 
    the eyes for $\tau_p$. Dashed curve represents an empirical fit of
    electron-electron scattering time as $\sim T^{-1}$ at high
    temperature. Solid curve sums these two contributions ($1/\tau_p$
    and the electron-electron scattering rate) as a guide
    curve to $\tau_p^{\ast}$. The data points of $\tau_p^\ast$ is
      obtained from Monte Carlo simulation of spin evolution. From
    Brand et al. \cite{PhysRevLett.89.236601}.} 
  \label{Harley_taup_Tdep}
\end{minipage}\hfill
  \begin{minipage}[h]{0.45\linewidth}
  \centering
  \includegraphics[height=5cm]{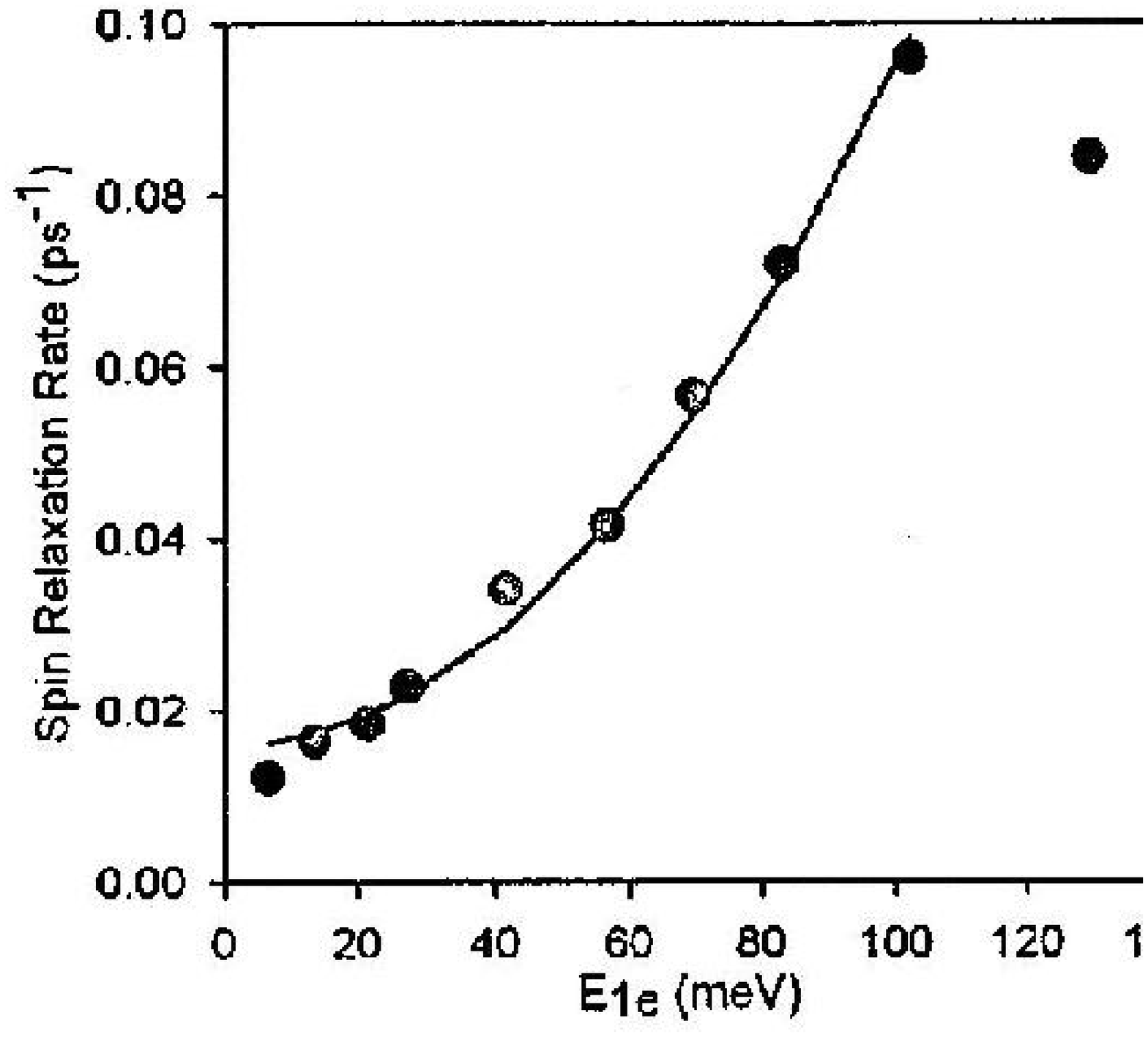}
  \caption{Spin relaxation rate as function of the subband
    quantization energy $E_{1e}$ in GaAs/Al$_x$Ga$_{1-x}$As quantum
    wells at room temperature. The solid curve is a fit to the
    data points of the form $\tau_s^{-1}=a+bE_{1e}^2$. From Malinowski
    et al. \cite{PhysRevB.62.13034}.}
  \label{Harley_width_dep}
\end{minipage}\hfill
\end{figure}

{\bf $\bullet$ Mobility dependence.}
Spin lifetime as function of mobility $\mu$ ($\propto \tau_p$) was studied in
Refs.~\cite{JJAP.38.2549,aleksiejunas:102105} at room temperature
where the qualitative relation $\tau_s\sim \mu^{-1}$ was observed which
signals the D'yakonov-Perel' spin relaxation mechanism. However, Brand
et al. found that in high mobility two-dimensional electron system at
low temperature ($T<100$~K), spin lifetime deviates from
$\tau_s\sim \mu^{-1}$ largely \cite{PhysRevLett.89.236601}. 
Electron-electron scattering is proved to be the key to understand the
observed results.\footnote{This was first predicted theoretically by
  Wu and Ning \cite{wu:epjb.18.373,JPSJ.70.2195} and later by Glazov and Ivchenko \cite{Glazov:jetp.75.403}.} In high mobility two-dimensional electron system at
low temperature, the electron-electron scattering is the dominant
scattering as other scatterings (the electron-impurity and
electron-phonon scatterings) are weak. As it randomizes the
momentum, the electron-electron scattering also contributes to the
D'yakonov-Perel' spin relaxation. However, since the electron-electron
scattering does not contribute to mobility, $\tau_s\sim \mu^{-1}$ does
not hold anymore. The total momentum scattering time with
the electron-electron scattering included, $\tau_p^{\ast}$, is then much
smaller than that deduced from mobility, $\tau_p$ (see
Fig.~\ref{Harley_taup_Tdep}). At room temperature,
the electron-longitudinal-optical-phonon scattering becomes the strongest
momentum scattering which also limits the mobility, hence the relation
$\tau_s\sim \mu^{-1}$ is recovered.

{\bf $\bullet$ Quantum well width dependence.}
The quantum well width dependence of spin relaxation was investigated in
Refs.~\cite{tackeuchi:797,tackeuchi:1131,britton:2140,JJAP.38.2549,JJAP.38.4680,0268-1242-14-3-002,Ohno2000817,PhysRevB.62.13034}.
All these studies were performed at room temperature in nondegenerate regime.
In this regime, by neglecting the cubic Dresselhaus term and the Rashba term and assuming
infinite well depth, spin lifetime is given by \cite{d'yakonov:110} 
$\tau_s = E_g /( 2\alpha^2E_{1e}^2k_BT \tau_p)$
with $\alpha = 2\gamma_{D}\sqrt{2m_e^3E_g}$ and
$E_{1e}$ being the electron subband quantization energy. It
hence gives $\tau_s\sim E_{1e}^{-2}$. Most experimental data
roughly agree with the relation. For example, Tackeuchi et
al. fitted $\tau_s\sim E_{1e}^{-2.2}$ for multiple GaAs/AlGaAs quantum
well \cite{tackeuchi:797}. However, there are some situations
where results deviate largely from the above relation \cite{JJAP.38.2549}.
Britton et al. reported that for small $E_{1e}$, the relation is
largely deviated whereas for large $E_{1e}$, the spin lifetime
generally shows the quadratic behavior \cite{britton:2140}. A similar
result was obtained by Malinowski et al. \cite{PhysRevB.62.13034}
which is shown in Fig.~\ref{Harley_width_dep}. It is seen that for
medium $E_{1e}$, the relation agrees well with experimental result. At
small $E_{1e}$ as the cubic Dresselhaus spin-orbit coupling plays
important role, the relation deviates from the data. At large $E_{1e}$
as the wavefunction penetration is unnegligible and the
infinite-depth-well assumption no longer holds, the relation also
deviates from experimental result.

\begin{figure}[bth]
  \centering
  \includegraphics[height=6.5cm]{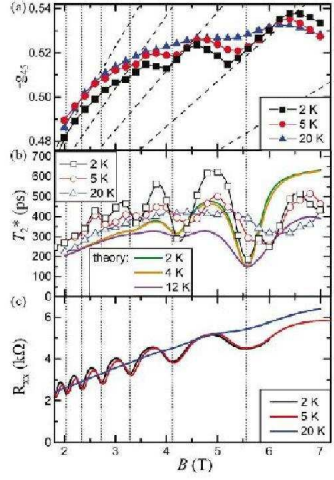}
  \caption{ Up panel: Spin dephasing time $T_2^\ast$ measured
    (symbols) as a function of magnetic field $B$ at $T=2$~K, 5 K, and
    20 K and calculated (curves) from a spin relaxation model at
    $T=2$~K, 4 K, and 12 K. Down panel: the longitudinal resistivity
    $R_{xx}$ as a function of B at $T=2$~K, 5 K, and 20 K. Measurement
    is performed in single $n$-modulation doped InGaAs/GaAs quantum
    well. Also shown are dotted curves indicating the
    position of $B_n$ for Landau-level index $n=16$, 14, 12, 10, 8 and
    6 ($n$ labels the filled Landau levels). From Sih et
    al. \cite{PhysRevB.70.161313}.} 
  \label{Sih_B_dep}
\end{figure}

{\bf $\bullet$ Magnetic field dependence.}
Spin lifetime has also been measured as function of magnetic
field in quantum wells in metallic regime in
Refs.~\cite{gerlovin:115330,studer:045302,stich:073309,korn,lau:142104,PhysRevB.70.161313}.
There are two special configurations of magnetic field: one that the
magnetic field lies in the quantum well plane (the Voigt configuration)
and the other that magnetic field is perpendicular to the well
plane (the Faraday configuration). In the Voigt configuration, the orbital
effect of magnetic field is negligible. In this case the magnetic
field has two consequences on spin relaxation: first it mixes the
in-plane and out-of-plane spin relaxations due to the Larmor precession;
second the Larmor spin precession slows down spin relaxation by a
factor of $(1+\omega_L^2\tau_p^2)$. In usual condition $\omega_L\tau_p\lesssim
0.1$ (e.g., $B$=2~T and $\tau_p=1$~ps in GaAs yield
$\omega_L\tau_p=0.05$), thus the second effect is weak. The first
effect is usually more important as the in-plane and out-of-plane spin
lifetimes differ largely. The effect of mixing 
saturates soon at a low magnetic
field around 0.1~T. This is consistent with experiments: the magnetic
field dependence is usually more pronounced at low field
\cite{studer:045302,stich:073309,lau:142104}, whereas at high field
the dependence becomes weak \cite{gerlovin:115330}. In asymmetric quantum
wells the in-plane spin relaxation can be quite anisotropic. In this
case the effect of magnetic field depends largely on its direction
\cite{averkiev:033305,liu:112111,stich:073309,larionov:033302,studer:045302}.
Interestingly, Stich et al. found that the magnetic field dependence
of spin lifetime for $B\parallel [110]$ exhibits a minimum
whereas the magnetic dependence for $B\parallel [1\bar{1}0]$ shows a
maximum for an asymmetric (001) quantum well \cite{stich:073309}. In
the Faraday configuration with magnetic field perpendicular to quantum
well plane, the orbital effect induces cyclotron motion which has
an important effect on spin relaxation. As the cyclotron frequency is
much larger than the Larmor frequency (in GaAs $\omega_c\simeq
70\omega_L$), the cyclotron motion effectively suppresses the spin
relaxation. This phenomena was observed by Sih et al. in InGaAs/GaAs
quantum well (See Fig.~\ref{Sih_B_dep}) \cite{PhysRevB.70.161313}.\footnote{Also observed in Si/Ge quantum well by Wilamowski and
  Jantsch \cite{PhysRevB.69.035328}.} At higher magnetic field and
low temperature, spin lifetime oscillates with magnetic field,
as the Landau level filling affects both spin precession and momentum
scattering. Finally in materials with strong energy dependence of the
$g$-factor (e.g., in narrow band-gap semiconductors) at
high magnetic field, the $g$-tensor inhomogeneity mechanism can be
important. In InGaAs quantum wells, spin relaxation was observed to
first increase and then decrease with increasing magnetic field for an
in-plane magnetic field \cite{Morita20041007}. The increase is due to
the mixing of out-of-plane and in-plane spin relaxations, whereas the
decrease can be attributed to the $g$-factor inhomogeneity spin
relaxation mechanism.

{\bf $\bullet$ Gate-voltage dependence.}
The spin relaxation can also be tuned by the gate-voltage
\cite{gerlovin:115330,larionov:033302}. The gate-voltage may have
several consequences on spin relaxation. First it changes electron
density. Second it changes the Rashba spin-orbit coupling. The
modification of the envelope function along the growth direction further
changes the linear Dresselhaus spin-orbit coupling. Third it may
change the mobility. Hence the underlying mechanism for the
gate-voltage dependence of the spin lifetime is complex.

{\bf $\bullet$ Initial spin polarization dependence.}
Initial spin polarization dependence was studied by Stich et al.
\cite{stich:176401,stich:205301,korn} where spin relaxation rate
decreases with initial spin polarization as predicted by Weng and Wu
\cite{PhysRevB.68.075312}. The underlying physics will be reviewed in
the next section on the kinetic spin Bloch equation approach
\cite{PhysRevB.61.2945,wu:epjb.18.373,PhysRevB.68.075312} where
the Coulomb Hartree-Fock term acts as an effective magnetic field in
spin precession \cite{PhysRevB.68.075312}. This effective magnetic
field is always along spin polarization direction. Realistic
calculations from the kinetic spin Bloch equations indicated that the
induced magnetic field can be as large as tens of Tesla
\cite{PhysRevB.68.075312}. Such large longitudinal effective magnetic
field largely suppresses the spin relaxation. Besides spin relaxation,
experimentalists also found evidence of the effective magnetic field
in the sign change of the Kerr rotation \cite{0295-5075-83-4-47006}.
Recently the Coulomb Hartree-Fock effective magnetic field was again
discussed in the issue of spin accumulation \cite{saarikoski:097204}.

{\bf $\bullet$ Excitation photon energy dependence and others.}
The excitation photon energy dependence of spin relaxation was
investigated in Ref.~\cite{leyland:205321} where the energy-dependent
momentum scattering time due to the electron-electron scattering was
revealed. Spin dynamics in higher subband was studied in
Refs.~\cite{0953-8984-16-4-013,0295-5075-83-4-47007,luo-2009},
indicating the important role of inter-subband momentum
scattering. Spin dynamics under microwave driving field was discussed
in Ref.~\cite{luo:192107}.

{\bf $\bullet$ Interface inversion asymmetry induced spin relaxation.}
The interface inversion asymmetry induced spin-orbit coupling and its
consequence to spin relaxation were studied in
Refs.~\cite{PhysRevB.58.R10179,PhysRevB.64.201301,PhysRevB.68.115311}.\footnote{The
  well width dependence of spin lifetime in narrow 
  quantum wells where the well and barrier materials do not share any
  common atom (e.g., in InGaAs/InP quantum well) may also contain the
  information of the interface inversion asymmetry
  \cite{tackeuchi:1131,Tackeuchi1999318}.} As was reviewed in
Section~2.4, in the case where the quantum well and the
barrier materials share no common atom, the interface inversion asymmetry
exhibits even in symmetric quantum wells. The interface inversion
asymmetry induced spin-orbit coupling is important in narrow quantum
wells. By comparing with the case with common atom of the same well
width, it was discovered that spin relaxation in quantum wells without
common atom is always shorter
\cite{tackeuchi:1131,PhysRevB.58.R10179,PhysRevB.64.201301}. This
effect is especially marked in narrow quantum wells. It can not be
explained by the D'yakonov-Perel' theory taking into account only the
Dresselhaus spin-orbit coupling \cite{tackeuchi:1131}. As in
[110]-grown quantum well, the two interfaces could be symmetric even
in the case without common atom, which gives a special case {\em
  without} the interface-induced spin-orbit coupling
\cite{PhysRevB.68.115311}. By comparing spin relaxation in [110]-grown
quantum well and [001]-grown quantum well with same well width, one
can also estimate the value of interface-induced spin-orbit coupling
\cite{PhysRevB.68.115311}.

{\bf $\bullet$ Spin relaxation in diluted nitride materials.}
Spin relaxation in diluted nitride materials such as GaAsN and InGaAsN
was studied in Refs.~\cite{lombez:252115,Lagarde.pssb.204.208,reith:211122}.
Interestingly, the spin lifetime increases with increasing temperature
up to room temperature for $T>40$~K, but decreases with increasing
temperature below 40~K. It was found that less than 1\% doping of
nitride in GaAs and InGaAs makes the spin lifetime at room temperature
increase by more than one order of magnitude in as-grown samples
before annealing \cite{lombez:252115}. After annealing the spin
lifetime is drastically droped \cite{lombez:252115}. The above
unusual behavior indicates that localized electrons bound to nitride
dopants play an important role. Even for delocalized electrons,
scattering with nitride dopants greatly reduces $\tau_p$ and hence
increases the spin lifetime. Annealing improves the crystal quality of
diluted nitride material which hence increases the mobility and reduces
the spin lifetime. Spin lifetime at room temperature as function of
subband quantization energy $E_{1e}$ gives $\tau_s\sim E_{1e}^{-1}$ in unannealed
InGaAsN/GaAs multiple quantum wells \cite{reith:211122}.

{\bf Single-particle theories for spin relaxation in (001) quantum wells}\\
\indent We now turn to review the single-particle theory of electron spin
relaxation in $n$-type and intrinsic two-dimensional electron
system. As in bulk system, the single-particle theory
assumes that for all ${\bf k}$ the strong scattering criteria
$\Omega_{\bf k}\tau_p\ll 1$ is fulfilled. It is also assumed that the
carrier-carrier scattering is irrelevant.
\footnote{However, it has been found that in high mobility
  two-dimensional system, electron-electron scattering is the dominant
  momentum scattering to the D'yakonov-Perel' spin relaxation
  \cite{PhysRevLett.89.236601,PhysRevB.68.075312,zhou:045305}.}
Within the elastic scattering approximation,\footnote{Although most of
  the single-particle theory is based on the elastic scattering
  approximation, some works go beyond that. Dyson and Ridley
  developed a method to calculate the momentum scattering time due to
  electron-longitudinal-optical-phonon scattering beyond the elastic
  scattering approximation. In bulk system, they showed that the
  elastic scattering approximation may have problem in treating the
  longitudinal-optical-phonon scattering as it is essentially
  inelastic \cite{PhysRevB.69.125211}. They further applied their
  method to study the D'yakonov-Perel' spin relaxation associated with
  the electron-longitudinal-optical-phonon scattering in quantum wells
  and wires in Ref.~\cite{PhysRevB.72.045326}.} spin lifetime
for system near the equilibrium (also implying that the spin
polarization is very small) can be calculated via
Eq.~(\ref{srt-single-2d}). Such paradigm has
been widely applied to study spin relaxation. Often the spin-orbit
coupling in the two-dimensional structure is calculated via the
${\bf k}\cdot{\bf p}$ method within the multiband envelope-function
approximation together with the Schr\"odinger-Poisson equation of the
heterostructure (see, e.g., Ref.~\cite{lau-2004}). The momentum
scattering times are calculated via the formulae developed for
calculating the mobility due to the electron-impurity and electron-phonon
scatterings. It should be mentioned that in heterostructures many
factors such as structure, doping and gate voltage can affect the
spin-orbit coupling. To quantitatively determine the spin lifetime,
one has to quantitatively determine the spin-orbit coupling. In
theoretical calculation, the spin-orbit coupling is determined by the
multiband envelope-function calculation with realistic structure
parameters. In experiments, the Rashba and linear Dresselhaus
spin-orbit couplings can be determined quantitatively via several
methods, such as electric-field-induced spin precession
\cite{lmeier:650,studer:045302} and current-induced modification of
$g$-factor \cite{wilamowski:prl187203}.

\begin{figure}[bth]
  \vskip -0.2cm
  \centering
  \includegraphics[height=6cm]{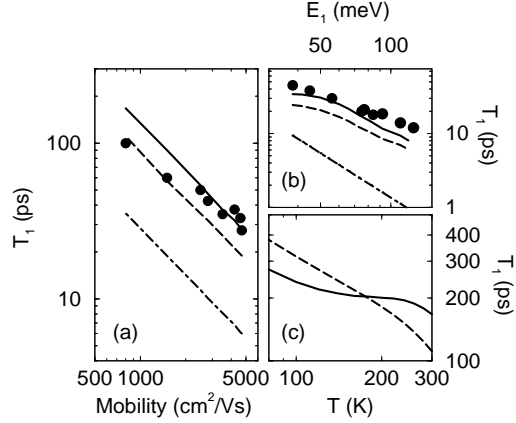}
  \caption{Electron-spin lifetime $T_1$ as a function of (a) mobility, (b)
    subband quantization energy $E_{1}$, and (c) temperature, for 75-\AA~ 
    GaAs/Al$_{0.4}$Ga$_{0.6}$As multiple quantum wells at room temperature. Dots
    represent the results of experiment (Ref.~\cite{JJAP.38.2549}). The
    theoretical results with electron-optical-phonon scattering
    (solid curves) and electron-neutral-impurity scattering (dashed curves) are
    shown, as well as the results from the D'yakonov-Kachorovskii
    theory (dot-dashed curves). From Lau et al. \cite{PhysRevB.64.161301}.}
  \label{Flatte_2d_cal}
\end{figure}

A good success of such paradigm is that it well reproduces the
subband quantization energy $E_{1e}$ dependence of the electron spin lifetime
\cite{PhysRevB.64.161301} which can not be explained by the
simple relation of $\tau_s\sim E_{1e}^{-2}$ from the
D'yakonov-Kachorovskii theory [see Fig.~\ref{Flatte_2d_cal}~(b)]. This is
due to the fact that this paradigm takes full account of spin-orbit
coupling via diagonalizing the multiband envelope-function equation.
In the same work Lau et al. also demonstrated that their calculation
achieved better agreement with experiments in the mobility dependence
of the spin lifetime compared with the D'yakonov-Kachorovskii theory (see
Fig.~\ref{Flatte_2d_cal}). They also demonstrated the role of the
electron-longitudinal-optical-phonon scattering to spin relaxation as
function of temperature compared with that of the impurity scattering
as depicted in Fig.~\ref{Flatte_2d_cal}.
Systematic calculation via such paradigm of various dependences of
electron spin lifetime can be found in Ref.~\cite{lau-2004}.

\begin{figure}[bth]
  \centering
  \includegraphics[height=6.5cm]{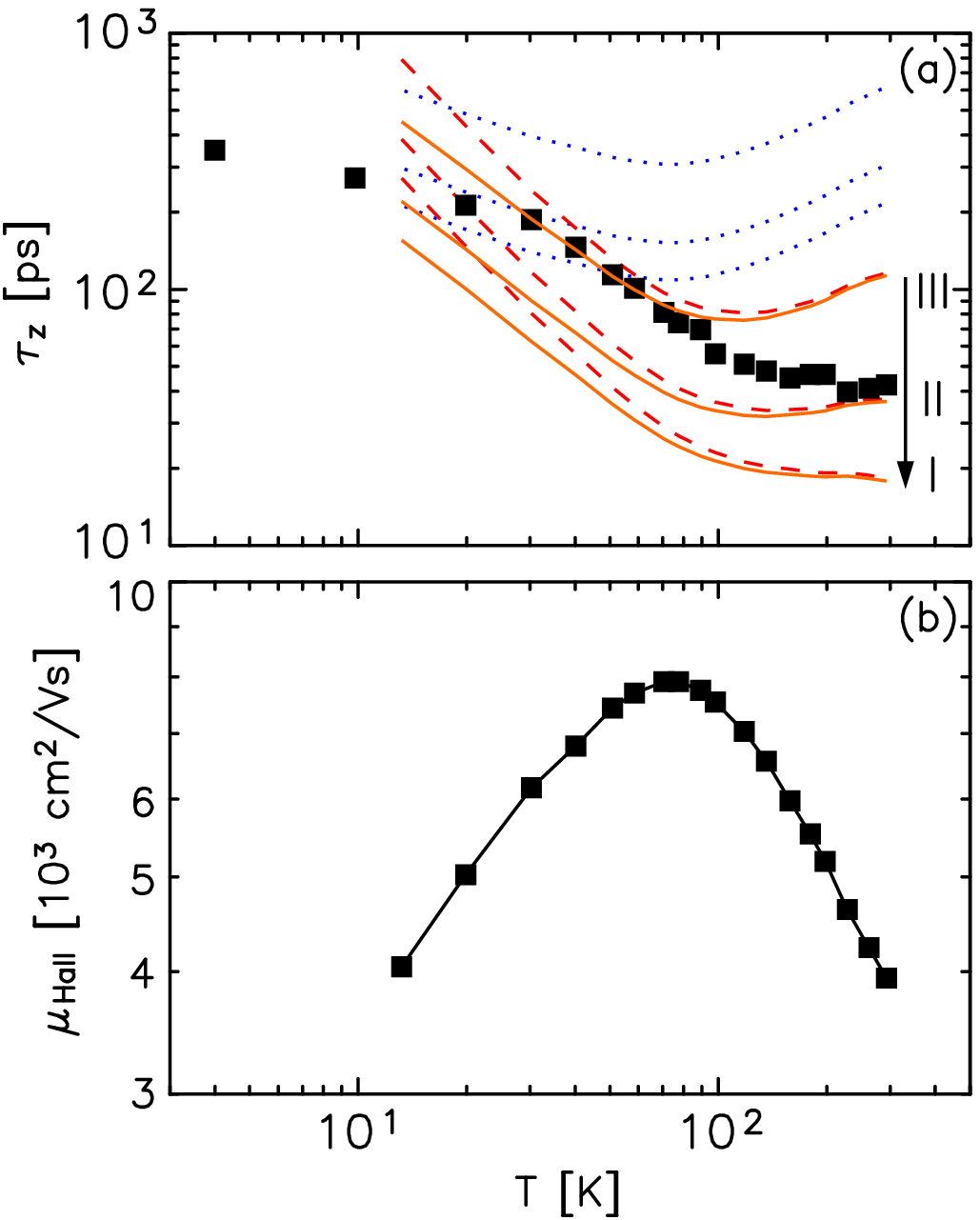}
  \caption{(a) Spin-lifetime $\tau_z$ and (b) measured Hall
    mobility $\mu_{\rm Hall}$ in experiments as a function of temperature $T$. In (a)
    three solid curves denote calculation with the mobility attributed
    to the three types of scattering (labeled as III, II and I in
      the figure.) The first type of scattering (I): phonon
      scatterings due to the deformation potential, 
    scattering by screened ionized impurities, neutral impurities, alloy
    and surface roughness; The second type of scattering (II): other kinds of
    phonon scatterings; The third type of scattering (III): scattering by
    weakly screened ionized impurities. The dotted curves refer to
    the calculation in the degenerate limit (zero temperature) 
    whereas the dashed curves denote the calculation in the
    nondegenerate limite (the Boltzmann statistics).
    From Kainz et al. \cite{PhysRevB.70.195322}.}
  \label{Kainz_T_dep}
\end{figure}

{\bf $\bullet$ Temperature dependence.}
The temperature dependence of the D'yakonov-Perel' spin relaxation was
calculated within such paradigm in
Refs.~\cite{PhysRevB.64.161301,Averkiev:36.91,PhysRevB.70.195322}.
Kainz et al. attempted to perform a microscopic calculations of
the temperature-dependent spin-relaxation rates with realistic system
parameters \cite{PhysRevB.70.195322}. By using the parameters from
experiments, they calculated the temperature dependence of spin relaxation
time and compared it with the experimental data by Ohno et
al. \cite{Ohno2000817} (see Fig.~\ref{Kainz_T_dep}). In their
calculation the momentum scattering time was {\em not} calculated
microscopically but inferred from the Hall mobility. They classified
three types of scatterings and assumed that the Hall mobility is solely
limited by each type of scattering. By doing so they obtained three
spin lifetimes. Finding that the experimental measured spin
lifetime falls into the region determined by the three
calculated ones, they concluded that the calculation agrees with the
experiments. By close examination, they found that the
electron-ionized-impurity scattering dominates at low temperature
($T<100$~K), whereas the electron-longitudinal-optical-phonon
scattering dominates at high temperature \cite{PhysRevB.70.195322}. The
temperature dependence was found to be more pronounced at low electron
density \cite{Averkiev:36.91,PhysRevB.70.195322}.

\begin{figure}[bth]
  \centering
  \includegraphics[height=6.5cm]{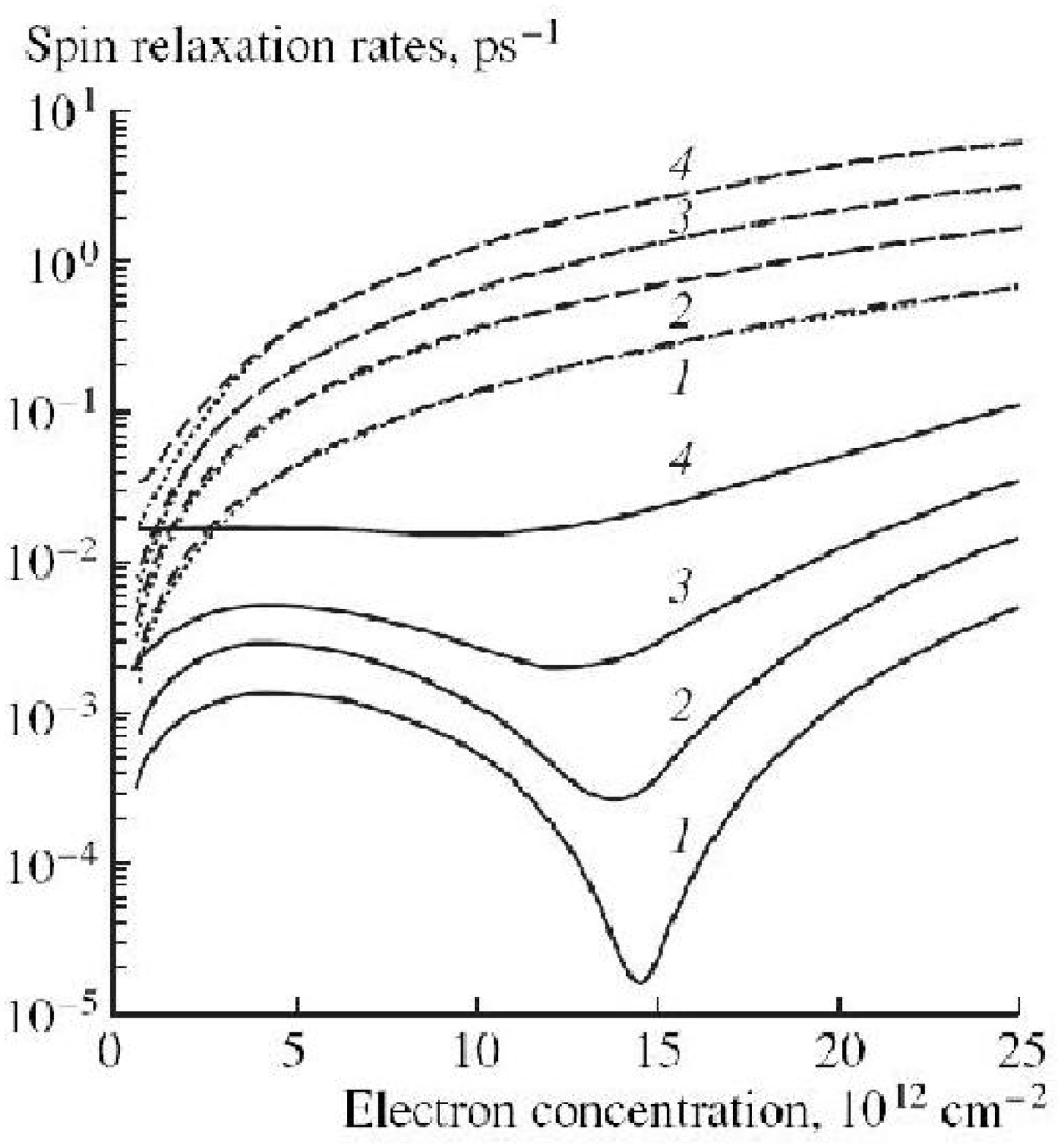}
  \caption{Spin-relaxation rates $1/\tau_{+}$ (solid curves),
    $1/\tau_{z}$ (dashed curves) and  $1/\tau_{-}$ (dotted curves)  as
    function of electron concentration for Boltzmann electron gas in
    GaAs/AlAs heterostructure at temperature $T$=(1) 30, (2) 77, (3)
    150 and (4) 300 K. From Averkiev et
    al. \cite{Averkiev:36.91}.}
  \label{Ne_dep_Golub}
\end{figure}

{\bf $\bullet$ Electron density dependence.}
The electron density dependence of spin relaxation was calculated in
Refs.~\cite{0957-4484-11-4-304,Averkiev:36.91,PhysRevB.68.075322,PhysRevB.70.195322,gilbertson:165335,li:153307}.
At zero temperature (degenerate regime), Golub et al. found that spin
relaxation rate increases monotonically with electron density
\cite{0957-4484-11-4-304}. Interestingly, Averkiev et al. found that
the ratio of spin lifetime $\tau_{-}/\tau_{+}$ [$\tau_{+}$
($\tau_{-}$) denotes spin lifetime for spin along $[110]$
($[1\bar{1}0]$) direction) has a peak in density dependence: at low
density the spin relaxation is dominated by $1/\tau_{-}$ whereas at high
density $1/\tau_{+}$ becomes more and more important for
$\alpha_R,\beta_D>0$  \cite{0957-4484-11-4-304,Averkiev:36.91}. From
Eq.~(\ref{srt-single-2d}), one can
understand the above results by noting that 
$\tilde{\beta}_D=\beta_D-\frac{1}{4}\gamma_Dk^2$ can become negative
at large $k$. Similar results have been obtained in temperature dependence
in nondegenerate regime \cite{Averkiev:36.91}. In heterojunction,
both the Rashba spin-orbit coupling and the linear Dresselhaus term
depend on the electron density as the built-in electric field and the
wave-function across the junction vary with electron density.
$\alpha_R$ and $\beta_D$ in heterojunction can be estimated as
$\alpha_R \simeq \alpha_0 n_e e^2 /(2\kappa_0\epsilon_0)$ and $
\beta_D \simeq \gamma_D \big[16.5\pi n_e e^2
    m^{\ast} /(8\kappa_0\epsilon_0)\big]^{\frac{2}{3}}$,
where $\alpha_0$ represents the Rashba coefficient, $\kappa_0$ is static
dielectric constant and $\epsilon_0$ stands for the vacuum dielectric
constant \cite{Averkiev:36.91}. Averkiev et al. found that the spin relaxation rate
$1/\tau_{+}$ has a minimum in the density dependence due to the
cancellation of the Rashba and linear Dresselhaus spin-orbit
coupling, whereas $1/\tau_{-}$ and $1/\tau_{z}$ increase with density
monotonically (see Fig.~\ref{Ne_dep_Golub}) \cite{Averkiev:36.91}. At the electron
density where $1/\tau_{+}$ has a minimum, the spin relaxation is
highly anisotropic. A more careful consideration of the same problem
within the multiband envelope-function approach was given by Kainz et
al. for various well width in
Refs.~\cite{PhysRevB.68.075322,PhysRevB.70.195322} where the
spin-orbit coupling was treated more carefully. Via similar approach
the Rashba and Dresselhaus spin-orbit couplings in $\delta$-doped
InSb/Al$_{x}$In$_{1-x}$Sb asymmetric quantum well was calculated in a
range of carrier densities \cite{gilbertson:165335}.\footnote{Density
  dependence of spin splitting in such structure was measured and
  compared with calculation in Ref.~\cite{gilbertson:235333}.} Based
on these results, the density dependence of spin lifetime was
calculated where a minimum in the density dependence of the in-plane
spin relaxation rate was also observed \cite{gilbertson:165335}. The
spin-orbit coupling and the density dependence of spin relaxation time
in such quantum well were also studied by Li et al. \cite{li:153307},
where the calculation was compared with the experimental results in
Ref.~\cite{1367-2630-8-4-049}. The authors also showed from the
eight-band ${\bf k}\cdot {\bf p}$ model that the spin-orbit coupling
deviates strongly from the linear-${\bf k}$ Rashba/Dresselhaus model
\cite{li:153307}.

\begin{figure}[bth]
  \begin{minipage}[h]{0.5\linewidth}
    \centering
    \includegraphics[height=7cm,angle=90]{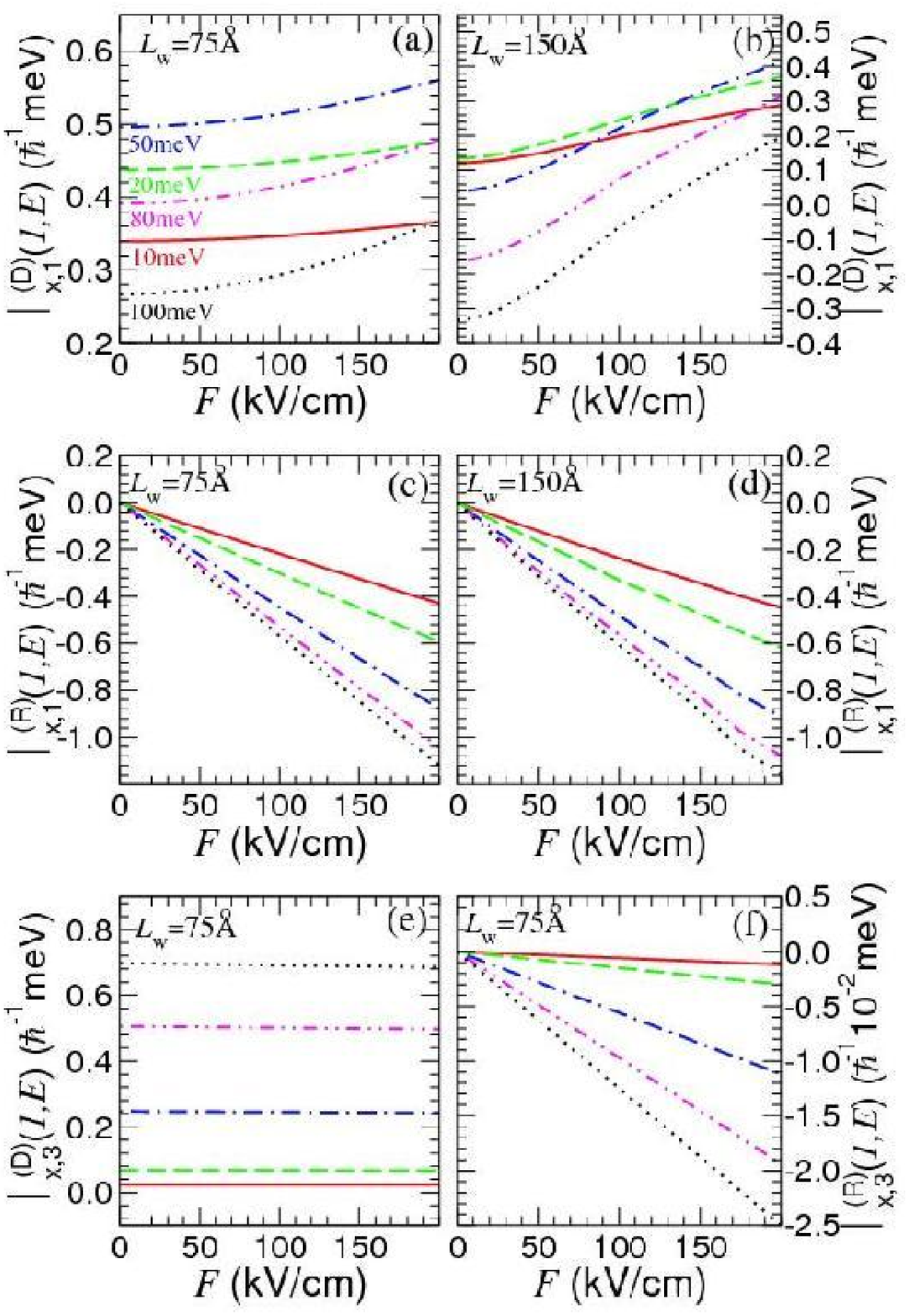}
  \end{minipage}\hfill
  \begin{minipage}[h]{0.5\linewidth}
    \centering
    \includegraphics[height=7cm,angle=90]{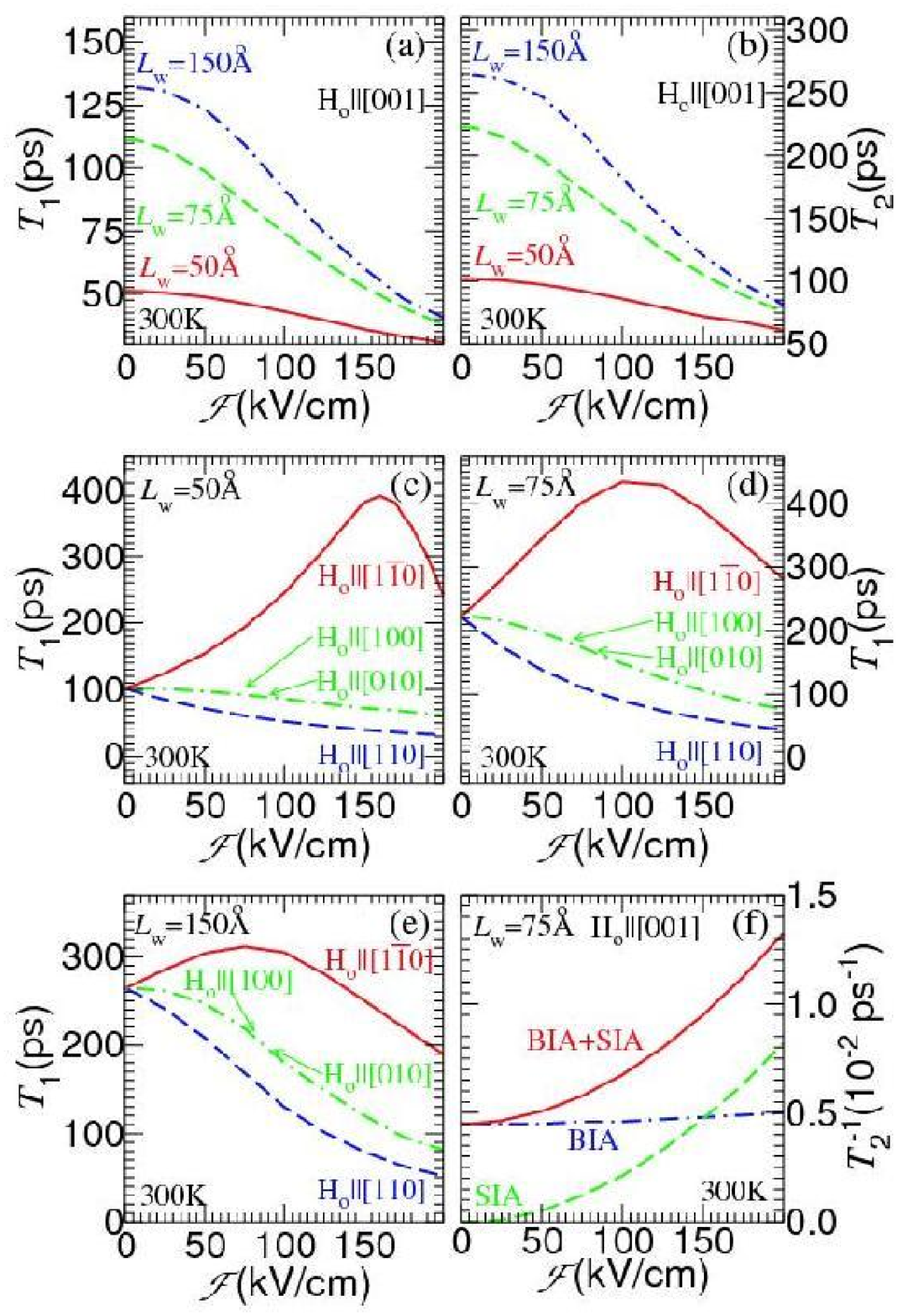}
  \end{minipage}\hfill
  \caption{Left figure: Electric field ${\cal{F}}$ (along the growth
    direction) dependence of electron spin precession vector for a
    75~\AA~ and a 150~\AA~ GaAs/Ga$_{0.6}$Al$_{0.4}$As quantum well at
    300 K for different energies $E=10$~meV (red solid curve), 20~meV
    (green dashed curve), 50~meV (blue dot-dashed curve), 80~meV
    (purple double-dot-dashed curve) and 100~meV (black dotted
    curve). $x$-component of the spin precession vector induced by
    the linear Dresselhaus term $\Omega^{({\rm D})}_{x,1}(1,E)$ for quantum wells with 
    well width (a) $L_w=75$~\AA~ and (b) $L_w=150$~\AA. $x$-component
    of the spin precession vector induced by the Rashba term
    $\Omega^{({\rm R})}_{x,1}(1,E)$ for (c) $L_w=75$~\AA~ and (d)
    $L_w=150$~\AA~ quantum wells. $x$-component of the spin precession
    vector induced by the cubic Dresselhaus term $\Omega^{({\rm
      D})}_{x,3}(1,E)$ for (e) $L_w=75$~\AA. $x$-component of the
     spin precession vector induced by higher order (cubic) Rashba term
  $\Omega^{({\rm R})}_{x,3}(1,E)$ for (f) $L_w=75$~\AA~ (Note that 
    the scale is two orders of magnitude smaller). Right figure:
    Electric field ${\cal{F}}$ (along the growth direction) dependence
    of electron spin relaxation time $T_1$, dephasing time $T_2$ and
    dephasing rate $1/T_2$ for $L_w=50$~\AA, $L_w=75$~\AA and
    $L_w=150$~\AA~ GaAs/Ga$_{0.6}$Al$_{0.4}$As quantum wells at 300~K with
    mobility $\mu=800$~cm$^2$/V~s. Note that the relative importance
    of the Dresselhaus (BIA) term and Rashba (SIA) term to 
    spin relaxation as function of the electric field along the growth
    direction is depicted in (f). From Lau and Flatt\'e
    \cite{PhysRevB.72.161311}.} 
  \label{Flatte_E_dep}
\end{figure}

{\bf $\bullet$ Gate-voltage dependence.}
The gate-voltage dependence of spin lifetime for a triangular
quantum well defined by a structure with an infinite height barrier at
left boundary and a constant electric field at right boundary was
studied by Averkiev et al. \cite{Averkiev:36.91}. They found that the
in-plane spin lifetime $\tau_{+}$ has a maximum when the Rashba
and linear Dresselhaus spin-orbit couplings cancel each other.
The gate-voltage dependence of the spin-orbit coupling in $\delta$-doped
InSb/In$_{1-x}$Al$_{x}$Sb asymmetric quantum well was calculated in
Ref.~\cite{gilbertson:165335}. It was also shown in
Ref.~\cite{lau:8682} that the spin lifetimes can be tuned
effectively via the electric field along the growth direction. Systematic study
on the dependence of spin lifetime on the electric field
across the GaAs/AlGaAs quantum well at room temperature was performed by
Lau and Flatt\'e \cite{PhysRevB.72.161311}. From a 14-band
envelope-function approach they calculated the Rashba and Dresselhaus
spin-orbit couplings as function of electric field for various well
widths (see Fig.~\ref{Flatte_E_dep}). They further calculated the
spin lifetime as function of electric field and reported that the
electric field effect is important at wide 
quantum well where the subband wavefunction is easily affected by the
electric field. Again the spin lifetime along the [110] direction
has a maximum as function of electric field. For a 7.5~nm quantum well
the contribution of the Rashba spin-orbit coupling in spin relaxation
exceeds that of the Dresselhaus one at 150~kV/cm. For
such narrow quantum well, the Dresselhaus spin-orbit coupling varies
little with electric field (up to 200~kV/cm), whereas the Rashba
spin-orbit coupling varies linearly with electric field. Similar
results were obtained by Yang and Chang \cite{yang:193314}.

{\bf $\bullet$ Nonlinear Rashba spin-orbit coupling to spin relaxation.}
Usually the multi-band envelope-function calculation includes the
nonparabolic effect in the spin-orbit coupling, e.g., at high energy the
Rashba spin-orbit coupling deviates from linearity
\cite{PhysRevB.72.161311,yang:193314}. The nonlinear effect in the Rashba
spin-orbit coupling was studied comprehensively by Yang and Chang
\cite{yang:193314}. They proposed a model to characterize such nonlinear
effect where the Rashba parameter is substituted by $\tilde{\alpha}_R
= \frac{\alpha_R}{1 + \zeta k^2}$, where $\zeta$ is a parameter
depending on quantum well structure and material. The model fits well
with the calculation. The nonlinearity is more pronounced in narrow
band-gap semiconductors. Within such model the spin relaxation as function
of electric field, well width, density and the ratio
$\alpha_R/\beta_D$ was investigated comprehensively \cite{yang:193314}.

{\bf $\bullet$ Magnetic field dependence.}
The magnetic field dependence of the spin relaxation in two-dimensional
system with both the Rashba and Dresselhaus spin-orbit couplings was
discussed by Glazov \cite{PhysRevB.70.195314}.\footnote{Similar study
  for spin relaxation in Si/Ge quantum well with only the Rashba
  spin-orbit coupling was given by Tahan et
  al. \cite{PhysRevB.71.075315}.} The main results are that the spin
relaxation tensor given by
Eq.~(\ref{srt-single-2d}) is extended to the
case with magnetic field along arbitrary direction. A special case is
that when the magnetic field is along the growth direction of the
two-dimensional structure, where Glazov gave
\be
\frac{1}{\tau_{z}} = 4\tilde{\tau}_1 k^2
\left[\frac{\alpha_R^2}{1+(\omega_L-\omega_c)^2\tilde{\tau}_1^2} +
  \frac{\beta_D^2}{1+(\omega_L+\omega_c)^2\tilde{\tau}_1^2} \right]
+
\frac{\tilde{\tau}_3k^6\gamma_D^2}{4}\frac{1}{1+(\omega_L-3\omega_c)^2\tilde{\tau}_3^2}.
\ee
Spin lifetime as function of the direction of magnetic field was
discussed and compared with experimental data
\cite{PhysRevB.70.195314}. Magnetic field effect on spin relaxation in
the weak scattering regime where the scattering frequency is
comparable with the spin precession frequency was studied by Glazov
\cite{Glazov2007531}. In such regime, spin polarization shows
zero-field oscillations due to the spin-orbit field induced spin
precession. Glazov found that the magnetic-field-induced cyclotron
rotation (which also rotates the ${\bf k}$-dependent spin-orbit field)
leads to fast oscillations of spin polarization around a non-zero
value and a strong suppression of spin relaxation \cite{Glazov2007531}. Recent
experiment confirmed such prediction in (001) quantum wells and showed
that the effects are absent in (110) quantum wells as the spin-orbit
field is along the growth direction which does not lead to any spin precession
\cite{2009arXiv0910.2847G}. Spin relaxation in the presence of both
electric and magnetic fields was discussed theoretically by Bleibaum
\cite{PhysRevB.71.195329,PhysRevB.71.235318}.\footnote{This has been reported
earlier by Weng et al. \cite{PhysRevB.69.245320} from the kinetic spin
Bloch equation approach (see next section).} In the presence of
spin-orbit coupling the electric field can induce an effective
magnetic field due to electron drifting
\cite{PhysRevB.69.245320}. Spin relaxation is modified by such
effective magnetic field and the real magnetic field due to
the induced spin precession. The above effects are in the Markovian
limit. Besides, there could emerge non-Markovian spin
dynamics under magnetic field. Glazov and Sherman considered the case
that the spin information can be stored in the closed orbits of
cyclotron motion and transferred to the open orbits \cite{0295-5075-76-1-102}.\footnote{The
  non-Markovian spin dynamics can also emerge in  insulating regime,
  where the nuclear spins stored the historical electron spin
  information.} As the closed orbit averaged out the spin-orbit
coupling, the electron spin lifetime in the closed orbits can be much
longer than that in the open orbits. In such a system, the fraction of spins in
the open orbits soon decays, whereas the fraction of spins in the
closed orbits decays slowly exhibiting a long-lived tail \cite{0295-5075-76-1-102}.\footnote{As
  spins in the closed orbits oscillate due to  spin-orbit coupling,
  the tail also oscillates.} The decay 
of the tail is governed by the scattering processes which transfer
electrons between closed and open orbits \cite{0295-5075-76-1-102}. Based on a
semiclassical model, the spin dynamics under weak/strong
scattering and weak/strong magnetic-field was
discussed \cite{0295-5075-76-1-102}. Particularly, they showed that at strong magnetic field $B$
along the growth direction, the tail disappears and the spin
lifetime is elongated as $\sim B^3$ \cite{0295-5075-76-1-102}. Under high magnetic field
in the quantum Hall regime, spin relaxation was studied in
Refs.~\cite{PhysRevB.43.14228,PhysRevB.46.4253,PhysRevLett.82.3324,dickman:jetp.lett.78.452,dikman:128,sherman:205335}.\footnote{Experimental
  studies can be found in
  Refs.~\cite{fukuoka:041304,Smet:nature415.281}.} It was shown that
at zero temperature with filling factor $\nu=1$, the spin relaxes
asymptotically with a power law rather than exponential
\cite{PhysRevLett.82.3324}. The spin relaxation is sensitive to the
filling factor and temperature. Experiments have shown that via tuning
the filling factor the effect of the hyperfine interaction can be
manipulated \cite{Smet:nature415.281}. As the magnetic field is high,
the $g$-tensor inhomogeneity mechanism can be important
\cite{sherman:205335}. In general, spin relaxation in quantum Hall
regime is quite different from that in the classical regime. The spin-flip
electron-phonon scattering plays important role. Besides, the
electron-electron interaction is crucial to the ground state as well
as to spin dynamics \cite{PhysRevLett.82.3324}.

{\bf $\bullet$ Weak localization effect and others.}
As in the D'yakonov-Perel' mechanism $\tau_s\sim 1/\tau_p$, the weak
localization correction to mobility will lead to correction in spin
relaxation as well. The weak localization correction to spin relaxation
was studied in
Refs.~\cite{PhysRevLett.76.3794,PhysRevB.70.205335,PhysRevLett.94.076406}.
The classical memory effect, which exists when the characteristic
scale of the disorder are comparable with the 
mean free path, leads to a non-Markovian spin dynamics
with nonexponential tail $\sim 1/t^2$ for quantum wells with
equal  Dresselhaus and Rashba spin-orbit coupling strengths
\cite{lyubinskiy:041301}. Pershin and Privman proposed
that the D'yakonov-Perel' spin relaxation in two-dimensional system
can be suppressed by a lattice of antidots \cite{PhysRevB.69.073310}.

{\bf $\bullet$ Spin relaxation in rolled-up two-dimensional electron gas.}
Spin dynamics in rolled-up two-dimensional electron gas was investigated
by Trushin and Schliemann \cite{1367-2630-9-9-346}. It was shown that
the symmetry of spin-orbit coupling varies with the radius rolled-up
structure \cite{1367-2630-9-9-346}. At certain radius, spin precession
and relaxation of a special spin component is completely quenched,
very similar to that in quantum wells with $\alpha_R=\beta_D$
(consider only the linear-${\bf k}$ spin-orbit coupling)
\cite{1367-2630-9-9-346}. 

\begin{figure}[bth]
  \centering
  \includegraphics[height=6.2cm]{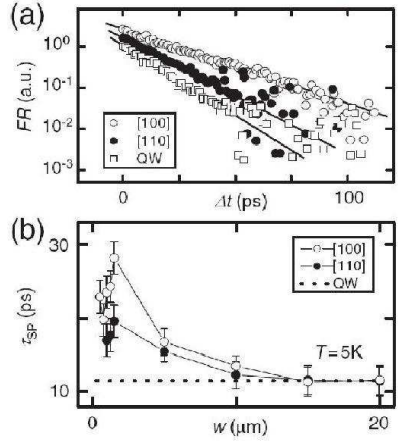}
  \caption{(a) Faraday rotation (FR) at 5 K for quantum well (open squares)
    and 750 nm wires patterned along [100] (open circles) and [110]
    (filled circles) as function of delay time $\Delta t$. Black
    curves are guides to the eyes, and the data are offset for
    clarity. (b) Width dependence $w$ of spin lifetime $\tau_{\rm SP}$
    for quantum wires patterned along [100] (open circles) and [110]
    (filled circles). The dotted line depicts the
    spin lifetime of the unpatterned quantum well. Measurements were  
    performed at $B=0$. From Holleitner et al. \cite{holleitner:036805}.}
  \label{Holleitner_width}
\end{figure}

{\bf $\bullet$ Crossover from two-dimension to one-dimension.}
An interesting problem is that how the spin
relaxation varies with the channel width of the two-dimensional
structure. This problem was investigated theoretically first by
Mal'shukov and Chao \cite{PhysRevB.61.R2413} and later by Kiselev and
Kim \cite{PhysRevB.61.13115,pssb.221.491}, showing that spin
relaxation for some nonuniform distributed spin polarization and the
uniform spin polarization along certain direction (direction of the
effective magnetic of the linear spin-orbit coupling with ${\bf k}$
along the unconstrained direction) is suppressed when the channel
width is smaller than the spin precession length
\cite{PhysRevB.61.R2413,PhysRevB.61.13115,pssb.221.491}. Originally,
these studies considered only the Rashba spin-orbit coupling. Recently
Kettemann extended the theory to include the Dresselhaus spin-orbit
coupling (both linear and cubic terms) \cite{kettemann:176808}.
After that experimental studies on submicron InGaAs wires observed
that spin lifetime first increases and then decreases with decreasing
channel width (see Fig.~\ref{Holleitner_width})
\cite{holleitner:036805,1367-2630-9-9-342}. The suppression of spin
relaxation was first explained by the theory of Mal'shukov and Chao, which,
however, is doubtful as the spin polarization distribution created by 
optical excitation is not of the type pointed out in their work. Detailed theoretical examination indicated
different explanations \cite{chang:125310}.

{\bf $\bullet$ Non-uniform system: Random spin-orbit coupling to spin relaxation.}
All the above studies are devoted to uniform system. For example, the
Rashba spin-orbit coupling is considered as
$H_R=\alpha_R(\sigma_xk_y-\sigma_yk_x)$ where $\alpha_R=\alpha_0
{\cal{E}}_z$ is proportional to a uniform electric field along thegrowth
direction. However, in genuine system, such electric field can not be
uniform due to imperfection. Such imperfection can arise from the
the fluctuation during the growth, making ${\cal{E}}_z$
position-dependent: ${\cal{E}}_z\to{\cal{E}}_z({\bf r}_{\parallel})$ with
${\bf r}_{\parallel}=(x,y)$. Spin relaxation due to this random Rashba
spin-orbit coupling can be important when other spin-relaxation
sources are ineffective \cite{PhysRevB.67.161303,sherman:209}.
This was studied comprehensively by Sherman and co-workers in
Refs.~\cite{PhysRevB.67.161303,sherman:209,PhysRevB.71.241312,dugaev:081301}.
A particular case is the symmetric (110) quantum wells, where the
average Rashba spin-orbit coupling is zero and the Dresselhaus
spin-orbit coupling does not lead to any spin relaxation due to
symmetry. Glazov and Sherman obtained that under weak and moderate
magnetic field along the growth direction at $r_c\gg l_d$ ($r_c$ is the
cyclotron radius and $l_d$ is characteristic length of random dopant
distribution) and $\alpha_Rk\tau_p\ll 1$
\cite{PhysRevB.71.241312},
\be
\tau_z^{-1} = 4\langle(\delta\alpha_R)^2\rangle
  k^2\tau_d +
\frac{4\langle\alpha_R\rangle^2k^2\tau_p}{(1+\omega_c^2\tau_p^2)},
\ee
where $\delta\alpha_R=\alpha_R-\langle\alpha_R\rangle$ is the random
Rashba coefficient with $\langle\alpha_R\rangle$ representing the
average Rashba coefficient. $\tau_d=l_d/v_{\bf k}$. The Zeeman interaction is
ignored as $\omega_L\ll \omega_c$. It is seen that both the random and the average
spin-orbit couplings lead to spin relaxation similar to the
D'yakonov-Perel' mechanism. As $\omega_c\tau_d\ll 1$ (as $r_c\gg
l_d$), the magnetic field does not suppress the spin relaxation due to
the random Rashba spin-orbit coupling. A simple estimation indicates
that in asymmetrically doped quantum well, the random spin-orbit coupling leads to a spin
relaxation rate two orders of magnitude smaller than that due to the
averaged one, leading to a spin lifetime of several tens of nanoseconds in GaAs
quantum wells \cite{sherman:209}. The spin relaxation due to the random
Rashba spin-orbit coupling may be an important source for (110) quantum
wells \cite{muller:206601}. At high magnetic field, $r_c\simeq l_d$,
the spin relaxation exhibits nonexponential tail due to the memory effect
as spin in closed orbits decays slowly
\cite{PhysRevB.71.241312}.

{\bf $\bullet$ Decay of non-uniformly distributed spin polarization.}
The decay of a standing wave of spin polarization in a two-dimensional
system with only Rashba spin-orbit coupling was studied by Pershin
\cite{PhysRevB.71.155317}. It was found that the spin relaxation
depends on the period of the standing wave. The coherent spin
precession of electrons moving in the same direction was shown be
responsible for such phenomena. In experiments such standing wave can
be generated by spin-grating technique \cite{PhysRevLett.76.4793}. In
a series of theoretical
\cite{bernevig:236601,stanescu:125307,weng:063714,bernevig:245123,shen_09}
and experimental
\cite{weber:076604,carter:136602,Koralek:nature458.610} works, spin
relaxation in spin-grating system was studied, where the spin
relaxation in such system is essentially related to the spin diffusion
limited by the D'yakonov-Perel' mechanism
\cite{bernevig:236601,weng:063714}. The spin relaxation in
spin-grating system or other nonuniform spin distributions will be
reviewed in Section~6 and Sec.~7.

\begin{figure}[bth]
  \begin{minipage}[h]{0.5\linewidth}
    \centering
    \includegraphics[height=6cm]{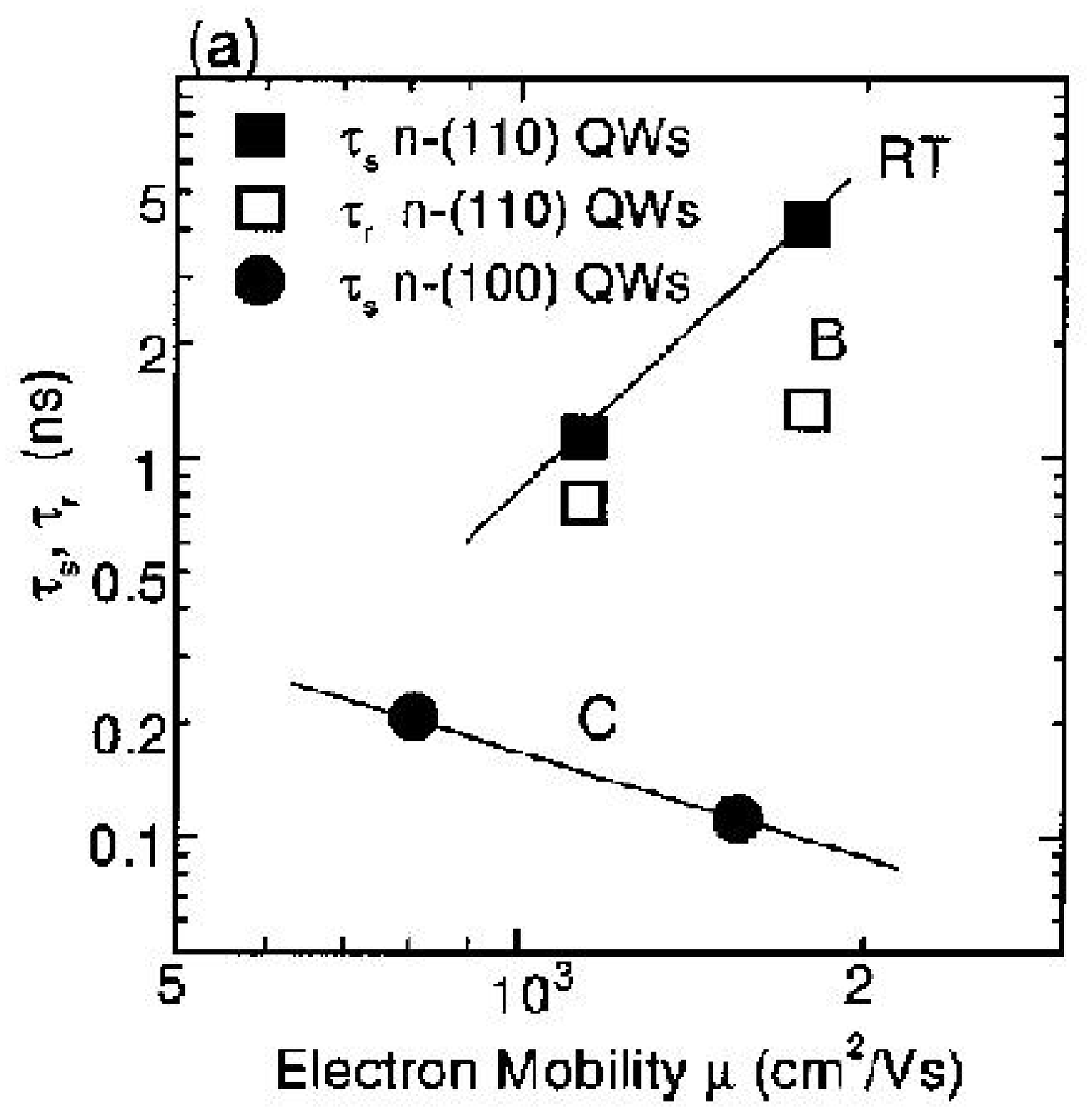}
  \end{minipage}\hfill
  \begin{minipage}[h]{0.5\linewidth}
    \centering
    \includegraphics[height=6cm]{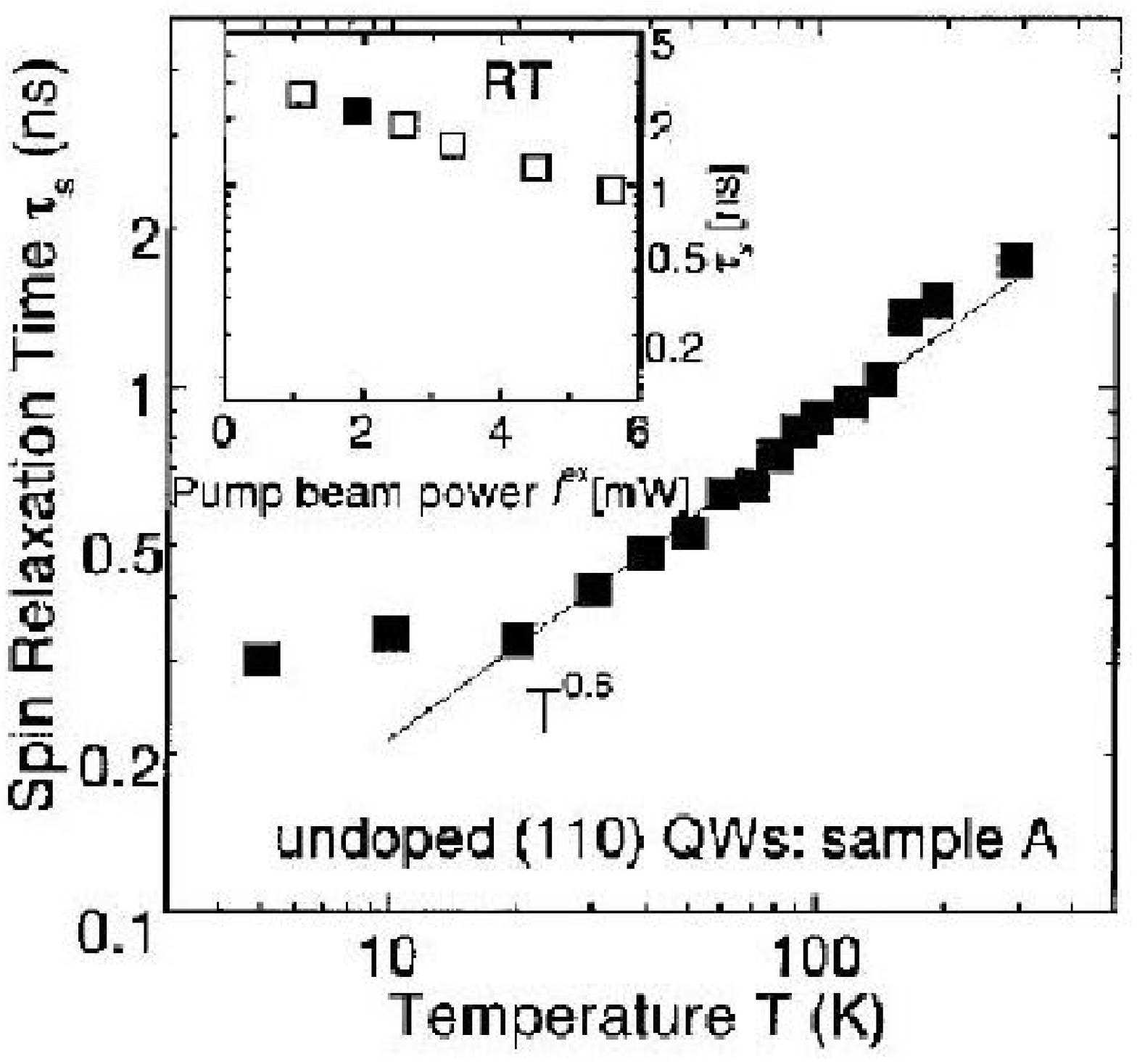}
  \end{minipage}\hfill
  \caption{Left: Spin lifetime $\tau_s$ as well as carrier
    recombination time $\tau_r$ in $n$-doped GaAs (110) and (100) 
    quantum wells as function of electron mobility $\mu$
    at room temperature. Right: Temperature $T$ dependence of spin
    relaxation time $\tau_s$ in undoped (110) quantum wells. The inset
    shows the excitation intensity $I^{\rm ex}$ dependence of $\tau_s$
    at room temperature. From Ohno et al. \cite{PhysRevLett.83.4196}.}
  \label{Ohno_110}
\end{figure}

{\bf Spin relaxation in (110) quantum wells: experiments and theories}\\
\indent We now turn to spin relaxation in two-dimensional structures grown
along the [110] direction. The bulk Dresselhaus spin-orbit coupling
becomes
\be
H_D = \gamma_D[(-k_x^2-2k_y^2+k_z^2)k_z, 4k_xk_yk_z,
k_x(k_x^2-2k_y^2-k_z^2)]\cdot\spin \big/ 2
\ee
where the three axises are ${\bf e}_x=\frac{1}{\sqrt{2}}(1,-1,0)$,
${\bf e}_y=(0,0,-1)$ and ${\bf e}_z=\frac{1}{\sqrt{2}}(1,1,0)$. When only
the lowest subband is considered, the effective spin-orbit coupling is
\be
H_{\rm SOC} = \gamma_D[0, 0, k_x(k_x^2-2k_y^2-\langle
\hat{k}_z^2\rangle)]\cdot\spin \big/ 2 + H_R
\ee
in which $\langle \hat{k}_z^2\rangle$ denotes the average over the lowest
subband and $H_R=\alpha_R(\sigma_xk_y-\sigma_yk_x)$ is the Rashba
spin-orbit coupling. In symmetric two-dimensional structures
$H_R=0$. It is noted that the effective magnetic field is then along the
[110] direction for all ${\bf k}$. The D'yakonov-Perel' spin
relaxation for spin polarization along the [110] direction is then
absent. The question arises that what kind of mechanism is now
responsible for spin relaxation along the [110] direction. To explore the
problem, Ohno et al. performed a systematic study of the dependence of
spin lifetime on the characteristic parameters such as the
subband quantization energy, electron mobility and the temperature (see
Fig.~\ref{Ohno_110}) \cite{PhysRevLett.83.4196,Adachi200136}. After a 
careful speculation on the observed dependences of spin relaxation
time, they concluded that the spin relaxation in undoped (110) quantum
wells is dominated by the Bir-Aronov-Pikus mechanism
\cite{PhysRevLett.83.4196}. However, the temperature dependence of
spin lifetime is anomalous and seems contradictory to the
Bir-Aronov-Pikus mechanism: the spin lifetime increases with
increasing temperature \cite{PhysRevLett.83.4196,Ohno2000817}. The
anomalous increase of spin lifetime with temperature can be
understood as following: the dissociation of excitons increases with
temperature rapidly which reduces the electron-hole exchange
interaction markedly and suppresses the Bir-Aronov-Pikus mechanism
\cite{PhysRevLett.83.4196}.\footnote{This is also confirmed by the
  increase of photo-carrier lifetime with temperature
  \cite{olbrich:245329}.} Similar effect was also found in ZnSe quantum
wells \cite{Hagele1999338}. However, the situation in $n$-doped
quantum wells is relatively obscure \cite{PhysRevLett.83.4196}, although
the observed mobility dependence suggests that the Elliott-Yafet
mechanism may dominate the spin relaxation at room temperature (see Fig.~\ref{Ohno_110})
\cite{PhysRevLett.83.4196}.

\begin{figure}[bth]
  \centering
  \includegraphics[height=5.5cm]{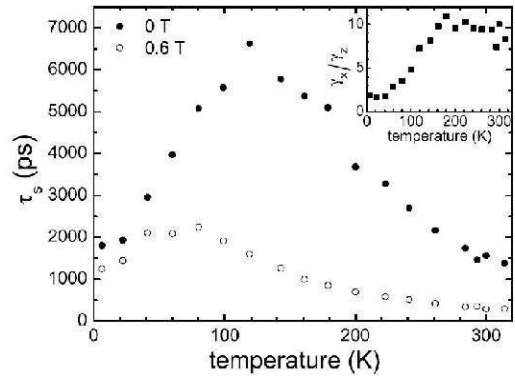}
  \caption{Temperature dependence of spin lifetime $\tau_s$ for $B=0$
    (filled circles) and $B=0.6$~T (open circles) in (110) GaAs
    quantum wells. Inset: Corresponding temperature dependence of spin
    relaxation anisotropy $\gamma_x/\gamma_z=\tau_z/\tau_x$. From D\"ohrmann et al. \cite{PhysRevLett.83.4196}.} 
  \label{Dohrmann_110}
\end{figure}

We now focus on the electron spin relaxation in $n$-doped symmetric
(110) quantum wells. Let us reexamine the three spin relaxation
mechanisms in symmetric (110) quantum wells more carefully. The
D'yakonov-Perel' mechanism leads to a spin relaxation tensor,
\be
\tau_z^{-1} = 0,\quad \quad
\tau_x^{-1} = \tau_y^{-1} = \langle\tau_p \beta_D^2  k^2 \rangle\big/2 
\ee
when only the linear spin-orbit coupling term is considered.
For the Elliott-Yafet
mechanism, usually the in-plane spin relaxation rate is larger than
the out-of-plane one \cite{jiang:155201}, but the two are generally
comparable \cite{0268-1242-23-11-114002,jiang:155201}. However, the Bir-Aronov-Pikus mechanism
is isotropic \cite{jiang:155201}. Therefore the
spin relaxation anisotropy will reflect the relevance of the
Bir-Aronov-Pikus mechanism. The experimental investigation was first
performed by D\"ohrmann et al. by measuring spin decay time in the
presence of an in-plane magnetic field \cite{PhysRevLett.93.147405}.
With an in-plane magnetic field, e.g., ${\bf B}\parallel {\bf e}_y$,
spin dynamics is governed by the following equations,
\be
\partial_t S_x = - S_x/\tau_x + \omega_L S_z, \quad \partial_t
S_z = - S_z/\tau_z - \omega_L S_x.
\ee
The solution is 
\be
S_z(t) =
S(0)e^{-\frac{1}{2}(\tau_x^{-1}+\tau_z^{-1})t}\cos(\omega
  t-\phi)/\cos(\phi)
\ee
for $2\omega_L>|\tau_z^{-1}-\tau_x^{-1}|$. Here $\tan\phi =
\frac{\tau^{-1}_x-\tau^{-1}_z}{2\omega}$ and $\omega =
\sqrt{\omega_L^2-(\tau^{-1}_x-\tau^{-1}_z)^2/4}$. The observed magnetic
field dependence is a step-function-like: for a moderate magnetic
field $B>0.5$~T, spin lifetime changes from $\tau_z$ to
$2/(\tau^{-1}_x+\tau^{-1}_z)$. The measured spin lifetimes at $B=0$
and $B=0.6$~T in GaAs (110) modulation $n$-doped quantum wells as
function of temperature are shown in Fig.~\ref{Dohrmann_110}.\footnote{Similar temperature dependence of spin lifetime in 
  (110) InGaAs quantum wells was studied in
  Ref.~\cite{schreiber:pssb244.2960}.} It is seen that at low
temperature the anisotropy is quite weak which indicates that the spin
relaxation is dominated by the Bir-Aronov-Pikus mechanism. With
increasing temperature, spin relaxation anisotropy increases as the
Bir-Aronov-Pikus mechanism becomes weaker and the D'yakonov-Perel'
mechanism for in-plane spin relaxation grows stronger.\footnote{In
  InGaAs quantum wells larger spin relaxation anisotropy was found
  \cite{morita:171905,schreiber:193304}, which is due to the much
  stronger Dresselhaus spin-orbit coupling in InGaAs.} The decrease
of $\tau_z$ at high temperature was explained by the intersubband
spin relaxation mechanism \cite{PhysRevLett.93.147405,Hagelebook}. The
intersubband spin relaxation mechanism can be understood as following:
when higher subbands are involved, the spin-orbit coupling
\be
H_D = \gamma_D [(-k_x^2-2k_y^2+k_z^2)k_z, 4k_xk_yk_z,
k_x(k_x^2-2k_y^2-k_z^2)]\cdot\spin \big/ 2
\ee
can enable intersubband spin-flip scattering as the first two terms
couple states with different spin and parity. At high temperature when
the higher subbands are populated, the intersubband spin relaxation
mechanism can be important \cite{PhysRevLett.93.147405}. This
mechanism also explains the mobility dependence of spin lifetime in
Fig.~\ref{Ohno_110}. Recently, Zhou and Wu proposed a virtual
intersubband spin relaxation, where the higher subbands need not to be
populated \cite{Zhou2009}. However, calculation indicated
that both the real and virtual intersubband spin relaxation mechanisms are
ineffective at low temperature for modulation $n$-doped quantum wells
as the scattering is suppressed \cite{Zhou2009,PhysRevLett.93.147405}.

The photo excitation of holes in the photoluminescence or Faraday/Kerr
rotation measurement makes the Bir-Aronov-Pikus mechanism become involved.
This masks the intrinsic electron spin relaxation in $n$-doped (110)
quantum wells at low temperature \cite{PhysRevLett.93.147405}. Recent
advancement in spin noise spectroscopy method
\cite{PhysRevLett.95.216603,romer:103903,starosielec:051116} enables
probing spin relaxation without photo-carrier excitation. The
Bir-Aronov-Pikus mechanism is then removed, and the intrinsic spin
lifetime can be approached. This method was applied to modulation
$n$-doped GaAs (110) quantum wells by M\"uller et al., where a much
longer spin lifetime of $\tau_z = 24$~ns at low temperature ($20$~K)
was obtained by careful analysis of the data \cite{muller:206601}.
The authors attributed the spin relaxation at such low temperature to
the random Rashba spin-orbit coupling mechanism
\cite{sherman:209,zhoupreprint}. Later, theoretical calculation
of the spin relaxation time limited by the random Rashba spin-orbit
coupling by Zhou and Wu via the fully microscopic kinetic spin Bloch
equation approach \cite{zhoupreprint} agrees well with
the experimental results by M\"uller et al. \cite{muller:206601}.

\begin{figure}[bth]
  \centering
  \includegraphics[height=6cm]{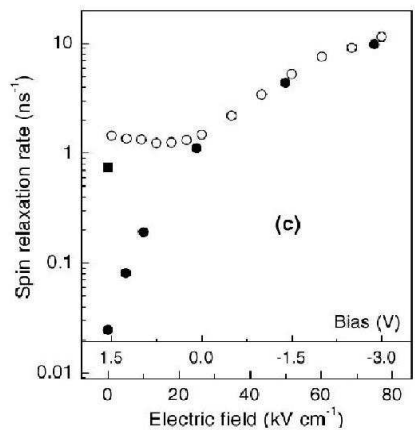}
  \caption{Measured spin relaxation rate {\sl vs.} bias voltage and
    corresponding electric field (open circles) compared with
    calculation for a symmetrical quantum well (filled circles).
    $T=170$~K. From Karimov et
    al. \cite{PhysRevLett.91.246601}.}
  \label{Harley_110}
\end{figure}

Other investigations include the achievement of high temperature gate
control of spin lifetime \cite{PhysRevLett.91.246601,Henini2004309,hall:202114},
which can be useful in spin switch devices or spin field-effect
transistors. The typical results of spin lifetime as function
of the electric field ${\cal{E}}$ along the growth direction are shown in
Fig.~\ref{Harley_110}. The low field deviation of $\tau_z\sim
{\cal{E}}^2$ indicates the relevance of spin relaxation mechanism
other than the D'yakonov-Perel' one associated with the Rashba spin-orbit
coupling. Bel'kov et al. studied the relation of symmetry to spin
relaxation in (110) quantum wells
\cite{bel'kov:176806,olbrich:245329}. The symmetry of the quantum well
was probed by the magnetic field induced photo-galvanic effect. At
certain configuration, the photocurrent is proportional to the Rashba
spin-orbit coupling coefficient. The authors observed that
photocurrent indeed vanishes for symmetric quantum wells. It was also
observed that the spin lifetime is longest in symmetric quantum wells.
Therefore the experiment demonstrated that the structure inversion
asymmetry can be tuned down to zero. As only the Rashba spin-orbit
coupling contributes to the D'yakonov-Perel' spin relaxation, the
(110) quantum wells can be used as a good platform to investigate the
Rashba spin-orbit coupling. Eldridge et al. presented an all-optical
measurements of the Rashba spin-orbit coupling: by measuring the spin
lifetime via polarized pump-probe reflection technique and
measuring the diffusion constant from spin-grating method (from which
$\tau_p$ is extracted), they extracted the Rashba coefficient
$\alpha_0=(2mk_BT\tau_s\tau_p)^{-1/2}/(e{\cal{E}})$ from
the D'yakonov-Perel' theory \cite{eldridge:125344}. At low
temperature, they found good quantitative agreement with
the ${\bf k}\cdot{\bf p}$ calculation and an unexpected temperature
dependence. In undoped (110) quantum wells, Eldridge et
al. discovered that the asymmetry of the band edge profile does {\em
  not} contribute to the Rashba spin-orbit coupling when the
electrostatic potential is absent \cite{2008arXiv0807.4845E}. These
findings confirm the theory of Lassnig \cite{PhysRevB.31.8076} (see
Sec.~2.3.3). Spin relaxation in (110) quantum wells under surface
acoustic waves was studied in Refs.~\cite{jr.:036603,jr.:153305}.

Theoretical investigation on spin relaxation in (110) quantum wells is
few up to now
\cite{Wu2002509,PhysRevB.72.115429,lau:8682,Zhou2009,tarasenko:165317,zhoupreprint,2009arXiv0911.1444G}.
Among these works, Wu and Gonokami first proposed that the
D'yakonov-Perel' mechanism can be effective for relaxation of spin
pointing along growth direction when an in-plane magnetic field is
exerted \cite{Wu2002509}. Later it was shown that the in-plane spin
relaxation rate can be tuned by strain which can be induced by the
mole fraction of gallium in an InGaAs/InP quantum well
\cite{PhysRevB.72.115429}. Effect of electric field across quantum
well on spin relaxation was studied in Ref.~\cite{lau:8682}. Recently,
Zhou and Wu proposed a virtual intrasubband spin-flip electron-phonon
and electron-impurity scattering due to the intersubband spin-orbit
coupling \cite{Zhou2009}. Tarasenko gave the spin relaxation tensor in
asymmetric (110) quantum wells with a finite Rashba spin-orbit
coupling \cite{tarasenko:165317}. He found that in the presence of
the Rashba spin-orbit coupling, the decay of electron spin initially
oriented along the growth direction is characterized by two spin
lifetimes.
Glazov et al. proposed a
symmetric multiple (110) GaAs quantum well structure to suppress the 
spin relaxation due to the random Rashba effect
\cite{2009arXiv0911.1444G}. In such structure, the donor Coulomb
potentials seen by electrons in the central quantum well is largely screened
by the electrons in other quantum wells, which hence strongly
suppresses the random Rashba spin-orbit coupling and spin relaxation
\cite{2009arXiv0911.1444G}. In this structure, spin-flip
scattering between electrons in different quantum wells, however,
leads to an additional spin relaxation. The spin-flip inter-well
electron-electron scattering is comparable with the random Rashba
spin-orbit coupling mechanism in the non-degenerate regime, but is
suppressed in the degenerate regime \cite{2009arXiv0911.1444G}.

\begin{figure}[bth]
  \centering
  \includegraphics[height=6cm]{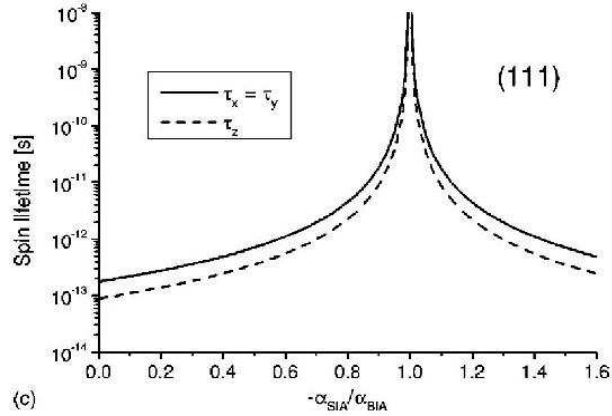}
  \caption{Calculated spin lifetime for (111) quantum wells as
    function of $-\alpha_{\rm SIA}/\alpha_{\rm BIA}$ ($\alpha_{\rm SIA}=\alpha_R$,
    $\alpha_{\rm BIA}= 2\gamma_D\langle\hat{k}_z^2\rangle/ \sqrt{3}$).
    From Cartoix\`a et al. \cite{PhysRevB.71.045313}.}
  \label{111_qwell}
\end{figure}

{\bf Spin relaxation in (111) quantum wells}\\
\indent Besides (110) quantum wells, spin relaxation in (111) quantum wells
also attracted much attention due to its particular symmetry. In (111)
quantum wells the linear Dresselhaus spin-orbit coupling is of the
same form of the Rashba one. The linear-${\bf k}$
spin-orbit coupling term is then
\be
H_{{\rm SO},1} = \alpha_{\rm IA} (\sigma_xk_y-\sigma_yk_x),
\ee 
where $\alpha_{\rm IA} =
\alpha_R + 2\gamma_D\langle\hat{k}_z^2\rangle /\sqrt{3}$. The
cubic-${\bf k}$ term reads
\be
H_{{\rm SO},3} = \frac{\gamma_D}{2\sqrt{3}}\left [k^2(-k_y\sigma_x + k_x\sigma_y)
+\sqrt{2}(3k_x^2-k_y^2)k_y\sigma_z\right ].
\ee
The spin lifetimes are then
\bea
\tau_x^{-1}=\tau_y^{-1}&=&
\left\langle k^2\tilde{\tau}_1 \left[12\alpha_{\rm IA}^2 -
    4\sqrt{3}\gamma_D\alpha_{\rm IA} k^2 + 
    (1+2\tilde{\tau}_3/\tilde{\tau}_1)\gamma_D^2 k^4 \right]\Big/6\right\rangle,\\
\tau_z^{-1}&=&\left\langle k^2\tilde{\tau}_1
  (\gamma_Dk^2-2\sqrt{3} \alpha_{\rm IA})^2\big/3\right\rangle.
\eea
It was first proposed by Cartoix\`a et
al. \cite{PhysRevB.71.045313,cartoixa:1462} that by tuning the
gate-voltage, the condition $\gamma_D\langle k^2\rangle
=2\sqrt{3}\alpha_{\rm IA}$ can be achieved, where spin
relaxation is suppressed for all spin components (see
Fig.~\ref{111_qwell}): $\tau_z=\infty$ and
$\tau_x=\tau_y=3/(\gamma_D^2\langle
  k^6\rangle\tilde{\tau}_3)$. Microscopic calculation using eight-band
${\bf k}\cdot{\bf p}$ method indicated the feasibility of such scheme
in an InAlAs/InGaAs/InP quantum well \cite{vurgaftman:053707}. Similar
scheme based on tuning spin-orbit coupling via strain was also
proposed \cite{PhysRevB.72.115429}.

{\bf Spin relaxation in arbitrarily oriented quantum wells}\\
\indent Accounting only the linear-${\bf k}$ Dresselhaus spin-orbit
coupling, spin relaxation tensor for an arbitrarily oriented quantum
well reads
\be
\tau_{ij}^{-1}=(\delta_{ij}{\rm
  Tr}\hat{\nu}-\nu_{ij})/\tau_s^0(\varepsilon_{\bf k}).
\ee
Here $\tau_s^0$ is the in-plane spin lifetime for (001) quantum
well. The tensor $\hat{\nu}$ depends on the orientation of the quantum
well ${\bf n}=(n_x,n_y,n_z)$\footnote{The $x$, $y$ and $z$ axis are
  taking along the [100], [010] and [001] directions respectively.} as
\cite{d'yakonov:110}
\be
\nu_{xx} = 4n_x^2(n_y^2+n_z^2)-(n_y^2-n_z^2)^2(9n_x^2-1), \quad\quad
\nu_{xy} = n_xn_y[9(n_x^2-n_z^2)(n_y^2-n_z^2)-2(1-n_z^2)],
\ee
with other components obtained by cyclic permutation of the $x$, $y$
and $z$ indexes.

\begin{figure}[bth]
  \centering
  \includegraphics[height=7cm]{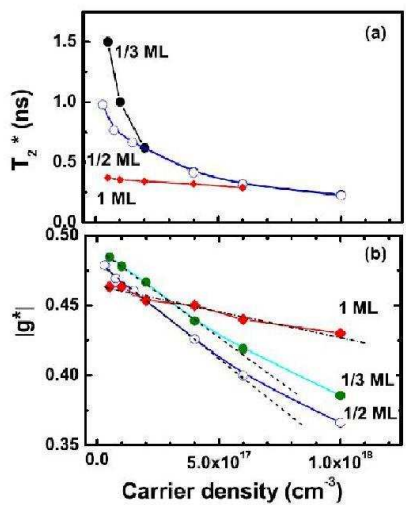}
  \caption{(a) Electron-spin dephasing time $T_2^{\ast}$
    and (b) effective $g$-factor $|g^\ast|$ for 1/3 monolayer (ML),
    1/2 ML and 1 ML InAs on (100) GaAs substrates with various carrier
    densities. The temperature is 77 K. From Yang et
    al. \cite{yang:035313}.}
  \label{Yang_submono}
\end{figure}

{\bf Spin relaxation in (sub)-monolayers: experiments}\\
\indent Spin relaxation in undoped sub-monolayer and monolayer InAs structures
grown in GaAs matrix was studied in Refs.~\cite{yang:035313,Sun2007158}. The 
monolayer structure on (001) surface can be regarded as an ideal
two-dimensional system. However, the (311) oriented monolayer forms
wire-like or disk-like microstructures on GaAs steps and facets.
In sub-monolayer InAs structures, such as 1/3 monolayer and 1/2 monolayer,
InAs was found to be organized as disk-like islands with lateral size of tens
of nanometers. However, the carrier system is still of two-dimensional
nature but with lower density of states and smaller mobility. It was
then found that the spin relaxation is suppressed by reducing the
layer thickness (see Fig.~\ref{Yang_submono}) \cite{yang:035313}. The boundaries and
deformation potentials are enhanced with decreased coverage, which
leads to the decrease of momentum scattering time and suppresses spin
relaxation \cite{yang:035313}. For monolayer structures, the (311) structure has a much
longer spin lifetime than the (001) one as the surface roughness
is much larger in the former \cite{yang:035313}. The $g$-factor and spin dephasing
time were also measured as function of excitation density for these
structures \cite{yang:035313}. The spin dephasing time decreases with excitation
density for all the 1/3, 1/2 and 1 monolayer structures \cite{yang:035313}. The decrease
of spin lifetime may be due to the enhancement of the
D'yakonov-Perel' mechanism as $\langle k^6 \rangle$ increases or due
to the enhancement of the Bir-Aronov-Pikus mechanism as the hole
density increases. However, via examination of the temperature
dependence of spin lifetime,  Yang et al. \cite{yang:035313} found that the
Bir-Aronov-Pikus mechanism can only be important at low excitation
density in 1/3 monolayer structure due to the strong electron-hole
exchange interaction between the spatially confined electrons and
holes \cite{Sun2007158}. The temperature dependence of spin lifetime
at higher
excitation density exhibits a peak \cite{yang:035313,Sun2007158}, which signals the D'yakonov-Perel'
spin relaxation associated with the electron-electron
scattering \cite{zhou:045305,jiang:125206}. Finally, the density
dependence of $g$-factor was used to
analysis the electronic density of states in these structures \cite{yang:035313}.

\begin{figure}[bth]
  \centering
  \includegraphics[height=7.5cm]{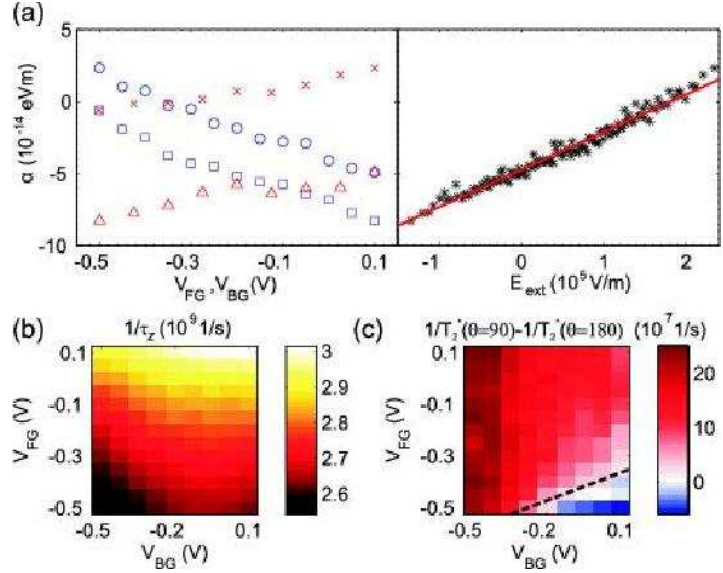}
  \caption{(001) parabolic quantum well. (a) Left
    panel: the Rashba spin-orbit coupling coefficient $\alpha$ as a
    function of the front-gate: FG (back-gate: BG) voltage with fixed
    BG (FG) voltage ($\triangle$: $V_{\rm FG}=0.1$~V; $\times$:
    $V_{\rm FG}=-0.5$~V; $\circ$: $V_{\rm BG}=0.1$~V; $\square$:
    $V_{\rm BG}=-0.5$~V). Right panel: $\alpha$ as a function of the
    external electric field $E_{\rm ext}$. The red line represents a
    least-squares fit. (b) Out-of-plane spin relaxation rate
    $1/\tau_z=1/T^\ast_2(\theta=90)+1/T^\ast_2(\theta=180)$.
    (c) In-plane spin-dephasing asymmetry
    $1/T^\ast_2(\theta=90)-1/T^\ast_2(\theta=180)$. $\theta=90$
    ($\theta=180$) denotes that the external
    magnetic field is along $[110]$ ($[\bar{1}10]$) direction. The
    dashed line marks the gate voltage where $\alpha=0$. From
    Studer et al. \cite{studer:027201}.} 
  \label{Studer_parabolic_qw}
\end{figure}

{\bf Spin relaxation in parabolic quantum wells: experiments}\\
\indent A good candidate for spintronic device structure is the parabolic
quantum wells, where the $g$-factor and spin lifetime can be
tuned efficiently by electrical means
\cite{gsalis:619,Salis200399,studer:027201}. Recently, the Rashba and
linear Dresselhaus spin-orbit couplings in parabolic quantum wells were
measured by monitoring the spin precession frequency of drifting
electrons via time-resolved Kerr rotation \cite{studer:027201}. It was
found that the Rashba spin splitting can be tuned significantly by the
gate biases, whereas the Dresselhaus spin-orbit coupling varies only
weakly. The spin relaxation was then tuned by gate-voltage (see
Fig.~\ref{Studer_parabolic_qw}). It was observed that the anisotropy of
spin relaxation vanishes when the Rashba spin-orbit coupling is tuned
to zero.

{\bf Spin relaxation in II-VI semiconductor two-dimensional
  structures: experiments}\\
\indent Spin relaxation in II-VI semiconductor two-dimensional structures has
also been widely studied. The first remarkable advancement is that the spin 
lifetime can be increased by several orders of magnitude via $n$-type
doping in ZnSe quantum wells \cite{J.M.Kikkawa08291997}. It was observed
that the spin lifetime is on the order of nanoseconds and is only weakly
temperature dependent \cite{J.M.Kikkawa08291997}. The spin lifetime in
undoped ZnSe quantum wells is on the order of 10~ps at low temperature
\cite{J.M.Kikkawa08291997,Hagele1999338}. In undoped quantum wells
the spin lifetime was observed to increase from less than 10~ps
at 20~K to 500~ps at 200~K \cite{Hagele1999338}. As the exciton
binding energy is large (19~meV), at low temperature spin dynamics is
governed by the exciton behavior. Experimental examination on the
relation between spin lifetime and momentum lifetime
reveals a motional narrowing nature $\tau_s\sim \tau_p^{-1}$, which
coincides with the picture of Maialle et al. \cite{PhysRevB.47.15776}:
the electron-hole exchange interaction serves as an effective magnetic
field depending on exciton momentum. The momentum dependent spin
precession leads to a spin relaxation similar to the D'yakonov-Perel'
one for electron spin. At high temperature the spin relaxation is
further suppressed with increasing temperature due to the ionization of
excitons which weakens the electron-hole exchange interaction
\cite{Hagele1999338}. The spin relaxations of electron, exciton and
trion were studied in $n$-doped single CdTe quantum wells with different
doping density \cite{PhysRevB.68.235316}. It was found that the
exciton spin lifetime (18-36~ps) is much shorter than the
electron one ($\sim$180~ps) and the trion spin
relaxation is governed by the fast hole spin relaxation
\cite{PhysRevB.68.235316}. Spin relaxation in (110) ZnSe quantum wells
was studied in Ref.~\cite{JS18.185} where behaviors quite different
from that in GaAs (110) quantum wells were found. Spin lifetime
was reported to decrease monotonically with increasing temperature in CdTe
quantum wells \cite{zhukov:205310}. The electron density dependence of
spin lifetime in CdTe quantum wells was found to be nonmotonic:
it exhibits a peak at $8\times 10^{10}$~cm$^{-2}$ for $T=5$~K \cite{pssb243.2290}.

\subsubsection{Electron spin relaxation in $p$-type III-V and II-VI
  semiconductor two-dimensional structures} 

The two main electron spin relaxation mechanisms in $p$-type
two-dimensional structures are the D'yakonov-Perel' and the
Bir-Aronov-Pikus mechanisms. It is believed that at low temperature
and/or high hole density the Bir-Aronov-Pikus mechanism dominates.
The Elliott-Yafet mechanism may also dominate spin relaxation in
heavily $p$-doped narrow bandgap semiconductors at very low
temperature where the Bir-Aronov-Pikus mechanism is suppressed by
Pauli blocking \cite{Titkovbook}. The D'yakonov-Perel' mechanism
determines spin relaxation in other regimes.

\begin{figure}[bth]
  \begin{minipage}[h]{0.5\linewidth}
    \centering
    \includegraphics[height=6cm]{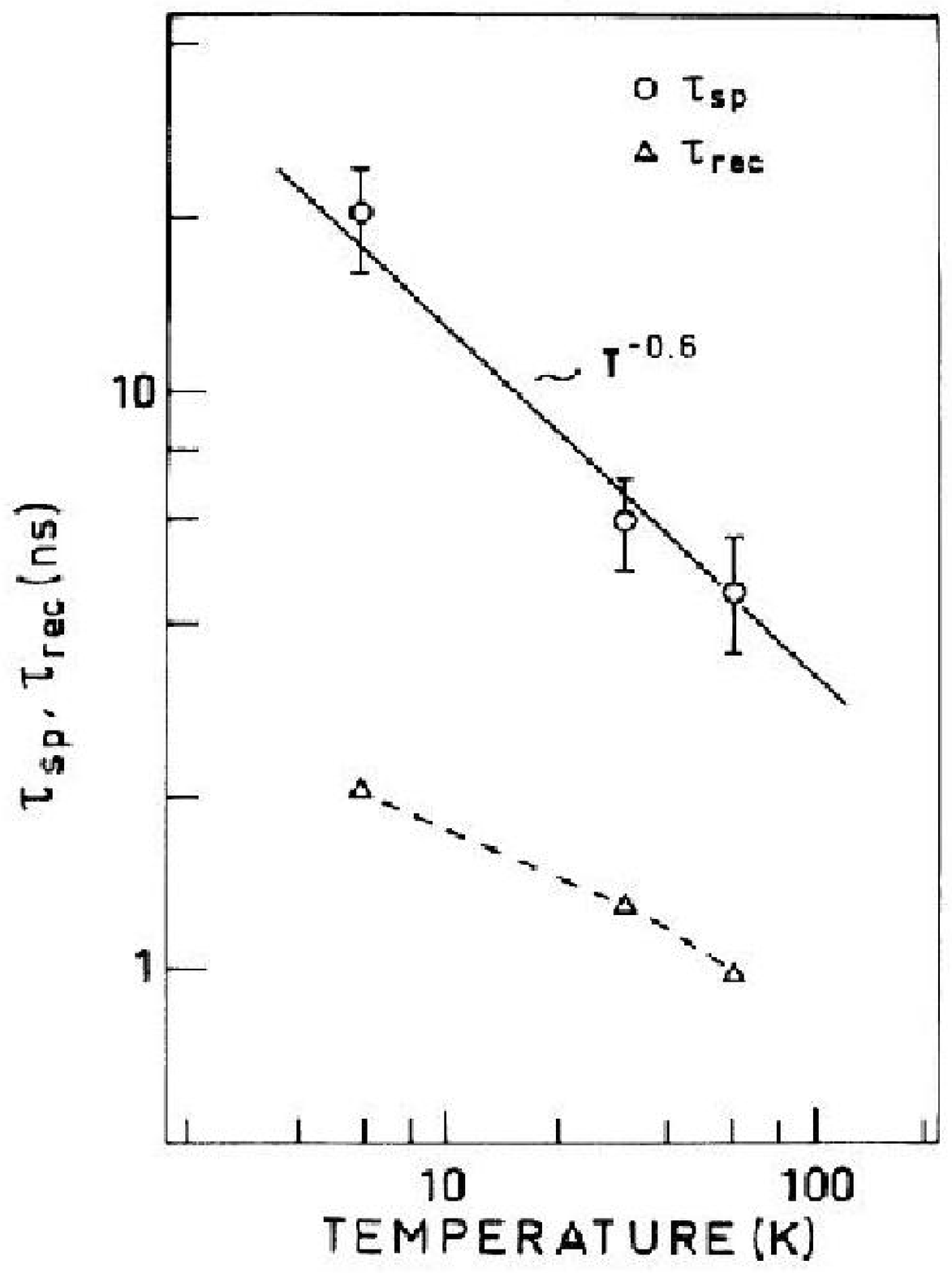}
  \end{minipage}\hfill
  \begin{minipage}[h]{0.5\linewidth}
    \centering
    \includegraphics[height=6cm]{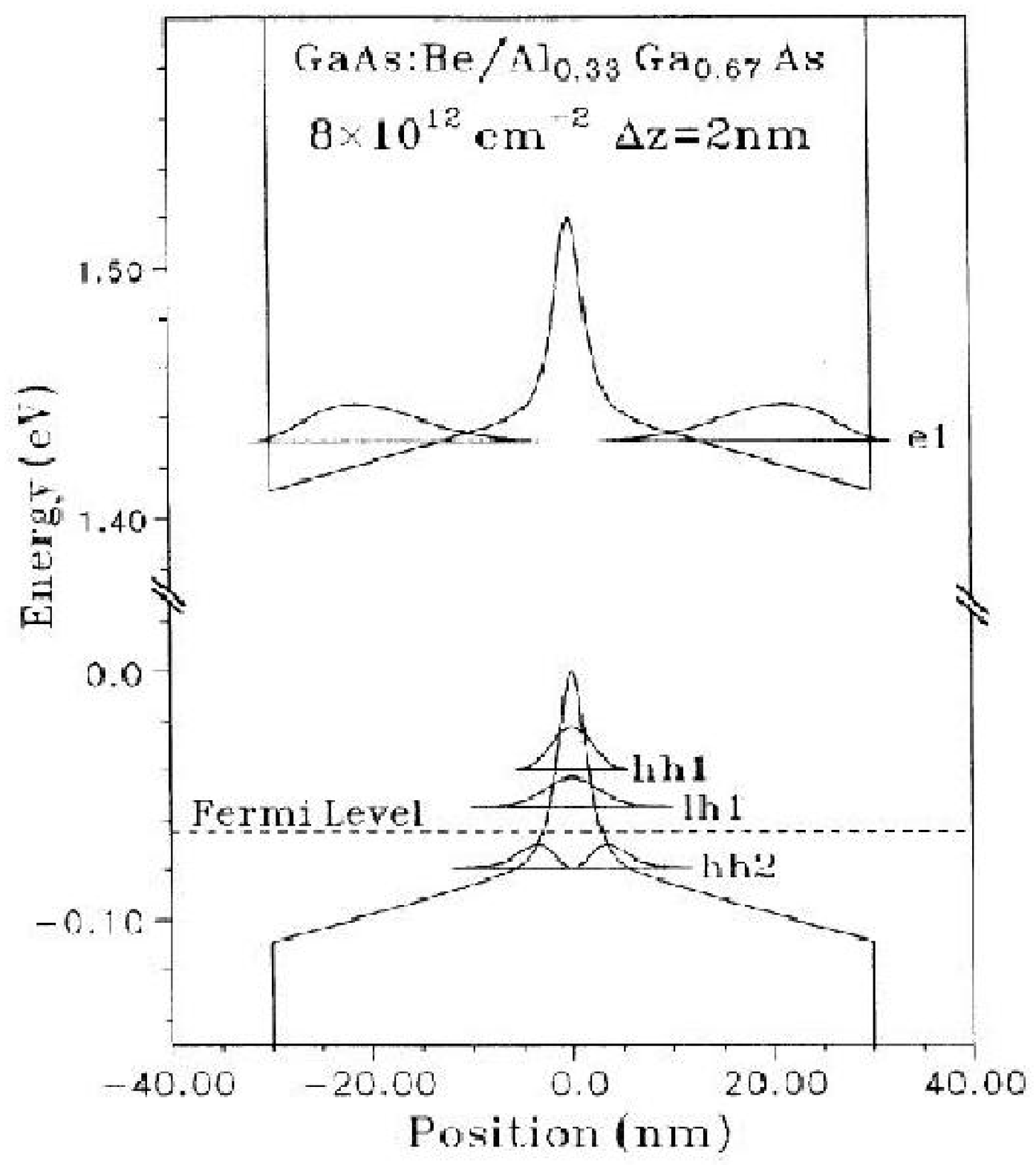}
  \end{minipage}\hfill
  \caption{Left: Temperature dependence of the electron-spin lifetime
    $\tau_{\rm sp}$ and luminescence recombination time $\tau_{\rm
      rec}$. The solid line indicates a $T^{-0.6}$ power law, the
    dashed line is drawn as a guide to the eyes. Right:
    Self-consistent potential profile of the structure
    investigated. Subband energies and probability densities at
    $k_{\parallel}=0$ are also shown. ``hh'' and ``lh'' denote
    heavy-hole and light-hole respectively; ``e1'' denotes the first
    electron subband. The original structure is a 60~nm wide
    GaAs/Al$_{0.33}$Ga$_{0.67}$As quantum well, where it is
    $\delta$-doped in its center with Be ($p$-type). From Wagner et 
    al. \cite{PhysRevB.47.4786}.}
  \label{Wagner_double_hetero}
\end{figure}

Experimental investigations on electron spin relaxation in $p$-type
quantum wells are scarce. In a 6~nm
$p$-modulation-doped GaAs multiple quantum well with hole density
$n_h=4\times 10^{11}$~cm$^{-2}$, Damen et al. measured an electron
spin lifetime of 150~ps at 10~K \cite{PhysRevLett.67.3432}.
In $\delta$-doped double heterostructures with a doping density
$8\times 10^{12}$~cm$^{-2}$, extremely long spin lifetimes up to
20~ns was observed at 6~K and a decrease of spin lifetime with
temperature $\tau_s\sim T^{-0.6}$ was found for $T=6$-60~K (see
Fig.~\ref{Wagner_double_hetero}) \cite{PhysRevB.47.4786}. The
extremely long spin lifetime was explained as suppression of
the electron-hole exchange interaction by spatially separated electrons
and holes in the $\delta$-doped double heterostructures. However, the
D'yakonov-Perel' spin relaxation should also be suppressed to achieve
such long spin lifetime, i.e., the momentum scattering should be
strong in the structure.\footnote{The electron-impurity scattering is
  the only possible candidate to suppress the D'yakonov-Perel' spin
  relaxation in the experimental condition, as other scatterings are
  limited at such low temperature and low electron density.} Later
Gotoh et al. measured spin relaxation in a device where the spacial
separation between electrons and holes can be tuned by a gate-voltage at
room temperature \cite{gotoh:3394}. They showed that the spin
relaxation is enhanced when the spacial electron-hole separation is
shortened. From the observed results, they concluded that electron
spin relaxation is dominated by the Bir-Aronov-Pikus mechanism in
their structures. However, the gate-voltage also modifies the Rashba
spin-orbit coupling, theses effects should also be taken into account
for a close examination. The temperature dependence of spin relaxation
at low temperature ($T\le 60$~K) in $p$-modulation-doped GaAs quantum
wells was investigated in Refs.~\cite{pssb.215.229,Potemski1999163,0953-8984-11-31-304}
where a decrease of spin lifetime was found and the spin relaxation
varies from 600~ps to 40~ps. Energy-resolved spin dynamics
at the sufaces of $p$-GaAs was studied by Schneider et al. using time-
and spin- resolved two-photon photoemission \cite{schneider:081302}. The spin lifetime
was found to increase with decreasing electron kinetic energy in
agreement with the corresponding theory \cite{PhysRevB.54.1967}. They
also observed a suppression of spin relaxation in (001) surfaces where
hole density is reduced by about an order of magnitude due to the
band-bending effect.

\begin{figure}[bth]
  \begin{minipage}[h]{0.5\linewidth}
    \centering
    \includegraphics[height=6.2cm]{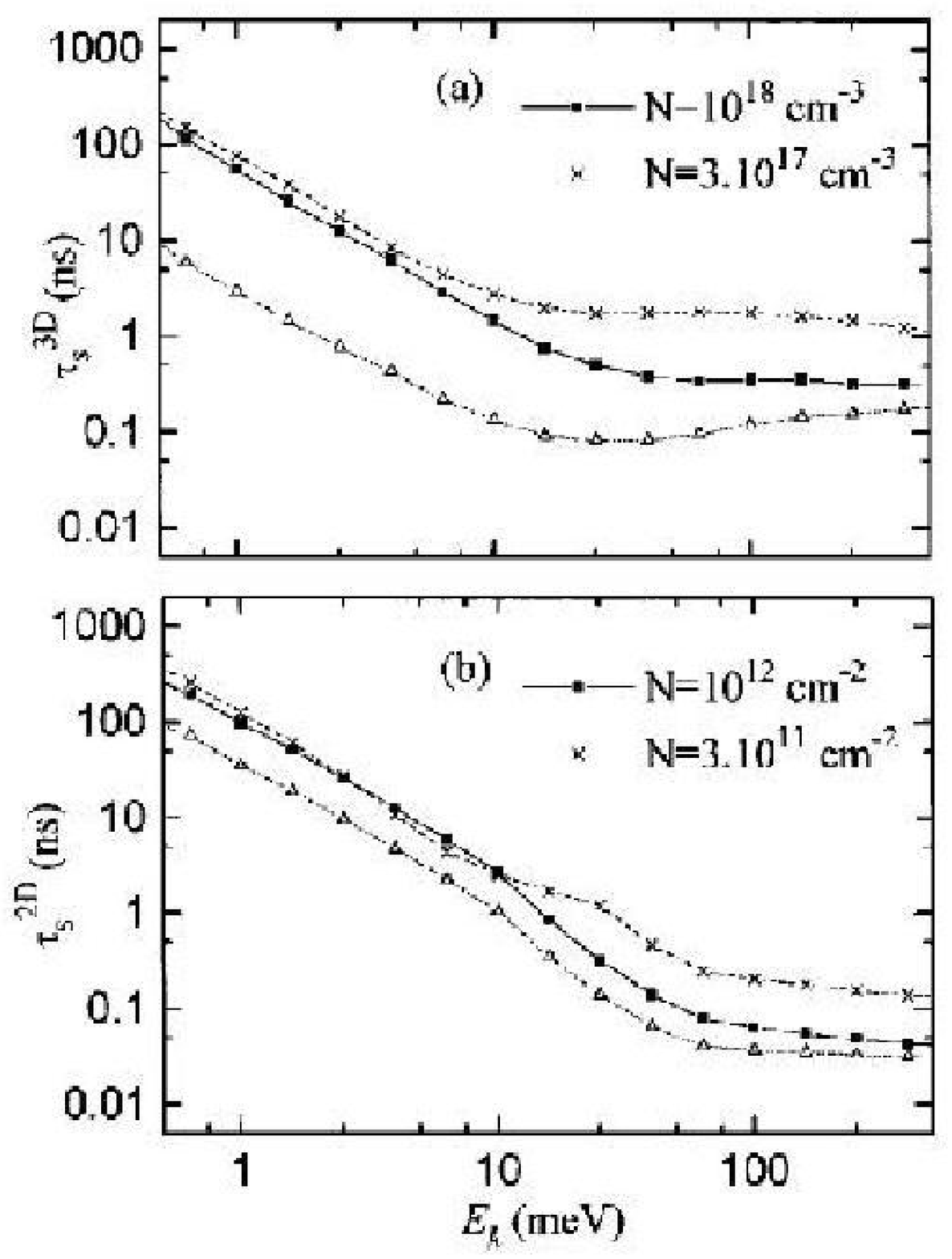}
  \end{minipage}\hfill
  \begin{minipage}[h]{0.5\linewidth}
    \centering
    \includegraphics[height=6.2cm]{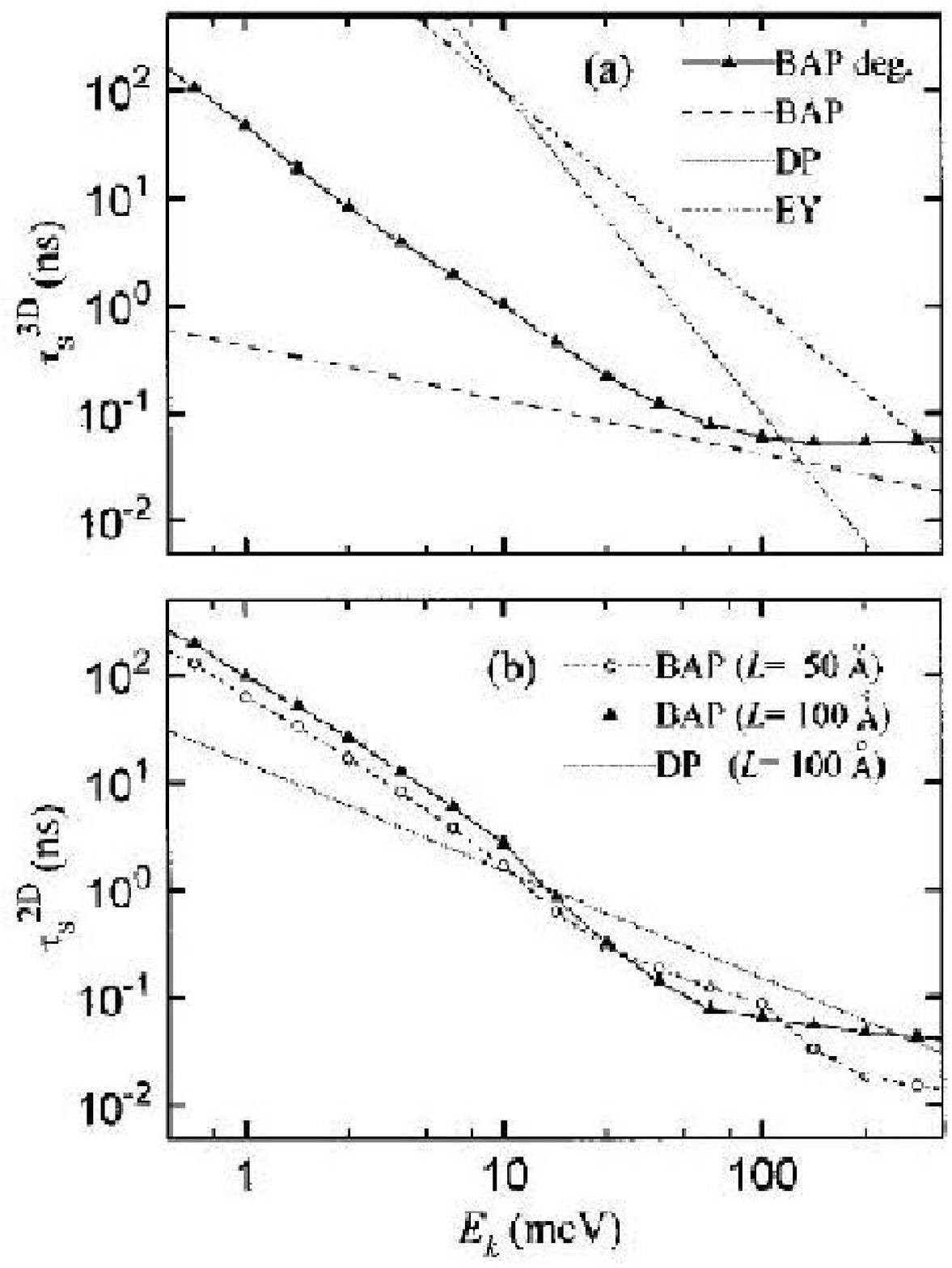}
  \end{minipage}\hfill
  \caption{Left: Electron spin lifetime
    $\tau_s$ due to the Bir-Aronov-Pikus mechanism in bulk GaAs
    and GaAs quantum wells as function of the
    electron kinetic energy for different hole densities $N$ as
      labeled in the figure. The dotted curves with open triangles
    are the same as the solid curves but including the Sommerfeld
    factor. Right: Calculated electron spin lifetimes $\tau_s$ 
    due to the D'yakonov-Perel' (DP), Elliott-Yafet (EY) and Bir-Aronov-Pikus
    (BAP) mechanisms as functions of the electron kinetic energy. The
    spin lifetimes due to the D'yakonov-Perel' and Elliott-Yafet
    processes are taken from Ref.~\cite{PhysRevB.16.820}, whereas the
    Bir-Aronov-Pikus (BAP) spin lifetimes are calculated by 
    assuming both nondegenerate holes (label in the figure
    as ``BAP'') and degenerate holes (label in the figure as ``BAP deg.'')
    with (a) hole densities $N=4\times 10^{18}$~cm$^{-3}$ in bulk GaAs and (b)
    $N=10^{12}$~cm$^{-2}$ in GaAs quantum wells with different well
    widths $L$. From Maialle \cite{PhysRevB.54.1967}.} 
  \label{Maialle_bap}
\end{figure}

Theoretically, the role of electron and hole spin relaxation dynamics
in time-resolved luminescence spectral was studied by Uenoyama and Sham
\cite{PhysRevB.42.7114,PhysRevLett.64.3070}. The ultrafast carrier and spin dynamics was then investigated
from a fully microscopic kinetic spin Bloch equation approach \cite{PhysRevB.61.2945,wu:epjb.18.373,PhysRevB.68.075312} including
also the carrier-carrier Coulomb scattering by Wu and Metiu
\cite{PhysRevB.61.2945}. Calculation of the Bir-Aronov-Pikus spin
relaxation in excitonic system and later in electron-hole plasma was
performed by Maialle et
al. \cite{PhysRevB.47.15776,PhysRevB.55.13771,PhysRevB.54.1967} where
the energy-dependent spin relaxation time and the effects of the
valence-band spin mixing as well as the electric field along the growth
direction were studied. These studies found that (i) spin lifetime
decreases with electron kinetic energy (see Fig.~\ref{Maialle_bap});
(ii) the valence-band spin mixing can lead to a factor of two
correction; and (iii) spin relaxation due to the Bir-Aronov-Pikus mechanism
can be tuned substantially by gate-voltage. The relative importance of
the Bir-Aronov-Pikus, Elliott-Yafet and D'yakonov-Perel' mechanisms
was also compared in GaAs quantum wells
\cite{PhysRevB.54.1967}. Maialle suggested that the Bir-Aronov-Pikus
mechanism is more important than the D'yakonov-Perel' mechanism at medium energy
in contrast to that the Bir-Aronov-Pikus mechanism is more important
at low energy in bulk GaAs from the same calculation (see
Fig.~\ref{Maialle_bap}).

\subsubsection{Electron spin relaxation in III-V and II-VI
  semiconductor one-dimensional structures}

There are much less works on electron spin relaxation in
one-dimensional structures compared with the electron spin
relaxation in two-dimensional structures. It is believed that
the D'yakonov-Perel' mechanism is the most efficient one, unless it is
suppressed in certain geometry. Due to the additional constrain,
the spin-orbit couplings in one-dimensional system is quite different from
those in the two-dimensional case. By denoting the unconstrained direction as
$\hat{z}$, the Rashba spin-orbit coupling for the lowest subband is then
\be
H_R = \alpha_0|e|^{-1} \left[\sigma_x\langle\partial_yV(x,y)\rangle
-\sigma_y \langle\partial_xV(x,y)\rangle\right] k_z ,
\ee
where $V(x,y)$ is the electrostatic potential and $\langle...\rangle$
stands for the average over the lowest subband wavefunction. The Dresselhaus spin-orbit
coupling depends on the growth direction of the quantum wire. If
$\hat{z}\parallel [001]$, then
\be
H_D = \gamma_D\langle (\hat{k}_x^2-\hat{k}_y^2) \rangle\sigma_z k_z.
\label{1D_dsoc}
\ee
One notices that spin precession direction is the same for {\em all}
$k_z$. Such symmetry leads to an infinite spin lifetimes for spin
polarization along the spin precession direction. However, spin
lifetime in other directions are still finite. By proper arrangement of
the growth direction and the confinement, the Rashba and Dresselhaus
spin-orbit couplings can be cancelled, where relaxation for all the
spin components due to the D'yakonov-Perel' mechanism can be inhibited
\cite{lu:073703,Liu.jsnm.9.525}.

\begin{figure}[bth]
  \centering
  \includegraphics[height=5.5cm]{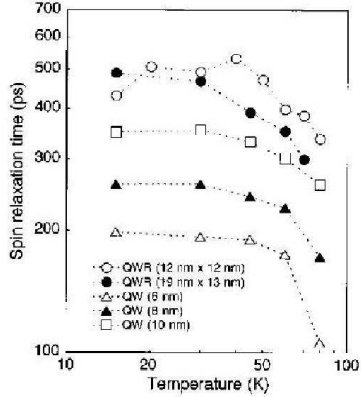}
  \caption{Temperature dependence of the spin relaxation time
    in quantum wires and quantum wells with various confinements (as
    labeled in the figure). From Sogawa et
    al. \cite{PhysRevB.58.15652}.}
  \label{Qwire_T_dep}
\end{figure}

Experimentally, spin relaxation in undoped rectangular GaAs/AlAs
quantum wires with small ($<20$~nm) lateral sizes was studied via
time-resolved photoluminescence \cite{PhysRevB.58.15652}. The measured
spin lifetime is shown in Fig.~\ref{Qwire_T_dep} together
with the spin lifetime in quantum wells for comparison. One finds
that spin lifetime in the narrower ($12$~nm$\times$12~nm) quantum
wire can be {\em larger} than that in the wider one
($19$~nm$\times$13~nm). This is in contrast to the fact that both the
D'yakonov-Perel' and Bir-Aronov-Pikus mechanisms increase with the
confinement, as also indicated by the spin lifetime in quantum
wells for different well widths. The temperature dependence of the wide
quantum wire is similar to that in quantum wells, which indicates the
multi-subband effect \cite{lu:073703}. In narrow quantum
wire, the temperature dependence is nonmonotonic, possibly indicating the
effect of the electron-hole Coulomb scattering
\cite{zhou:075318,jiang:125206}.
Wire width dependence of the spin relaxation in wide quantum wires was
studied in Refs.~\cite{holleitner:036805,1367-2630-9-9-342}, where the
crossover from two dimension to  one dimension was discussed. Transport
studies of spin lifetime was reported in Ref.~\cite{kunihashi:226601}.

\begin{figure}[bth]
  \begin{minipage}[h]{0.45\linewidth}
    \centering
    \includegraphics[height=5cm]{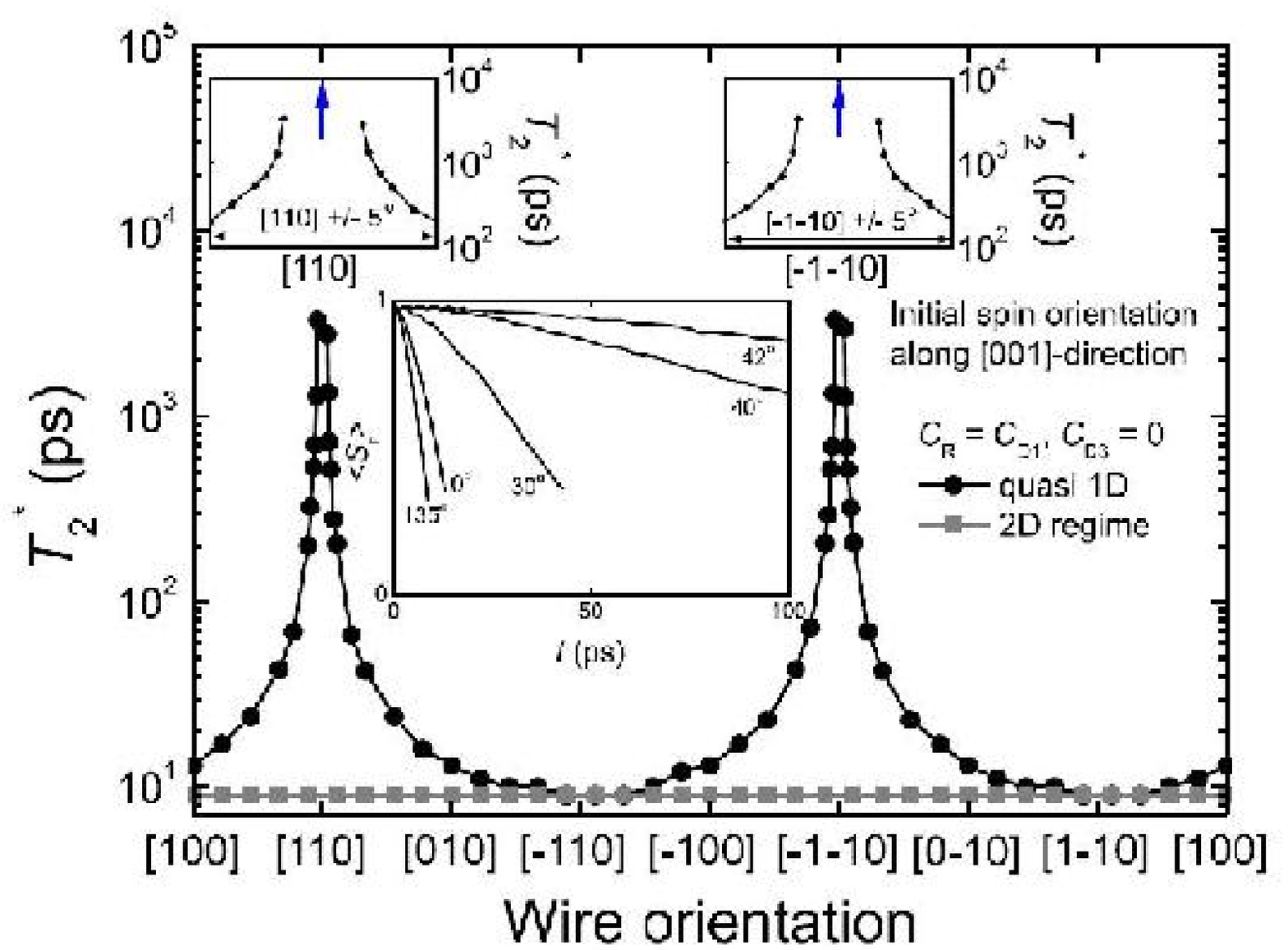}
    \caption{The spin-dephasing time $T_2^{\ast}$ of a spin ensemble
      is plotted as a function of the wire orientation with respect to
      the [100] lattice direction. The spin polarization is initially oriented
      along the [001] direction. $T_2^{\ast}$ is strongly enhanced for a
      quasi-one-dimensional wire oriented in the [110] direction ({\sl black}) as
      compared to an ensemble in a two-dimensional system ({\sl
        gray}). In the calculation, the Rashba and Dresselhaus
      spin-orbit coupling strengths are taken to be equal, and
      the cubic Dresselhaus term is ignored. The two upper insets 
show the results near the
      two peaks. The lower inset shows the evolution of the spin
      polarization for various wire orientation (with respect to the
      [100] direction). From Liu et al. \cite{Liu.jsnm.9.525}.}
    \label{Liu_qwr}
  \end{minipage}\hfill
  \begin{minipage}[h]{0.45\linewidth}
    \centering
    \includegraphics[height=3.5cm]{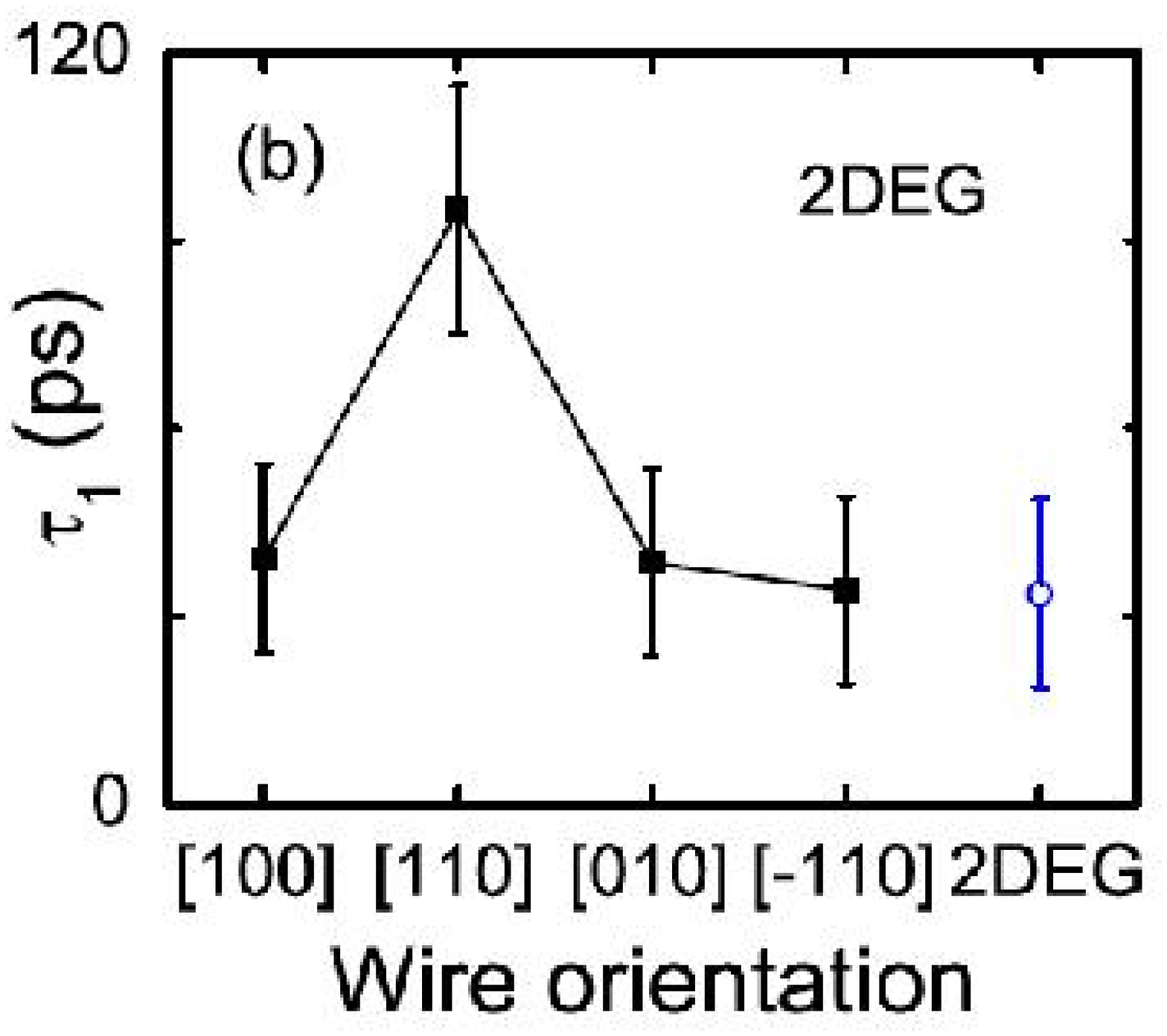}
    \caption{Spin-dephasing time $\tau_1$ measured via Kerr
      rotation in quasi-one-dimensional structures with different wire
      orientation together with spin relaxation in two-dimensional
      electron gas (2DEG). From Denega et al. \cite{2009arXiv0910.2336D}.} 
    \label{Denega_qw_or_exp}
  \end{minipage}\hfill
\end{figure}

Theoretically, the study of spin relaxation in quantum wires focused
on the D'yakonov-Perel' mechanism. Pramanik et al. studied spin
relaxation in a wide quantum wire (4~nm$\times$30~nm) using a
semiclassical approach (via the Monte Carlo simulation)
\cite{PhysRevB.68.075313}. In their work spin relaxation was found to
be anisotropic as analyzed above. They showed that electric fields
which drive electrons to high $k$ states lead to faster spin
relaxation \cite{PhysRevB.68.075313}.
Later Dyson
and Ridley studied electron spin relaxation in quantum wires due to
the electron-longitudinal-optical-phonon scattering \cite{PhysRevB.72.045326}. Unlike previous
studies concerning the electron-longitudinal-optical-phonon scattering, no elastic scattering
approximation was made. They gave analytical expressions for the energy
dependent scattering time where the inelastic nature of the collision
was fully taken into account. They found that the spin relaxation rate
increases with confinement asymmetry as indicated by
Eq.~(\ref{1D_dsoc}) where 
the Dresselhaus spin-orbit coupling is proportional to $\langle
(\hat{k}_x^2-\hat{k}_y^2)\rangle$.
Spin lifetime due to the electron-longitudinal-optical-phonon
scattering was also found to increase with temperature \cite{PhysRevB.72.045326}. In the
presence of both the Rashba and Dresselhaus spin orbit couplings, the spin
relaxation becomes anisotropic. Consider a special case, where the
strength of the
Rashba spin-orbit coupling is equal to that of the linear Dresselhaus
spin-orbit coupling: In two-dimensional case, this leads to a zero
spin-splitting for ${\bf k}$ along [110] direction, whereas a maximum
spin-splitting for ${\bf k}$ along [1$\bar{1}$0]
direction. By supperimposing additional constrainment to form quantum wires
with the growth direction along [110] direction and keeping the condition
that the strengths of the Rashba and linear Dresselhaus spin-orbit couplings are equal,
it is easy to understand that the spin-orbit field is zero and
spin relaxation is inhibited. Such condition holds for narrow
quantum wires where only the lowest subband is relevant. Recently, Liu
et al. showed that it also holds for quasi-one-dimensional quantum
wires where the wire width is comparable to the spin precession length
or mean free path (see Fig.~\ref{Liu_qwr}) \cite{Liu.jsnm.9.525}. The
underlying physics is that electrons with a large transverse momentum
component to the wire orientation almost do not contribute to the
spin-dephasing because of motional narrowing as they suffer strong
boundary scattering. Only electrons with a large momentum component
parallel to wire orientation contribute significantly to spin
relaxation. Hence spin relaxation depends largely on the wire
orientation for equal Rashba and Dresselhaus spin-orbit coupling strengths when
the cubic Dresselhaus term is neglected\footnote{Note that the
  dependence of the spin relaxation on quantum wire orientation
  was also shown by Holleitner et
  al. \cite{holleitner:036805,1367-2630-9-9-342} where a shorter spin
  lifetime in wires along [110] direction than that in wires
  along [100] direction was found. This may be because of an opposite
  sign of the Rashba spin-orbit coupling in those wires according to the
  theory by Liu et al. \cite{Liu.jsnm.9.525}.} (see
Fig.~\ref{Liu_qwr}) \cite{Liu.jsnm.9.525}. The authors also considered
the realistic conditions of imperfect orientation along [110]
direction, the cubic Dresselhaus term and inequality of the Rashba and
linear Dresselhaus spin-orbit coupling strengths \cite{Liu.jsnm.9.525}. Recent
experimental results confirmed the above theoretical predictions (see
Fig.~\ref{Denega_qw_or_exp}) \cite{2009arXiv0910.2336D}. In the
experiment, the strength of the Rashba and linear Dresselhaus spin-orbit
coupling is not equal. From the wire orientation dependence, the
authors estimated that the ratio of the strengths of the Rashba and
linear Dresselhaus spin-orbit coupling is about 2.

\subsubsection{Hole spin relaxation in metallic regime}

Because of the complexity of the valence bands, hole spin relaxation
is quite different from the electron spin relaxation. In bulk III-V and
II-VI semiconductors, due to the strong spin-orbit coupling and the
heavy-light-hole mixing, the D'yakonov-Perel' spin relaxation
is very efficient, yielding very short spin lifetime $\tau_s\sim
100$~fs \cite{PhysRevLett.89.146601,krauss:256601}. In nanostructures,
the heavy-light-hole degeneracy at ${\bf k}=0$ is lifted and the
effect of the spin-orbit coupling is reduced. Spin relaxation is hence
slowed down. In quantum wells, hole spin relaxation is in the
picosecond regime, usually longer at low temperature
\cite{syperek:187401,yugova:167402,2009arXiv0909.3711K}. Hole spin relaxation in nanostructures is more
complicated, as hole spin-orbit coupling and hole subband structure
can vary largely with the geometry (e.g., size and growth direction) of
the nanostructures \cite{winklerbook}. Besides the D'yakonov-Perel'
mechanism, other mechanisms may also be important. For example,
consider a unstrained symmetric $p$-type quantum well with only the
lowest heavy-hole subband being relevant. The Rashba spin-orbit
coupling vanishes as the structure is symmetric. The Luttinger
Hamiltonian only induces hole-spin-mixing but no zero-field spin
splitting in the lowest heavy-hole subband due to its space-inversion
symmetry. The zero-field spin splitting can only originate from the
Dresselhaus spin-orbit coupling, which is much weaker than the
spin-orbit coupling in bulk system due to the Luttinger
Hamiltonian. The D'yakonov-Perel' mechanism is then largely
suppressed. On the other hand, the spin-mixing induced by the
Luttinger Hamiltonian leads to an effective ``spin''-flip scattering
associated with any momentum scattering (the Elliott-Yafet-type
mechanism), such as hole-phonon scattering
\cite{PhysRevB.42.7114,PhysRevB.43.9687,Bastard1992335,PhysRevLett.64.3070,d'yakonov:1072}.
Such ``spin''-flip processes can be very efficient, especially near a
subband anti-crossing point \cite{lue:165321}. In general both the
D'yakonov-Perel' and Elliott-Yafet-type mechanisms should be
considered. Ferreira and Bastard compared the two spin relaxation
mechanisms and found that the D'yakonov-Perel' mechanism is usually
more important in asymmetric quantum wells
\cite{0295-5075-23-6-010}. Finally, in heavily $n$-doped samples, the
electron-hole exchange interaction (the Bir-Aronov-Pikus mechanism) may
also be relevant.

\begin{figure}[bth]
  \centering
  \includegraphics[height=7cm]{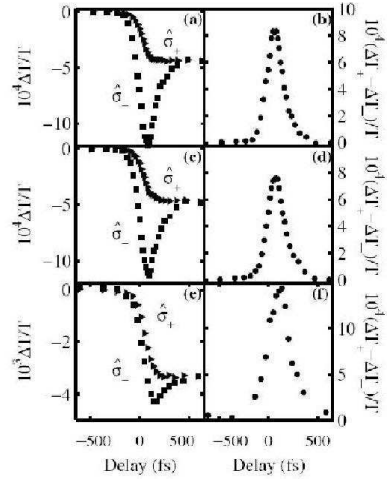}
  \caption{Pump-probe measurements of hole spin dynamics in bulk
    undoped GaAs at different probe wavelengths 3200 nm (top), 3000 nm
    (middle) and 3800 nm (bottom). $T$: transmision, $\Delta T$:
    differential transmision. Difference in differential transmision
    probed by right ($\Delta T_{+}$) and left ($\Delta T_{-}$)
    circular polarizations $(\Delta T_{+}-\Delta T_{-})/T$ is related
    to hole spin polarization. From Hilton and Tang
    \cite{PhysRevLett.89.146601}.}
  \label{Hilton_hole_bulk}
\end{figure}

Below we first review hole spin relaxation in bulk materials.
Experiments in undoped GaAs revealed that hole spin
lifetime is about $100$~fs at room temperature
\cite{PhysRevLett.89.146601}.\footnote{Hole spin relaxation in
  split-off bands was studied by Kauschke et
  al. \cite{PhysRevB.35.3843}.}  Theretical calculation of the
ultrafast nonequilibrium dynamics under optical excitation including
the spin-orbit coupling as well as hole-phonon and hole-hole
scatterings repeated the observed results \cite{krauss:256601}. Near
the equilibrium, calculation reported a spin lifetime about 200~fs for
heavy-hole and less than 100~fs for light-hole at room temperature
\cite{PhysRevB.71.245312}. Hole spin relaxation was also considered in 
the framework of non-Markovian stochastic theory
\cite{PhysRevB.71.233202}. Spin lifetime of the $\Gamma_7$ band
hole in wurtzite GaN was measured to be 120~ps
\cite{PhysRevB.72.121203}, much longer than that in GaAs. This is
because unlike the zinc-blende structures, valence bands in wurtzite semiconductors
are splitted into three {\em separated} bands, $\Gamma_9$, $\Gamma_7$
and $\Gamma_{7^\prime}$. The band separation reduces the effect of
the spin-orbit coupling and suppresses the D'yakonov-Perel spin relaxation.

Hole spin relaxation in bulk semiconductor under strain was studied by
D'yakonov and Perel' \cite{yakonov_strain}. It was found that the hole
spin along the strain axis relaxes much slower than the unstrained
samples. This is because strain lifts the heavy-hole and light-hole
degeneracy at $\Gamma$ point and largely reduces the
spin-orbit effect in both heavy- and light-hole bands.

\begin{figure}[bth]
  \begin{minipage}[h]{0.4\linewidth}
    \centering
    \includegraphics[height=5.5cm]{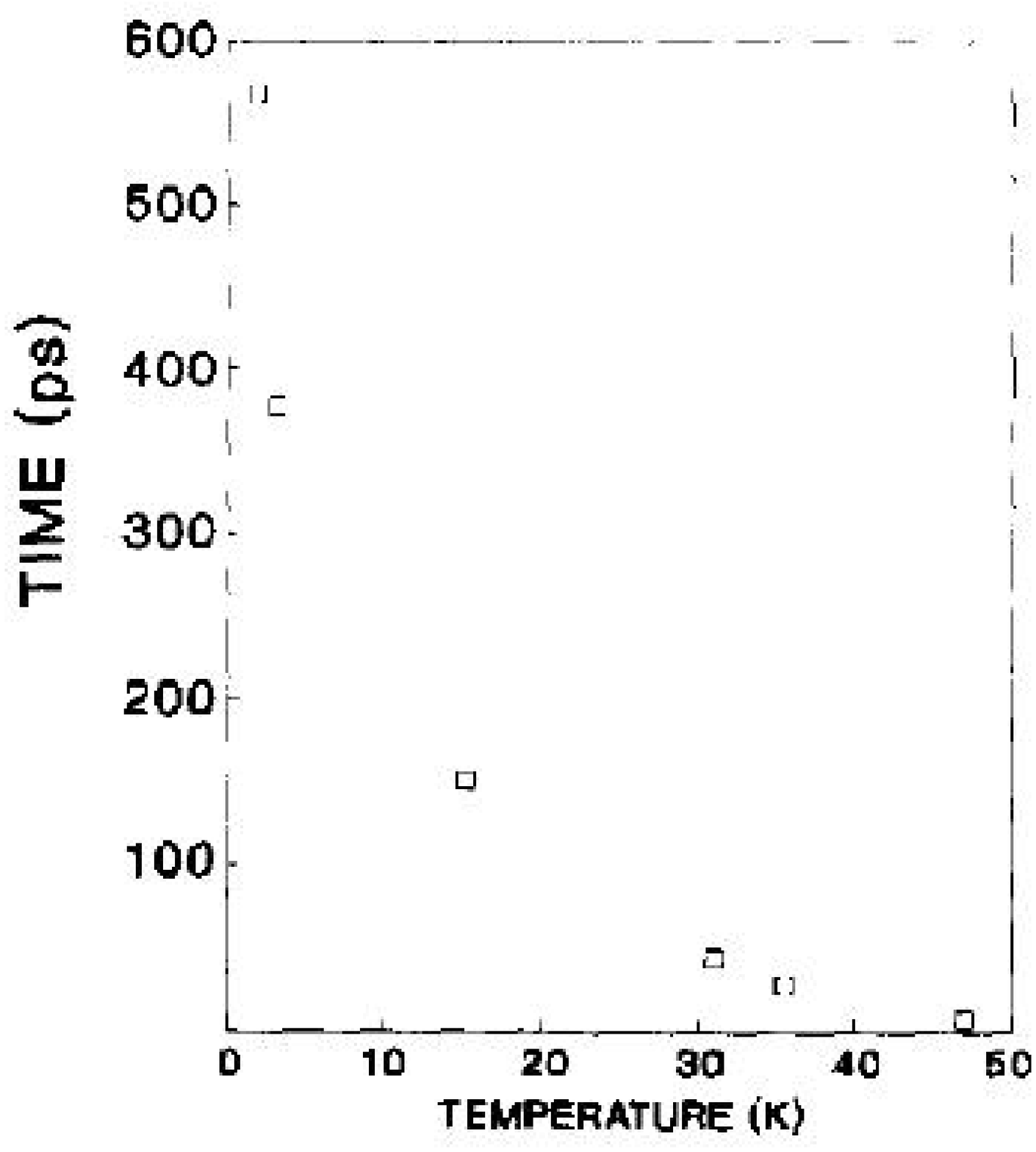}
    \caption{Temperature dependence of hole-spin lifetime in
      $n$-doped GaAs/AlGaAs quantum wells. Reproduced from Baylac et
      al. \cite{Baylac199557}.} 
    \label{Baylac_Tdep_ntype}
  \end{minipage}\hfill
  \begin{minipage}[h]{0.5\linewidth}
    \centering
    \includegraphics[height=5.5cm]{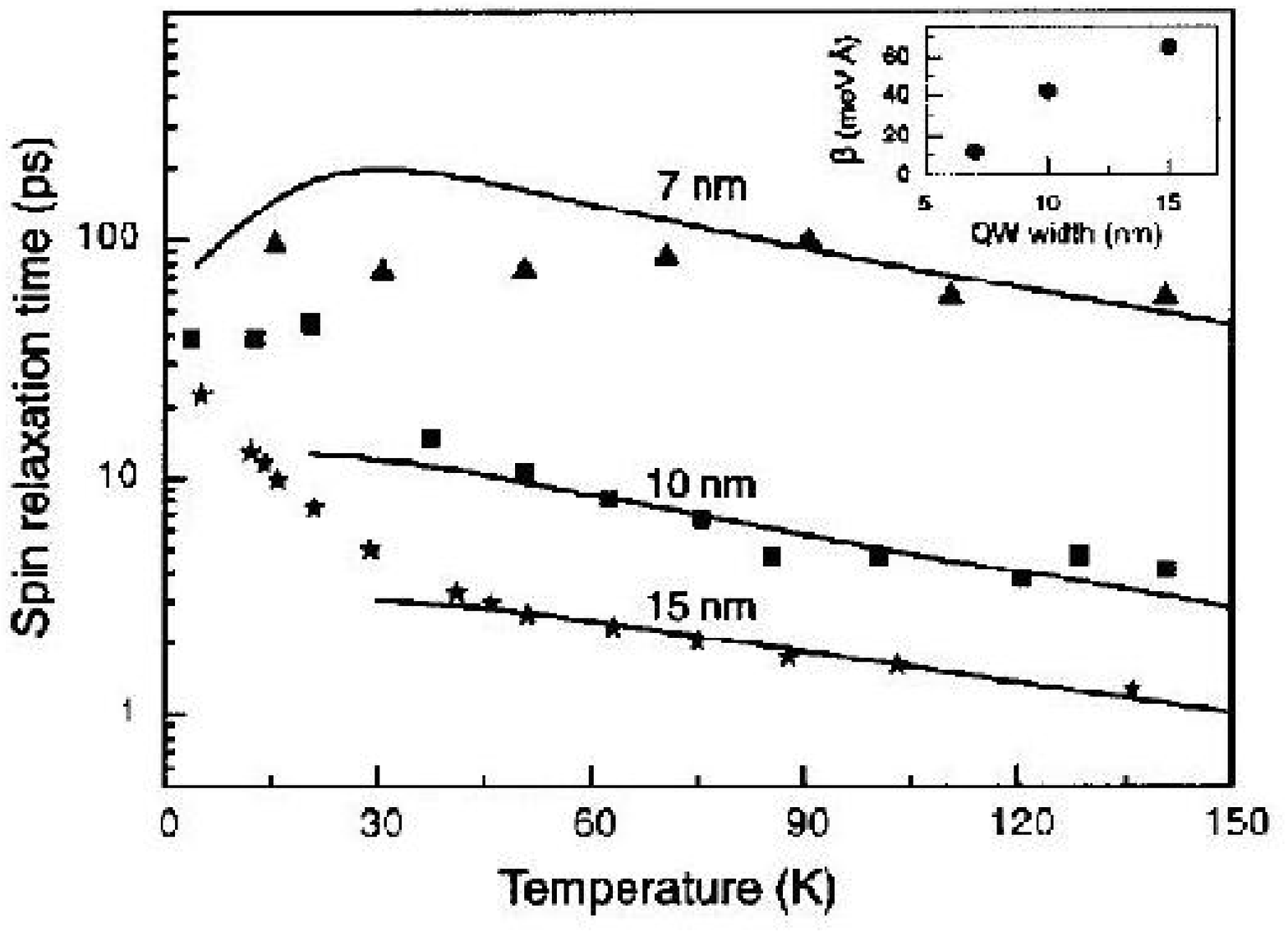}
    \caption{Temperature dependence of hole-spin relaxation time
    in $p$-doped GaAs/AlGaAs quantum wells. The curves are the
    calculated results. The inset shows the hole spin-orbit coupling
    constant $\beta$ (in meV \AA) as function of quantum well width.
    From Schneider
    et al. \cite{schneider:420}.}
  \label{hole_p_GaAsqw_T}
  \end{minipage}\hfill
\end{figure}

More studies were devoted to hole spin relaxation in two-dimensional
structures. Experimentally, Ganichev et al. measured hole spin
relaxation by spin-sensitive bleaching of intersubband hole spin
orientation in a $p$-type quantum well with well width $L_w=15$~nm. At
low temperature ($T<50$~K) hole spin lifetime varies as
$\sim T^{-1/2}$ for hole density $=2\times 10^{11}$~cm$^{-2}$
\cite{PhysRevLett.88.057401}. Subsequent studies via the same method
extended the data to $T<140$~K and to different well widths (see
Fig.~\ref{hole_p_GaAsqw_T}) \cite{schneider:420}. Strikingly, spin
relaxation is more efficient in wider quantum wells, in contrast to
the inverse tendency in electron case. This is a specific feature of
two-dimensional hole systems where the spin-orbit coupling is
determined by heavy-light hole mixing, which is stronger in wider
quantum wells [see Eq.~(\ref{alpha_hole}), the Rashba coefficient is
proportional to the inverse of the hole subband splitting, i.e., the
Rashba coefficient decreases with increasing well width]
\cite{0268-1242-23-11-114017}. Minkov et al. found that the hole spin
lifetime (deduced from the weak antilocalization measurements)
decreases rapidly with hole density at very low temperature $T=0.44$~K
for $3<n_h<10\times10^{11}$~cm$^{-2}$ but varies slowly with
temperature for $T<5$~K at $n_h\simeq 8 \times 10^{11}$~cm$^{-2}$ in
$p$-InGaAs/GaAs quantum wells \cite{PhysRevB.71.165312}. This is
consitent with the D'yakonov-Perel' spin relaxation in degenerate
regime, where temperature (hole density) changes both hole
distribution and hole-momentum scattering marginally (markedly). In
undoped 1.5 monolayer InAs/GaAs quantum well at 10~K, Li et
al. observed long hole spin lifetime ($\tau_s^h\simeq 200$~ps)
which can be comparable to the electron spin lifetime
$\tau_s^e\simeq 350$~ps in the same structure \cite{cpl.26.057303}.
This may be because that the hole spin-orbit coupling is weak in such
ultrathin layers, as hole spin-orbit coupling decreases with quantum
well width \cite{0268-1242-23-11-114017}. Interestingly, it was
  found that the hole spin lifetime has a peak in the temperature
  dependence \cite{Li2010}, which resembles the temperature dependence of electron
  spin lifetime in similar structure \cite{yang:035313,Sun2007158}. This is
  consistent with the D'yakonov-Perel' spin relaxation due to the
  hole-hole Coulomb scattering \cite{zhang:155311}.
Baylac et al. observed that the hole
spin lifetime decreases with photo-excitation energy at very
low temperature (1.7~K) \cite{Baylac1995161}. The result was explained
by the fact that hole spin relaxation rate increases with hole kinetic energy
\cite{Baylac1995161}. Both the D'yakonov-Perel' and Elliott-Yafet
mechanisms give such dependence
\cite{PhysRevB.43.9687,0295-5075-23-6-010,lue:125314}. In a designed
double quantum well structure, Lu et al. demonstrated that hole spin
relaxation in a narrow well with width $L_w=4.5$~nm at room
temperature can be suppressed ($\tau_s^h\simeq 100$~ps) if electrons
tunnel out to an adjacent quantum well rapidly \cite{Lu.5.326}. This
gives an evidence that the Bir-Aronov-Pikus mechanism plays important
role in hole spin relaxation. Hole spin relaxation time in $n$-doped
quantum wells was first measured by Damen et al., finding that
$\tau_s^h\simeq 4$~ps at 10~K \cite{PhysRevLett.67.3432}. Long hole
spin lifetime up to nanoseconds was observed in later experiments
in $n$-doped quantum wells at low temperature
\cite{PhysRevB.60.5811,PhysRevB.46.7292,Roussignol1994263,Baylac199557,yugova:167402}
which is assigned to localized holes. Hole spin relaxation in $n$-type
quantum wells was reported to decrease with magnetic field
\cite{PhysRevB.60.5811} which is because that the $g$-tensor
inhomogeneity mechanism is important in two-dimensional hole system
\cite{PhysRevB.66.113302}. Also hole spin lifetime decreases
drastically with increasing temperature in $n$-type quantum wells
\cite{Baylac199557,yugova:167402} (see Fig.~\ref{Baylac_Tdep_ntype})
as spin relaxation rate increases with increasing hole kinetic
energy. Hole spin relaxation in type-II quantum wells was studied by
Kawazoe et al. \cite{PhysRevB.47.10452} where two spin decay components
were observed: a faster one with $\tau_s^h\sim 20$-100~ps and a slower
one with $\tau_s^h\sim 20$~ns. The slower decay was attributed to the
localized holes \cite{PhysRevB.47.10452}. In II-VI semiconductor
quantum wells, such as CdTe quantum wells, hole spin lifetime
is on the order of tens of ps \cite{zhukov:155318,chen:115320,PhysRevB.68.235316}.

Theoretically, hole spin relaxation and its relation to
photoluminescence spectra were studied by Uenoyama and Sham in
$n$-doped, undoped and $p$-doped quantum wells
\cite{PhysRevB.42.7114,PhysRevLett.64.3070}, where the hole-phonon
scattering was examined. Specific calculations of ``spin''-flip
scattering, which leads to the Elliott-Yafet-like spin relaxation, were
performed by Ferreira and Bastard
\cite{PhysRevB.43.9687,0295-5075-23-6-010} and by 
Vervoort et al. \cite{PhysRevB.56.R12744}, taking into account of the 
interface inversion asymmetry. The D'yakonov-Perel' spin relaxation
due to the ``spin''-conserving scattering via standard single particle 
paradigm was given by Ferreira and Bastard \cite{0295-5075-23-6-010}.
The role of the Bir-Aronov-Pikus mechanism in hole spin relaxation in
quantum wells was discussed by Maialle \cite{0268-1242-13-8-004}. It was
found that in narrow quantum wells, the Bir-Aronov-Pikus mechanism is
comparable with other mechanisms.

Finally, it should be mentioned that up till now the hole spin relaxation
in one-dimensional structures was only investigated by L\"u et
al. theoretically \cite{lue:165321}. They studied the topic
comprehensively via the kinetic spin Bloch equation approach
\cite{PhysRevB.61.2945,wu:epjb.18.373,PhysRevB.68.075312}. Their
work will be introduced in Section~5.

\subsection{Carrier spin relaxation in III-V and II-VI paramagnetic diluted magnetic
  semiconductors}

Semiconductors dilutedly doped with magnetic impurities, such as Mn,
are interesting materials because the magnetism and electrics can
be incorporated together. Understanding carrier spin dynamics in
such kind of materials is important for the application. Also the
spin dynamics of the magnetic impurity or the magnetization dynamics is closely
related to the carrier spin dynamics. Most of the works in the literature focus on Mn
doped III-V and II-VI diluted magnetic semiconductors. In this
subsection we review carrier spin dynamics in these materials. We
restrict our review on spin dynamics in the paramagnetic phase, where
the description is much easier.

\begin{figure}[bth]
\begin{minipage}[h]{0.45\linewidth}
  \centering
  \includegraphics[height=5.4cm]{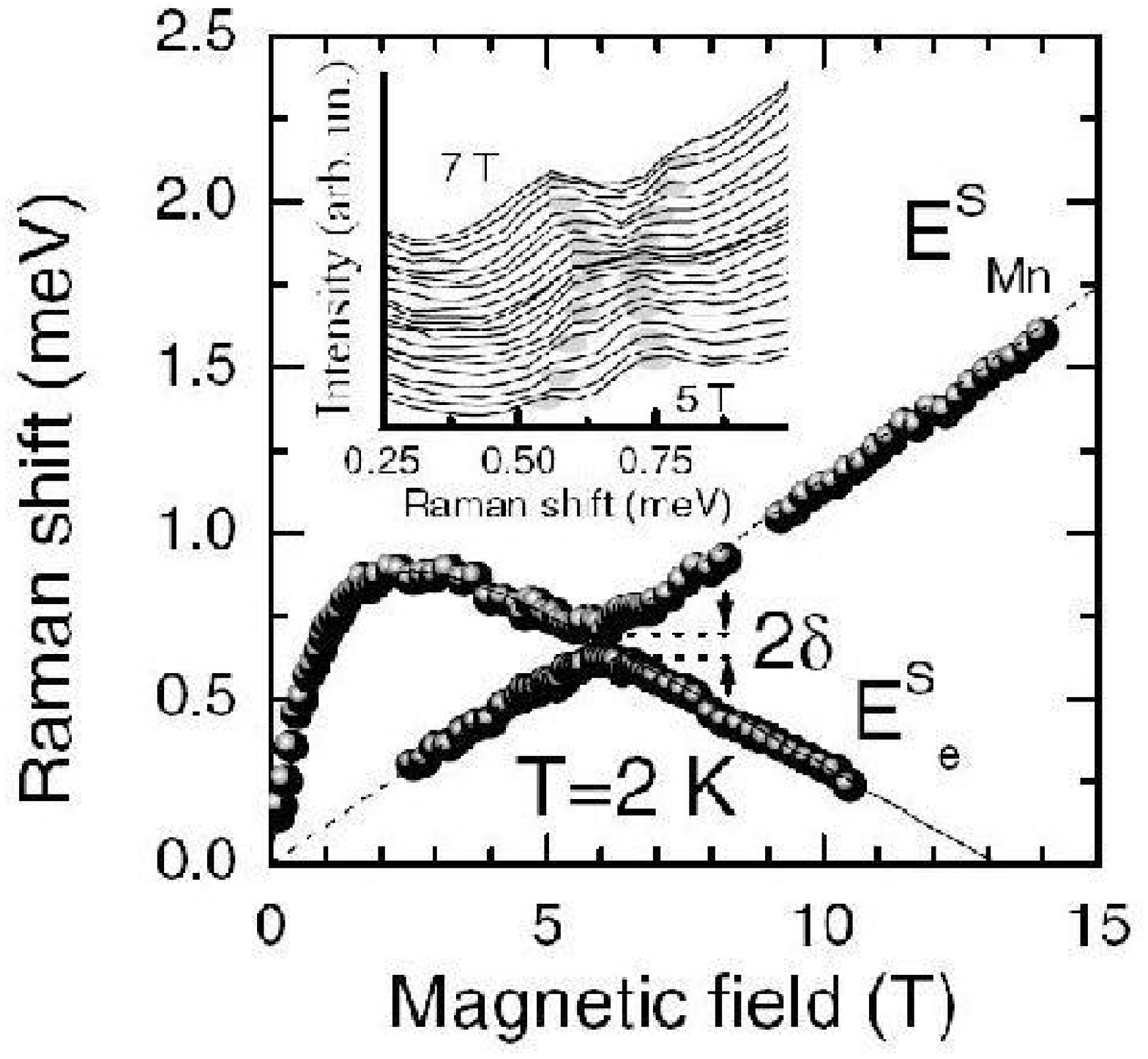}
  \caption{Raman shift for the spin-flip transitions for electron and
    Mn ion as function of magnetic field. The expected behavior for
    noninteracting subsystems calculated is indicated by the solid and
    dashed curves. $E^S_{\rm Mn}$ and $E^S_{e}$ denote Mn and electron
    spin splitting respectively. The anti-crossing splitting is
    $2\delta$. The temperature is $T=2$~K. Inset shows the Raman
    spectra (every 0.1~T) in the region of the anti-crossing. From
    Teran et al. \cite{PhysRevLett.91.077201}.}
  \label{Teran_sd_anticros}
\end{minipage}\hfill
  \begin{minipage}[h]{0.5\linewidth}
    \centering
    \includegraphics[height=4.5cm]{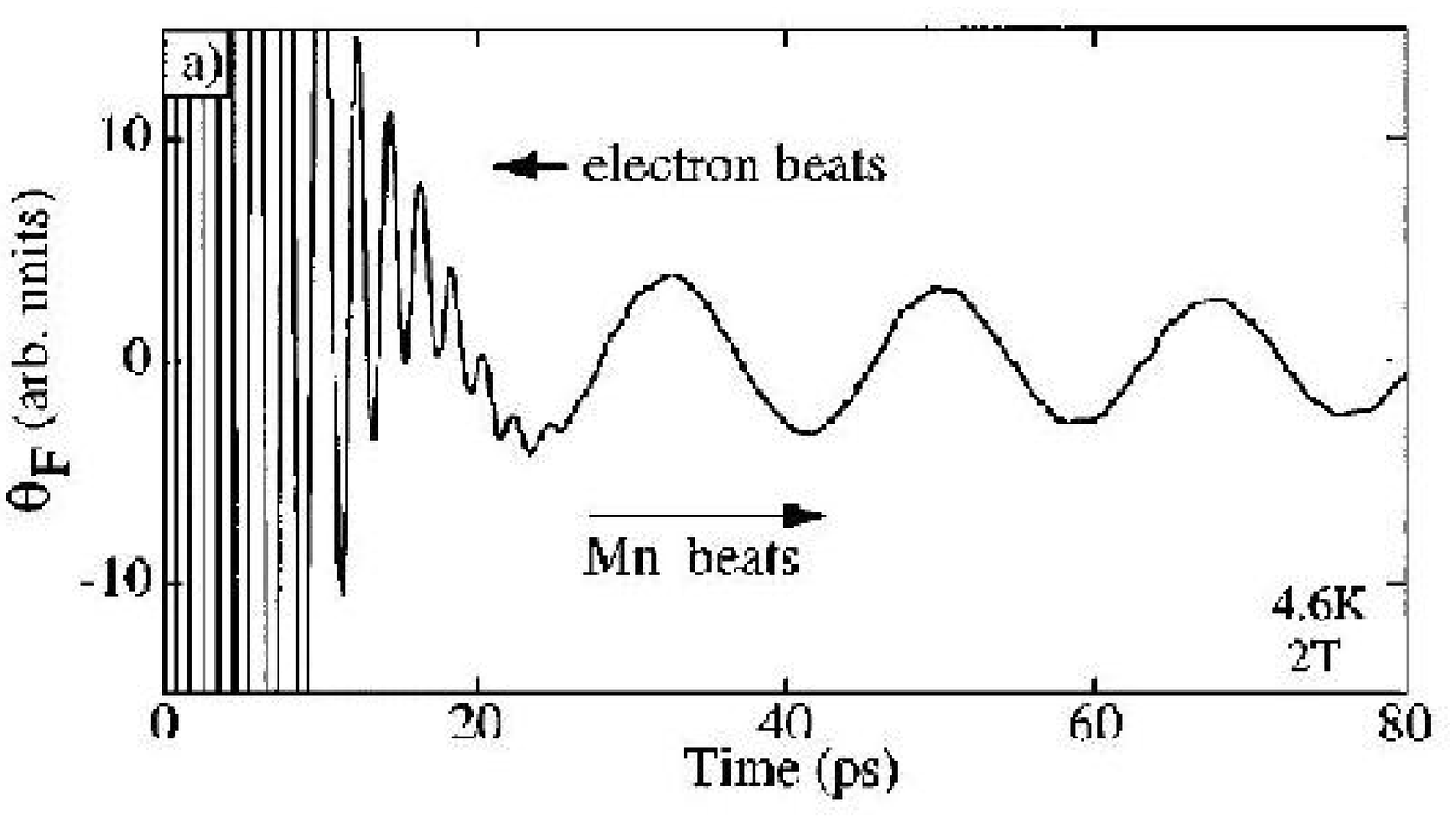}
    \caption{Induced Faraday rotation $\Theta_{\rm F}$, showing the
      final oscillations of the electron spins superimposed on an
      induced precession of the Mn spins. From Crooker et
      al. \cite{PhysRevLett.77.2814}.} 
    \label{Crooker_Mn_beats}
\end{minipage}\hfill
\end{figure}

A direct consequence of the $s(p)$-$d$ exchange interaction is that it
contributes to the carrier spin-flip scattering and hence carrier spin
relaxation. The $s(p)$-$d$ exchange interaction is believed to
dominate spin relaxation in (II,Mn)-VI semiconductors
\cite{PhysRevB.64.085331}. Besides spin relaxation, the $s(p)$-$d$
exchange interaction also leads to spin precession: carrier spin
precesses in the mean field of the $s(p)$-$d$ exchange
interaction. Under external magnetic field, carrier spins precess in
the coaction of the external field and the mean field of Mn spins. For
example, ${\bf B}\parallel \hat{{\bf x}}$, electron spin Larmor precession
frequency is then
\be
\omega_L = E_Z^e = |g_e\mu_BB-xN_0J_{\rm sd}\langle S^d_{x}\rangle|.
\ee
$\langle S^d_{x}\rangle=-\frac{5}{2}B_{5/2}[5g_{\rm Mn}\mu_BB/(2k_BT)]$
is the average Mn spin polarization at equilibrium where $B_{5/2}(x)$ is
the spin-5/2 Brillouin function and $g_{\rm Mn}$ is the Mn spin
$g$-factor. When the Mn density is high, the exchange interaction
can be much larger than the Zeeman interaction $g_e\mu_BB$, leading to
the giant Zeeman splitting. In ZnSe/MnSe heterostructures, a large
$g$-factor of 400 was observed by Crooker et
al. \cite{PhysRevLett.75.505}. The contribution from the exchange
interaction to spin precession leads to nonlinear magnetic field
dependence of Larmor frequency, which can be used to determine the
strength of the exchange coupling in experiment
\cite{PhysRevLett.95.017204,PhysRevLett.91.077201,PhysRevB.72.235313,stern:045329}.
The exchange mean field is largely reduced at elevated
temperature. This leads to a drastic decrease in spin precession
frequency with increasing temperature
\cite{PhysRevLett.95.017204,PhysRevB.72.235313,ronnburg:117203}.
The precession of carrier spins and Mn spins can be strongly coupled
when their frequencies match (see Fig.~\ref{Teran_sd_anticros})
\cite{PhysRevLett.91.077201,vladimirova:081305}. Besides,
nonequilibrium carrier spin polarization can be transferred to the Mn
spin system via exchange interaction
\cite{rcmyers:203,PhysRevLett.77.2814,PhysRevB.56.7574} and {\it vice
versa}
\cite{PhysRevB.59.2050,PhysRevB.65.035313,kneip:035306,kneip:045305,akimov:165328,tsitsishvili:155305}.
Mn spin beats were observed by Crooker et al. (see Fig.~\ref{Crooker_Mn_beats})
\cite{PhysRevLett.77.2814,PhysRevB.56.7574} and theoretically studied by
Linder and Sham \cite{Linder1998412}.

Historically, carrier spin dynamics was first studied in (II,Mn)VI
diluted magnetic semiconductors. In these materials no charge doping is invoked,
as Mn$\Rightarrow$Mn$^{2+}+2e^{-}$. At high doping density and low
temperature the Mn-Mn interaction leads to a spin-glass
order. However, at high temperature or low doping density it is
paramagnetic. The bandgap also changes with Mn doping. For example in
CdMnTe, the bandgap increases with Mn doping density. A widely studied
system is Cd$_{1-x}$Mn$_x$Te/Cd$_{1-y}$Mn$_y$Te quantum well,
where the layer with higher Mn density serves as barrier. A particular
case is $x=0$, the $s(p)$-$d$ exchange interaction is then incorporated
through the wavefunction penetrated into the barrier layers. The
$s(p)$-$d$ exchange interaction can then be engineered via tuning the
structure \cite{PhysRevLett.64.2430}. Another example of such
structure is to insert MnSe layers into ZnCdSe quantum wells
\cite{PhysRevLett.75.505}. Crooker et al. found that the $s(p)$-$d$
exchange interaction changes significantly with the number of MnSe
layers [in their experiment, they used 1($\times 3$ atomic layer),
3($\times1$ atomic layer) and 24 ($\times 1/8$ atomic layer) layers of
MnSe] in ZnCdSe quantum well while keeping the overall Mn doping
density unchanged
\cite{PhysRevLett.75.505}. The modification of the $s(p)$-$d$ exchange
interaction comes from tuning the overlap between the 
confined electrons/holes and Mn ions by structure engineering. Another reason is
that the Mn-Mn antiferromagnetic coupling between neighboring Mn spins
which ``locks'' the Mn spins is largely suppressed in $1/8$ MnSe
sub-monolayers as the average distance between Mn spins increases.
Interestingly Smyth et al. found that the spin subband of the
ZnMnSe/ZnSe superlattice can be significantly modified by external
magnetic field \cite{PhysRevLett.71.601}: Due to the giant Zeeman
splitting, the spin-down subband is lowered and spin-up
subband is raised. At the threshold magnetic field $B_{c}$, spin-up
subband energy is as high as the barrier energy. Above $B_{c}$ spin-up
electrons switch into the ZnSe layers. The spin and electronic
structure is then tuned by the external magnetic field. In parabolic
ZnSe/ZnCdSe quantum well with MnSe monolayers inserted, Myers et
al. demonstrated that the $s$-$d$ exchange interaction can be
manipulated by gate voltage \cite{PhysRevB.72.041302}. As the
$s(p)$-$d$ exchange interaction is 
tuned, the carrier spin dynamics is also manipulated
\cite{PhysRevLett.75.505,PhysRevLett.71.601,PhysRevB.72.041302,PhysRevB.56.9726}.

\begin{figure}[bth]
  \centering
  \includegraphics[height=7.5cm,angle=90]{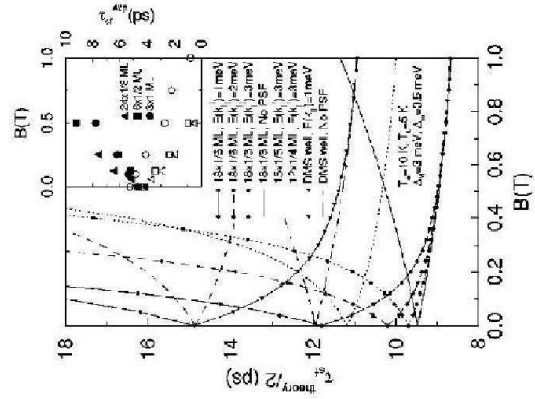}
  \caption{Calculated electron spin-flip times $\tau_{\rm
        sf}^{\rm theory}/2$ as a function of the
    magnetic field $B$. The inclusion of phase-space-filling (PSF)
    enhances the two-branch feature of the $\tau_{sf}$ {\sl vs.} $B$ curves.
    Different structures are considered: heterostructures with
    $18\times1/6$ monolayer (ML), $15\times1/5$ ML, $12\times1/4$ ML
    and diluted magnetic semiconductor (DMS) quantum well.
    Inset: experimental results by Crooker et
    al. \cite{PhysRevLett.75.505} with various heterostructures:
    $24\times 1/8$ ML (triangle), $6\times 1/2$ ML (square) and
    $3\times 1$ ML (circle). The two branches of spin-flip time
    correspond to $\tau_{\up\to\down}$ and $\tau_{\down\to \up}$.
    The calculation is not intended to
    compare quantitatively with the experimental results.
    From Egues and Wilkins \cite{PhysRevB.58.R16012}.}
  \label{Egues_B_tau}
\end{figure}

Femtosecond pump-probe measurements were first performed by Freeman et
al. \cite{PhysRevLett.64.2430}, revealing a very short spin relaxation
time of $\sim$3~ps in Cd$_{1-x}$Mn$_x$Te/Cd$_{1-y}$Mn$_y$Te quantum wells
\cite{PhysRevLett.64.2430,freeman:5102}. After that, Bastard and
Chang \cite{PhysRevB.41.7899}, Bastard and Ferreira
\cite{Bastard1992335} presented theoretical studies on spin relaxation
due to the $s$-$d$ exchange interaction in such kind of quantum
wells. They obtained comparable spin lifetime with
experiments. They also predicted well width dependence of spin
lifetime and proposed a bias double-quantum-well structure to
manipulate the spin lifetime. Interesting behavior was observed
under magnetic field. As the neighbor Mn ions are coupled
antiferromagnetically, Mn ions can link together to form clusters
  which have small spin momentums and become ineffective in
spin-flip scattering. Hence, only {\em isolated} Mn 
ions count. Bastard and Chang took an effective Mn density as
$n_{\rm eff}=x(1-x)^{12}N_0$ to warrant only Mn ions without any nearest
neighbors. Akimoto et al. \cite{PhysRevB.56.9726} found that electron
spin relaxation rate in CdTe/CdMnTe quantum wells is proportional to
the probability of finding electron in the barrier layer of CdMnTe, which
is consistent with the theory of Bastard and Chang
\cite{PhysRevB.41.7899}. As stated in previous paragraph, spin-up
subband is raised higher than the spin-down one. Hence spin-flip
scattering from up-spin to down-spin is more favorable than the
opposite one at low temperature. Smyth et al. reported that
spin-flip time of spin-up electron decreases with increasing magnetic
field whereas that of spin-down electron increases: the spin-flip
times vary with magnetic field as two branches
\cite{PhysRevLett.71.601}. Similar behavior was observed in other
diluted magnetic heterostructures (see inset of Fig.~\ref{Egues_B_tau})
\cite{PhysRevLett.75.505,PhysRevB.50.10851,PhysRevB.50.7689,PhysRevB.71.165203,488685,baumberg:6199}.
Such two-branch behavior is weakened with increasing excess
photo-carrier energy or decreasing exchange interaction
\cite{PhysRevB.50.10851,488685}. Both electron and hole spin
lifetimes were found to be smaller in diluted magnetic
heterostructures compared to the nonmagnetic ones
\cite{PhysRevLett.77.2814}, which indicates the relevance of the
$s(p)$-$d$ exchange interaction to the spin relaxation.

\begin{figure}[bth]
  \centering
  \includegraphics[height=5cm]{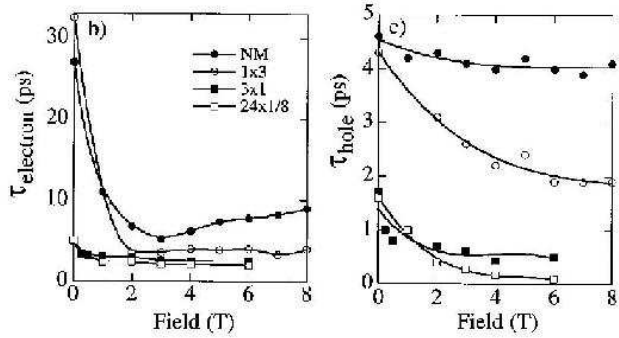}
  \caption{Electron (left) and hole (right) spin lifetime as
    function of transverse magnetic field for nonmagnetic (NM)
    (Zn$_{0.77}$Cd$_{0.23}$Se/ZnSe) quantum well as well as that for
    quantum well with $1\times 3$, $3\times 1$ and $24\times 1/8$ MnSe
    monolayers inserted (for the details of the heterostructures see
    Ref.~\cite{PhysRevLett.77.2814}). $T=4.6$~K. From 
    Crooker et al. \cite{PhysRevLett.77.2814}.}
  \label{Crooker_tau_B}
\end{figure}

Theoretically, spin relaxation rate can be calculated from the Fermi
Golden rule \cite{J.Kossut.pssb,PhysRevB.39.10918},
\be
\Gamma_{{\bf k}\uparrow,{\bf k}^{\prime}\downarrow} =
2\pi |\langle {\bf k}\uparrow|H_{s-d}|{\bf
  k}^{\prime}\downarrow\rangle|^2 \delta(\varepsilon_{{\bf
    k}\uparrow}-\varepsilon_{{\bf k}^{\prime}\downarrow}).
\ee
The spin-flip rates are
\be
\tau^{-1}_{\up\to \down} = \langle \Gamma_{{\bf k}\uparrow,{\bf
    k}^{\prime}\downarrow}(1-f_{{\bf  k}^{\prime}\downarrow})\rangle,\quad\quad
\tau^{-1}_{\down\to \up} = \langle \Gamma_{{\bf k}\downarrow,{\bf
    k}^{\prime}\uparrow}(1-f_{{\bf  k}^{\prime}\uparrow})\rangle.
\ee
Here ${\up\to \down}$ (${\down\to \up}$) denotes the spin-up to spin-down
(spin-down to spin-up) transition. $\langle...\rangle$ means average
over the electron ensemble. The total spin relaxation rate is
$\tau^{-1}_s = \tau^{-1}_{\up\to \down} + \tau^{-1}_{\down\to \up}$.
For example in quantum wells at zero magnetic field
\cite{jiang:155201,PhysRevB.67.115319,tsitsishvili:195402},
\be
\tau_s^{-1} = I_s xN_0 J_{\rm sd}^2m_e \langle\langle S^d_{-}S^d_{+} + S^d_{+}S^d_{-} \rangle\rangle/(4L_w),
\ee
where $J_{\rm sd}$ is the $s$-$d$ exchange constant, $x$ stands for
the mole fraction of Mn and $N_0$ denotes the density of the unit
cells. ${\bf S}^d=(S^d_x,S^d_y,S^d_z)$ is Mn spin operator and
$S^d_{\pm}=S^d_{x}\pm iS^{d}_{y}$. $\langle\langle...\rangle\rangle$
stands for average over Mn spin distribution. Here $I_s=L_w\int
|\psi_e^{\uparrow}(z)|^2|\psi_e^{\downarrow}(z)|^2 {\rm d}{z}$ with
$\psi^{\uparrow(\downarrow)}_e$ being the spin-resolved electron
subband wavefunction and $L_w$ denoting the well width. At equilibrium state,
$\langle\langle S^d_{-}S^d_{+} + S^d_{+}S^d_{-}
\rangle\rangle=\frac{4}{3}S_d(S_d+1)$. As to the two spin-flip
transition rates, $\tau^{-1}_{\up\to \down}\propto \langle\langle
S^{d}_{-}S^{d}_{+}\rangle\rangle=S_d(S_d+1)-\langle\langle
(S^{d}_{z})^2\rangle\rangle-\langle\langle S^{d}_{z}\rangle\rangle$ 
whereas $\tau^{-1}_{\up\to \down}\propto \langle\langle
S^{d}_{+}S^{d}_{-}\rangle\rangle=S_d(S_d+1)-\langle\langle 
(S^{d}_{z})^2\rangle\rangle+\langle\langle S^{d}_{z}\rangle\rangle$.
In the presence of a magnetic field along the $z$ axis, the Mn spin
polarization $\langle\langle S^{d}_{z}\rangle\rangle<0$. This enhances
the spin-up to spin-down transition but suppresses the inverse
process. The calculation of Egues and Wilkins indicated that the
phase-space-filling effect (Pauli blocking factor $(1-f_{{\bf
    k}^{\prime}\uparrow(\downarrow)})$) further enhances the
difference between the two spin-flip scattering rates (see Fig.~\ref{Egues_B_tau})
\cite{PhysRevB.58.R16012}. It is noted that the total spin relaxation
rate is proportional to $S_d(S_d+1)-\langle\langle
(S^{d}_{z})^2\rangle\rangle$ which decreases with increasing magnetic
field. The underlying physics is that the magnetic field pins the Mn spin
and then suppresses the spin-flip processes and electron spin relaxation
\cite{murayama:261105,pssc.3.1109}. Magnetic field effect on
electron/hole/exciton spin-flip scattering rates at low temperature in
CdMnTe quantum wells was investigated comprehensively by Tsitsishvili and Kalt \cite{tsitsishvili:195402}.
In most cases the magnetic field dependence of spin relaxation rate is
weak. Unlike longitudinal magnetic field, a transverse magnetic field
always leads to faster spin relaxation (for both electrons and holes)
(see Fig.~\ref{Crooker_tau_B})
\cite{PhysRevLett.72.717,PhysRevB.70.115307,Kikkawa1998394,PhysRevLett.77.2814,PhysRevB.56.7574,PhysRevB.64.085331}.
This is because that the magnetic field pinning of Mn spins along
transverse direction increases the factor $S_d(S_d+1)-\langle\langle
(S^{d}_{z})^2\rangle\rangle=\langle\langle
(S^{d}_{x})^2+(S^{d}_{y})^2\rangle\rangle$
\cite{jiang:155201,PhysRevB.67.115319}.
Electron spin relaxation is enhanced for exciton bound electrons as
the center-of-mass effective mass is larger for exciton. This was first
predicted by Bastard and Ferreira \cite{Bastard1992335} and then
confirmed experimentally by Smits et al. \cite{PhysRevB.70.115307}.
Temperature dependence of electron, hole and exciton spin relaxation
has been studied in
Refs.~\cite{zhu:142109,zhou:053901,ronnburg:117203,PhysRevB.70.115307}.
Smits et al. presented a systematic experimental study on excitonic
enhancement of electron/hole spin relaxation in CdMnTe based quantum
well structures \cite{PhysRevB.70.115307}.

\begin{figure}[bth]
  \centering
  \includegraphics[height=5.5cm]{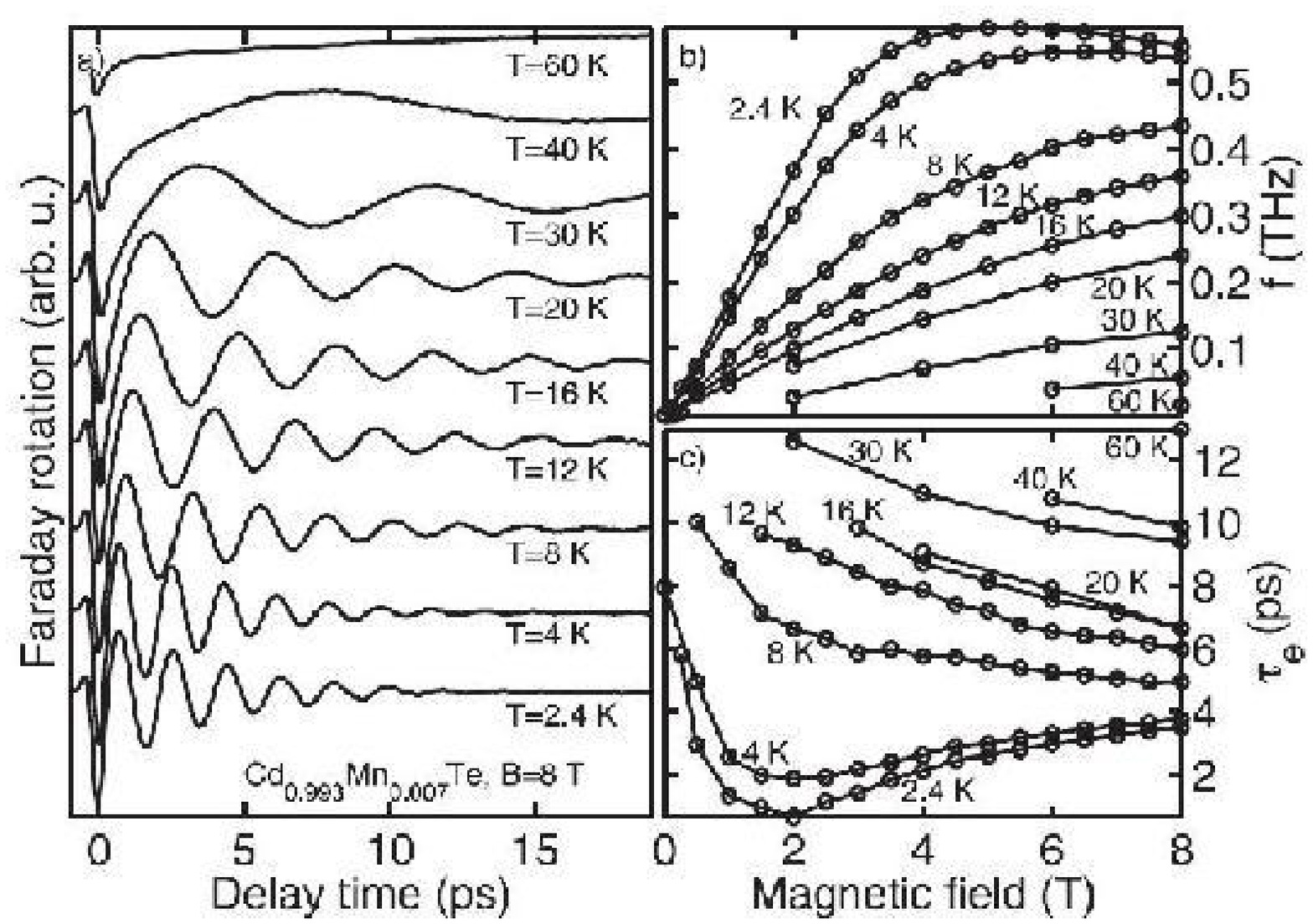}
  \caption{Results of time-resolved Faraday rotation experiments on 
    Cd$_{1-x}$Mn$_x$Te (x=0.0073) at 8 T and for various temperatures:
    (a) Temperature dependence of the Faraday-rotation transients;
    (b) temperature and magnetic-field dependence of the electron
    spin precession frequency extracted from the transients; (c)
    fitted electron spin dephasing time $\tau_e$ as a function of
    temperature and magnetic field. From R\"{o}nnburg et
    al. \cite{ronnburg:117203}.}
  \label{Ronnburg_CdMnTe_B_T_dep}
\vskip 0.2cm
 \begin{minipage}[h]{0.45\linewidth}
\centering
\includegraphics[height=5.5cm]{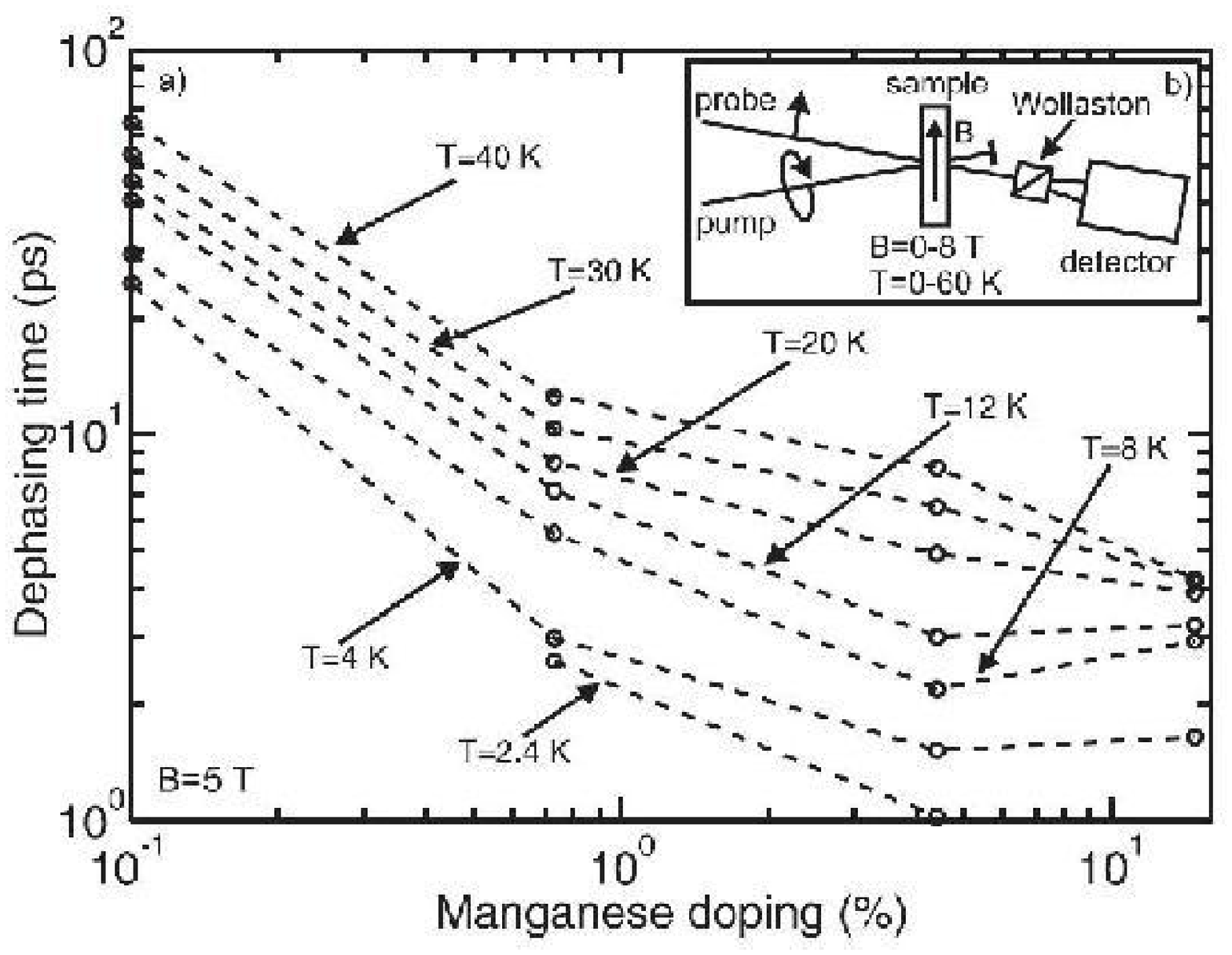}
 \end{minipage}\hfill
 \begin{minipage}[h]{0.45\linewidth}
   \centering
  \includegraphics[height=5.5cm]{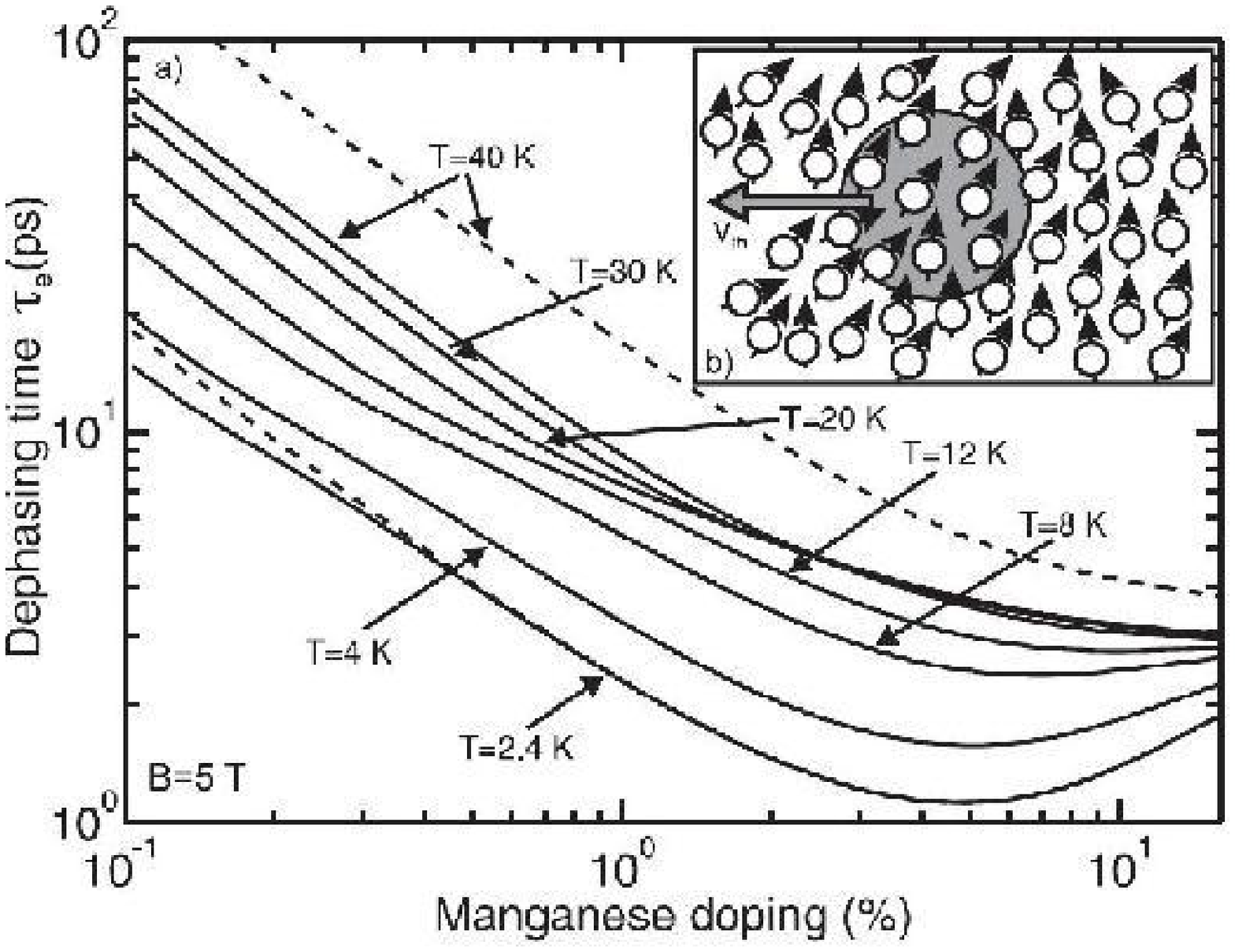}
\end{minipage}\hfill
  \caption{Left: experiment. Right: theory. Doping density and
    temperature dependence of electron spin dephasing time. Magnetic
    field is fixed at 5~T. Inset in right figure: the picture of
    exciton moving through the local moments of magnetic dopants. 
    Inset in the left figure: experimental set-up.
    From R\"onnburg et al. \cite{ronnburg:117203}.}
  \label{Ronnburg_CdMnTe_x_T_dep}
\end{figure}

Recently, R\"onnburg et al. presented a systematic experimental
investigation on electron spin relaxation in bulk CdMnTe. Their major
results are shown in Figs.~\ref{Ronnburg_CdMnTe_B_T_dep} and
\ref{Ronnburg_CdMnTe_x_T_dep}. In the experiment, the photon energy was
chosen to be centered on the 1s exciton absorption lines and the
photo-excitation intensity was kept low in order to minimize sample
heating and carrier-density-dependent effects. A salient feature is
that electron spin lifetime increases with temperature, which
signals a motional-narrowing spin relaxation mechanism. This is quite
different from the previously considered $s$-$d$ exchange scattering
mechanism where electron spin relaxation rate in bulk system is
proportional to $\langle k\rangle$, which increases with temperature.
R\"onnburg et al. then presented a new theory of spin relaxation based
on the thermal and spacial fluctuation of the $s$-$d$ mean field.
The transverse spin relaxation rate is given by
\be
\frac{1}{T_2} = \gamma^2\tau_0\left[\langle(\delta M_z)^2\rangle + \frac{\left(\langle(\delta M_x)^2\rangle +\langle(\delta M_y)^2\rangle\right)/2}{1+(g_e\mu_BB+\gamma\langle\delta M_z\rangle)^2\tau_0^2}\right],
\ee
where $\gamma=J_{\rm sd}/( g_{\rm Mn}\mu_B)$, ${\bf
  M}=(M_x,M_y,M_z)$ is the (Mn spin) magnetization and $\delta M_i$
($i=x,y,z$) denotes the fluctuation of magnetization felt by optically
excited electrons. $\tau_0$ stands for the correlation time of the
fluctuations. $g_e$ is the intrinsic $g$-factor of the itinerant
electrons. By working out the thermal and spacial fluctuations, the
spin relaxation rate is given explicitly as \cite{ronnburg:117203}
\be
\frac{1}{T_2} = \frac{\gamma^2\tau_0}{N_{\rm
    ex}}\left[n_0xk_BT\chi(B,T)+\langle M_z\rangle^2 +
\frac{n_0xk_BT[3\chi(0,T)-\chi(B,T)]-\langle
  M_z\rangle^2}{2+2(g_e\mu_BB+\gamma\langle\delta
  M_z\rangle)^2\tau_0^2}\right]. 
\ee
Here $\chi(B,T)=\frac{\partial\langle M_z(B,T)\rangle}{\partial B}$ is
the susceptibility. $\langle ...\rangle$ denotes the thermal average
of Mn spins. By assuming the spacial fluctuation to be a Poisson
distribution, it gives the amplitude of the spacial fluctuation
$\langle (\delta M_i)^2\rangle_{\rm sf}=\langle M_i\rangle^2$ ($i=x,y,z$).
According to the central limit theorem, the fluctuation is reduced by
the number of Mn spins felt by the electron wave-packet $N_{\rm ex}$.
In intrinsic bulk CdMnTe at low temperature electrons and holes are
bound together to form excitons. The spacial extension of
exciton-bound electron wave-packet is controlled by two factors: the
exciton Bohr radius and the thermal length $L_{\rm th}$ of the
center-of-mass motion. The latter is estimated as the inverse of the
thermal fluctuation of the center-of-mass wave-vector, 
$L_{\rm th}=0.37/\sqrt{mk_BT}$ with $m=m_e+m_h$. Taking into account
these two factors, one obtains the volume of the wave-packet $V_{\rm ex}$ and then
the number of Mn spins within the volume $N_{\rm ex}=xn_0V_{\rm ex}$.
The correlation time also consists of two contributions,
$1/\tau_0=1/\tau_{\rm prop}+1/\tau^{\rm Mn}_{\rm sf}$. $\tau^{\rm
  Mn}_{\rm sf}$ is the Mn spin-flip time which is at least several
hundreds of picoseconds \cite{PhysRevB.56.7574} and hence
ineffective. $\tau_{\rm prop}$ denotes the exciton propagation time,
which is the time for an exciton to see a new environment of Mn spins.
It is approximated as the time for an exciton to propagate a distance
equal to its spacial extension with a thermal average velocity. In
such a way $\tau_0$ is determined. By taking parameters in the
literature, R\"onnburg et al. calculated the transverse spin
lifetime and found that the calculated results agree remarkably
well with experimental data both qualitatively and quantitatively (see
Fig.~\ref{Ronnburg_CdMnTe_x_T_dep}). The Mn doping $x$ dependence of
$T_2$ at small $x$ shows $1/x$ behavior as $M_z$, $\chi$ and $N_{\rm
  ex}$ are proportional to $x$. At high Mn doping, the
antiferromagnetic interaction between neighboring Mn ions should be
taken into account with a correction $x_{\rm eff}=x(1-x)^{12}$
\cite{PhysRevB.58.R16012}. This leads to the observed increase of
$T_2$ at large $x$ in Fig.~\ref{Ronnburg_CdMnTe_x_T_dep}. The magnetic
field dependence also agrees well with the calculation (not shown). The
intial decrease of $T_2$ with magnetic field is mainly due to the increase
of $\langle M_z\rangle^2$. At high magnetic field, $\chi(B,T)$ and
$\langle M_z\rangle$ saturate. The magnetic field dependence then
mainly comes from the factor $1/[2+2(g_e\mu_BB+\gamma\langle
\delta M_z\rangle)^2\tau_0^2]$ which makes $T_2$ increase with
$B$ (Similar behavior was also observed in the experiment of
Cronenberger et al. \cite{Cronenberger2008427}). The magnetic field dependence is
more pronounced at low temperature where $\tau_0$ is larger and the
magnetic field dependence of $\langle M_z\rangle$ is stronger.

\begin{figure}[bth]
  \begin{minipage}[h]{0.4\linewidth}
    \centering
  \includegraphics[height=5.1cm]{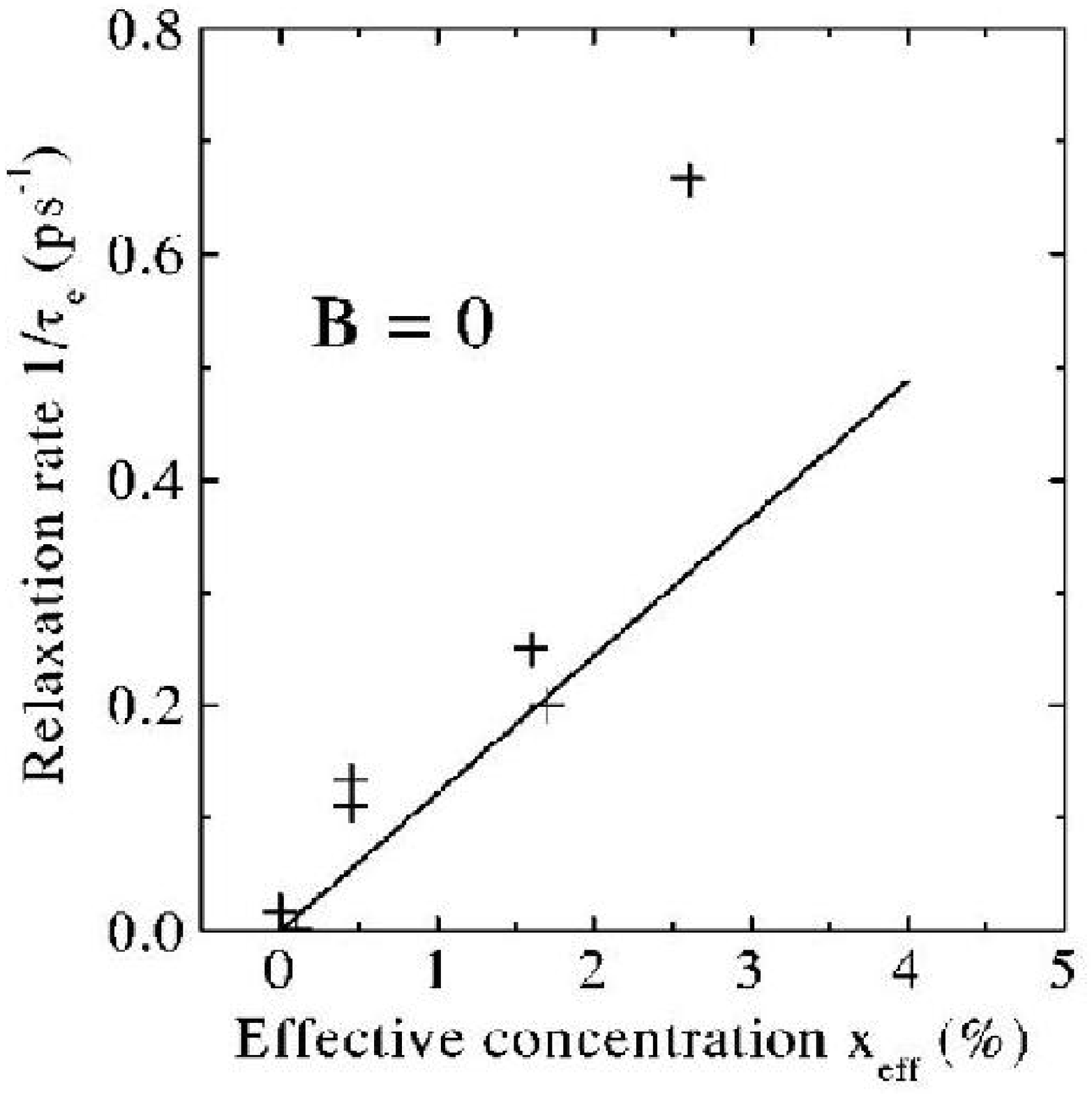}
  \caption{Electron spin relaxation rates {\sl vs.} Mn effective content
    $x_{\rm eff}$ for a QW width $L_w=8$~nm: experimental data (crosses),
    calculated values (solid line). From
    Camilleri et al. \cite{PhysRevB.64.085331}.} 
  \label{Camilleri_CdMnAs_xdep}
  \end{minipage}\hfill
  \begin{minipage}[h]{0.5\linewidth}
  \centering
  \includegraphics[height=5cm]{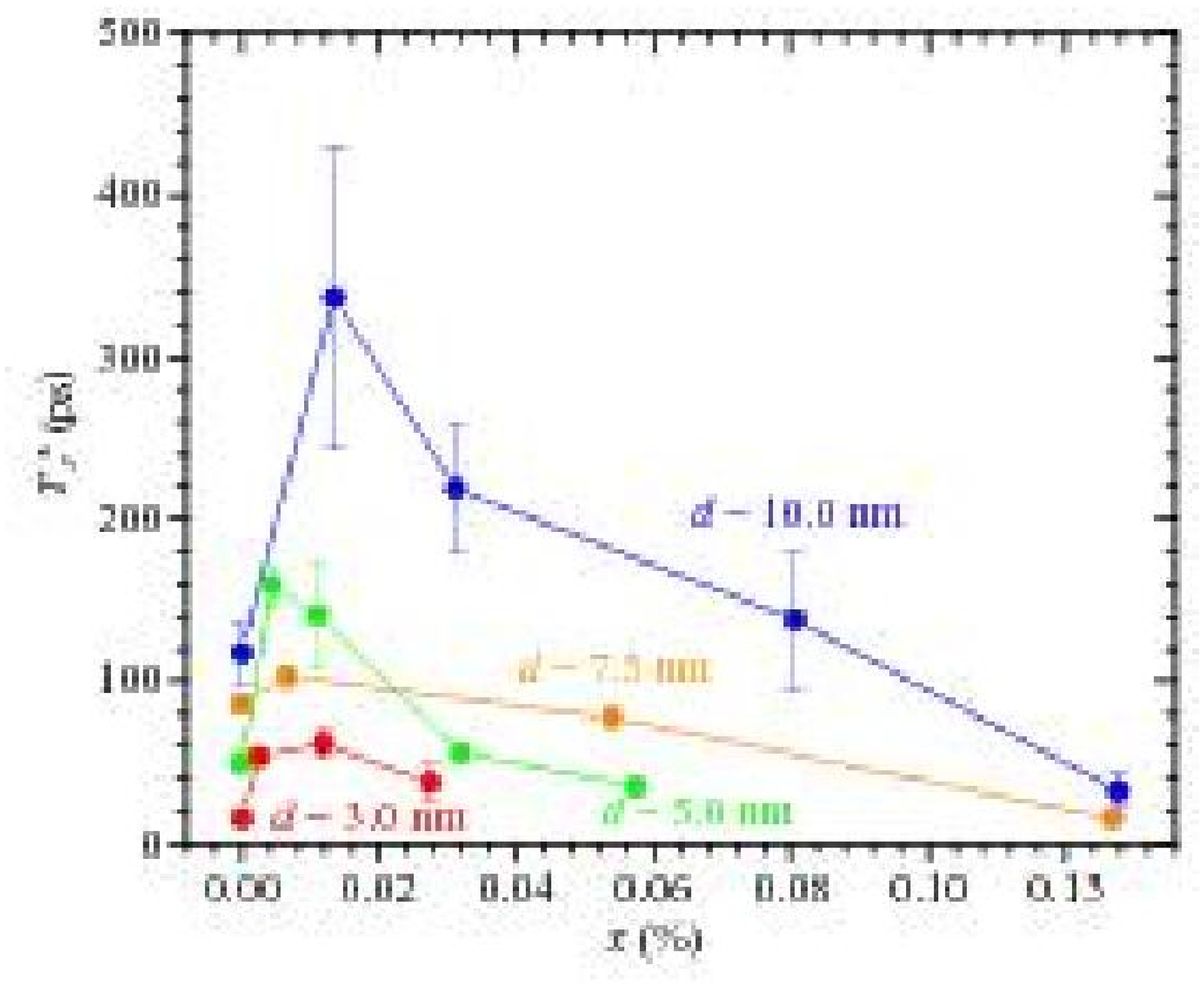}
  \caption{The transverse electron spin lifetime $T_2^{\ast}$ at
    $T=5$~K versus the percentage of Mn $x$ for four quantum well sample
    sets of varies widths. The plotted values of $T_2^{\ast}$ are the mean
    values from magnetic field $B=0$ to 8 T, and the error bars
    represent the standard deviations. From Poggio et
    al. \cite{PhysRevB.72.235313}.} 
  \label{Poggio_GaMnAs_xdep}
\end{minipage}\hfill
\end{figure}

The Mn density dependence is informative and reveals the underlying
spin relaxation mechanisms. In CdMnTe quantum wells, spin relaxation
rate was found to be proportional to $x$ for $x>4\times 10^{-3}$,
indicating spin relaxation due to the $s$-$d$ exchange scattering
(see Fig.~\ref{Camilleri_CdMnAs_xdep})
\cite{PhysRevB.64.085331}. However, in GaMnAs quantum wells Poggio et
al. found that spin lifetime first increases and then decreases with
increasing Mn density. A peak appears around $x\sim 10^{-4}$ for well width from
3 to 10~nm (see Fig.~\ref{Poggio_GaMnAs_xdep})
\cite{PhysRevB.72.235313}. It was speculated that the dominant spin
relaxation mechanisms for Mn densities below and above the peak density
are different. However, calculation based on realistic parameters is
needed to determine the relevance of various mechanisms. Such
calculation has been done by Jiang et al. recently \cite{jiang:155201}.

Relaxation dynamics of spin-polarized electrons excited at higher
energy subband was revealed by probe-energy dependence of the spin
lifetime in Ref.~\cite{saito:262111}. The probe-energy
dependence is understood by the intraband electron energy relaxation
and subsequent spin relaxation. Energy-resolved spin relaxation in
wurtzite magnetic semiconductor CdMnSe was studied in
Ref.~\cite{QE.24.315}.

\begin{figure}[bth]
  \begin{minipage}[h]{0.5\linewidth}
    \centering
    \includegraphics[height=6cm]{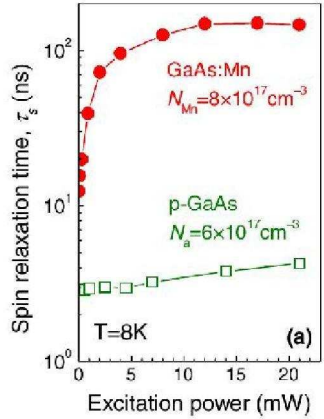}
  \end{minipage}\hfill
  \begin{minipage}[h]{0.5\linewidth}
    \centering
    \includegraphics[height=4.5cm]{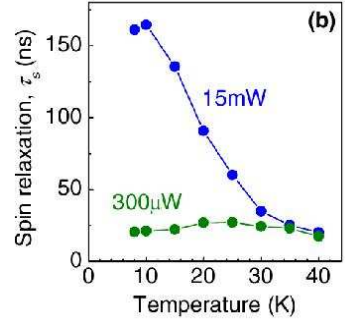}
  \end{minipage}\hfill
  \caption{(a) Spin relaxation time $\tau_s$ {\sl vs.} excitation power for
    GaAs:Mn ($N_{\rm Mn}=8\times10^{17}$~cm$^{-3}$) (dots)
    and for p-GaAs containing nonmagnetic acceptors
    ($N_{a}=6\times10^{17}$~cm$^{-3}$) (open squares).
    (b) Spin lifetimes {\sl vs.} temperature for different
    excitation powers in GaAs:Mn. Curves are to guide the eyes.
    From Astakhov et al. \cite{astakhov:076602}.}
  \label{Astakhov_GaMnAs}
\end{figure}

At low temperature, Mn ion can bind a hole to form a neutral
center. Hence optically excited electrons interact with the neutral
centers rather than the Mn ions and holes separately. The ground state
of the neutral center is a two-fold degenerate state with angular
momentum $m_J=\pm 1$ as the $p$-$d$ exchange interaction is
anti-ferromagnetic.\footnote{Such a spin structure can be utilized to
  establish optical initialization and readout of Mn spin
  \cite{rcmyers:203} as circularly polarized light (with spin
  $\sigma_J=\pm 1$) selectively excites the $m_J=\pm 1$ state due to the
  optical selection rules.} The exchange interaction between
electron and the neutral center consists of two origins, the $s$-$d$
exchange interaction and the electron-hole exchange
interaction. Interestingly, both interactions are
ferromagnetic. Due to opposite Mn and hole spin orientations, the
whole exchange interaction is weakened.\footnote{This leads to a
  deviation of the $s$-$d$ exchange constant deduced from the measured
  electron spin precession frequency from the genuine $s$-$d$ exchange
  constant \cite{sliwa:165205}.} The cancellation of the two exchange
interaction leads to suppression of spin relaxation. As GaMnAs is
partially compensated semiconductors, i.e., interstitial Mn's act as
donors whereas substitutional Mn's act as acceptors, hole density is
smaller than the Mn acceptor density. Hence there are still some
charged Mn acceptors. Recently, Astakhov et al. found that the exchange
interaction can be further reduced by generating more holes via optical
pumping. They found that below some threshold excitation power, spin
relaxation is suppressed by increasing the optical excitation power
(see Fig.~\ref{Astakhov_GaMnAs}) \cite{astakhov:076602}. A picture of
localized electrons was presented in their paper to explain the
results at low temperature. In such picture, the spin relaxation is
due to random spin precession when electrons pass by the neutral
centers. The spin relaxation rate is
$1/\tau_s=\frac{2}{3}\langle\omega^2\rangle\tau_c$ where $\tau_c$ is
the correlation time. In the presence of a longitudinal magnetic field,
spin relaxation is suppressed by a factor of $1/[1+(\omega_L\tau_c)^2]$.
By measuring the longitudinal magnetic field dependence, Akimov et
al. found that $\tau_c$ in GaMnAs is much longer than that in GaAs:Ge
with similar acceptor concentration. They explained that the strong
exchange coupling between Mn acceptor and hole protects hole spin from
dissipation and the correlation time of random electron spin
precession is only limited by electron hopping, leading to a large
$\tau_c$ \cite{akimov:081203}.

Hole spin relaxation in diluted magnetic quantum wells and
heterostructures was studied in
Refs.~\cite{PhysRevLett.77.2814,PhysRevB.56.7574,PhysRevB.70.115307,PhysRevB.64.085331}.
For heavy-holes, the $p$-$d$ exchange interaction is 
not able to flip hole spin unless facilitated by heavy-light hole
mixing \cite{PhysRevB.43.9687}. It was reported that spins of
exciton-bound holes relax faster than free holes due to enhanced
heavy-light hole mixing in excitons \cite{PhysRevB.70.115307}. Also
the mixing is increased by the mean field of the $p$-$d$
interaction. Camilleri observed a relation for hole spin relaxation
rate $1/\tau_{s}^{h}\simeq a(x_{\rm eff}\langle M\rangle)^2+b$ where
$x_{\rm eff}$ is the effective (isolated) Mn mole fraction, $\langle M\rangle$
denotes the magnetization and $b$ stands for the heavy-light hole
mixing independent of $\langle M\rangle$ and other spin relaxation sources
\cite{PhysRevB.64.085331}. Hole spin relaxation always decreases with
increasing magnetic field
\cite{PhysRevLett.77.2814,PhysRevB.56.7574,PhysRevB.70.115307,PhysRevB.64.085331}.
Theoretical investigation of hole spin relaxation is presented in
Refs.~\cite{PhysRevB.43.9687,tsitsishvili:195402}.

\subsection{Spin relaxation in Silicon and Germanium in metallic regime}

Spin relaxation in silicon and germanium is an old topic. Spin
relaxation of donor bound electrons has been studied intensively since the
middle of last century
\cite{PhysRev.107.491,PhysRev.109.319,PhysRev.114.1245,PhysRev.118.1523,PhysRev.118.1534,PhysRevB.2.2429,PhysRev.134.A265}
and revisited in recent decades in the context of quantum computation
\cite{PhysRevB.66.035314,PhysRevB.68.193207,PhysRevB.72.235201,PhysRevB.68.115322,PhysRevB.72.161306,witzel:035322}.
In this subsection, we focus on spin relaxation in silicon and
germanium in the metallic regime.

\begin{figure}[bth]
  \centering
  \includegraphics[height=6cm]{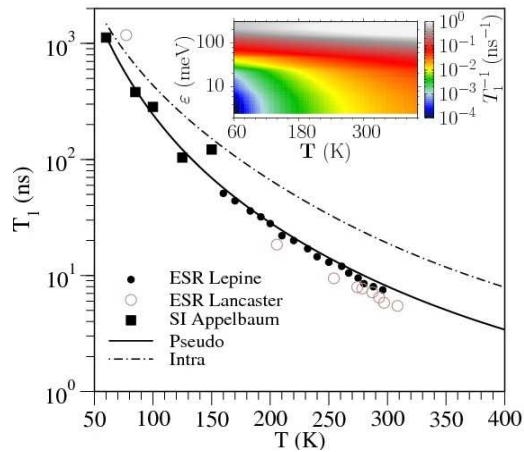}
  \caption{ Spin relaxation time $T_1$ in bulk silicon as a function of
    temperature $T$. The solid curve is the calculation (Pseudo), the symbols are 
    the spin injection (SI Appelbaum) \cite{Appelbaum:447.295} and the
    electron spin resonance (ESR Lepine \cite{PhysRevB.2.2429} and ESR
    Lancaster \cite{0370-1328-84-1-304}) data (see
    Ref.~\cite{Fabianbook}). The dashed-dotted curve is the spin
    lifetime calculated with only the intravalley scattering (Intra).
    The inset shows the contour plot of the spin relaxation rate
    $T_1^{-1}$ of  hot electrons, as a function of the electron energy
    $\varepsilon$ and lattice temperature $T$.
    From Cheng et al. \cite{2009arXiv0906.4054C}.}
  \label{Cheng_si}
\end{figure}

Let us first start with bulk silicon.\footnote{There is few study
  in bulk germanium in metallic regime.} As silicon possesses space
inversion symmetry, the spin-orbit coupling term of the conduction band is
zero. Hence the D'yakonov-Perel' mechanism is absent. In $n$-doped
silicon, the only relevant spin relaxation mechanism is the
Elliott-Yafet mechanism. Experimentally, electron spin relaxation in
bulk silicon was mostly studied via electron spin resonance as the
optical orientation and detection method is forbidden due to its
indirect band gap nature. Most of the investigations focused on the low
temperature regime
\cite{PhysRevB.36.6198,PhysRevB.5.1716,PhysRevB.3.4285,PhysRevB.3.4232},
whereas only a few covered the high temperature regime
\cite{PhysRevB.2.2429,0370-1328-84-1-304}. Among these works, a
systematic investigation was performed by L\'epine
\cite{PhysRevB.2.2429}. Analysis indicated that the spin relaxation
scenario varies with temperature.\footnote{The analysis has been summerized in
  Ref.~\cite{zuticrmp}.} At high temperature ($T>150$~K) most of
electrons are in extended states and spin relaxation is dominated by
the band electrons other than donors. Theoretically, conduction band electron
spin relaxation in silicon was first studied by Elliott \cite{PhysRev.96.266}
and Yafet \cite{yafetbook} separately. The Elliott-Yafet spin relaxation consists
of both the Elliott process where spin-flip is due to the spin-mixing
of conduction band and the Yafet process where spin-flip is caused by
direct spin-phonon coupling. By taking
into account of the Yafet process due to intravalley electron-acoustic
phonon scattering, Yafet gave a qualitative relation of
$\tau_s\sim T^{-\frac{5}{2}}$ \cite{yafetbook}. Recently, Cheng et al. found that the
Elliott process and the Yafet process interfere destructively in
silicon. The total spin relaxation rate is much slower than that
predicted by the individual Yafet or Elliott process
\cite{2009arXiv0906.4054C}. They gave new qualitative relation of
$\tau_s\sim T^{-3}$, which agrees quite well with experiments
\cite{2009arXiv0906.4054C} [see Fig.~\ref{Cheng_si}].

Two-dimensional structure can introduce structure and/or interface
inversion asymmetry which induces the spin-orbit coupling in conduction
band. Spin-orbit coupling in silicon or germanium quantum wells was
studied in
Refs.~\cite{PhysRevB.71.075315,PhysRevB.69.115333,nestoklon:235334,nestoklon:155328,2009arXiv0908.2417P}.
Correspondingly, the D'yakonov-Perel' spin relaxation in
two-dimensional structure was investigated in
Ref.~\cite{PhysRevB.71.075315}. Experimentally, Tyryshkin et
al. measured spin echoes and deduced a spin coherence time up to
3~$\mu$s for a high mobility two-dimensional electron system
formed in Si/SiGe quantum wells \cite{PhysRevLett.94.126802}. The
angular dependence of spin relaxation indicated that the
D'yakonov-Perel' mechanism due to the Rashba spin-orbit coupling is
the relevant one in high mobility samples
\cite{PhysRevLett.94.126802,PhysRevB.66.195315,Wilamowski2002439}.
Like that in bulk silicon, spin relaxation in two-dimensional silicon
structures was also studied mostly via electron spin resonance
\cite{PhysRevB.56.R4359,Sandersfeld2000312,Jantsch2002504,Wilamowski2002439,Wilamowski2003111,PhysRevB.66.195315,PhysRevB.69.035328,pssb.210.643,malissa:1739,Truittbook,2009arXiv0912.3037S}.
Recently, electrically detected spin resonance has also been
developed and applied to silicon two-dimensional structures
\cite{PhysRevB.59.13242,PhysRevLett.91.246603,matsunami:066602}.

The dependence of spin lifetime on the orientation of
external magnetic field was studied in
Refs.~\cite{Jantsch2002504,Wilamowski2002439,PhysRevB.69.035328},
revealing the relevance of the D'yakonov-Perel' spin relaxation
mechanism in high mobility samples. Examination of the relation
between the spin lifetime and the momentum scattering rate
indicated that in low mobility samples the spin relaxation is dominated by
the Elliott-Yafet mechanism \cite{Wilamowski2003111}.\footnote{The
  relation between spin lifetime and momentum scattering time
  was also studied in Refs.~\cite{PhysRevB.69.035328,PhysRevB.66.195315}.} Temperature
and density dependences of spin relaxation rate were studied in
Refs.~\cite{PhysRevB.66.195315,PhysRevB.69.035328,PhysRevB.56.R4359}
and \cite{PhysRevB.66.195315} respectively. Spin relaxation can be
suppressed by cyclotron motion as demonstrated by Wilamowski et
al. \cite{PhysRevB.69.035328}. Spin relaxation in quantum Hall regime
was studied by Matsunami et al. \cite{matsunami:066602}.

Theoretically, Sherman pointed out that even in nominally symmetric
silicon quantum wells, the doping inhomogeneity can cause a random
spin-orbit coupling \cite{sherman:209}. Such effect may play an important
role in spin relaxation in symmetric silicon quantum wells
\cite{PhysRevB.67.161303,sherman:209}. Similarly, the surface
roughness can also induces a random spin-orbit coupling due to
interface inversion asymmetry \cite{PhysRevB.69.115333}. Hole
spin-orbit coupling in silicon quantum wells was discussed in
Refs.~\cite{PhysRevB.71.035321,zhang:155311}. The effect of
carrier-carrier scattering on the D'yakonov-Perel' spin
relaxation in $n$-type silicon quantum wells
\cite{0295-5075-87-5-57005} and $p$-type silicon (germanium) quantum
wells \cite{zhang:155311} has also been reported in the recent literature.

\subsection{Magnetization dynamics in magnetic semiconductors}

Magnetic semiconductors, especially the Mn doped III-V magnetic
semiconductors, have attracted much attention for decades. The
invention of ferromagnetic III-V semiconductors opens the perspective
of integrating the magnetic recording and electronic circuit on one
chip and has inspired a lot
of studies \cite{H.Ohno08141998,1005423,nmat1325,nmat883,jungwirth:809}.
As the ferromagnetic order in III(Mn)-V magnetic semiconductors originates from the carrier mediated Mn spin-spin interactions via the
strong $s(p)$-$d$ exchange interactions \cite{jungwirth:809}, the
manipulation of carriers would lead to efficient control over the
magnetization. The efficient manipulation of magnetization via optical
\cite{PhysRevLett.95.167401,wang:217401,wang:235308,wang:021101,hall:032504,astakhov:152506,Kondo:pssc3.4263,PhysRevLett.88.137202,PhysRevLett.78.4617,qi:112506,rozkotova-2008-93,hashimoto:202506,rozkotova-2008-92,rozkotova-2008-44,kobayashi:07C519,PhysRevB.69.033203,qi:085304,hashimoto:067202,astakhov:187401,PhysRevB.68.193203,PhysRevLett.92.237203}
and electrical \cite{2008arXiv0812.3160C,408944a0,li:112513,Chiba:nature455.515}
means have been realized.

In this subsection we review recent studies on optically induced
magnetization dynamics in III(Mn)-V ferromagnetic
semiconductors. Historically, the optically induced magnetization
dynamics was first discovered in ferromagnetic metal (nickel)
\cite{PhysRevLett.76.4250}. The studies in metal have some relation
with those in ferromagnetic semiconductors, which hence should also be
mentioned. After more than a decade study, the picture of the
underlying physics has been discovered and some usefull models have
been raised. However, studies which can quantitatively relate the
microscopic carrier dynamics with the observed macroscopic
magnetization dynamics are still absent.

\begin{figure}[bth]
  \centering
  \includegraphics[height=5.cm]{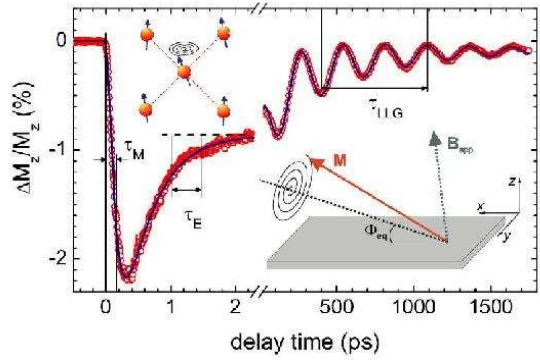}
  \caption{Main: experimental development of the induced perpendicular
   component of ${\bf M}$, ${\rm \Delta M_z/M_z}$, after laser heating
   a nickel thin film at $t=0$. Insets: precession of an individual
   spin in the exchange field (left) and of ${\bf M}$ in the effective
   field (right). $\tau_{\rm M}$, $\tau_{\rm E}$ and $\tau_{\rm LLG}$
   denote the demagnetization time, the energy relaxation time and the
   Landau-Lifschitz-Gilbert damping time, respectively. The three
   stages of the magnetization dynamics is clearly seen. From Koopmans et al. \cite{PhysRevLett.95.267207}.}
  \label{FM_dem}
\end{figure}

We first review optically induced magnetization dynamics in
ferromagnetic metals. Let us start from the observed phenomena. In
experiments, strong femtosecond laser pulses (usually $\sim
1$~mJ~cm$^{-2}$) are used to pump the ferromagnetic materials. The
magnetization is then monitored by probe pulses via magneto-optical
Kerr effect. The relative magnetization $M(t)/M(0)$ is assumed to be
proportional to the relative Kerr rotation $\theta(t)/\theta(0)$ where
$\theta(0)$ [$M(0)$] is the Kerr rotation (magnetization) before
pumping. A typical curve is shown in Fig.~\ref{FM_dem}. The magnetization
dynamics is characterized by three stages: (i) an ultrafast
demagnetization within 1~ps; (ii) a recovery of magnetization via
equilibration processes such as electron-phonon scattering from 1~ps
to $\sim 20$~ps; (iii) after $\sim 20$~ps a damped precession of the
magnetization due to the effective exchange field
\cite{PhysRevLett.85.844,PhysRevLett.95.267207}. The characteristic
time scale of these stages are the ultrafast demagnetization time
$\tau_M$, the energy lifetime $\tau_E$ and the
Landau-Lifschitz-Gilbert damping time $\tau_{\rm LLG}$. In Nickel,
$\tau_M\sim 100$~fs, $\tau_E\sim 0.5$~ps, and $\tau_{\rm LLG}\sim
700$~ps \cite{PhysRevLett.95.267207}. Many works have been attributed to
study the magnetization dynamics by various methods to reveal the
underlying mechanism and the dependences of the dynamics on the
external conditions
\cite{PhysRevLett.76.4250,PhysRevLett.78.4861,PhysRevLett.85.844,PhysRevLett.95.267207,cinchetti:177201,Stamm:NatMater.6.740,2009arXiv0904.4399A,carpene:174422}.
It was found that the magnetization in stage (ii) is governed by the
hot-electron temperature, which agrees well with the static $M(T)$
curve \cite{PhysRevLett.78.4861}. The damped precession of the
magnetization in stage (iii) was found to be due to the deviation of
the magnetization direction from the equilibrium one caused by the
laser pumping \cite{PhysRevLett.95.267207}. Among these processes, the
ultrafast demagnetization and the magnetization precession have
attracted a lot of attention to uncover the underlying microscopic
mechanisms.

The ultrafast demagnetization in ferromagnetic metals is the first
story which closely grasped the attention of the society. It was
first discovered by Beaurepaire et al. in nickel in 1996
\cite{PhysRevLett.76.4250}. The magnetization decreases by 50~\%
within $\lesssim 100$ femtoseconds, which is the fastest magnetization
dynamics ever found. The large ``instantaneous'' demagnetization has
been puzzling researchers, as previous measurement yielded a spin-lattice
lifetime of $\sim 100$~ps \cite{PhysRevB.46.5280}. Purely
electronic mechanisms such as electron-magnon scattering was also
suggested to explain the ultrafast demagnetization. However,
although the pure electronic processes can flip individual electron
spins rapidly \cite{PhysRevLett.79.5158}, they conserve the total spin
of the electron system and hence can not lead to demagnetization
\cite{cinchetti:177201,PhysRevLett.85.844}. Koopmans et al. first
suggested that the measured optical signal at the first picosecond does
not reflect the magnetization dynamics, but only reflects the optical
dynamics \cite{PhysRevLett.85.844}. They found that the relative
change in the Kerr rotation does not coincide with that in Kerr
ellipticity within the first 0.5~ps, which distincts with the
magnetization induced Kerr effect. Hence the ``instantaneous'' decrease
of the Kerr rotation or ellipticity can not be directly related to
the demagnetization \cite{PhysRevLett.85.844}. A close examination
indicated that the photo-excited electrons thermalize also within the
first picosecond \cite{PhysRevLett.85.844}. A more direct and
reliable measurement of the spin dynamics via X-ray magnetic circular
dichroism observed a demagnetization time of $\sim 120$~fs
\cite{Stamm:NatMater.6.740}. In the same 
experiment, the photo-electron thermalization time was identified as
$\sim 640$~fs. At this stage, the time scale of the magnetization
dynamics is faithfully characterized. However, the demagnetization
mechanism is still unclear. Although several mechanisms and models have
been proposed
\cite{PhysRevLett.76.4250,PhysRevLett.85.3025,PhysRevLett.95.267207,atxitia:057203,2009arXiv0906.5104K},
they have not yet been confirmed by experiments unambiguously.

The focus of this subsection is the photo-induced magnetization dynamics in
ferromagnetic III(Mn)-V semiconductors. One of the significant difference
is that the magnetization is provided by the Mn spin, hence the
magnetization and carrier degrees of freedom are separated. 
Furthermore, unlike the complex energy bands in metals, the energy
bands in ferromagnetic III(Mn)-V semiconductors are well described by
the Kane model with the $s(p)$-$d$ exchange interaction
\cite{jungwirth:809}, which greatly simplifies the theoretical study.
Theoretical studies can then help to identify the underlying
mechanisms of the photo-induced magnetization dynamics
\cite{cywinski:045205,chovan:085321,PhysRevB.69.085209,tserkovnyak:5234}.

\begin{figure}[bth]
  \centering
  \includegraphics[height=6.5cm]{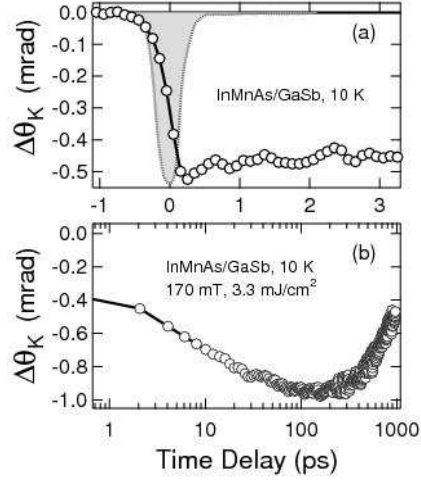}
  \caption{(a) The first 3~ps of demagnetization dynamics in
    InMnAs. The shadowed region denotes the cross correlation of the
    pump and probe pulses. (b) Demagnetization dynamics covering the
    entire time range of the experiment. $\Delta\theta_{\rm K}$ is the
    derived difference Kerr rotation which is supposed to be
    proportional to the induced magnetization change along the growth
    direction. From Wang et al. \cite{PhysRevLett.95.167401}.}
  \label{JWang_InMnAs}
\end{figure}

Experimentally, laser-induced demagnetization in (III,Mn)V
ferromagnetic semiconductors was first studied by Kojima et
al. \cite{PhysRevB.68.193203} in GaMnAs. The magnetization
dynamics was monitored via the magneto-optical Kerr rotation. They
observed a slow demagnetization time of $\sim 500$~ps, which was
attributed to the possible thermal isolation between the charge and
spin system in GaMnAs \cite{PhysRevB.68.193203}. The sub-picosecond
ultrafast demagnetization in ferromagnetic semiconductors was first
demonstrated by Wang et al. \cite{PhysRevLett.95.167401} in InMnAs with much
higher intensity $\gtrsim 1$~mJ/cm$^{-2}$ per pump pulse than the one in
the experiment of Kojima et al. (45~$\mu$J/cm$^{-2}$). A typical
trace of the time-resolved Kerr rotation is illustrated in
Fig.~\ref{JWang_InMnAs}. The phenomena are similar to that in
ferromagnetic metals: a sub-picosecond ($\sim 200$~fs) 
demagnetization and a long time recovery. However, it is noted that
the magnetization further decreases after the first 1~ps, in contrast
with the rapid recovery in ferromagnetic metals
\cite{PhysRevLett.95.267207}. In InMnAs, this slow demagnetization
process can last for several hundreds of picosecond at low pump
fluence. With increased pump fluence the duration of the slow
demagnetization process decreases whereas the decay rate
increases. Meanwhile the degree of demagnetization due to the fast
process increases. At high pump fluence $\gtrsim 10$~mJ/cm$^2$, the
magnetization is completely quenched after the fast process and the
slow demagnetization diminishes. The slow demagnetization process is
assigned to the Mn spin-lattice relaxation in the existing literature
\cite{PhysRevLett.95.167401}. The arguments are as follows: In
III(Mn)-V ferromagnetic semiconductors the large hole spin
polarization blocks the carrier-Mn spin-flip scattering, hence the Mn
spin system is thermally isolated \cite{PhysRevB.68.193203}. On the
other hand, the carrier system is efficiently coupled to phonon bath
via the carrier-phonon interaction at a time scale of $\lesssim 1$~ps.
The thermalized phonon bath then leads to a Mn spin-lattice relaxation
at the time scale of 100~ps, which results in the slow demagnetization
process \cite{PhysRevB.68.193203}.

\begin{figure}[bth]
  \centering
  \includegraphics[height=5.2cm]{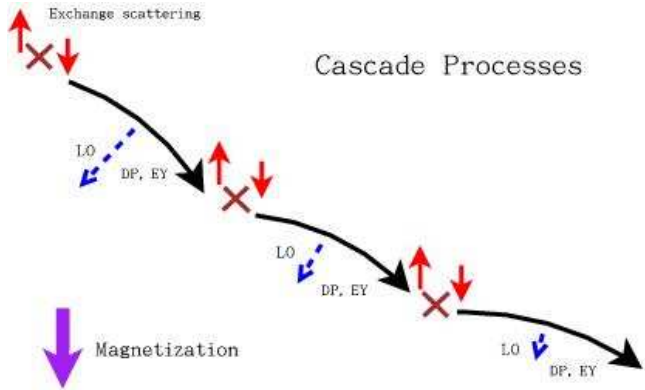}
  \caption{Illustration of the microscopic process of
    ultrafast demagnetization. ${\mbox{\boldmath
          $\times$\unboldmath}}$ denotes the $p$-$d$ exchange
    scattering. Thin arrow denotes the spin polarization direction of
    photo-excited holes. Bold arrow represents the initial
    magnetization direction. Each $p$-$d$ exchange scattering reduces
    the magnetization and transfers the spin polarization into the
    hole system. Holes lose their spin polarization quickly through the
    D'yakonov-Perel' (DP) and Elliott-Yafet (EY) mechanisms. During these
    processes, the efficient hole-longitudinal-optical-phonon
    scattering [denoted as blue arrow (LO) in the figure] relaxes
    the excess energy of the photo-excited holes. After the energy
    relaxation, hole spin relaxation slows down (the
    hole-longitudinal-optical-phonon scattering also slows down) and
    becomes less efficient than the $p$-$d$ exchange scattering. Hole
    spin polarization then saturates and the cascade processes
    terminate.} 
  \label{FScasacade}
\end{figure}

We now focus on the fast demagnetization process. The underlying
physics of the ultrafast demagnetization is illustrated in
Fig.~\ref{FScasacade}. The key process is that via the strong $p$-$d$
coupling the Mn spin polarization is efficiently transferred to the
hole spin polarization, which is similar to the inverse Overhauser
effect \cite{cywinski:045205}. However, such processes is suppressed
if the hole spin relaxation rate is smaller than the rate at which the
spin polarization is injected from the Mn spin system via the $p$-$d$
exchange scattering. Efficient hole spin relaxation is required for a
ultrafast demagnetization. It is known that hole spin relaxation is
very fast in bulk $p$-GaAs, at a time scale of $100$~fs, due to the strong
spin-orbit coupling \cite{PhysRevLett.89.146601,krauss:256601}.
However, in III(Mn)-V ferromagnetic semiconductors, the large spin
splitting due to the 
mean $p$-$d$ exchange field removes the $\Gamma$
point degeneracy and suppresses the effect of the spin-orbit coupling
at low kinetic energy
\cite{opt-or,zuticrmp}. Fortunately for photo-excited holes, of which
the kinetic energy is high \cite{PhysRevLett.95.167401}, the
spin-orbit splitting is larger than the exchange splitting. The hole
spin relaxation can again be very efficient. The fast spin
relaxation of photo-excited holes facilitates the ``inverse Overhauser
effect'' and leads to the ultrafast demagnetization. Due to the
hole-phonon and hole-hole scattering, the photo-excited holes soon
lose their energy and thermalize within 1~ps. After the energy
relaxation process, the ultrafast demagnetization is terminated. Such
scenario has been laid out in recent studies
\cite{cywinski:045205,wang:235308}.\footnote{Note that a similar
    picture of Mn ion spin relaxation in ZnMnSe/ZnBeSe paramagnetic
    quantum well was also depicted in Ref.~\cite{akimov:165328}.}

In Ref.~\cite{cywinski:045205}, Cywinski et al. developed a mean field
theory to describe the above scenario of ultrafast demagnetization. In
their theory, the magnetization was described by the mean Mn spin
polarization which was then characterized by a single Mn spin density
matrix. The Mn-Mn correlations beyond the mean field level were
argued to be averaged out in such a strong carrier excitation
regime. The instantaneous Mn and hole spin splitting was provided by
the mean field of the $p$-$d$ exchange interaction. The hole dynamics
was then described by the spin-resolved distribution function 
$f_{n{\bf k}}$ with $n$ denoting the (spin) band index. The
magnetization dynamics was obtained, in principle, by solving the
coupled kinetic equations of holes and Mn spin. The ``inverse
Overhauser effect'', i.e., the transfer of spin polarization from Mn
spin system to hole spin system due to the nonequilibrium hole
distribution triggered by the photo-excitation, was then included in
the kinetic equation. Within such a model, the Mn spin-flip rate due
to the ``inverse Overhauser effect'' (from $m$ state to $m\pm 1$ state
with $-5/2\le m\le 5/2$ being the Mn spin index) can be written as
\cite{cywinski:045205}
\be
W_{m,m\pm 1} = \frac{\beta^2}{4} 2\pi |\langle
m|S^{\mp}|m+1\rangle|^2\sum_{nn^{\prime}}\int \frac{d{\bf
    k}}{(2\pi)^3}\frac{d{\bf k}^{\prime}}{(2\pi)^3} |\langle
n^{\prime}{\bf k}^{\prime}|s^{\pm}|n{\bf k}\rangle|^2 f_{n{\bf
    k}}(1-f_{n^{\prime}{\bf
    k}^{\prime}})\delta(\tilde{\varepsilon}_{n{\bf
    k}}-\tilde{\varepsilon}_{n^{\prime}{\bf
    k}^{\prime}}\pm\Delta_{M}), 
\ee
where $\beta$ is the $p$-$d$ exchange constant and $S^{\pm}$ ($s^{\pm}$)
are the Mn (hole) spin ladder operators. $\Delta_M$ represents the Mn
instantaneous mean field spin splitting produced by the holes. 
$\tilde{\varepsilon}_{n{\bf k}}$ is the instantaneous band energy with
the mean field exchange interaction included. By considering the hole
energy- and spin-relaxation in a phenomenological way, the magnetization
dynamics was solved within a simplified two band hole approximation
\cite{cywinski:045205}. The results reproduced the main features of
the ultrafast demagnetization: (i) the process is terminated at the
time scale of hole energy equilibration time $\tau_E$
\cite{cywinski:045205}; (ii) the demagnetization rate and the degree of the
magnetization change increase with the pump fluence and hole spin
relaxation rate \cite{cywinski:045205}. The obtained demagnetization
time is on the order of 1~ps \cite{cywinski:045205}. The feature
  of the magnetization dynamics also qualitatively
agrees with the observation in 
experiments \cite{cywinski:045205,wang:235308}. Further
calculations based on the realistic band structure and microscopic
interactions, such as what has been done in
paramagnetic phase in Ref.~\cite{jiang:155201}, is needed to
elucidate the underlying physics unambiguously and to compare
with the experimental results.

At low pump fluence, such as $\sim 10$~$\mu$J~cm$^{-2}$ (compared to
the fluence $\gtrsim 1$~mJ~cm$^{-2}$ used in demagnetization
measurements), rich magnetization dynamics is triggered by photo
illumination. Damped precession of the magnetization was observed,
where the driving force was attributed to the photo-induced change in
the magnetic anisotropy energy
\cite{oiwa.js.9.18,wang:233308,qi:112506,rozkotova-2008-92,rozkotova-2008-44,hashimoto:067202,qi:085304,zhu:142109,kobayashi:07C519}.
It was also discovered that the magnetization precession can be coherently
manipulated by photo pulses
\cite{hashimoto:202506,rozkotova-2008-93}. Besides, with circularly
polarized light, the magnetization is rotated due to the $p$-$d$
exchange mean field produced by the photo-injected hole spin
polarization \cite{PhysRevB.69.033203,qi:112506,zhu:142109}.
The photo-induced change in the magnetic anisotropy energy has been
argued to originate from: (i) the change of the temperature of the
hole gas due to hot photo-excited holes \cite{rozkotova-2008-92}; (ii) the
change of the hole density \cite{oiwa.js.9.18}; (iii) the nonthermal
effect \cite{hashimoto:067202}. The dependence of the precession
frequency (closely related to the magnetic anisotropy energy)
and damping constant (the Gilbert constant) were hence studied for
various photo intensity, temperature and hole density, and compared
with theoretical results \cite{PhysRevB.69.085209,tserkovnyak:5234,kapetanakis:097201}.
It should be mentioned that there is other mechanism which can also trigger such
magnetization precession, such as the photo-induced spin precession
due to the interband polarization proposed by Chovan and Perakis
\cite{chovan:085321}. Very recently Kapetanakis et
al. \cite{PhysRevLett.103.047404} developed a nonequilibrium theory
based on the original idea of Ref.~\cite{chovan:085321} and well
explained the experimental results of magnetization precession
\cite{wang:021101}. In fact, within the first picosecond, the hot 
photo-excited holes can also modify the 
magnetization direction besides changing the magnetization amplitude 
(demagnetization) due to the fact that in the presence of hole spin-orbit 
coupling the total (hole+Mn)
spin polarization is not conserved. Once the magnetization direction
deviates from the equilibrium one, a damped precession is
motivated by the effective magnetic field due to the external and
internal (anisotropy) magnetic fields \cite{PhysRevLett.85.844}.

\section{Spin relaxation and  dephasing based on kinetic spin Bloch 
equation approach}

In this section, we introduce a fully microscopic many-body approach
named kinetic spin Bloch equation approach developed by Wu
et al. \cite{PhysRevB.61.2945,PhysRevB.68.075312,JPSJ.70.2195,wu:pssb.222.523}. Unlike the approaches
reviewed in the previous sections which treat scatterings using the
relaxation time approximation, the kinetic spin Bloch equation approach treats all scatterings
explicitly and self-consistently to all orders. Especially, the
carrier-carrier Coulomb scattering is explicitly included in the
theory. This allows them to study spin dynamics not only near but also
far away from the equilibrium, for example, the spin dynamics in the
presence of high electric field (hot-electron
condition) \cite{PhysRevB.69.245320} and with large initial spin
polarizations \cite{PhysRevB.68.075312,stich:176401,stich:205301}.
This section only focuses on the results in spacial uniform systems
while those in the spacial non-uniform  systems are given in
Section~7. We organize this section as follows: In Section~5.1 we 
first introduce the kinetic spin Bloch equation approach. Then we address 
the effect of electron-electron Coulomb scattering on the spin dynamics 
in Section~5.2. The general kinetic spin Bloch equations in
$n$- or $p$-doped confined structures are given in Section~5.3 
where we even include the case of spacial gradient.\footnote{For the case
of spin diffusion and transport, see Section~7.} The results of
 spin relaxation and
dephasing in quantum wells, quantum wires, and bulk samples
 are reviewed in Sections~5.4, 5.5 and 5.6, respectively.

\subsection{Kinetic spin Bloch equation approach}

\subsubsection{Model and Hamiltonian}
We first use a four-spin-band model to demonstrate the establishment
of the kinetic spin Bloch equations and apply these equations to investigate the spin
precession and relaxation/dephasing. By considering a (001) zinc
blende quantum well with its growth axis in the $z$-direction and a
moderate magnetic field along the $x$-direction (Voigt configuration),
we write the Hamiltonian of electrons in the conduction and valence
bands as
\begin{equation}
H=\sum_{\mu{\bf k}\sigma}\varepsilon_{\mu{\bf k}}c_{\mu{\bf
    k}\sigma}^\dagger c_{\mu{\bf k}\sigma}+\sum_{\mu{\bf
    k}\sigma\sigma^\prime}g\mu_B[{\bf B}+{\bf \Omega}_\mu({\bf
  k})]\cdot{\bf S}_{\mu\sigma\sigma^\prime}c_{\mu{\bf
    k}\sigma}^\dagger c_{\mu{\bf k}\sigma^\prime}+H_{E}+H_{I},
\label{eq5.1.1-1}
\end{equation}
with $\mu=c$ and $v$ standing for the conduction and valence bands,
respectively. Here we only include the lowest subbands of the
conduction and valence bands. This is valid for quantum wells
with sufficiently small well widths. $\varepsilon_{v{\bf
    k}}=-E_g/2-k^2/2m_h=-E_g/2-\varepsilon_{hk}$ and $\varepsilon_{c{\bf
    k}}=E_g/2+k^2/2m_e=E_g/2+\varepsilon_{ek}$ with $m_h$ and $m_e$
denoting effective masses of the hole and electron, respectively. $E_g$
is the band gap. For quantum well without strain, the heavy hole band
is above the light hole one and hence the valence band to be
considered is the heavy-hole band with ${\bf S}_{v\sigma\sigma^\prime}$
representing spin 3/2 matrices and $\sigma=\pm3/2$. ${\bf
  S}_{c\sigma\sigma^\prime}$ stands for the spin 1/2 matrices for
conduction electrons. ${\bf \Omega}_\mu({\bf k})$ represents the effective
magnetic field from the D'yakonov-Perel' term of the $\mu$-band.

$H_{E}$ in Eq.~(\ref{eq5.1.1-1}) denotes the dipole coupling with the
light field $E_{-\sigma}(t)$ with $\sigma=\pm$ representing the
circularly polarized light. Due to the selection rule
\begin{equation}
  H_{E}=-d\sum_{\bf k}[E_-(t)c_{c{\bf k}\frac{1}{2}}^\dagger c_{v{\bf
      k}\frac{3}{2}}+{\rm H.c.}]-d\sum_{\bf k}[E_+(t)c_{c{\bf
      k}-\frac{1}{2}}^\dagger c_{v{\bf k}-\frac{3}{2}}+{\rm H.c.}],
\label{eq5.1.1-2}
\end{equation}
in which $d$ denotes the optical-dipole matrix element and
$E_\sigma(t)=E_\sigma^{(0)}(t)\cos\omega t$ with $\omega$ being the
central frequency of the light pulse. $E_\sigma^{(0)}(t)$ denotes a
Gaussian pulse with pulse width $\delta t$. $H_{I}$ stands for the
interaction Hamiltonian. It contains both the spin-conserving
scattering, such as the electron-electron Coulomb, electron-phonon and
electron-impurity scatterings, and the spin-flip scattering, such as
the scattering due to the Bir-Aronov-Pikus and Elliott-Yafet mechanisms.

\begin{figure}[h]
\begin{center}
\includegraphics[width=7cm]{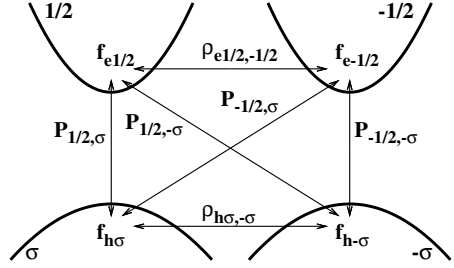}
\caption{Illustration of the four-spin-band model: two conduction
  bands with spin $\pm 1/2$ and two valence bands with spin
  $\pm\sigma$. The spin coherence $\rho$, optical and forbidden
  optical coherences $P$ as well as carrier distributions of the four
  bands are also illustrated.}
\label{fig5.1.1-1}
\end{center}
\end{figure}

It can be seen from $H_{E}$ that the laser pulse introduces the optical
coherence between the conduction and valence bands $P_{{\bf
    k}\frac{1}{2}\frac{3}{2}}\equiv e^{i\omega t}\langle c_{v{\bf
    k}\frac{3}{2}}^\dagger c_{c{\bf k}\frac{1}{2}}\rangle$ and $P_{{\bf
    k}-\frac{1}{2}-\frac{3}{2}}\equiv e^{i\omega t}\langle c_{v{\bf
    k}-\frac{3}{2}}^\dagger c_{c{\bf k}-\frac{1}{2}}\rangle$. In the
meantime, due to the presence of the magnetic field in the Voigt
configuration, these optical coherences may further transfer coherence
to $P_{{\bf k}-\frac{1}{2}\frac{3}{2}}\equiv e^{i\omega t}\langle c_{v{\bf
    k}\frac{3}{2}}^\dagger c_{c{\bf k}-\frac{1}{2}}\rangle$ and $P_{{\bf
    k}\frac{1}{2}-\frac{3}{2}}\equiv e^{i\omega t}\langle c_{v{\bf
    k}-\frac{3}{2}}^\dagger c_{c{\bf k}\frac{1}{2}}\rangle$, of which the
direct optical transition is forbidden. Due to the presence of the
magnetic field and/or effective magnetic field, if there is any spin
polarization, i.e., $f_{\mu{\bf k}\sigma}\equiv\langle c_{\mu{\bf
    k}\sigma}^\dagger c_{\mu{\bf k}\sigma}\rangle\neq f_{\mu{\bf
    k}-\sigma}$, electrons can flip from $\sigma$-band to
$-\sigma$-band, and hence induce the correlation between two spin
bands, i.e., $\rho_{\mu{\bf k}\sigma-\sigma}\equiv\langle c_{\mu{\bf
    k}-\sigma}^\dagger c_{\mu{\bf k}\sigma}\rangle\equiv \rho_{\mu{\bf
    k}-\sigma\sigma}^\ast$. Similar to the optical coherence which is
represented by $P_{{\bf k}\sigma\sigma^\prime}$, the spin coherence
can be well represented by $\rho_{\mu{\bf k}\sigma-\sigma}$. In the
following we call it spin coherence. These coherences as well as
electron/hole distributions illustrated in Fig.~\ref{fig5.1.1-1} are
the quantities to be determined self-consistently through the kinetic spin Bloch equations.  

\subsubsection{Kinetic spin Bloch equations}
We construct the kinetic spin Bloch equations using the nonequilibrium Green function
method \cite{haugjauho} as follows \cite{PhysRevB.61.2945}
\begin{equation}
\dot{\rho}_{\mu\nu,{\bf
    k},\sigma\sigma^\prime}=\dot{\rho}_{\mu\nu,{\bf
      k},\sigma\sigma^\prime}|_{\rm coh}+\dot{\rho}_{\mu\nu,{\bf
      k},\sigma\sigma^\prime}|_{\rm scat}.
\label{eq5.1.2-3}
\end{equation}
Here $\rho_{\mu\nu,{\bf k},\sigma\sigma^\prime}$ represents the
single-particle density matrix with $\mu$ and $\nu$ $=$ $c$ or
$v$. ($\mu$ and $\nu$  can be subband indices and/or valley indices also.) 
The diagonal elements describe the carrier distribution functions
$\rho_{\mu\mu,{\bf k},\sigma\sigma}=f_{\mu{\bf k}\sigma}$ and the
off-diagonal elements either represent the optical coherence
$\rho_{cv,{\bf k},\sigma\sigma^\prime}=P_{{\bf
    k}\sigma\sigma^\prime}e^{-i\omega t}$ for the inter conduction
band and valence band polarization, or the spin coherence
$\rho_{\mu\mu,{\bf k},\sigma-\sigma}$ for the inter spin-band
correlation. The coherent terms $\dot{\rho}_{\mu\nu,{\bf
      k},\sigma\sigma^\prime}|_{\rm coh}$ are composed of the contributions 
from the energy spectrum, electric pumping field, magnetic and/or
effective magnetic field as well as the Coulomb Hartree-Fock term.
The scattering terms $\dot{\rho}_{\mu\nu,{\bf
      k},\sigma\sigma^\prime}|_{\rm scat}$ consist of the spin-conserving scatterings
such as the electron-electron Coulomb, electron-phonon and 
electron--non-magnetic-impurity scatterings and/or the spin-flip scatterings
such as the scatterings due to the Bir-Aronov-Pikus and/or Elliott-Yafet mechanisms and the 
hyperfine interaction. The detailed expressions of these terms will be  
given in the following subsections for different specific problems.

 By solving the kinetic spin Bloch equations (\ref{eq5.1.2-3}) with the initial
conditions:
\begin{equation}
  \rho_{\mu\nu,{\bf k},\sigma\sigma^\prime}(0)=\left\{
      \begin{array}{ll}
        f_{\mu{\bf k}\sigma}(0) &\mbox{with $\mu=\nu$ and $\sigma=\sigma^\prime$}\\
        0 &\mbox{with $\mu\neq\nu$ and/or $\sigma\neq\sigma^\prime$}
      \end{array}
    \right.,
    \label{eq5.1.2-4}
  \end{equation}
one obtains the time evolution of the density matrix.

\subsubsection{Faraday rotation angle, spin dephasing and spin
  relaxation}
The Faraday rotation angle can be calculated for two degenerate
Gaussian pulses with a delay time $\tau$. The first pulse (pump) is
circularly polarized, e.g., $E_{\rm pump}^0(t)=E_\pm^0(t)$, and travels in the
${\bf k}_1$-direction. The second pulse (probe) is linear polarized
and is much weaker than the first one, e.g.,
$E_{\rm probe}^0(t)\equiv\chi[E_-^0(t-\tau)+E_+^0(t-\tau)]$ with $\chi\ll
1$. The probe pulse travels in the ${\bf k}_2$ direction. The Faraday
rotation angle is defined as \cite{Linder1998412,PhysRevLett.74.4698}
\begin{equation}
\Theta_{\rm F}(\tau)=C\sum_{\bf k}\int \mbox{Re}[\bar{P}_{{\bf
    k}\frac{1}{2}\frac{3}{2}}(t){E_-^0}^\ast(t-\tau)-\bar{P}_{{\bf
    k}-\frac{1}{2}-\frac{3}{2}}(t){E_+^0}^\ast(t-\tau)]dt,
\label{eq5.1.2-5}
\end{equation}
with $\bar{P}_{{\bf k}\sigma\sigma^\prime}$ standing for the optical
transition in the probe direction. $C$ is a constant. The spin
relaxation/dephasing can be determined from the slope of the envelope
of the Faraday rotation angle $\Theta_{\rm F}(\tau)$. However,
calculation of this quantity is quite time consuming. In reality, we 
use the slope of the envelope of 
\be
\Delta N_\mu = \sum_{\bf k}(f_{\mu{\bf k}|\sigma|}-f_{\mu{\bf
      k}-|\sigma|}),\quad\quad
\rho_\mu = \sum_{\bf k}\left|\rho_{\mu\mu,{\bf
      k},|\sigma|-|\sigma|}\right|,\quad\quad 
\rho_\mu^\ast = \Big|\sum_{\bf k}\rho_{\mu\mu,{\bf
      k},|\sigma|-|\sigma|}\Big|,
\label{eq5.1.3}
\ee
to substract the spin relaxation time $T_1$, spin
dephasing time $T_2$ and ensemble spin dephasing time $T_2^\ast$,
respectively \cite{Lue501,PhysRevB.61.2945}. Similarly, the optical dephasing is
described by 
the incoherently summed polarization \cite{haugkoch,haugjauho},
\begin{equation}
P_{\sigma\sigma}(t)=\sum_{\bf k}\left|P_{{\bf k},\sigma\sigma}(t)\right|.
\end{equation}
The later was first introduced by Kuhn and Rossi \cite{PhysRevLett.69.977}. It is
understood that both true dissipation and the interference among the
${\bf k}$ states may contribute to the decay. The incoherent summation
is therefore used to isolate the irreversible decay from the decay
caused by interference.
   
\subsection{Electron-electron Coulomb scattering to spin dynamics}
\label{sec5.2}
It has long been believed that the spin-conserving electron-electron
Coulomb scattering does not contribute to the spin
relaxation/dephasing \cite{flatte}. It was first pointed out by Wu
and Ning \cite{wu:epjb.18.373} that in the presence of inhomogeneous broadening
in spin precession, i.e., the spin precession frequencies are ${\bf
  k}$-dependent, any scattering, including the spin-conserving
scattering, can cause irreversible spin dephasing. This inhomogeneous
broadening can come from the energy-dependent $g$-factor
\cite{wu:epjb.18.373,PhysRevB.47.6807,PhysRev.154.737,PhysRevB.4.1285,PhysRevLett.51.130,PhysRevB.32.6965,PhysRevB.44.9048}, the D'yakonov-Perel' term
\cite{JPSJ.70.2195}, the random spin-orbit coupling \cite{sherman:209}, and even the
momentum dependence of the spin diffusion rate along the spacial
gradient \cite{PhysRevB.66.235109}. This can be illustrated from the kinetic spin Bloch equations of
a four-spin band model in a quantum well with the lowest valence band
being the heavy-hole band and $\rho_{cv,{\bf
    k},\sigma\sigma^\prime}=P_{{\bf k}\sigma\sigma^\prime}e^{-i\omega
  t}$. When the D'yakonov-Perel' term ${\bf \Omega}({\bf k})=0$, the coherent parts of the kinetic spin Bloch equations
are given by \cite{wu:epjb.18.373}
\bea
\left.\frac{\partial f_{e{\bf k}\sigma}}{\partial
  t}\right|_{\rm coh} &=& 2\sum_{\bf q}V_{\bf
  q}\mbox{Im}\big(\sum_{\sigma^\prime}P_{{\bf
      k+q}\sigma\sigma^\prime}^\ast P_{{\bf
      k}\sigma\sigma^\prime}+\rho_{cc,{\bf
      k+q},-\sigma\sigma}\rho_{cc,{\bf k},\sigma-\sigma}\big)
  -g\mu_BB\mbox{Im}\rho_{cc,{\bf k},\sigma-\sigma},
\label{eq5.2-1}
\\\nonumber
\left.\frac{\partial \rho_{cc,{\bf k},\sigma-\sigma}}{\partial
  t}\right|_{\rm coh}&=&i\sum_{\bf q}V_{\bf q}\big[(f_{e{\bf k+q}\sigma}-f_{e{\bf
    k+q}-\sigma})\rho_{cc,{\bf k},\sigma-\sigma}-(f_{e{\bf k}\sigma}-f_{e{\bf
    k}-\sigma})\rho_{cc,{\bf k+q},\sigma-\sigma}\\  &&\quad +P_{{\bf
    k+q}\sigma\sigma_1}P_{{\bf k}-\sigma\sigma_1}^\ast-P_{{\bf
    k+q}-\sigma\sigma_1}^\ast P_{{\bf
    k}\sigma\sigma_1}\big]+\frac{i}{2}g\mu_BB(f_{e{\bf
  k}\sigma}-f_{e{\bf k}-\sigma}),
\label{eq5.2-2}
\\\nonumber
\left.\frac{\partial P_{{\bf k}\sigma\sigma^\prime}}{\partial
  t}\right|_{\rm coh}&=&-i\delta_{\sigma\sigma^\prime}({\bf k})P_{{\bf
    k}\sigma\sigma^\prime}-\frac{i}{2}g\mu_BBP_{{\bf
    k}-\sigma\sigma^\prime}-i\sum_{\bf q}V_{\bf q}\big[P_{{\bf
    k+q}\sigma\sigma^\prime}(1-f_{h{\bf k}\sigma^\prime}\\ &&\quad -f_{e{\bf
    k}\sigma})-P_{{\bf k+q}-\sigma\sigma^\prime}\rho_{cc,{\bf
    k},\sigma-\sigma}+P_{{\bf k}-\sigma\sigma^\prime}\rho_{cc,{\bf
    k+q},\sigma-\sigma}\big],
\label{eq5.2-3}
\end{eqnarray}
for electron distribution function $f_{e{\bf k}\sigma}\equiv 
f_{c{\bf k}\sigma}$, spin coherence and optical
coherence, respectively. $f_{h{\bf k}\sigma}\equiv 1-f_{v{\bf k}\sigma}$
is the hole distribution function. The first term on the right hand side of
Eq.~(\ref{eq5.2-1}) is the Fock term from the Coulomb scattering with
$V_{\bf q}=\frac{2\pi e^2}{q}$ denoting the Coulomb matrix element. The first term of
Eq.~(\ref{eq5.2-3}) gives the free evolution of the polarization
components with the detuning
\begin{equation}
\delta_{\sigma\sigma^\prime}({\bf k})=\varepsilon_{eh{\bf
    k}}-\Delta_0-\sum_{\bf q}V_{\bf q}(f_{e{\bf k+q}\sigma}+f_{h{\bf
    k+q}\sigma^\prime}),
\label{eq5.2-4}
\end{equation}
in which $\varepsilon_{eh{\bf k}}=\varepsilon_{e{\bf
    k}}+\varepsilon_{h{\bf k}}$ and $\Delta_0=\omega-E_g$. $\Delta_0$
is the detuning of the center frequency of the light pulses with
respect to the unrenormalized band gap. The last term in
Eq.~(\ref{eq5.2-3}) describes the excitonic correlations whereas the
first term in Eq.~(\ref{eq5.2-2}) describes the Hartree-Fock
contributions to the spin coherence. 

For spin-conserving scattering,
\begin{equation}
\sum_{\bf k}\left.\frac{\partial \rho_{\mu\nu},{\bf
    k},\sigma\sigma^\prime}{\partial t}\right|_{\rm scat}=0.
\label{eq5.2-5}
\end{equation}
By performing the summation over ${\bf k}$ from both sides of the
kinetic spin Bloch equations and further noticing that the Hartree-Fock contributions from the
Coulomb interaction in the coherent parts of the kinetic spin Bloch equations satisfy
\begin{equation}
\sum_{\bf k}\left.\frac{\partial \rho_{\mu\nu},{\bf
    k},\sigma\sigma^\prime}{\partial t}\right|_{\rm coh}^{\rm HF}=0,
\label{eq5.2-6}
\end{equation}
one has
\begin{eqnarray}
\frac{\partial^2}{\partial t^2}\mbox{Im}\sum_{\bf k}\rho_{cc,{\bf
    k},\sigma-\sigma}&=&-g^2\mu_B^2B^2\mbox{Im}\sum_{\bf k}\rho_{cc,{\bf
    k},\sigma-\sigma},\\
\label{eq5.2-7}
\frac{\partial}{\partial t}\mbox{Re}\sum_{\bf k}\rho_{cc,{\bf
    k},\sigma-\sigma}&=&0.
\label{eq5.2-8}
\end{eqnarray}
This shows that there is no spin dephasing in the absence of
inhomogeneous broadening and spin-flip scattering. For optical
dephasing, $\delta_{\sigma\sigma^\prime}(\bf k)$ in
Eq.~(\ref{eq5.2-3}) is ${\bf k}$-dependent, i.e., here exists the inhomogeneous
broadening. Therefore any optical-dipole conserving scattering can lead to
irreversible optical dephasing \cite{haugjauho}. Similarly, if there is
inhomogeneous broadening in the spin precession, any spin-conserving
scattering can lead to irreversible spin dephasing. As
electron-electron Coulomb scattering is a spin-conserving scattering,
of course it contributes to the spin dephasing in the presence of the
inhomogeneous broadening \cite{wu:epjb.18.373}.

Wu and Ning first showed that with the energy-dependent $g$-factor as
an inhomogeneous broadening, the Coulomb scattering can lead to
irreversible spin dephasing \cite{wu:epjb.18.373}. In [001]-grown $n$-doped
quantum wells, the importance of the Coulomb scattering for spin
relaxation/dephasing was proved by Glazov and Ivchenko
\cite{Glazov:jetp.75.403} by using perturbation theory and by Weng and Wu
\cite{PhysRevB.68.075312} from the fully microscopic kinetic spin Bloch equation approach. In a
temperature-dependent study of the spin dephasing in [001]-oriented
$n$-doped quantum wells, Leyland et al. experimentally verified
the effects of the electron-electron Coulomb scattering
\cite{leyland:165309,PhysRevLett.89.236601}. Later 
Zhou et al. even predicted a peak from the Coulomb scattering in
the temperature dependence of the spin relaxation time in a
high-mobility low-density $n$-doped [001] quantum well
\cite{zhou:045305}. This was later demonstrated by Ruan et al.
experimentally \cite{ruan:193307}.

\subsection{Kinetic spin Bloch equations in $n$- or 
$p$-doped confined structures}
\label{sec5.3}
In this section we present the spin dynamics in $n$- or $p$-type confined
semiconductor structures. In this case, we only need to consider a 
simplified single-particle density matrix $\rho_{{\bf k}}$ which
consists only the electron/hole distribution functions and the inter-spin-band polarization (spin coherence). In the effective-mass
approximation, the Hamiltonian of electrons in the confined system
reads
\begin{eqnarray}
H&=&\sum_i[H_{e}({\bf R}_i)+e{\bf E}\cdot{\bf r}_i]+H_{I},
\label{eq5.3-1}
\\
H_{e}({\bf R})&=&\frac{{\bf P}^2}{2m^\ast}+H_{\rm so}({\bf P})+V({\bf
  r}_c)+\frac{1}{2}g^\ast\mu_B{\bf B}\cdot{\bf\sigma},
\label{eq5.3-2}
\end{eqnarray}
in which ${\bf r}_c$ represents the coordinate along the confinement
and ${\bf R}=({\bf r},{\bf r}_c)$. ${\bf P}=-i\hbar\nabla-e{\bf A}({\bf
R})$ is the momentum with ${\bf B}=\nabla\times{\bf A}$. $H_{\rm so}$ in
Eq.~(\ref{eq5.3-2}) is the Hamiltonian of the spin-orbit coupling which
consists the Dresselhaus/Rashba term as well as the strain-induced
spin-orbit coupling. $V({\bf r}_c)$ represents the confinement
potential. The energy spectrum of single-particle Hamiltonian
(\ref{eq5.3-2}) reads
\begin{equation}
H_{e}|{\bf k},n\rangle=\varepsilon_n({\bf k})|{\bf k},n\rangle
\label{eq5.3-3}
\end{equation}
with $n$ denoting the index for both subband [subjected to the
confinement $V({\bf r}_c)$] and spin. Hence $|{\bf k},n\rangle=|{\bf
  k}\rangle|n\rangle_{\bf k}$.

Due to the spin-orbit coupling, $_{\bf k}\langle
n|n^\prime\rangle_{{\bf k}^\prime}$ usually is not diagonal if ${\bf
  k}^\prime\neq{\bf k}$. The Hilbert space spanned by $|{\bf
  k},n\rangle$ is helix space \cite{cheng:032107,cheng:083704}. The kinetic spin Bloch equations in
this space are very complicated. Usually we use the wavefunctions
$|n\rangle_{\bf k}$ at a fixed ${\bf k}_0$ as a complete basis, i.e.,
$\{|n\rangle_0\}$. Therefore, the Hilbert space becomes $\{|{\bf
  k}\rangle|n\rangle_0\}$ which is complete and orthogonal. We call this
space collinear space. The eigenfunction $|{\bf k},n\rangle$ can be
expanded in this space as 
\begin{equation}
|{\bf k},n\rangle=\sum_m{}_0\langle m|\langle{\bf k}|{\bf
  k},n\rangle|{\bf k}\rangle|m\rangle_0.
\label{eq5.3-4}
\end{equation}
The matrix element of the Hamiltonian $H_{e}$ in the collinear space
reads
\begin{equation}
E_{\bf k}^{nn^\prime}={}_0\langle n|\langle {\bf k}|H_{e}|{\bf
  k}\rangle|n^\prime\rangle_0.
\label{eq5.3-5}
\end{equation}
As $|{\bf k}\rangle|n\rangle_0$ in general is not the eigenfunction of
$H_{e}$, $E_{\bf k}^{nn^\prime}$ is not diagonal. $e{\bf E}\cdot{\bf r}$
in Eq.~(\ref{eq5.3-1}) is the driving term of the electric field along
the confinement-free directions. $H_{I}$ is the interaction Hamiltonian.

In the collinear space, the density operator can be written as
\begin{equation}
\rho({\bf Q})=\sum_{{\bf k},n_1n_2}I_{n_1n_2}({\bf q}_c)c_{{\bf
    k+q}n_1}^\dagger c_{{\bf k}n_2},
\label{eq5.3-6}
\end{equation}
in which $I_{n_1n_2}({\bf q}_c)={}_0\langle n_1|e^{i{\bf q}_c\cdot{\bf
    r}_c}|n_2\rangle_0$ is the form factor. The interaction
Hamiltonian $H_{I}$ reads
\begin{eqnarray}
H_{\rm ee}&=&\sum_{{\bf Qkk^\prime},n_1n_2n_3n_4}V({\bf
  Q})I_{n_1n_2}({\bf q}_c)I_{n_3n_4}({\bf q}_c)c_{{\bf
    k^\prime+q}n_3}^\dagger c_{{\bf k-q}n_1}^\dagger c_{{\bf
    k}n_2}c_{{\bf k^\prime}n_4},
\label{eq5.3-7}
\\
H_{\rm ep}&=&\sum_{{\bf Qk},n_1n_2,\lambda}M_\lambda({\bf
  Q})I_{n_1n_2}({\bf q}_c)c_{{\bf k+q}n_1}^\dagger c_{{\bf
    k},n_2}(a_{{\bf Q}\lambda}+a_{-{\bf Q}\lambda}^\dagger),
\label{eq5.3-8}
\\
H_{\rm ei}&=&\sum_{{\bf Qk},n_1n_2}V_i({\bf Q})I_{n_1n_2}({\bf
  q}_c)\rho_i({\bf Q})c_{{\bf k+q}n_1}^\dagger c_{{\bf k}n_2},
\label{eq5.3-9}
\end{eqnarray}
for electron-electron, electron-phonon and electron-impurity
interactions, respectively.

By solving the non-equilibrium Green function with generalized
Kadanoff-Baym Ansatz and the gradient expansion \cite{haugjauho}, the
kinetic spin Bloch equations read
\begin{equation}
\frac{\partial \rho_{\bf k}({\bf r},t)}{\partial
  t}=\left.\frac{\partial}{\partial t}\rho_{\bf k}({\bf
  r},t)\right|_{\rm dr}+\left.\frac{\partial}{\partial t}\rho_{\bf k}({\bf
  r},t)\right|_{\rm dif}+\left.\frac{\partial}{\partial t}\rho_{\bf k}({\bf
  r},t)\right|_{\rm coh}+\left.\frac{\partial}{\partial t}\rho_{\bf k}({\bf
  r},t)\right|_{\rm scat}.
\label{eq5.3-10}
\end{equation}
The first term on the right hand side of the above equation is 
\begin{equation}
\left.\frac{\partial}{\partial t}\rho_{\bf k}({\bf
  r},t)\right|_{\rm dr}=\frac{1}{2}\{\nabla_{\bf r}\bar{\varepsilon}_{\bf k}({\bf
  r},t),\nabla_{\bf k}\rho_{\bf k}({\bf r},t)\}
\label{eq5.3-11}
\end{equation}
with $\{A, B\}=AB+BA$ representing the anti-commutator. $(\bar{\varepsilon}_{\bf
    k}({\bf r},t))_{n_1n_2}=E_{\bf k}^{n_1n_2}+e{\bf E}\cdot{\bf
    r}\delta_{n_1n_2}+\Sigma_{\rm HF}^{n_1n_2}({\bf k},{\bf r},t)$, in which
  $\Sigma_{\rm HF}^{n_1n_2}({\bf k},{\bf r},t)=-\sum_{\bf Q}I({\bf
    q}_c)\rho_{\bf k-q}({\bf r},t)V({\bf Q})I(-{\bf q}_c)$ is the
  Hartree-Fock term. The second term describes the diffusion
\begin{equation}
\left.\frac{\partial}{\partial t}\rho_{\bf k}({\bf
  r},t)\right|_{\rm dif}=-\frac{1}{2}\{\nabla_{\bf k}\bar{\varepsilon}_{\bf k}({\bf
  r},t),\nabla_{\bf r}\rho_{\bf k}({\bf r},t)\}.
\label{eq5.3-12}
\end{equation}
The coherent term is given by
\begin{equation}
\left.\frac{\partial}{\partial t}\rho_{\bf k}({\bf
  r},t)\right|_{\rm coh}=-i[\bar{\varepsilon}_{\bf k}({\bf r},t),\rho_{\bf
    k}({\bf r},t)],
\label{eq5.3-13}
\end{equation}
where $[A, B]=AB-BA$ is the commutator. The scattering term in
Eq.~(\ref{eq5.3-10}) reads
\begin{equation}
\left.\frac{\partial}{\partial t}\rho_{\bf k}({\bf
  r},t)\right|_{\rm scat}=-\{S_{\bf k}(>,<)-S_{\bf k}(<,>)+S_{\bf k}(>,<)^\dagger -S_{\bf k}(<,>)^\dagger\},
\label{eq5.3-14} 
\end{equation}
with
\begin{eqnarray}\nonumber
S_{\bf k}(>,<)&=& \  \sum_{\bf Q}N_i\int_{-\infty}^td\tau V_i({\bf Q})I({\bf
  q}_c)e^{-iE_{\bf k-q}(t-\tau)}\rho_{\bf k-q}^>(\tau)V_i(-{\bf
  Q})I(-{\bf q}_c)\rho_{\bf k}^<(\tau) e^{iE_{\bf k}(t-\tau)}
\\\nonumber && \hspace{-0.35cm}+\sum_{\bf
  Q}\int_{-\infty}^td\tau M_\lambda({\bf Q})I({\bf
  q}_c)e^{-iE_{\bf k-q}(t-\tau)}\rho_{\bf k-q}^>(\tau)M_\lambda(-{\bf
  Q}) I(-{\bf q}_c)\rho_{\bf k}^<(\tau) e^{iE_{\bf k}(t-\tau)}
[N^<({\bf
  Q})e^{i\omega_{\bf Q}(t-\tau)}+N^>({\bf Q})e^{-i\omega_{\bf
    Q}(t-\tau)}]\\ \nonumber &&\hspace{-0.35cm} +\sum_{{\bf Q}{\bf k}^\prime {\bf
    q}_c^\prime}\int_{-\infty}^td\tau V({\bf Q})I({\bf
  q}_c)e^{-iE_{\bf k-q}(t-\tau)}\rho_{\bf k-q}^>(\tau)V(-{\bf
  q}_c^\prime)I(-{\bf q}^\prime_c)\rho_{\bf k}^<(\tau) e^{iE_{\bf
    k}(t-\tau)} \mbox{Tr}[I(-{\bf q}_c)e^{-iE_{\bf k}(t-\tau)}\rho_{{\bf
k}^\prime}^>(\tau)\\  && \quad\times I({\bf q}_c^\prime)\rho_{{\bf k}^\prime-{\bf
q}}^<(\tau)e^{-iE_{{\bf k}^\prime-{\bf q}}(t-\tau)}].
\label{eq5.3-15}
\end{eqnarray}
 Here $N^<({\bf Q})=N({\bf Q})$ is the phonon distribution, $N^>({\bf
   Q})=N({\bf Q})+1$, $\rho_{\bf k}^>=1-\rho_{\bf k}$, and $\rho_{\bf
   k}^<=\rho_{\bf k}$. $N_i$ is the impurity density. As $E_{\bf k}=\sum_{n}\varepsilon_n({\bf k})T_{{\bf
     k},n}$ with $T_{{\bf k},n}=|{\bf k},n\rangle\langle{\bf k},n|$,
 $e^{iE_{\bf k}t}=\sum_n e^{i\varepsilon_n({\bf k})t}T_{{\bf
     k},n}$. By further assuming the spin precession period is much
 longer than the average momentum scattering time and applying the
 Markovian approximation, i.e., replacing $\rho_{\bf k}(\tau)$ in the
integrand of Eq.~(\ref{eq5.3-15}) by
\begin{equation}
\rho_{\bf k}(\tau)\approx e^{iE_{\bf k}(t-\tau)}\rho_{\bf k}(t)e^{-iE_{\bf
    k}(t-\tau)},
\label{eq5.3-16}
\end{equation}
the time integral can be carried out and one arrives at the energy
conservation:
\begin{eqnarray}\nonumber
S_{\bf k}(>,<)&=&\ \pi \sum_{{\bf Q},n_1n_2} N_i V_i({\bf Q})I({\bf
  q}_c)\rho_{\bf k-q}^>(t)T_{{\bf k-q},n_1}V_i(-{\bf
  Q})I(-{\bf q}_c)T_{{\bf k},n_2}\rho_{\bf
  k}^<(t) \delta(\varepsilon_{n_1}({\bf k-q})-\varepsilon_{n_2}({\bf
  k}))\\\nonumber && +\pi\sum_{{\bf
  Q},n_1n_2}M_\lambda({\bf Q})I({\bf
  q}_c)\rho_{\bf k-q}^>(t)T_{{\bf k-q},n_1} M_\lambda(-{\bf
  Q})I(-{\bf q}_c)T_{{\bf k},n_2}\rho_{\bf k}^<(t) \\\nonumber &&\quad\times 
[N^<({\bf
  Q})\delta(\varepsilon_{n_1}({\bf k-q})-\varepsilon_{n_2}(\bf
k)-\omega_{\bf Q}) + N^>({\bf Q})\delta(\varepsilon_{n_1}({\bf k-q})-\varepsilon_{n_2}({\bf
k})+\omega_{\bf Q})]\\\nonumber && +\pi\sum_{{\bf Q}{\bf k}^\prime {\bf
    q}_c^\prime,n_1n_2n_3n_4}V({\bf Q}) I({\bf
  q}_c)\rho_{\bf k-q}^>(t)T_{{\bf k-q},n_1}V(-{\bf Q})I(-{\bf
  q}^\prime_c)T_{{\bf k},n_2}\rho_{\bf k}^<(t)\\ &&\quad \times \mbox{Tr}[I(-{\bf q}_c)\rho_{{\bf
k}^\prime}^>(t) T_{{\bf k}^\prime,n_3}I({\bf q}_c^\prime)T_{{\bf
k}^\prime-{\bf q},n_4}\rho_{{\bf k}^\prime-{\bf
q}}^<(t)]\delta(\varepsilon_{n_1}({\bf k-q})-\varepsilon_{n_2}({\bf
k})+\varepsilon_{n_3}({\bf k}^\prime) -\varepsilon_{n_4}({\bf
  k}^\prime-{\bf q})).
\label{eq5.3-17}   
\end{eqnarray}
The expression of $S_{\bf k}(<,>)$ can be obtained by interchanging $<$ and $>$ in Eqs.~(\ref{eq5.3-16}) and (\ref{eq5.3-17}).
 
\subsection{Spin relaxation and dephasing in $n$- or $p$-type quantum wells}
\label{sec5.4}
In this section, we review the spin dynamics in $n$- or $p$-type
quantum wells under various conditions. We first write the kinetic spin Bloch equations in
both the helix and collinear spin spaces in quantum wells in
Sec.~\ref{sec5.4.1}. Then we review the spin dynamics near the equilibrium and far
away from the equilibrium in Sec.~\ref{sec5.4.2} and \ref{sec5.4.3}
respectively. We highlight the effect of the Coulomb scattering to the
spin relaxation and dephasing in Sec.~\ref{sec5.4.4}. After that
we review the non-Markovian effect of hole spin dynamics in
Sec.~\ref{sec5.4.5}. We review the electron spin relaxation due to the
Bir-Aronov-Pikus mechanism in Sec.~\ref{sec5.4.6}. The spin dynamics in the presence
of a strong THz laser field is reviewed in Sec.~\ref{sec5.4.7}. Spin
relaxation in paramagnetic GaMnAs quantum wells,\footnote{Recently Shen {\em
et al.} have further extended the kinetic spin Bloch equation
theory to study the Gilbert damping in ferromagnetic semiconductors
\cite{2010arXiv1001.4576S}.} GaAs (110) quantum
wells, Si/SiGe and Ge/SiGe quantum wells and wurtzite ZnO (001)
quantum wells are reviewed in Secs.~\ref{sec5.4.8}-\ref{sec5.4.11},
respectively. 
\subsubsection{Kinetic spin Bloch equations in (001) quantum wells}
\label{sec5.4.1}
We first consider a quantum well with a small well width so that only
the lowest subband is needed. The electron Hamiltonian of
Eq.~(\ref{eq5.3-2}) now reads
\begin{equation}
H_{e}=\frac{{\bf k}^2}{2m^\ast}+{\bf h}({\bf k})\cdot{\bgreek
  \sigma}+\varepsilon_0.
\label{eq5.4.1-1}
\end{equation}
Here $\varepsilon_0$ is the energy of the lowest subband. It can be
calculated from the Schr\"odinger equation with confinement potential
$V({\bf r}_c)=V(z)$. ${\bf h}({\bf k})=\frac{1}{2}g\mu_B[{\bf B}+{\bf
  \Omega}^{2D}({\bf k})]$ with ${\bf \Omega}^{2D}({\bf k})$ being the
Dresselhaus and/or Rashba terms. The eigenenergy and eigenfunction
of Eq.~(\ref{eq5.4.1-1}) are $\varepsilon_\xi({\bf
  k})=\frac{k^2}{2m^\ast}+\xi|{\bf h}({\bf k})|+\varepsilon_0$ and
$|{\bf k},\xi=\pm\rangle=\frac{T_{{\bf
    k},\xi}|\uparrow\rangle}{\langle\uparrow|T_{{\bf
    k},\xi}|\uparrow\rangle}$, respectively. The projector operator
reads $T_{{\bf k},\xi}=\frac{1}{2}[1+\xi\frac{{\bf h}({\bf k})}{|{\bf
    h}({\bf k})|}\cdot{\bgreek\sigma}]$. The space spanned by $\{|{\bf
k},\xi\rangle\}$ is the helix space. As $|{\bf k},\xi\rangle$ is ${\bf
k}$-dependent, the kinetic spin Bloch equations are more complicated in this
space. Meanwhile, one may use the space spanned by $\{|{\bf
k}\rangle|n\rangle_0\}$. Here we choose ${\bf k}_0=0$, i.e., the
$\Gamma$-point and then $|n\rangle_0$ is the eigenvector of
$\sigma_z$, $|\sigma\rangle=|\uparrow\rangle$ or $|\downarrow\rangle$, which is ${\bf
k}$-independent. This space is the collinear space. In the collinear
spin space, the density matrix is 
\begin{equation}
\rho_{\bf k}=\left(\begin{array}{cc} f_{{\bf k}\uparrow} & \rho_{{\bf k}\uparrow\downarrow}\\ \rho_{{\bf k}\downarrow\uparrow} & f_{{\bf k}\downarrow}\end{array} \right).
\label{eq5.4.1-2}
\end{equation}
When an electric field ${\bf E}$ is applied along the quantum well,
the kinetic spin Bloch equations are given by \cite{PhysRevB.69.245320,cheng:073702}
\begin{equation}
\frac{\partial \rho_{\bf k}(t)}{\partial t}=e{\bf E}\cdot\nabla_{\bf
  k}\rho_{\bf k}(t)-i[{\bf h}({\bf k})\cdot{\bgreek\sigma}+\Sigma_{\rm HF}({\bf k},t), \rho_{\bf k}(t)]+\left.\frac{\rho_{\bf
      k}(t)}{\partial t}\right|_{\rm scat}.
\label{eq5.4.1-3}
\end{equation}
Here $\Sigma_{\rm HF}({\bf k},t)=-\sum_{\bf q}V_{\bf q}\rho_{\bf
  k-q}(t)$, and $S_{\bf k}(>,<)$ in the scattering terms
[Eq.~(\ref{eq5.3-17})] reads
\begin{eqnarray}\nonumber
S_{\bf k}(>,<)&=&\ \pi N_i\sum_{{\bf q},\eta_1\eta_2}|U_{\bf q}|^2\rho_{\bf k-q}^>(t)T_{{\bf k-q},\eta_1}T_{{\bf k},\eta_2}\rho_{\bf k}^<(t)\delta(\varepsilon_{\eta_1}({\bf k-q})-\varepsilon_{\eta_2}({\bf
  k}))\\\nonumber &&+\pi\sum_{{\bf
  Q},\eta_1\eta_2}|g_{{\bf Q},\lambda}|^2\rho_{\bf k-q}^>(t)T_{{\bf k-q},\eta_1}T_{{\bf k},\eta_2}\rho_{\bf k}^<(t)[N^<({\bf
  Q})\delta(\varepsilon_{\eta_1}({\bf k-q}) -\varepsilon_{\eta_2}(\bf
k)-\omega_{\bf Q})\\\nonumber &&\quad +N^>({\bf Q})\delta(\varepsilon_{\eta_1}({\bf k-q})-\varepsilon_{\eta_2}(\bf
k)+\omega_{\bf Q})]\\\nonumber &&+\pi\sum_{{\bf q}{\bf
    k}^\prime,\eta_1\eta_2\eta_3\eta_4}V_{\bf q}^2\rho_{\bf k-q}^>(t)T_{{\bf k-q},\eta_1}T_{{\bf k},\eta_2}\rho_{\bf k}^<(t)\mbox{Tr}[\rho_{{\bf
k}^\prime}^>(t)T_{{\bf k}^\prime,\eta_3}T_{{\bf
k}^\prime-{\bf q},\eta_4}\rho_{{\bf k}^\prime-{\bf
q}}^<(t)]\\ &&\quad \times\delta(\varepsilon_{\eta_1}({\bf k-q})-\varepsilon_{\eta_2}(\bf
k)+\varepsilon_{\eta_3}({\bf k}^\prime)-\varepsilon_{\eta_4}({\bf
  k}^\prime-{\bf q})),
\label{eq5.4.1-4}
\end{eqnarray}
with $|U_{\bf q}|^2=\sum_{q_z}V_i({\bf Q})V_i({-\bf Q})I(q_z)I(-q_z)$
and $|g_{{\bf Q},\lambda}|^2=M_\lambda({\bf Q})I(q_z)M_\lambda({-\bf
  Q})I(-q_z)$ and $V_q=\sum_{q_z}V({\bf Q})I(q_z)I(-q_z)$ being the
matrix elements with the form factors. It is noted that
$I_{\sigma_1\sigma_2}(q_z)$ is diagonal in the collinear spin space $|
\sigma\rangle$. In the calculation, one either uses the static
screening or the screening under the random-phase approximation
\cite{zhou:045305} for $V_{\bf q}$ and $U_{\bf q}$ \cite{jetp.99.1279},
according to the different conditions of investigation. In the
collinear spin space, the screened Coulomb potential and
electron-impurity interaction potential in the random-phase
approximation read
\begin{eqnarray}
V_{\bf q}&=&\sum_{q_z}v_{\bf Q}|I(iq_z)|^2/\epsilon({\bf q}),
\\
|U_{\bf q}|^2&=&\sum_{q_z}u_{\bf Q}^2|I(iq_z)|^2/\epsilon({\bf q})^2,
\end{eqnarray}
where $v_{\bf Q}=\frac{4\pi e^2}{{\bf Q}^2}$ is the bare Coulomb
potential, $u_{\bf Q}^2=Z_i^2v_{\bf Q}^2$ with $Z_i$ the charge number of
impurity, and 
\begin{equation}
\epsilon({\bf q})=1-\sum_{q_z}v_{\bf Q}|I(iq_z)|^2\sum_{{\bf k}\sigma}\frac{f_{{\bf k+q}\sigma}-f_{{\bf
      k}\sigma}}{\epsilon_{\bf k+q}-\epsilon_{\bf k}}.
\label{rpa}
\end{equation}
Eq.~(\ref{eq5.4.1-3}) is valid in both the collinear and helix spin
spaces. By performing a unitary transformation $\rho_{\bf k}^h=U_{\bf
  k}^\dagger\rho_{\bf k}^cU_{\bf k}$, one may transfer the density matrix from the collinear space
$\rho_{\bf k}^c$ to the helix one $\rho_{\bf k}^h$, with
\begin{equation}
U_{\bf k}=\left(\begin{array}{cc} \langle\uparrow|{\bf k},+\rangle &
    \langle\uparrow|{\bf k},-\rangle \\ \langle\downarrow|{\bf
      k},+\rangle & \langle\downarrow|{\bf k},-\rangle\end{array}
\right).
\label{eq5.4.1-6}
\end{equation}
Finally, we point out that the energy spectrum $\varepsilon_\eta({\bf
  k})$ in the scattering Eq.~(\ref{eq5.4.1-4}) contains the spin-orbit
coupling. When the coupling is much smaller than the Fermi energy, one
may neglect the coupling and hence
\begin{equation}
\delta(\varepsilon_\eta({\bf q})-\varepsilon_{\eta^\prime}({\bf
  k}))\approx\delta(\varepsilon({\bf q})-\varepsilon({\bf k})).
\label{eq5.4.1-7}
\end{equation}
By further utilizing the relation
\begin{equation}
\sum_\eta T_{{\bf k},\eta}=1,
\label{eq5.4.1-8}
\end{equation}
the scattering becomes
\begin{eqnarray}\nonumber
S_{\bf k}(>,<)&=&\ \pi N_i\sum_{{\bf q}}|U_{\bf q}|^2\rho_{\bf k-q}^>(t)\rho_{\bf k}^<(t)\delta(\varepsilon({\bf k-q})-\varepsilon({\bf
  k}))\\\nonumber &&+\pi\sum_{{\bf
  Q}}|g_{{\bf Q},\lambda}|^2\rho_{\bf k-q}^>(t)\rho_{\bf k}^<(t)[N^<({\bf
  Q})\delta(\varepsilon({\bf k-q})-\varepsilon(\bf
k)-\omega_{\bf Q}) +N^>({\bf Q})\delta(\varepsilon({\bf k-q})-\varepsilon(\bf
k)+\omega_{\bf Q})]\\\nonumber && + \pi\sum_{{\bf q}{\bf
    k}^\prime}V_{\bf q}^2\rho_{\bf k-q}^>(t)\rho_{\bf k}^<(t)\mbox{Tr}[\rho_{{\bf
k}^\prime}^>(t)\rho_{{\bf k}^\prime-{\bf
q}}^<(t)]\delta(\varepsilon({\bf k-q})-\varepsilon(\bf
k)+\varepsilon({\bf k}^\prime)-\varepsilon({\bf
  k}^\prime-{\bf q})).
\label{eq5.4.1-9}
\end{eqnarray}
This is the form used in the electron systems \cite{PhysRevB.66.235109,zhou:045305,PhysRevB.68.075312,PhysRevB.69.245320,weng:410,PhysRevB.70.195318,PhysRevB.72.033311,jiang:113702,cheng:073702,Lue501,stich:176401,cheng:205328,stich:073309,stich:205301,weng:063714,zhang:235323,jiang:prb125309,0295-5075-84-2-27006,zhang:075303,jiang:155201}.
               
Before discussing the results from the full kinetic spin Bloch equations, we first show a
simplest case by keeping only the electron-impurity scattering in the
scattering term, where one can obtain an analytical solution in the
spacial homogeneous system by further neglecting the Coulomb Hartree-Fock
contribution \cite{lue:125314}. The kinetic spin Bloch equations then read
\be
\frac{\partial}{\partial t}\rho_{\bf
  k}(t) = -i[\alpha(k_y\sigma_x-k_x\sigma_y),\rho_{\bf
  k}(t)]-2\pi N_i\sum_{\bf q}|U_{\bf q}|^2\delta(\varepsilon_{\bf
  k-q}-\varepsilon_{\bf k}) [\rho_{\bf k}(t)-\rho_{\bf k-q}(t)].
\label{eq5.4.1-10}
\ee
By expanding $\rho_{\bf k}$ as $\rho_{\bf k}=\sum_l\rho_l(k)e^{i\theta_k l}$, one has
\begin{equation}
\frac{\partial}{\partial
  t}\rho_l(k,t)=\alpha k\big\{[S^+,\rho_{l+1}(k,t)]-[S^-,\rho_{l-1}(k,t)]\big\}-|U_l(k)|^2\rho_l(k,t),
\label{eq5.4.1-11}
\end{equation}
in which $|U_l(k)|^2=\frac{m^\ast
    N_i}{2\pi\hbar^2}\int_0^{2\pi}d\theta |U(\sqrt{2k^2(1-\cos 
  \theta)})|^2(1-\cos l\theta)$
with $|U_0(k)|^2=0$ and $|U_{-1}(k)|^2=|U_1(k)|^2\equiv
\frac{1}{\tau_p(k)}$. By multiplying $\frac{1}{2}{\bgreek\sigma}$ and
then performing trace from both sides of Eq.~(\ref{eq5.4.1-11}), one
comes to
\begin{equation}
\frac{\partial}{\partial
  t}{\bf S}_l(k,t)=\alpha k[{\cal F}^\dagger{\bf S}_{l+1}(k,t)-{\cal
  F}{\bf S}_{l-1}(k,t)]-|U_l(k)|^2{\bf S}_l(k,t),
\label{eq5.4.1-12}
\end{equation}
with ${\bf S}_l(k,t)=\frac{1}{2}\mbox{Tr}[\rho_l(k,t){\bgreek\sigma}]$
and ${\cal F}=\left(\begin{array}{ccc} 0 & 0 & 1\\ 0 & 0 & -i \\ -1 & i & 0\end{array}
  \right)$. In deriving Eq.~(\ref{eq5.4.1-12}), we have used the relation
$\frac{1}{2}\mbox{Tr}([S^+,\rho_l]{\bgreek\sigma})={\cal F}^\dagger{\bf
  S}_l(k,t)$ and $\frac{1}{2}\mbox{Tr}([S^-,\rho_l]{\bgreek\sigma})={\cal F}{\bf
  S}_l(k,t)$. By keeping only the lowest three orders of ${\bf S}_l$ in the
strong scattering limit ($x_k\ll 1$), 
i.e., $l=0,\pm
1$, one has 
\begin{equation}
\left[\frac{\partial}{\partial t}-\alpha k\left(\begin{array}{ccc}
    -\frac{1}{\alpha k\tau_p(k)} & -{\cal F} & 0\\ {\cal F}^\dagger &
    0 & -{\cal F} \\ 0 & {\cal F}^\dagger  &  -\frac{1}{\alpha
      k\tau_p(k)} \end{array}\right)\right]\left(\begin{array}{c}
    {\bf S}_1 \\ {\bf S}_0 \\ {\bf S}_{-1} \end{array}\right)=0.
\label{eq5.4.1-14}
\end{equation} 
The solution of Eq.~(\ref{eq5.4.1-14}) reads
\begin{eqnarray}
{\bf S}_1(k,t)&=&-\frac{x_k}{2\sqrt{1-x_x^2}}e^{-\frac{t}{2\tau_p(k)}}\sinh\frac{t}{2\tau_p(k)/\sqrt{1-x_k^2}}\left(\begin{array}{c} 1 \\ i \\ 0 \end{array}\right)f(\varepsilon_k-\mu),\label{eq5.4.1-15}\\
{\bf S}_0(k,t)&=&e^{-\frac{t}{2\tau_p(k)}}\left[\frac{\sinh\frac{t}{2\tau_p(k)/\sqrt{1-x_k^2}}}{\sqrt{1-x_k^2}}+\cosh\frac{t}{2\tau_p(k)/\sqrt{1-x_k^2}}\right]\left(\begin{array}{c} 0 \\ 0 \\ 1 \end{array}\right)f(\varepsilon_k-\mu),\label{eq5.4.1-16}\\
{\bf S}_{-1}(k,t)&=&-\frac{x_k}{2\sqrt{1-x_x^2}}e^{-\frac{t}{2\tau_p(k)}}\sinh\frac{t}{2\tau_p(k)/\sqrt{1-x_k^2}}\left(\begin{array}{c} 1 \\ -i \\ 0 \end{array}\right)f(\varepsilon_k-\mu),\label{eq5.4.1-17}
\end{eqnarray}
with the initial conditions being ${\bf S}_1(k,0)={\bf S}_{-1}(k,0)=0$
and ${\bf S}_0(k,0)=f(\varepsilon_k-\mu){\bf e}_z$, i.e., the initial
spin polarization being along the $z$-axis. $x_k$ in
Eqs.~(\ref{eq5.4.1-15}-\ref{eq5.4.1-17}) is $x_k=4\alpha
k\tau_p(k)$. From Eqs.~(\ref{eq5.4.1-15}-\ref{eq5.4.1-17}), electron
spin at momentum ${\bf k}$ reads
\begin{equation}
{\bf S}_{\bf k}(t)={\bf S}_0(k,t)+{\bf S}_1(k,t)e^{i\theta_k}+{\bf
  S}_{-1}(k,t)e^{-i\theta_k},
\label{eq5.4.1-18}
\end{equation}
and hence the component along the $z$-axis is given by
\be
S_{\bf
  k}^z(t)=S_0^z(k,t)=\frac{1}{2}f(\varepsilon_k-\mu)\left[\left(1+1\big/\sqrt{1-x_k^2}\right)e^{-\frac{t}{2\tau_p(k)}(1-\sqrt{1-x_k^2})}
  + \left(1-1\big/\sqrt{1-x_k^2}\right)e^{-\frac{t}{2\tau_p(k)}(1+\sqrt{1-x_k^2})}\right].
\label{eq5.4.1-19}
\ee
From Eq.~(\ref{eq5.4.1-19}), one can see the different time
evolutions of the spin polarization at different regimes. When
$x_k>1$, i.e., $\alpha k\tau_p(k)>1/4$, the system is in the weak
scattering regime and the terms $e^{\pm
  \frac{t}{2\tau_p(k)}\sqrt{1-x_k^2}}$ give just the spin
oscillations. Hence the spin relaxation is given by 
\begin{equation}
1/\tau_s({\bf k})=1/[2\tau_p(k)].
\label{eq5.4.1-20}
\end{equation}
However, the spin oscillation frequency is $\omega_{\bf
  k}=\sqrt{(2\alpha k)^2-1/[2\tau_p(k)]^2}$. One can see from
$\omega_{\bf k}$ that the scattering tends to suppress the spin
oscillation frequency. This indicates the counter-effect of the
scattering to the inhomogeneous broadening. When $x_k<1$, the spin
polarization decays according to
\begin{equation}
1/\tau_{s,\pm}({\bf k})=\left(1\pm\sqrt{1-[4\alpha
  k\tau_p(k)]^2}\right)\big/[2\tau_p(k)].
\label{eq5.4.1-21}
\end{equation}
In the strong scattering limit, i.e., $x_k\ll 1$,
\begin{eqnarray}
1/\tau_{s,+}({\bf k})&=&1/\tau_p(k),\label{eq5.4.1-22}\\
1/\tau_{s,-}({\bf k})&=&(2\alpha k)^2\tau_p(k).\label{eq5.4.1-23}
\end{eqnarray}
As $x_k\ll 1$, $\tau_{s,-}({\bf k})\gg\tau_{s,+}({\bf k})$ and the
spin relaxation is determined by $\tau_{s,-}({\bf k})$. It is noted that $\tau_{s,-}(\bf k)$ is exactly the result in the
literature \cite{dp,opt-or}.

Finally, one can see from Eqs.~(\ref{eq5.4.1-21}) and
(\ref{eq5.4.1-23}) that in the weak scattering regime, a stronger
scattering leads to a faster spin relaxation. Nevertheless, in the
strong scattering regime, a stronger scattering leads to a weaker
spin relaxation. This can be understood as following. In the presence
of inhomogeneous broadening, the scattering has dual effects to the
spin relaxation: (i) It gives a spin relaxation channel; (ii) It
has a counter-effect to the inhomogeneous broadening by making the
system more homogeneous. In the weak scattering limit, the counter
effect is weak and hence adding a new scattering always leads to an additional
relaxation channel and hence a fast spin relaxation. In the strong
scattering limit, the counter effect becomes significant and hence
adding a new scattering always leads to a longer spin relaxation time.

\subsubsection{Spin relaxation and dephasing
 in $n$-type (001) GaAs quantum wells near the equilibrium}
\label{sec5.4.2}
By numerically solving the kinetic spin Bloch equations with all the relevant scatterings
included, Weng and Wu studied the spin dephasing in $n$-type GaAs
quantum wells at high temperature ($\ge$ 120~K), where the electron-acoustic
phonon scattering is unimportant, first for small well width
\cite{PhysRevB.68.075312} with only the lowest subband and then for
large well width \cite{PhysRevB.70.195318} with mutisubband effect
considered. They further investigated the hot-electron effect in spin
dephasing by applying a large in-plane electric field
\cite{PhysRevB.69.245320}, where the hot-electron effect is investigated. By
further increasing the in-plane electric field, electrons can populate
higher subband and/or higher valleys. These effects were investigated by Weng
and Wu \cite{PhysRevB.70.195318} for mutisubband case and Zhang et al.
\cite{zhang:235323} for multivalley case. The spin relaxation at low
temperature was first investigated by Zhou et al.
\cite{zhou:045305} from kinetic spin Bloch equation approach by including the electron-acoustic
phonon scattering. Jiang and Wu studied the effect of strain on spin
relaxation \cite{PhysRevB.72.033311}. Spin relaxation for system with
competing Dresselhaus and Rashba terms was investigated theoretically
by Cheng and Wu \cite{cheng:083704} and both experimentally and theoretically by Stich
et al. \cite{stich:073309}. Spin relaxation with large initial spin
polarization was first studied by Weng and Wu \cite{PhysRevB.68.075312} and
many predictions were verified experimentally with good agreement
between the experimental data and theoretical calculations by Stich
et al. \cite{stich:176401,stich:205301}. The density dependence
of the spin relaxation was also investigated both theoretically and
experimentally \cite{0295-5075-84-2-27006}. In this and next sections we review the main
results of the above investigations.

It was revealed that the spin relaxation time based on the D'yakonov-Perel'
mechanism has a rich temperature dependence depending on different
impurity densities, carrier densities and well widths.

 \begin{figure}[htb]
 \begin{center}
 \centerline{\hspace{1.2cm} \includegraphics[width=6cm]{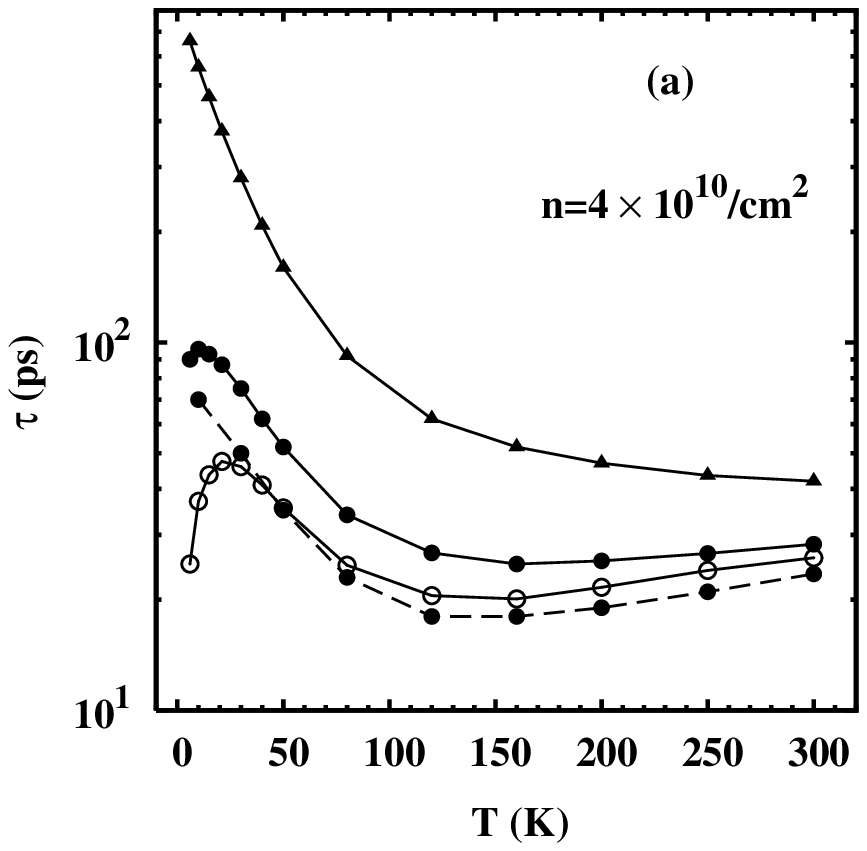}\hspace{-1.7cm}\includegraphics[width=6cm]{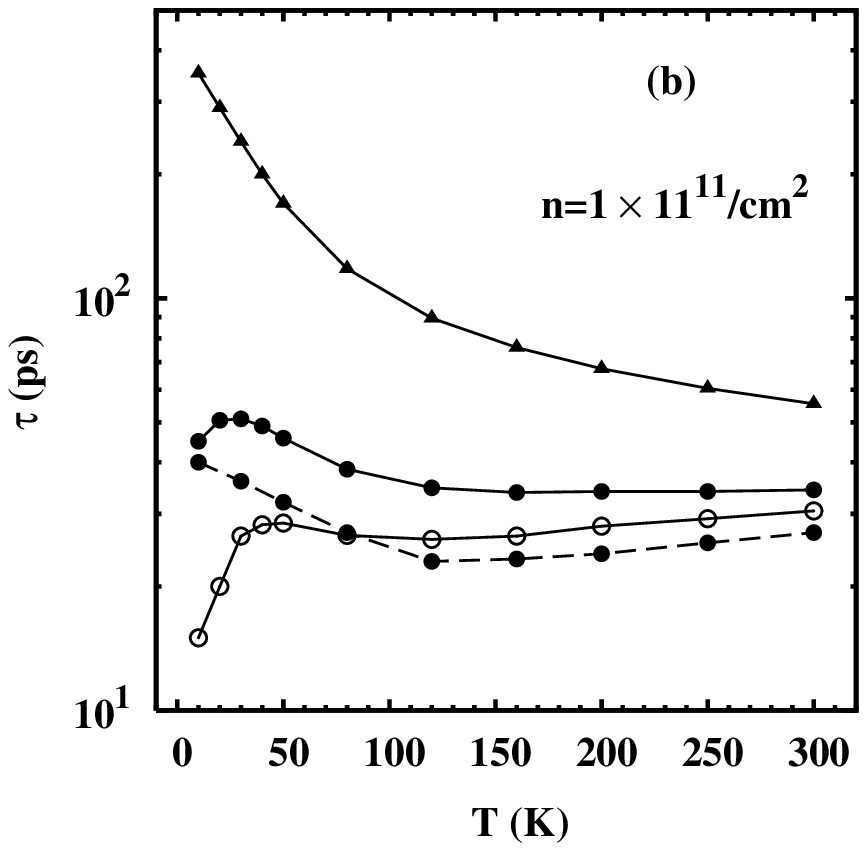}\hspace{-1.7 cm}\includegraphics[width=6cm]{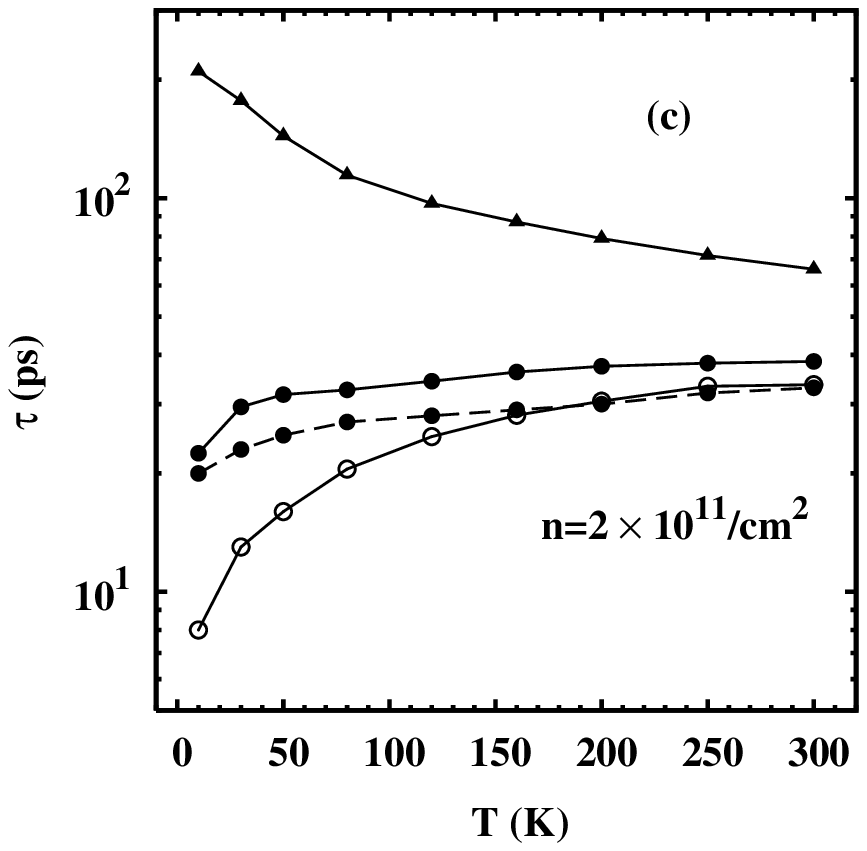}}
 \caption{Spin relaxation time $\tau$ {\sl vs.} the temperature $T$ in GaAs
(001) quantum well with well
   width $a=7.5$~nm and electron density $n$ being (a) $4\times
   10^{10}$~cm$^{-2}$, (b) $1\times 10^{11}$~cm$^{-2}$, and (c) $2\times
   10^{11}$~cm$^{-2}$, respectively. Solid curves with triangles:
   impurity density $n_i=n$; solid curves with dots: $n_i=0.1n$; solid
   curves with circles: $n_i=0$; dashed curves with dots: $n_i=0.1n$ and
   no Coulomb scattering. From Zhou et al. \cite{zhou:045305}.}
 \label{fig5.4.2-1}
 \end{center}
 \end{figure}

 \begin{figure}[htb]
 \begin{center}
 \centerline{\hspace{1.2cm} \includegraphics[width=6cm]{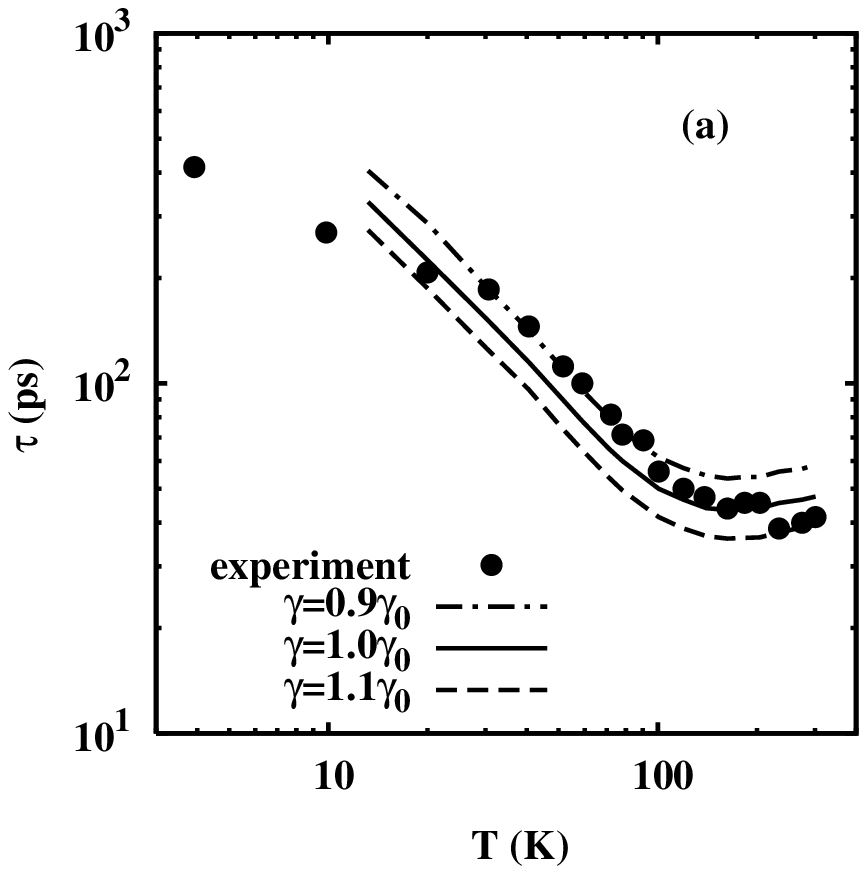}\hspace{-.7cm}\includegraphics[width=6cm]{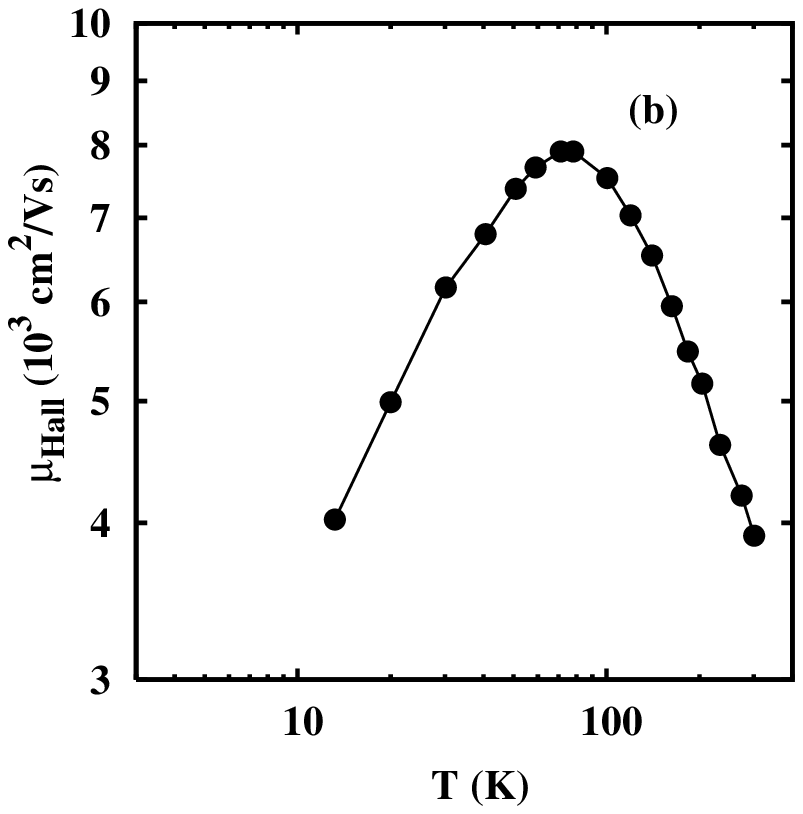}}
 \caption{(a) Spin relaxation time $\tau$ {\sl vs.} temperature $T$ for GaAs (001)
   quantum well with well width $a=7.5$~nm and electron density 
   $n=4\times10^{10}$~cm$^{-2}$ at three different spin-splitting
   parameters. Dots: experimental data; dot-dashed curve:
   $\gamma=0.9\gamma_0$; solid curve: $\gamma=\gamma_0$; dashed curve:
   $\gamma=1.1\gamma_0$. $\gamma_0=0.0114$~eV$\cdot$nm$^3$. (b) Hall
   mobility $\mu_{\rm Hall}$ {\sl vs.} temperature $T$
   (Ref.~\cite{Ohno2000817}), from which the temperature dependence of
   the impurity density is determined. From Zhou et
   al. \cite{zhou:045305}.} 
 \label{fig5.4.2-2}
 \end{center}
 \end{figure}

\begin{figure}[htb]
\begin{center}
\includegraphics[width=6cm]{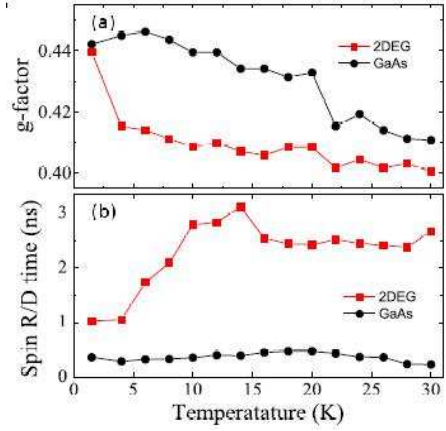}
\caption{ (a) Measured electron $g$-factor as a function of temperature for
  two-dimensional electron gas (squares) in 
GaAs/Al$_{0.35}$Ga$_{0.65}$As heterostructure  and bulk GaAs (circles). (b)
Measured electron spin relaxation/dephasing time as a function of temperature
  for two-dimensional electron gas (squares) and bulk GaAs (circles). All of
  the data were taken at $B=0.5$~T and powers of
  pump:probe=200:20~$\mu$W. From Ruan et al. \cite{ruan:193307}.}
\label{fig5.4.2-3}
\end{center}
\end{figure}

Figure~\ref{fig5.4.2-1} shows the temperature dependence of the spin
relaxation time of a 7.5~nm GaAs/Al$_{0.4}$Ga$_{0.6}$As quantum well at different electron and
impurity densities \cite{zhou:045305}. For this small well width, only
the lowest subband is needed in the calculation. It is shown in the
figure that when the electron-impurity scattering is dominant, the
spin relaxation time decreases with increasing temperature monotonically. This is
in good agreement with the experimental finding as shown in
Fig.~\ref{fig5.4.2-2}, where the dots are experimental data from Ohno
et al. \cite{Ohno2000817}, and the theoretical calculation
based on the kinetic spin Bloch equations well reproduces the experimental results from 20~K
to 300~K \cite{zhou:045305}. However, it is shown that for
sample with high mobility, i.e., low impurity density, when the
electron density is low enough, there is a peak at low
temperature. This peak, located around the Fermi temperature of
electrons $T_F^e=E_F/k_B$, is identified to be solely due to the Coulomb
scattering \cite{zhou:045305,PhysRevB.70.245210}. It disappears when the Coulomb
scattering is switched off, as shown by the dashed curves in the
figure. This peak also disappears at high impurity densities. It is also noted in
Fig.~\ref{fig5.4.2-1}(c) that for electrons of high density so that
$T_F^e$ is high enough and the contribution from the electron--longitudinal optical-phonon
scattering becomes marked, the peak disappears even for sample with no
impurity and the spin relaxation time increases with temperature
monotonically. The physics leading to the peak is due to the crossover
of the Coulomb scattering from the degenerate to the non-degenerate
limit. At $T<T_F^e$, electrons are in the degenerate limit  and the
electron-electron scattering rate $1/\tau_{\rm ee}\propto T^{2}$. At
$T>T_F^e$, $1/\tau_{\rm ee}\propto T^{-1}$ \cite{jetp.99.1279,giuliani_05}. Therefore, at
low electron density so that $T_F^e$ is low enough and the electron-acoustic
phonon scattering is very weak comparing with the electron-electron
Coulomb scattering, the Coulomb scattering is the dominant scattering
for high mobility sample. Hence the different temperature dependence
of the Coulomb scattering leads to the peak. It is noted that the peak
is just a feature of the crossover from the degenerate to the
non-degenerate limit. The location of the peak also depends on the
strength of the inhomogeneous broadening. When the inhomogeneous
broadening depends on momentum linearly, the peak tends to appear at
the Fermi temperature. A similar peak was predicted in the electron spin
relaxation in $p$-type GaAs quantum well and the hole spin
relaxation in (001) strained asymmetric Si/SiGe quantum well, where
the electron and hole spin relaxation times both show a peak at the
hole Fermi temperature $T_F^h$ \cite{zhang:155311,zhou0905.2790}. When the
inhomogeneous broadening depends on momentum cubically, the peak tends
to shift to a lower temperature. It was predicted that a peak in the
temperature dependence of the electron spin relaxation time appears at
a temperature in the range of ($T_F^e/4$, $T_F^e/2$) in the intrinsic
bulk GaAs \cite{jiang:125206} and a peak in the temperature dependence
of the hole spin relaxation time at $T_F^h/2$ in $p$-type Ge/SiGe
quantum well \cite{zhang:155311}. Ruan et al. demonstrated the
peak experimentally in a high-mobility low-density
GaAs/Al$_{0.35}$Ga$_{0.65}$As heterostructure \cite{ruan:193307} as
shown in Fig.~\ref{fig5.4.2-3} where a peak appears
at $T_F^e/2$ in the spin relaxation time versus temperature
curve.

\begin{figure}[htb]
\begin{center}
\centerline{
\hspace {1.2 cm} 
\includegraphics[width=6cm]{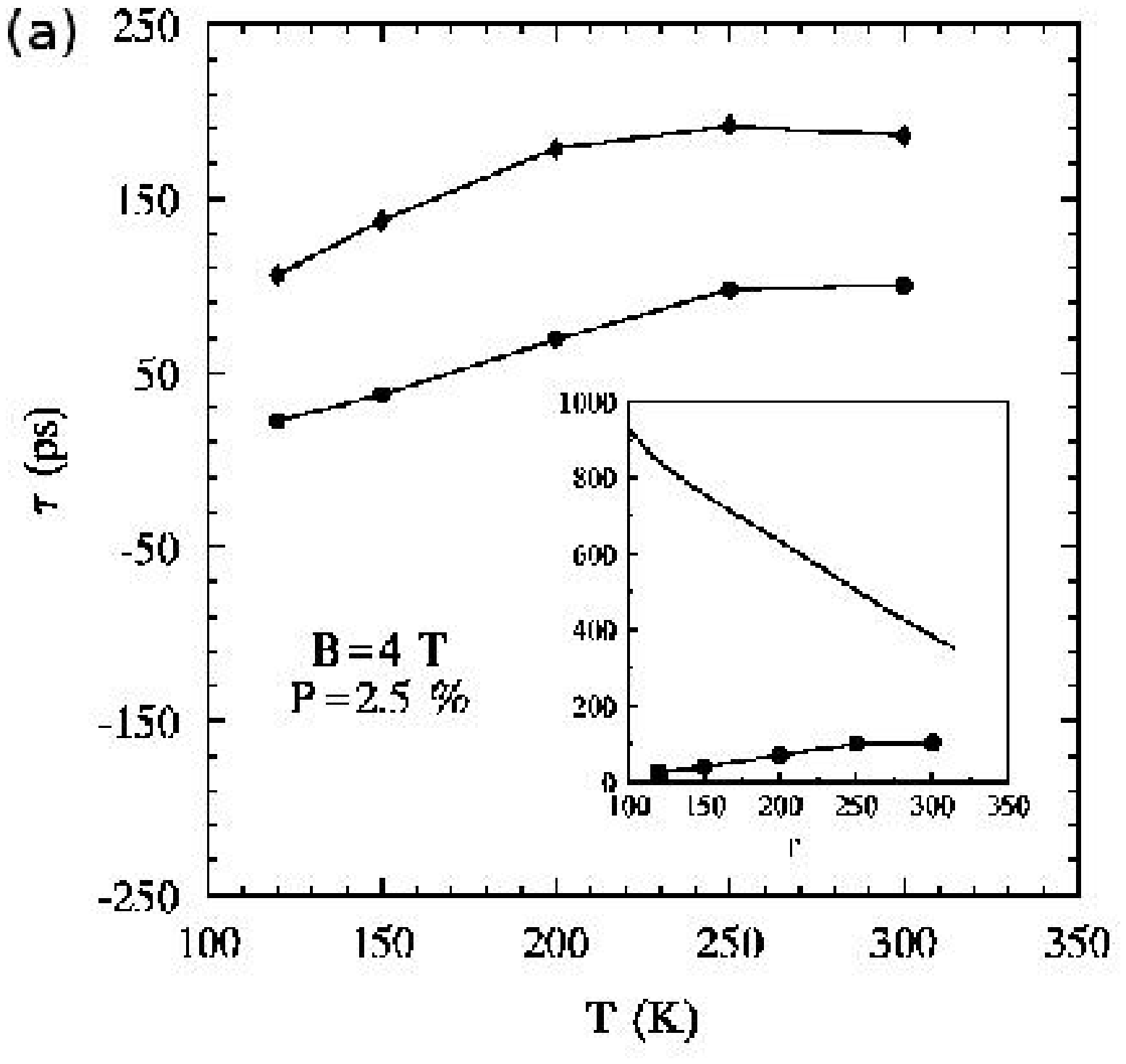}\hspace {0 cm}
\includegraphics[width=6cm]{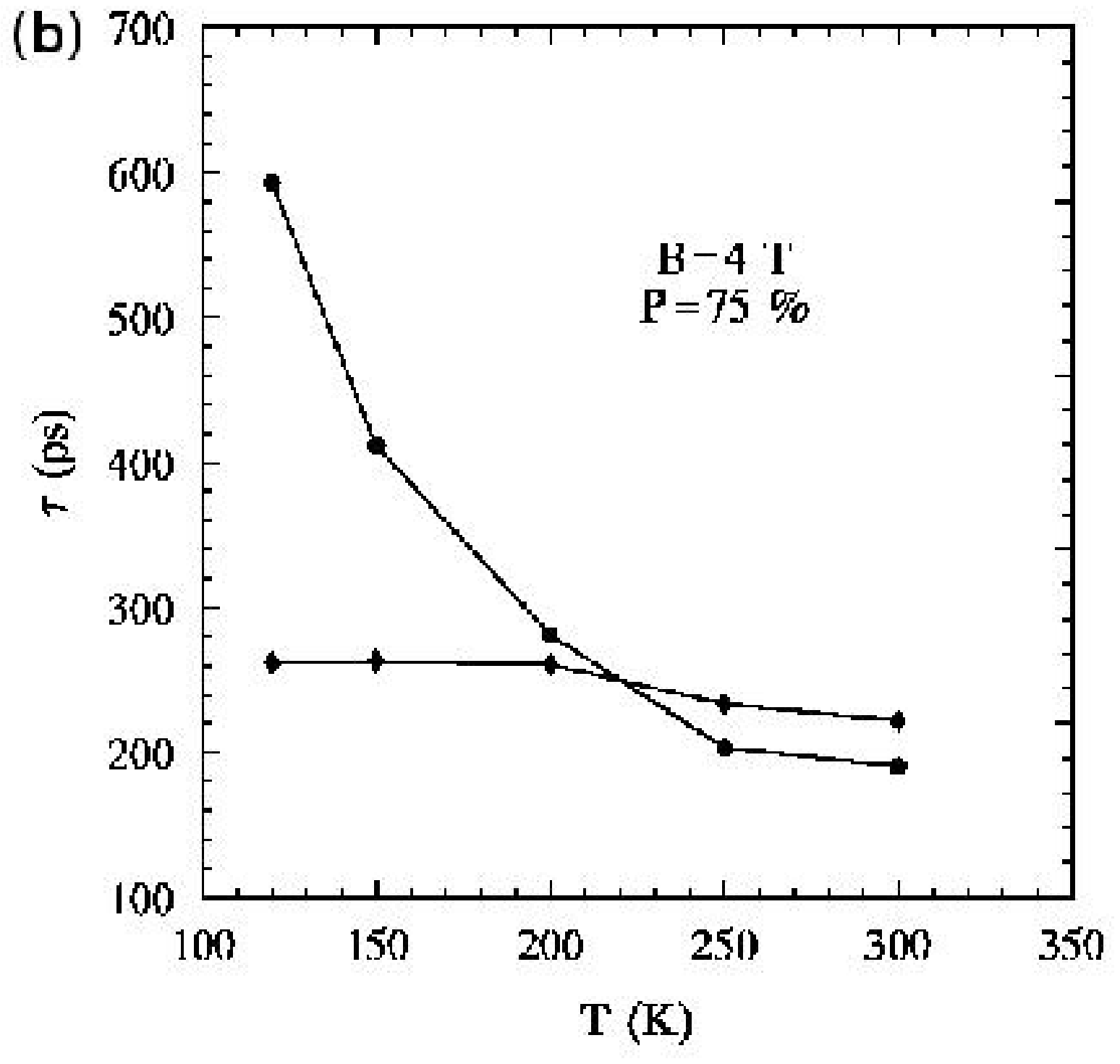}}
\caption{Spin dephasing time $\tau$ {\sl vs.} temperature $T$ with spin
  polarization $P=2.5$~\% (a) and $P=75$~\% (b) under two different 
impurity levels in  GaAs (001) quantum well. Curves
  with dots: $N_i=0$; curves with squares: $N_i=0.1N_e$. The spin
  dephasing times predicted by the simplified treatment of D'yakonov-Perel' term
  (solid curve) and the kinetic spin Bloch equation approach (dots) for $N_i=0$ are plotted
  in the inset for comparison. From Weng and Wu \cite{PhysRevB.68.075312}.}
\label{fig5.4.2-4}
\end{center}
\end{figure}

\begin{figure}[htb]
\begin{center}
\includegraphics[width=6cm]{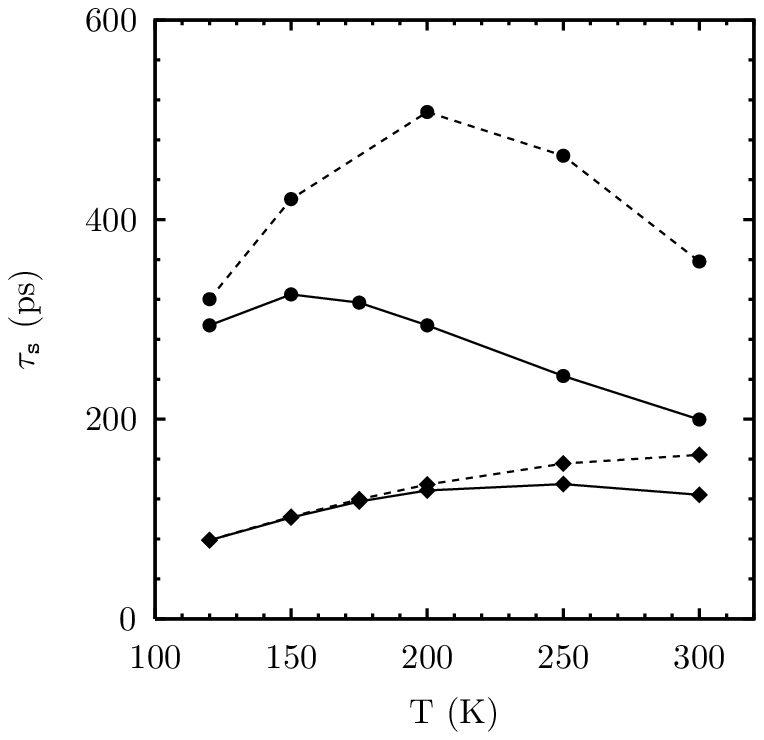}
\caption{Spin dephasing time $\tau_{s}$ {\em vs}. the background temperature
$T$ for two GaAs (001) quantum wells with width
$a=17.8$~nm ($\bullet$) and $12.7$~nm ($\blacklozenge$). The solid
 curves are the spin dephasing time calculated 
with the lowest two subbands included and
  the dashed curves are those calculated with only the lowest
  subband. From Weng and Wu \cite{PhysRevB.70.195318}.}
\label{fig5.4.2-5}
\end{center}
\end{figure}

It is also seen from Fig.~\ref{fig5.4.2-1} that at high temperature
($>100$~K), except for the case where the electron-impurity scattering
dominates the scattering, the spin relaxation time increases with
increasing temperature. Weng and Wu also showed this in a 15~nm
quantum well with a magnetic field $B=4$~T in the Voigt configuration
\cite{PhysRevB.68.075312}. They even compared the spin relaxation times
obtained from the kinetic spin Bloch equation approach and from the single-particle model
\cite{0953-8984-14-12-202,PhysRevB.64.161301} $1/\tau=\int_0^{\infty}dE_{\bf k}(f_{{\bf
    k}\frac{1}{2}}-f_{{\bf k}-\frac{1}{2}})\Gamma({\bf k})/\int_0^{\infty}dE_{\bf k}(f_{{\bf
    k}\frac{1}{2}}-f_{{\bf k}-\frac{1}{2}})$, in which $\Gamma({\bf
  k})$ is the spin relaxation rate. It is seen from the inset of
Fig.~\ref{fig5.4.2-4} that the spin relaxation time based on the
previous single-particle model decreases rather than increases with increasing temperature. As the impurity density
is set to zero in both computations, the main difference comes
from the Coulomb scattering which is missing in the single particle
calculation. 

However, even for sample without any impurity, the temperature
dependence of the spin relaxation time can also decrease with
increasing temperature at high temperatures. This was shown by Weng and
Wu in GaAs (001) quantum well with wider well width \cite{PhysRevB.70.195318}. In
the calculation, they considered two well widths, i.e., 17.8~nm and
12.7~nm, and showed the spin relaxation times as function of
temperature with both the lowest subband and multisubband effects
considered. One can see from the dashed curves (only the lowest
subband is calculated) in Fig.~\ref{fig5.4.2-5} that for small well width, the spin relaxation
time increases with increasing temperature but for larger well width, 
the spin relaxation time first increases and then decreases with
temperature. It is also noted that when the multisubband effect is
included (solid curves), similar results are also obtained. 

\begin{figure}[htb]
\begin{center}
\includegraphics[width=3.5cm]{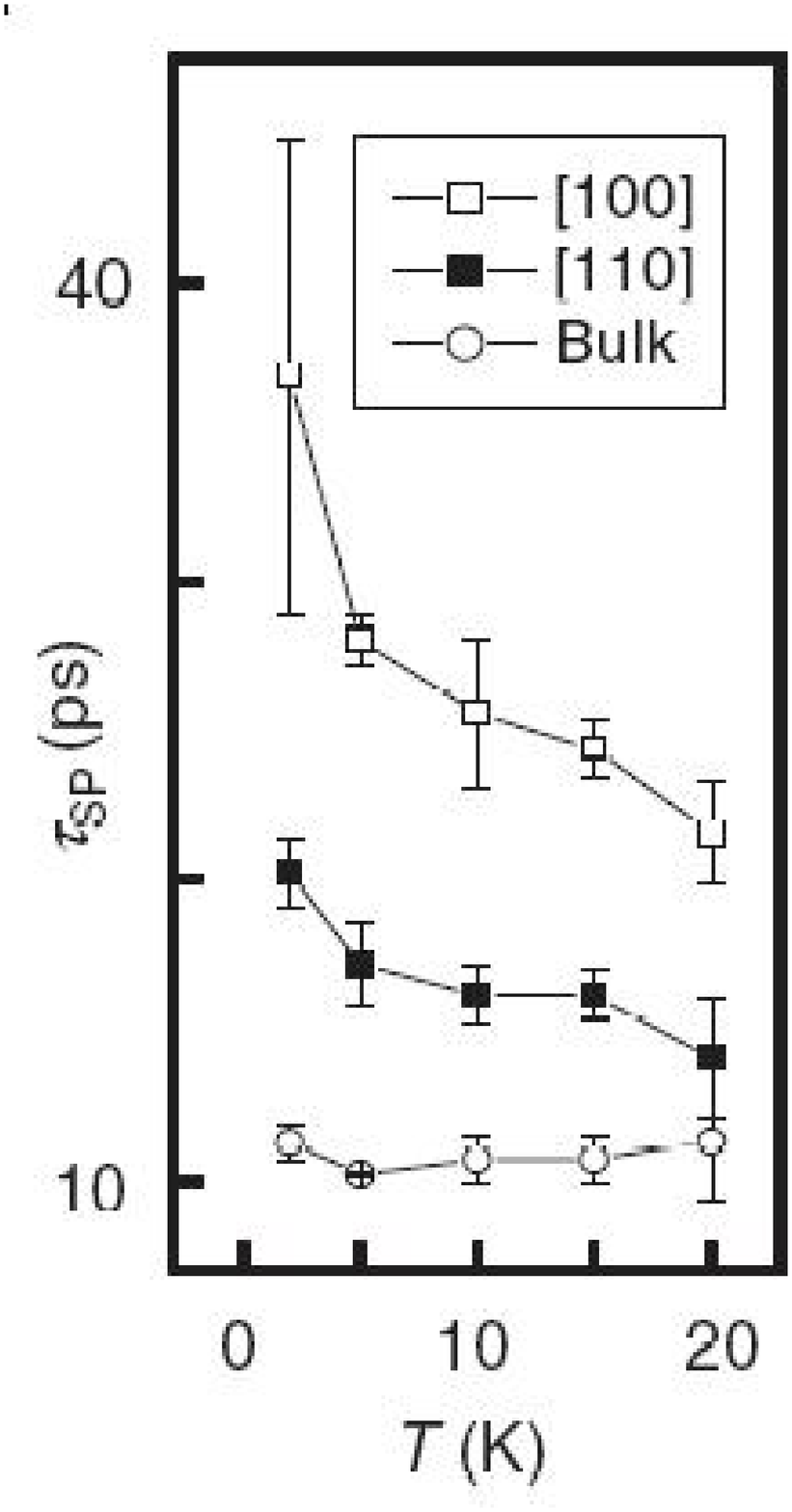}
\caption{Measured temperature dependence of the spin relaxation time for
 two-dimensional $n$-InAsGa channels with a width of 1.5~$\mu$m,
  where the cubic in ${\bf k}$ terms of the Dresselhaus term dominate
  the spin relaxation. Open and filled squares represent data of
  channels along [100] and [110], while the open circles depict the
  spin relaxation time of the unstructured quantum well. From
  Holleitner et al. \cite{1367-2630-9-9-342}.}
\label{fig5.4.2-6}
\end{center}
\end{figure}

\begin{figure}[htb]
\begin{center}
\centerline{
\hspace {1.2 cm} 
\includegraphics[width=6cm]{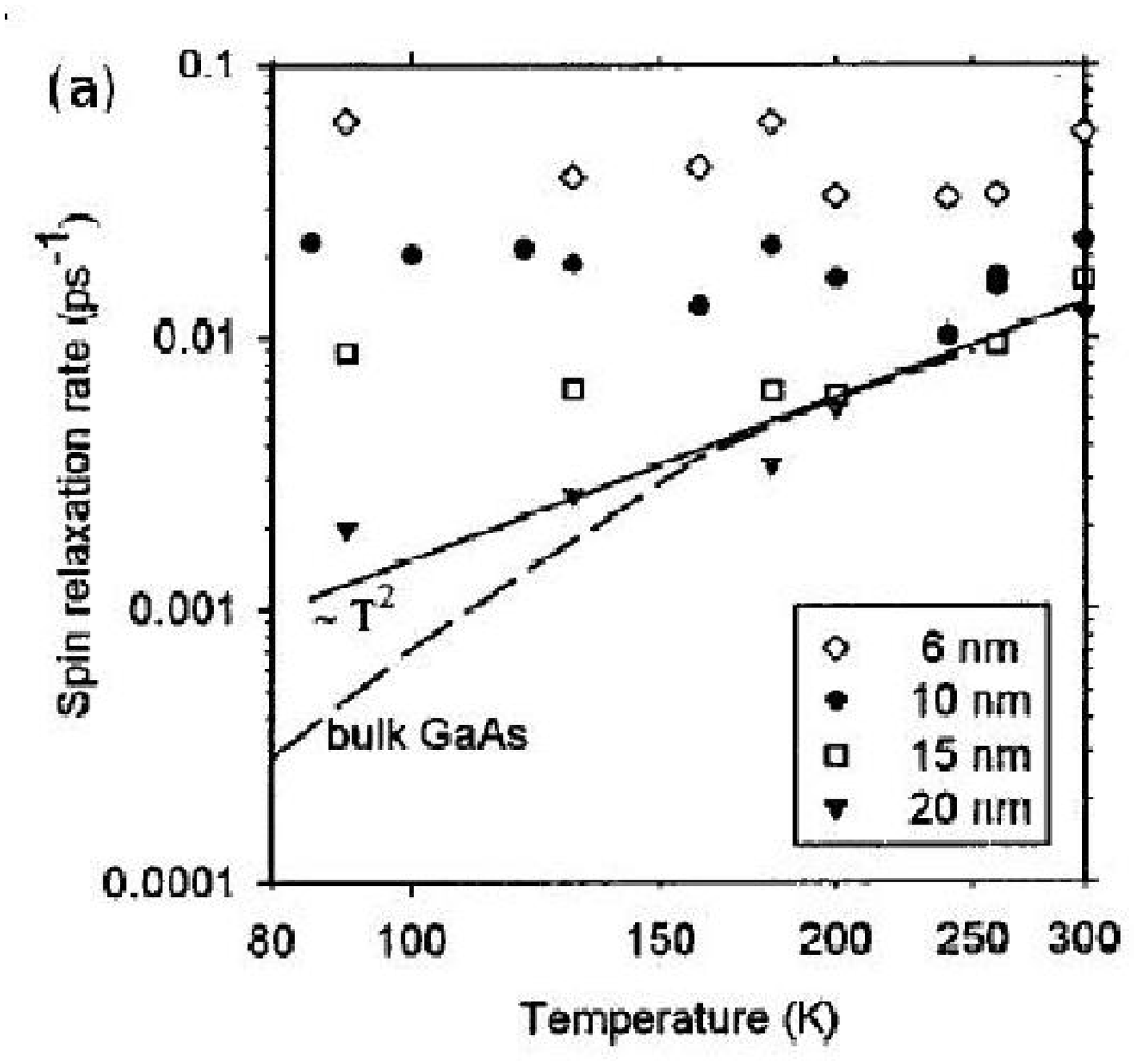}\hspace {0 cm}
\includegraphics[width=6.7cm,height=5.4cm]{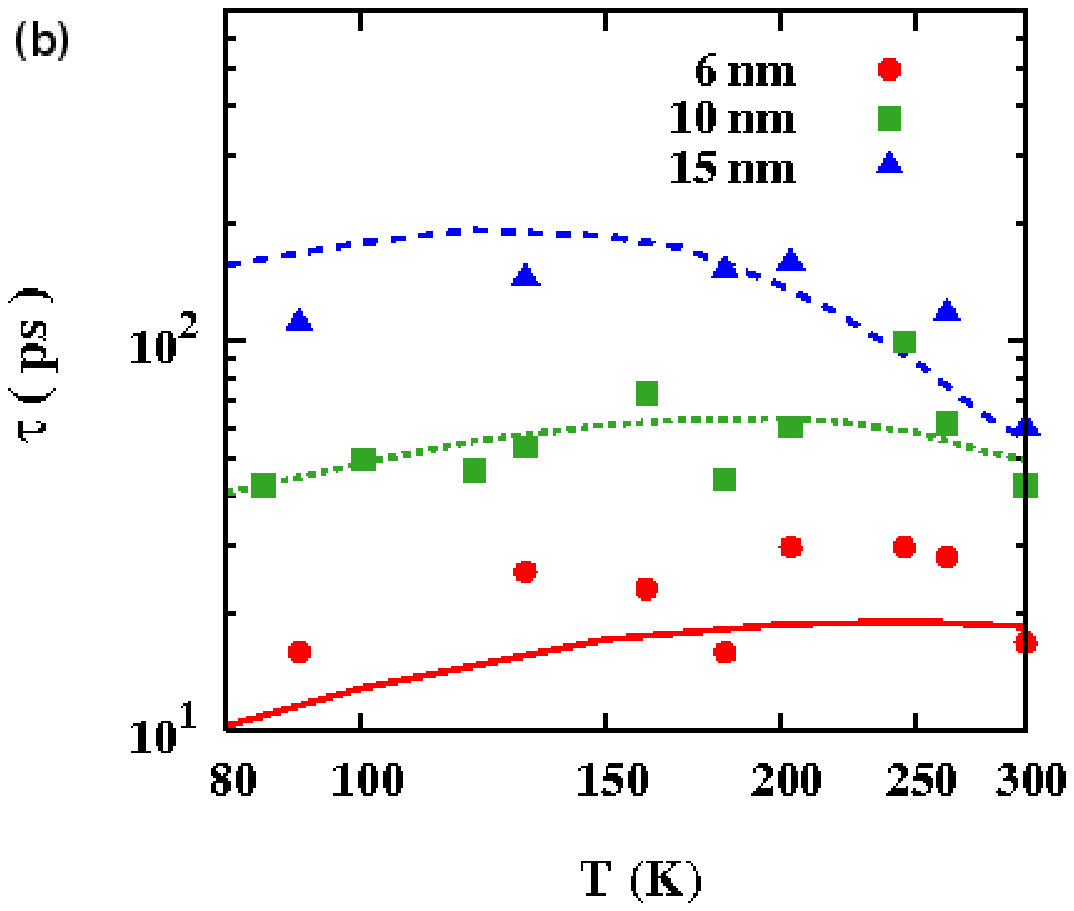}}
\caption{ Electron spin relaxation in intrinsic GaAs/Al$_{0.35}$Ga$_{0.65}$As 
quantum wells of a variety of widths. (a) Experimental results from
Malinowski et al. \cite{PhysRevB.62.13034}. The solid line
represents a quadratic dependence of relaxation rate on temperature. The dashed curve is data on
  electron spin relaxation in bulk GaAs from Maruschak et al.
  (as reproduced by Meier and Zakharchenya \cite{opt-or}). From Malinowski et al. \cite{PhysRevB.62.13034}. (b) Theoretical calculation via the kinetic spin Bloch equation
    approach (curves). The dots are data from (a). The 
parameters are $N_e=N_h=4\times
    10^{11}$~cm$^{-2}$, $N_i=0.1~N_e$, $\gamma_{\rm
      D}=0.0162$~eV$\cdot$nm$^{3}$. Here $N_e$ and $N_h$ are the
    electron and hole densities respectively, $N_i$ is the impurity
    density, and $\gamma_{\rm D}$ is the spin-orbit splitting parameter.}
\label{fig5.4.2-7}
\end{center}
\end{figure}

The physics leading to above rich behaviors originates from the
competition between the inhomogeneous broadening and the
scattering. With the increase of temperature, electrons are distributed to
higher momentum states. This leads to a larger inhomogeneous broadening
from the D'yakonov-Perel' term and hence a shorter spin relaxation time. In the
meantime, higher temperature also causes scattering (especially the
electron-phonon scattering) stronger. The scattering tends to make
electrons distribute more homogeneously and suppresses the
inhomogeneous broadening. Therefore the scattering tends to cause a
longer spin relaxation time with the increase of temperature. When the
electron-impurity scattering dominates the whole scattering, the spin
relaxation time decreases with increasing temperature monotonically,
thanks to the increase of the inhomogeneous broadening together with the weak
temperature dependence of the electron-impurity scattering. For samples with
high mobility, at low temperature where the scattering is determined by 
the electron-electron Coulomb scattering, a peak appears due to 
the crossover from the degenerate to non-degenerate limits. At high
 temperature where electrons are in non-degenerate
limit, the scattering is determined by
the electron-electron and electron-phonon scatterings. The calculations
show that for small well width when only the linear Dresselhaus term is
dominant, the temperature dependence of the scattering is stronger and
the spin relaxation time increases with temperature. However, for wide
quantum well, at certain temperature the cubic Dresselhaus term
becomes dominant. The fast increase of the inhomogeneous broadening
from the cubic term overcomes the effect from the scattering and the
spin relaxation time decreases with temperature. Jiang and
Wu further introduced strain to change the relative importance of the
linear and cubic D'yakonov-Perel' terms and showed the different temperature
dependences of the spin relaxation time \cite{PhysRevB.72.033311}. This
prediction has been realized experimentally by Holleitner et al.
in Ref.~\cite{1367-2630-9-9-342}, where they showed that in $n$-type two-dimensional InGaAs
channels, when the linear D'yakonov-Perel' term is suppressed, the spin relaxation
time decreases with temperature monotonically, as shown in
Fig.~\ref{fig5.4.2-6}. It is noted for the unstructured quantum well
in Fig.~\ref{fig5.4.2-6}, where the linear term is important, the spin
relaxation time increases with temperature. Similar findings were
reported by Malinowski et al. \cite{PhysRevB.62.13034}, who
measured the spin relaxation in intrinsic GaAs/Al$_x$Ga$_{1-x}$As
quantum wells with different well widths, for temperature higher than
80~K as shown in Fig.~\ref{fig5.4.2-7}(a). One clearly sees an
increase of the spin relaxation time with increasing temperature for small well
width and a decrease of the spin relaxation time for large well
width. Figure~\ref{fig5.4.2-7}(b) shows the theoretical calculation
based on the kinetic spin Bloch equations for three small well widths, where the lowest
subband approximation is valid for the small well widths (6 and 10~nm)
and barely valid for the 15~nm well width. All the relevant
scatterings, such as the electron-electron, electron-hole, electron-phonon
and electron-impurity scatterings are included in the computation. The
peaks in the cases of 10 and 15~nm well widths originate from the
competition of the linear and cubic D'yakonov-Perel' terms addressed above.

\begin{figure}[ht]
\begin{center}
\hspace{1 cm}\includegraphics[width=6.5cm]{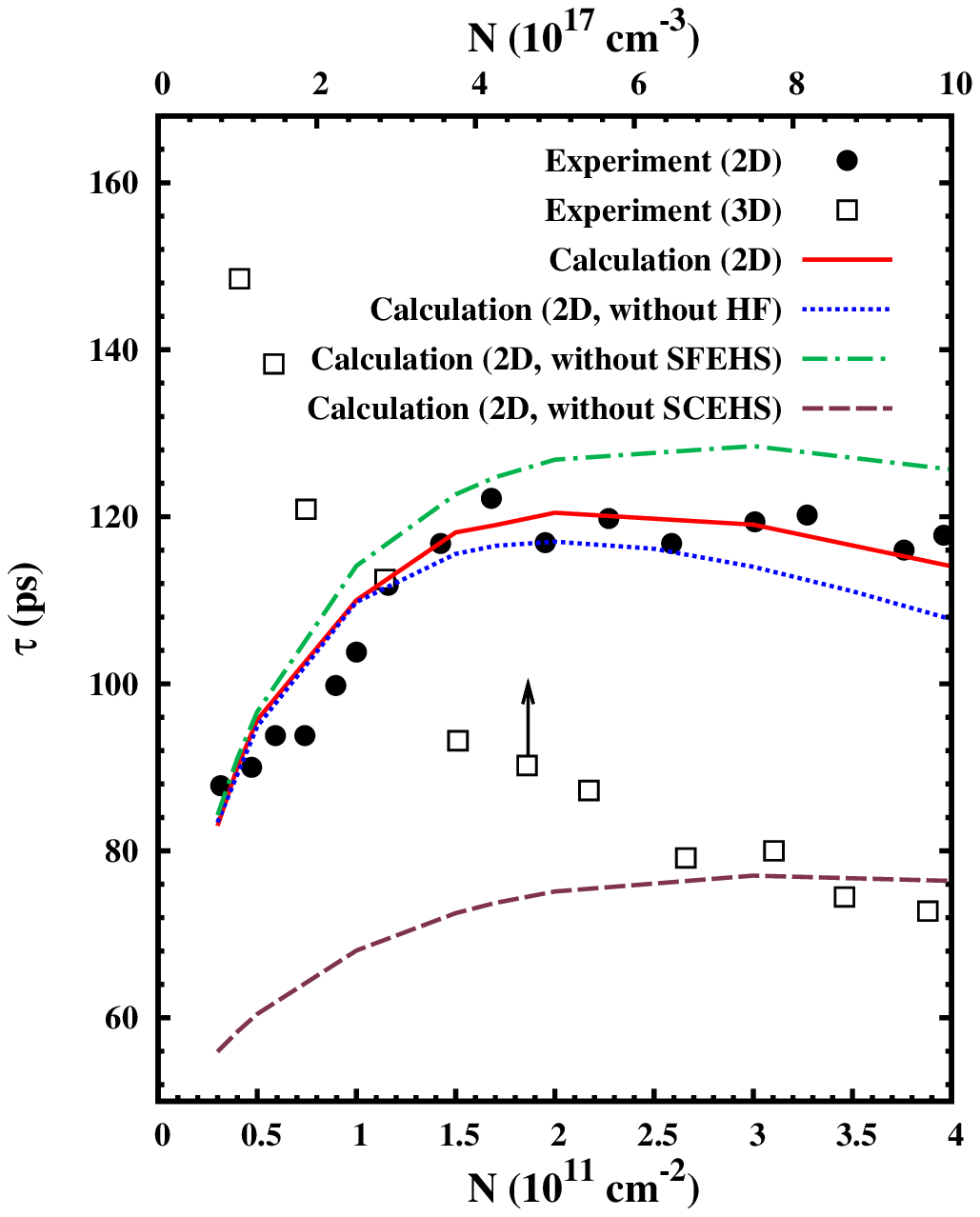}
\caption{ Carrier density dependence of electron
  spin relaxation time $\tau$ at room temperature. Dots: experimental
  data in a GaAs (001) quantum well (2D) with a width of 10~nm; 
open squares: experimental data in
  bulk material (3D). Solid curve: full theoretical
  calculation; chain curve: theoretical calculation without
  the spin-flip electron-hole scattering (SFEHS); dashed curve: theoretical calculation without
  the spin-conserving electron-hole scattering (SCEHS); dotted
  curve: theoretical calculation without the Coulomb Hartree-Fock term. Note the scale
  of the bulk data is on the top frame of the figure. From Teng et al. \cite{0295-5075-84-2-27006}.}
\label{fig5.4.2-8}
\end{center}
\end{figure}

Similar situation also happens in the density dependence of the spin
relaxation time. A peak in the density dependence of the spin
relaxation time was predicted theoretically and realized
experimentally \cite{0295-5075-84-2-27006}, as shown in Fig.~\ref{fig5.4.2-8} where
the spin relaxation time is plotted against the photoexcited carrier
density in a GaAs quantum well at room temperature. This peak can be easily understood from
the relation $\tau_s\propto[\langle\Omega^2({\bf
  k})\rangle\tau_p^\ast]^{-1}$, where $\tau_p^\ast$ should include the effect
from the Coulomb scattering
\cite{wu:epjb.18.373,Glazov:jetp.75.403,leyland:165309}. When the system is in the
non-degenerate limit, the average of $\langle\Omega^2({\bf k})\rangle$
is performed at the Boltzmann distribution and is therefore
independent of the carrier density. Consequently $\tau_s$ increases
with increasing carrier density as $\tau_p^\ast$ decreases with the 
density. However, when the carrier density is high enough and the
average should be performed using the Fermi distribution,
$\langle\Omega^2({\bf k})\rangle$ becomes density
dependent. Especially in the strong degenerate limit, $\langle k
\rangle\sim k_F\propto\sqrt{N}$ and $\langle\Omega^2({\bf k})\rangle$
becomes strongly density dependent. Moreover, the carrier-carrier
Coulomb scattering decreases with increasing density at strong
degenerate limit. As a result, the spin relaxation time decreases with
increasing carrier density. The peak position depends on the competition
between the inhomogeneous broadening and the scattering. Also the
linear and cubic D'yakonov-Perel' terms influence the peak position.

\begin{figure}[htb]
\begin{center}
\hspace{1 cm}\includegraphics[width=8.cm]{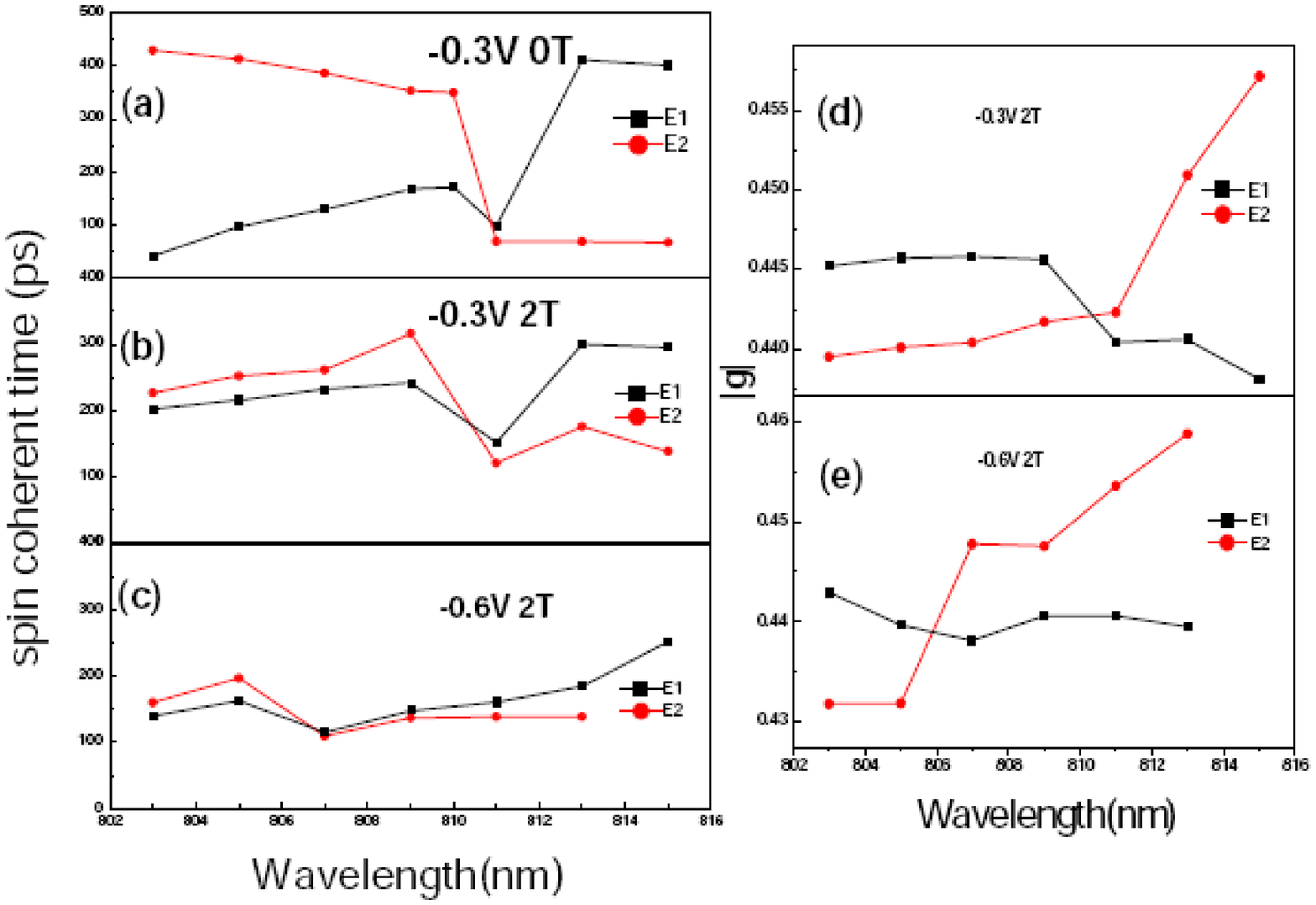}
\caption{ (a), (b) and (c) Measured spin coherence times for the 1st and 2nd
  subbands of a single-barrier GaAs/AlAs heterostructure, 
as the wavelength varies for three different conditions:
  $-0.3$~V, 0~T; $-0.3$~V, 2~T and $-0.6$~V, 2~T. The points are experimental
  results, and the curves are best fittings to the points. (d) and (e)
  Effective $g^\ast$ factors for the 1st and 2nd subbands plotted as a function of
  wavelength. From Zhang et al. \cite{0295-5075-83-4-47007}.}
\label{fig5.4.2-9}
\end{center}
\end{figure}

By including the intersubband scattering, especially the intersubband
Coulomb scattering, Weng and Wu studied the multisubband effect in
spin relaxation \cite{PhysRevB.70.195318}. It was discovered that although the
higher subband has a much larger inhomogeneous broadening, the spin
relaxation times of two subbands are identical thanks to the strong
intersubband Coulomb scattering \cite{PhysRevB.70.195318}. This prediction was
later verified experimentally by Zhang et al. \cite{0295-5075-83-4-47007}, who
studied the spin dynamics in a single-barrier heterostructure by
time-resolved Kerr rotation. By applying a gate voltage, they
effectively manipulated the confinement of the second subband and the
measured spin relaxation times of the first and second subbands are
almost identical at large gate voltage as shown in
Fig.~\ref{fig5.4.2-9}(c). The big deviations at small gate voltages
are because that the wavefunctions of the second subband are extended due
to the weak confinement and hence the intersubband Coulomb scattering
becomes weaker.

\begin{figure}[htb]
\begin{center}
\centerline{
\includegraphics[width=8.cm]{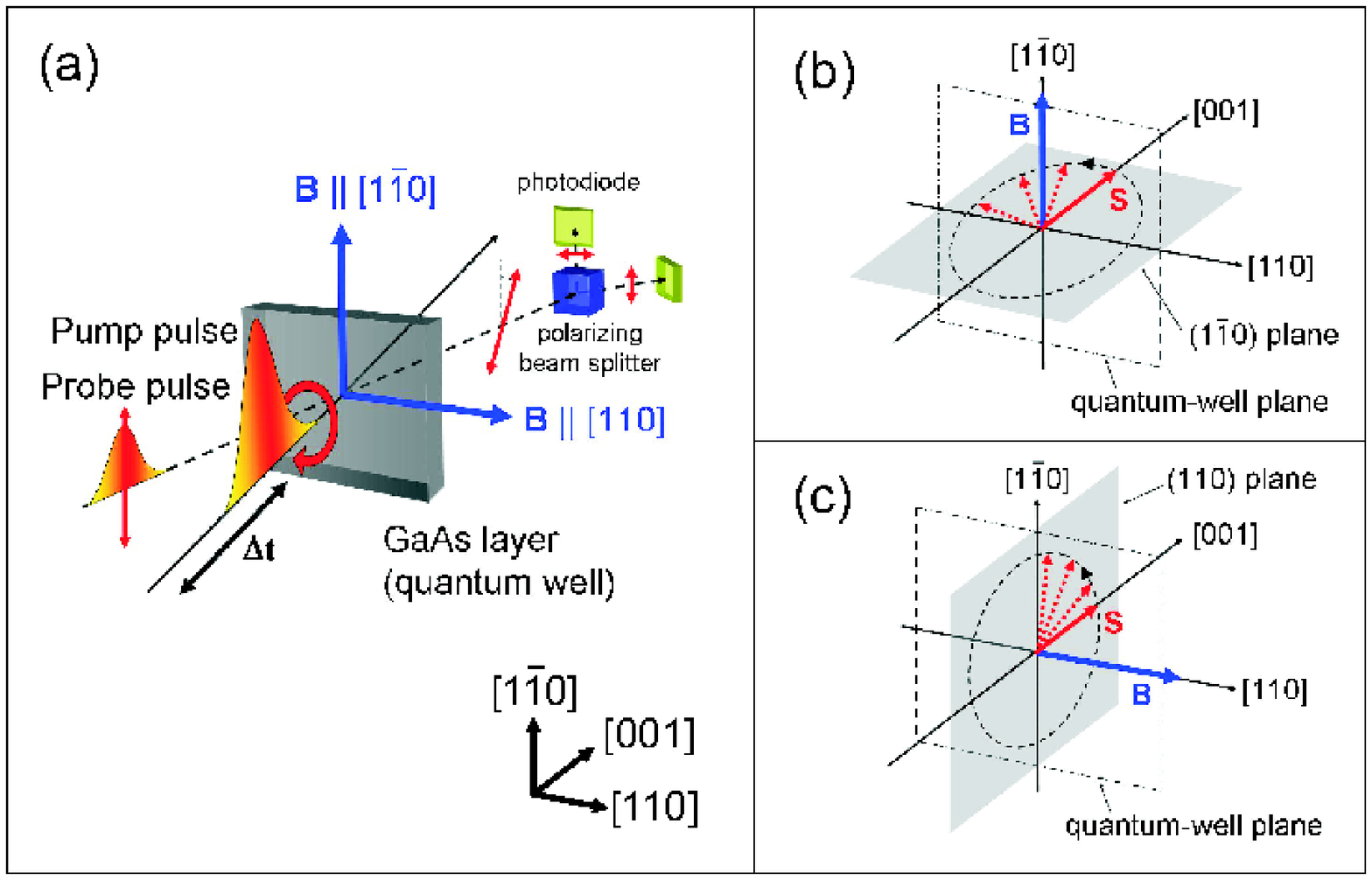}\hspace{0.5 cm}\includegraphics[width=4.cm]{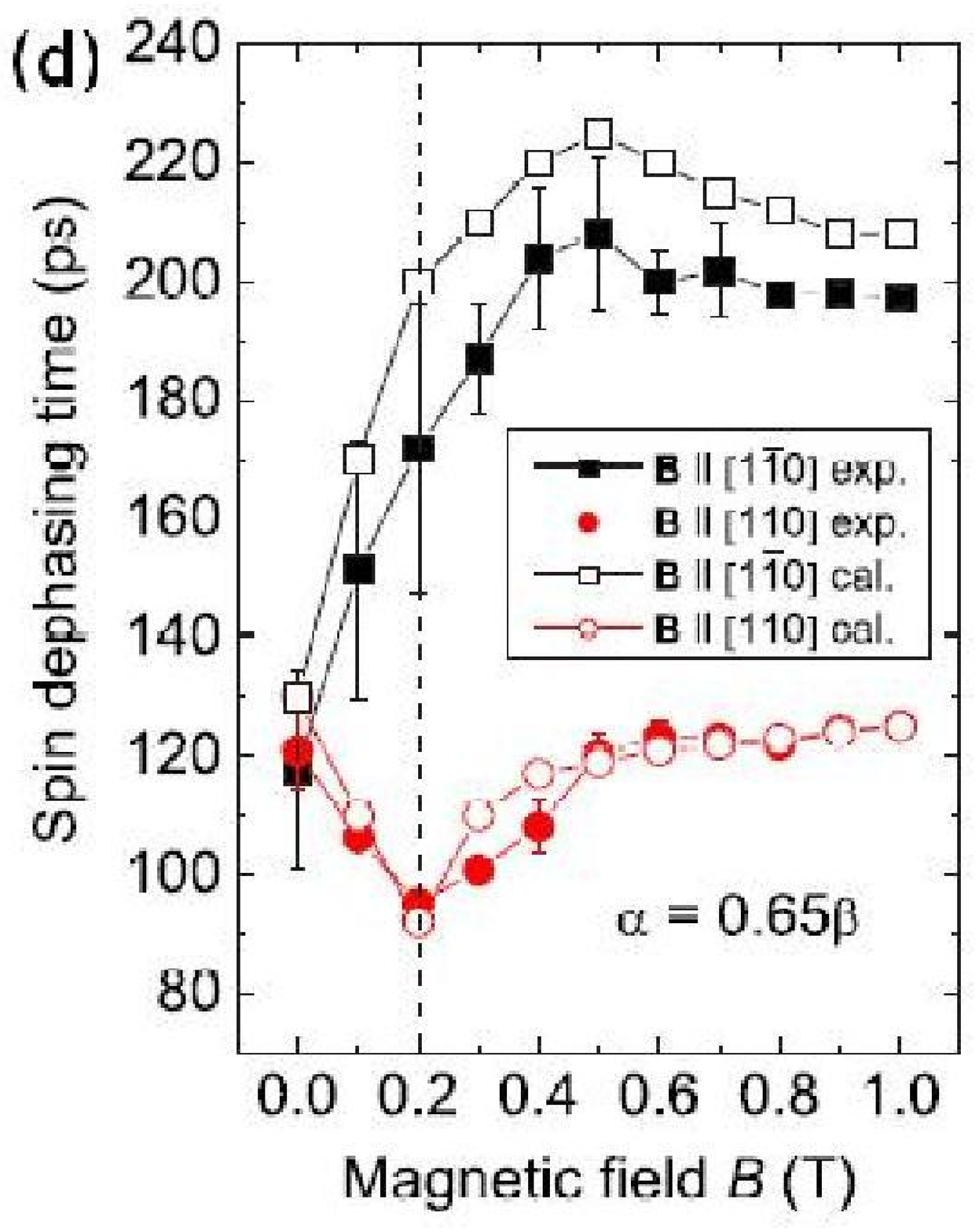}
}
\caption{ (a) Schematic of the time-resolved Faraday
  rotation experiment. In-plane magnetic fields are applied either in
  the [110] or [1$\bar{1}$0] directions. (b) Sketch of the
  precession of optically oriented spins about a [1$\bar{1}$0]
  in-plane magnetic field. (c) Sketch of the
  precession of optically oriented spins about a [110]
  in-plane magnetic field. (d) Comparison of the experimental (solid
  symbols) and theoretically calculated (open symbols) spin dephasing
  times for different in-plane magnetic-field directions
in GaAs (001) quantum well. $\alpha$ and
  $\beta$ are the Rashba and Dresselhaus spin-orbit coefficients,
  respectively. From Stich et al. \cite{stich:073309}.} 
\label{fig5.4.2-10}
\end{center}
\end{figure}

Finally we address spin relaxation in systems with both the
Dresselhaus and Rashba spin-orbit couplings. It has been first pointed
out by Averkiev and Golub \cite{PhysRevB.60.15582} that for (001) quantum
well with identical Dresselhaus and Rashba spin-orbit coupling
strengths, when the cubic Dresselhaus term is ignored, the effective
magnetic field becomes fixed and is aligned along the (110) or
($\bar{1}$10) direction depending on the sign of the Rashba
field. Therefore one obtains infinite spin lifetime if the spin
polarization is parallel to the effective magnetic field direction. In reality, due to
the presence of the cubic Dresselhaus term and/or the difference
between the Dresselhaus and Rashba spin-orbit coupling strengths, the spin life
time along the (110) or ($\bar{1}$10) direction is still finite. But
there is strong anisotropy of the spin lifetime along different spin
polarizations. Cheng and Wu studied this anisotropy under identical
linear Dresselhaus and Rashba coupling strengths, but with the cubic
Dresselhaus term included \cite{cheng:083704}. Stich et al. applied
a magnetic field parallel to the (110) and ($\bar{1}$10) directions and
obtained large magnetoanisotropy of electron spin dephasing in a high
mobility (001) GaAs quantum well \cite{stich:073309}. The initial spin
polarization was obtained by optical pumping and is therefore along
the growth direction of the quantum well. The experimental setup is
illustrated in Fig.~\ref{fig5.4.2-10}(a). Due to the mixing of the
different anisotropic spin orientations by the magnetic field, they
observed different magnetic field dependences of the spin dephasing
time along the (110) and ($\bar{1}$10)
directions, as shown in Fig.~\ref{fig5.4.2-10}(d). The maximum and
minimum are determined by the relative strengths of the Dresselhaus
and Rashba terms. It is also seen that calculation based on the fully
microscopic kinetic spin Bloch equations can well reproduce the
experimental findings. 

\subsubsection{Spin relaxation and dephasing in $n$-type (001) GaAs
  quantum wells far away from the equilibrium}
\label{sec5.4.3}
Due to the full account of the Coulomb scattering, the kinetic spin
Bloch equation approach can be applied to study spin system far away
from the equilibrium. By so called far away from the equilibrium, one
refers to the spin dynamics with large spin polarization and/or with
large in-plane electric field where the hot-electron effect becomes
significant. 

\begin{figure}[htb]
\begin{center}
\centerline{
\includegraphics[width=6.cm]{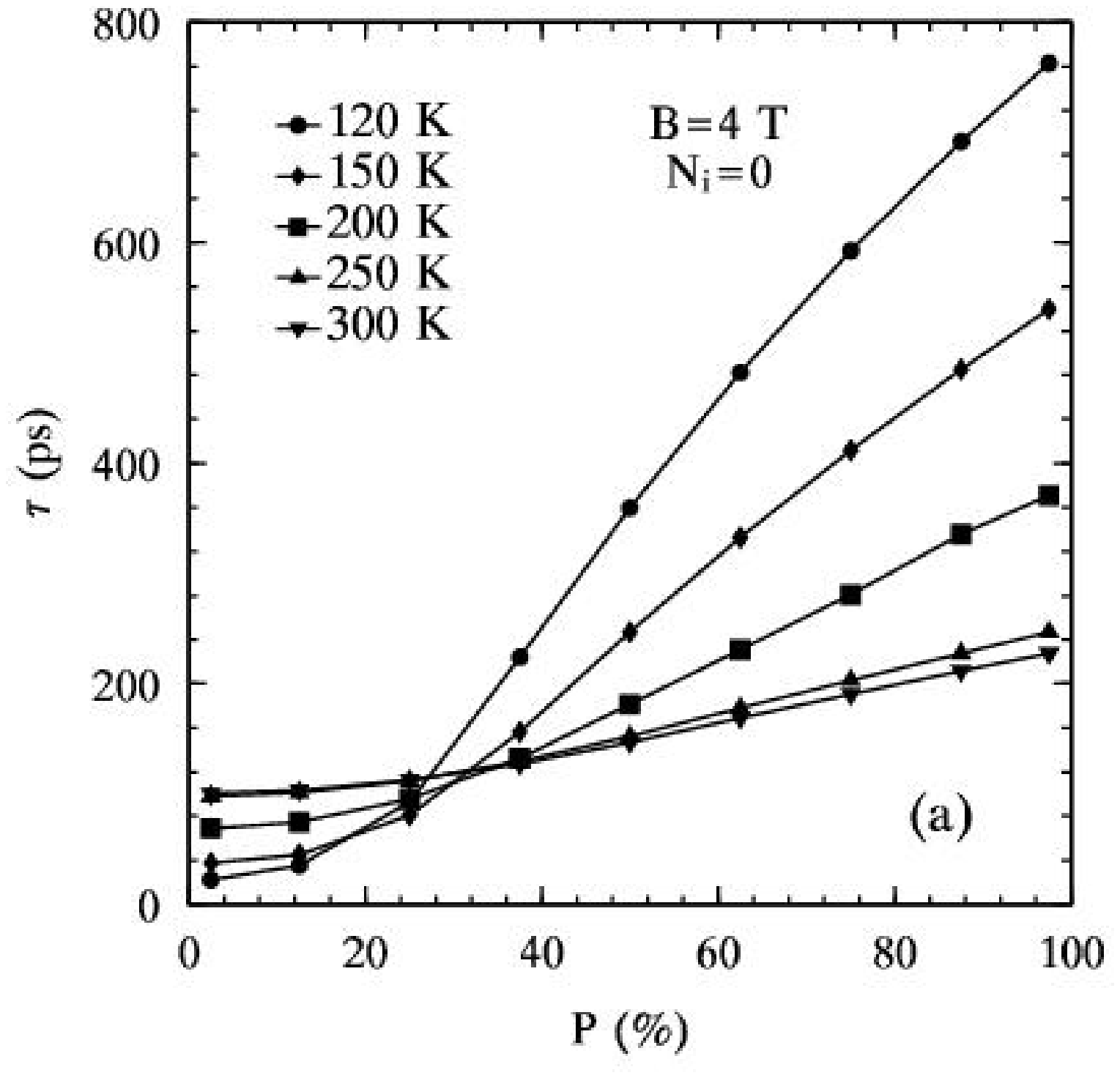}\hspace{0.5 cm}\includegraphics[width=6.cm]{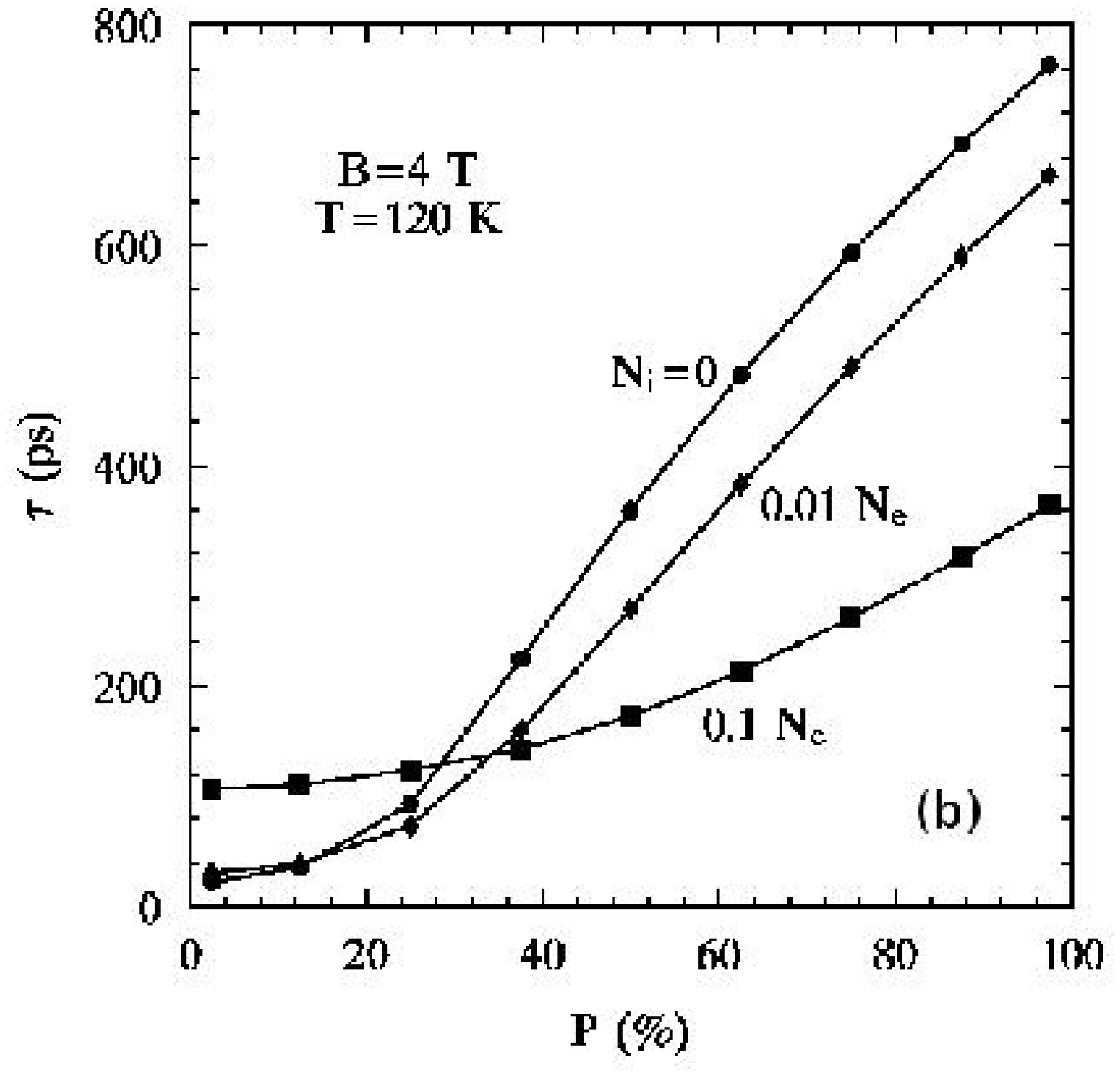}
}
\caption{(a) Spin dephasing time $\tau$ {\sl vs.} the initial spin
  polarization $P$ of electrons in GaAs (001) quantum well
 at different temperatures. The impurity density
  $N_i=0$. (b) Spin dephasing time $\tau$ {\sl vs.} the initial spin
  polarization $P$ at $T=120$~K for different impurity levels. Dots ($\bullet$): $N_i=0$;
    Diamonds ($\blacklozenge$): $N_i=0.01N_e$;
    Squares ($\blacksquare$): $N_i=0.1N_e$. The curves are plotted for
    the aid of eyes. From Weng and Wu \cite{PhysRevB.68.075312}.} 
\label{fig5.4.3-1}
\end{center}
\end{figure}

\begin{figure}[htb]
\begin{center}
\includegraphics[width=8.cm]{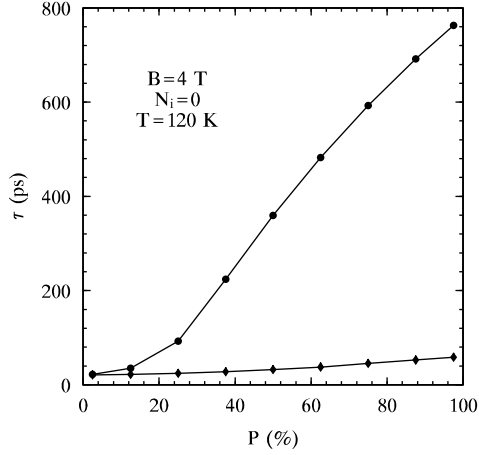}
\caption{Spin dephasing time $\tau$ {\sl vs.} the initial spin
    polarization $P$ for $T=120$~K and $N_i=0$
 in GaAs (001) quantum well.
    Dots ($\bullet$): With the longitudinal component of the Hartree-Fock term
    included; Diamonds ($\blacklozenge$): Without the longitudinal
    component of the Hartree-Fock term. The curves are plotted for the aid of eyes. From Weng and Wu \cite{PhysRevB.68.075312}.} 
\label{fig5.4.3-2}
\end{center}
\end{figure}

\begin{figure}[htb]
\begin{center}
\centerline{
\begin{minipage}[]{13cm}
\parbox[t]{4cm}{
  \includegraphics[width=4cm]{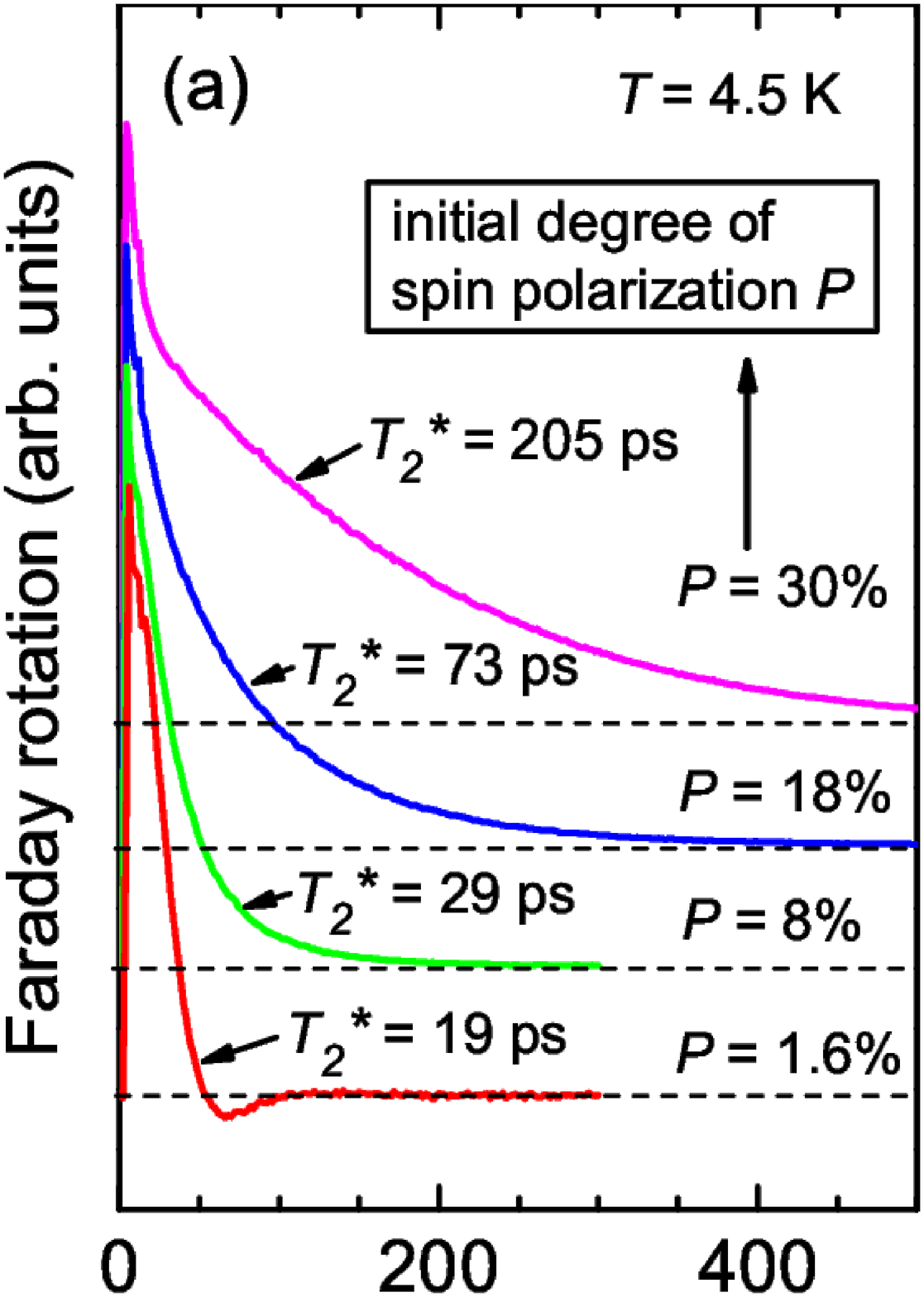}}
\parbox[t]{8cm}{
\vskip -5.5 cm
  \includegraphics[width=8.cm]{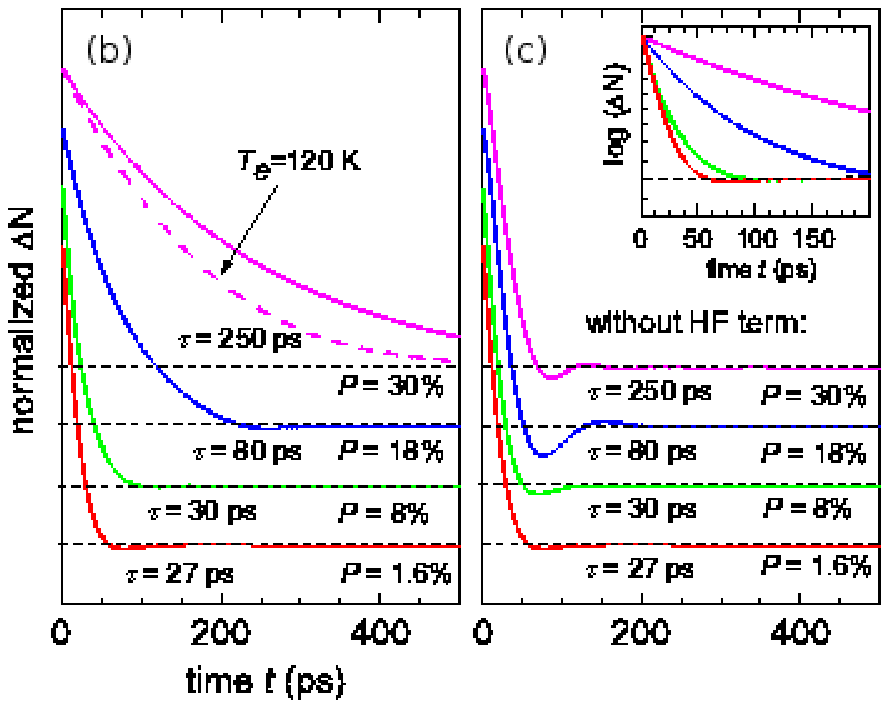}
}
\end{minipage}
}
\caption{ (a) Normalized time-resolved Faraday rotation traces for different
degrees of initial electron spin polarization, $P$ in GaAs (001) 
quantum well. The total
densities of electrons, $n_{tot}=n_e+n_{ph}^{tot}$, are 2.19, 2.66,
3.83, 8.39 (in units of 10$^{11}$ cm$^{-2}$) for $P=1.6\% $, $8\%$,
$18\%$, $30\%$, respectively. (b) Calculated spin decay curves for the
experimental parameters, like initial spin polarization, total
electron densities, electron mobility, and temperature $T=4.5$ K
(solid lines). In the calculation, a phenomenological decay time
 is incorporated as a single fitting
parameter $\tau$ (see text). The dashed curve
 for $P=30\%$ is calculated for a hot electron temperature
of $T_e=120$\ K. (c) Same as (b) but calculated without the Hartree-Fock term.
In particular for large $P$ the decay is much faster than in the
experiments [cf.~(a)]. The inset shows the data, displayed in
(b), in a semilogarithmic plot. At low $P$ the zero-field
oscillations are superimposed.  From Stich et al. \cite{stich:176401}.} 
\label{fig5.4.3-3}
\end{center}
\end{figure}

\begin{figure}[htb]
\begin{center}
\includegraphics[width=7.cm]{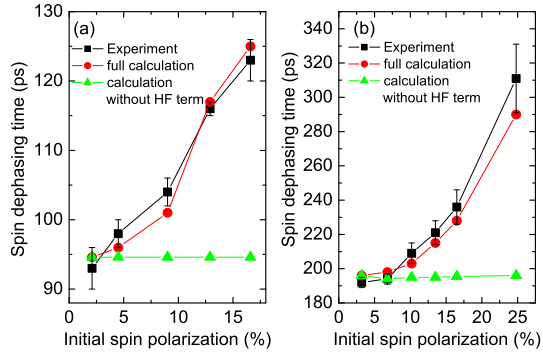}
\caption{GaAs (001) quantum well: (a) The spin dephasing times as a function of initial spin
polarization for constant, \emph{low} excitation density and
variable polarization degree of the pump beam. The measured spin
dephasing times are compared to calculations with and without the
Hartree-Fock (HF) term, showing its importance. (b) The spin dephasing times measured and calculated
for constant, \emph{high} excitation density and variable
polarization degree.  From Stich et al. \cite{stich:205301}.} 
\label{fig5.4.3-4}
\end{center}
\end{figure}

\begin{figure}[htb]
\begin{center}
\centerline{
\includegraphics[width=6.5cm]{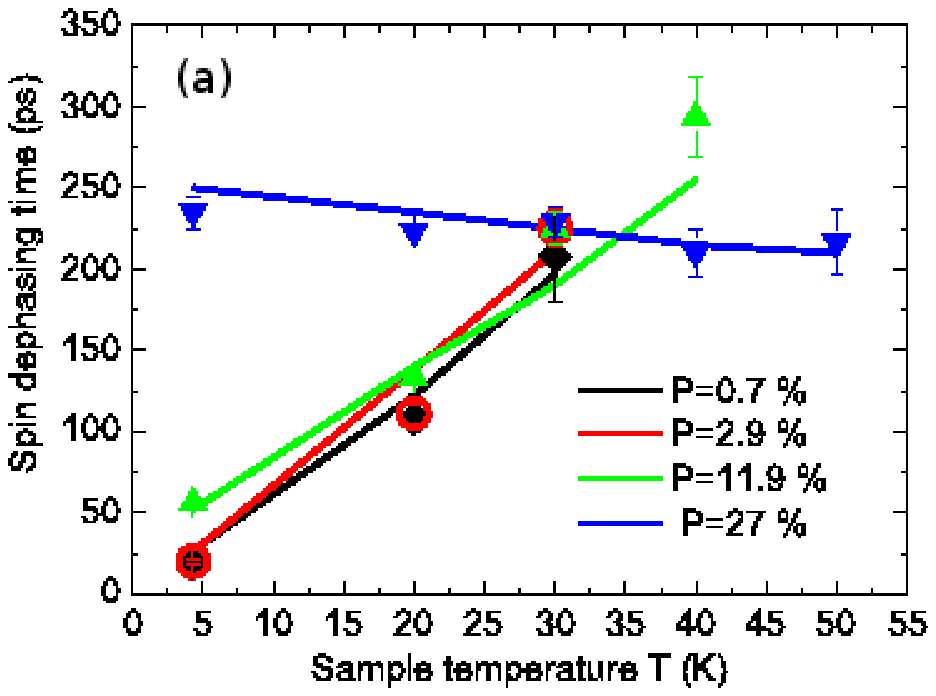}\hspace{0.5 cm}\includegraphics[width=6.2cm]{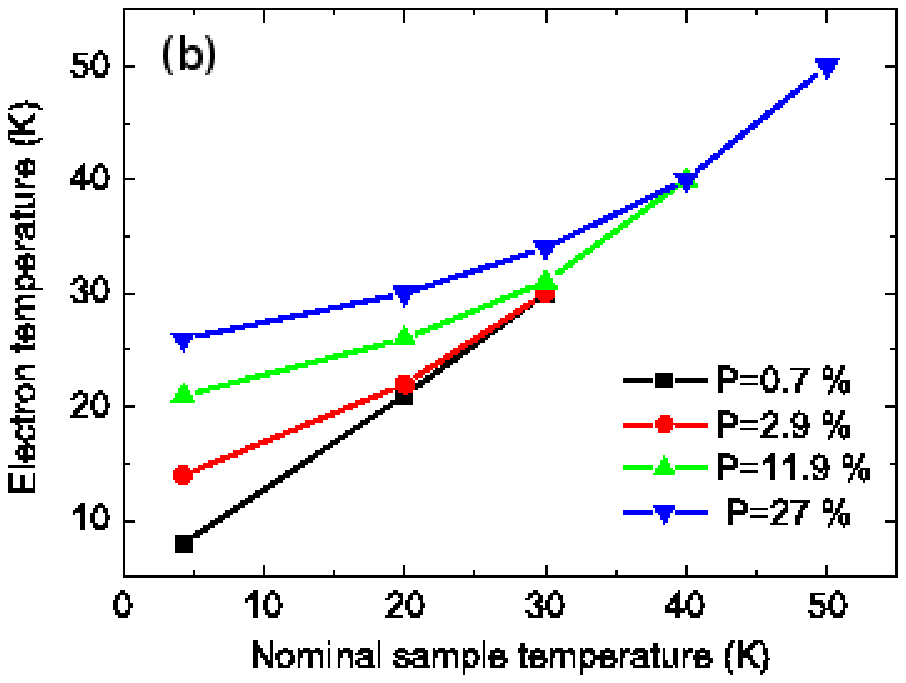}
}
\caption{ GaAs (001) quantum well: 
(a) Spin dephasing time as a function of sample temperature, for
 different initial spin polarizations. The measured data points are represented by
 solid points, while the calculated data are represented
 by  lines of the same colour. (b) Electron temperature determined
 from intensity-dependent photoilluminance measurements as
 a function of the nominal sample temperature, for different pump beam fluence
 and initial spin polarization, under experimental conditions corresponding to
 the measurements shown in (a). The measured data points are represented by
 solid points, while the curves serve as guide to the eye. From Stich
 et al. \cite{stich:205301}.} 
\label{fig5.4.3-5}
\end{center}
\end{figure}

\begin{figure}[htb]
\begin{center}
\includegraphics[width=12.cm]{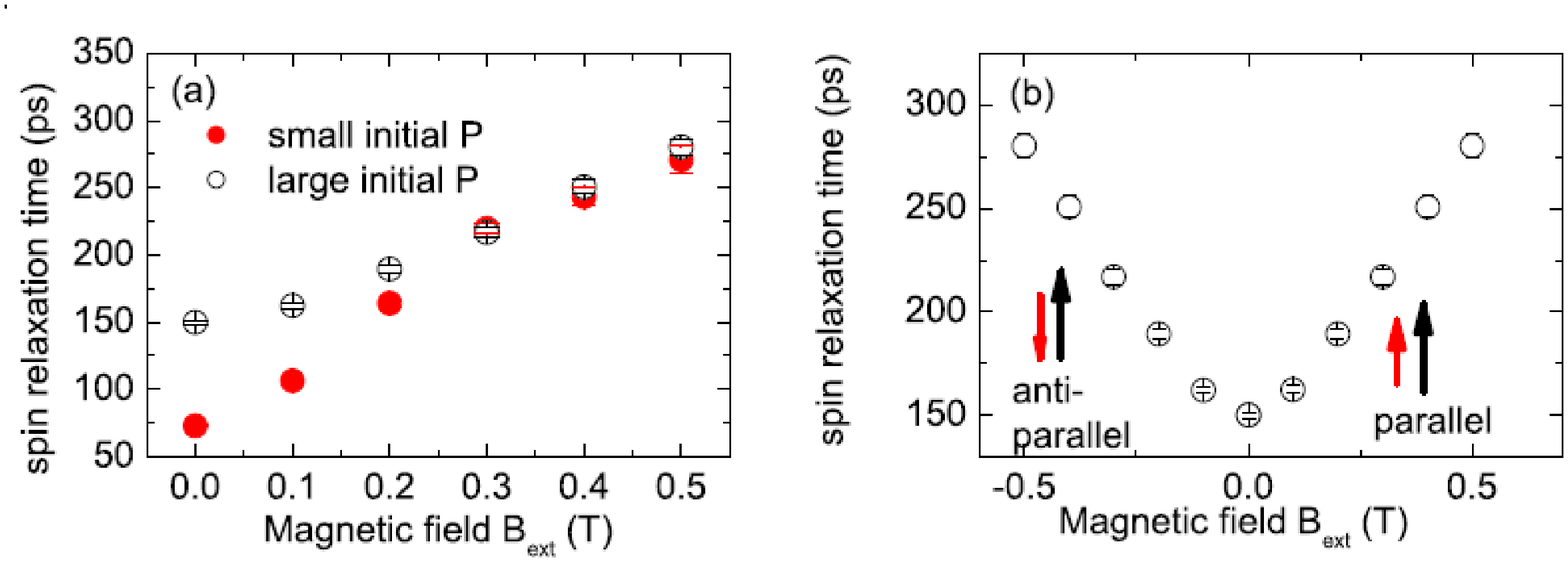}
\caption{ (a) Spin dephasing times as a function of an
  external magnetic field perpendicular to the quantum well plane for
  small and large initial spin polarization in GaAs (001) quantum well. 
(b) Same as (a) for large
  initial spin polarization and both polarities of the external
  magnetic field. From Korn et al. \cite{korn}.} 
\label{fig5.4.3-6}
\end{center}
\end{figure}

Weng and Wu first studied the spin relaxation with large initial spin
polarization \cite{PhysRevB.68.075312} and discovered marked increase of the spin
relaxation/dephasing time with increasing spin polarization as shown in
Fig.~\ref{fig5.4.3-1}(a). It is noted that the plotted spin dephasing
time is the inverse of the spin dephasing rate. Therefore one expects
a markedly increased total spin lifetime. This enhancement is more
pronounced in samples with larger mobility, as shown in
Fig.~\ref{fig5.4.3-1}(b). The physics leading to such an increase was
identified from the Coulomb Hartree-Fock term. In the presence of the
spin polarization, the Coulomb Hartree-Fock term provides an effective
magnetic field along the $z$-axis:
\begin{equation}
B_z^{\rm HF}=\frac{1}{g\mu_B}\sum_{\bf q}V_{\bf q}(f_{{\bf
    k+q}\frac{1}{2}}-f_{{\bf k+q}-\frac{1}{2}}).
\label{eq5.4.3-1}
\end{equation}
This effective magnetic field can be as large as 20~T and effectively
blocks the spin precession as the energies of the spin-up and -down
electrons are no longer detuned. This can be seen clearly from
Fig.~\ref{fig5.4.3-2} that by removing the longitudinal component of
the Hartree-Fock term, the polarization dependence of the spin
relaxation time becomes pretty mild. It is noted that unlike a real
magnetic field which breaks the time-reversal symmetry, the Coulomb
interaction does not. This can be seen from the fact that the
Hartree-Fock term cancels each other after performing the summation over all
the {\bf k} states [Eq.~(\ref{eq5.2-6})]. Therefore the spin relaxation time
with a large effective magnetic field is still finite. As the
effective magnetic field decreases with temperature, Weng and 
Wu pointed out that the spin relaxation time should decrease with
increasing temperature at large spin polarization, in contrast to the
case with small spin polarization [as shown in
Fig.~\ref{fig5.4.2-4}(b)].

These predictions have been realized experimentally by Stich et al. \cite{stich:176401,stich:205301} and Zhang et al.
\cite{0295-5075-83-4-47006}. By changing the intensity of the circularly polarized
lasers, Stich et al. measured the spin dephasing time in a high
mobility $n$-type GaAs quantum well as a function of initial spin
polarization as shown in Fig.~\ref{fig5.4.3-3}(a). Indeed they
observed an increase of the spin dephasing time with the increased
spin polarization, and the theoretical calculation based on the
kinetic spin Bloch equations nicely reproduced
the experimental findings when the Hartree-Fock term was included
[Fig.~\ref{fig5.4.3-3}(b)]. It was also shown that when the
Hartree-Fock term is removed, one does not see any increase of the
spin dephasing time [Fig.~\ref{fig5.4.3-3}(c)]. Later, they further
improved the experiment by replacing the circular-polarized laser
pumping with the elliptic polarized laser pumping. By doing
so, they were able to vary the spin polarization without changing the
carrier density. Figure.~\ref{fig5.4.3-4} shows the measured
spin dephasing times as function of initial spin polarization under
two fixed pumping intensities, together with the theoretical calculations with
and without the Coulomb Hartree-Fock term. Again the spin dephasing
time increases with the initial spin polarization as predicted and the
theoretical calculations with the Hartree-Fock term are in good
agreement with the experimental data. Moreover, Stich et al.
also confirmed the prediction of the temperature dependences of the
spin dephasing time at low
and high spin polarizations \cite{PhysRevB.68.075312}. Figure~\ref{fig5.4.3-5}(a) shows
the measured temperature dependences of the spin dephasing time at
different initial spin polarizations. As predicted, the spin
dephasing time increases with increasing temperature at small spin
polarization but decreases at large spin polarization. The theoretical
calculations also nicely reproduced the experimental data. The
hot-electron temperatures in the calculation were taken from the
experiment [Fig.~\ref{fig5.4.3-5}(b)]. The effective magnetic field
from the Hartree-Fock term has been measured by Zhang et al.
\cite{0295-5075-83-4-47006} from the sign switch of the Kerr signal and the phase reversal
of Larmor precessions with a bias voltage in a GaAs heterostructure. 

Korn et al. \cite{korn} also estimated the average effect by applying an
external magnetic field in the Faraday configuration, as shown in
Fig.~\ref{fig5.4.3-6}(a) for the same sample reported above
\cite{stich:176401,stich:205301}. They compared the spin dephasing times of both large and
small spin polarizations as function of
external magnetic field. Due to the effective magnetic field from the
Hartree-Fock term, the spin relaxation times are different under small
external magnetic field but become identical when the magnetic field
becomes large enough. From the merging point, they estimated the mean
value of the effective magnetic field is below 0.4~T. They further
showed that this effective magnetic field from the Hartree-Fock term
cannot be compensated by the external magnetic field, because it does
not break the time-reversal symmetry and is therefore not a genuine
magnetic field, as said above. This can be seen from
Fig.~\ref{fig5.4.3-6}(b) that the spin relaxation time at large spin
polarization shows identical external magnetic field dependences when
the magnetic field is parallel or antiparallel to the growth
direction.

The spin dynamics in the presence of a high in-plane electric field
was first studied by Weng et al. \cite{PhysRevB.69.245320} in GaAs quantum well
with only the lowest subband by solving the kinetic spin Bloch
equations. To avoid the ``runaway'' effect \cite{Dmitriev2000565,dmitriev:3793}, the electric
field was calculated upto 1~kV/cm. Then Weng and Wu further
introduced the second subband into the model and the in-plane electric
field was increased upto 3~kV/cm \cite{PhysRevB.70.195318}. Zhang et al. 
included $L$ valley and the electric field was further increased upto
7~kV/cm \cite{zhang:235323}. 

The effect of in-plane electric field to the spin relaxation in system
with strain was investigated by Jiang and Wu \cite{PhysRevB.72.033311}. Zhou
et al. also investigated the electric-field effect at low
lattice temperatures \cite{zhou:045305}. 

\begin{figure}[htb]
\begin{center}
\centerline{
\includegraphics[width=6.5cm,height=4.5cm]{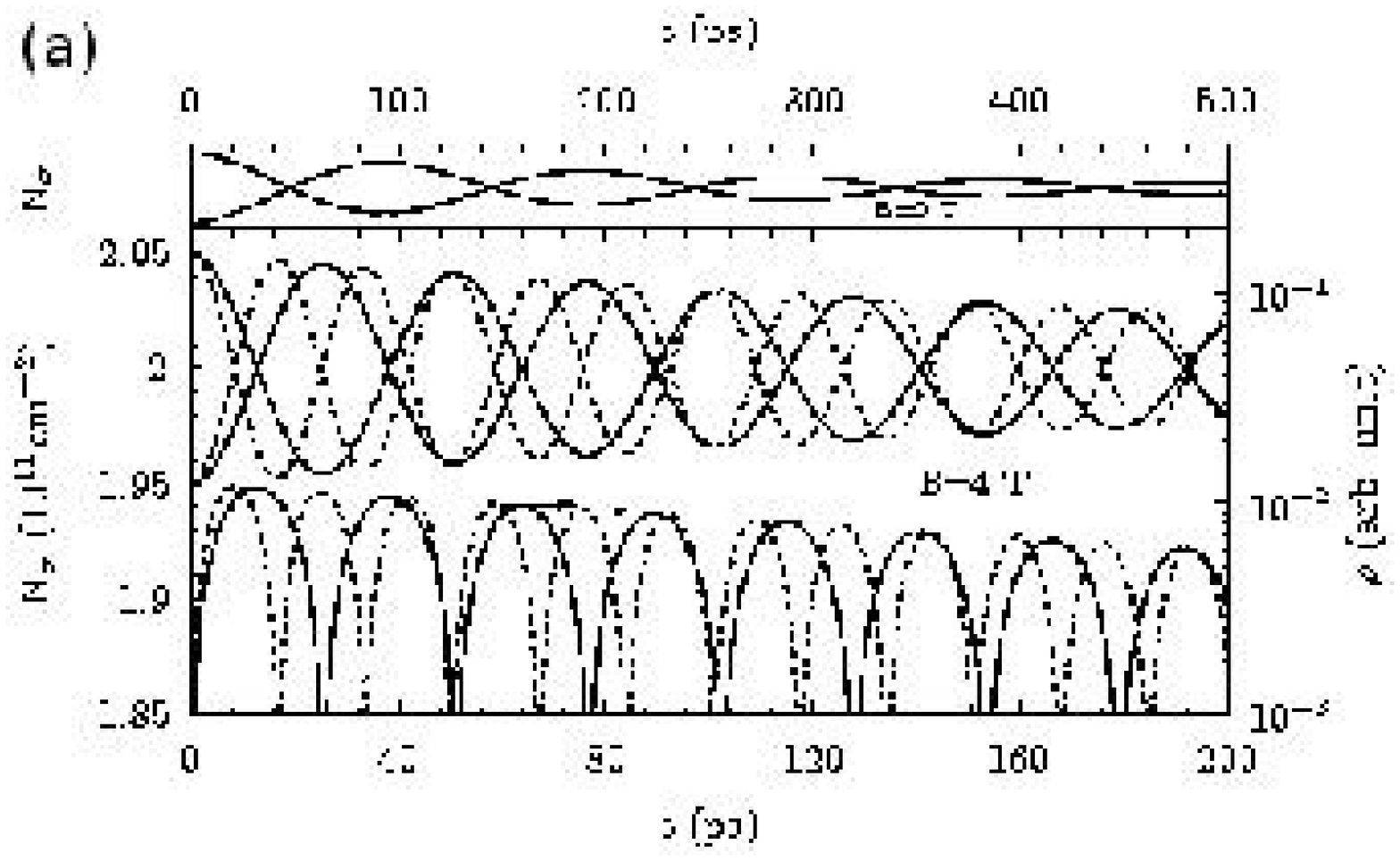}\hspace{0.5 cm}\includegraphics[width=6.2cm,height=4.5cm]{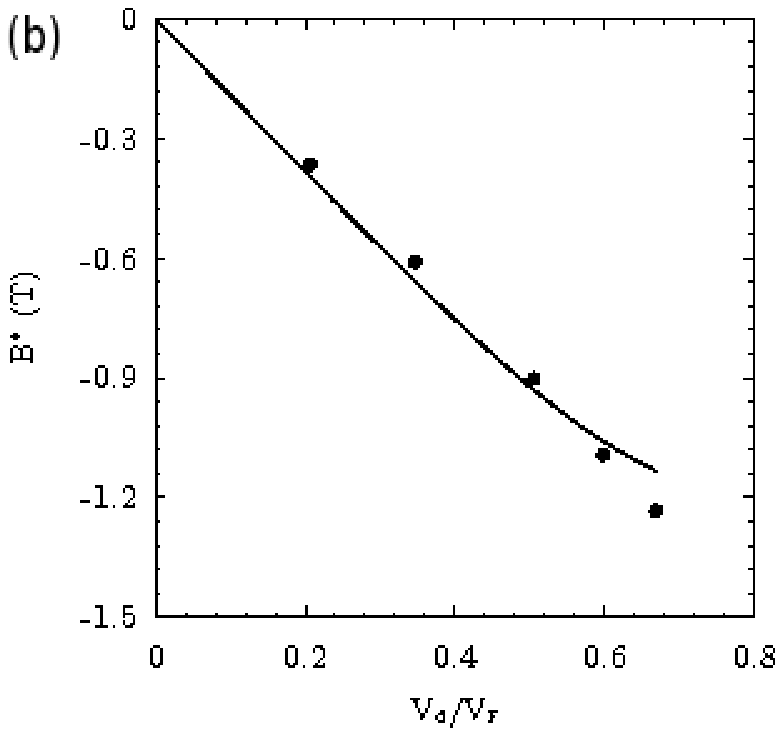}
}
\caption{(a) Electron densities of spin-up and -down electrons 
and the incoherently summed spin coherence $\rho$ {\em vs}. time $t$ for 
 GaAs (001) quantum well (the well
  width is 15~nm) with initial spin polarization $P=2.5$~\% under different electric fields
  $E=0.5$~kV/cm (solid curves) and $E=-0.5$~kV/cm (dashed curves).
  Top panel: $B=0$\ T; bottom panel: $B=4$\ T.
  Note the scale of the spin coherence is on the right side of the
  figure and the scale of the top panel is different from that
of the bottom one. (b) Net effective magnetic field $B^{\ast}$
    from the D'yakonov-Perel' term {\em vs}. the drift velocity $V_d$ at $T=120$~K
 with impurity density $N_i=0$. The solid curve is  the corresponding
 result from Eq.~(\ref{eq5.4.3-3}). From Weng et al. \cite{PhysRevB.69.245320}.} 
\label{fig5.4.3-7}
\end{center}
\end{figure}

The in-plane electric field leads to two effects. (i) It shifts the
center-of-mass of electrons to ${\bf k}_d=m^\ast{\bf v}_d=m^\ast\mu{\bf E}$ with
$\mu$ representing the mobility, which further induces an effective
magnetic field via the D'yakonov-Perel' term \cite{PhysRevB.69.245320}. The induced
effective magnetic field can be estimated by 
\begin{equation}
{\bf B}_{\rm tot}={\bf B}+{\bf B}^\ast={\bf B}+\frac{1}{g\mu_B}\int d{\bf k}(f_{{\bf
    k}\frac{1}{2}}-f_{{\bf k}-\frac{1}{2}}){\bf \Omega}({\bf k})\Big/\int d{\bf k}(f_{{\bf
    k}\frac{1}{2}}-f_{{\bf k}-\frac{1}{2}}).
\label{eq5.4.3-2}
\end{equation}
(ii) The in-plane electric field also leads to the hot-electron effect
\cite{conwell}. By taking the electron distribution function as the
drifted Fermi function $f_{{\bf k}\sigma}=\{\exp[(({\bf k}-m^\ast{\bf
  v}_d)^2/2m^\ast-\mu_\sigma)/k_BT_e]+1\}^{-1}$ in the case with only
the lowest subband included, the effective magnetic field for small
spin polarization can be roughly estimated as \cite{PhysRevB.69.245320}
\begin{equation}
B^\ast=\gamma_{\rm D}{m^\ast}^2v_d\{E_f/[2(1-e^{-E_f/k_BT_e})]-E_c\}/g\mu_B,
\label{eq5.4.3-3}
\end{equation}
with $E_f$ and $E_c$, the Fermi and confinement energies of the
quantum well, respectively. $T_e$ is the hot-electron temperature. Figure
\ref{fig5.4.3-7}(a) shows the spin precession in the presence of a 
high electric field in a 15~nm quantum well at $T=120$~K. One finds
that the spin precesses even in the absence of any external magnetic
field and the spin precession frequency changes with the direction of
the electric field in the presence of an external magnetic field. The
effective magnetic field deduced from the spin precessions in
Fig.~\ref{fig5.4.3-7}(a) is plotted in Fig.~\ref{fig5.4.3-7}(b), which
is in good agreement with the corresponding result from
Eq.~(\ref{eq5.4.3-3}).

\begin{figure}[htb]
\begin{center}
\centerline{
\includegraphics[width=5cm,height=4cm]{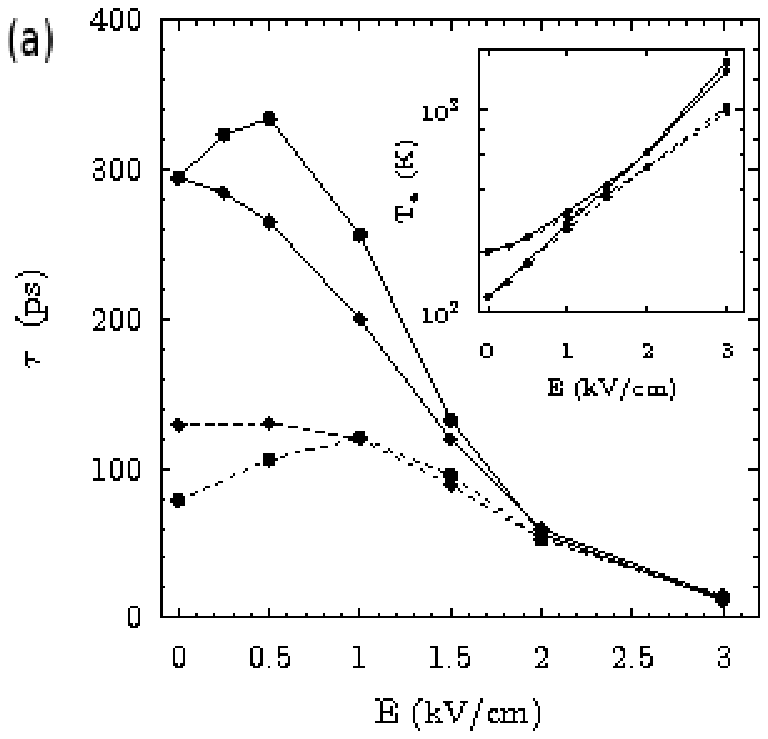}\hspace{0.5 cm}\includegraphics[width=5.5cm,height=4.4cm]{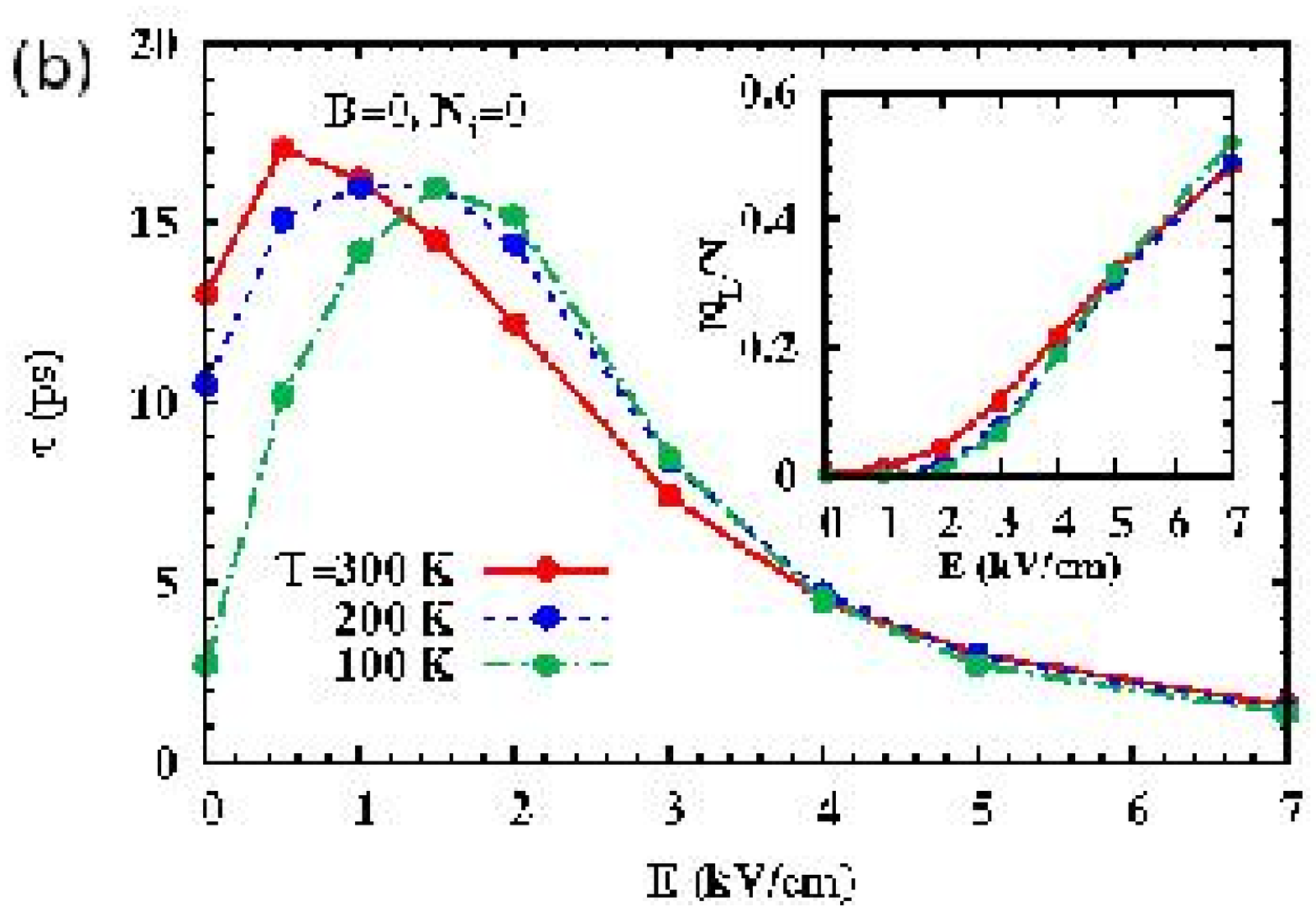}
}
\caption{GaAs (001) quantum wells:
 (a) Spin dephasing time {\em vs}. the applied electric
    field $E$ at different temperatures and  well widths:
    $\bullet$, $T=120$~K;  $\blacklozenge$, $T=200$~K;
Solid curves, $a=17.8$~nm; Dashed
 curves, $a=12.7$~nm. Inset: The corresponding
    electron temperature $T_e$ as a function of the electric
    field. From Weng and Wu \cite{PhysRevB.70.195318}. (b) Spin relaxation time $\tau$ {\em vs}. electric filed $E$ when
temperature $T$=300 (solid curves), 200 (dashed curves) and
100~K (chain curves) with $B=N_i=0$. Inset: fraction of electrons in $L$
      valleys against electric field. From Zhang et al. \cite{zhang:235323}.} 
\label{fig5.4.3-8}
\end{center}
\end{figure}

The spin relaxation time can be effectively manipulated by the
in-plane electric field. When only the lowest subband is considered,
the electric field influences the spin relaxation via concurrent
effects of the increase of the inhomogeneous broadening by driving
electrons to higher momentum states and the increase/decrease
(depending on the type of scattering and also the
degenerate/nondegenerate limit) of scattering from the hot-electron
temperature. Weng et al. reported rich electric field dependence
of the spin relaxation time under various conditions
\cite{PhysRevB.69.245320}. When the electric field is high enough so that
higher subbands and/or valleys are involved, due to the different
spin-orbit coupling strengths and/or effective masses in different
subbands and valleys, the spin relaxation time can be manipulated more
effectively \cite{PhysRevB.70.195318,zhang:235323}. This can be seen from
Fig.~\ref{fig5.4.3-8}(a) and (b) where spin relaxation times are
plotted against the in-plane electric field in GaAs quantum wells, with
the second subband and $L$ valley being populated at high electric
field, respectively. The inhomogeneous broadening comes from the
Dresselhaus terms. In $\Gamma$ valley, it reads
\begin{equation}
{\bf h}_{\Gamma}({\bf k}_{\Gamma})=\frac{\gamma}{2}\big(k_{\Gamma x}(k_{\Gamma y}^{2}-\langle
k_{\Gamma z}^{2}\rangle),k_{\Gamma y}(\langle k_{\Gamma
  z}^{2}\rangle-k_{\Gamma x}^{2}), \langle k_{\Gamma z}\rangle(k_{\Gamma
  x}^{2}-k_{\Gamma y}^{2})\big),
\label{eq5.4.3-4}
\end{equation}
and in $L$-valley, it reads \cite{aronov,Fu20082890}
\begin{equation}
{\bf h}_{L_{i}}({\bf
  k}_{L_{i}})=\beta\big(k_{L_{i}x},k_{L_{i}y},\langle
k_{L_{i}z}\rangle\big)\times\hat{{\bf n}}_{i}. 
\label{eq5.4.3-5}
\end{equation}
Here $\hat{\bf n}_i$ is the unit vector along the longitudinal
principle axis of $L_i$ valley. Note the coordinate system of above
equations has been given in Ref.~\cite{Ivchenkoref}. ${\bf k}_\Gamma={\bf k}$
and ${\bf k}_{L_i}={\bf k}-{\bf K}_{L_i}^0$ with ${\bf
  K}_{L_i}^0=\frac{\pi}{a_0}(1,\pm 1,\pm 1)$. $a_0$ is the lattice
constant. $\langle k_{\lambda z}\rangle$ ($\langle k_{\lambda
  z}^2\rangle$) represents the average of the operator $-i\partial/\partial
z-K_{\lambda z}^{0}$ [$(-i\partial/\partial z-{K}_{\lambda
  z}^{0})^{2}$] over the electron states in $\lambda$-valley. Under the
infinite-depth assumption, $\langle n|k_{\lambda
  z}^{2}|n\rangle=\frac{n^2\pi^2}{a^2}$ with $n$, the subband index and
$a$, the well width. Therefore, when the linear-$k$ terms are dominant,
the inhomogeneous broadening of the second subband in ${\bf
  h}_\Gamma({\bf k}_\Gamma)$ is four times as large as the first
subband. Also a tight-binding calculation by Fu et al. \cite{Fu20082890}
indicates $\beta=0.026$~eV$\cdot$nm, which is much larger than
$\frac{1}{2}\gamma_{\rm D}\langle k_{\Gamma z}^2\rangle\approx
0.001$~eV$\cdot$nm when only the lowest subband of the $\Gamma$-valley
is considered and the well width is set to be 7.5~nm
\cite{zhang:235323}. Therefore, the inhomogeneous broadenings of the
higher subband and valley are much larger. It is therefore easy to
understand that the spin relaxation time decreases with increasing
electric field when more electrons are excited to higher subband
and/or valley.

\begin{figure}[htb]
\begin{center}
\includegraphics[width=6.5cm]{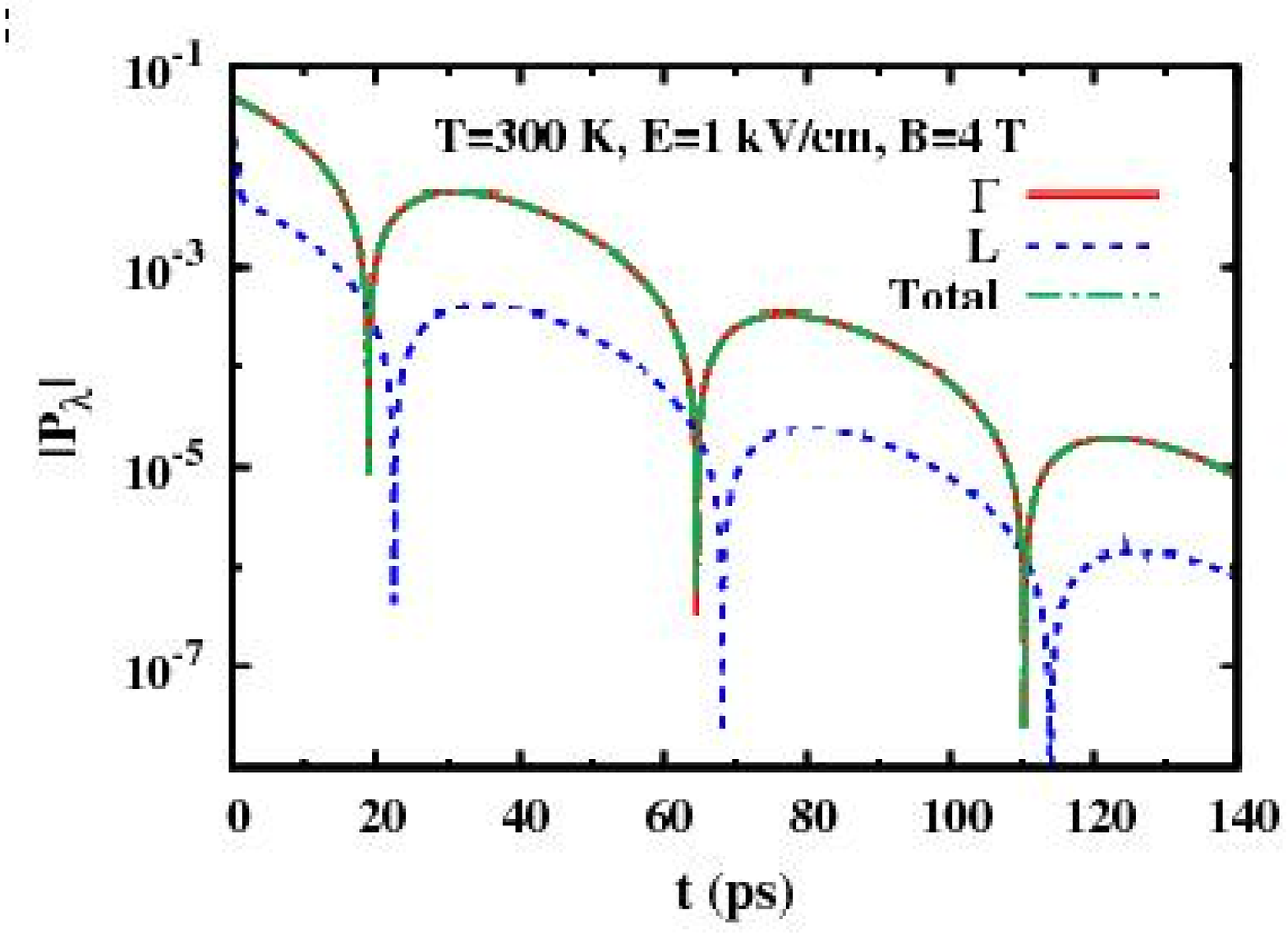}
\caption{ Temporal evolution of spin polarization
  $|P_{\lambda}|$ in $\lambda$ valley ($\lambda=\Gamma$, $L$) of
GaAs (001) quantum well with electric field $E=1$~kV/cm and 
magnetic field $B=4$~T. From  Zhang et al. \cite{zhang:235323}.} 
\label{fig5.4.3-9}
\end{center}
\end{figure}

\begin{figure}[htb]
\begin{center}
\includegraphics[width=6.cm]{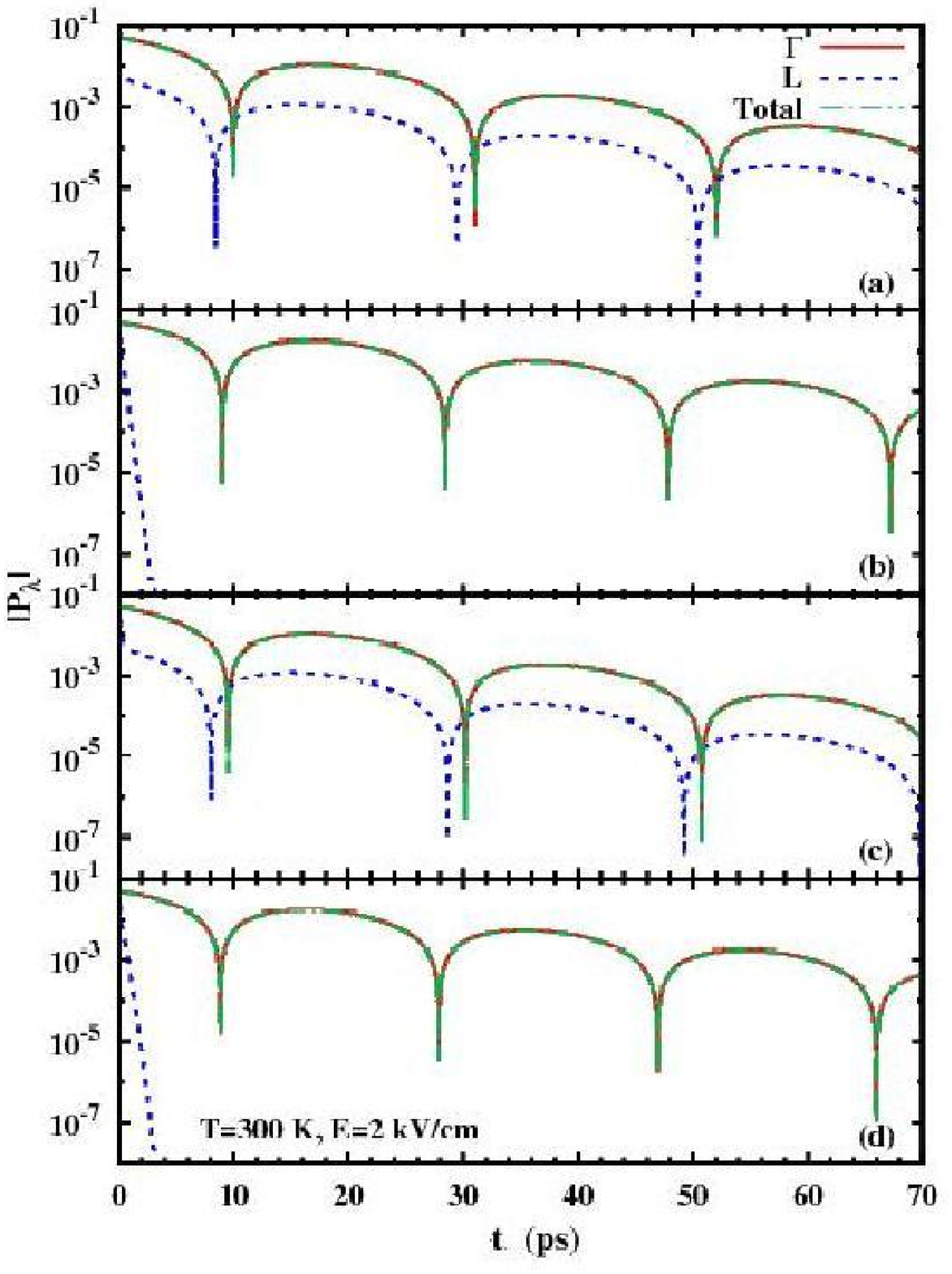}
\caption{ Spin polarization  $|P_{\lambda}|$ {\em vs}. $t$ 
under electric field $E=2$~kV/cm in GaAs (001) quantum well. Solid curve:
      $\Gamma$ valley; dashed
      curve: $L$ valley; and chain curve: the total. (a):
      with both the intervalley electron-phonon scattering ($H_{\rm
        \Gamma-L}^{\rm e-p}$) and intervalley electron-electron
      Coulomb scattering ($H_{\Gamma-L}^{\rm e-e}$);
      (b): without $H_{\rm \Gamma-L}^{\rm e-p}$; (c): without
      $H_{\rm \Gamma-L}^{\rm e-e}$; and (d): without $H_{\rm \Gamma-
        L}^{\rm e-e}$ and $H_{\rm \Gamma-L}^{\rm e-p}$. $T=300$~K and
      $N_i=B=0$. From Zhang et al. \cite{zhang:235323}.} 
\label{fig5.4.3-10}
\end{center}
\end{figure}

By simply looking at the inhomogeneous broadenings of $\Gamma$ and $L$
valleys, one may naively come to the conclusion that the spin
relaxation rate of electrons in $L$ valley should be much faster than
that in $\Gamma$ valley. Also by noticing the $g$-factor of $L$ valley
$g_L=1.77$ \cite{shen:063719}, which is also much larger than that of
$\Gamma$ valley, $g_\Gamma=-0.44$ \cite{madelung}, one may also
speculate that the spin precession under a magnetic field in the Voigt
configuration should be much faster in the $L$ valley than in the
$\Gamma$ valley. Zhang et al. showed both are in fact incorrect
\cite{zhang:235323}. They calculated the spin precessions of electrons
in $\Gamma$ and $L$ valleys in GaAs quantum well with both high
in-plane electric field and a magnetic filed in the Voigt configuration,
and discovered that the spin precession frequencies as well as spin
dephasing times of both valleys are identical to each other, as shown
in Fig.~\ref{fig5.4.3-9}, and the ``effective'' $g$-factor of the
$L$ valley from the spin precession is 0.44, which is in the same magnitude as that of the
$\Gamma$ valley. The physics leading to these unexpected results was
revealed due to the strong intervalley electron-phonon scattering. As
shown in Fig.~\ref{fig5.4.3-10} where the spin precessions of an
$n$-type (001) GaAs quantum well under electric field $E=2$~kV/cm are
plotted with all the scatterings included (a), without the intervalley
electron-phonon scattering (b), without the intervalley
electron-electron Coulomb scattering (c), and without intervalley
scattering (d), respectively. It is seen that when the intervalley
electron-phonon scattering is switched off, the spin polarization in
the $L$ valleys decays pretty fast. This is in contrast to the
multi-subband case addressed in the previous subsection where the
inter-subband Coulomb scattering is the cause of the identical spin
relaxation times of each subband.

\subsubsection{Effect of the Coulomb scattering to the spin dephasing
  and relaxation}
\label{sec5.4.4}
\begin{figure}[htb]
\begin{center}
\centerline{
\includegraphics[width=4.cm,height=4.4cm]{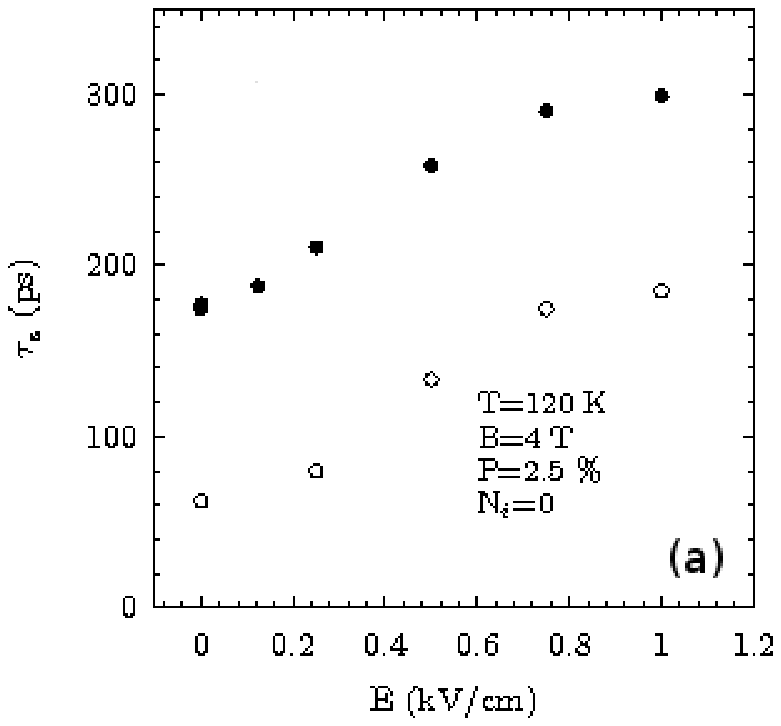}\hspace{0.9 cm}\includegraphics[width=5cm,height=4.4cm]{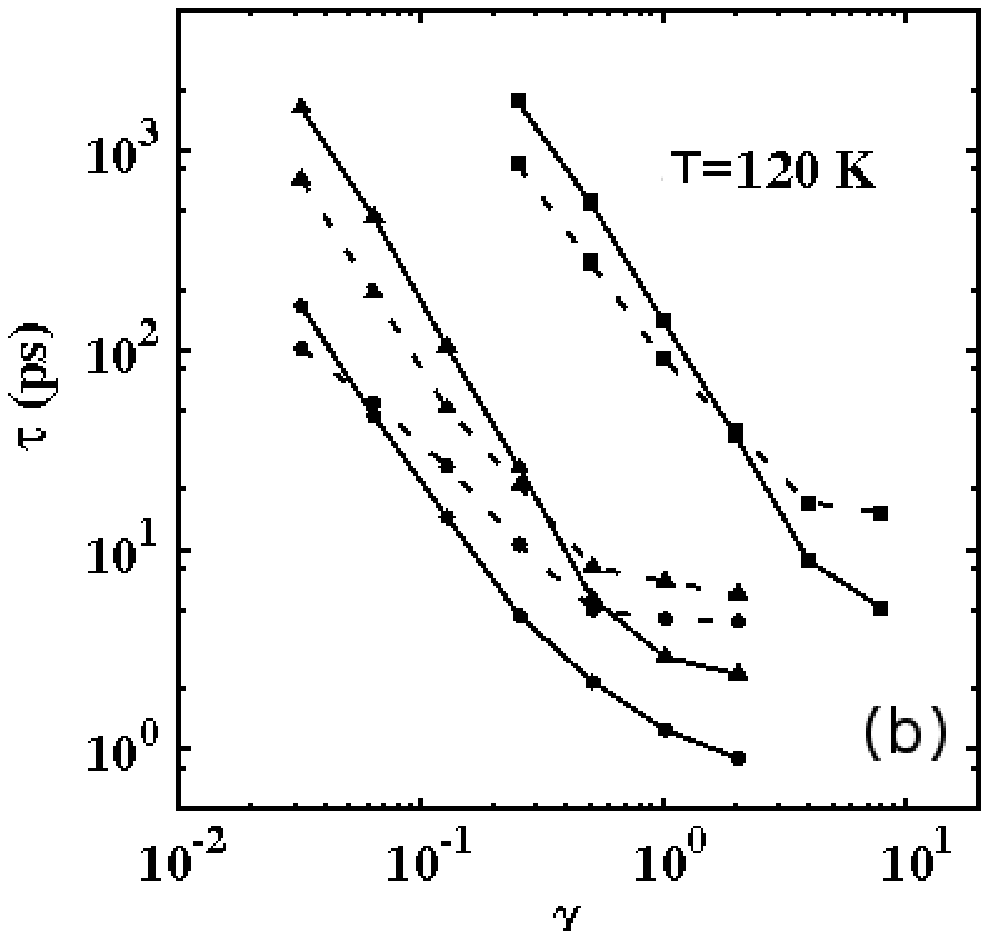}\includegraphics[width=5cm,height=4.4cm]{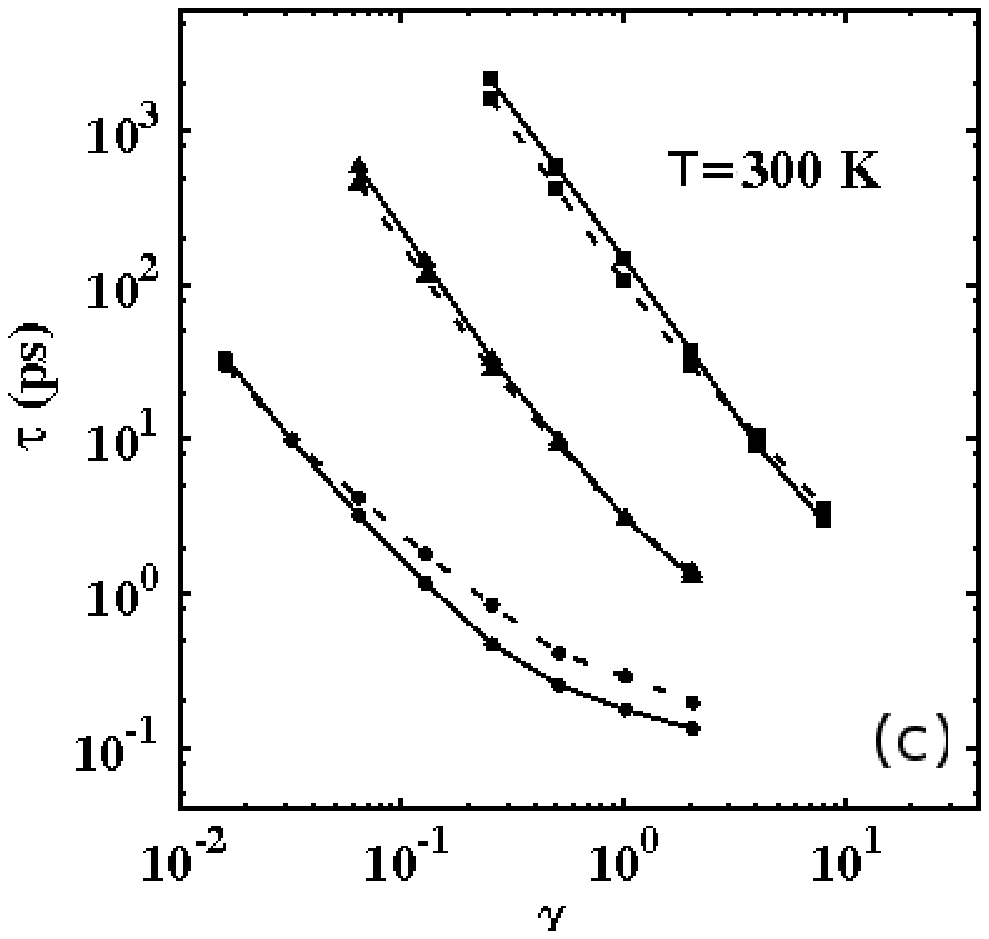}
}
\caption{(a) Spin dephasing time in GaAs (001) quantum wells with (close
  circles) and without (open circles) the Coulomb scattering. From
  Weng et al. \cite{PhysRevB.69.245320}. (b) and (c): Spin dephasing time {\sl vs.}
  the scaling coefficient of the D'yakonov-Perel' term $\gamma$ for $T=120$~K and
  300~K, respectively. The solid  (dashed) curves are the results with (without) the Coulomb scattering.
  {\tiny $\blacksquare$}: Electrons; {$\blacktriangle$}: light holes;
  {$\bullet$}: heavy holes. The impurity density $N_i=0$ and magnetic
  field $B=0$. From L\"{u} et al. \cite{lue:125314}.} 
\label{fig5.4.4-1}
\end{center}
\end{figure}

As addressed in the previous subsections, the Coulomb scattering can
contribute to the spin dephasing and relaxation due to the D'yakonov-Perel'
mechanism. A natural question would be how the Coulomb scattering
changes the spin dephasing/relaxation time. Glazov and Ivchenko
pointed out that the Coulomb scattering can prolong the spin
dephasing/relaxation time limited by the D'yakonov-Perel'
mechanism. Weng et al. compared the electron spin dephasing times in (001) GaAs
quantum well with and without the Coulomb scattering in the presence
of an in-plane electric field \cite{PhysRevB.69.245320}. As shown in
Fig.~\ref{fig5.4.4-1}(a), in which the spin dephasing times with and
without the Coulomb scattering are plotted against in-plane electric
field. It is seen that adding Coulomb scattering always leads to a
longer spin dephasing time both near (at $E\sim0$) and far
away from (with hot-electron effect) the equilibrium. This looks quite
contradictory to the previous understanding of optical dephasing
\cite{haugjauho} where adding a Coulomb scattering always leads to a
shorter optical dephasing time. Also as stated in Sec.~\ref{sec5.2}
that in the presence of the inhomogeneous broadening, adding a new
scattering leads to a new dephasing channel. How can the additional
``channel'' leads to a longer spin dephasing time? To understand this,
L\"u et al. \cite{lue:125314} put a scaling coefficient $\gamma$ in
front of the D'yakonov-Perel' term and compared the spin dephasing times of
electrons, heavy holes and light holes in (001) GaAs quantum well with
and without the Coulomb scattering as function of $\gamma$. As shown
in Fig.~\ref{fig5.4.4-1}(b) and (c), for larger $\gamma$, i.e., in the
weak scattering limit, adding Coulomb scattering always leads to a
shorter spin dephasing time. However, for small $\gamma$, i.e., in the
strong scattering limit, adding Coulomb scattering leads to a longer
spin dephasing time. This can be understood as follows: In the weak
scattering limit, the counter effect of the scattering to the
inhomogeneous broadening is pretty weak, and the only effect of the
scattering is an additional spin dephasing channel. However, in the 
strong scattering limit, the counter effect of the scattering is more
pronounced and hence the spin dephasing time is increased by adding
the Coulomb scattering. It is noted that $\gamma=1$ corresponds to
the genuine case and from the figure one notices that electrons
happen to be in the strong scatterign limit. Therefore adding the
Coulomb scattering leads to a longer spin dephasing time. However,
both heavy and light holes are in the weak scattering
limit. Consequently adding the Coulomb scattering makes the spin
dephasing time shorter. In fact, for optical dephasing problem, the
inhomogeneous broadening in Eq.~(\ref{eq5.2-3}) happens to be in the weak
scattering limit. Therefore, the Coulomb scattering always brings a
shorter optical dephasing time \cite{haugjauho}. 

The strong/weak scattering limit can be measured by $\xi\equiv|g^\ast\mu_B{\bf
  \Omega}({\bf k})|\tau_p^\ast$. When $\xi\gg 1$ ($\ll1$), the system is
in the weak (strong) scattering limit. Here $\tau_p^\ast$ is the
effective momentum scattering time from not only the carrier-phonon
and carrier-impurity scatterings, but also from the Coulomb scattering
\cite{Glazov:jetp.75.403,leyland:165309}. It is also noted that the regime
of strong and weak scattering can be changed even for the same sample
by temperature. It has been shown by Harley group
\cite{leyland:165309,PhysRevLett.89.236601} and also later by Stich
et al. \cite{stich:176401,stich:073309} that at low temperature
electrons in high mobility (001) GaAs quantum well can be in the weak
scattering limit. Also for holes in $p$-type (001) GaAs quantum wells,
L\"u et al. have shown that the system can be changed from the
weak scattering limit to the strong one by impurity density\footnote{
It is noted that the impurity density should be 1/4 of the value
presented in Ref.~\cite{lue:125314}.} and temperature,
etc. \cite{lue:125314}. In fact, in the strong (weak) scattering limit,
besides the Coulomb scattering, adding any scattering, can cause a
longer (shorter) spin dephasing/relaxation time.

\subsubsection{Non-Markovian effect in the weak scattering limit}
\label{sec5.4.5}
As stated in Sec.~\ref{sec5.3}, the scattering terms of the kinetic spin Bloch equations are
given by Eqs.~(\ref{eq5.3-14}) and (\ref{eq5.3-15}), which are
time-integrals. By applying the Markovian approximation
[Eq.~(\ref{eq5.3-16})], one can carry out the time-integral and the
scattering terms are simplified to Eq.~(\ref{eq5.3-17}), where one has
the energy conservation [$\delta$-functions in Eq.~(\ref{eq5.3-16})]
and the trace of the history disappears. This approximation is valid
only in the strong scattering limit where the spin precession between
two adjacent scatterings is negligible. However, in the weak
scattering limit, electron spin can experience many precessions
between two scattering events, and the Markovian approximation is
invalid. In this circumstance one has to trace back the history by carrying out the time
integral. Hence the dynamics becomes non-Markovian.

\begin{figure}[ht]
  \centerline{\includegraphics[width=4.5cm]{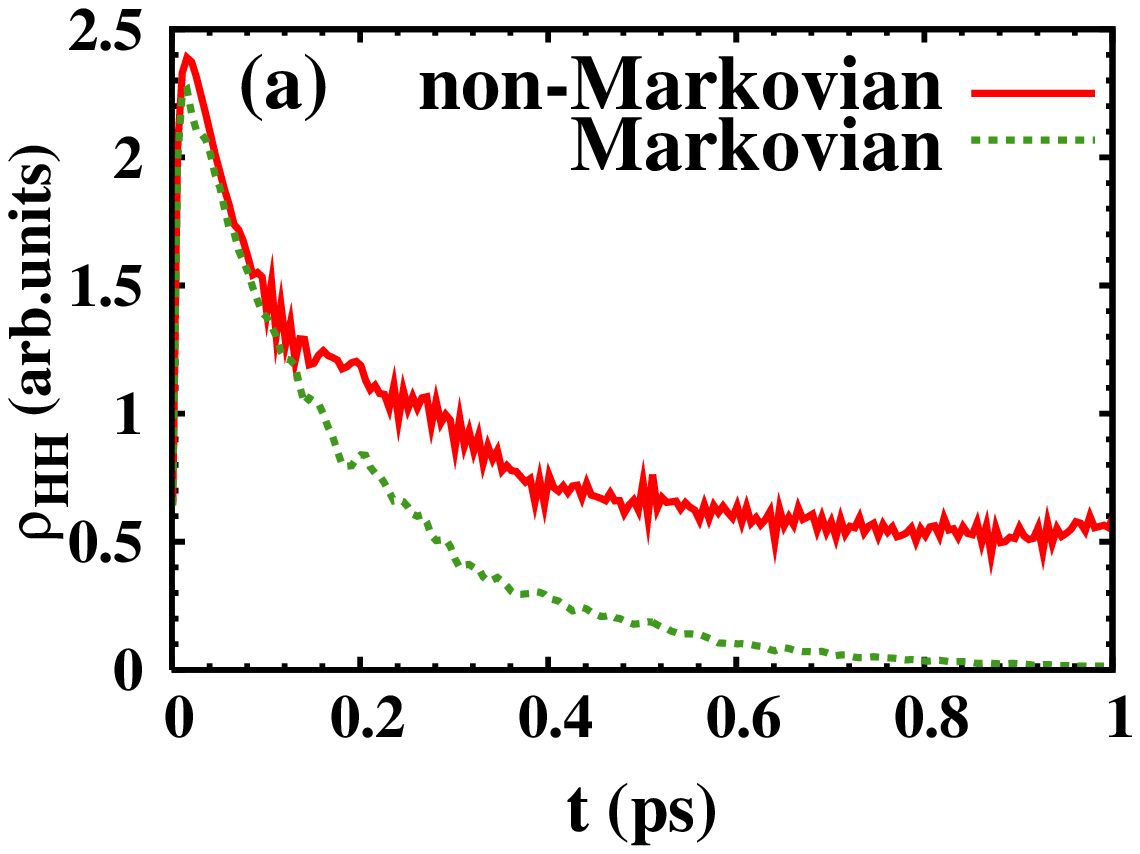}\includegraphics[width=4.5cm]{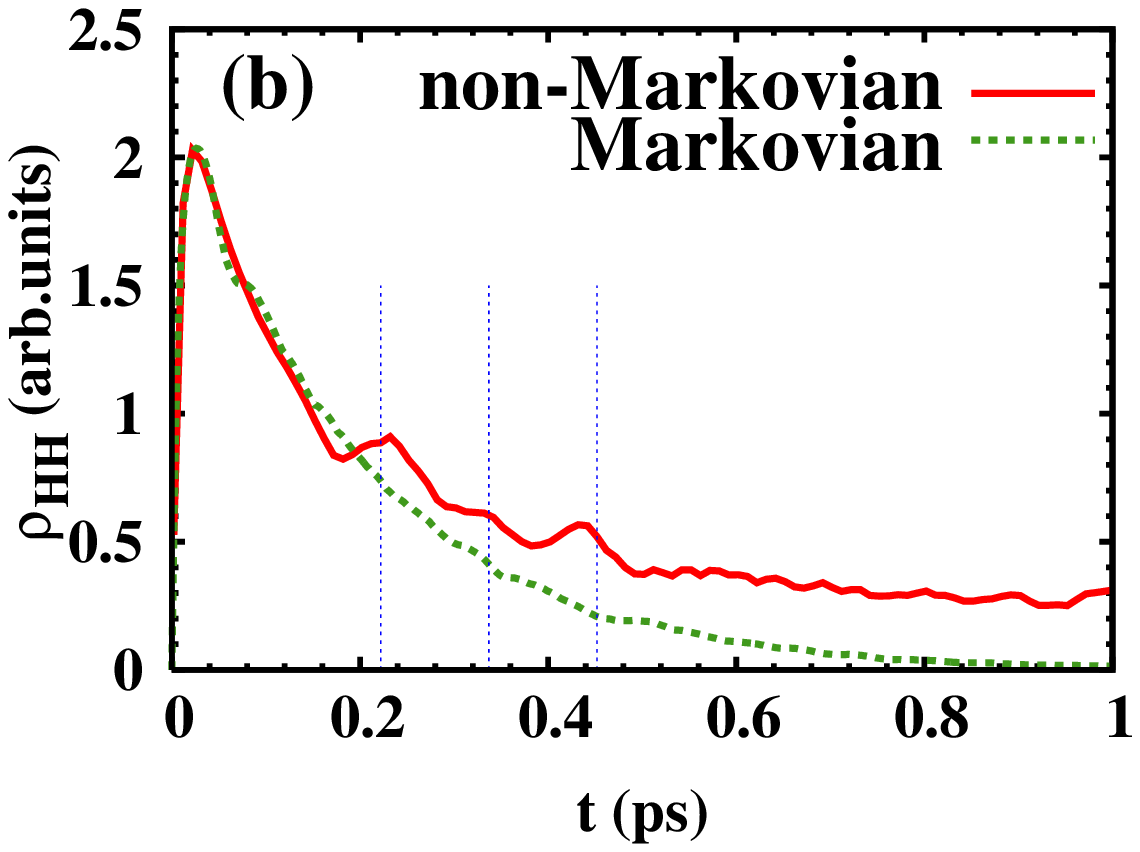}} 
  \centerline{\includegraphics[width=4.5cm]{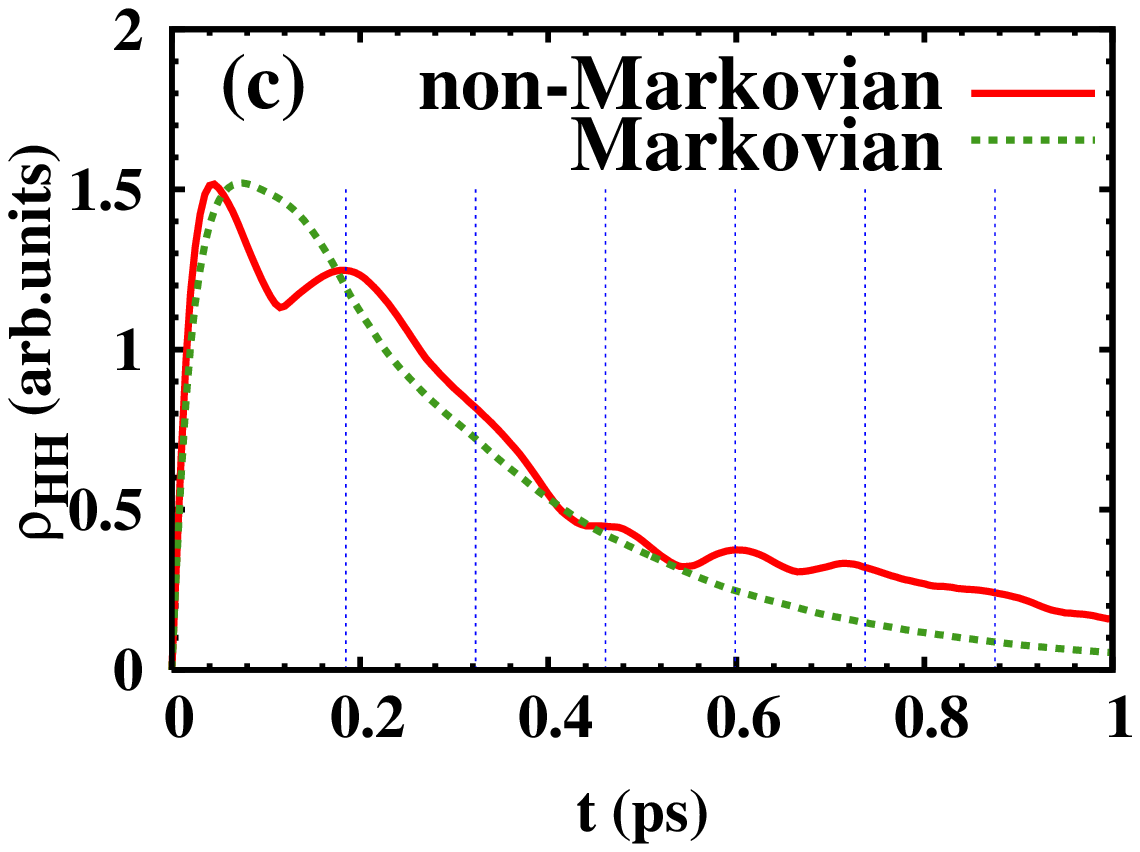}\includegraphics[width=4.5cm]{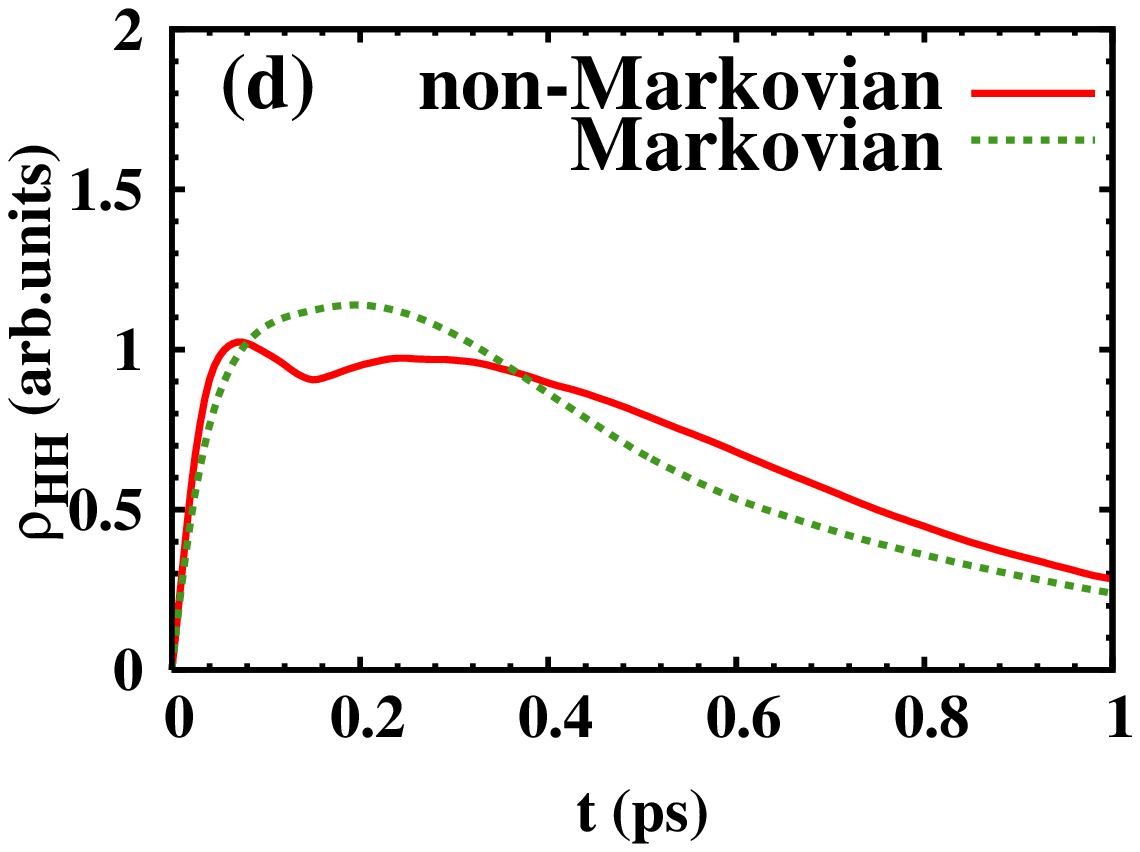}}
  \caption{Time evolutions of the incoherently
    summed spin coherence  $\rho_{\rm HH}$ of heavy holes in
GaAs (001) quantum wells with Rashba spin-orbit coupling
    strength (a) $\gamma_{54}^{7h7h}E_zm_0/\hbar^2$ =6.25~nm, (b)
    2.5~nm, (c) 0.625~nm and (d) 0.25~nm,
    respectively. $\gamma_{54}^{7h7h}$ is the Rashba
    coefficient and $E_z$ is the electic field from the gate voltage
    \cite{lue:125314}. Correspondingly, the mean spin
    precession time ${\Omega}^{-1}$ is (a) 0.046~ps, (b) 0.116~ps, 
    (c) 0.463~ps and (d) 1.160~ps. Solid curve: in non-Markovian
    limit; Dashed curve: in Markovian limit. The well width $a=5$~nm
    and the temperature $T=300$~K. From Zhang and Wu \cite{zhang:193312}.}
\label{fig5.4.5-1.eps}
\end{figure}

Glazov and Sherman first investigated the electron spin relaxation in
GaAs quantum wells under a strong magnetic field by the Monte Carlo
simulation \cite{0295-5075-76-1-102}. They reported a longer non-Markovian
spin relaxation time \cite{0295-5075-76-1-102}. Zhang and Wu studied the
non-Markovian hole spin dynamics in $p$-type GaAs quantum wells using
kinetic spin Bloch equation approach \cite{zhang:193312}. By performing time derivative of the
time integral in Eq.~(\ref{eq5.3-15}), they transferred the
integrodifferential kinetic spin Bloch equations in non-Markovian limit to a larger set of
differential equations. They compared the time evolutions of the
incoherently summed spin coherence of heavy holes in both the
Markovian and non-Markovian limits at different Rashba strengths. From
the decay of the incoherently summed spin coherence, one may extract
$T_2$.\footnote{Instead of $T_2^\ast$ which is measured from the Faraday/Kerr
rotation experiment.} As shown in Fig.~\ref{fig5.4.5-1.eps}(d), in the
strong scattering limit, both approaches yield the same spin dephasing
time. Nevertheless, in the weak scattering limit, the non-Markovian
kinetics gives a longer hole spin dephasing time, which is in good
agreement with the prediction of Glazov and Sherman
\cite{0295-5075-76-1-102}. The physics behind this phenomena is easily
understood from the nature of non-Markovian scattering, which keeps
some memory of the coherence during the scattering. What is  most
important is that they predicted
quantum spin beats when the spin precession time is comparable to the
momentum scattering time. The beats are solely from the non-Markovian
effect (or memory effect), with the beating frequency being exactly
the longitudinal optical-phonon frequency [see Fig.~\ref{fig5.4.5-1.eps}(b) and
(c)]. The quantum spin beats are similar to the longitudinal optical-phonon quantum
beats in optical four wave mixing signal \cite{PhysRevLett.75.2188}. However, a
measurement of these quantum spin beats cannot be performed by the
regular Faraday/Kerr rotation measurement. A possible way is through
the spin echo measurement. 

\subsubsection{Electron spin relaxation due to the Bir-Aronov-Pikus mechanism in
  intrinsic and $p$-type GaAs quantum wells}
\label{sec5.4.6}
It has long been believed in the literature that the Bir-Aronov-Pikus mechanism is
dominant at low temperature in $p$-type samples and has important
contribution to intrinsic sample with high photo-excitation
\cite{PhysRevB.56.R7076,PhysRevB.72.245202,boggess:1333,PhysRevB.47.4786,PhysRevLett.67.3432,gotoh:3394,schneider:081302}.
These conclusions were made based on Eq.~(\ref{BAPes}) from the
Fermi Golden rule, which in (001) quantum well reads \cite{zhou:075318}
\begin{equation}
\frac{1}{\tau_{\rm BAP}({\bf k})}=4\pi\sum_{{\bf k}^\prime {\bf
    q}}\delta(\varepsilon_{{\bf k-q}}^e-\varepsilon_{{\bf
    k}}^e+\varepsilon_{{\bf k^\prime}}^h-\varepsilon_{{\bf
    k^\prime-q}}^h)|M({\bf K-q})|^2f_{\bf k^\prime-q}^h(1-f_{\bf
  k}^h)
\label{taubap}
\end{equation}
with ${\bf K=k+k^\prime}$ and
\begin{eqnarray}
|M({\bf k-q})|^2 =\frac{9\Delta E_{LT}^2}{16|\phi_{3D}(0)|^4}\left[\sum_{q_z}\frac{f_{ex}(q_z)({\bf K-q})^2}{q_z^2+({\bf K-q})^2}\right]^2. 
\end{eqnarray}
Here $\Delta E_{LT}$ is the longitudinal-transverse splitting in
bulk. $|\phi_{3D}(0)|^2=1/(\pi a_0^3)$ denotes the bulk exciton state
at zero relative distance. $f_{ex}(q_z)$ is the form factor
\cite{zhou:075318,PhysRevB.47.15776}. In the meantime, the spin relaxation time from the D'yakonov-Perel'
mechanism used for comparison was calculated without the Coulomb
scattering which has been demonstrated to be important in the previous
sections.

Zhou and Wu reexamined the problem using the fully microscopic kinetic spin Bloch equation
approach \cite{zhou:075318}. They constructed the kinetic spin Bloch equations in intrinsic and $p$-type
(001) GaAs quantum wells with the electron-phonon, electron-impurity,
electron-electron Coulomb and electron-heavy hole Coulomb scatterings
explicitly included. The electron-heavy hole Coulomb scattering
includes the spin-conserving scattering and the spin-flip scattering
(the Bir-Aronov-Pikus term). They also extended the screening under
the random phase approximation Eq.~{(\ref{rpa}) to
  include the contribution from heavy holes:
\begin{equation}
\epsilon({\bf q})=1-\sum_{q_{z}}v_{Q}f_e(q_{z})\sum_{{\bf k},\sigma}
\frac{f_{{\bf k}+{\bf q},\sigma}
-f_{{\bf k},\sigma}}{\varepsilon^{e}_{\bf k+q}-\varepsilon^e_{\bf
  k}}
\mbox{}-\sum_{q_{z}}v_{Q}f_h(q_{z})\sum_{{\bf k^{\prime}},\sigma}\frac{f^h_{{\bf k^{\prime}}+{\bf q},\sigma}
-f^h_{{\bf k^{\prime}},\sigma}}{\varepsilon^{h}_{\bf
    k^{\prime}+q}-\varepsilon^h_{\bf k^{\prime}}},
\end{equation}
where $f_e(q_z)$ and $f_h(q_z)$ are the form factors \cite{zhou:075318}. The
scattering terms from the spin-flip electron-heavy hole exchange
interaction (Bir-Aronov-Pikus term) read
\begin{eqnarray}\nonumber
\left.\frac{\partial f_{{\bf k},\sigma}}{\partial t}\right|_{\rm BAP}&=&-2\pi\sum _{{\bf k^{\prime},q}}
\delta(\varepsilon_{\bf k-q}^e-\varepsilon_{\bf k}^e
+\varepsilon_{\bf k^{\prime}}^h-\varepsilon_{\bf k^{\prime}-q}^h)
|M({\bf K-q})|^2
\\ &&\quad \times [(1-f_{{\bf k^{\prime}},\sigma}^h)f_{{\bf
    k^{\prime}-q},-\sigma}^h
 f_{{\bf k},\sigma}(1-f_{{\bf k-q},-\sigma})
-f_{{\bf k^{\prime}},\sigma}^h(1-f_{{\bf k^{\prime}-q},-\sigma}^h)
(1-f_{{\bf k},\sigma})f_{{\bf k-q},-\sigma}]
\label{BAPscat1} ,\\
\left. \frac{\partial \rho_{\bf k}}{\partial t}\right |_{\rm BAP}&=&-\pi\sum_{{\bf k^{\prime},q},\sigma}
  \delta(\varepsilon_{\bf k-q}^e-\varepsilon_{\bf k}^e
+\varepsilon_{\bf k^{\prime}}^h-\varepsilon_{\bf k^{\prime}-q}^h)
|M({\bf K-q})|^2 \\ &&\quad \times [(1-f_{{\bf
    k^{\prime}},\sigma}^h)f_{{\bf k^{\prime}-q},-\sigma}^h (1-f_{{\bf k-q},-\sigma})\rho_{\bf k}+
  f_{{\bf k^{\prime}},\sigma}^h(1-f_{{\bf k^{\prime}-q},-\sigma}^h)
  f_{{\bf k-q},\sigma}\rho_{\bf k}]. 
\label{BAPscat}
\end{eqnarray}
It is noted from Eq.~(\ref{BAPscat1}) that there are quadratic terms in
the sense of the electron distribution function $f_{{\bf
    k}\sigma}$. This immediately implies that one cannot recover
Eq.~(\ref{taubap}) from the Fermi Golden rule. The only way to recover
Eq.~(\ref{taubap}) is under the elastic scattering approximation:
$\varepsilon_{\bf k-q}^e\approx\varepsilon_{\bf k}^e$, and
$\varepsilon_{{\bf k}^\prime}^h\approx\varepsilon_{\bf k^\prime-q}^h$
where the quadratic terms exactly cancel each other and the spin
relaxation time is given by Eq.~(\ref{taubap}). It is well known that
the elastic scattering approximation is valid only in the
non-degenerate regime at high temperatures. When the temperature is
low enough so that electrons are degenerate, the elastic scattering
approximation is not valid any more. Therefore the quadratic terms
become unnegligible. In fact, the terms $(1-f_{{\bf k-q},-\sigma})$ and
$(1-f_{{\bf k}\sigma})$ in Eq.~(\ref{BAPscat1}) come from the Pauli
blocking, which reduces the spin relaxation rate. Therefore the Bir-Aronov-Pikus
mechanism becomes less important at low temperatures when the Pauli
blocking becomes important. This effect was missing in the literature
\cite{PhysRevB.56.R7076,PhysRevB.72.245202,boggess:1333,PhysRevB.47.4786,PhysRevLett.67.3432,gotoh:3394,schneider:081302}.
In Fig.~\ref{fig5.4.6-1}(a), the spin relaxation times due to 
the Bir-Aronov-Pikus mechanism in intrinsic GaAs quantum well calculated from the
full spin-flip electron-hole exchange scattering
[Eq.~(\ref{BAPscat1})] are compared with those without the Pauli blocking at
different pumping densities. It is seen that the previous calculations
based on Eq.~(\ref{taubap}) always overestimate the importance of
the Bir-Aronov-Pikus mechanism at low temperatures. In Fig.~\ref{fig5.4.6-1}(b),
the spin relaxation times of intrinsic GaAs quantum wells due to the
Bir-Aronov-Pikus mechanism and the D'yakonov-Perel' mechanism are compared at different
photo-excitations. It is seen that instead of being important at low
temperature, the Bir-Aronov-Pikus mechanism is even negligible at low temperature
when the Pauli blocking becomes important. Two errors in the previous
treatment may cause the incorrect conclusion that the Bir-Aronov-Pikus mechanism is
dominant at low temperature: (i) The previous theory overlooked the Pauli blocking
and hence overestimated the importance of the Bir-Aronov-Pikus mechanism; (ii) It
overlooked the contribution of the Coulomb scattering to the D'yakonov-Perel'
mechanism and thus underestimated the importance of the D'yakonov-Perel' mechanism. It is also
noted that the temperature at which the Bir-Aronov-Pikus may have some contribution
is around the hole Fermi temperature. Similar conclusions were also
obtained in $p$-type GaAs quantum wells \cite{zhou:075318}.

\begin{figure}[ht]
  \centerline{\includegraphics[width=6cm]{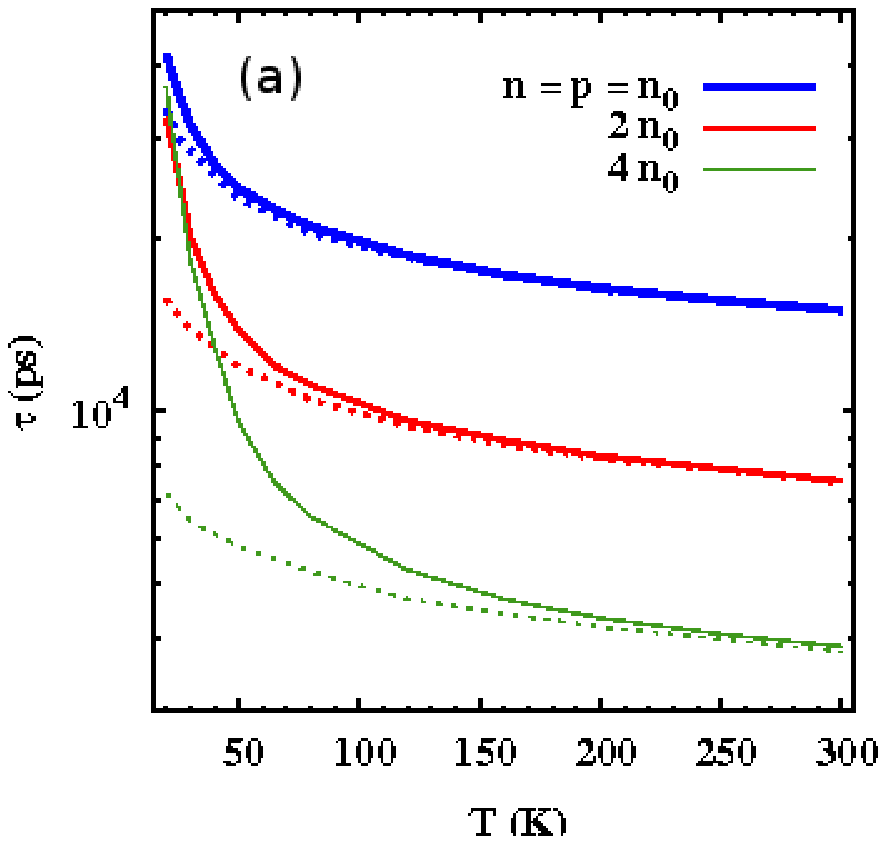}\includegraphics[width=6cm]{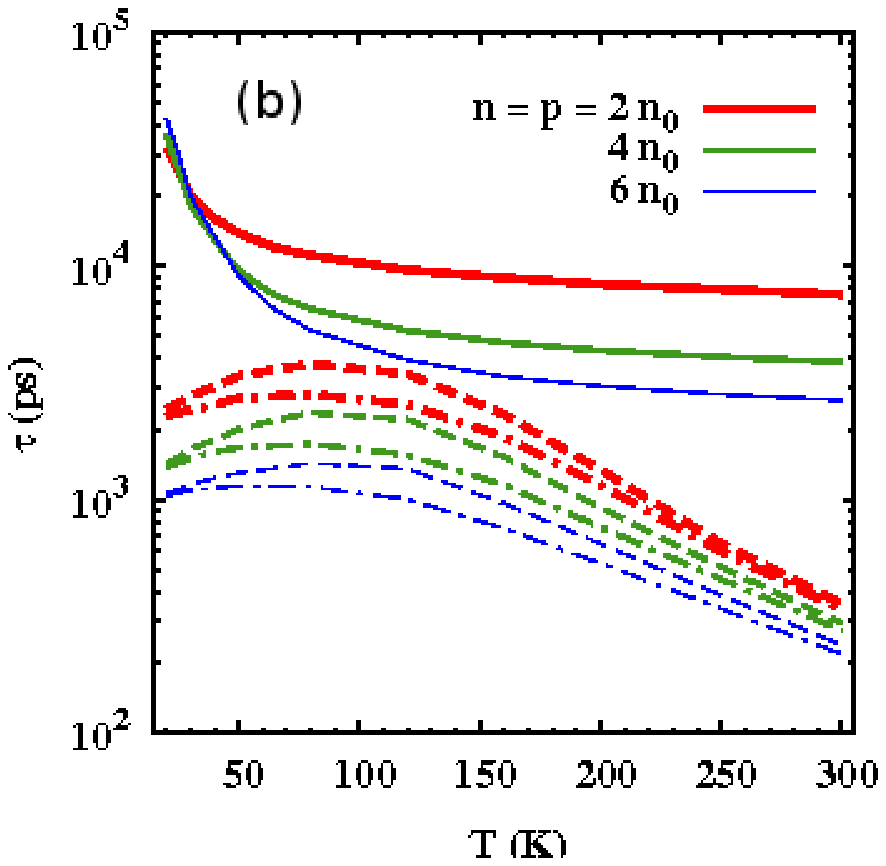}}
  \caption{GaAs (001) quantum wells:
 (a) Spin relaxation time due to the Bir-Aronov-Pikus mechanism 
with full spin-flip
electron-hole exchange scattering (solid curves) and
with only the linear  terms in the spin-flip
electron-hole exchange scattering (dotted curves) at 
different electron densities against temperature $T$.  $n_0 = 10^{11}$\ cm$^{-2}$. (b) Spin relaxation time due to the Bir-Aronov-Pikus (solid curves) and
D'yakonov-Perel' (dashed curves) mechanisms and
the total spin relaxation time (dash-dotted curves) {\it vs.} temperature $T$
in intrinsic quantum wells
at different densities ($n =p=2$, 4, $6n_0$)
when well width $a = 20$~nm and impurity density $n_i = n$. $n_0 =
10^{11}$~cm$^{-2}$. From Zhou and Wu \cite{zhou:075318}.}
\label{fig5.4.6-1}
\end{figure}           

\begin{figure}[ht]
  \centerline{\includegraphics[width=6cm]{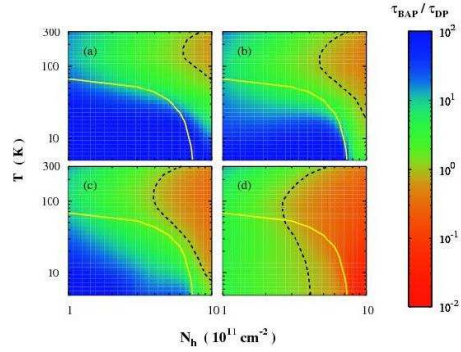}}
  \caption{ Ratio of the spin relaxation time due to the Bir-Aronov-Pikus
 mechanism to that due to the D'yakonov-Perel' mechanism, 
$\tau_{\rm BAP}/\tau_{\rm DP}$, in GaAs (001) quantum wells
     as function of temperature and hole density with
     (a) $N_i=0$, $N_{ex}=10^{11}$~cm$^{-2}$;
     (b) $N_i=0$, $N_{ex}=10^{9}$~cm$^{-2}$; (c) $N_i=N_h$,
     $N_{ex}=10^{11}$~cm$^{-2}$; (d) $N_i=N_h$, $N_{ex}=10^{9}$~cm$^{-2}$.
     The black dashed curves indicate the cases satisfying
     $\tau_{\rm BAP}/\tau_{\rm DP}=1$. Note the smaller the
     ratio $\tau_{\rm BAP}/\tau_{\rm DP}$ is, the more
     important the Bir-Aronov-Pikus mechanism becomes.
     The yellow solid curves indicate the cases satisfying
     $\partial_{\mu_h}[N_{{\rm LH}^{(1)}}+N_{{\rm HH}^{(2)}}]/\partial_{\mu_h}N_h=0.1$.
     In the regime above the yellow curve the multi-hole-subband
     effect becomes significant. From Zhou et al. \cite{zhou0905.2790}.}
\label{fig5.4.6-2}
\end{figure}           

\begin{figure}[ht]
  \centerline{\includegraphics[width=6cm]{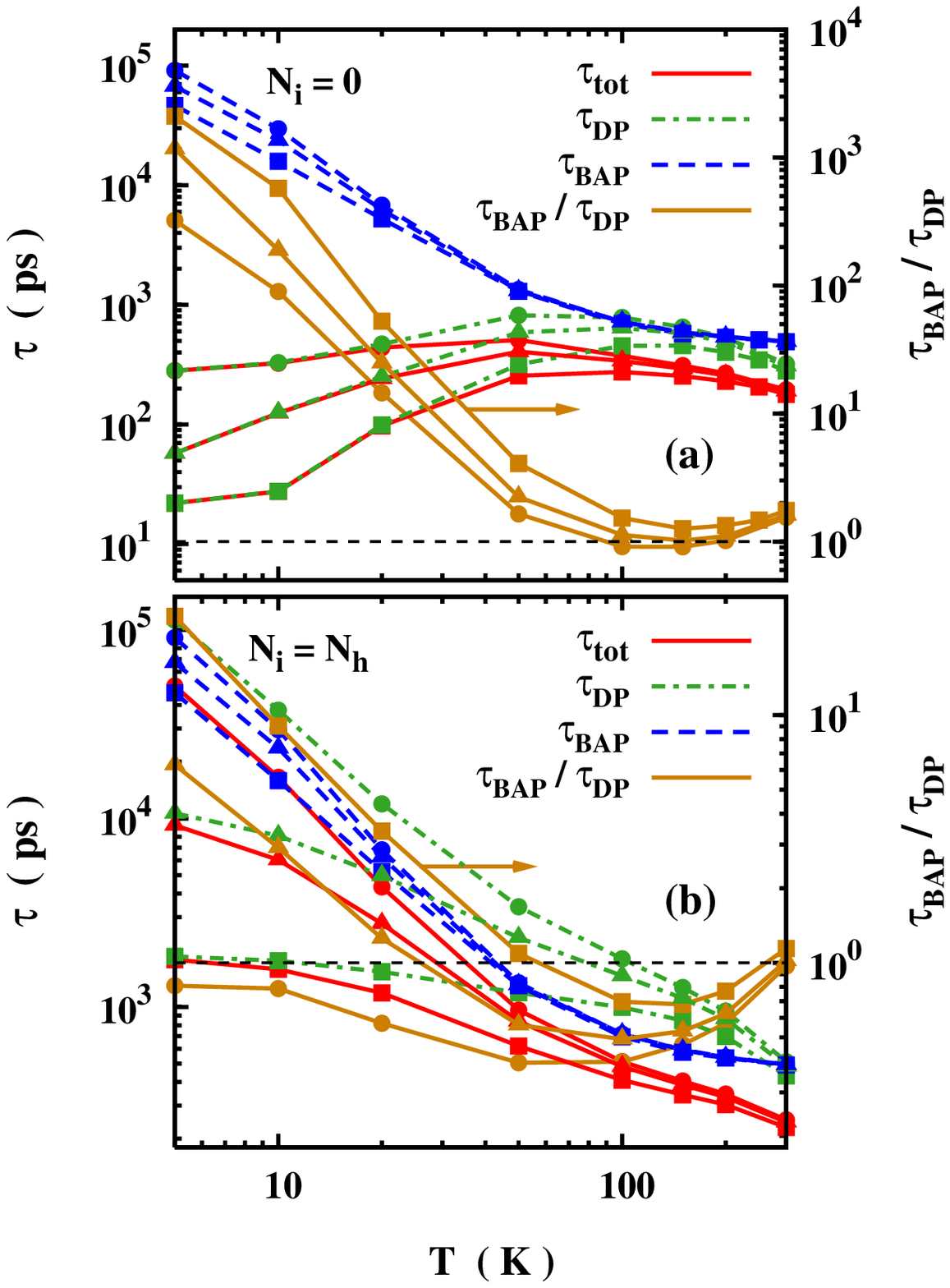}}
  \caption{  Spin relaxation times of electrons in GaAs (001) quantum wells
 due to the D'yakonov-Perel' and Bir-Aronov-Pikus mechanisms, the
    total spin relaxation time together with the ratio $\tau_{\rm BAP}/\tau_{\rm DP}$
   {\it vs}. temperature $T$ for $N_{\rm ex}=10^{9}$~cm$^{-2}$ (curves with
    $\bullet$), $3\times 10^{10}$~cm$^{-2}$ (curves with
    $\blacktriangle$) and $10^{11}$~cm$^{-2}$ (curves with
    $\blacksquare$) with hole density $N_{h}=5 \times 10^{11}$~cm$^{-2}$ and
    impurity densities (a) $N_i=0$ and (b) $N_i=N_h$.
    The electron Fermi temperatures for those excitation
    densities are $T_{\rm F}^e=0.41$, $12.4$ and $41.5$~K, respectively. The
    hole Fermi temperature is $T_{\rm F}^h=124$~K. Note the scale of
    $\tau_{\rm BAP}/\tau_{\rm DP}$ is on the right-hand side
    of the frame. From Zhou et al. \cite{zhou0905.2790}.}
\label{fig5.4.6-3}
\end{figure}

Later, Zhou et al. performed a thorough investigation of
electron spin relaxation in $p$-type (001) GaAs quantum wells by
varying impurity, hole and photo-excited electron
densities over a wide range of values \cite{zhou0905.2790}, under the idea that very
high impurity density and very low photo-excited electron density may
effectively suppress the importance of the D'yakonov-Perel' mechanism and the Pauli
blocking. Then the relative importance of the Bir-Aronov-Pikus and D'yakonov-Perel'
mechanisms may be reversed. This indeed happens as shown in the
phase-diagram-like picture in Fig.~\ref{fig5.4.6-2} where the relative
importance of the Bir-Aronov-Pikus and D'yakonov-Perel' mechanisms is plotted as function of hole
density and temperature at low and high impurity densities and
photo-excitation densities. For the situation of high hole density
they even included multi-hole subbands as well as the light hole band. It
is interesting to see from the figures that at relatively high
photo-excitations, the Bir-Aronov-Pikus mechanism becomes more important than the
D'yakonov-Perel' mechanism only at high hole densities and high temperatures (around
hole Fermi temperature) when the impurity is very low [zero in
Fig.~\ref{fig5.4.6-2}(a)]. Impurities can suppress the D'yakonov-Perel' mechanism and
hence enhance the relative importance of the Bir-Aronov-Pikus mechanism. As a
result, the temperature regime is extended, ranging from  the hole
Fermi temperature to the electron Fermi temperature for high hole
density. When the photo-excitation is weak so that the Pauli blocking
is less important, the temperature regime where the Bir-Aronov-Pikus mechanism is
important becomes wider compared to the high excitation case. In
particular, if the impurity density is high enough and the
photo-excitation is so low that the electron Fermi temperature is
below the lowest temperature of the investigation, the Bir-Aronov-Pikus mechanism
can dominate the whole temperature regime of the investigation at sufficiently high hole
density, as shown in Fig.~\ref{fig5.4.6-2}(d). The corresponding spin
relaxation times of each mechanism under high or low impurity and
photo-excitation densities are demonstrated in
Fig.~\ref{fig5.4.6-3}. They also discussed the density dependences of
spin relaxation with some intriguing properties related to the high
hole subbands \cite{zhou0905.2790}.

\begin{figure}[ht]
  \centerline{\includegraphics[width=6cm]{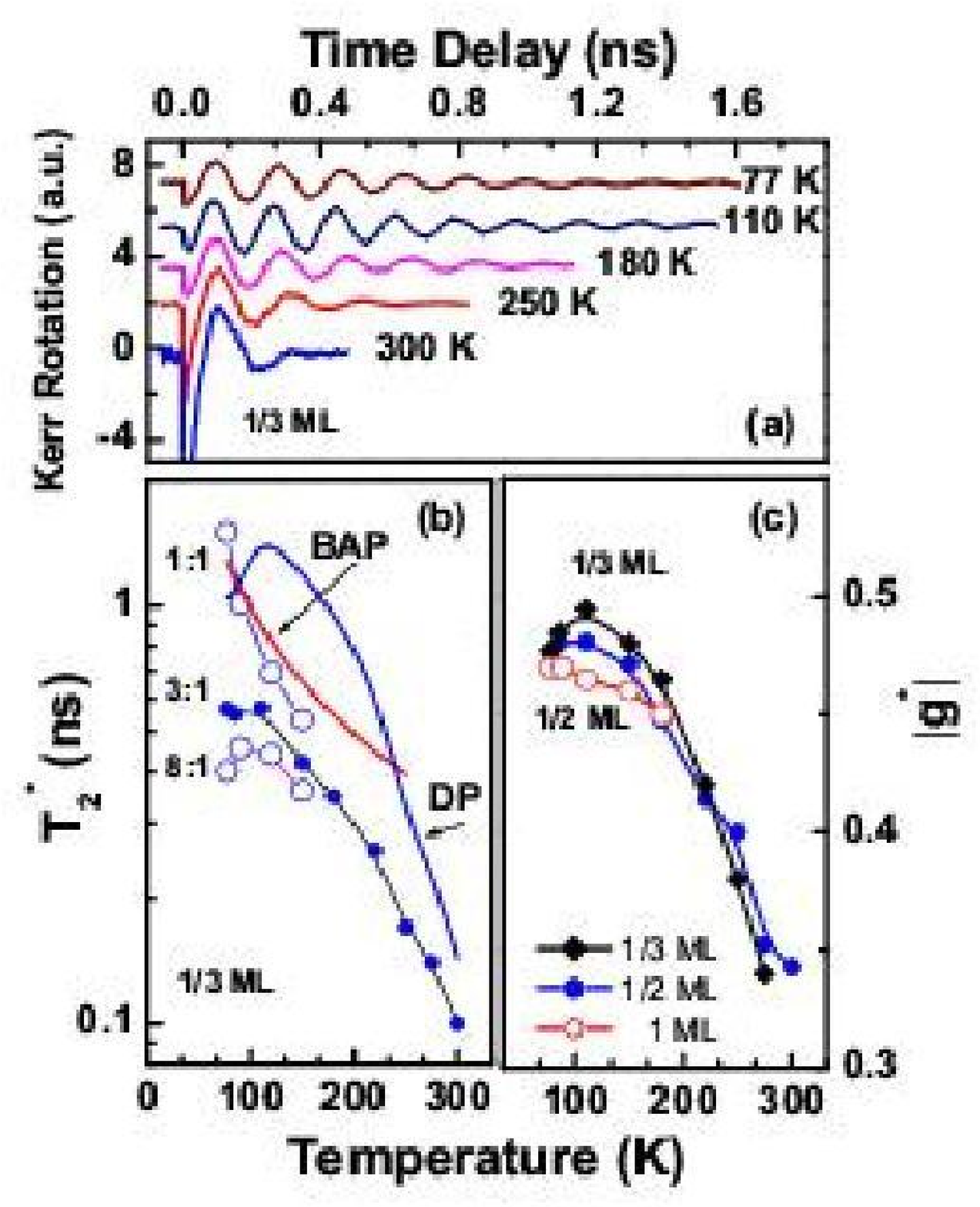}}
  \caption{  Kerr rotation of the 1/3 monolayer InAs sample at different
temperatures measured at $B=0.82$~T. The pumping density is
about 1.5 $\times$10$^{{17}}$/cm$^{{3}}$. The temperature
dependence of the electron spin decoherence time $T_{2}^{*}$ and
the effective $g$-factor are shown in (b) and (c), respectively.
The $T_{2}^{*}$ of 1/3 monolayer InAs sample with various pumping density
at low temperature as specified is also shown for comparison. From Yang et al. \cite{yang:035313}.}
\label{fig5.4.6-4}
\end{figure}       

The predicted Pauli-blocking effect in the Bir-Aronov-Pikus mechanism has been
partially demonstrated experimentally by Yang et al. \cite{yang:035313}, as
shown in Fig.~\ref{fig5.4.6-4}. They showed by increasing the pumping
density, the temperature dependence of the spin dephasing time
deviates from the one from the Bir-Aronov-Pikus mechanism and the peaks at high
excitations agree well with those predicted by Zhou and Wu \cite{zhou:075318}.

\subsubsection{Spin dynamics in the presence of a strong THz laser
  field in quantum wells with strong spin-orbit coupling}
\label{sec5.4.7}
The influence of a strong THz field, which can effectively change the
electron orbital momentums and significantly modify the electron density
of states, has been extensively studied in spin-unrelated problems
such as the dynamic Franz-Keldysh effect and optical side-band effect \cite{Rashbajsc,tokura:047202,PhysRevLett.91.126405,rashba:apl5295,cheng:032107,efros:prb165325}. So
far, the application of strong THz field to the spin systems is very
limited and only in theory. Johnsen is the first one who investigated
the optical sideband generation in systems with spin-orbit coupling
and suggested the formation of odd sidebands which are otherwise
absent without the spin-orbit coupling \cite{PhysRevB.62.10978}. Cheng and Wu showed
theoretically that in InAs quantum wells, a strong in-plane THz
electric field can induce a large spin polarization oscillating at the
same frequency of the THz driving field \cite{cheng:032107}. Later Jiang et al. and
Zhou suggested similar effects in single charged
InAs quantum dots \cite{jiang:jap063709} and
in $p$-type GaAs (001) quantum wells \cite{Zhou:PE1386}. These calculations in Refs.~\cite{PhysRevB.62.10978,cheng:032107,jiang:jap063709,Zhou:PE1386} are
performed without any dissipation. However, in reality, due to the
scattering there is spin dephasing and relaxation. Whether the large
spin polarization oscillations are still kept in the presence of the
dissipation is unanswered in Refs.~\cite{cheng:032107,jiang:jap063709,Zhou:PE1386}. Also the THz field can effectively
affect the density of states, and also cause the hot-electron effect,
which in turn should have pronounced effect on the spin
relaxation/dephasing. To answer these questions, Jiang et al.
extended the kinetic spin Bloch equations to study the spin kinetics in the presence of
strong THz laser fields via the Floquet-Markov approach, first in
quantum dots \cite{jiang:035307} then in quantum wells \cite{jiang:prb125309}. Their investigation suggested that
the dissipation does not block large THz spin polarization
oscillations and the spin relaxation and dephasing can be effectively
manipulated by the THz field. 

In the Coulomb gauge ${\bf A}(t)={\bf
  E}\cos(\Omega t)/\Omega$ and the scalar potential $\Phi=0$, the
electron Hamiltonian in a (001) quantum well with small well width
along the $z$-axis reads 
\begin{equation}
  H_{e} = \sum_{{\bf k}\sigma\sigma^{\prime}}
  H_0^{\sigma\sigma^{\prime}}({\bf k},t) \hat{\sl c}^{\dagger}_{{\bf
      k}\sigma} \hat{\sl c}_{{\bf k}\sigma^{\prime}},
  \label{eq5.4.7-1}
\end{equation}
with
\bea
\hat{H}_0({\bf k},t) &=& \frac{[{\bf k} +e{\bf A}(t)]^2}{2m^{\ast}} \mbox{\boldmath
    $\hat{1}$\unboldmath} +
  \alpha_{\rm R}\{\hat{\sigma}_xk_y-\hat{\sigma}_y[k_x+eA(t)]\}
  \nonumber\\ &=&
\{\varepsilon_{\bf k} +\gamma_E
  k_x\Omega\cos(\Omega t) + E_{\rm em}[1+\cos(2\Omega t)]\} \mbox{\boldmath
    $\hat{1}$\unboldmath}+
  \alpha_{\rm R}(\hat{\sigma}_xk_y-\hat{\sigma}_yk_x) -
  \alpha_{\rm R}\hat{\sigma}_yeE\cos(\Omega t)/\Omega.
  \label{eq5.4.7-2}
\end{eqnarray}
Here $\varepsilon_{\bf k}=\frac{{\bf k}^2}{2m^{\ast}}$,
$\gamma_{E}=\frac{eE}{m^{\ast}\Omega^2}$ and
$E_{\rm em}=\frac{e^2E^2}{4m^{\ast}\Omega^2}$. $\Omega$ is the Thz
frequency and $\alpha_{\rm R}$ is the Rashba coefficient. It is noted that the last
term manifests that the THz electric field acts as a THz magnetic field
along the $y$ axis
\begin{equation}
  B_{\mbox{eff}}= -2\alpha_{\rm R} eE\cos(\Omega t)/\left(g\mu_B\Omega\right),
\label{beff}
\end{equation}
where $g$ is the electron $g$-factor. We will show later that this
THz-field-induced
effective magnetic field has many important effects on  spin
kinetics. The term proportional to $E_{\rm em}$ is responsible for the
dynamical Franz-Keldysh effect \cite{cheng:032107,PhysRevB.57.8860,PhysRevLett.76.4576}. This term does
  not contain any dynamic variable of the electron system and thus
  has no effect on the kinetics of the electron system.
Usually, the largest time-periodic term is the term $\gamma_E
k_x\Omega\cos(\Omega t)$, where the sideband effect mainly comes from.
Under an intense THz field, this term can be comparable to or larger than
 $\varepsilon_{\bf k}$. The Hamiltonian of the scattering remains all
 the same as before, e.g., Eqs.~(\ref{eq5.3-7}-\ref{eq5.3-9}).

The Schr\"odinger equation for electron with momentum ${\bf k}$ reads
\begin{equation}
i\partial_{t}\Psi_{{\bf k}}(t) = \hat{H}_0({\bf k},t) \Psi_{{\bf
    k}}(t).
\label{sch}
\end{equation}
According to the Floquet theory \cite{PhysRev.138.B979}, the solution to the above
equation is
\be
 \Psi_{{\bf k}\eta}(t)
= e^{i{\bf k}\cdot{\bf r}-i\varepsilon_{\bf k}t}\phi_{1}(z)
\xi_{{\bf k}\eta}(t) e^{-i[\gamma_E k_x \sin(\Omega t) + E_{\rm em}t+E_{\rm em}
\frac{\sin(2\Omega t)}{2\Omega}]}
\equiv e^{i{\bf k}\cdot{\bf r}}\Phi_\eta(z) e^{-i[\gamma_E k_x \sin(\Omega t) + E_{\rm em}t+E_{\rm em}
  \frac{\sin(2\Omega t)}{2\Omega}]},
\label{Floquet}
\ee
with $\eta=\pm$ denoting the spin branch and $\phi_{1}(z)$ being the
wavefunction of the lowest
subband. $\xi_{{\bf k}\eta}(t)=e^{-iy_{{\bf
    k}\eta}t}\sum_{n\sigma}\upsilon_{n\sigma}^{{\bf
    k}\eta}e^{in\Omega t}\chi_{\sigma}$
where $y_{{\bf k}\eta}$ and $\upsilon_{n\sigma}^{{\bf k}\eta}$
are the eigen-values and eigen-vectors of the  equation
\begin{eqnarray}
 (y_{{\bf k}\eta}-n\Omega) \upsilon_{n\sigma}^{{\bf k}\eta} &=&
   \frac{i\sigma}{2\Omega}\alpha_{\rm R} eE (\upsilon_{n-1,-\sigma}^{{\bf
      k}\eta} +\upsilon_{n+1,-\sigma}^{{\bf k}\eta})  + \alpha_{\rm R}(k_y+i\sigma k_x)\upsilon_{n,-\sigma}^{{\bf k}\eta}.
\label{floquet}
\end{eqnarray}
This equation is equivalent to Eq.~(2) in Ref.~\cite{cheng:032107}.
For each ${\bf k}$, the spinors \{$|\xi_{{\bf k}\eta}(t)\rangle$\} at
any time $t$ form a complete-orthogonal basis of the spin
space \cite{Grifoni1998229,jiang:035307,Kohler2005379}. The time evolution operator
for state ${\bf k}$ can be written as
\begin{eqnarray}
\hat{U}_0^e({\bf
  k},t,0)&=&\sum_{\eta}|\xi_{{\bf k}\eta}(t)\rangle\langle \xi_{{\bf
    k}\eta}(0)|e^{-i[\varepsilon_{\bf k}t+\gamma_E k_x \sin(\Omega t)]} e^{-i[E_{\rm em}t+E_{\rm em}\sin(2\Omega t)/(2\Omega)]}.
\end{eqnarray}

At zero temperature and in the absence of any scattering, the spectral
function is \cite{cheng:032107}
\begin{equation}
A({\bf k};t_1,t_2)=\sum_{\eta =\pm 1}\Phi_\eta({\bf
  k},t_1)\Phi_\eta^\dagger({\bf k},t_2),
\label{eq5.4.7-3}
\end{equation}
from which the density matrix reads 
\begin{equation}
\rho(t_1, t_2)=\frac{1}{(2\pi)^2}\int d{\mathbf k} A({\mathbf k};t_1,
t_2).
\label{eq5.4.7-4}
\end{equation}
Note $\rho(t_1,t_2)$ is a 2$\times$2 matrix in the spin space. In the
collinear spin space,
$\rho_{\uparrow\uparrow}=\rho_{\downarrow\downarrow}$. The average
magnetic moment from the THz field (along the $x$-axis) and the Rashba
field is given by
\begin{equation}
{\mathbf M}(T)=\left(0,
-\frac{g\mu_B}{n_{\uparrow}+n_{\downarrow}}
\int_{-\infty}^{E_f(T)}d\omega\ 
\mbox{Im}\rho_{\uparrow,\downarrow}(\omega, T), 0\right),
\label{eq5.4.7-5}
\end{equation}
with $E_f(T)$ determined from $n_{\sigma} =
\frac{1}{2\pi}\int_{-\infty}^{E_f(T)}d\omega\rho_{\sigma,\sigma}(\omega,
T)$. Here $T=(t_1+t_2)/2$ and $\omega$ is Fourier transferred from
$t=t_1-t_2$. Figure~\ref{fig5.4.7-1} shows the average magnetic moment
$M_y$ versus the time $T$ at different electric fields and THz
frequencies. It is seen that a THz magnetic signal is effectively
induced by the THz electric one.

\begin{figure}[htb]
\centerline{\includegraphics[width=6cm]{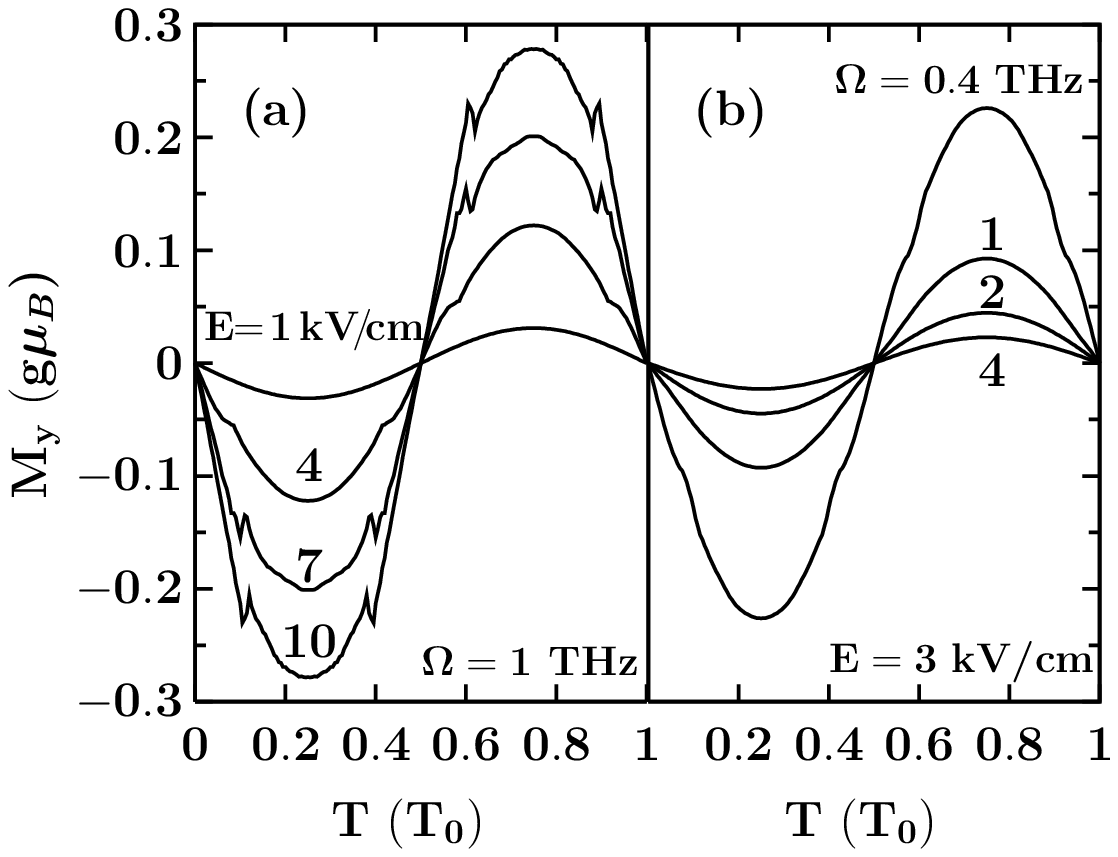}}
\caption{The average magnetic moment $M_y$ {\it vs}. the time of
InAs (001) quantum well at
    $E=1$, 4, 7 and 10 kV/cm for fixed $\Omega=1$\ THz (a) at
$\Omega=0.4$, 1, 2 and 4\ THz for fixed $E=3$\ kV/cm (b).
The electron density is $N=10^{11}$ cm$^{-2}$. From Cheng and Wu \cite{cheng:032107}.}
\label{fig5.4.7-1}
\end{figure}

To extend the kinetic spin Bloch equations to the situation with an intense THz field, Jiang
et al. pointed out \cite{jiang:prb125309} that it is insufficient to include this field
only in the driving term as in the previous investigation of the
hot-electron effect where a static electric field is applied \cite{PhysRevB.69.245320,zhou:045305}. The
correct way is to evaluate the collision integral with the Floquet
wavefunctions, i.e., the solution of the time-dependent Schr\"odinger
equation Eq.~(\ref{sch}) \cite{PhysRevE.55.300}. Moreover, the Markovian approximation should be made with
respect to the spectrum determined by the Floquet wavefunctions \cite{PhysRevE.55.300}. These
improvements constitute the Floquet-Markov
approach \cite{PhysRevE.55.300,Grifoni1998229}, which works well 
when the driven system is in dynamically stable regime and the
system-reservoir coupling can be treated perturbatively. With the
Floquet-Markov approach and by projecting the density matrix in the
Floquet picture $\hat{\rho}^{F}_{\bf k}(t) = \hat{U}_0^e{}^\dagger({\bf k},t,0)
  \hat{\rho}_{\bf k}(t) \hat{U}_0^{e}({\bf k},t,0)$, the kinetic spin Bloch equations read 
\begin{equation}
\label{kinetic spin Bloch equation}
  \partial_{t}\hat{\rho}^F_{{\bf k}}(t) =
  \left.\partial_{t}\hat{\rho}^F_{{\bf k}}(t)\right|_{\rm coh} +
  \left.\partial_{t}\hat{\rho}^F_{{\bf k}}(t)\right|_{\rm scat},
\end{equation}
where $\partial_{t}\hat{\rho}^F_{{\bf k}}(t)|_{\rm coh}$ and
$\partial_{t}\hat{\rho}^F_{{\bf k}}(t)|_{\rm scat}$ are the coherent and
scattering terms, respectively. The coherent terms, which describe the coherent precession determined
by the electron Hamiltonian $H_{e}$ and the Hartree-Fock
contribution of the electron-electron Coulomb interaction, can be written as
\begin{eqnarray}
\left. \partial_{t}\hat{\rho}^{F}_{{\bf k}}(t)\right|_{\rm coh} &=&
i\sum_{{\bf
      k}^{\prime},q_z} V_{{\bf k}-{\bf
      k}^{\prime},q_z}|I(iq_z)|^2 \big{[}\hat{S}_{{\bf k},{\bf
      k}^{\prime}}(t,0)\hat{\rho}^{F}_{{\bf k}^{\prime}}(t)
    \hat{S}_{{\bf k}^{\prime},{\bf k}}(t,0),\ \hat{\rho}^{F}_{{\bf
    k}}(t)\big{]}.
\end{eqnarray}
Here $\hat{\Sigma}_{\rm HF}^F({\bf k},t)=-\sum_{{\bf
    k}^{\prime},q_z}V_{{{\bf k}-{\bf
      k}^{\prime}},q_z}|I(iq_z)|^2\hat{\rho}^F_{{\bf k}^{\prime}}(t)$ is the Coulomb
 Hartree-Fock self-energy. The scattering terms are
composed of terms due to the electron-impurity
($\left.\partial_{t}\rho_{{\bf k}}^F\right|_{\rm ei}$), electron-phonon
($\left.\partial_{t}\rho_{{\bf k}}^F\right|_{\rm ep}$) and
electron-electron ($\left.\partial_{t}\rho_{{\bf k}}^F\right|_{\rm ee}$) scatterings, respectively.
Under the generalized Kadanoff-Baym Ansatz \cite{haugjauho}, 
these scattering terms read \cite{jiang:prb125309}
\begin{eqnarray}\nonumber
 \left. \partial_{t}\rho^{F(\eta\eta^{\prime})}_{{\bf
       k}}(t)\right|_{\rm ei}
 &=& -
\sum_{{\bf k}^{\prime},q_z,n}^{\eta_1\eta_2\eta_3} \pi n_i U^2_{{\bf k}-{\bf
      k}^{\prime},q_z} |I(iq_z)|^2  \bigg{[} \Big{\{} S^{(\eta\eta_1)}_{{\bf k},{\bf
          k}^{\prime}}(t,0)S^{(n)(\eta_2\eta_3)}_{{\bf
      k}^{\prime},{\bf k}} \delta(n\Omega+\bar{\varepsilon}_{{\bf
      k}^{\prime}\eta_2} -\bar{\varepsilon}_{{\bf k}\eta_3} )\\
  &&\quad\times
  \Big{(}\rho^{>F(\eta_1\eta_2)}_{{\bf
 k}^{\prime}}(t) \rho^{<F(\eta_3\eta^{\prime})}_{{\bf k}} (t)
  -\rho^{<F(\eta_1\eta_2)}_{{\bf
 k}^{\prime}}(t) \rho^{>F(\eta_3\eta^{\prime})}_{{\bf k}}(t) \Big{)} \Big{\}}
  +\Big{\{}\eta\leftrightarrow\eta^{\prime}\Big{\}}^{\ast} \bigg{]},
\label{scatei}
\end{eqnarray}
\begin{eqnarray}\nonumber
 \left. \partial_{t}\rho^{F(\eta\eta^{\prime})}_{{\bf
       k}}(t)\right|_{\rm ep}
&=& -{\sum_{{\bf k}^{\prime},q_z,n,\lambda,\pm}^{\eta_1\eta_2\eta_3} \pi
  |M_{\lambda, {\bf k}-{\bf k}^{\prime},q_z}|^2 |I(iq_z)|^2 }
  \bigg{[} \Big{\{}
  S^{(\eta\eta_1)}_{{\bf k},{\bf k}^{\prime}}(t,0)
  S^{(n)(\eta_2\eta_3)}_{{\bf k}^{\prime},{\bf k}}
  e^{\mp it\omega_{\lambda, {\bf k}-{\bf k}^{\prime},q_z}}
  \delta(\pm\omega_{\lambda, {\bf k}-{\bf
      k}^{\prime},q_z}+n\Omega + \bar{\varepsilon}_{{\bf
      k}^{\prime}\eta_2} - \bar{\varepsilon}_{{\bf k}\eta_3} )
  \\ &&\quad \times \Big{(}
   N^{\pm}_{\lambda, {\bf k}-{\bf k}^{\prime},q_z}\rho^{>F(\eta_1\eta_2)}_{{\bf k}^{\prime}}(t)
   \rho^{<F(\eta_3\eta^{\prime})}_{\bf k}(t) -N^{\mp}_{\lambda, {\bf k}-{\bf k}^{\prime},q_z} \rho^{<F(\eta_1\eta_2)}_{{\bf
       k}^{\prime}}(t) \rho^{>F(\eta_3\eta^{\prime})}_{\bf k}(t)
   \Big{)} \Big{\}} +
{\Big{\{}\eta\leftrightarrow\eta^{\prime}\Big{\}}^{\ast}} \bigg{]},
\label{scatep}
\end{eqnarray}
\begin{eqnarray}\nonumber
\left.\partial_{t}\rho^{F(\eta\eta^{\prime})}_{{\bf k}}(t)\right|_{\rm
  ee}
&=& -  \sum_{{\bf k}^{\prime},{\bf k}^{\prime\prime},n,n^{\prime}}^{\eta_1...\eta_7} \pi \Big{[}\sum_{q_z} V_{{\bf k}-{\bf k}^{\prime},q_z}
  |I(iq_z)|^2 \Big{]}^2
  \bigg{[} \Big{\{}
    T^{(\eta\eta_1)}_{{\bf k},{\bf k}^{\prime}}(t,0)
    T^{(n^{\prime})(\eta_2\eta_3)}_{{\bf k}^{\prime},{\bf
        k}} T^{(n-n^{\prime})(\eta_4\eta_5)}_{{\bf k}^{\prime\prime},{\bf
        k}^{\prime\prime}-{\bf k}+{\bf k}^{\prime}} T^{(\eta_6\eta_7)}_{{\bf
        k}^{\prime\prime}-{\bf k}+{\bf k}^{\prime},{\bf
    k}^{\prime\prime}}(t,0) \\\nonumber
&&\quad \times 
\delta(n\Omega+
\bar{\varepsilon}_{{\bf k}^{\prime}\eta_2}-\bar{\varepsilon}_{{\bf
    k}\eta_3}+\bar{\varepsilon}_{{\bf k}^{\prime\prime}\eta_4}
  -\bar{\varepsilon}_{{\bf k}^{\prime\prime}-{\bf k}+{\bf
    k}^{\prime}\eta_5} )\Big{(} \rho^{>F(\eta_1\eta_2)}_{{\bf k}^{\prime}}(t)
\rho^{<F(\eta_3\eta^{\prime})}_{\bf k}(t)
\rho^{<F(\eta_5\eta_6)}_{{\bf
        k}^{\prime\prime}-{\bf k}+{\bf k}^{\prime}}(t)
\rho^{>F(\eta_7\eta_4)}_{{\bf k}^{\prime\prime}}(t)\\ && \quad -\rho^{<F(\eta_1\eta_2)}_{{\bf k}^{\prime}}(t)
\rho^{>F(\eta_3\eta^{\prime})}_{\bf k}(t)
\rho^{>F(\eta_5\eta_6)}_{{\bf
        k}^{\prime\prime}-{\bf k}+{\bf k}^{\prime}}(t)
\rho^{<F(\eta_7\eta_4)}_{{\bf k}^{\prime\prime}}(t) \Big{)} \Big{\}}
    +{\Big{\{}\eta\leftrightarrow\eta^{\prime}\Big{\}}^{\ast}} \bigg{]}.
\label{scatee}
\end{eqnarray}
In these equations, $N^{\pm}_{\lambda, {\bf k}-{\bf k}^{\prime},q_z}=N_{\lambda, {\bf k}-{\bf
    k}^{\prime},q_z}+\frac{1}{2}(1 \pm 1)$ stands for the phonon number, $n_i$ is the
impurity density, $\hat{\rho}^{>}_{\bf k} = \hat{{\bf
    1}} - \hat{\rho}_{\bf k}$,
$\hat{\rho}^{<}_{\bf k} = \hat{\rho}_{\bf
    k}$,  and $\bar{\varepsilon}_{{\bf k}\eta}=\varepsilon_{{\bf
    k}} + y_{{\bf k}\eta}$.
\be
 S^{(\eta_1\eta_2)}_{{\bf k}^{\prime},{\bf
      k}}(t,0)=\langle \xi_{{\bf
    k}^{\prime}\eta_1}(t)|\xi_{{\bf k}\eta_2}(t)\rangle e^{i[(\varepsilon_{{\bf k}^{\prime}}-\varepsilon_{\bf
      k})t+\gamma_E\sin(\Omega t)(k_x^{\prime}-k_x)]}
 = \sum_{n} S^{(n)(\eta_1\eta_2)}_{{\bf
    k}^{\prime},{\bf k}}
e^{i t(n\Omega+\bar{\varepsilon}_{{\bf
      k}^{\prime}\eta_1}-\bar{\varepsilon}_{{\bf k}\eta_2}) },
\ee
with
\begin{equation}
S^{(n)(\eta_1\eta_2)}_{{\bf k}^{\prime},{\bf k}} =
\sum_{m\sigma}F^{{\bf k}^{\prime}\eta_1 \ast}_{m\ \sigma}\ F^{{\bf k}\eta_2}_{n+m\ \sigma}.
\end{equation}
Here $F^{{\bf k}\eta}_{n\ \sigma}=\sum_{m} \upsilon^{{\bf
    k}\eta}_{n+m\ \sigma} J_{m}(\gamma_E k_x)$ with $J_{m}(x)$ standing
    for the $m$-th order Bessel function.
\begin{eqnarray}
 T^{(\eta_1\eta_2)}_{{\bf k}^{\prime},{\bf k}}(t,0)
 &=&\langle \xi_{{\bf
    k}^{\prime}\eta_1}(t)|\xi_{{\bf k}\eta_2}(t)\rangle
e^{i(\varepsilon_{{\bf k}^{\prime}}-\varepsilon_{\bf k})t}  = \sum_{n} T^{(n)(\eta_1\eta_2)}_{{\bf
    k}^{\prime},{\bf k}}
e^{i t(n\Omega+\bar{\varepsilon}_{{\bf
      k}^{\prime}\eta_1}-\bar{\varepsilon}_{{\bf k}\eta_2}) },
\end{eqnarray}
with
\begin{equation}
T^{(n)(\eta_1\eta_2)}_{{\bf k}^{\prime},{\bf k}} =
\sum_{m\sigma}\upsilon^{{\bf k}^{\prime}\eta_1 \ast}_{m\ \sigma}\
\upsilon^{{\bf k}\eta_2}_{n+m\ \sigma}.
\end{equation}
$\{\eta\leftrightarrow\eta^{\prime}\}$ stands for the
same terms as in the previous $\{\}$ but with the interchange
$\eta\leftrightarrow\eta^{\prime}$. The term of the
electron-electron scattering is
quite different from those of the electron-impurity  and
electron-phonon scattering, as the momentum
  conservation eliminates the term of  $e^{i\gamma_E\sin(\Omega t)k_x}$.

The above equations clearly show the sideband effects, i.e., $n\Omega$
in the $\delta$-functions. The extra energy, $n\Omega$, is provided by
the THz field during each scattering process. This makes transitions
from the low-energy states (small $k$) to high-energy ones (large
$k$) become possible, even through the elastic electron-impurity
scattering. These processes are the sideband-modulated scattering
processes.

The kinetic spin Bloch equations are solved numerically. After that,
$\rho^{F(\eta\eta^{\prime})}_{{\bf k}}(t)$ for each ${\bf k}$ is
obtained. From
\begin{equation}
\rho^{F(\eta\eta^{\prime})}_{{\bf
    k}}(t) = \langle \xi_{{\bf k}\eta}(0) |\hat{U}_0^{e\ \dagger}({\bf k},t,0)
  \hat{\rho}_{\bf k}(t) \hat{U}_0^{e}({\bf k},t,0)
  |\xi_{{\bf k}\eta^{\prime}}(0)\rangle=\langle \xi_{{\bf k}\eta}(t) | \hat{\rho}_{\bf
  k}(t) |\xi_{{\bf k}\eta^{\prime}}(t)\rangle,
\end{equation}
by performing a unitary transformation, one comes to
the single particle
density matrix $\hat{\rho}_{\bf k}(t)$ in the collinear basis
\{$|\sigma\rangle$\} which is composed by the
eigen-states of $\hat{\sigma}_{z}$. In this
spin space, the spin polarization along any direction can be obtained
readily, e.g., $S_z=\sum_{{\bf k}}\frac{1}{2}(\rho_{\bf
  k}^{\uparrow\uparrow}-\rho_{\bf k}^{\downarrow\downarrow})$,
$S_x=\sum_{{\bf k}}\mbox{Re}\{\rho_{\bf k}^{\uparrow\downarrow}\}$ and 
$S_y=-\sum_{{\bf k}}\mbox{Im}\{\rho_{\bf k}^{\uparrow\downarrow}\}$.
From the temporal evolution of $S_z$, the spin relaxation time is
extracted.

\begin{figure}[htb]
\centerline{\includegraphics[width=8cm]{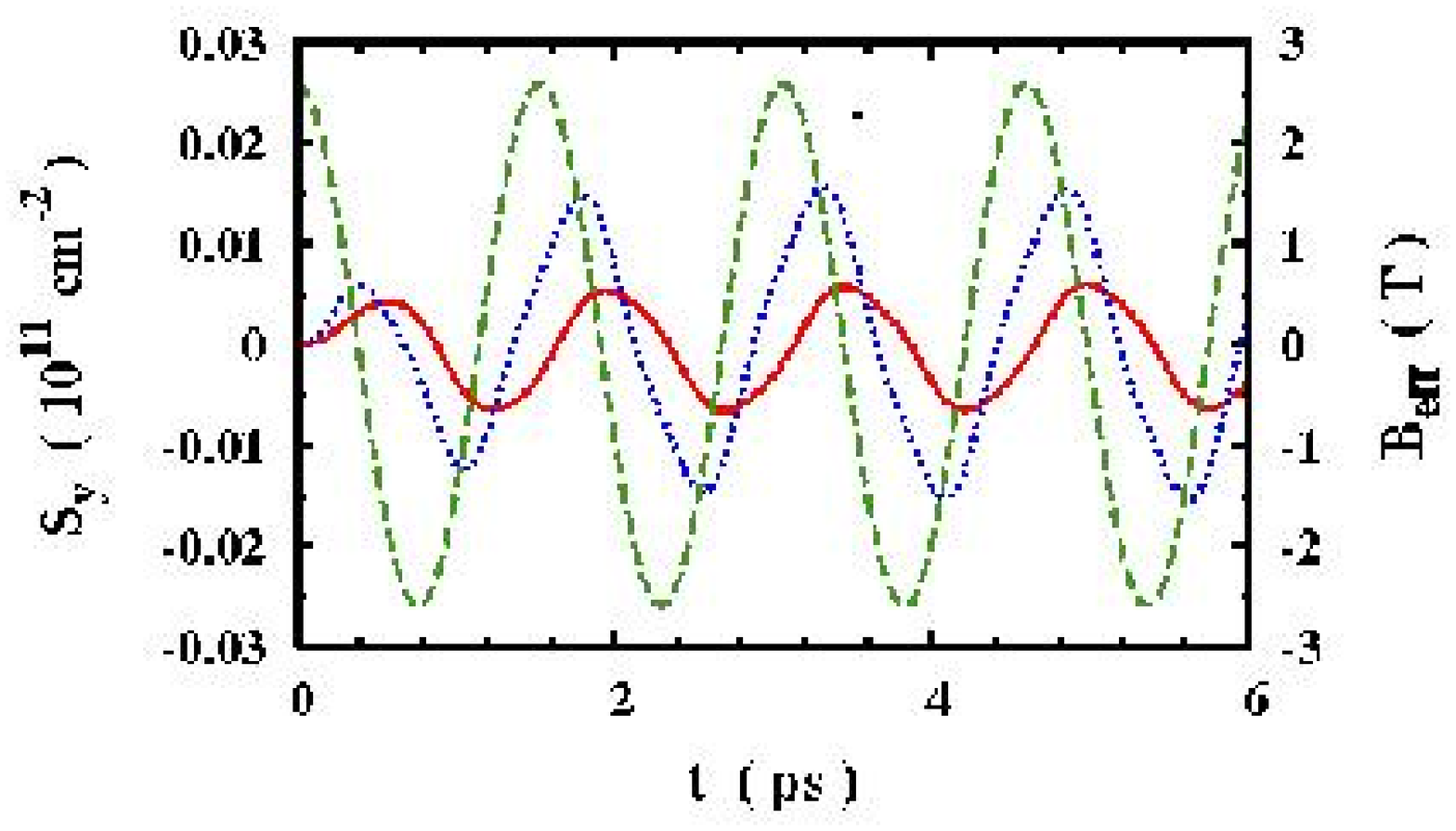}}
  \caption{ Spin polarization along $y$ axis,
$S_y$, as function of time with zero initial spin polarization for $E=0.5$\ kV/cm (solid curve) and
1.0\ kV/cm (dotted curve) in InAs (001) quantum well. $T=50$\ K and
$N_i=0.05 N_e$. The dashed curve is the
  THz-field-induced effective magnetic field $B_{\mbox{eff}}$ with $E=1.0$ kV/cm. Note
  that the scale of the dashed curve is on the right-hand side of the
  frame. From Jiang et al. \cite{jiang:prb125309}.}
\label{fig5.4.7-2}
\end{figure}

By numerically solving the kinetic spin Bloch equations with all the scattering included,
Jiang et al. showed that with dissipation, the THz field can
still pump a large (several percent) spin polarization which
oscillates at the same frequency with the THz field, as shown in
Fig.~\ref{fig5.4.7-2}. This feature coincides with what predicted by
Cheng and Wu in the dissipation-free investigation \cite{cheng:032107}. What differs from
the previous case is shown in Fig.~\ref{fig5.4.7-2} that there is a
delay of $S_y$ with respect to the THz-field induced magnetic field
$B_{\rm eff}$. This delay is due to the retarded response of the spin
polarization to the spin pumping caused by the THz field. The amplitude
of the steady-state spin polarization $S_y^0$ (the peak value of
$S_y$) depends on the THz field strength and the THz frequency. These
features have been addressed in detail in Ref.~\cite{jiang:prb125309}.

\begin{figure}[htb]
\centerline{\includegraphics[width=4.5cm]{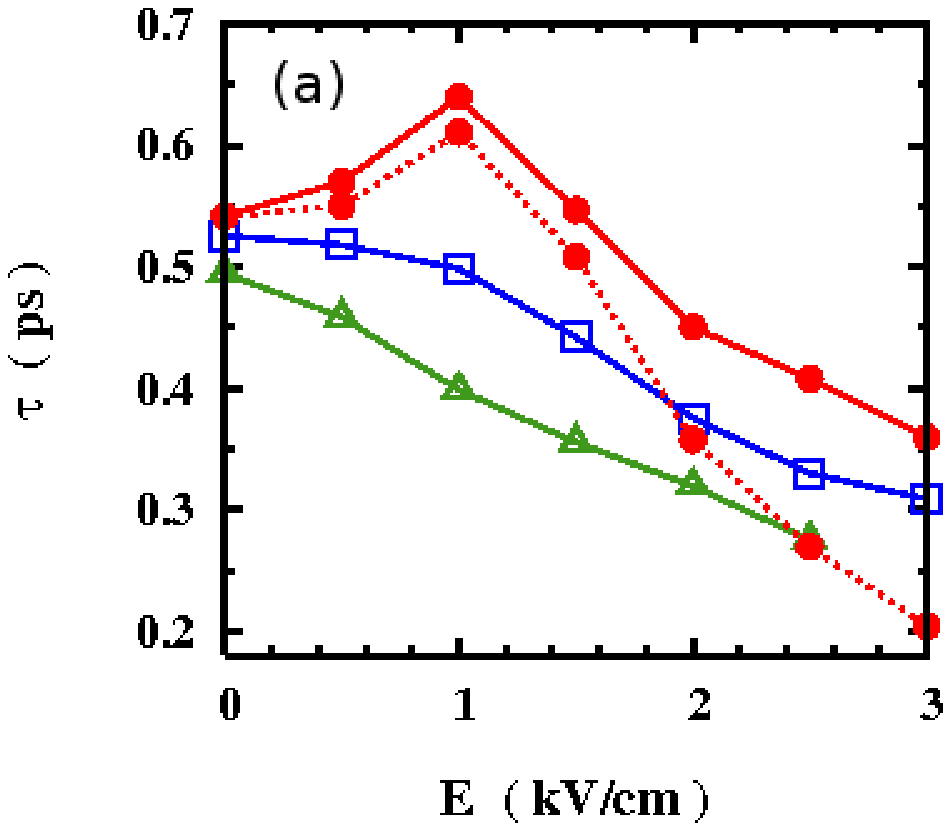}\includegraphics[width=4.5cm]{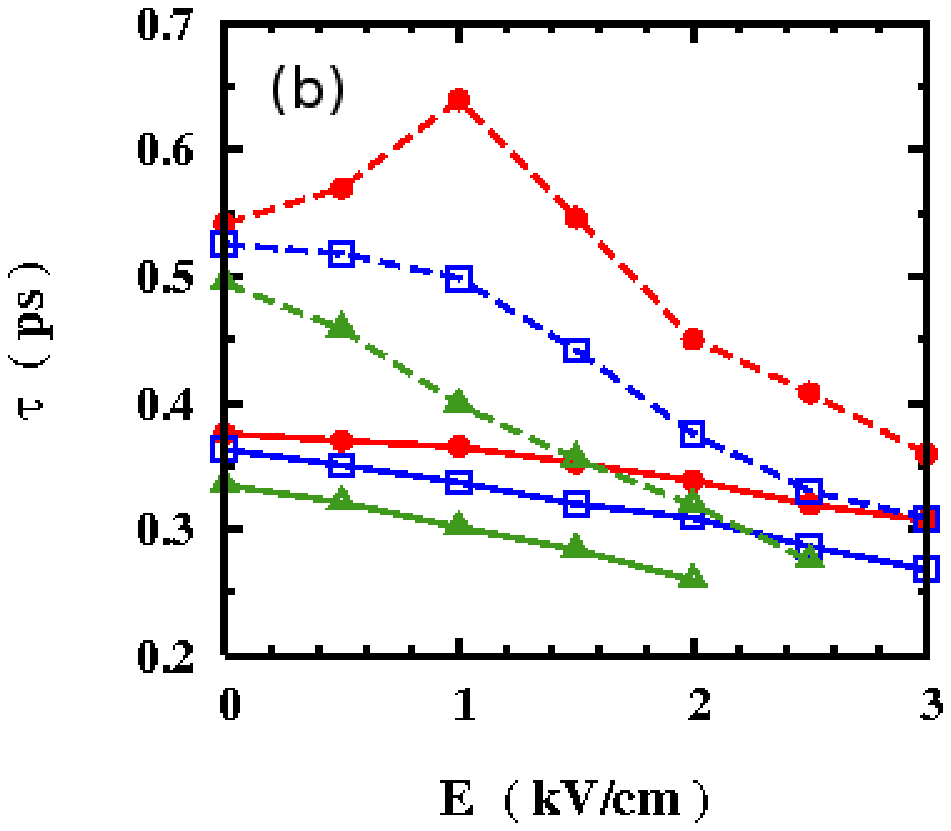}\includegraphics[width=4.5cm]{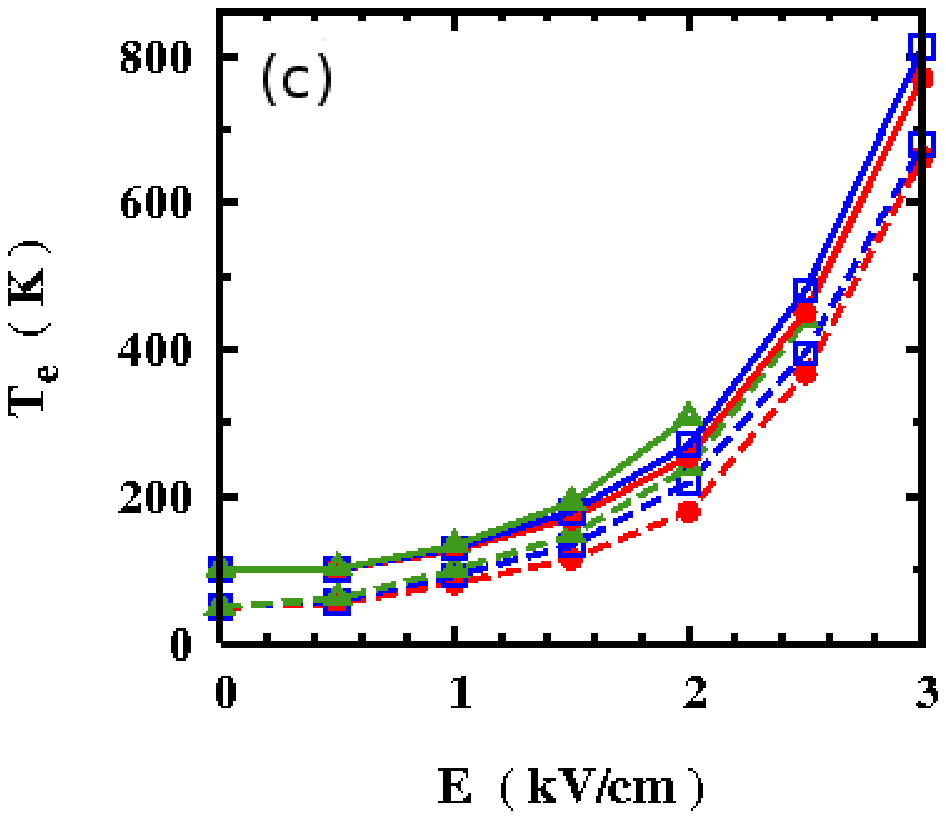}}
  \caption{ InAs (001) quantum wells: 
(a) Dependence of the spin relaxation time $\tau$
  on THz field strength for impurity densities: $N_i=0$ ($\bullet$);
  $N_i=0.02N_e$ ($\square$); $N_i=0.05N_e$ ($\triangle$). Solid
  curves: from full calculation; Dotted curve: from the calculation
 without the THz-field-induced effective magnetic field
 $B_{\mbox{eff}}$. (b) Dependence of the spin relaxation time $\tau$
  on THz field strength for impurity densities: $N_i=0$ ($\bullet$);
$N_i=0.02N_e$ ($\square$); and $N_i=0.05N_e$ ($\triangle$). $T=100$~K
  (solid curves) and $50$~K (dashed curves). (c) Dependence of the hot-electron temperature $T_e$
  on THz field strength for different impurity densities: $N_i=0$ ($\bullet$
  ); $N_i=0.02N_e$ ($\square$); $N_i=0.05N_e$ ($\triangle$). $T=100$~K
  (solid curves) and $T=50$~K (dashed curves). From Jiang et al.
  \cite{jiang:prb125309}.}
\label{fig5.4.7-3}
\end{figure}

Apart from pumping THz spin polarizations, the THz field can
effectively manipulate the electron density of states and cause the
hot-electron effect. Both in turn can lead to the manipulation of the
spin relaxation and dephasing. Figure~\ref{fig5.4.7-3} shows the
dependence of the spin relaxation time on the THz field strength at
different impurity densities (a) and temperatures (b), together with
the corresponding hot-electron temperatures (c). Two consequences of
the THz field lead to the rich behaviors in the figure: (i) the total
THz field-induced effective magnetic field $B$ [$B=B_{\rm eff}+B_{\rm av}$
with $B_{\rm av}(t)=2\alpha_{\rm R}\langle k_x\rangle/(|g|\mu_B)$ from the
Rashba spin-orbit coupling] and (ii) the hot-electron effect. Effect (i)
can give a magnetic field as large as several tesla [2.6~T per 1 kV/cm
THz field with $\nu\equiv\Omega/(2\pi)=0.65$~THz]. This effective magnetic
field blocks the inhomogeneous broadening from the Rashba spin-orbit
coupling and thus elongates the spin relaxation time. Effect (ii)
leads to the enhancement of momentum scattering as well as the
inhomogeneous broadening, while the enhancement of the inhomogeneous
broadening tends to shorten the spin relaxation time, the boost of the
scattering tends to increase (or decrease) the spin relaxation time in
the strong (or weak) scattering limit as discussed in the previous
sections. Similarly the spin relaxation time can also be manipulated
by the THz frequency as shown in Fig.~\ref{fig5.4.7-4}.

\begin{figure}[htb]
\centerline{\includegraphics[height=4.5cm]{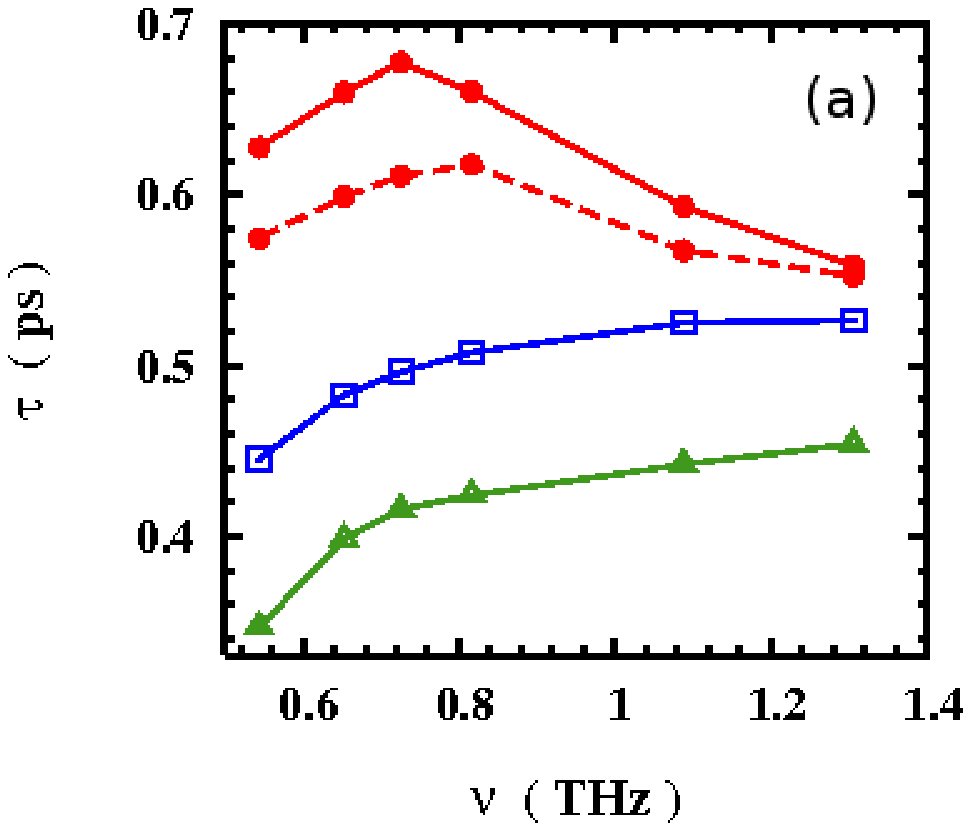}\includegraphics[height=4.5cm]{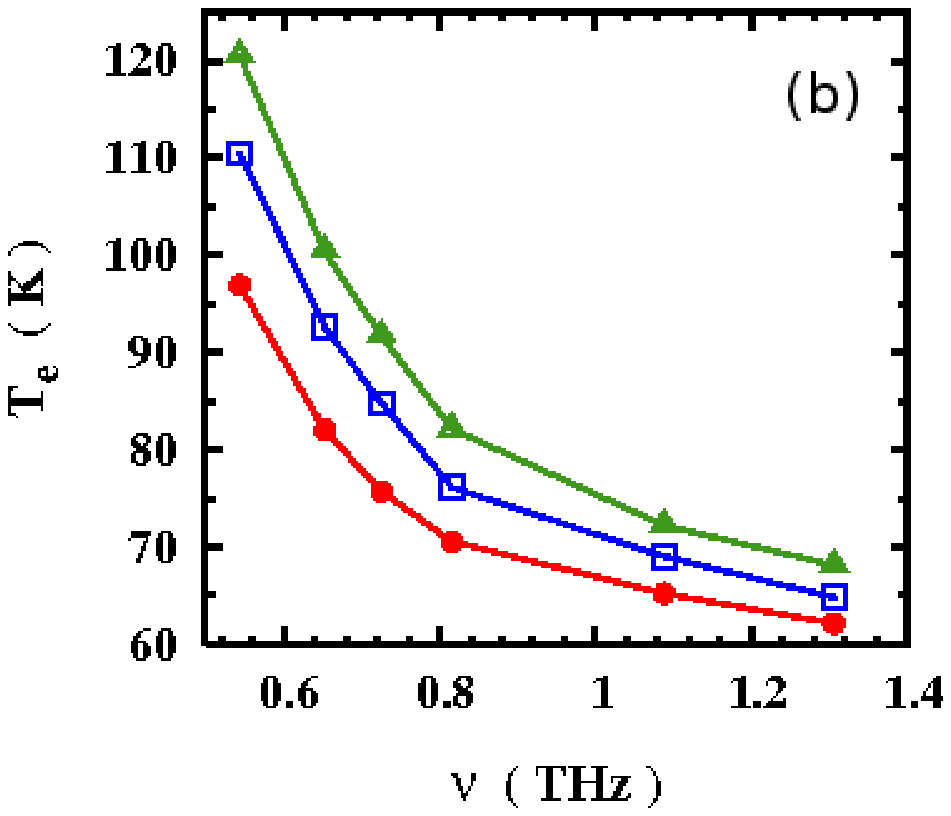}}
\caption{InAs (001) quantum wells:
 (a) Dependence of the spin relaxation time $\tau$
  on THz frequency for impurity densities: $N_i=0$
  ($\bullet$); $N_i=0.02N_e$ ($\square$); and $N_i=0.05N_e$
  ($\triangle$). Solid curves: from full calculation; Dotted curve:
  from the calculation without the THz-field-induced effective
  magnetic field $B_{\mbox{eff}}$. $E=1$~kV/cm and $T=50$~K. (b)
  Dependence of the hot-electron temperature $T_e$ 
  on THz frequency for impurity densities: $N_i=0$
  ($\bullet$); $N_i=0.02N_e$ ($\square$); and $N_i=0.05N_e$
  ($\triangle$). $E=1$~kV/cm and $T=50$~K. From Jiang et al. \cite{jiang:prb125309}.}
\label{fig5.4.7-4}
\end{figure}

\subsubsection{Spin relaxation and dephasing in GaAs (110) quantum wells}
\label{sec5.4.8}
In symmetric GaAs (110) quantum wells, the D'yakonov-Perel' term mainly comes from
the Dresselhaus term which reads
\begin{eqnarray}
g\mu_B\Omega_x^{nn^\prime}({\bf k})&=&\gamma_{\rm D}[-(k_x^2+2k_y^2)\langle
n|k_z|n^\prime\rangle+\langle n|k_z^3|n^\prime\rangle],\\
g\mu_B\Omega_y^{nn^\prime}({\bf k})&=&4\gamma_{\rm D}k_xk_y\langle n|k_z|n^\prime\rangle,\\
g\mu_B\Omega_z^{nn^\prime}({\bf k})&=&\gamma_{\rm D}(k_x^2-2k_y^2-\langle n|k_z^2|n^\prime\rangle)\delta_{nn^\prime},
\end{eqnarray}
where $\langle n| k_z^m|n^\prime\rangle=\int
dz\phi_n^\ast(z)(-i\partial/\partial z)^m\phi_{n^\prime}(z)$ with
$\phi_n(z)$ representing the envelope function of the electron in $n$-th
subband. As $\langle n|k_z|n\rangle=\langle n|k_z^3|n\rangle=0$, when
only the lowest subband is relevant, the effective magnetic field
${\bf \Omega}({\bf k})$ is along the $z$-axis. Therefore, it was
proposed both theoretically and experimentally \cite{PhysRevB.64.161301,boggess:1333,PhysRevLett.83.4196,Adachi200136} that the D'yakonov-Perel'
mechanism can not cause any spin relaxation if the spin polarization
is along the $z$-axis. 

Wu and Kuwata-Gonokami first pointed out that if a magnetic field in
the Voigt configuration is applied in the system, there is spin
relaxation and dephasing due to the D'yakonov-Perel' mechanism \cite{Wu2002509}. This can be
seen from the coherent term of the kinetic spin Bloch equations \cite{Wu2002509}:
\begin{eqnarray}
&&\hspace{-0.59cm}\left.\frac{\partial f_{{\bf k}\sigma}}{\partial t}\right|_{\rm coh}=-g\mu_BB\mbox{Im}\rho_{{\bf k},\sigma-\sigma}
+2\mbox{Im}\sum_{\bf q}V_{\bf q}\rho_{{\bf k},\sigma-\sigma}\rho_{{\bf k}+{\bf q},-\sigma
\sigma}, \label{eq5.4.8-1}
\\ &&\hspace{-1cm}\left.\frac{\partial \rho_{{\bf k},\sigma-\sigma}}{\partial t}\right|_{\rm coh}=\frac{i}{2}g\mu_BB(f_{{\bf k}\sigma}
-f_{{\bf k}-\sigma})-i\sigma\omega({\bf k})\rho_{{\bf
    k},\sigma-\sigma} +i\sum_{\bf q}V_{\bf q}(f_{{\bf k}+{\bf q}\sigma}-f_{{\bf k}+{\bf q}
-\sigma}) \rho_{{\bf k},\sigma-\sigma} -i\sum_{\bf q}V_{\bf q}(f_{{\bf
k}\sigma}-f_{{\bf k}-\sigma})\rho_{{\bf k}+{\bf q},\sigma-\sigma},\nonumber\\
\label{eq5.4.8-2}
\end{eqnarray}
where $B$ is along the $x$-axis and $\omega({\bf
  k})=-g\mu_B\Omega_z^{11}({\bf k})$. It is seen that in the presence
of the magnetic field, the spin coherence $\rho_{{\bf
    k},\sigma-\sigma}$ is involved and $\omega({\bf k})$ in
Eq.~(\ref{eq5.4.8-2}) provides the inhomogeneous broadening which leads to
the spin relaxation and dephasing \cite{Wu2002509}. Experimentally D\"ohrmann
et al. \cite{PhysRevLett.93.147405,Hagelebook} observed a ``turn on'' of the spin relaxation by
switching on the magnetic field.

The spin relaxation in the absence of the magnetic field in symmetric
GaAs (110) quantum wells with small well width is an interesting
problem and has been studied extensively \cite{Ohno2000817,Adachi200136,PhysRevLett.93.147405,Hagelebook,PhysRevLett.91.246601,bel'kov:176806,olbrich:245329,muller:206601}. In most of these works,
the main reason limiting the spin relaxation time is attributed to the
Bir-Aronov-Pikus mechanism. In the spin noise spectroscopy measurements by M\"uller
et al. \cite{muller:206601}, the excitation of the semiconductor is
negligible and hence the Bir-Aronov-Pikus mechanism is avoided. They reported spin
relaxation about 24~ns and attributed it to the D'yakonov-Perel' mechanism due to
the random Rashba fields caused by fluctuations in the donor density
first proposed by Sherman \cite{sherman:209,PhysRevB.67.161303}. Zhou
and Wu further investigated this effect using the kinetic spin Bloch equations, where the contribution of
the Coulomb scattering originally missed in Refs.~\cite{sherman:209,PhysRevB.67.161303} is
included \cite{zhoupreprint}. The method to incorporate the effect of the Random
Rashba spin-orbit coupling in the kinetic spin Bloch equation approach is to calculate the
time evolutions of the kinetic spin Bloch equations under sufficient Rashba coefficients
$\alpha_{\rm R}$, which are assumed to satisfy the Gaussian distribution
$P(\alpha_{\rm R})=\frac{1}{\sqrt{2\pi}\Delta}e^{-\alpha_{\rm R}^2/(2\Delta^2)}$. The
spin relaxation time is obtained by the slope of the coherently summed
spin polarization
\begin{equation}
S_z=\int d\alpha_{\rm R} P(\alpha_{\rm R})\sum_{\bf k}\frac{1}{2}(f_{{\bf
    k}\uparrow}^{\alpha_{\rm R}}-f_{{\bf k}\downarrow}^{\alpha_{\rm R}}).
\end{equation}
A calculation based on the kinetic spin Bloch equations can nicely reproduce the experimental
findings by M\"uller et al. \cite{muller:206601} in
Fig.~\ref{fig5.4.8-1}(a). They also showed that at low impurity
density, the Coulomb scattering has a strong influence to the spin
relaxation, as shown in Fig.~\ref{fig5.4.8-1}(b), in which a peak due
to the Coulomb scattering in the temperature dependence of the spin
relaxation time was predicted.

Besides the Random Rashba spin-orbit coupling field, Zhou and Wu
proposed a virtual intersubband spin-flip spin-orbit coupling induced
spin relaxation in GaAs (110) quantum wells and showed this mechanism
becomes important for samples with high impurity density \cite{Zhou2009}.

\begin{figure}[htb]
\centerline{\includegraphics[height=4.5cm]{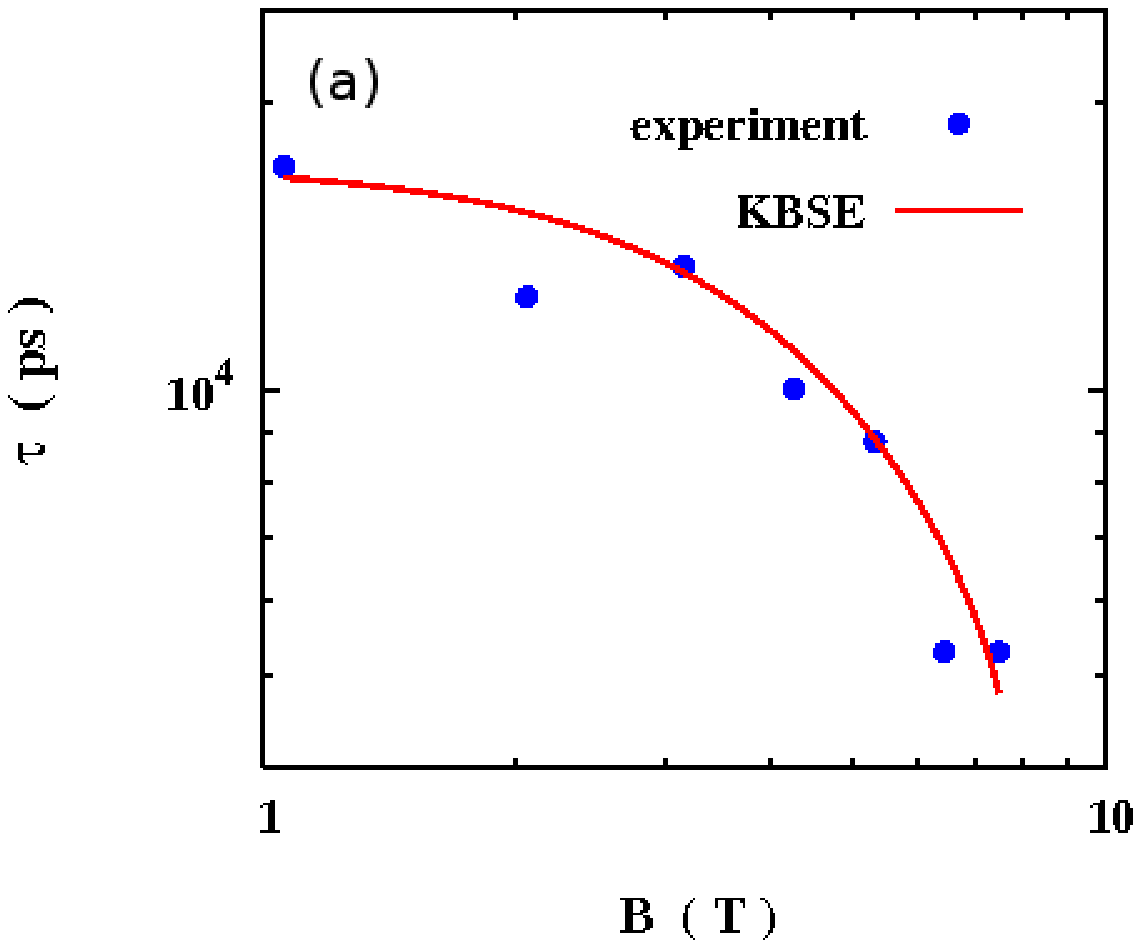}
\includegraphics[height=4.5cm]{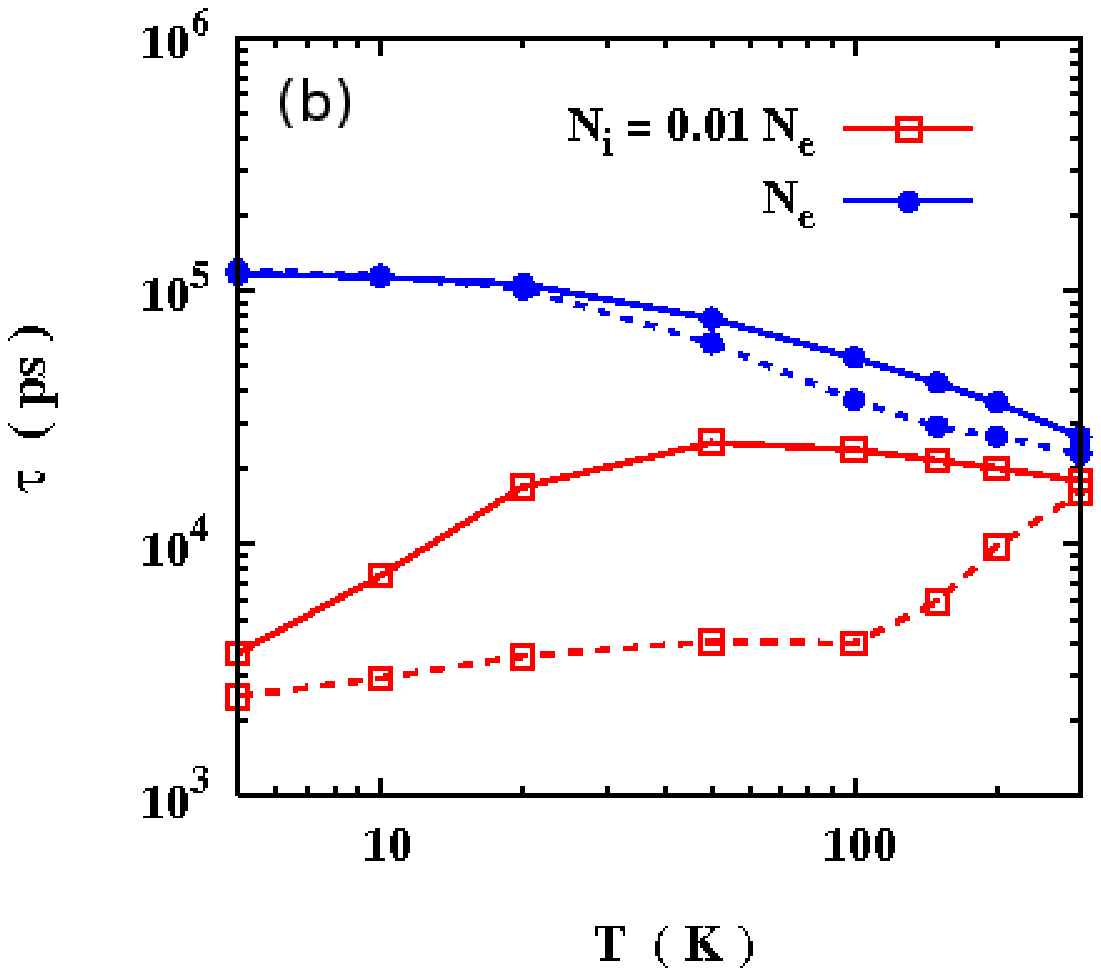}}
\caption{GaAs (110) quantum wells:
 (a) Spin relaxation times {\sl vs.} the in-plane magnetic field
    strength from the KBSE approach and from the experimental data in
    Ref.~\cite{muller:206601}
    with temperature $T=20$~K, well width $a=16.8$~nm, electron density
    $N_e=1.8\times 10^{11}$~cm$^{-2}$ and impurity density
    $N_i=0.01~N_e$. The fitting parameters are $E_w=4.2$~kV/cm,
    $\gamma_{\rm D}=5.02$~eV$\cdot${\AA}$^3$. (b) Spin relaxation
    times due to the random Rashba spin-orbit coupling induced D'yakonov-Perel' mechanism {\sl vs.}
    temperature $T$ for impurity densities $N_i=N_e$ and
    $0.01~N_e$ with $B=0$, $a=16.8$~nm and
    $N_e=1.8\times 10^{11}$~cm$^{-2}$. The corresponding Fermi
    temperature is $75$~K. The solid (dashed) curves represent the
    results calculated with (without) the electron-electron Coulomb
    scattering. From Zhou and Wu \cite{Zhou2009}.}
\label{fig5.4.8-1}
\end{figure}

\subsubsection{Spin dynamics in paramagnetic Ga(Mn)As quantum wells}
\label{sec5.4.9}
Jiang et al. further extended the kinetic spin Bloch equations to study the electron
spin relaxation in paramagnetic Ga(Mn)As quantum wells
\cite{jiang:155201}. Besides the D'yakonov-Perel' \cite{dp,DP2} and Bir-Aronov-Pikus \cite{BAP,aronov} mechanisms, they further included the
spin relaxation mechanism due to the exchange coupling of the
electrons and the localized Mn spins (the s-d exchange scattering
mechanism) \cite{jungwirth:809} and the Elliott-Yafet mechanism \cite{PhysRev.85.478,PhysRev.96.266} due to the presence of
the heavily doped Mn. In Ga(Mn)As, the Mn dopants can be either
substitutional or interstitial; the substitutional Mn accepts one
electron whereas the interstitial Mn releases two. Direct doping in
low-temperature molecular-beam epitaxy growth gives
rise to more substitutional Mn ions than interstitial ones, which
makes the Ga(Mn)As a $p$-type semiconductor \cite{jungwirth:809,furdyna,balk,PhysRevB.72.235313}. In GaAs quantum
wells near Ga(Mn)As layer, the Mn dopants can diffuse into the GaAs
quantum wells, where the Mn ions mainly take the interstitial
positions, making the quantum well $n$-type \cite{PhysRevLett.92.037201,Schulz20082163,wuarxiv}.

The kinetic spin Bloch equations are the same as those in GaAs quantum wells, except the
additional terms in the coherent and scattering terms associated with
the Mn spin ${\bf S}$ and the Elliott-Yafet mechanism. The additional coherent
term reads 
\begin{eqnarray}
  \left. \partial_t \hat{\rho}_{\bf k}\right|^{\rm coh}_{\rm Mn} = - i [
    \hat{H}_{\rm mf}^{\rm sd}, \hat{\rho}_{\bf k}],
\end{eqnarray}
with $\hat{H}_{\rm sd}^{\rm mf}=-N_{\rm Mn}\alpha\langle{\bf
  S}\rangle\cdot\frac{\spin}{2}$. Here $\langle{\bf S}\rangle$ is the
average spin polarization of Mn ions and $\alpha$ is the $s$-$d$
exchange coupling constant. For simplicity, Mn ions are assumed to be
uniformly distributed within and around the quantum wells with a bulk
density $N_{\rm Mn}$. The additional scattering terms are those from
the $s$-$d$ exchange scattering and the Elliott-Yafet mechanism. The $s$-$d$
exchange scattering term is given by 
\be
  \left. \partial_t \hat{\rho}_{\bf k}\right|_{\rm sd}^{\rm scat} =
  -\pi N_{\rm Mn} \alpha^2 I_s \sum_{\eta_1 \eta_2 {\bf
      k}^{\prime}} G_{\rm Mn}(-\eta_1{-\eta_2})
  \delta(\varepsilon_{\bf k}-\varepsilon_{{\bf k}^{\prime}})
  (\hat{s}^{\eta_1} \hat{\rho}^>_{{\bf k}^{\prime}}
    \hat{s}^{\eta_2} \hat{\rho}^<_{\bf k}
    - \hat{s}^{\eta_2}
    \hat{\rho}^<_{{\bf k}^{\prime}}
    \hat{s}^{\eta_1}
    \hat{\rho}^>_{\bf k} + {\rm H.c.}
    ),
\ee
with $G_{\rm  Mn}(\eta_1\eta_2)=\frac{1}{4}{\rm
  Tr}(\hat{S}^{\eta_1}\hat{S}^{\eta_2}\hat{\rho}_{\rm
  Mn})$. $\hat{S}^{\eta}$ and $\hat{s}^{\eta}$ ($\eta=0,\pm 1$) are
the spin ladder operators. The equation of motion for Mn spin density matrix consists of three parts
$\partial_t \hat{\rho}_{\mathrm{Mn}} = \left. \partial_t
\hat{\rho}_{\mathrm{Mn}}\right|_{\rm coh}  + \left. \partial_t
\hat{\rho}_{\mathrm{Mn}}\right|_{\rm scat} + \left. \partial_t
\hat{\rho}_{\mathrm{Mn}}\right|_{\rm rel}$. The first part describes the
coherent precession around  the external mangetic field and the
$s$-$d$ exchange mean field, $\left. \partial_t
\hat{\rho}_{\mathrm{Mn}}\right|_{\rm coh} = - i \left[ g_{\rm Mn} \mu_{\rm B}
  {\bf B}\cdot \hat{{\bf S}}  - \alpha\sum_{{\bf k}}{\rm
    Tr} (\frac{\hat{\mbox{\boldmath$\sigma$\unboldmath}}}{2} \hat{\rho}_{\bf
    k}) \cdot \hat{{\bf S}},\ \hat{\rho}_{\mathrm{Mn}}\right]$.
The second part represents the $s$-$d$ exchange scattering with
electrons $\left. \partial_t \hat{\rho}_{\mathrm{Mn}}\right|_{\rm scat}
= -\frac{\pi \alpha^2}{4} \sum_{\eta_1 \eta_2 {\bf k}} 
\delta(\varepsilon_{\bf k}-\varepsilon_{{\bf k}^{\prime}})
{\rm Tr}(\hat{s}^{-\eta_2} \hat{\rho}^<_{{\bf k}^{\prime}}
\hat{s}^{-\eta_1} \hat{\rho}^>_{\bf k} )
\Big[ (\hat{S}^{\eta_1}\hat{S}^{\eta_2} \hat{\rho}_{\mathrm{Mn}}
  - \hat{S}^{\eta_1}\hat{\rho}_{\mathrm{Mn}}\hat{S}^{\eta_2}) 
  + {\rm H.c.}  \Big]$. The third part characterizes the Mn spin
relaxation due to other mechanisms, such as the $p$-$d$ exchange
interaction with holes or Mn-spin--lattice interaction, with a
relaxation time approximation,
$ \left. \partial_t \hat{\rho}_{\mathrm{Mn}}\right|_{\rm rel} = -\left(
\hat{\rho}_{\mathrm{Mn}}-\hat{\rho}_{\mathrm{Mn}}^{0}
\right)/\tau_{\mathrm{Mn}}$. Here $\hat{\rho}_{\mathrm{Mn}}^{0}$
represents the equilibrium Mn spin density
matrix. $\tau_{\mathrm{Mn}}$ is the Mn spin relaxation time, which
is typically 0.1$\sim$10~ns \cite{rcmyers:203}.

After incorporating the Elliott-Yafet mechanism, besides the ordinary
spin-conserving terms, there are spin-flip terms. For
electron-impurity scattering these additional terms are
\begin{eqnarray}
  \left. \partial_t \hat{\rho}_{\bf k}\right|_{\rm ei}^{\rm EY} &=& -\pi
  n_{i} \sum_{{\bf k}^{\prime}} \delta(\varepsilon_{\bf
  k}-\varepsilon_{{\bf k}^{\prime}}) \Big[ U^{(1)}_{{\bf k}-{\bf
  k}^\prime} \big( \hat{\Lambda}^{(1)}_{{\bf k},{\bf k}^\prime}
  \hat{\rho}^>_{{\bf k}^{\prime}}
  \hat{\Lambda}^{(1)}_{{\bf k}^\prime,{\bf k}}
  \hat{\rho}^<_{\bf k} - \hat{\Lambda}^{(1)}_{{\bf k},{\bf k}^\prime}
  \hat{\rho}^<_{{\bf k}^{\prime}} \hat{\Lambda}^{(1)}_{{\bf k}^\prime,{\bf k}}
    \hat{\rho}^>_{\bf k} \big)\nonumber \\ && \quad
    + U^{(2)}_{{\bf k}-{\bf k}^\prime}
    \big( \hat{\Lambda}^{(2)}_{{\bf k},{\bf k}^\prime} \hat{\rho}^>_{{\bf k}^{\prime}}
   \hat{\Lambda}^{(2)}_{{\bf k}^\prime,{\bf k}}
      \hat{\rho}^<_{\bf k} - \hat{\Lambda}^{(2)}_{{\bf k},{\bf k}^\prime}
    \hat{\rho}^<_{{\bf k}^{\prime}} \hat{\Lambda}^{(2)}_{{\bf k}^\prime,{\bf k}}
    \hat{\rho}^>_{\bf k} \big)
     + {\rm H.c.} \Big],
   \end{eqnarray}
where $n_i = N_{\rm Mn}^{\rm S}+4N_{\rm Mn}^{\rm I} +
n_{i0}$ with $N_{\rm Mn}^{\rm S}$, $N_{\rm Mn}^{\rm I}$ and $n_{i0}$
representing the densities of substitutional Mn, interstitial Mn and non-magnetic
impurities, respectively. $U^{(1)}_{{\bf k}-{\bf
    k}^\prime} = \frac{\lambda_c^2}{4}  \sum_{q_z} V^2_{{\bf k}-{\bf
    k}^{\prime},q_z} |I(iq_z)|^2
q_z^2$ and $U^{(2)}_{{\bf k}-{\bf k}^\prime} = - \lambda_c^2
\sum_{q_z} V^2_{{\bf k}-{\bf k}^{\prime},q_z} |I(iq_z)|^2$.
Here
$\lambda_c=\frac{\eta(1-\eta/2)}{3m_cE_g(1-\eta/3)}$ with
$\eta=\frac{\Delta_{\rm SO}}{\Delta_{\rm SO} + E_g}$. $E_g$ and $\Delta_{\rm SO}$
are the band-gap and the spin-orbit splitting of the valence band,
respectively \cite{opt-or}.
The spin-flip matrices are given by
$\hat{\Lambda}^{(1)}_{{\bf k}^\prime,{\bf k}} =
[({\bf k}+{\bf k}^{\prime},0)\times
  \hat{\mbox{\boldmath${\sigma}$\unboldmath}}]_z$ and
$\hat{\Lambda}^{(2)}_{{\bf k},{\bf k}^\prime} = [({\bf k},0)\times({\bf
    k}^{\prime},0)]\cdot
\hat{\mbox{\boldmath${\sigma}$\unboldmath}}$.
It is noted that $\hat{\Lambda}^{(1)}_{{\bf
    k}^\prime,{\bf k}}$ and $\hat{\Lambda}^{(2)}_{{\bf k}^\prime,{\bf k}}$
contribute to the out-of-plane and in-plane spin relaxations,
respectively. They are generally different and therefore the spin
relaxation due to the Elliott-Yafet mechanism in quantum wells is anisotropic.
The Elliott-Yafet mechanism can be incorporated into other scatterings
similarly \cite{jiang:125206}. 

By solving the kinetic spin Bloch equations, Jiang et al. \cite{jiang:155201} studied the spin
relaxation in both $n$- and $p$-type Ga(Mn)As quantum wells. For
$n$-type sample, the total electron density is given by
$N_e=N_e^i+N_e^{\rm Mn}+N_{\rm ex}$ with $N_e^{\rm Mn}$, $N_e^i$ and
$N_{\rm ex}$ representing the densities from Mn donors, other donors and
photo excitation, respectively. The calculated spin relaxation times
due to various mechanisms are shown as function of Mn concentration in
$n$-type samples with $N_e^i=0$ (a) and $N_e^i=10^{11}$~cm$^{-2}$ (b)
respectively (Fig.~\ref{fig5.4.9-1}). It is interesting to see that even for the strong
$s$-$d$ exchange coupling taken in the calculation, the spin
relaxation due to the $s$-$d$ exchange coupling is still much weaker
than that due to the D'yakonov-Perel' mechanism. Also the Bir-Aronov-Pikus and Elliott-Yafet mechanisms are
irrelevant to the spin relaxation. The spin relaxatin is solely
determined by the D'yakonov-Perel' mechanism. Moreover, they predicted a peak in the
Mn concentration dependence of the spin relaxation time. The physics
leading to the peak is the same as the carrier-density dependence of
the spin relaxation time in $n$-type GaAs quantum wells addressed in
Sec.~\ref{sec5.4.2}, with the peak being around the electron Fermi temperature
$T_F^e$. 
\begin{figure}[htbp]
  \begin{center}
    \includegraphics[height=6cm]{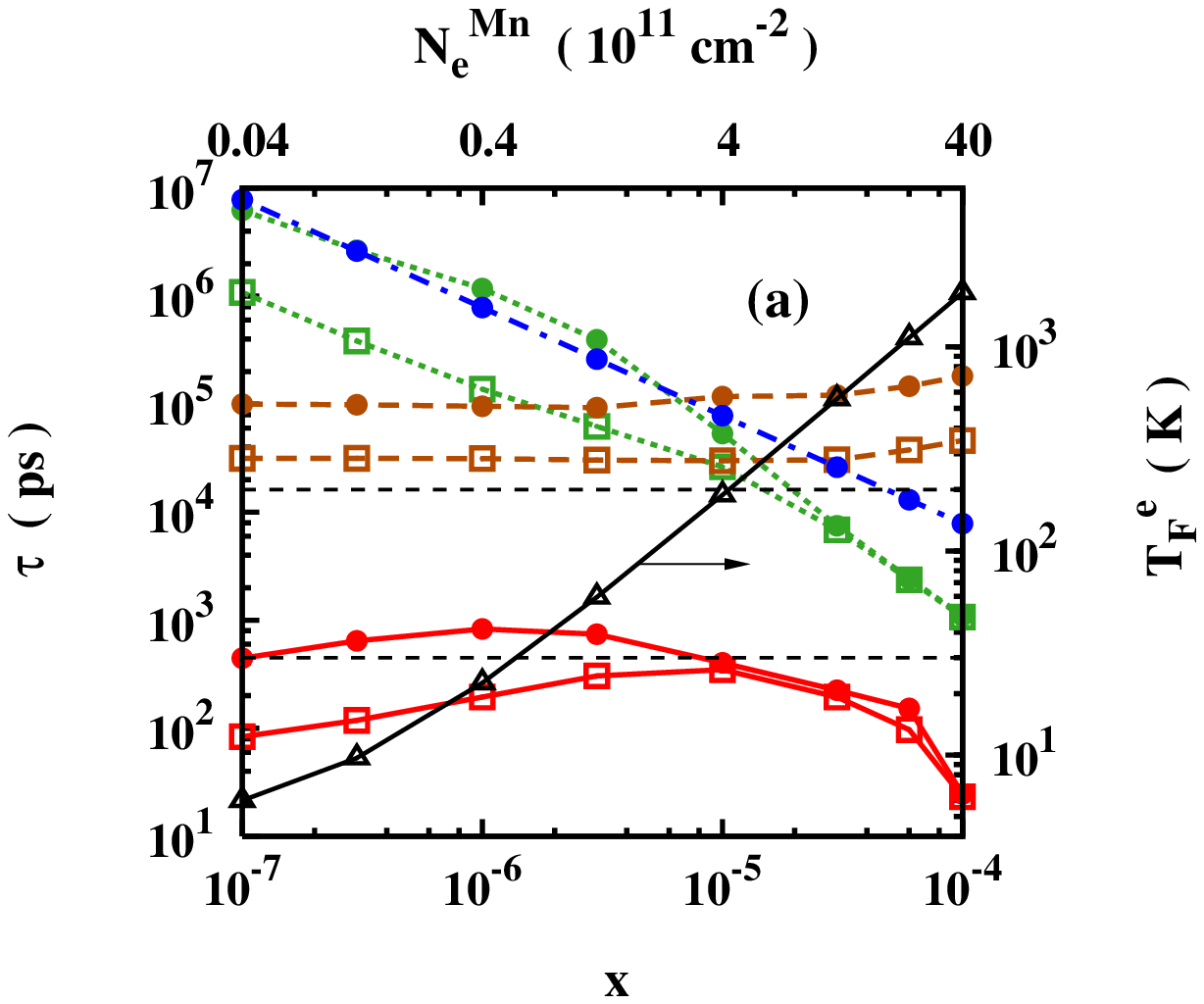}\includegraphics[height=6cm]{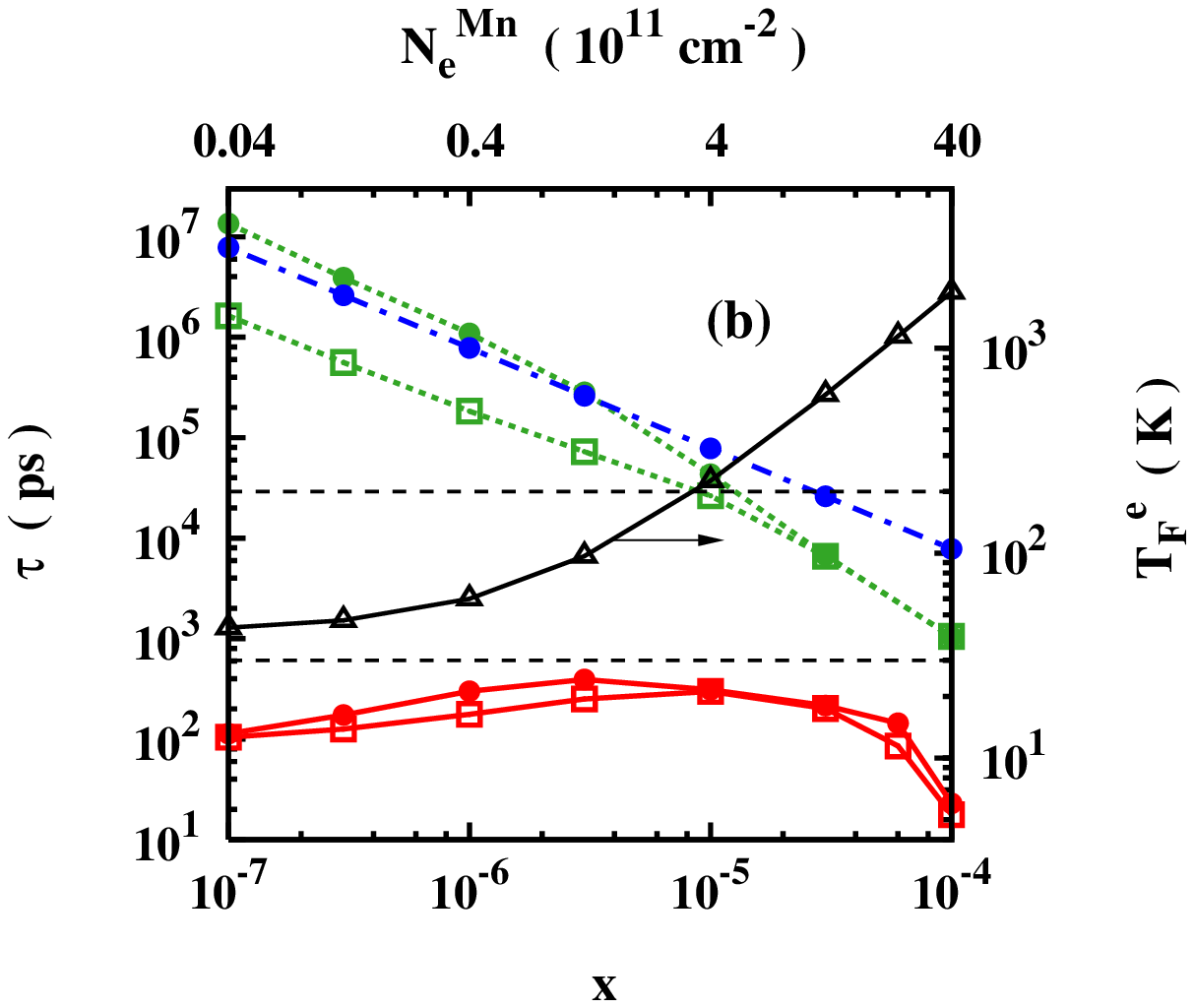}
  \end{center}
  \caption{  Spin relaxation time $\tau$ due to various mechanisms in
$n$-type Ga(Mn)As quantum wells which are (a) undoped or (b) $n$-doped
before Mn-doping as function of Mn concentration $x$ at 30~K ($\bullet$)
    and 200~K ($\square$). Red solid curves: the spin relaxation time due to the D'yakonov-Perel'
    mechanism $\tau_{\rm DP}$; Green dotted curves: the spin relaxation time due to the
Elliott-Yafet mechanism $\tau_{\rm EY}$; Brown dashed curves: the spin relaxation time due to
    the Bir-Aronov-Pikus mechanism $\tau_{\rm BAP}$; Blue chain curve: the spin relaxation time due
    to the $s$-$d$ exchange scattering mechanism $\tau_{\rm sd}$.
    The Fermi temperature of electrons $T_{\rm F}^e$ is ploted as black
    curve with $\triangle$ (the scale of $T_{\rm F}^e$ is on the right hand
    side of the frame) and $T_{\rm F}^e=T$ for both $T=30$ and $200$~K
    cases are plotted as black dashed curves. The scale
    of the electron density from Mn donors $N_e^{\rm Mn}$ is also
    plotted on the top of the frame. From Jiang et al. \cite{jiang:155201}.}
  \label{fig5.4.9-1}
\end{figure}

\begin{figure}[htb]
  \begin{center}
    \includegraphics[height=6cm]{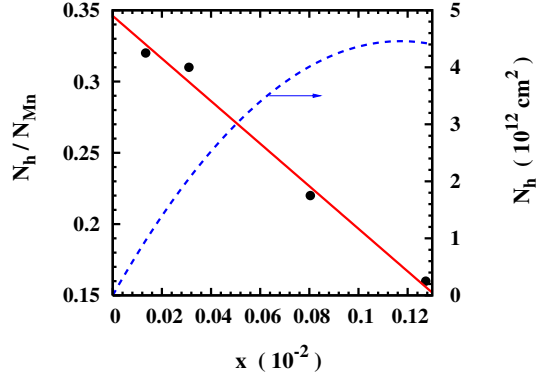}
  \end{center}
  \caption{Ratio of the hole density to the Mn
    density $N_h/N_{\rm Mn}$ {\sl vs.} the Mn concentration $x$ in
    $p$-type Ga(Mn)As quantum wells. The black dots represent the
    experimental data. The red solid curve is the fitted one. The hole
    density $N_{h}$ is also plotted (the blue dashed curve). Note that
    the scale of $N_h$ is on the right hand side of the frame. From
    Jiang et al. \cite{jiang:155201}.}
  \label{fig5.4.9-2}
\end{figure}

\begin{figure}[htb]
  \begin{center}
    \includegraphics[height=4.5cm]{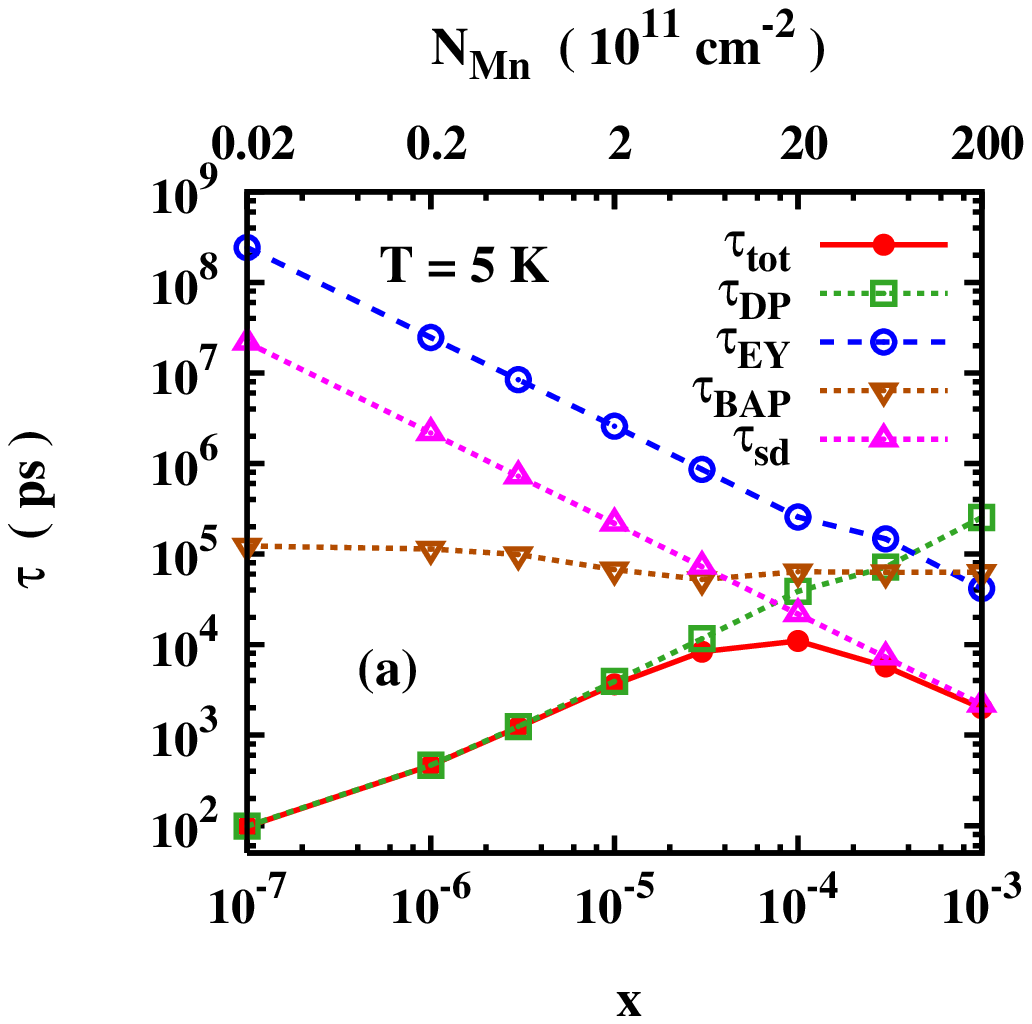}
\includegraphics[height=4.5cm]{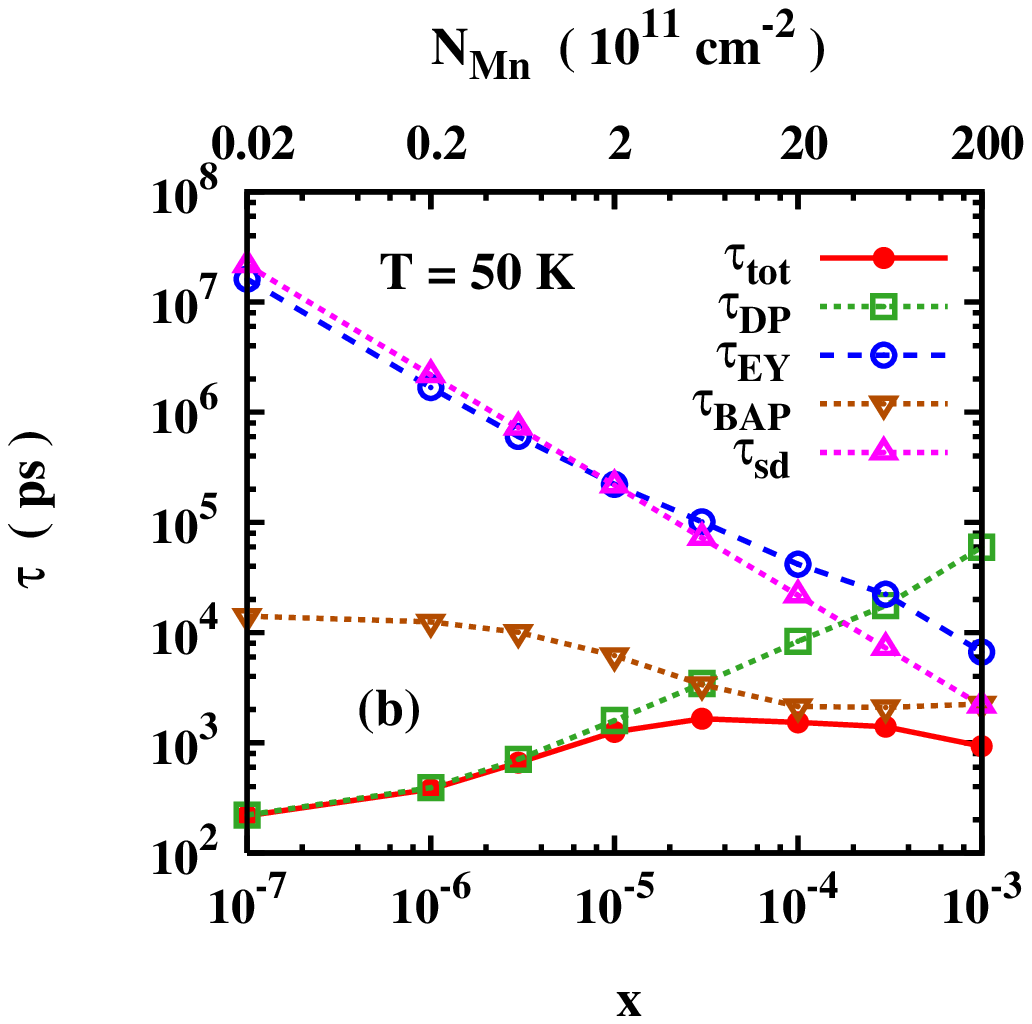}
\includegraphics[height=4.5cm]{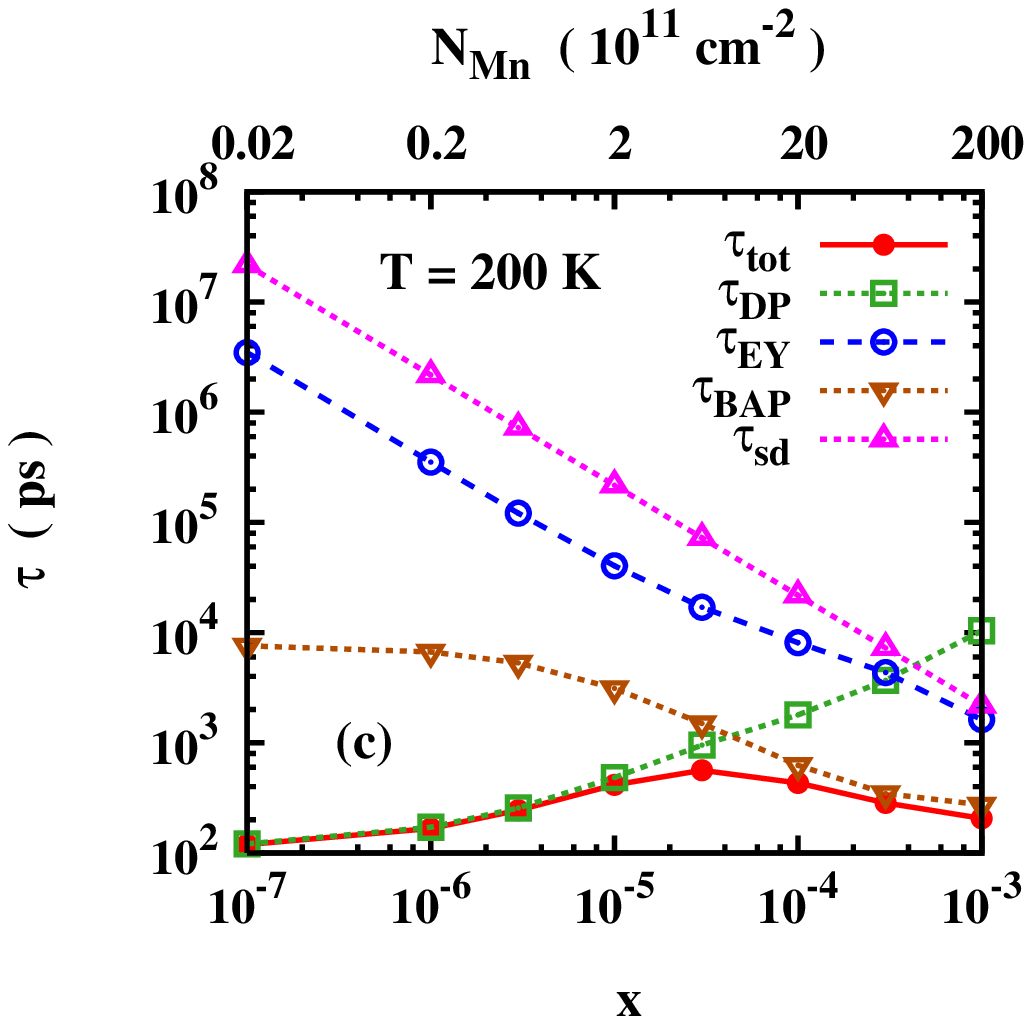}
  \end{center}
  \caption{  Spin relaxation time $\tau$ due to various mechanisms and
    the total spin relaxation time in $p$-type Ga(Mn)As quantum wells
against the Mn
    concentration $x$ at (a) $T=5$, (b) 50 and (c) $200$~K. The scale
    of $N_{\rm Mn}$ is also plotted on the top of the frame. From Jiang et al. \cite{jiang:155201}.}
  \label{fig5.4.9-3}
\end{figure}

For $p$-type Ga(Mn)As quantum wells, both substitutional and
interstitial Mn ions exist in the system. For simplicity, all the
holes are assumed free. Due to the presence of interstitial Mn ions,
the ratio of the hole density $N_h$ to the Mn density  N$_{\rm Mn}$ was 
obtained by fitting the experimental data in Ref.~\cite{PhysRevB.72.235313}, as shown in
Fig.~\ref{fig5.4.9-2}. The electron spin relaxations due to various mechanisms
were calculated as function of Mn concentration at different
temperatures by Jiang et al. \cite{jiang:155201}, as shown in
Fig.~\ref{fig5.4.9-3}. Unlike the case of $n$-type samples, due to the presence
of large hole density at high Mn concentration, the Bir-Aronov-Pikus and $s$-$d$
exchange scattering mechanisms can be important. At very low
temperature [Fig.~\ref{fig5.4.9-3}(a)], due to the Pauli blocking addressed in
Sec.~\ref{sec5.4.6}, the Bir-Aronov-Pikus mechanism is negligible. The spin relaxation is
determined by the D'yakonov-Perel' mechanism at low Mn concentration and the $s$-$d$
exchange scattering mechanism at high Mn concentration. At medium
temperature [Fig.~\ref{fig5.4.9-3}(b)], both the Bir-Aronov-Pikus and $s$-$d$ exchange scattering
mechanisms determine the spin relaxation at high Mn concentrations. At
high temperature [Fig.~\ref{fig5.4.9-3}(c)], the spin relaxation is determined
by the Bir-Aronov-Pikus mechanism. As the D'yakonov-Perel' mechanism determines the spin
relaxation at low Mn concentrations, the spin relaxation time limited
by it increases with increasing Mn concentration (increasing
electron-impurity scattering). Moreover, both the Bir-Aronov-Pikus and $s$-$d$
exchange scattering mechanisms increase with increasing Mn density. As
a result, there is a peak in the Mn density dependence of the electron
spin relaxation time at any temperature (except for the extremely low
temperature where the localization becomes important). 

The temperature, photo-excitation density and magnetic field
dependences of the spin relaxation in paramagnetic Ga(Mn)As quantum
wells have also been investigated in detail in
Ref.~\cite{jiang:155201}. 

\subsubsection{Hole spin dynamics in (001) strained asymmetric Si/SiGe
and Ge/SiGe quantum wells}
\label{sec5.4.10}
Among different kinds of hosts for spintronics devices, Silicon appears to
be a particularly promising one, partly due to the high possibility of
eliminating hyperfine couplings by isotropic purification and well
developed microfabrication technology \cite{0268-1242-12-12-001}. Many investigations
have been carried out to understand the electron spin relaxation in
bulk Silicon and its nanostructures \cite{PhysRevB.71.075315,PhysRevB.66.195315,Jantsch2002504,Appelbaum:447.295,PhysRevB.2.2429,0370-1328-84-1-304}. The study of hole spin
relaxation in Silicon is very limited. Glavin and Kim presented a first
calculation of the spin relaxation of two-dimensional holes in
strained asymmetric Si/SiGe (Ge/SiGe) quantum wells
\cite{PhysRevB.71.035321} by means
of the single-particle approximation where the important effect of the
Coulomb scattering to the spin relaxation is absent. More seriously,
as pointed out by Zhang and Wu \cite{zhang:155311}, the nondegenerate perturbation
method with only the lowest unperturbed subband of each hole state
considered in the calculation of the subband energy spectrum and
envelope functions by Glavin and Kim \cite{PhysRevB.71.035321} is inadequate in
converging the calculation. However, when more unperturbed subbands are
included as basis functions, the nondegenerate perturbation method
even fails. By applying the exact diagonalizing method to obtain the energy
spectrum and envelope functions, Zhang and Wu studied the hole spin
relaxation in (001) strained asymmetric Si/Si$_{0.7}$Ge$_{0.3}$
(Ge/Si$_{0.3}$Ge$_{0.7}$) quantum wells in the situation with only the
lowest hole subband being relevant by means of the kinetic spin Bloch equation approach
\cite{zhang:155311}. 

The structures of SiO$_{2}$/Si/Si$_{0.7}$Ge$_{0.3}$ and
SiO$_{2}$/Ge/Si$_{0.3}$Ge$_{0.7}$ (001) quantum wells are illustrated
in Fig.~\ref{fig5.4.10-1}, with the confining potential $V(z)$
approximated by the triangular potential due to the large gate
voltage. The subband envelope functions are obtained by solving the
eigen-equation of the 6$\times$6 Luttinger Hamiltonian
including the heavy hole, light hole and split-off hole states \cite{PhysRev.97.869,PhysRev.102.1030,strainbook}
under the confinement $V(z)$, with sufficient basis functions included
\cite{zhang:155311}. Due to the biaxial strain
\cite{PhysRev.97.869,PhysRev.102.1030,strainbook,PhysRevB.46.4110}, the
lowest subband in Si/SiGe quantum wells is light hole-like, which is an admixture of
the light hole and split-off hole states, whereas that in Ge/SiGe
quantum wells is a pure heavy hole state. By using the L\"owdin
partition method \cite{winklerbook}, the
effective Hamiltonian of the lowest hole subband in Si/SiGe (Ge/SiGe) 
quantum wells can be written as \cite{zhang:155311}
\begin{equation}
  H^{(l,h)}_{\rm eff}=-\frac{{\bf k}^2}{2m^{(l,h)}}-\frac{1}{2}{\bgreek\sigma}\cdot{\bf
    \Omega}^{(l,h)}(k_x,k_y),
\label{eqa3}
\end{equation}
where ${\bf k}$ is the in-plane momentum, $m^{(l)}$
[$m^{(h)}$] is the in-plane effective mass of the lowest light (heavy)
hole subband in Si/SiGe (Ge/SiGe) quantum wells,
$\bgreek{\sigma}$ are the Pauli matrices, and ${\bf
  \Omega}^{(l)}$ [${\bf \Omega}^{(h)}$] is the Rashba term of
 the lowest light hole (heavy hole) subband in Si/SiGe (Ge/SiGe) quantum wells. ${\bf\Omega}^{(l)}$ has both the
linear and cubic dependences on momentum, whereas ${\bf\Omega}^{(h)}$
has only the cubic dependence. For the lowest light hole subband in Si/SiGe quantum wells,
\begin{eqnarray}
m^{(l)}&=&m_0[A-B(\lambda^{(l1l1)}_{00}/2-\sqrt{2}\lambda^{(l1l2)}_{00})]^{-1},\label{ml}\\
{\bf \Omega}^{(l)}&=&{\bf \Omega}^{(l)}_1+{\bf \Omega}^{(l)}_3,\\
\Omega_{1x,y}^{(l)}&=&\Xi k_{x,y},\\
\Omega_{3x}^{(l)}&=&\Pi Bk_x(k_x^2+k_y^2)+\Theta [3Bk_x(k_x^2-k_y^2)+2\sqrt{3(3B^2+C^2)}k_y^2k_x],\\
\Omega_{3y}^{(l)}&=&\Pi Bk_y(k_x^2+k_y^2)+\Theta [3Bk_y(k_y^2-k_x^2)+2\sqrt{3(3B^2+C^2)}k_x^2k_y],
\end{eqnarray}
with
\begin{eqnarray}
\Xi&=&\frac{\hbar}{m_0}\sqrt{6(3B^2+C^2)}\kappa_{00}^{(l1l2)},\label{xi}\\
  \Pi&=&-\frac{\hbar^3}{2m_0^2}\sqrt{\frac{3(3B^2+C^2)}{2}}\sum_{\alpha=l,s}\sum_{n=0}^{\infty}(1-\delta_{l\alpha}\delta_{0n})\frac{\kappa_{0n}^{(l1\alpha
      2)}-\kappa_{0n}^{(l2\alpha 1)}}{E_0^{(l)}-E_n^{(\alpha)}}[\sqrt{2}(\lambda_{0n}^{(l1\alpha 2)}+\lambda_{0n}^{(l2\alpha 1)})-\lambda_{0n}^{(l1\alpha 1)}],\label{pi}\\
\Theta&=&-\frac{\hbar^3}{2m_0^2}\sqrt{\frac{3B^2+C^2}{3}}\sum_{n=0}^{\infty}\frac{\kappa_{0n}^{(l1h)}-\frac{1}{\sqrt{2}}\kappa_{0n}^{(l2h)}}{E_0^{(l)}-E_n^{(h)}}(\sqrt{2}\lambda_{0n}^{(l2h)}+\lambda_{0n}^{(l1h)})\label{theta}.
\end{eqnarray}
For the lowest heavy hole subband in Ge/SiGe quantum wells,
\begin{eqnarray}
m^{(h)}&=&m_0(A+B/2)^{-1},\label{mh}\\
{\bf \Omega}^{(h)}&=&{\bf \Omega}^{(h)}_3,
\label{rashba1}\\
\Omega_{3x}^{(h)}&=&\Lambda
[3Bk_x(k_x^2-k_y^2)-2\sqrt{3(3B^2+C^2)}k_y^2k_x],
\label{rashba2}\\
\Omega_{3y}^{(h)}&=&\Lambda
[3Bk_y(k_x^2-k_y^2)+2\sqrt{3(3B^2+C^2)}k_x^2k_y],
\label{rashba3}
\end{eqnarray}
with
\begin{eqnarray}
\Lambda
&=&-\frac{\hbar^3}{2m_0^2}\sqrt{\frac{3B^2+C^2}{3}}\sum_{\alpha=l,s}\sum_{n=0}^{\infty}\frac{\frac{1}{\sqrt{2}}\kappa_{0n}^{(h\alpha
    2)}-\kappa_{0n}^{(h\alpha 1)}}{E_0^{(h)}-E_n^{(\alpha)}}(\sqrt{2}\lambda_{0n}^{(h\alpha 2)}+\lambda_{0n}^{(h\alpha
  1)}).
\label{lambda}
\end{eqnarray}
Here $A$, $B$ and $C$ are the valence band
parameters, which relate to the Luttinger parameters
$\gamma_1$,
$\gamma_2$ and $\gamma_3$ through $A=\gamma_1$, $B=2\gamma_2$ and
$\sqrt{3B^2+C^2}=2\sqrt{3}\gamma_3$. $E_n^{(\alpha)}$ ($\alpha$=$h$,
$l$, $s$) are the  subband energy
levels. $\lambda_{nn^\prime}^{(\alpha\beta)}$
and $\kappa_{nn^\prime}^{(\alpha\beta)}$ are defined as
$\lambda_{nn^\prime}^{(\alpha\beta)}=\int_{-\infty}^{+\infty}dz\chi_n^{(\alpha)}(z)\chi_{n^\prime}^{(\beta)}(z)$
and
$\kappa_{nn^\prime}^{(\alpha\beta)}=\int_{-\infty}^{+\infty}dz\chi_n^{(\alpha)}(z)\frac{d\chi_{n^\prime}^{(\beta)}(z)}{dz}$,
with $\chi_n^{(\alpha)}$ the envelope functinos \cite{zhang:155311}. 
The calculated spin-orbit coupling coefficients of the lowest subband
of Si/SiGe quantum well ($\Xi$, $\Pi$ and $\Theta$) and Ge/SiGe quantum well
($\Lambda$) are given in Fig.~\ref{fig5.4.10-2}. 

\begin{figure}[htb]
\begin{center}
    {\includegraphics[width=7cm]{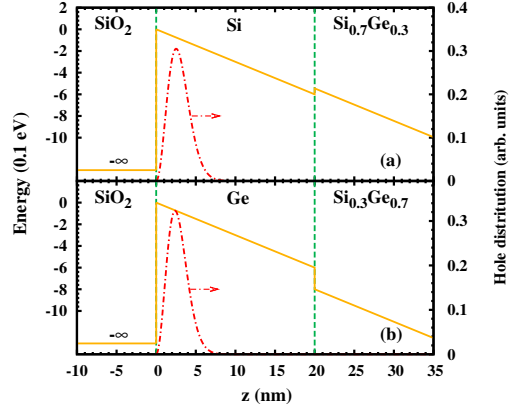}}
\end{center}
 \caption{ Schematics of the
 SiO$_2$/Si/Si$_{0.7}$Ge$_{0.3}$ quantum well structure (a) and
  SiO$_2$/Ge/Si$_{0.3}$Ge$_{0.7}$ quantum well structure (b). 
Two vertical dashed lines in each figure represent the two interfaces. The solid curves
represent the confining potential $V(z)$ with electric field $E=300$~kV/cm. 
The valence band discontinuities
at the Si/Si$_{0.7}$Ge$_{0.3}$ and Ge/Si$_{0.3}$Ge$_{0.7}$ interfaces 
are neglected in the triangular potential approximation. The chain 
curves with their  scale on the right hand side of the
   frame respectively represent the lowest light hole and heavy hole distributions in Si/Si$_{0.7}$Ge$_{0.3}$
and Ge/Si$_{0.3}$Ge$_{0.7}$ quantum wells along the $z$-axis. From
Zhang and Wu \cite{zhang:155311}.}
  \label{fig5.4.10-1}
\end{figure}

 \begin{figure}[htb]
   \begin{center}
    {\includegraphics[width=7cm]{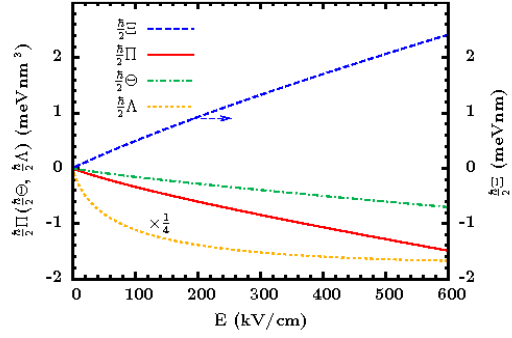}}
  \end{center}
    \caption{ Spin-orbit coupling coefficients
      $\frac{\hbar}{2}\Xi$, $\frac{\hbar}{2}\Pi$ and 
      $\frac{\hbar}{2}\Theta$ for the lowest light hole subband in Si/SiGe quantum wells and
      $\frac{\hbar}{2}\Lambda$ for the lowest heavy hole subband in Ge/SiGe 
      quantum wells against the electric field $E$. The
      scale of $\frac{\hbar}{2}\Xi$ is on the right hand side of the
      frame. From Zhang and Wu \cite{zhang:155311}.}    
  \label{fig5.4.10-2}
\end{figure}

The hole spin relaxation time is calculated by solving the kinetic spin Bloch equations, with
the hole-deformation optical/acoustic phonon \cite{bufler:2626}, hole-impurity
and hole-hole Coulomb scatterings explicitly included \cite{zhang:155311}. The
typical results are shown in Fig.~\ref{fig5.4.10-3} for Si/SiGe
quantum wells, where a peak appears in both the temperature dependence (located
around the hole Fermi temperature) and the density dependence (located
around the crossover from the degenerate to the nondegenerate regimes) of hole
spin relaxation time. It is also shown that the hole-hole Coulomb scattering plays
an essential role in the spin relaxation. As shown in
Fig.~\ref{fig5.4.10-4}, by switching off the hole-hole Coulomb
scattering, the spin relaxation times differ dramatically. Similar
behavior also happens in Ge/SiGe quantum wells, except that the peak
in the temperature dependence of the hole spin relaxation time is
located around half of the hole Fermi temperature. The shift of the
peak to the lower temperature is suggested to be due to the cubic momentum
dependence of the Rashba term [Eqs.~(\ref{rashba1}-\ref{rashba3})], in contrast to the
linear one which is important in the case of Si/SiGe quantum
wells. The effects of impurity and gate voltage on the hole spin
relaxation are discussed in detail in Ref.~\cite{zhang:155311}.

\begin{figure}[htb]
  \begin{center}
    \includegraphics[height=5cm]{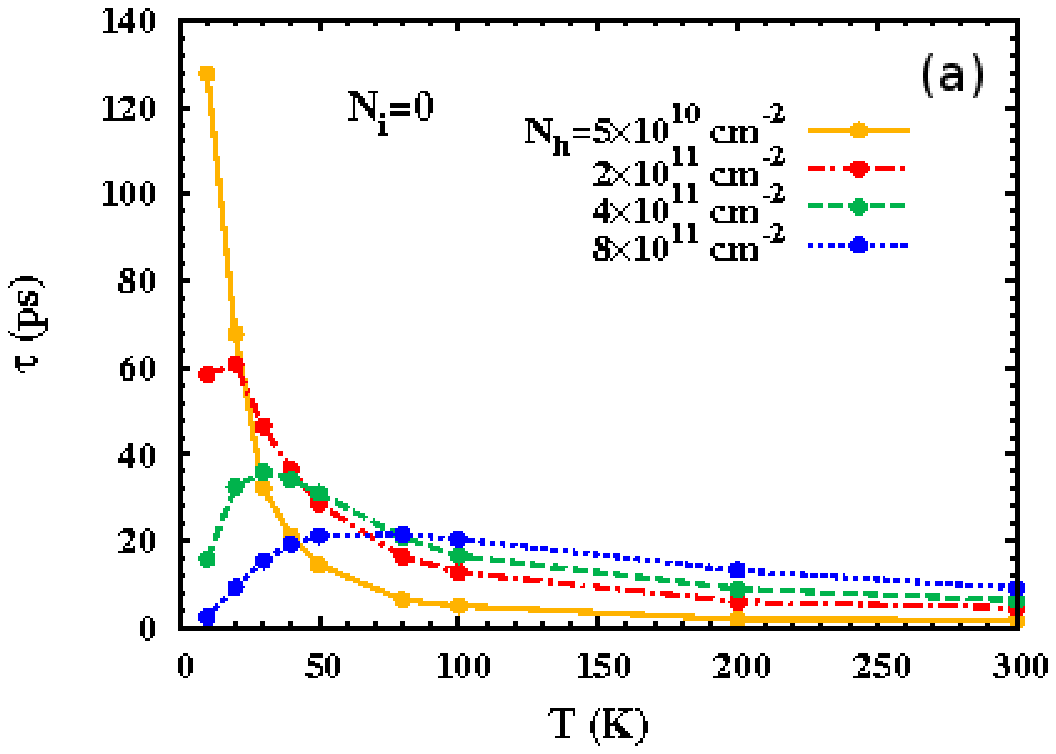}\includegraphics[height=5cm]{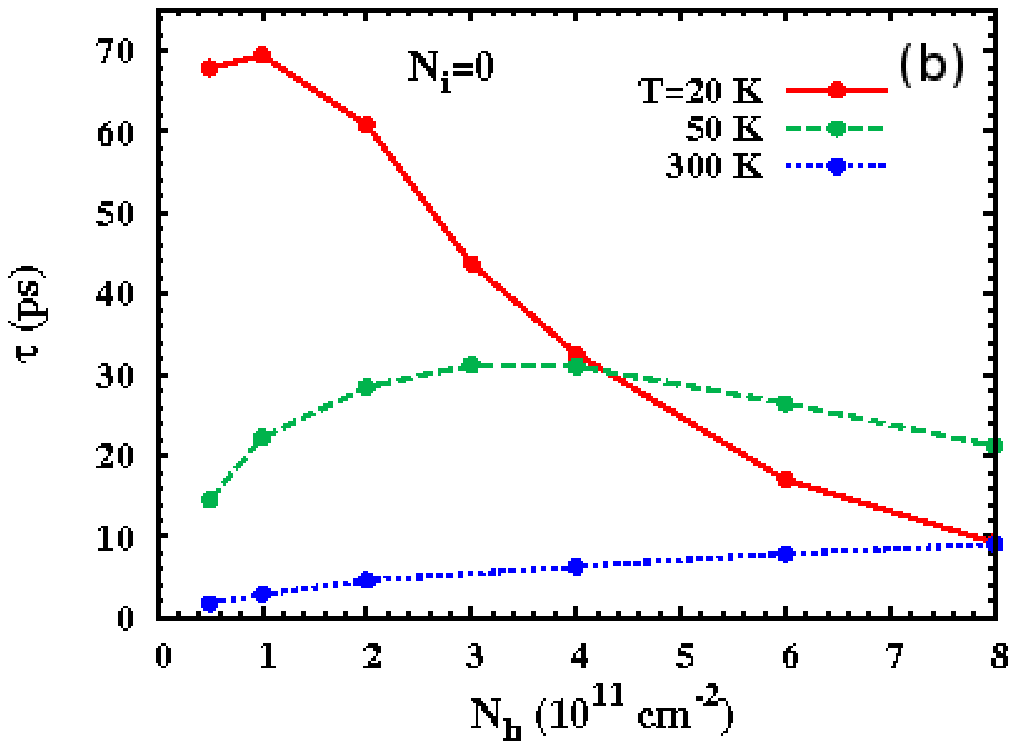}
  \end{center}
  \caption{   (a) Spin relaxation time $\tau$ against
      temperature $T$ with different hole densities. (b) Spin relaxation time $\tau$ against
      hole density $N_h$ with different temperatures. For both cases the impurity
      density $N_i=0$ and the electric field $E=300$~kV/cm. From Zhang
      and Wu \cite{zhang:155311}.}
  \label{fig5.4.10-3}
\end{figure}

\begin{figure}[htb]
\begin{center}
    {\includegraphics[width=7cm]{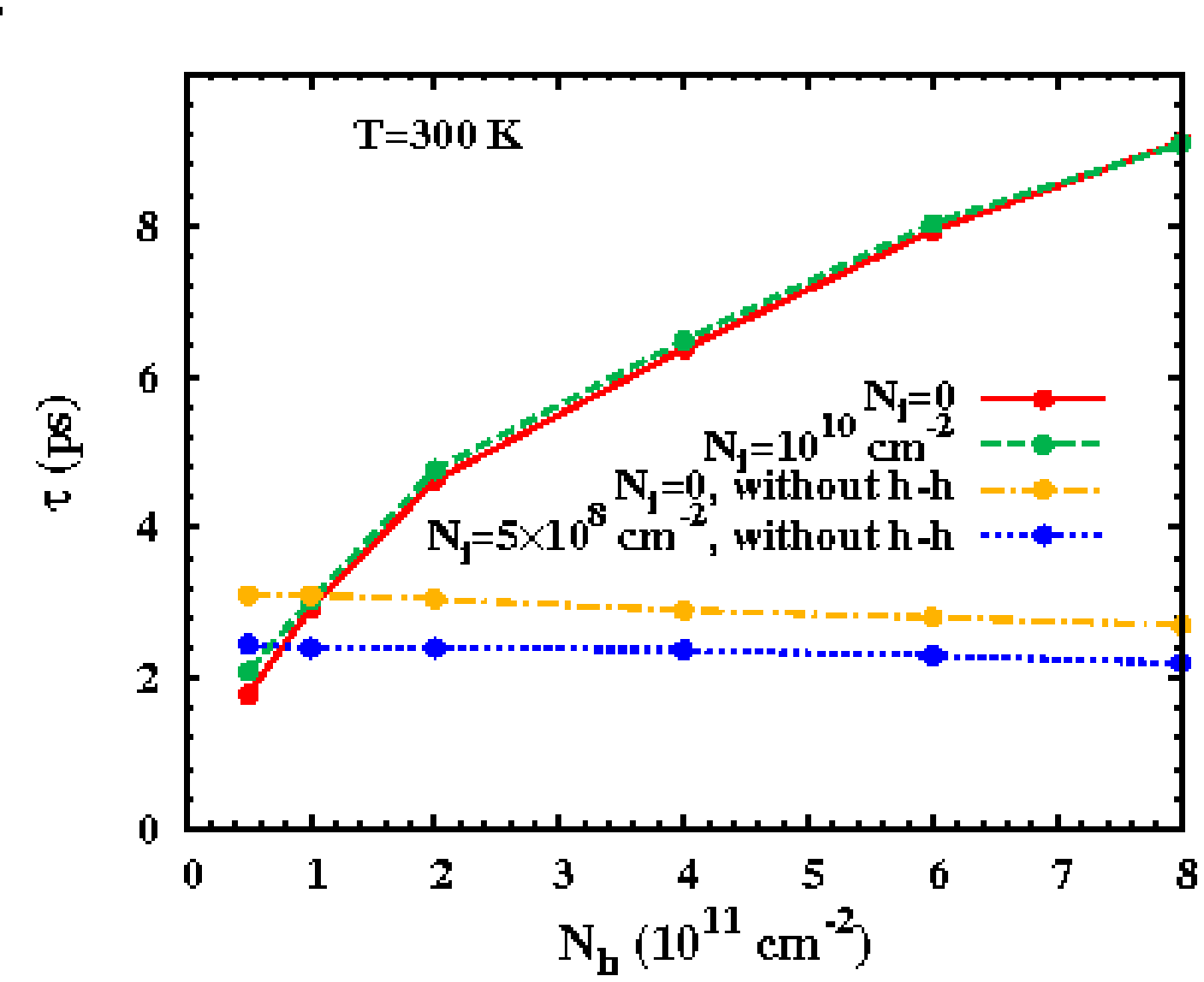}}
\end{center}
    \caption{Spin relaxation time $\tau$ against
      hole density $N_h$ with different scatterings
      included in Si/SiGe quantum well. 
$T=300$~K and $E=300$~kV/cm. Solid curve: with the
      hole-hole Coulomb (h-h), the hole-optical phonon 
      (h-op) and the hole-acoustic phonon (h-ac) scatterings. Dashed curve: same as the solid curve
      with the additional hole-impurity scattering
      ($N_i=10^{10}$~cm$^{-2}$) included; Chain curve: with the h-op and
      h-ac scatterings; Dotted curve: same as the chain curve
      with the additional hole-impurity scattering ($N_i=5\times 10^{8}$~cm$^{-2}$)
     included. From Zhang and Wu \cite{zhang:155311}.}    
  \label{fig5.4.10-4}
\end{figure}

Finally, it is noted that unlike holes in GaAs quantum wells where the
system is in the weak scattering limit \cite{lue:125314}, holes in Si/SiGe
quantum wells are generally in the strong scattering limit thanks to
the strong hole-hole Coulomb scattering and weak Rashba spin-orbit
coupling. However, holes in Ge/SiGe quantum wells can be in the weak
scattering limit with high density at low temperature due to the
larger Rashba term \cite{zhang:155311}. In any case, adding impurities can shift
the system from weak scattering limit to the strong one \cite{lue:125314,zhang:155311}.

\subsubsection{Spin relaxation and dephasing in $n$-type wurtzite ZnO 
(0001) quantum  wells}
\label{sec5.4.11}
While there are a lot of studies on the spin dynamics in cubic
zinc-blende semiconductors, much attention has also been devoted to
the spin properties of zinc oxide (ZnO) with wurtzite structures
\cite{ozgur:041301,ghosh:232507,ghosh:162109,liu:186804,lagarde:045204,lagarde:033203,T.Dietl02112000,harmon:115204}.
Harmon et al. calculated the spin relaxation time in
bulk material in the framework of single particle approach
\cite{harmon:115204}. L\"u and Cheng \cite{0268-1242-24-11-115010} applied the kinetic spin Bloch equation approach and
calculated the spin relaxation in $n$-type ZnO (0001) quantum
wells under various conditions, including well width, impurity
density and external electric field (hot-electron effect). In wurtzite
structure, the D'yakonov-Perel' term comes from the Rashba term due to the intrinsic
wurtzite structure inversion asymmetry ${\bf \Omega}_{\rm R}({\bf k})$ and
the Dresselhaus term ${\bf \Omega}_{\rm D}({\bf k})$ which can be written as
\cite{fu:093712}:
\begin{eqnarray}
g\mu_B{\bf \Omega}_{\rm R}({\bf k})&=&\alpha_e(k_y,-k_x,0),\\
g\mu_B{\bf \Omega}_{\rm D}^n({\bf
  k})&=&\gamma_e(b\langle k_z^2\rangle_n-k_{||}^2)(k_y,-k_x,0),
\end{eqnarray}
with $\alpha_e$, $\gamma_e$ and $b$ standing for the spin-orbit
coupling coefficients. $\langle k_z^2\rangle_n=n^2\pi^2/a^2$ is
the subband energy in the hard wall potential approximation. By taking
the lowest two subbands into account, L\"u and Cheng solved the
kinetic spin Bloch equations. The typical results are shown in Fig.~\ref{fig5.4.11-1}.

\begin{figure}[htb]
  \begin{center}
    \includegraphics[height=4cm]{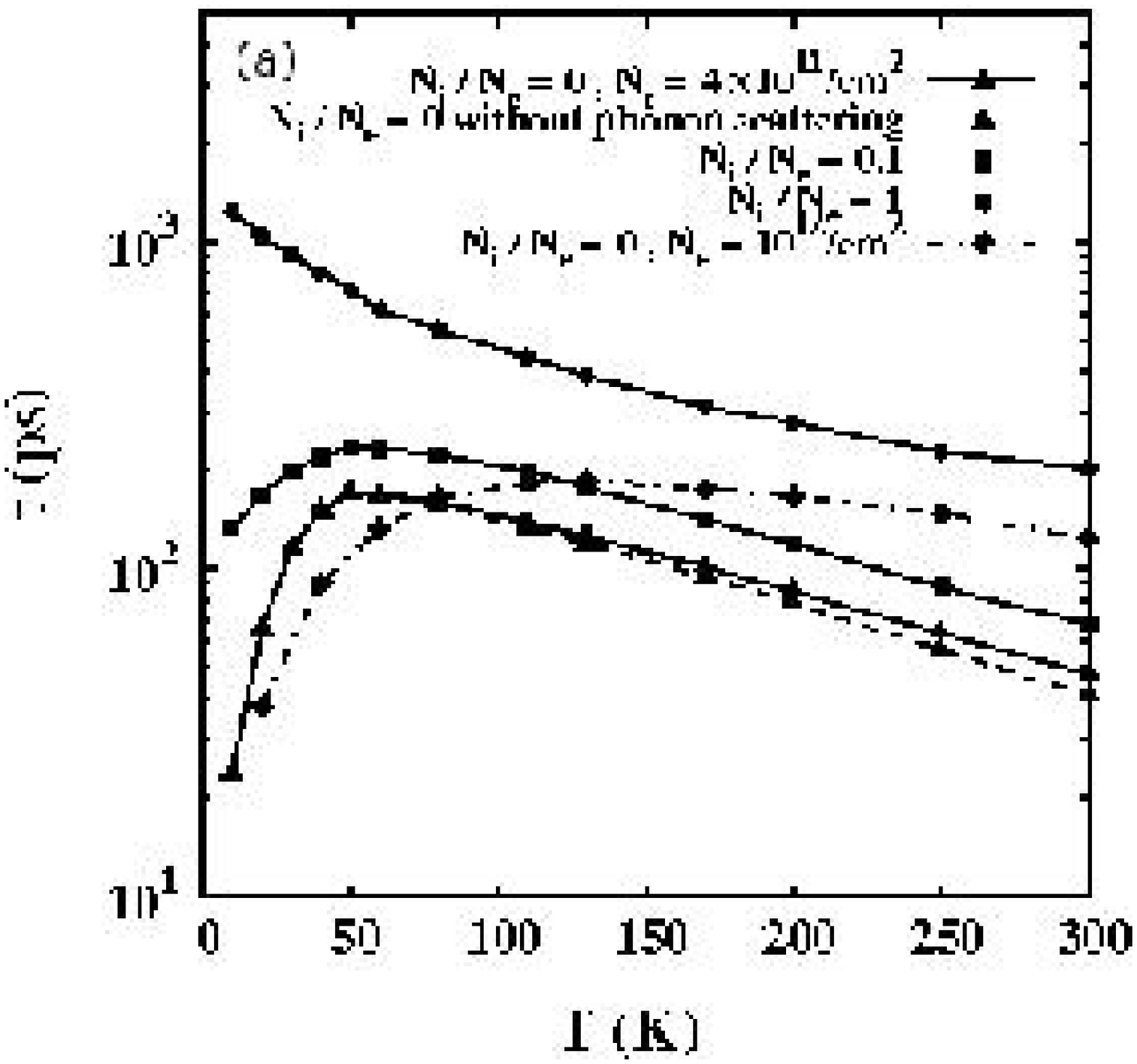}\includegraphics[height=4.3cm]{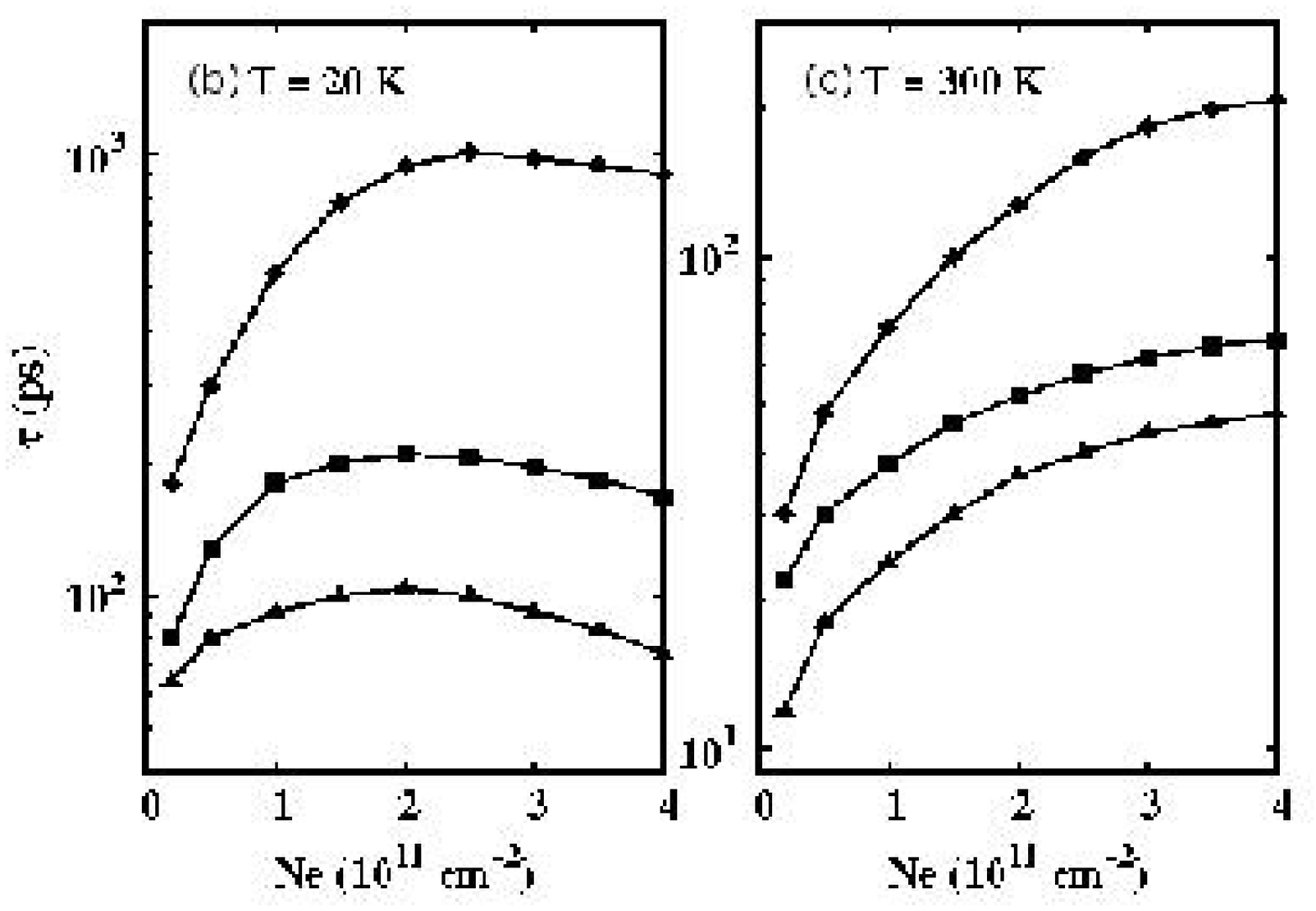}
  \end{center}
  \caption{ZnO (0001) quantum wells:
 (a) Spin relaxation time $\tau$ {\it vs}. temperature $T$ at
different impurity densities. The dashed curve is obtained from the
calculation of excluding the electron-phonon scattering. (b) and (c):
Spin relaxation time {\sl vs.} the electron density with different
impurity densities and temperatures (20~K and 300~K respectively). {$\blacktriangle$}: $N_i/N_e = 0$; {$\blacksquare$}:
$N_i/N_e = 0.1$; {$\blacklozenge$}: $N_i/N_e = 1$. From L\"{u} and Cheng \cite{0268-1242-24-11-115010}.}
  \label{fig5.4.11-1}
\end{figure}

Basically the properties of the spin relaxation in ZnO quantum wells
are all the same with those in GaAs ones. One still observes the peak
in the temperature and density dependences (the former requires low
impurity density) of the spin
relaxation time. The difference is that for GaAs quantum wells, the temperature
peak of the spin relaxation time can only be
observed at low electron density (i.e., low transition
  temperature) and low impurity density \cite{zhou:045305,zhou:075318,ruan:193307}, because the
electron-phonon scattering becomes strong enough to destroy the nonmonotonic
$T$ dependence of the scattering time induced by the electron-electron
scattering. Such case can be avoided in ZnO quantum wells, in which the electron-phonon 
scattering is always pretty weak due to the large optical phonon energies
($\sim800$~K). Thus the temperature peak can be found even for high
electron density samples.

\subsection{Spin relaxation in quantum wires}
In recent years, progress in nanofabrication and growth techniques has
made it possible to produce high-quality quantum wires and investigate
physics in these nanostructures \cite{pfeiffer:073111,danneau:012107,klochan:092105,hof:115325,farhangfar:205437,lehnen:205307,agrawal:245405,csontos:155323}. The energy spectrum of
quantum wire systems with strong spin-orbit coupling has been studied
both experimentally \cite{danneau:026403,goldoni:2965,csontos:155323} and theoretically \cite{pramanik:155325,zhang:155316,csontos:073313,csontos:155323,PhysRevB.43.4732,PhysRevB.40.5507,PhysRevB.55.7726,PhysRevB.70.155302,csontos:023108}. Unlike quantum
wells discussed in Sec.~\ref{sec5.4}, there is additional confinement in
quantum wires and the remaining free degree of freedom is along the
wire growth direction. Therefore, there are stronger subband effect
and also marked anisotropy along the growth directions of
wires. Consequently this gives more choices for the manipulation of
the spin degree of freedom. 

For quantum wires, the spin relaxation was calculated in the framework
of single particle approach
\cite{PhysRevB.68.075313,PhysRevB.72.045326,schwab:155316,PhysRevB.58.15652}
and Monte Carlo simulations 
\cite{ohno:241308,PhysRevB.68.075313,Liu.jsnm.9.525,PhysRevB.61.13115}. Cheng et al. first applied the kinetic spin Bloch equations to study
the spin dynamics in (001) oriented InAs quantum wires with only the
lowest subband involved \cite{Cheng2003365}. L\"u et al. studied the
influence of higher subbands and wire orientations on spin
relaxation in $n$-type InAs quantum wires \cite{lu:073703}. They reported that
the intersubband Coulomb scattering can make an important contribution
to the spin relaxation. Also due to intersubband scattering in
connection with the spin-orbit coupling, spin relaxation in quantum
wires can show different characteristics from those in bulk and in quantum
wells. Hole spin relaxation in $p$-type GaAs quantum wires was also
investigated by L\"u et al. from the kinetic spin Bloch equation approach \cite{lue:165321}.

\subsubsection{Electron spin relaxation in $n$-type InAs quantum
  wires}
By modeling the InAs quantum wire by a rectangular confinement
potential of infinite well depth along the $x$-$y$  plane ($|x|\le
a_x$, $|y|\le a_y$), one can write the Rashba and Dresselhaus
Hamiltonian by replacing $k_x$, $k_x^2$, $k_y$, and $k_y^2$ in the bulk
Rashba and Dresselhaus Hamiltonian by
$\langle\psi_{nx}|-i\partial/\partial x|\psi_{n^\prime x}\rangle$,
$\langle\psi_{nx}|(-i\partial/\partial x)^2|\psi_{n^\prime x}\rangle$,
$\langle\psi_{ny}|-i\partial/\partial y|\psi_{n^\prime y}\rangle$ and
$\langle\psi_{ny}|(-i\partial/\partial y)^2|\psi_{n^\prime y}\rangle$, respectively. Here
  $\psi_{nx}=\sqrt{\frac{2}{a_x}}\sin\frac{n_x\pi x}{a_x}$ and
  $\psi_{ny}=\sqrt{\frac{2}{a_y}}\sin\frac{n_y\pi y}{a_y}$. The bulk
  Rashba term reads 
\be
  \label{Rashba_100}
    H_{\rm R}(\mathbf{k}) = {\gamma^{6c6c}_{41}}
    \mbox{\boldmath$\sigma$\unboldmath} \cdot \mathbf{k}
  \times \mathbf{\mathcal{E}} = {\gamma^{6c6c}_{41}}[\sigma_x
  (  k_y  {\cal E}_z - 
  k_z {\cal E}_y) + \sigma_y (k_z {\cal E}_x -  k_x 
  \mathcal{E}_z) + \sigma_z ( k_x  {\cal E}_y -   k_y  {\cal E}_x)].
\ee
For a (100) InAs quantum wire, the $x$, $y$ and $z$ axes correspond to the
 [100], [010] and [001] crystallographic directions, respectively, and
 the bulk Dresselhaus term can be written as \cite{winklerbook}:
\begin{eqnarray}
  \label{Dresselhaus_100} 
   H^{100}_{\rm D} &=& {b^{6c6c}_{41}} \{ \sigma_x [ k_x ( k_y^2  -
  k_z^2)] + \sigma_y [ k_y (k_z^2 -
   k_x^2 )] + \sigma_z [k_z( k_x^2  -
   k_y^2 )] \}.
\end{eqnarray}
For a (110) quantum wire, the $x$, $y$ and $z$ directions correspond to the
[$\bar{1}$10], [001] and [110] 
crystallographic directions, and one has
\begin{equation}
  \label{Dresselhaus_110} 
  H^{110}_{\rm D} = {b^{6c6c}_{41}} \{ \sigma_x [-\frac{1}{2} k_z  (
  k_x^2   -  k_z^2  + 2 k_y^2 )] 
 + 2 \sigma_y  k_x   k_y   k_z + \sigma_z [\frac{1}{2}  
  k_x   (  k_x^2   -  k_z^2  - 2 k_y^2 )]  \}.
\end{equation}
For a (111) quantum wire, the $x$, $y$ and $z$ directions correspond to the
[11$\bar{2}$], [$\bar{1}$10], and [111] crystallographic
directions, and one has
\begin{eqnarray}
  \label{Dresselhaus_111} \nonumber
   H^{111}_{\rm D} &=& {b^{6c6c}_{41}} \{ \sigma_x [- \frac{\sqrt{2}}{\sqrt{3}} 
  k_x  k_y  k_z - \frac{1}{2\sqrt{3}} 
  k_y^3   - \frac{1}{2\sqrt{3}}  k_y 
  k_x^2  + \frac{2}{\sqrt{3}}  k_y  k_z^2 -
  \frac{\sqrt{2}}{3}  k_y^2  k_z] \\  &&\quad +
  \sigma_y[ \frac{1}{2\sqrt{3}}  k_x^3 +
  \frac{1}{2\sqrt{3}}   k_x   k_y^2 
  -\frac{1}{\sqrt{6}}   k_x^2  k_z  - 
  \frac{1}{\sqrt{6}} k_z ( k_x^2   +  k_y^2
  )]  + \sigma_z [\frac{\sqrt{3}}{\sqrt{2}} k_x^2
    k_y  - \frac{1}{\sqrt{6}}  k_y^3
   - \frac{2}{3} k_z  k_y^2 ]\}.
\end{eqnarray}

The kinetic spin Bloch equations read $\dot{\bf { \rho}}_{{ k}} = \dot{\bf { \rho}}_{{
      k}}|_{\rm coh} +   \dot{\bf { \rho}}_{{ k}}|_{\rm scat}$, with
  the coherent terms being
\begin{eqnarray}
  \dot{\rho_{k}}\Big|_{\rm coh} = i \Big[
  \sum_{\mathbf{Q}} V_{\mathbf{Q}}  
  I_{\mathbf{Q}} {\rho}_{k-q} I_{-\mathbf{Q}} ,  \rho_k \Big]
  -i \Big[ H_{e}(k),  \rho_k   \Big]. 
 \label{coh1}
\end{eqnarray}
Here $\mathbf{Q} \equiv
(q_x, q_y, q)$.  $I_{\mathbf{Q}} $ is the form factor. In $(s,s')$
space [$s=(n_x,n_y,\sigma)$], it reads 
\begin{eqnarray}
  I_{\mathbf{Q},s_1,s_2} =\langle s_1 | e^{i\mathbf{Q} \cdot
   \mathbf{r}} | s_2 \rangle  
   = \delta_{\sigma_1,\sigma_2}F(m_1,m_2,q_y,a_y)F(n_1,n_2,q_x,a_x),
\end{eqnarray}
where
\be
    F(m_1 , m_2,q,a) = 2 i a q [e^{i a q} \cos{\pi (m_1 -
      m_2)} -1 ]  \left[\frac{1}{\pi^2(m_1 - m_2)^2 -
        a^2 q^2} - \frac{1}{\pi^2(m_1 + m_2)^2 - a^2 q^2}\right].
    \label{Ffunction}
\ee
The first term in Eq.~(\ref{coh1}) is the Coulomb Hartree-Fock term, and
the second term comes from the single particle Hamiltonian $H_{e} =
\frac{{\mathbf P}^2}{2 m^*} + H_{\rm R} + H_{\rm D} + V_c(\mathbf{ r})$ in $(s,s')$
space. For small spin polarization, the contribution from the Hartree-Fock
term in the coherent term is negligible \cite{PhysRevB.68.075312,stich:176401,stich:205301} and the
coherent spin dynamics 
is essentially due to the spin precession around the effective
internal fields described by Eqs.~(\ref{Rashba_100})-(\ref{Dresselhaus_111}).

The scattering contributions to the dynamic equation of the
spin-density matrix include the electron-nonmagnetic impurity,
electron-phonon and electron-electron
scatterings:
 \begin{eqnarray}
   \frac{\partial \rho_{{k}}}{\partial t}\Big|_{\rm scat} &=&
   \frac{\partial \rho_{{k}}}{\partial t}\Big|_{\rm im} + 
   \frac{\partial \rho_{{k}}}{\partial t}\Big|_{\rm ph} +
   \frac{\partial \rho_{{k}}}{\partial t}\Big|_{\rm ee}, \nonumber \\ 
   \frac{\partial \rho_{{k}}}{\partial t}\Big|_{\rm im} &=& \pi N_i \sum_{\mathbf{
       Q}}^{s_1,s_2} |U^i_{\mathbf{Q}}|^2
   \delta(E_{s_1,k-q} - E_{s_2,k})  I_{\mathbf{Q}} 
   [(1-{ \rho}_{k-q})T_{s_1}
   I_{-{\mathbf{Q}}} T_{s_2} {
     \rho}_k -
   \rho_{k-q} T_{s_1} I_{-{\mathbf{Q}}} T_{s_2} (1-\rho_k)]  + {\rm H.c.}, \nonumber  \\
   \frac{\partial \rho_{{k}}}{\partial t}\Big|_{\rm ph} &=&  \pi \sum_{{\mathbf{Q}},\lambda}^{s_1,s_2}
   |M_{{\mathbf{Q}},\lambda}|^2 
   I_{\mathbf{Q}} \big\{
   \delta(E_{s_1,k-q} - E_{s_2,k} +
   \omega_{{\mathbf{Q}},\lambda} ) [(N_{{\mathbf{Q}},\lambda} +1)(1-{
     \rho}_{k-q})   T_{s_1}I_{-{\mathbf{Q}}} T_{s_2} { \rho}_k
    - N_{\mathbf{
       Q},\lambda} \rho_{k-q}  T_{s_1} I_{-{\mathbf{Q}}} T_{s_2}(1-\rho_k)]
   \nonumber \\ && \quad +\ \delta(E_{s_1,k-q} - E_{s_2,k} -
   \omega_{{\mathbf{Q}},\lambda} )  [N_{{\mathbf{Q}},\lambda}(1-{
     \rho}_{k-q})  T_{s_1}I_{-{\mathbf{Q}}}
   T_{s_2}{ \rho}_k - (N_{\mathbf{
       Q},\lambda}+1) \rho_{k-q}  T_{s_1}I_{-{\mathbf{Q}}}
   T_{s_2} (1-\rho_k)]
   \big\}+ {\rm H.c.},  \nonumber \\
   \frac{\partial \rho_{{k}}}{\partial t}\Big|_{\rm ee} &=&  \pi
   \sum_{{\mathbf{Q}},k^{\prime}}^{s_1,s_2,s_3,s_4} V_{\mathbf{Q}}^2   \delta(E_{s_1,{
       k}-q} - E_{s_2,k} + E_{s_3,k^{\prime}} -
   E_{s_4, k^{\prime} -q})I_{\mathbf{Q}}
  \big\{(1-{\rho}_{k-q})  T_{s_1}
   I_{-{\mathbf{Q}}} T_{s_2}{   \rho}_k
   \mbox{Tr}[(1-\rho_{k^{\prime}})T_{s_3}
   I_{\mathbf{Q}} T_{s_4}\rho_{k^{\prime} -q}
   I_{-\mathbf{ Q}}] \nonumber \\
 && \quad -\ \rho_{\mathbf{
       k}-q} T_{s_1} I_{-\mathbf{ Q}} T_{s_2}(1-\rho_k)
   \mbox{Tr}[\rho_{k^{\prime}}T_{s_3}
   I_{\mathbf{Q}}  T_{s_4} (1-\rho_{
       k^{\prime}-q})
   I_{-\mathbf{ Q}} ]\big\} + {\rm H.c.},
 \label{scat1}
 \end{eqnarray}
 in which $T_{s_1,s,s'} =  \delta_{s_1, s} \delta_{s_1,s'}$. 
 The statically screened Coulomb potential in the random-phase
 approximation reads \cite{haugjauho}
 \begin{equation}
   \label{V_q}
   V_{q} = {\sum_{q_x,q_y}v_{Q}|I_{\mathbf{ Q}}|^2}/{\kappa(q)},
 \end{equation}
 with the bare Coulomb potential $v_{Q} = 4 \pi e^2 / Q^2 $ and 
 \begin{equation}
   \label{RPA}
   \kappa(q) = 1- {\sum_{q_x,q_y}v_{Q}\sum_{ks}|I_{\mathbf{ Q},s,s}|^2}
   \frac {f_{k+q,s} - f_{k,s}}{E_{s,k+q} - E_{s,k}}.
 \end{equation}
 In Eq.~(\ref{scat1}), $N_i$ is the density of impurities, and $
 |U^{i}_{\mathbf{Q}}|^2 $ is the impurity potential. Furthermore, $|M_{\mathbf{
     Q},\lambda}|^2$ and $N_{\mathbf{ Q},\lambda} =
 [\mbox{exp}(\omega_{\mathbf{ Q},\lambda} /k_B T) - 1]^{-1}$ are the
 matrix elements of the electron-phonon interaction and the Bose
 distribution function, respectively. The phonon energy spectrum for
 phonon with mode $\lambda$ and wavevector $\mathbf{Q}$ is denoted by
 $\omega_{\mathbf{ Q},\lambda}$.

By numerically solving the kinetic spin Bloch equations, L\"u et al. \cite{lu:073703}
investigated the influence of the wire size, orientation, doping
density, and temperature on the spin relaxation time. They also showed
that the Coulomb scattering makes marked contribution to the spin
relaxation. Some typical results that the spin relaxation time can be
effectively manipulated in quantum wires with different orientations
are summarized in Fig.~\ref{fig5.5.1-1}. For (100) quantum wires
[Fig.~\ref{fig5.5.1-1}(a)], due to the competition of the longitudinal
and transverse effective magnetic fields from the Dresselhaus and
Rashba terms addressed above, the spin relaxation time can be
efficiently manipulated by the wire size. Specifically, there is a
minimum in the spin relaxation time when $a_x=a_y$, thanks to the
cancellation of the longitudinal component from the Dresselhaus term
($\langle k_x^2\rangle-\langle k_y^2\rangle=0$) when only the lowest
subband is populated. For (110) quantum wire with $a_x=a_y=30$~nm, the electronic population is mainly in the lowest subband. In
the presence of an electric field of the form ($E_x$, $E_y$, 0), the
relevant Rashba and Dresselhaus terms are 
\begin{eqnarray}
  \label{Rashba_110_1band}
  &&  H^{110}_{\rm R} = {\gamma^{6c6c}_{41}}(-\sigma_x {
    E}_y k_z + \sigma_y E_x k_z), \\
  \label{Dresselhaus_110_1band} 
  &&  H^{110}_{\rm D} =  -\frac{1}{2} {b^{6c6c}_{41}} \sigma_x k_z ( \langle
  k_x^2 \rangle  -  k_z^2  + 2\langle k_y^2 \rangle ).
\end{eqnarray}
The effective magnetic field formed by the Dresselhaus term is along
the $x$-direction, which corresponds to the [$\bar{1}$10]
crystallographic direction, and the effective magnetic field formed by
the Rashba term is in the $x$-$y$ plane. If the direction of the total
effective magnetic field formed by the spin-orbit coupling is tuned to be exactly the
direction of the initial spin polarization, then one can expect an
extremely long spin relaxation time as pointed out in
Refs.~\cite{Liu.jsnm.9.525,cheng:083704}. This is exactly the case as shown in
Fig.~\ref{fig5.5.1-1}(b). However, for wider wire size, the spin
relaxation time is much smaller due to the contribution of higher
subbands. For (111) quantum wires, again in the case with only the lowest
subband being relevant, the Rashba and Dresselhaus terms read 
\begin{eqnarray}
  \label{Rashba_111_1band}
   H^{111}_{\rm R} &=& {\gamma^{6c6c}_{41}}(-\sigma_x {
    E}_y k_z + \sigma_y E_x k_z), \\
  \label{Dresselhaus_111_1band} 
   H^{111}_{\rm D} &=& {b^{6c6c}_{41}} ( -  \frac{\sqrt{2}}{3} \sigma_x k_z  \langle
  k_y^2 \rangle  -
  \frac{1}{\sqrt{6}}\sigma_y k_z (\langle k_x^2 \rangle  + \langle k_y^2
  \rangle) -\frac{2}{3} \sigma_z 
    k_z \langle k_y^2 \rangle  ).
\end{eqnarray}
Similar to the case of (110) quantum wires, one expects very long spin
relaxation time if the total effective magnetic field points into the
directions of initial spin polarization. For a numerical example of
this effect, L\"u et al. chose $E_y$ such that $\gamma^{6c6c}_{41}
E_y + (\sqrt{2}/3)b^{6c6c}_{41} \langle k_y^2 \rangle = 0 $ for a
small wire width $a_x = a_y = 10$~nm, so that the $x$ component of the
total effective magnetic field is zero. For an initial spin
polarization along the $z$-direction, which corresponds to the [111]
crystallographic direction, the spin relaxation time as a function of
$E_x$ is shown in Fig.~\ref{fig5.5.1-1}(c). It is seen that when $a_x
= a_y = 10$~nm, there is a pronounced maximum of the 
spin relaxation time at $E_x = 70$~kV/cm, which fulfills the relation
${\gamma^{6c6c}_{41}} E_x + \frac{1}{\sqrt{6}}{b^{6c6c}_{41}}(\langle
k_x^2 \rangle + \langle k_y^2 \rangle ) \approx 0$.  Consequently, for
this field strength, the direction of the total effective magnetic
field is exactly along the direction of the initial spin polarization
and this leads to a very long spin relaxation time.

\begin{figure}[htb]
  \begin{center}
   \hspace{-0.5 cm} \includegraphics[width=5cm]{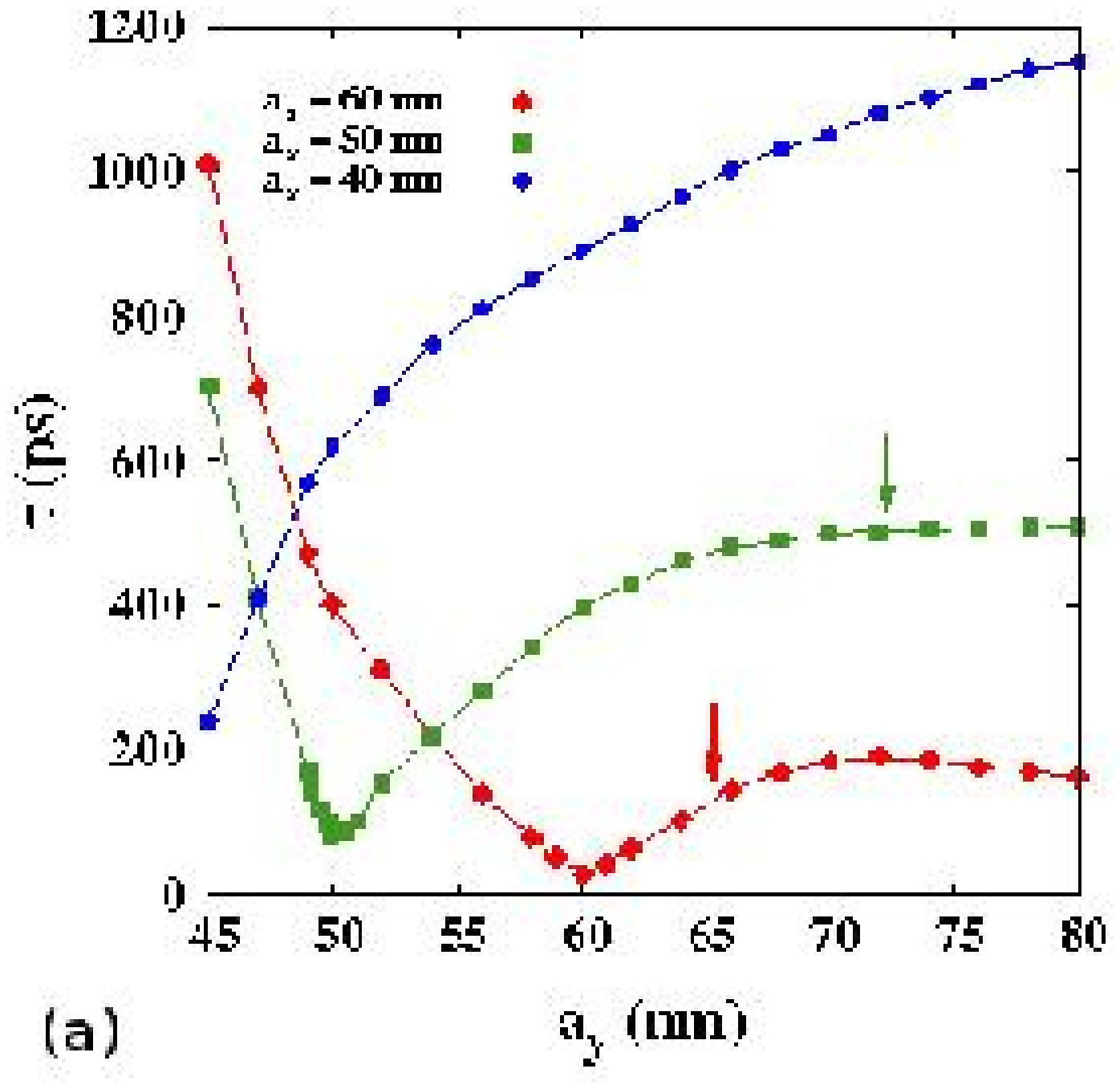}\hspace{-0.5 cm}\includegraphics[width=5cm]{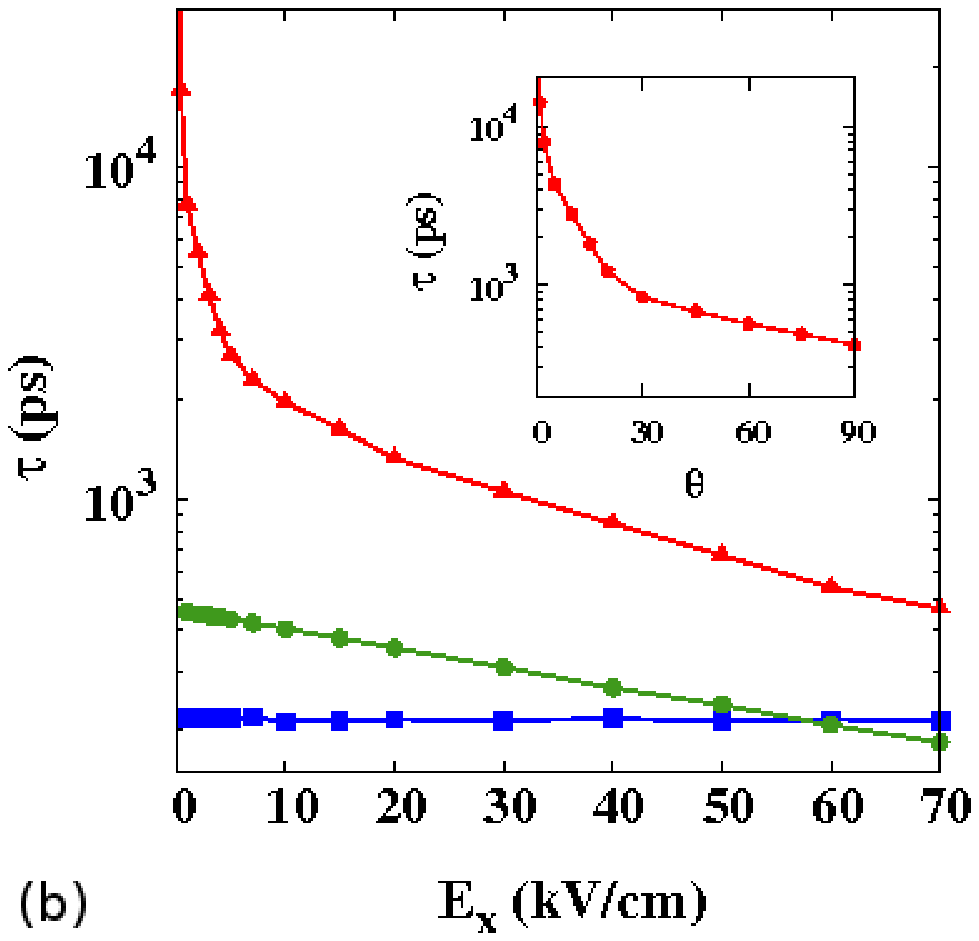}\hspace{-0.5 cm}\includegraphics[width=5cm]{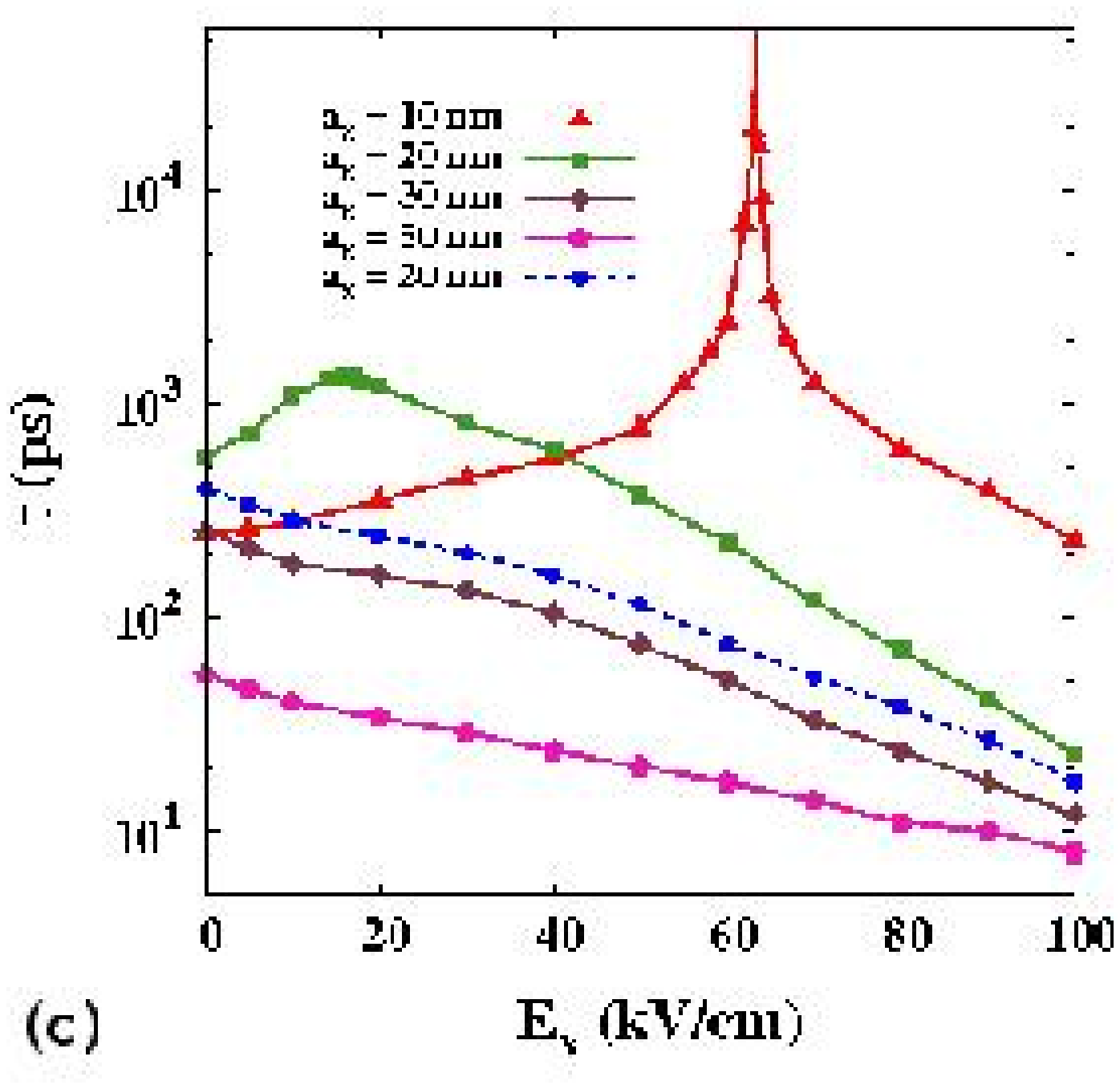}
  \end{center}
  \caption{ (a) Spin relaxation time $\tau$  {\it vs}.
    the quantum wire width in $y$ direction $a_y$ for InAs (100) quantum
    wires at different $a_x$. The electron density is $N = 4 \times 10^5$~cm$^{-1}$
    and $T = 100$~K. The arrows mark the densities at which the electron
    populations in the second and higher subbands are
    approximately 30~\%. (b) Spin relaxation time $\tau$ {\sl vs.}
    $E_x$ for (110) quantum wires at $T = 50$~K and $N_e = 4 \times
    10^5$~cm$^{-1}$. {$\blacktriangle$}:  $a_x = a_y = 30$~nm with an
    initial spin polarization along the $x$-direction; $\blacksquare$:
    $a_x = a_y = 30$~nm with an initial spin polarization along
    the $y$-direction; {$\bullet$}: $a_x = a_y = 50$~nm with an
    initial spin polarization along the $x$-direction. (c) Spin
    relaxation time $\tau$ {\sl vs.} $E_x$ for different wire
    sizes $a_x = a_y$ at $T = 50$~K and $N_e = 4 \times
    10^5$~cm$^{-1}$. The growth direction of the quantum wire is along the
      [111] crystallographic direction. The solid curves are the
    results with the initial spin polarization along $z$-direction and the dashed curve is the
    result with the initial spin polarization along
    $x$-direction. From L\"{u} et al. \cite{lu:073703}.}
  \label{fig5.5.1-1}
\end{figure}

\subsubsection{Hole spin relaxation in $p$-type (001) GaAs quantum wires}
Investigation on hole spin relaxation in quantum wires is very
limited \cite{PhysRevB.58.15652}. Unlike the hole spins in bulk III-V materials which
relax very fast due to the mixture of the heavy-hole and light-hole
states, in confined structures such as quantum wells \cite{lue:125314} and
quantum dots \cite{PhysRevB.71.075308}, the degeneracy of the heavy and light hole bands
is lifted and the mixture of these bands can be tuned by strain
\cite{PhysRevB.71.075308}. Therefore the spin lifetime of holes can be much longer than
that in the bulk \cite{PhysRevB.71.075308,lue:125314}. In quantum wires, similar situation also
happens. L\"u et al. performed a systematic investigation on the
spin relaxation of $p$-type GaAs quantum wires by numerically solving
the kinetic spin Bloch equations \cite{lue:165321}. They reported that the quantum wire size influences
the spin relaxation time effectively by modulating the energy spectrum
and the heavy-hole--light-hole mixing of wire states.

By considering a $p$-doped (001) GaAs quantum wire with rectangular
confinement and hard wall potential, L\"u et al. first obtained
the subband structure by diagonalizing the hole Hamiltonian including
the quantum confinement. Here the light-hole admixture is dominant in
the lowest spin-split subband, but the heavy-hole admixture becomes
also important in higher subbands due to the heavy-hole--light-hole
mixing. Then they investigated the time evolution of holes
by numerically solving the fully microscopic
kinetic spin Bloch equations in the obtained subbands, with all the
scatterings,
particularly the Coulomb scattering,
explicitly included. They found that the quantum wire size influences the
spin relaxation time effectively because the spin-orbit coupling and the subband
structure in quantum wires depend strongly on the confinement. When the
quantum wire size increases, the lowest spin-split subband and the
second-lowest spin-split subband can be very close to each other at
an anticrossing point. If the anticrossing is close to the Fermi surface,
the contribution from the spin-flip scattering reaches a maximum and
correspondingly the spin relaxation time reaches a minimum. Moreover,
they showed that the dependence of 
the spin relaxation time on confinement size in quantum wires behaves oppositely to the trend found in
quantum wells. It was also found that,
 when the quantum wire size is very small, the
spin relaxation time can  either increase or decrease with increasing hole density, depending on
the spin mixing of the subbands. However, the behavior of holes in
quantum wires 
where the spin relaxation time increases or decreases with hole density is
quite different from the one of light holes in quantum wells with small
well width \cite{lue:125314}. These features originate from the subband
structure of the quantum wires and
the spin mixing which give rise to the spin-flip scattering.
The spin mixing and inter-subband
scattering  are  modulated more dramatically in quantum wires by changing the
hole distribution in different subbands.
They also investigated the effects of temperature and
initial spin polarization, showing that the inter-subband scattering and
the Coulomb Hartree-Fock contribution can make
a marked contribution to the spin relaxation.

By assuming the growth direction of the quantum wire along the
$z$-axis ([001] crystallographic direction), the Hamiltonian of holes in the basis of
spin-3/2 projection ($J_z$) eigenstates with quantum numbers $+\frac{3}{2}$,
$+\frac{1}{2}$, $-\frac{1}{2}$ and $-\frac{3}{2}$ can be written as \cite{winklerbook}
\begin{equation}
  \label{Luttinger_Hamiltonian}
  H_{h} = \left( \begin{array}{cccc} H_{hh} & S & R &
      0 \\ S^\dagger & H_{lh} & 0 & R \\ R^\dagger & 0 &
      H_{lh}& -S \\ 0 & R^\dagger & -S^\dagger & H_{hh}  \end{array}
  \right)+  H_{8v8v}^r +   H_{8v8v}^b + V_c({\bf r}),
\end{equation}
where $V_c({\bf r})$ is the hard-wall confinement potential in $x$ and
$y$ directions and

\begin{eqnarray}
  \label{hh}
  &&  H_{hh} = \frac{1}{2m_0} [(\gamma_1 + \gamma_2)[P_x^2
  + P_y^2] +(\gamma_1 -2 \gamma_2) P_z^2,\\
  \label{lh}
  &&  H_{lh} = \frac{1}{2m_0} [(\gamma_1 - \gamma_2)[P_x^2
  + P_y^2] +(\gamma_1 + 2 \gamma_2) P_z^2,\\
  \label{S}
  &&  S = -  \frac{\sqrt{3} \gamma_3}{m_0} P_z[P_x -  iP_y], 
  R = - \frac{\sqrt{3}}{2m_0} \{\gamma_2 [P_x^2 - P_y^2]-
  2 i \gamma_3  P_xP_y \},\\
  \label{Rashba}
  &&  H_{8v8v}^r = \frac{\gamma^{8v8v}_{41}}{\hbar} [J_x (P_y {\cal E}_z -
  P_z {\cal E}_y) + J_y (P_z {\cal E}_x - P_x{\cal E}_z)+ J_z (P_x {\cal E}_y - P_y {\cal E}_x)], \\
  \label{Dresselhaus}
  &&  H_{8v8v}^b = \frac{b^{8v8v}_{41}}{\hbar^3} \{ J_x [P_x(P_y^2 -
  P_z^2)] + J_y [P_y(P_z^2 -
  P_x^2)] + J_z [P_z(P_x^2 -
  P_y^2)] \}.
\end{eqnarray}
In these equations, $m_0$  denotes the free electron mass,
$\gamma_1$, $\gamma_2$ and $\gamma_3$ are the Luttinger
parameters, ${\cal E}$ is the  electric field and $J_i$ are
  spin-3/2 angular momentum matrices \cite{PhysRev.102.1030}. $H_{8v8v}^r$ is the
  spin-orbit coupling arising from the
 structure inversion asymmetry and $H_{8v8v}^b$ is the
 spin-orbit coupling from the bulk inversion asymmetry. As shown
 in Ref.~\cite{lue:165321}, these two terms turn out to be one or
two orders of magnitude smaller than the intrinsic spin-orbit coupling from the
four-band Luttinger-Kohn Hamiltonian [the first term in
Eq.~(\ref{Luttinger_Hamiltonian})]. Moreover, from the first term in Eq.~(\ref{Luttinger_Hamiltonian}), one can
 see that the light hole spin-up
 states can be directly mixed with the heavy hole states by $S$ and
  $R$, but the mixing between light hole spin-up states and light hole spin-down
 states has to be mediated by the heavy hole states.
All the mixing is related to the confinement. When the
 confinement decreases, the mixing increases due to the
 decrease of the energy gap between the light hole and heavy hole states.

The kinetic spin Bloch equations $\dot{\bf { \rho}}_{{ k}} + \dot{\bf { \rho}}_{{
      k}}|_{\rm coh} +   \dot{\bf { \rho}}_{{ k}}|_{\rm scat} = 0$
  can be written in either the collinear spin space which is
constructed by basis $\{s\}$, with $\{s\}$  obtained from the eigenfunctions of
the diagonal part of $H_{h}(k)$.
$|s\rangle = |m,n \rangle |\sigma \rangle $ with $ \langle r
|m,n \rangle = \frac{2}{\sqrt{a_x a_y}} \sin(\frac{m\pi
  y}{a_y}) \sin(\frac{n\pi x}{a_x}) e^{i {k}z}$ and $|\sigma
\rangle$ standing for the eigenstates of 
 $J_z$. Then the matrix elements in the
collinear spin space $\rho^c_{k,s_1,s_2}$ is
written as $\rho^c_{k,s_1,s_2} = \langle s_1 |
\rho_k | s_2 \rangle $. Here the superscript ``$c$''
denotes the quantum number distinguishing states in the collinear spin
space. One can also project $\rho_k$ in the ``helix'' spin space which is
constructed by basis $\{\eta\}$ with $\eta$ being the eigenfunctions of
$H_{h}(k)$:
\begin{equation}
\label{spec}
H_{h}(k)|\eta\rangle=E_{\eta,k}|\eta\rangle.
\end{equation}
This basis function is a mixture of
light-hole and heavy-hole states and is $k$ dependent.
Then the matrix elements in the helix spin
space $\rho^h_{{k},\eta,\eta^{\prime}}$ can be written as $\rho^h_{{
    k},\eta,\eta^{\prime}} = \langle \eta |\rho_k|
\eta^{\prime} \rangle$, with the superscript ``$h$''
denoting  the helix spin
space. The density matrix in the helix spin space can be
transformed from that in the collinear one by a unitary
transformation: $ \rho^h_k = U^{\dagger}_k\rho_{k}^c U_k $,
where $U_k(i,\alpha) = \eta^i_{\alpha}(k)$ with
$\eta^i_{\alpha}(k)$ being the $i$th element of the $\alpha$th eigenvector
after the diagonalization of $H_h(k)$. 

In helix spin space, the coherent terms read \begin{eqnarray}
  \label{coh}
\dot{\rho}^h_{{k}}|_{\rm coh} =  i \Big[ \sum_{\bf Q} V_{\bf Q}
U^{\dagger}_{k}I_{\bf Q} U_{k-q} { \rho}^h_{k-q}
U^{\dagger}_{k-q}I_{\bf -Q} U_{k},  \rho^h_k
\Big]-i \Big[ U^{\dagger}_{k}H_{h}(k)U_{k} ,  \rho^h_k
\Big],
\end{eqnarray}
where $I_{\bf Q} $  is the form factor in the collinear spin space with wave
vector ${\bf Q} \equiv (q_x, q_y, q)$.
The first term in Eq.~(\ref{coh}) is the
Coulomb Hartree-Fock term and the second term is the contribution
from the intrinsic spin-orbit coupling from the  Luttinger-Kohn Hamiltonian.
$I_{\bf Q} $ can be written as  $I_{{\bf Q},s_1,s_2} = \langle s_1 | e^{i
  {\bf Q} \cdot {\bf r}} | s_2 \rangle  = \delta_{\sigma_1,\sigma_2}
F(m_1,m_2,q_y,a_y)F(n_1,n_2,q_x,a_x)$, with $F(m_1 , m_2,q,a)$ being
expressed as Eq.~(\ref{Ffunction}). For small spin polarization, the contribution from the
Hartree-Fock term in the coherent term is negligible \cite{PhysRevB.68.075312,stich:176401,stich:205301}
 and the spin precession  is determined
by the spin-orbit coupling,
$\dot{\rho}^h_{{k},\eta,\eta^{\prime}}|_{\rm coh} = -i\rho^h_{{k},\eta,\eta^{\prime}}(E_{\eta,k} -
  E_{\eta^{\prime},k})$, which is proportional to the energy gap between
  $\eta$ and $\eta^{\prime}$ subbands.

The scattering terms include the
hole-nonmagnetic-impurity,  hole-phonon
and hole-hole Coulomb scatterings. In the helix spin space,
the scattering terms are given by

\begin{eqnarray}
  \dot{\rho}^h_{{k}}|_{\rm scat} &=& \pi N_i \sum_{{\bf
      Q},\eta_1,\eta_2} |U^i_{\bf Q}|^2
  \delta(E_{\eta_1,k-q} - E_{\eta_2,k})  U^{\dagger}_{k}I_{\bf Q} U_{k-q}
  [(1-{ \rho}^h_{k-q})T_{k-q,\eta_1}
  U^{\dagger}_{k-q}I_{-{\bf Q}} 
U_{k}T_{k,\eta_2} {
    \rho}^h_k \nonumber\\ && \quad  -
  \rho^h_{k-q} T_{k-q,\eta_1} U^{\dagger}_{{k-q}}I_{-{\bf Q}} U_{k}T_{k,\eta_2} (1-\rho^h_k)]
  \nonumber \\ && + \pi \sum_{{\bf Q},\eta_1,\eta_2,\lambda} |M_{{\bf Q},\lambda}|^2
  U^{\dagger}_{k}I_{\bf Q} U_{k-q} \{
  \delta(E_{\eta_1,k-q} - E_{\eta_2,k} +
  \omega_{{\bf Q},\lambda} )[(N_{{\bf Q},\lambda} +1) (1-{
    \rho}^h_{k-q})  T_{k-{q},\eta_1}U^{\dagger}_{k-q}I_{-{\bf Q}} U_{k}T_{{
      k},\eta_2} { \rho}^h_k \nonumber\\ && \quad - N_{{\bf Q},\lambda} \rho^h_{k-q}  T_{{
      k}-q,\eta_1} U^{\dagger}_{{\bf
      k-q}}I_{-{\bf Q}} U_{k}T_{k,\eta_2}(1-\rho^h_k)] +\delta(E_{\eta_1,k-q} - E_{\eta_2,k} -
  \omega_{{\bf Q},\lambda} ) [N_{{\bf Q},\lambda}(1-{
    \rho}^h_{k-q})  T_{{k}-q,\eta_1} \nonumber \\ && \quad \times U^{\dagger}_{k-q}I_{-{\bf Q}}
U_{{k}}T_{k,\eta_2}{ \rho}^h_k - (N_{{\bf
 Q},\lambda}+1) \rho^h_{k-q} T_{{
      k}-q,\eta_1}U^{\dagger}_{k-q}I_{-{\bf Q}} U_{k}T_{k,\eta_2}(1-\rho^h_k)]
  \}  \nonumber \\ && + \pi \sum_{{\bf Q},k^{\prime}}
  \sum_{\eta_1,\eta_2,\eta_3,\eta_4} V_{\bf Q}^2 U^{\dagger}_{k}I_{\bf Q} U_{k-q}
 \delta(E_{\eta_1,{
      k}-q} - E_{\eta_2,k} + E_{\eta_3,k^{\prime}} -
  E_{\eta_4, k^{\prime} -q})
  \{
  (1-{\rho}^h_{k-q})T_{k-q,\eta_1}
  U^{\dagger}_{k-q}I_{-{\bf Q}} U_{k} T_{{
      k},\eta_2}{   \rho}^h_k
  \nonumber \\
&&\quad \times\mbox{Tr}[(1-\rho^h_{k^{\prime}})T_{\eta_3,k^{\prime}}
  U^{\dagger}_{k}I_{\bf Q} U_{k-q} T_{k^{\prime} -q,\eta_4}\rho^h_{k^{\prime}
    -q}
  U^{\dagger}_{k-q}I_{-{\bf
Q}} U_{k}] - \rho^h_{{\bf
      k}-q} T_{k-q,\eta_1} U^{\dagger}_{k-q}I_{-{\bf Q}} U_{k} T_{{
      k},\eta_2}(1-\rho^h_k)
\nonumber \\ &&\quad \times\mbox{Tr}[\rho^h_{k^{\prime}}T_{\eta_3,k^{\prime}}
  U^{\dagger}_{k}I_{\bf Q} U_{k-q} T_{k^{\prime} -q,\eta_4} (1-\rho^h_{
      k^{\prime}-q})
  U^{\dagger}_{k-q}I_{-{\bf Q}} U_{k}]\} + {\rm H.c.},
\label{scat}
\end{eqnarray}
in which $T_{k,\eta}(i,j) =  \delta_{\eta i} \delta_{\eta j}
$. $V_{\bf Q}$ in Eq.~(\ref{scat}) reads $ V_{\bf Q} =
4 \pi e^2 /[\kappa_0 (q^2 + q_{\|}^2 + \kappa^2)]$, with
$\kappa_0$ representing the static dielectric constant and $\kappa^2 =
4\pi e^2 N_h /(k_B T \kappa_0 a^2)$ standing for the Debye screening
constant. $N_i$ in  Eq.~(\ref{scat}) is the impurity density and  $
|U^{i}_{\bf Q}|^2 = \{ 4\pi Z_i  e^2 /[\kappa_0 (q^2 + q_{\|}^2 +
\kappa^2)]\}^2$ is the impurity potential with $Z_i$ standing for the
charge number of the impurity.  $|M_{{\bf Q},\lambda}|^2$ and $N_{{\bf
    Q},\lambda} = [\mbox{exp}(\omega_{{\bf Q},\lambda} /k_B T) - 1]^{-1}$ are
the matrix element of the hole-phonon interaction and the Bose
distribution function  with phonon energy spectrum $\omega_{{\bf
    Q},\lambda}$ at phonon mode $\lambda$ and wave vector ${\bf Q}$,
respectively. Here the hole-phonon scattering includes the
hole--longitudinal optical-phonon  and hole--acoustic-phonon scatterings with the
explicit expressions of $|M_{{\bf Q},\lambda}|^2$ can be found in
Refs.~\cite{PhysRevB.68.075312,PhysRevB.70.195318,zhou:045305}. It is noted that the energy spectrum $E_{\eta,k}$ in
the scattering term contains the spin-orbit coupling which cannot be
ignored due to the strong coupling. By solving the kinetic spin Bloch equations, one obtains
hole spin relaxation.

\begin{figure}[thb]
  \begin{center}
    \includegraphics[width=9cm]{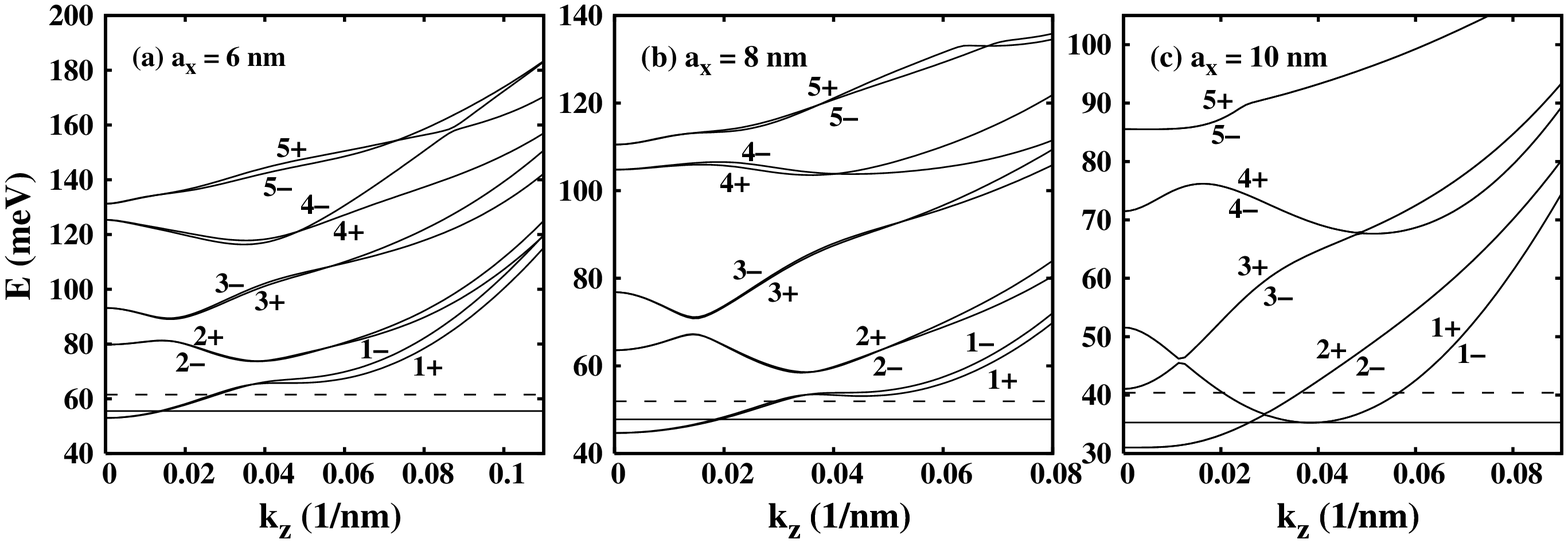}
  \end{center}
  \begin{center}
    \includegraphics[width=9cm]{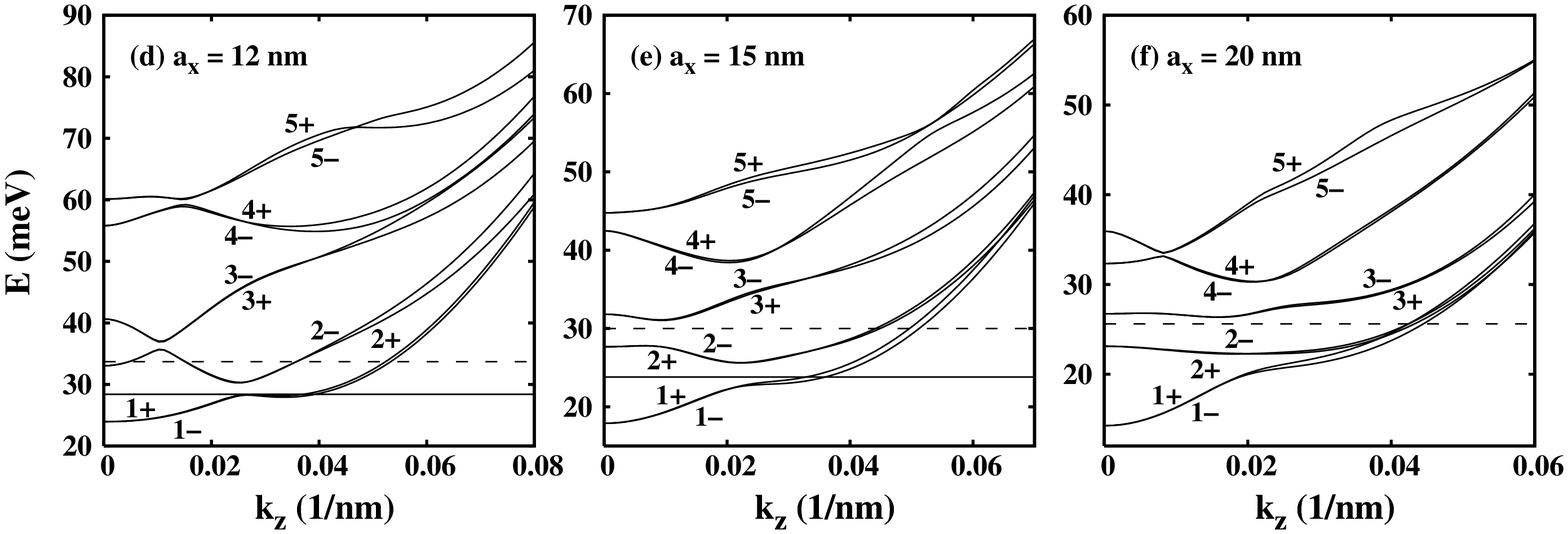}
  \end{center}
  \caption{Typical energy spectra of holes in GaAs (001)
quantum wires for (a) $a_x=6$~nm; (b) $a_x=8$~nm; (c)
    $a_x=10$~nm; (d) $a_x=12$~nm; (e)  $a_x=15$~nm; and (f)
    $a_x=20$~nm. $a_y=10$~nm.
    ${\langle E\rangle}$ at $T = 20$~K is also
    plotted: solid line for $N_h = 4 \times 10^5 $~cm$^{-1}$ and
    dashed line for $N_h = 2 \times 10^6 $~cm$^{-1}$. From L\"{u} et al. \cite{lue:165321}.
 }
\label{fig5.5.2-1}
\end{figure}

The typical subband structure is shown in Fig.~\ref{fig5.5.2-1} for
different confinements. Each subband is denoted as 
$l+$ ($l-$) if the dominant spin component  is the
spin-up (-down) state.
One can see from Fig.~\ref{fig5.5.2-1} that $1+$ and $1-$ subbands are very close
to each other, so are the subbands $2\pm$. The spin-splitting
between them is mainly caused by the spin-orbit coupling arising from
the bulk inversion asymmetry, because that the
spin-splitting caused by the spin-orbit coupling arising from the
structure inversion asymmetry is three orders of
magnitude smaller
than the diagonal terms in Eq.~(\ref{Luttinger_Hamiltonian}) and
can not be seen in Fig.~\ref{fig5.5.2-1}. The spin-splitting caused by
the bulk inversion asymmetry is
proportional to $(P_x^2 - P_y^2)$, which disappears when the
confinement in $x$ and $y$ directions are symmetrical. Therefore,
$l\pm$ are almost degenerate when $a_x = a_y =
10$~nm. If one excludes the spin-orbit coupling from the bulk inversion asymmetry and structure inversion asymmetry, $l\pm$
are always degenerate because of the Kramers degeneracy.
One also observes that when $a_x$ gets larger, the subbands are closer to
each other. Especially, in the case of
$a_x=a_y=10$~nm, there are anticrossing points due to the
heavy-hole--light-hole mixing in the Luttinger Hamiltonian.
When $a_x$ keeps on increasing,
the anticrossing
point at small $k$ between the $1\pm$ and $2\pm$ gradually
 disappears. However, at large $k$ region,
the lowest two subbands become very close to each other. These will lead
to significant effect on spin relaxation time. In
Fig.~\ref{fig5.5.2-1}, a quantity $\langle E\rangle$, with 
\begin{equation}
{\langle E\rangle} =\frac{\sum_{l}\int^{+\infty}_{-\infty} dk
(\rho^h_{k,l+,l+}-\rho^h_{k,l-,l-}) (E_{l+,k}+E_{l-,k})}{2\sum_{l}\int^{+\infty}_{-\infty} dk
  (\rho^h_{k,l+,l+}-\rho^h_{k,l-,l-})},
\end{equation}
is introduced to represent the energy region where spin precession and
relaxation between the $+$ and $-$ bands mainly take place. It is seen
from the figure that for $N_h=4\times 10^5$~cm$^{-1}$ and $2\times
10^6$~cm$^{-1}$, $\langle E\rangle$ only intersects with $1\pm$ and
$2\pm$ subbands. It is also seen that the dominant spin
component in $1+$ ($1-$) state is the spin-up (spin-down) light-hole
state. 

There are three mechanisms leading to spin relaxation. First, the
spin-flip scattering, which includes the scattering between $l+$ and
$l-$ subbands and the scattering between $l+$ and $l^{\prime}-$
subbands ($l\ne l^{\prime}$), can cause spin relaxation. The spin
relaxation time decreases with the spin-flip scattering, with the scattering strength
being proportional to the spin mixing of the helix subbands. Second,
because of the coherent term $\dot{\rho}^h_{{k}}|_{\rm coh}$, there is a
spin precession between different subbands. The frequency of this spin
precession depends on $k$ and this dependence serves as inhomogeneous
broadening. As shown in 
Refs.~\cite{PhysRevB.61.2945,wu:epjb.18.373,JPSJ.70.2195,PhysRevB.68.075312,PhysRevB.70.195318,PhysRevB.69.245320}, in the presence of the
inhomogeneous broadening, even the spin-conserving scattering can
cause irreversible spin relaxation.  As a result, the spin-conserving
scattering, i.e., the scattering between $l+$ and $l^{\prime}+$ and
the scattering between $l-$ and $l^{\prime}-$, can cause spin
relaxation along with the inhomogeneous broadening. At last, the
spin-flip scattering along with the inhomogeneous broadening can also cause an
additional spin relaxation.  

It is seen from Fig.~\ref{fig5.5.2-1}(a) that when $N_h = 4 \times 10^5 $~cm$^{-1}$
and $a_x = 6$~nm,
$\langle E \rangle$ only intersects  with the $1\pm$ subbands and is
far away from the $2\pm$ subbands. Therefore,
holes populate the $1\pm$  subbands only.
 As pointed out before, the coherent term
 $\dot{\rho}^h_{k,1+,1-}|_{\rm coh}$ is
proportional to $(E_{1+,k}-E_{1-,k})$. As holes are only
populating states in the small $k$ region where the
spin splitting between $1\pm$ is negligible,  the spin
precession between these two states, and thus the inhomogeneous
broadening, is very small. Consequently  the main spin-relaxation mechanism
is due to the spin-flip scattering, i.e., the scattering between
$1\pm$ subbands. 

In the case of larger $a_x$ and $N_h$ as shown in Fig.~\ref{fig5.5.2-1}(c)-\ref{fig5.5.2-1}(f),
where $\langle E \rangle$ is close to or intersects with the $2\pm$ subbands,
holes populate both the $1\pm$ and $2\pm$ subbands. The spin-flip
scattering here includes the scattering between 
$1\pm$ states, the scattering between $2\pm$ states and the spin-flip
scattering between $1\pm$ and $2\pm$ subbands. 
This spin-flip scattering is still found to be the main spin
relaxation mechanism. Besides, differing from the case of
Fig.\ 2(a), the coherent term $\dot{\rho}^h_{k,1\pm,2\pm}|_{\rm coh}$ 
is proportional to the energy gap between $1\pm$ and $2\pm$, and it is
much larger than $\dot{\rho}^h_{k,1+,1-}|_{\rm coh}$. As a result, there
is a much stronger spin precession between $1\pm$ and $2\pm$ subbands
with a frequency depending on $k$, and the inhomogeneous broadening
caused by this precession along
with both the spin-conserving scattering and the spin-flip scattering
can make a considerable contribution to the spin relaxation.

A typical hole spin relaxation time as a function of wire width is
given in Fig.~\ref{fig5.5.2-2} at different temperatures and hole
densities. The underlying physics can be well understood from the
corresponding subband mixing in Fig.~\ref{fig5.5.2-1}. Similarly the
temperature, hole density, and spin polarization dependences of spin
relaxation were also
discussed in detail in Ref.~\cite{lue:165321}.

\begin{figure}[thb]
  \begin{center}
    \includegraphics[width=7cm]{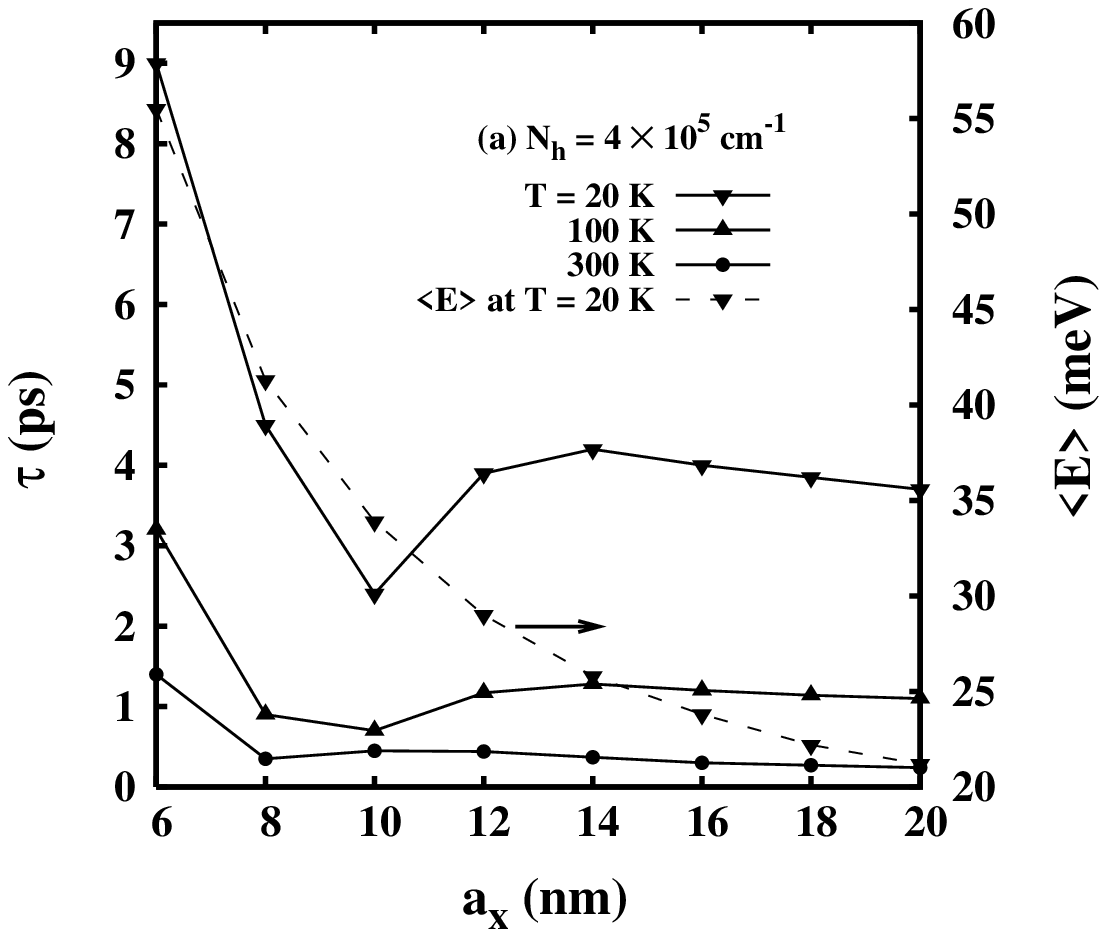}\includegraphics[width=7cm]{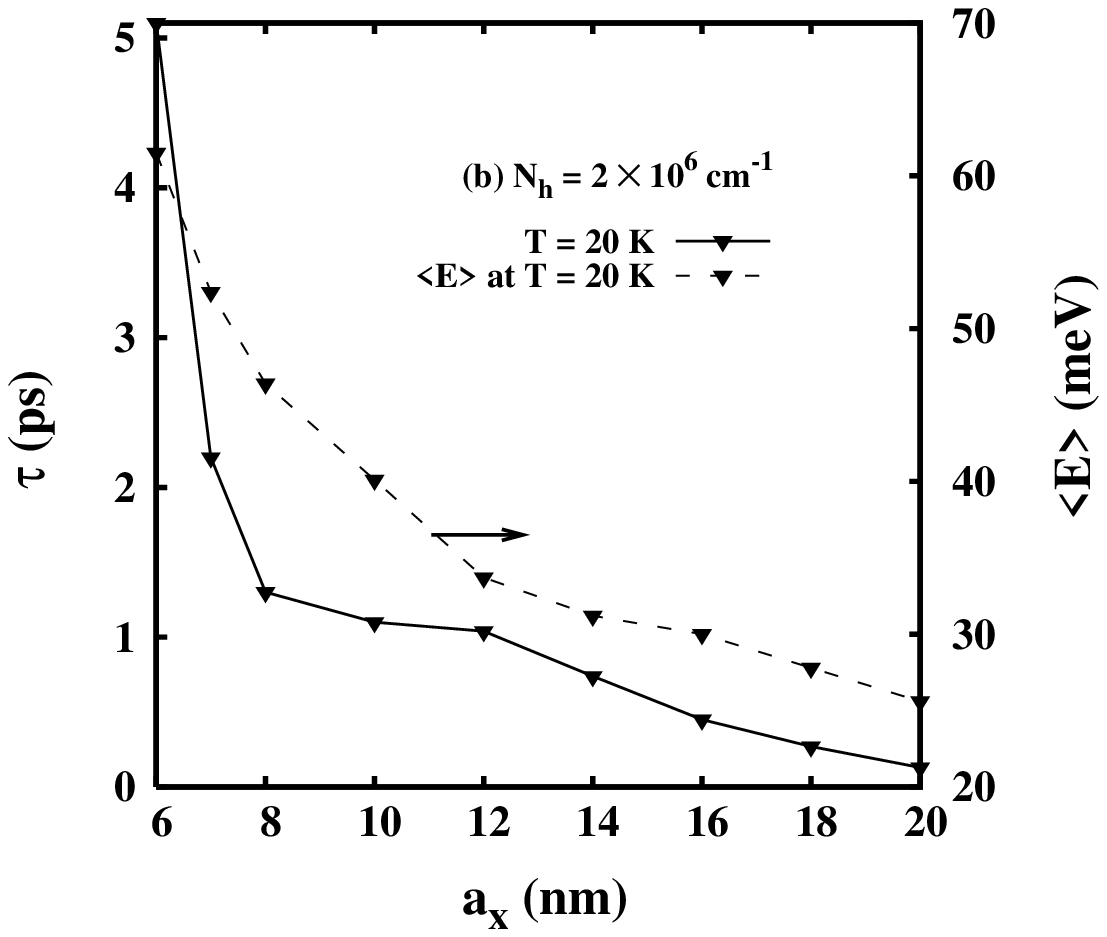}
  \end{center}
  \caption{Spin relaxation time $\tau$ {\it vs}. the quantum wire width in $x$
   direction $a_x$ for (a) $N_h = 4 \times
    10^5$~cm$^{-1}$ at different temperatures and (b) $N_h = 2 \times
    10^6$~cm$^{-1}$ at $T = 20$~K. $a_y=10$~nm.
 From L\"{u} et al. \cite{lue:165321}.}
\label{fig5.5.2-2}
\end{figure}

\subsection{Spin relaxation  in bulk III-V semiconductors}
\label{jiang-wu-bulk}

The study of spin dynamics in bulk III-V semiconductors has a long
history. The theoretical study of spin relaxation and the
systematic experimental investigation started as early as 1970s
\cite{opt-or}. The early studies have been reviewed comprehensively in the
book {\em ``Optical Orientation''} \cite{opt-or}. In the past
decade, the topic gained renewed interest in the context of
semiconductor spintronics \cite{S.A.Wolf11162001}. Differing from
early experimental studies which mainly focus on electron spin
relaxation in $p$-type bulk III-V semiconductors, experimental
investigations in the past decade mainly focus on $n$-type and
intrinsic bulk III-V semiconductors \cite{zuticrmp}. Theoretically,
Song and Kim systematically calculated the density and temperature
dependences of electron spin relaxation time in bulk $n$-type and
$p$-type GaAs, InAs, GaSb and InSb by including the D'yakonov-Perel',
Elliott-Yafet and Bir-Aronov-Pikus mechanisms \cite{PhysRevB.66.035207}.
However, their approach was based on the approximate formulae in the
book {\em ``Optical Orientation''} \cite{opt-or} where the momentum
scattering time is calculated via the approximate formulae for
mobility \cite{PhysRevB.66.035207}. A key mistake is that they used
the formulae that can only be used in the nondegenerate regime. This
makes their results in the low-temperature and/or high-density regime
questionable. Moreover, the single-particle approach they used limits
the validity of their results. Other theoretical investigations have
similar problems
\cite{krishnamurthy:1761,PhysRevB.64.161301,PhysRevB.71.245312,PhysRevB.54.1967,harmon:115204,barry:3686}.
Therefore a systematic many-body investigation from the fully
microscopic kinetic spin Bloch equation approach is needed.

In this subsection, we review the comprehensive study on the topic via
the fully microscopic kinetic spin Bloch equation approach by Jiang and Wu
\cite{jiang:125206} where many important predictions and results that
can not be achieved via the single-particle approach were
obtained.\footnote{Review of the experimental studies and the
  single-particle theories on electron spin relaxation in bulk
  $n$-type and intrinsic III-V semiconductors in metallic regime is
  presented in Sec.~4.2.1 and 4.2.2, respectively. Studies on electron
  spin relaxation in bulk $p$-type III-V semiconductors was reviewed
  in the book {\em ``Optical Orientation''} \cite{Titkovbook}.}

The system considered is the bulk III-V semiconductors, where the spin
interactions have been introduced in Sec.~2. Specifically, the spin-orbit
coupling consists of the Dresselhaus term,
\be
{\bf \Omega}({\bf k})=2\gamma_{D}[k_x(k_y^2-k_z^2),
k_y(k_z^2-k_x^2), k_z(k_x^2-k_y^2)]
\ee
and the strain-induced term which is linear in ${\bf k}$. The
electron-hole exchange interaction is given by Eq.~(\ref{bapitot})
which consists of both the short-range interaction [see
Eq.~(\ref{bapisr})] and the long-range one [see
Eq.~(\ref{bapilr})]. The Elliott-Yafet mechanism is included in the
electron-impurity, electron-phonon, electron-electron and
electron-hole scatterings in the kinetic spin Bloch equations \cite{jiang:125206}.

\begin{figure}[htb]
\begin{center}
\includegraphics[height=4.65cm]{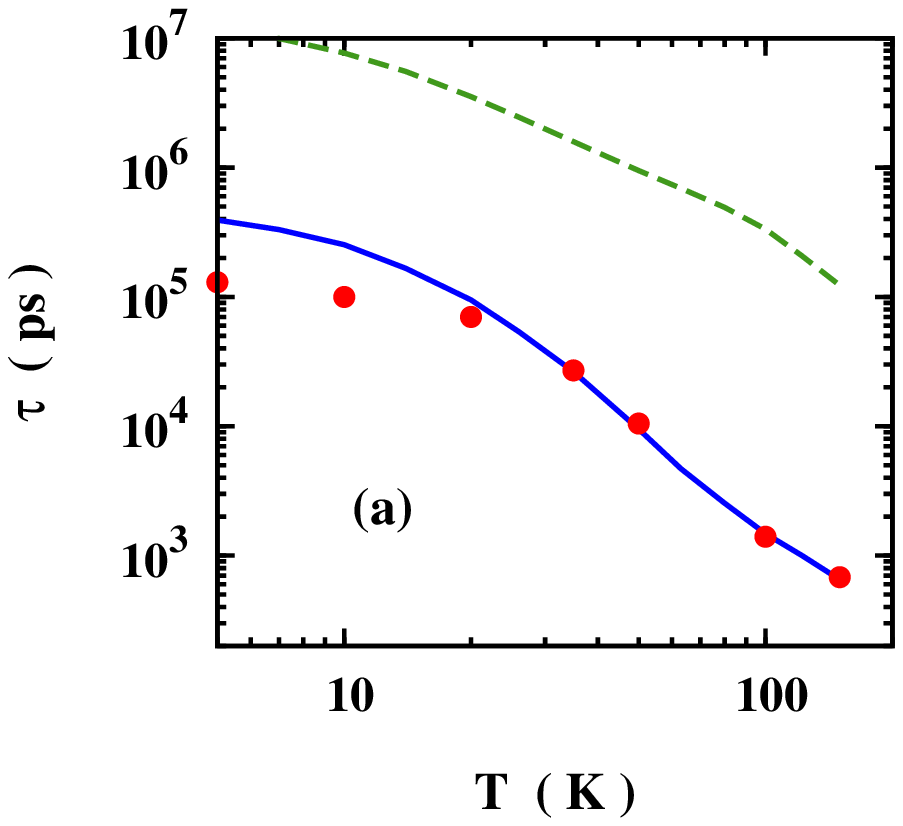}\includegraphics[height=4.65cm]{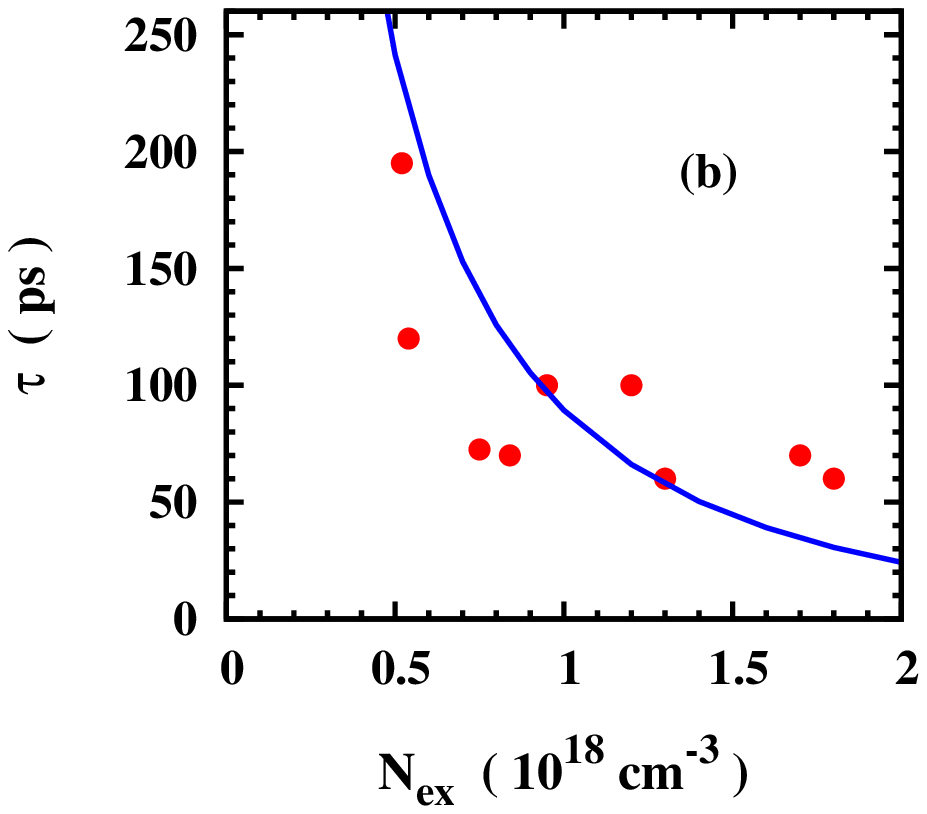}\includegraphics[height=4.65cm]{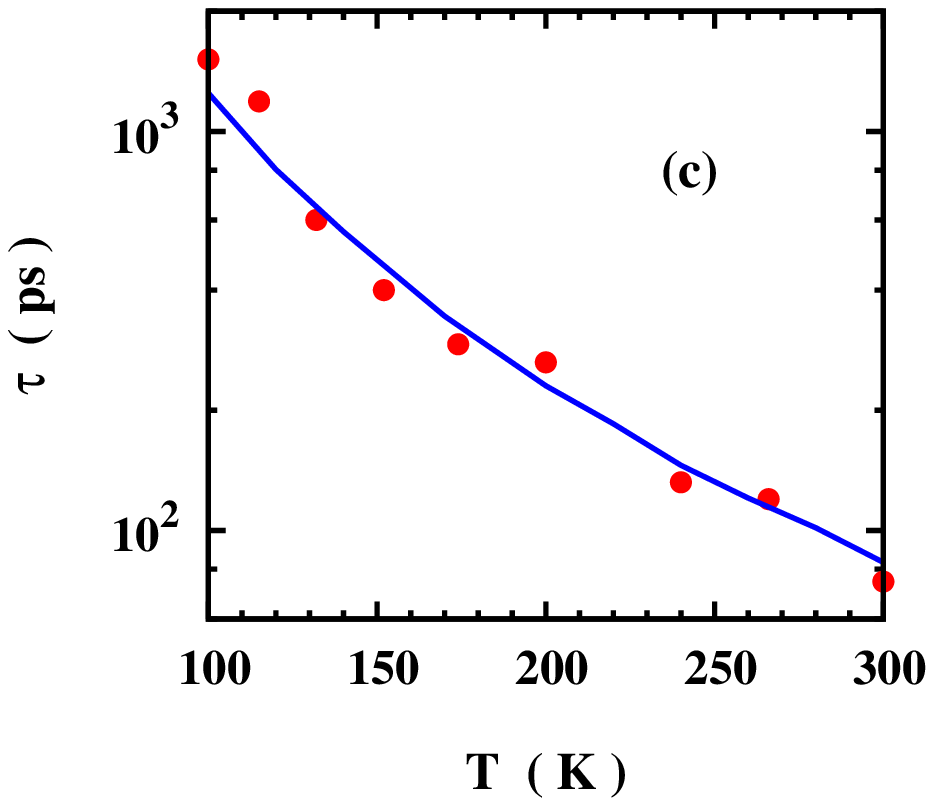}
\end{center}
\caption{ (a) $n$-GaAs. Spin relaxation times $\tau$ from the
  experiment in Ref.~\cite{PhysRevLett.80.4313} ($\bullet$) and
  from the calculation via the kinetic spin Bloch equation approach with only the D'yakonov-Perel'
  mechanism (solid curve) as well as that with only the Elliott-Yafet
  mechanism (dashed curve). $n_{e}=10^{16}$~cm$^{-3}$, $n_i=n_e$ and
  $N_{\rm ex}=10^{14}$~cm$^{-3}$. $\gamma_{
  D}=8.2$~eV$\cdot$\AA$^{3}$. (b) $p$-GaAs. Spin relaxation times $\tau$ from the 
  experiment in Ref.~\cite{PhysRevB.24.3623} ($\bullet$) and
  from the calculation via the kinetic spin Bloch equation approach (solid
  curve). $n_h=6\times 10^{16}$~cm$^{-3}$, $n_i=n_h$ and
  $T=100$~K. $\gamma_{D}=8.2$~eV$\cdot$\AA$^{3}$. (c)
  $p$-GaAs. Spin relaxation times $\tau$ from the experiment in
  Ref.~\cite{PhysRevB.37.1334} ($\bullet$) and from the calculation via
  the kinetic spin Bloch equation approach (solid curve). $n_h=1.6\times 10^{16}$~cm$^{-3}$,
  $n_i=n_h$ and $N_{\rm ex}=10^{14}$~cm$^{-3}$. $\gamma_{
  D}=10$~eV$\cdot$\AA$^{3}$. From Jiang and Wu \cite{jiang:125206}.} 
\label{fig_bulk_exp}
\end{figure}

\subsubsection{Comparison with experiments}

Jiang and Wu compared their calculation via the kinetic spin Bloch
equation approach \cite{jiang:125206} with experimental results
measured in GaAs in Refs.~\cite{PhysRevLett.80.4313,PhysRevB.24.3623,PhysRevB.37.1334}.
The results are presented in Fig.~\ref{fig_bulk_exp}(a)-(c) where the
calculated spin lifetimes are plotted as solid curves and the
experimental results as red dots. It is seen that the calculation
agrees quite well with experimental data for both $n$- and $p$-type
GaAs in a wide temperature and density regimes. The deviation
in the low-temperature regime ($T<20$~K) in
Fig.~\ref{fig_bulk_exp}(a) is due to the rising of electron
localization. Good quantitative agreement with experimental data for
intrinsic GaAs at room temperature was also achieved by Jiang and Wu
in Ref.~\cite{jiang-2009}. In those calculations, {\em all} the material
parameters are taken from the standard handbook of
{\em Landolt-B\"ornstein} \cite{madelung}. The only free parameter is
the Dresselhaus spin-orbit coupling constant $\gamma_{D}$ which has
not been unambiguously determined by experiment or
theory. Nevertheless the parameter $\gamma_{D}$ for GaAs used in the
calculation is close to the value from recent {\it ab initio}
calculation with GW approximation
($\gamma_{D}=8.5$~eV$\cdot$\AA$^{3}$) \cite{chantis:086405} and that
from the recent fitting of the magnetotransport in chaotic GaAs quantum dots
($\gamma_{D}=9$~eV$\cdot$\AA$^{3}$) \cite{krich:226802}. The
good agreement with experimental data indicates that the calculation
has achieved {\em quantitative accuracy}.

\subsubsection{Electron-spin relaxation in $n$-type bulk III-V
  semiconductors}

\begin{figure}[htb]
  \begin{minipage}[h]{0.5\linewidth}
   \centering
  \includegraphics[height=5.5cm]{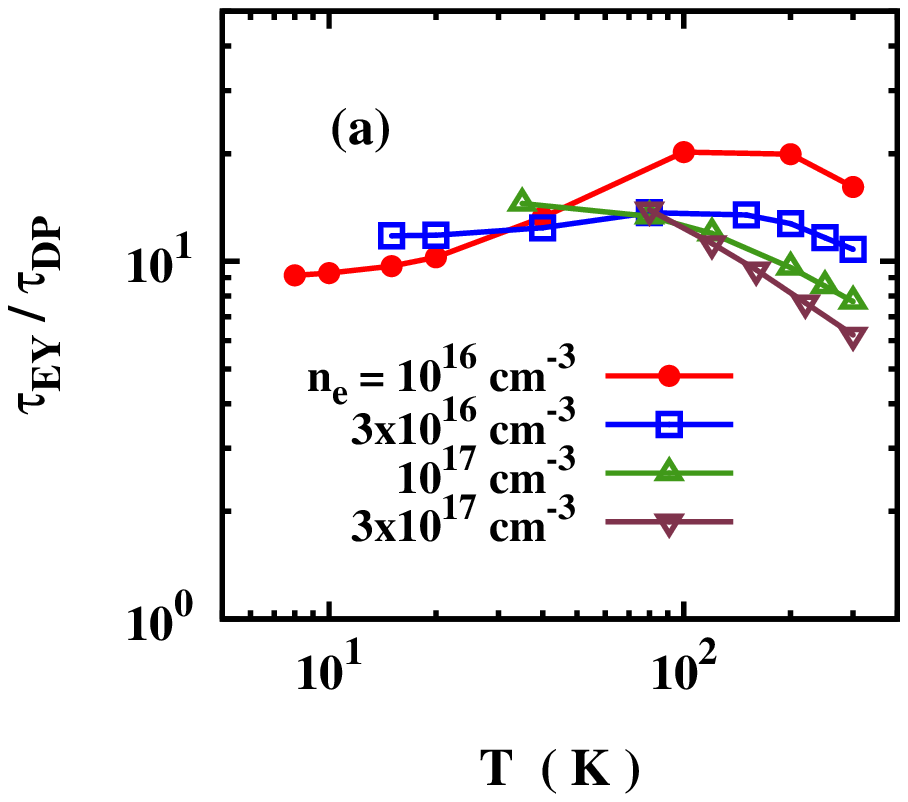}
  \end{minipage}\hfill
  \begin{minipage}[h]{0.5\linewidth}
  \centering
  \includegraphics[height=5.5cm]{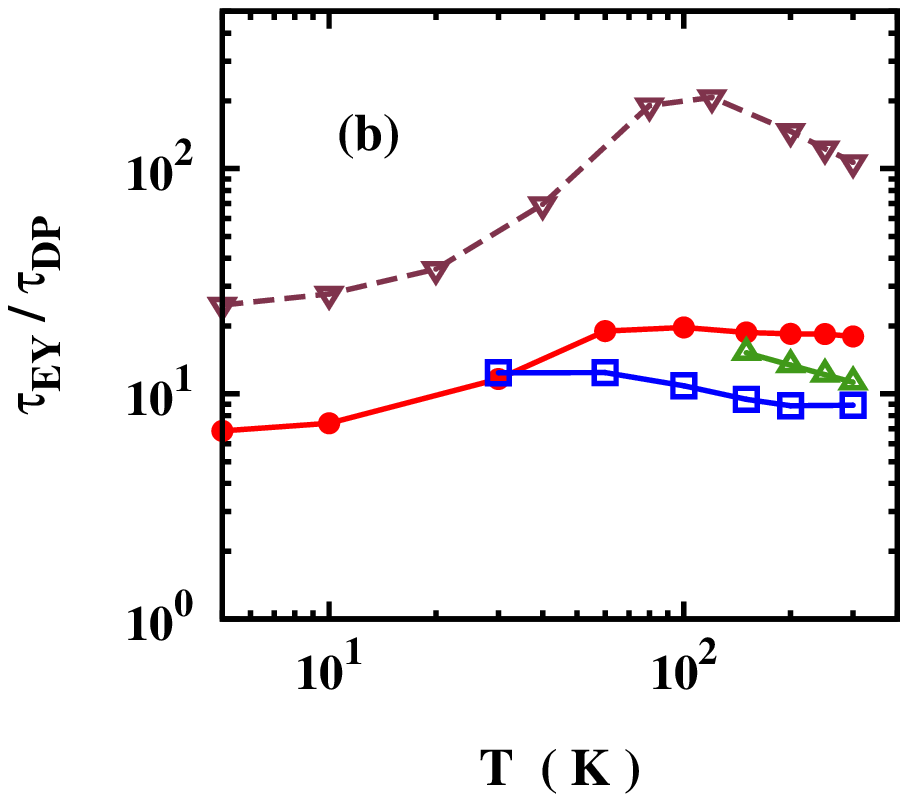}
  \end{minipage}\hfill
  \caption{ Ratio of the Elliott-Yafet spin relaxation time $\tau_{\rm EY}$ to the
    D'yakonov-Perel' one $\tau_{\rm DP}$ for (a) $n$-InSb (b) $n$-InAs and $n$-GaAs
    as function of temperature  for various electron densities. In Fig.(b):
  $n_{e}=10^{16}$~cm$^{-3}$ (curve with $\bullet$), $2\times
  10^{17}$~cm$^{-3}$ (curve with $\square$), $10^{18}$~cm$^{-3}$
  (curve with $\triangle$) for InAs, and $n_{e}=10^{16}$~cm$^{-3}$
  (curve with $\triangledown$) for GaAs. From Jiang and Wu
  \cite{jiang:125206}.}
\label{fig_ey_dp}
\end{figure}

{\bf Comparison of different spin relaxation mechanisms.}
In $n$-type bulk III-V semiconductors at low photo-excitation density,
the Bir-Aronov-Pikus mechanism is irrelevant as the hole density is low. The
mechanisms left are then the Elliott-Yafet and D'yakonov-Perel' mechanisms. Previously, it
was believed that the Elliott-Yafet mechanism is important in narrow bandgap
semiconductors, such as InSb and InAs. Jiang and Wu compared the
relative efficiency of the two mechanisms at various conditions for
InSb and InAs \cite{jiang:125206}. The results are shown in
Fig.~\ref{fig_ey_dp}. In contrast to previous understanding, they
found that the Elliott-Yafet mechanism is much {\em less} efficient than the D'yakonov-Perel'
one in both InSb and InAs. Although the low-temperature results
for the high density cases are absent in Fig.~\ref{fig_ey_dp}, it was
found that in such regime the ratio $\tau_{\rm EY}/\tau_{\rm DP}$
varies slowly with temperature \cite{jiang:125206}. For GaAs, the Elliott-Yafet
mechanism is still less important as indicated by
Fig.~\ref{fig_ey_dp}(b).

To give a qualitative picture of the relative importance of the Elliott-Yafet
mechanism in other III-V semiconductors, Jiang and Wu analyzed the
problem by using the approximate formulae for the D'yakonov-Perel' and Elliott-Yafet spin
relaxation time [Eqs.~(\ref{eyapp}) and (\ref{dpapp})]. From those
equations,
\begin{equation}
  \frac{\tau_{\rm EY}}{\tau_{\rm DP}} = \frac{2Q}{3A}\langle \varepsilon_{\bf k}\rangle
  \tau_p^2 \Theta.
\label{EYDP2}
\end{equation}
Here $\Theta= 8\gamma_D^2m_e^3E_g^2 (1-\eta/3)^2
/[(1-\eta/2)^2\eta^2]$. $Q$ and $A$ are numerical factors around
unity. The factor $\Theta$, which is solely determined by the material
parameters, is listed in Table~\ref{tab_dpey} for various III-V
semiconductors. The spin-orbit coupling parameter $\gamma_D$ is fitted
from experiments (except that $\gamma_D$'s for InP and GaSb are
from the ${\bf k}\cdot{\bf p}$ calculation in
Ref.~\cite{PhysRevB.72.193201}), whereas other parameters are from 
{\em Landolt-B\"ornstein} \cite{madelung}. One notices from the table
that the factor $\Theta$ is much smaller for InAs and InSb than other
III-V semiconductors. According to this, the Elliott-Yafet mechanism should be
much less efficient than the D'yakonov-Perel' one in GaSb and InP. Actually, one
notices that $\Theta\propto E_g^2$ which decreases rapidly with
decreasing bandgap. In commonly used III-V semiconductors, InSb has
smallest bandgap. However, even for InSb the Elliott-Yafet mechanism is much less
important than the D'yakonov-Perel' mechanism in the metallic regime. Therefore, in
other III-V semiconductors, the Elliott-Yafet mechanism is also {\em
  unimportant}.\footnote{This is also true for intrinsic III-V
  semiconductors and in most cases for $p$-type semiconductors as well
  as for some II-VI semiconductors (such as CdTe, see
  Ref.~\cite{2009arXiv0910.1506J}).}

\begin{center}
\begin{table}[htbp]
\caption{{The $\Theta$ factor for III-V semiconductors. From Jiang and
    Wu \cite{jiang:125206}.}} 
\vskip 0.2cm 
\centering
\begin{tabular}{lllllllllllll}\hline\hline
 \mbox{}      &\mbox{}\mbox{}\mbox{}&  GaAs  &\mbox{}\mbox{}&  GaSb
  &\mbox{}\mbox{}&  InAs  &\mbox{}\mbox{}&   InSb &\mbox{}\mbox{}&   InP \\
\hline
$\Theta$ (eV) &\mbox{}\mbox{}\mbox{}&  2.7$\times 10^{-2}$  &\mbox{}\mbox{}&  0.12
  &\mbox{}\mbox{}&  2.0$\times 10^{-3}$  &\mbox{}\mbox{}&   9.2$\times 10^{-4}$  &\mbox{}\mbox{}&  0.27\\  
\hline\hline
\end{tabular}
\label{tab_dpey}
\end{table}
\end{center}

Very recently, Litvinenko et
  al. studied the magnetic field dependence of spin lifetime in
  $n$-type InSb and InAs experimentally \cite{litvinenko:111107}. They
  found in $n$-InSb that the spin lifetime increases significantly with
  increasing magnetic field in the Faraday configuration (magnetic field
  parallel to spin polarization) [see Fig.~\ref{Ben_Murdin_EYDP}]. As
  the Elliott-Yafet spin relaxation has little magnetic field
  dependence, this result implies that the dominant spin relaxation
  mechanism in $n$-InSb is the D'yakonov-Perel' mechanism at low
  magnetic field. At high magnetic field the Elliot-Yafet spin
  relaxation dominates as the  D'yakonov-Perel' mechanism is strongly suppressed
by the longitudinal magnetic field. They found similar results in $n$-InAs. These
 results confirmed the previous prediction by Jiang and Wu that the 
  Elliott-Yafet mechanism is {\em less} efficient than the
  D'yakonov-Perel' one in $n$-type InSb and InAs \cite{jiang:125206}.

\begin{figure}[htb]
    \centering
    \includegraphics[height=5.cm,angle=90]{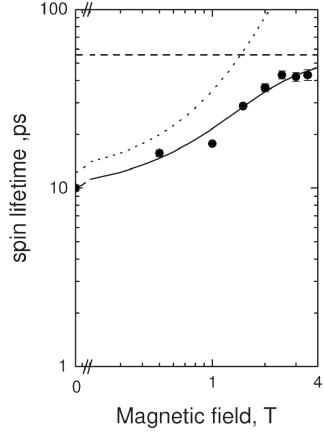}
    \caption{Spin lifetime in $n$-InSb as function of magnetic
      field at 100~K. The magnetic field is parallel to the spin
      polarization direction (Faraday configuration). Theoretical
      dependences of the Elliott-Yafet (dashed curve),
      D'yakonov-Perel' and the total (solid curve) spin lifetimes are
      also shown. Reproduced from Litvinenko et al. \cite{litvinenko:111107}.}
    \label{Ben_Murdin_EYDP}
\end{figure}

{\bf The D'yakonov-Perel' spin relaxation.}
As both the Bir-Aronov-Pikus and Elliott-Yafet mechanisms are unimportant in bulk $n$-type
III-V semiconductors in metallic regime, the only relevant one is the
D'yakonov-Perel' mechanism. Although the D'yakonov-Perel' mechanism has been studied for about
forty years, the understanding on it in bulk III-V semiconductors is
yet adequate. For example, in the previous literature, the
electron-electron scattering has long been believed to be irrelevant in bulk
III-V semiconductors. Jiang and Wu showed that the electron-electron
scattering is important for spin relaxation in $n$-GaAs in the
{\em nondegenerate} regime except when the
electron-longitudinal-optical-phonon dominates momentum
scattering. The same conclusion should also hold for other bulk
$n$-type III-V semiconductors.

{\sl The D'yakonov-Perel' spin relaxation: density dependence.} 
In Fig.~\ref{fig_ne_dep}(a), spin relaxation time as function of
electron density is plotted for $n$-GaAs at 40~K. The density of
impurity is taken as the same as that of electron, $n_i=
n_e$. Remarkably, one notices that the density dependence is 
{\em nonmonotonic} with a {\em peak} around
$n_e=10^{16}$~cm$^{-3}$. Previously, the nonmonotonic density
dependence of spin lifetime was observed in low-temperature 
($T\lesssim$5~K) experiments, where the localized electrons play a
crucial role and the electron system is in the insulating regime or
around the metal-insulator transition point. Jiang and Wu found, for
the first time, that the spin lifetime in {\em metallic} regime is also
{\em nonmonotonic}. Moreover, they pointed out that it is a {\em
  universal} behavior for {\em all} bulk III-V semiconductors at {\em
  all} temperature where the peak is located at $T_F\sim T$ with $T_F$
being the electron Fermi temperature. From the piont of view of spintronic
device application, as devices are more favorable to operate in the
metallic regime, this prediction gives the important information that the
longest spin lifetime in metallic regime is at $T_F\sim T$. The
underlying physics for the nonmonotonic density dependence in metallic
regime is elucidated below.

To understand the D'yakonov-Perel' spin relaxation qualitatively, let us first
recall the widely used approximate formulae \cite{Titkovbook}, $\tau_{\rm DP}\simeq
1/[\langle |{\bf \Omega}({\bf k})|^2-\Omega_z^2({\bf k})\rangle \tau_p^{\ast}]$
where $\langle...\rangle$ denotes the ensemble
average.\footnote{$\langle...\rangle=\frac{\int d{\bf k}~(f_{\bf k}^{\uparrow}-f_{\bf
    k}^{\downarrow}) ... }{\int d{\bf  k}~(f_{\bf k}^{\uparrow}-f_{\bf
    k}^{\downarrow})}$.} 
The expression contains two key factors of the D'yakonov-Perel' spin relaxation: (i)
the inhomogeneous broadening from the ${\bf k}$-dependent transverse
spin-orbit field $\sim \langle |{\bf \Omega}({\bf k})|^2
-\Omega_z^2({\bf k})\rangle$; (ii) the momentum scattering time
$\tau_p^{\ast}$ [including the contributions of the electron-impurity,
electron-phonon, electron-electron and electron-hole (whenever holes
exist) scatterings]. The D'yakonov-Perel' spin relaxation time increases with
increasing momentum scattering rate, but
decreases with increasing inhomogeneous broadening.

To elucidate the underlying physics, Jiang and Wu plotted the spin
relaxation times calculated with only the electron-impurity,
electron-electron and electron-phonon scatterings in
Fig.~\ref{fig_ne_dep}(a) respectively. It is seen that the spin relaxation time with
only one kind of scattering is smaller than that with all scatterings,
which is a consequence of the motional narrowing nature of the D'yakonov-Perel'
mechanism $\tau_s\propto 1/\tau_p$. One notices that the
electron-electron scattering gives {\em important} contribution to spin
relaxation in the nondegenerate (low density) regime. Interestingly,
both the electron-electron and electron-impurity scatterings lead to
nonmonotonic density dependence of spin relaxation time. One notices that the
electron-phonon scattering is much weaker (the corresponding spin
relaxation time is much shorter as $\tau_s\propto 1/\tau_p$) as the
temperature is low.

Let us first look at the density dependence of the electron-electron
scattering time. In fact, the density and temperature dependences of
the electron-electron scattering time have been widely investigated in
spin-unrelated problems (see, e.g., Ref.~\cite{giuliani_05}). From the
previous works \cite{giuliani_05,jetp.99.1279}, after some
approximation, the asymptotic density and temperature dependences of
the electron-electron scattering time $\tau_p^{ee}$ in the degenerate
and nondegenerate regimes are given by,
\bea
 \tau_p^{ee} \propto n_{e}^{\frac{2}{3}}/ T^2 \quad \quad {\rm for}\quad  T\ll T_{F}, 
 \label{taupee1} \\
  \tau_p^{ee} \propto T^{\frac{3}{2}} / n_{e} \quad\quad {\rm for}\quad  T\gg T_{F}.
 \label{taupee2}
\eea
From the above equations, one notices that the electron-electron
scattering time in the nondegenerate and degenerate regimes has
different density dependence. In the nondegenerate (low density)
regime, the electron-electron scattering time decreases with electron
density [see Eq.~(\ref{taupee2})], where the inhomogeneous broadening $\sim \langle |{\bf
  \Omega}({\bf k})|^2-\Omega_z^2({\bf k})\rangle\propto \langle
\varepsilon_{\bf k}^3\rangle$ ($\varepsilon_{\bf k}$ is electron
kinetic energy) varies slowly with density as the electron
distribution is close to the Boltzmann distribution in nondegenerate
regime. The spin relaxation time thus increases with electron
density. In degenerate (high density) regime, both the electron-electron
scattering time [see Eq.~(\ref{taupee1})] and the inhomogeneous
broadening increase with electron density. Therefore, the spin
relaxation time $\tau_s\simeq 1/[\langle |{\bf \Omega}({\bf
  k})|^2-\Omega_z^2({\bf k}) \rangle \tau_p^{ee}]$ decreases with electron density.

For spin relaxation associated with the electron-impurity scattering,
the scenario is similar: In nondegenerate regime, the decrease of the
electron-impurity scattering time with electron density
($1/\tau_p^{ei}\propto n_i=n_e$) leads to the increase of the spin relaxation time with
increasing electron density. In the degenerate regime, the
inhomogeneous broadening increases with increasing electron density,
as $\langle |{\bf \Omega}({\bf k})|^2-\Omega_z^2({\bf k}) \rangle
\propto k_{\rm F}^6 \propto n_{e}^2$. On the other hand, the
electron-impurity scattering time varies slowly with electron density
because $1/\tau_p^{ei} \sim n_i V_{k_{F}}^2 k_F \sim n_{e}k_F/k_{F}^4
\propto n_{e}^{0}$. Consequently, the spin relaxation time decreases with increasing
electron density. For spin relaxation related to the electron-phonon
scattering, the situation, however, is different: In nondegenrate
regime both the inhomogeneous broadening and the electron-phonon
scattering rate vary slowly with electron density as electron
distribution is close to the Boltzmann distribution. In degenerate
regime, the increase of the inhomogeneous broadening is faster
than the variation of the electron-phonon scattering, which hence
leads to the decrease of spin relaxation time with increasing electron
density.

In summary, the electron density depdence of spin relaxation
time is nonmonotonic as the momentum scattering time and the
inhomogeneous broadening have different {\em qualitative} density
dependences in the nondegenrate and degenerate regimes. The spin relaxation
time increases (decreases) in the nondegenrate (degenerate) regime
with increasing electron density. A peak hence locates in the
crossover regime, where $T_F$ is around $T$. Such a scenario should hold for
{\em all} the III-V semiconductors at {\em all} temperature and the
nonmonotonic density dependence is thus a {\em universal}
behavior.\footnote{The nonmonotonic density dependence should also
  exist in bulk II-VI semiconductors with zinc-blende structures which
has similar band structure with III-VI semiconductors.} Furthermore,
in Fig.~\ref{fig_ne_dep}(b), Jiang and Wu showed that the nonmonotonic
density dependence of spin relaxation time also exists even when the
spin-orbit coupling is dominated by the linear-${\bf k}$
term.\footnote{From this point, the nonmonotonic density dependence
  should also exist in bulk wurtzite semiconductors with bulk
  inversion asymmetry, such as GaN, AlN
  and ZnO.} Subsequently, similar behavior was found in
two-dimensional system
\cite{jiang:155201,0268-1242-24-11-115010,zhang:155311}, where the
underlying physics is similar. The predicted peak was later observed
by Krauss et al. \cite{krauss0902.0270} (see, also \cite{shenpeak}).

\begin{figure}[htb]
\begin{minipage}[h]{0.5\linewidth}
    \centering
    \includegraphics[height=5cm]{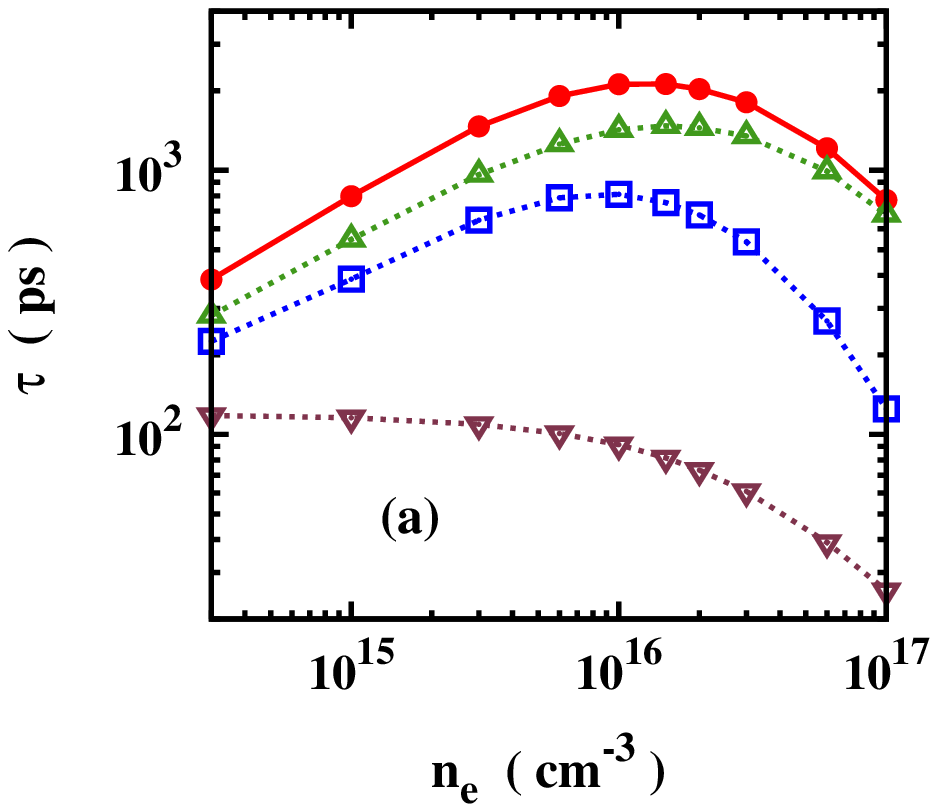}
  \end{minipage}\hfill
  \begin{minipage}[h]{0.5\linewidth}
    \centering
    \includegraphics[height=5cm]{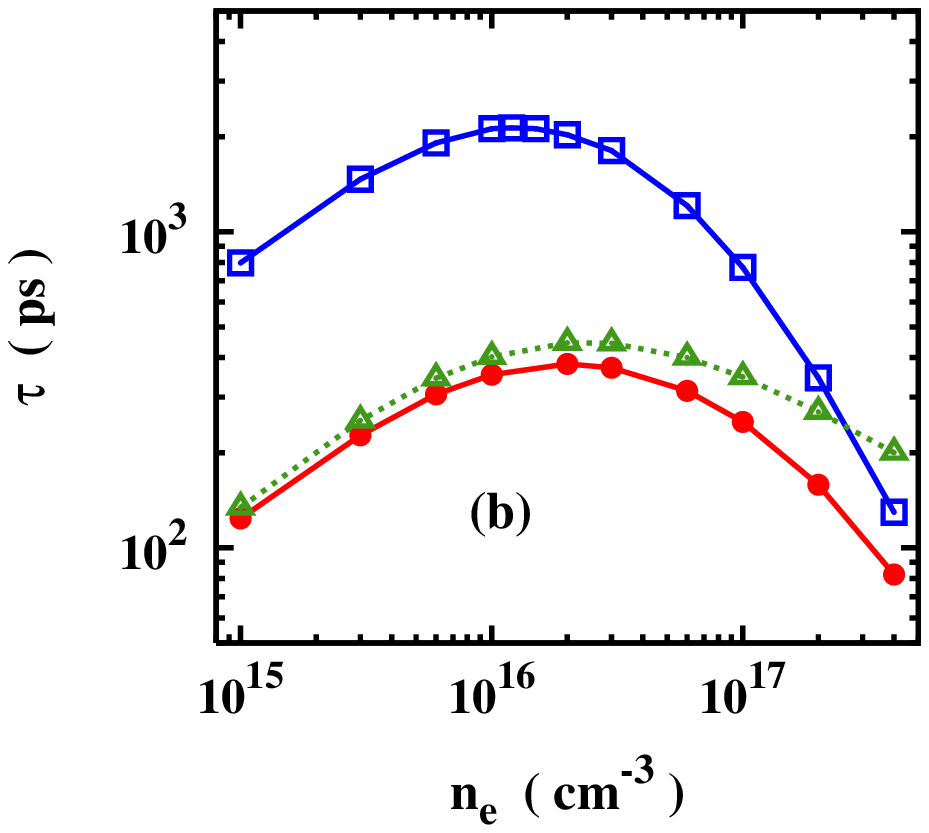}
  \end{minipage}\hfill
  \caption{ $n$-GaAs at $T=40$~K. (a) spin relaxation time $\tau$ as
  function of electron density $n_{e}$ ($n_i=n_{e}$) from full calculation
  (curve with $\bullet$), from calculation with only the electron-electron
  scattering (curve with $\square$), with only the
  electron-impurity scattering (curve with $\triangle$), and with
  only the electron-phonon scattering (curve with $\triangledown$);
  (b) spin relaxation time $\tau$ as function of electron density $n_{e}$ ($n_i=n_{e}$)
  for the case with strain-induced spin-orbit coupling (curve with
  $\bullet$: with both the linear- and the cubic-${\bf k}$ spin-orbit coupling; curve
  with $\triangle$: with only the linear-${\bf k}$ spin-orbit coupling) and the case
  without strain (curve with $\square$). From Jiang and Wu \cite{jiang:125206}.}
  \label{fig_ne_dep}
\end{figure}

\subsubsection{Electron-spin relaxation in intrinsic bulk III-V
  semiconductors}

In intrinsic semiconductors, the carriers are generated by
photo-excitation where the electron density is equal to the hole
density $n_e=n_h=N_{\rm ex}$ ($N_{\rm ex}$ denotes the excitation
density). As the impurity density is very low (one can take $n_i=0$),
the carrier-carrier  scattering is dominant unless at high
temperature where the electron-longitudinal-optical-phonon scattering becomes more
important. Intrinsic bulk semiconductors thus offer a good
platform to study the many-body effect to electron spin relaxation.

{\bf Comparison of different mechanisms.}
As both the Bir-Aronov-Pikus and D'yakonov-Perel' mechanisms contribute to electron spin
relaxation, the relative efficiency of the Bir-Aronov-Pikus and D'yakonov-Perel' mechanisms
should be compared. In Fig.~\ref{fig_intrinsic1}(a) the Bir-Aronov-Pikus and D'yakonov-Perel'
spin relaxation times as function of temperature are plotted. It is noted that the Bir-Aronov-Pikus
spin relaxation time is over one order of magnitude larger than
the D'yakonov-Perel' one, which indicates that the Bir-Aronov-Pikus mechanism is much less
efficient than the D'yakonov-Perel' one in intrinsic bulk III-V
semiconductors. Systematic calculation for various temperatures and
excitation densities confirms that the Bir-Aronov-Pikus mechanism is
{\em irrelevant} in intrinsic bulk GaAs in the metallic regime. Such
conclusion also holds for GaSb. Recent experiments arrived at the same
conclusion for intrinsic InSb \cite{murdin:096603}. Therefore, the Bir-Aronov-Pikus
mechanism is unimportant in intrinsic GaAs, GaSb and InSb.

\begin{figure}[htb]
\begin{center}
\includegraphics[height=4.5cm]{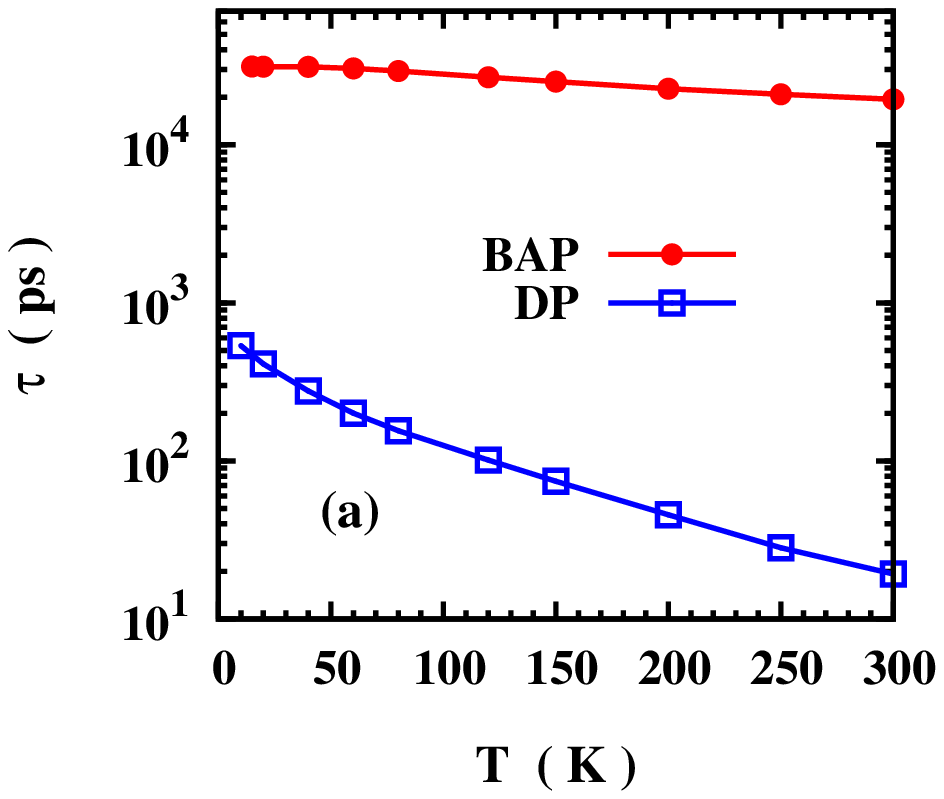}\includegraphics[height=4.5cm]{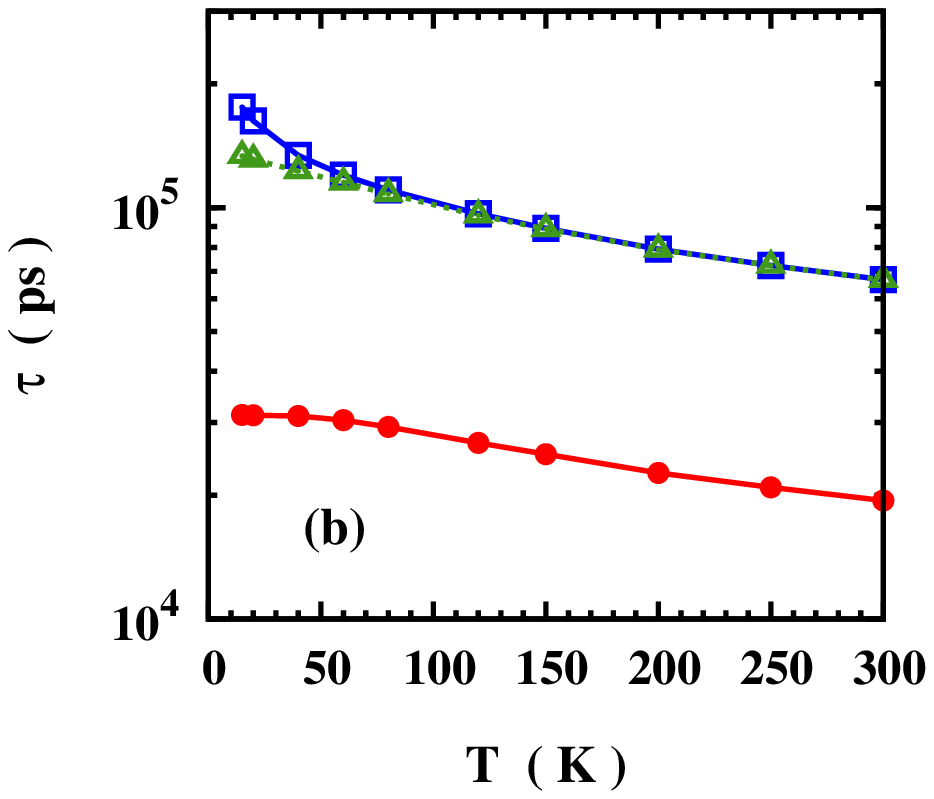}\includegraphics[height=4.5cm]{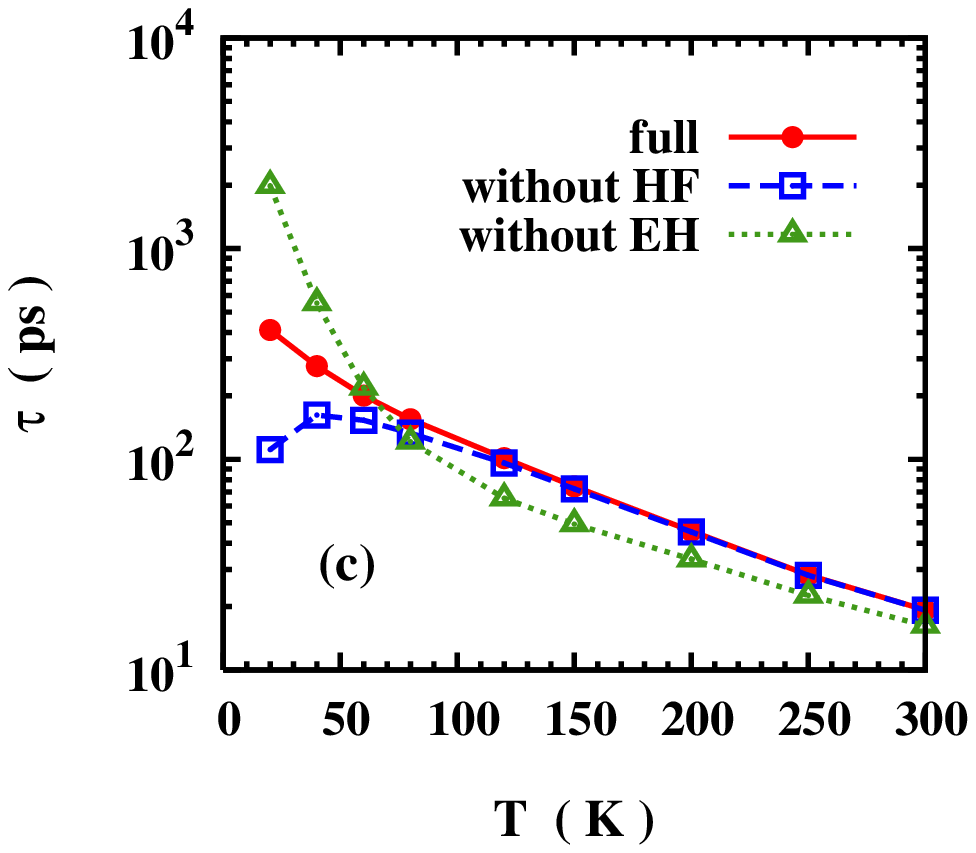}
\end{center}
  \caption{ Spin relaxation time $\tau$ for intrinsic
    GaAs with excitation density $N_{\rm ex}=10^{17}$~cm$^{-3}$. The
    initial spin polarization is $P=50\%$. (a) spin relaxation time $\tau$ due to
    the Bir-Aronov-Pikus (BAP) and D'yakonov-Perel' (DP) mechanisms as function of temperature. (b) The Bir-Aronov-Pikus
    spin relaxation time calculated from Eq.~(\ref{BAPelastapp1}) (dotted curve with
    $\triangle$), from the kinetic spin Bloch equation approach with both long-range and
    short-range exchange scatterings (solid curve with $\bullet$) as
    well as from the kinetic spin Bloch equation approach with only the short-range exchange
    scattering (solid curve with $\square$). (c) The D'yakonov-Perel' spin relaxation time from full
    calculation (solid curve with $\bullet$), from the calculation
    without the Coulomb Hartree-Fock term (dashed curve with $\square$), and
    from the calculation without the electron-hole Coulomb scattering
    (dotted curve with $\triangle$). From Jiang and Wu \cite{jiang:125206}.}
  \label{fig_intrinsic1}
\end{figure}

{\bf The Bir-Aronov-Pikus spin relaxation: short-range {\sl vs.} long-range interaction
  and the Pauli blocking.}
As to the Bir-Aronov-Pikus spin relaxation, it should be mentioned that the
long-range part of the electron-hole exchange interaction [see
Eq.~(\ref{bapilr})] has always been ignored in the literature
\cite{zuticrmp,opt-or,PhysRevB.66.035207,PhysRevB.16.820}. Jiang and
Wu examined the relative contribution of the long-range and
short-range interactions to the Bir-Aronov-Pikus spin relaxation. In
Fig.~\ref{fig_intrinsic1}(b) the Bir-Aronov-Pikus spin relaxation time
calculated with both the long-range and short-range exchange
interactions as well as that without the long-range exchange
interaction are plotted. It is seen that the spin relaxation time increases by about
three times when the long-range exchange interaction is removed. This
indicates that the long-range interaction is {\em more} important than
the short-range one in GaAs and hence can not be neglected. Moreover,
in the previous literature, the Bir-Aronov-Pikus spin relaxation time was calculated via
the Fermi Golden rule \cite{opt-or,PhysRevB.54.1967},
\be
\frac{1}{\tau_{\rm BAP}({\bf k})} = 4\pi\hspace{-3pt} \sum_{{\bf
    q},{\bf k}^{\prime}\atop m,m^{\prime}}\hspace{-3pt}
\delta(\varepsilon_{{\bf k}} +\varepsilon^{h}_{{\bf k}^{\prime}
  m^{\prime}}-\varepsilon_{{\bf k}-{\bf q}} - \varepsilon^{h}_{{\bf
    k}^{\prime}+{\bf q}m}) |{\cal{J}}^{(+)\ {\bf k}^{\prime}
  m^{\prime}}_{{\bf k}^{\prime}+{\bf q}m}|^2 f^{h}_{{\bf k}^{\prime}
  m^{\prime}} (1 - f^{h}_{{\bf k}^{\prime}+{\bf q}m}).
\label{BAPelastapp1}
\ee
As pointed out by Zhou and Wu in the two-dimensional system
\cite{zhou:075318} (see Sec.~5.4.6), such approach, which ignores the
Pauli-blocking effect of electron distribution, fails at low
temperature (in degenerate regime). The spin relaxation time calculated from
Eq.~(\ref{BAPelastapp1}) with only the short-range exchange
interaction is plotted in Fig.~\ref{fig_intrinsic1}(b). One notices
that the results from Eq.~(\ref{BAPelastapp1}) indeed deviate from
the exact results via the kinetic spin Bloch equation approach at low temperature.

{\bf The D'yakonov-Perel' spin relaxation: temperature dependence and the effects
  of the Coulomb Hartree-Fock term and the electron-hole scattering.}
In Fig.~\ref{fig_intrinsic1}(c) the D'yakonov-Perel' spin relaxation time as function of temperature
is plotted. The initial spin polarization is $P=50\%$, i.e., ideal
circularly polarized light excitation. To elucidate the role of the
Coulomb Hartree-Fock term, the spin relaxation time calculated without the
Coulomb Hartree-Fock term is also plotted. It is seen that the Coulomb Hartree-Fock term
largely affects the spin relaxation at low temperature, whereas at high
temperature it is ineffective. The underlying physics is as follows:
The spin relaxation time under the Hartree-Fock effective magnetic
field is estimated as [similar to Eq.~(\ref{dp-B})]
\begin{equation}
  \tau_s(P) = \tau_s(P=0)[1+(g\mu_{\rm B} B_{\rm HF}\tau_p^{\ast})^2]
\label{HFeq}
\end{equation}
where $B_{\rm HF}$ is the averaged effective magnetic field and
$\tau_p^{\ast}$ stands for the momentum scattering including the
carrier-carrier scattering. At low temperature, the carrier-carrier
[see Eq.~(\ref{taupee1})] and electron-phonon scatterings are
suppressed. Therefore, $\tau_p^{\ast}$ is large and the Hartree-Fock magnetic field has
strong effect on spin relaxation time. However, at high temperature
the momentum scattering is strong and $\tau_p^{\ast}$ is small. Consequently,
the Hartree-Fock magnetic field has little effect on spin relaxation.

It is noted that without the Coulomb Hartree-Fock term the spin relaxation time has {\em nonmonotonic}
temperature dependence. Usually, the spin relaxation time without the Coulomb Hartree-Fock term is close
to the spin relaxation time at small initial spin polarization. To confirm that, Jiang
and Wu plotted the spin relaxation time at the same condition but for $P=2\%$ in
Fig.~\ref{fig_int_T_P}. It is seen that the spin relaxation time is indeed {\em
  nonmonotonic} in temperature dependence.\footnote{The nonmonotonic
  temperature dependence of spin relaxation time has been observed in bulk
  intrinsic GaAs recently \cite{2009arXiv0910.1714R}, which confirms
  the prediction by Jiang and Wu \cite{jiang:125206}.} In contrast,
the spin relaxation time decreases with increasing temperature in $n$-type III-V
semiconductors as the electron-impurity scattering dominates at low
temperature. The nonmonotonic temperature dependence of the spin relaxation time
originates from the nonmonotonic temperature dependence of the
electron-electron and electron-hole scattering times as noted from
Eqs.~(\ref{taupee1}) and (\ref{taupee2}).\footnote{The electron-hole
  scattering time has similar temperature dependence as that of the
  electron-electron scattering.} The peak is then located in
the crossover regime. Systematic calculation indicates that the peak
temperature is around $T_F/3$ and lies in the range of $(T_F/4,T_F/2)$
for various carrier density in both GaAs and InAs \cite{jiang:125206}.

The spin relaxation time calculated without the electron-hole scattering is also
plotted in Fig.~\ref{fig_intrinsic3}(c) to indicate the contribution
of the electron-hole scattering. It is seen that at high temperature
($T>60$~K), the spin relaxation time becomes smaller without the electron-hole
scattering. However, at low temperature ($T<60$~K), the spin relaxation time becomes
larger. The decrease of the spin relaxation time at high temperature indicates the
importance of the electron-hole scattering according to the motional
narrowing $\tau_s\propto 1/\tau_p$. However, at low temperature, the
Coulomb Hartree-Fock term complicates the behavior. According to
Eq.~(\ref{HFeq}), if the Coulomb Hartree-Fock term plays significant role [i.e.,
$(g\mu_{\rm B} B_{\rm HF}\tau_p^{\ast})^2>1$], then $\tau_s\sim
[(|{\bf \Omega}|^2-\Omega_z^2)\tau_p^{\ast}]^{-1}(g\mu_{\rm B} B_{\rm
  HF}\tau_p^{\ast})^2\sim \tau_p^{\ast}$. Therefore removing the
electron-hole scattering leads to longer spin relaxation
time. These results
demonstrate that the electron-hole scattering plays an important role in
both low and high temperature regimes.

\begin{figure}[htb]
\centering
\includegraphics[height=5.5cm]{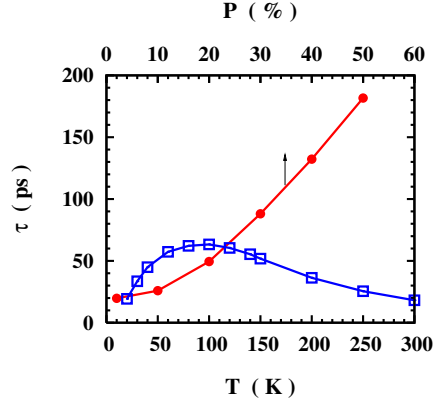}
\caption{ Intrinsic GaAs with $N_{\rm ex}=2\times
  10^{17}$~cm$^{-3}$. Spin relaxation time $\tau$ as function of temperature for
  $P=2~\%$ (curve with $\square$) and the spin relaxation time as function of initial
  spin polarization $P$ for $T=20$~K (curve with $\bullet$) (note that
  the scale of $P$ is on the top of the frame). From Jiang and
  Wu \cite{jiang:125206}.}
\label{fig_int_T_P}
\end{figure}

{\bf The D'yakonov-Perel' spin relaxation: initial spin polarization dependence.}
The effect of the Coulomb Hartree-Fock term is also reflected in the initial
polarization dependence of the spin relaxation time
\cite{PhysRevB.68.075312,stich:176401,stich:205301,korn}, which is
plotted in Fig.~\ref{fig_int_T_P}. It is seen that the spin relaxation time increases
by about one order of magnitude when the initial spin polarization
increases from $2\%$ to $50\%$. Without the Coulomb Hartree-Fock term the
increment of the spin relaxation time is negligible (not shown). It was also
found that, in contrast to the two-dimensional case
\cite{PhysRevB.68.075312,stich:176401,stich:205301,korn}, the initial
spin polarization dependence in $n$-type III-V semiconductors is very
weak. This is because the momentum scattering in $n$-type III-V
semiconductors is strong even at low temperature as the impurity
density is high ($n_i=n_e$) \cite{jiang:125206}.

{\bf The D'yakonov-Perel' spin relaxation: density dependence.} 
In Fig.~\ref{fig_intrinsic3}(a), the density dependence of the spin relaxation time is
plotted. It is seen again that the Bir-Aronov-Pikus mechanism is much less
efficient then the D'yakonov-Perel' mechanism. Remarkably, the density dependence is
{\em nonmonotonic} and a peak exhibits. The underlying physics is
similar with that for the $n$-type case except that the
electron-impurity scattering is substituted by the electron-hole
scattering. As the electron-hole scattering has similar density
dependence with that of the electron-electron scattering, the density
dependence of the spin relaxation time is then nonmonotonic. The peak is also located
in the crossover regime, where $T_F$ is around $T$. Such behavior is
also {\em universal} for {\em all} bulk intrinsic III-V semiconductors
at {\em all} temperature.\footnote{Such density dependence should also
hold for strained III-V semiconductors, wurtzite semiconductors with bulk
  inversion asymmetry and bulk II-VI semiconductors.} 

To elucidate the role of the electron-hole scattering and the Coulomb
Hartree-Fock term in spin relaxation, the spin relaxation times calculated without these terms
are plotted together with the spin relaxation time from the full calculation in
Fig.~\ref{fig_intrinsic3}(b). The importance of the electron-hole
scattering is obvious in a wide density range. The Coulomb Hartree-Fock term is shown to be
important only in the high density regime.\footnote{The Hartree-Fock effective
  magnetic field can be estimated as $B_{\rm HF} = \tilde{V}_{q} n_{e}
  P/(g\mu_B)$ where $\tilde{V}_{q}$ describes the average Coulomb
  interaction. Hence the Hartree-Fock effective magnetic field is strong at high
  electron density.}

\begin{figure}[htb]
 \begin{minipage}[h]{0.5\linewidth}
   \centering
    \includegraphics[height=4.65cm]{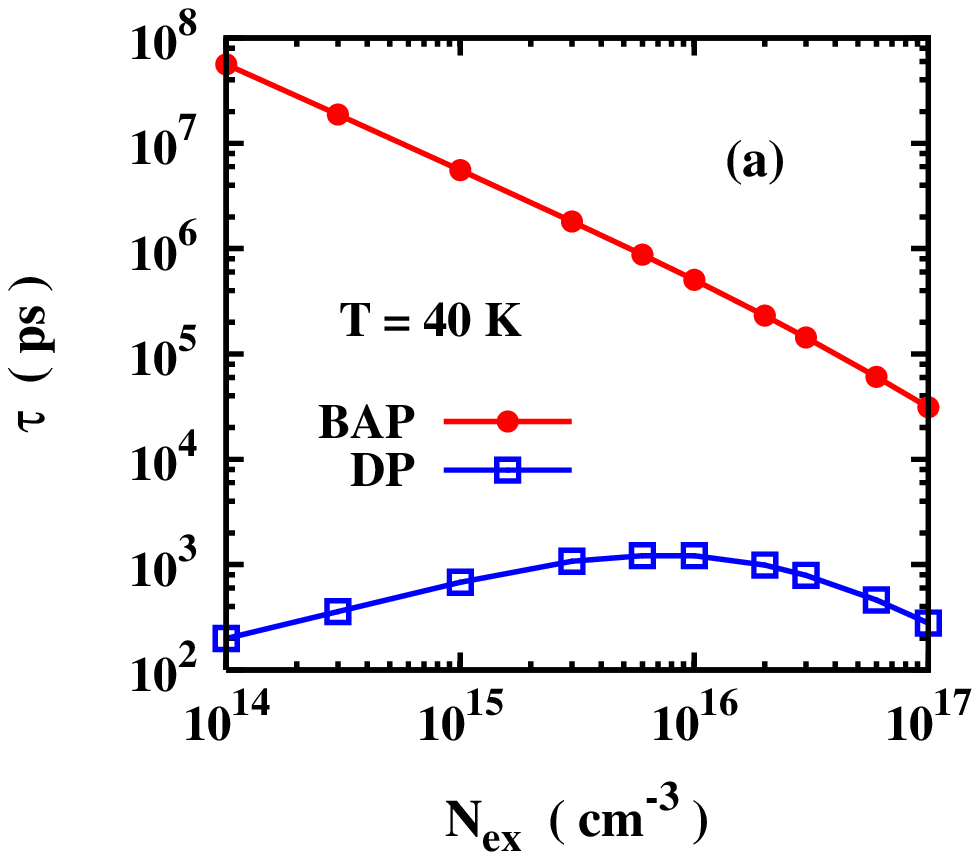}
  \end{minipage}\hfill
  \begin{minipage}[h]{0.5\linewidth}
    \centering
    \includegraphics[height=4.65cm]{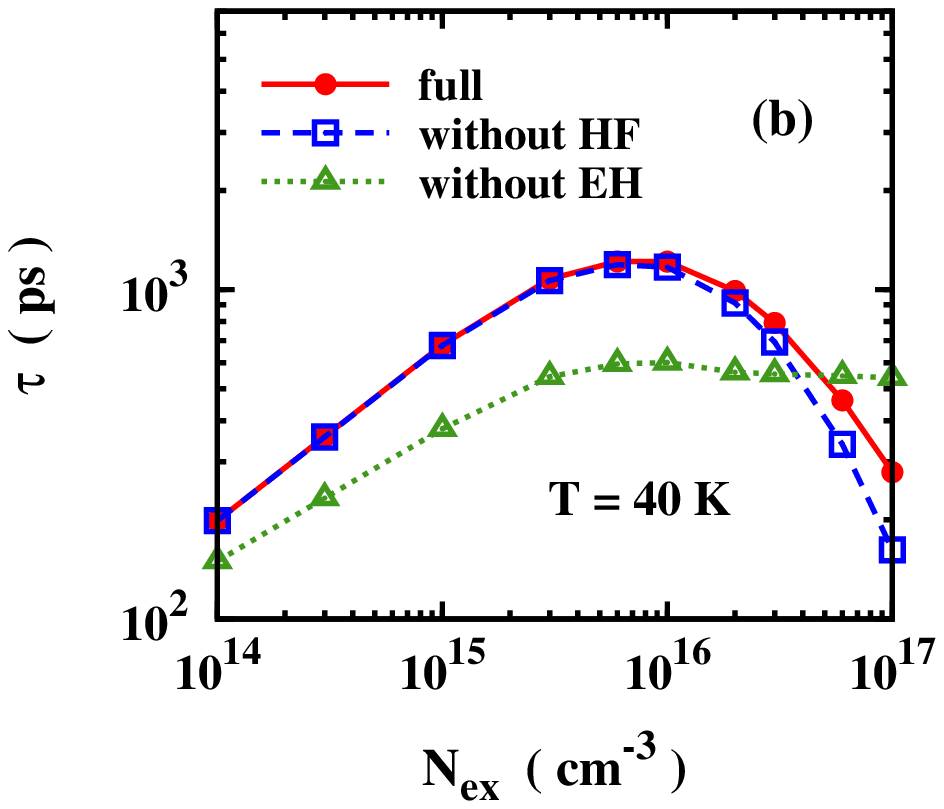}
  \end{minipage}\hfill
  \caption{ Intrinsic GaAs with $P=50\%$ and
    $T=40$~K. (a) The Bir-Aronov-Pikus (BAP) and D'yakonov-Perel' (DP) spin relaxation times $\tau$ as function of
    photo-excitation density $N_{\rm ex}$. (b) The D'yakonov-Perel' spin relaxation time from the
    full calculation (curve with $\bullet$), from the calculation
    without the Coulomb Hartree-Fock term (curve with $\square$), and from the
    calculation without the electron-hole Coulomb scattering (curve
    with $\triangle$). From Jiang and Wu \cite{jiang:125206}.}
  \label{fig_intrinsic3}
\end{figure}

\subsubsection{Electron-spin relaxation in $p$-type bulk III-V
  semiconductors}

{\bf Comparison of the D'yakonov-Perel' and Bir-Aronov-Pikus
  mechanisms in GaAs}. The main sources of spin relaxation have been
recognized as the Bir-Aronov-Pikus 
and D'yakonov-Perel' mechanisms \cite{opt-or}.\footnote{The Elliott-Yafet
  mechanism was checked to be unimportant in metallic regime for both
  $p$-GaAs and $p$-GaSb by Jiang and Wu \cite{jiang:125206}.} An
important issue is the relative efficiency of the two mechanisms under
various conditions. This was studied comprehensively by Jiang and Wu
in Ref.~\cite{jiang:125206}. Below we review their results.

{\sl Low photo-excitation case.}
Jiang and Wu first discussed the low photo-excitation case. In this
case, the electron density is low and the electron system is
nondegenerate. The ratio of the Bir-Aronov-Pikus spin relaxation time to the D'yakonov-Perel' one is plotted in
Fig.~\ref{fig:T_p}(a) for various hole densities as function of
temperature. It is seen that the D'yakonov-Perel' mechanism dominates at high
temperature, whereas the Bir-Aronov-Pikus mechanism dominates at low
temperature. This is consistent with the common belief in the literature
\cite{aronov,PhysRevB.66.035207,PhysRevB.16.820,zuticrmp,opt-or,Fabianbook}.

The temperature dependences of the Bir-Aronov-Pikus and D'yakonov-Perel' spin relaxation times are plotted in
Fig.~\ref{fig:T_p}(b) for a typical case with 
$n_h=3\times 10^{18}$~cm$^{-3}$. It is seen that both the D'yakonov-Perel' spin
relaxation time and the Bir-Aronov-Pikus one decrease with temperature. As
electron system is nondegenerate, the inhomogeneous broadening varies
as $\sim \langle |{\bf \Omega}|^2-\Omega_z^2\rangle \propto T^3$, which leads
to the rapid decrease of the D'yakonov-Perel' spin relaxation time. The Bir-Aronov-Pikus spin relaxation time decreases with
temperature partly because of the Pauli blocking of holes. To
elucidate this, the Bir-Aronov-Pikus spin relaxation time without the Pauli blocking of holes is
plotted as dotted curve in Fig.~\ref{fig:T_p}(b). The results indicate
that the Pauli blocking of holes effectively suppresses the Bir-Aronov-Pikus spin
relaxation at low temperature and makes the temperature dependence of
the Bir-Aronov-Pikus spin relaxation time stronger.

\begin{figure}[htb]
  \begin{minipage}[h]{0.45\linewidth}
    \centering
    \includegraphics[height=5cm]{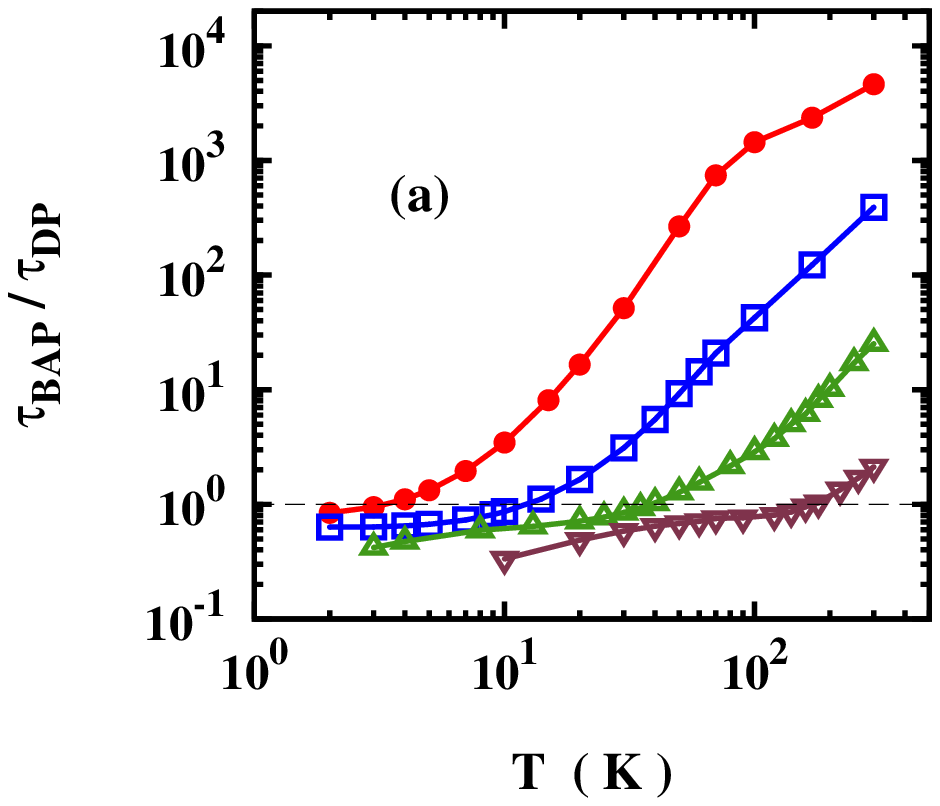}
  \end{minipage}\hfill
  \begin{minipage}[h]{0.45\linewidth}
    \centering
    \includegraphics[height=5cm]{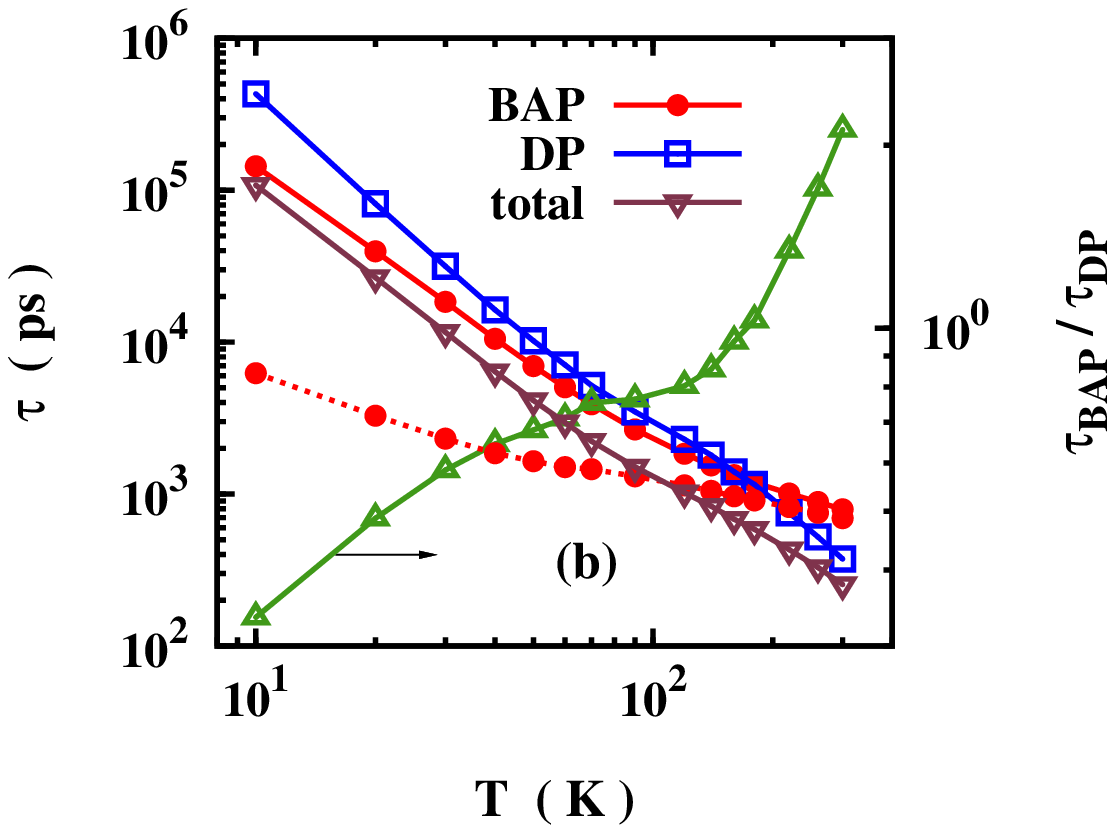}
  \end{minipage}\hfill
\caption{ $p$-GaAs. Ratio of the Bir-Aronov-Pikus spin
  relaxation time to the D'yakonov-Perel' one $\tau_{\rm BAP}/\tau_{\rm DP}$
  as function of temperature for various hole densities with $N_{\rm
    ex}=10^{14}$~cm$^{-3}$ and $n_i=n_h$. (a): $n_h=3\times
  10^{15}$~cm$^{-3}$ (curve with $\bullet$), $3\times
  10^{16}$~cm$^{-3}$ (curve with $\square$), 
  $3\times 10^{17}$~cm$^{-3}$ (curve with $\triangle$), and 
  $3\times 10^{18}$~cm$^{-3}$ (curve with $\triangledown$).
  (b): The spin relaxation times due to the Bir-Aronov-Pikus (BAP) and
  D'yakonov-Perel' (DP) mechanisms, the total spin relaxation time and
  the ratio $\tau_{\rm BAP}/\tau_{\rm DP}$ (curve with $\triangle$)
  {\sl vs.} the temperature for $n_h=3\times 10^{18}$~cm$^{-3}$. The
  dotted curve represents the Bir-Aronov-Pikus spin relaxation time 
  calculated without the Pauli blocking of holes. Note the scale of
  $\tau_{\rm BAP}/\tau_{\rm DP}$ is on the right hand side of the
  frame. From Jiang and Wu \cite{jiang:125206}.}
\label{fig:T_p}
\end{figure}

{\sl High photo-excitation case.}
The high photo-excitation case was also discussed by Jiang and Wu. The
excitation density was taken as $N_{\rm ex}=0.1n_h$. The ratio of the
Bir-Aronov-Pikus spin relaxation time to the D'yakonov-Perel' one as function of temperature for various hole
densities is plotted in Fig.~\ref{fig:T_p2}(a). Interestingly,
the ratio is nonmonotonic and has a minimum roughly around the Fermi
temperature of electrons, $T\sim T_{\rm F}$. In contrast to the low
photo-excitation  case, the Bir-Aronov-Pikus mechanism no longer dominates the low
temperature regime. To understand such behavior, the spin relaxation times due to the
two mechanisms and their ratio are plotted in Fig.~\ref{fig:T_p2}(b)
for a typical case $n_h=3\times 10^{18}$~cm$^{-3}$. Before looking at
the figure, one should note that the only difference for the two cases
it that the electron density is much larger in the high excitation
density. In the figure, one notices that, quite differently, the
D'yakonov-Perel' spin relaxation time saturates at low temperature, which is the reason why the
ratio of the two spin relaxation times increases at decreasing temperature. The
underlying physics for the saturation of the D'yakonov-Perel' spin relaxation time at low temperature
is as follows: As the electron density is high, at low temperature the
electron system enters into the degenerate regime, where the
inhomogeneous broadening and the momentum scattering (dominated by
the electron-impurity scattering) varies slowly with temperature. This
thus leads to the saturation of the D'yakonov-Perel' spin relaxation time at low temperature.

\begin{figure}[htb]
\begin{minipage}[h]{0.45\linewidth}
    \centering
    \includegraphics[height=5cm]{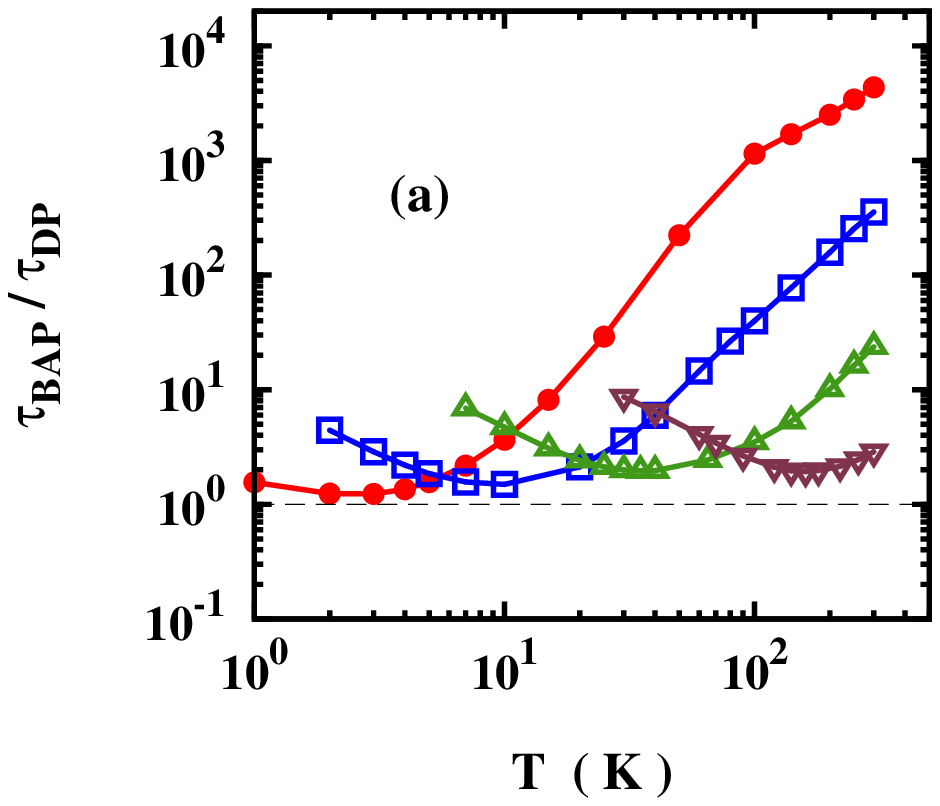}
  \end{minipage}\hfill
  \begin{minipage}[h]{0.45\linewidth}
    \centering
    \includegraphics[height=5cm]{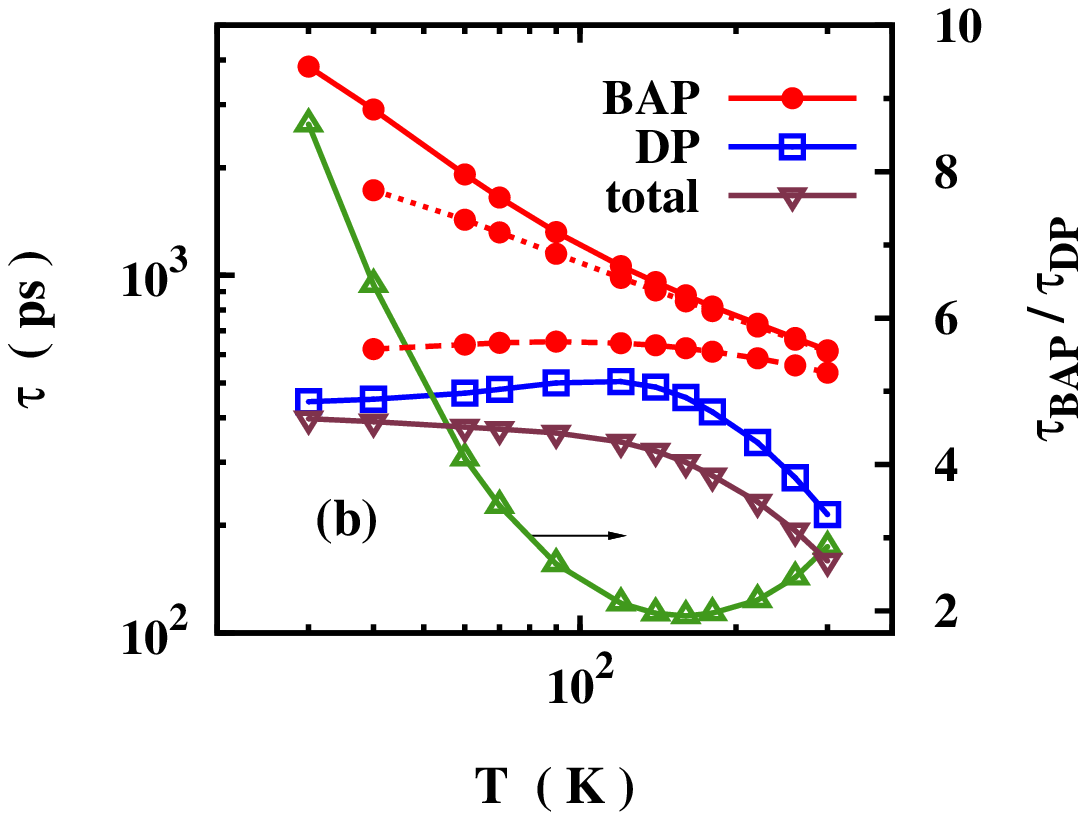}
  \end{minipage}\hfill
\caption{ $p$-GaAs. Ratio of the Bir-Aronov-Pikus spin
  relaxation time to the D'yakonov-Perel' one $\tau_{\rm BAP}/\tau_{\rm DP}$
  as function of temperature for various hole densities with $N_{\rm
    ex}=0.1n_h$ and   $n_i=n_h$. (a): $n_h=3\times 10^{15}$~cm$^{-3}$
  (curve with $\bullet$), $3\times 10^{16}$~cm$^{-3}$ (curve with
  $\square$), $3\times 10^{17}$~cm$^{-3}$ (curve with $\triangle$),
  and  $3\times 10^{18}$~cm$^{-3}$ (curve with $\triangledown$). The
  hole Fermi temperatures for these densities are $T_{\rm F}^{h}=1.7$,
  7.7, 36, and 167~K, respectively. The electron Fermi temperatures
  are $T_{\rm F}=2.8$, 13, 61, and 283~K, respectively. (b): The spin
  relaxation times due to the Bir-Aronov-Pikus (BAP) and
  D'yakonov-Perel' (DP) mechanisms, the total spin relaxation time and
  the ratio $\tau_{\rm BAP}/\tau_{\rm DP}$ (curve with $\triangle$) {\sl vs.} the
  temperature for $n_h=3\times 10^{18}$~cm$^{-3}$. The dotted (dashed)
  curve represents the Bir-Aronov-Pikus spin relaxation time
  calculated without the Pauli blocking of electrons (holes).  Note 
  the scale of $\tau_{\rm BAP}/\tau_{\rm DP}$ is on the right hand
  side of the frame. From Jiang and Wu \cite{jiang:125206}.} 
\label{fig:T_p2}
\end{figure}

{\bf Role of screening on D'yakonov-Perel' spin relaxation.}
One may find from Fig.~\ref{fig:T_p2}(b) that the D'yakonov-Perel' spin relaxation time has
{\em nonmonotonic} temperature dependence. Unlike the nonmonotonic
temperature dependence in intrinsic semiconductors, here the
behavior is not caused by the carrier-carrier scattering. The
underlying physics is a little bit more complex: it is related to the
temperature dependence of screening. To elucidate the underlying
physics, Jiang and Wu plotted the D'yakonov-Perel' spin relaxation times calculated with the
Thomas-Fermi screening \cite{haugkoch} [which applies in the
degenerate (low temperature) regime], the Debye-Huckle screening
\cite{haugkoch} [which applies in the nondegenerate (high temperature)
regime] and the screening with the random-phase approximation
\cite{haugkoch} (which applies in the whole temperature regime) in
Fig.~\ref{fig:T_peak_screen}(a). It is noted that with the 
Thomas-Fermi screening (which is temperature-independent) the peak
disappears, whereas with the Debye-Huckle screening the peak
remains. This indicates that the increase of screening with decreasing
temperature is crucial for the appearance of the peak. The scenario is
as follows: With decreasing 
temperature, the electron system gradually enters into the degenerate
regime and the temperature dependence of the inhomogeneous broadening
$\sim \langle |{\bf \Omega}|^2-\Omega_z^2\rangle$ becomes
mild. However, as the hole Fermi energy is smaller than the electron
one (due to its large effective mass), there is a temperature interval
where the hole system is still in the nondegenerate regime. The
screening, which mainly comes from holes (again, due to its large
effective mass), still increases with decreasing temperature
significantly, $\kappa^2\sim 1/T$. Therefore, the electron-impurity
scattering (the dominant one at such temperatures) time,
$\tau_p^{ei}\propto 1/[\langle V_q^2\rangle \propto \langle
 (q^2+\kappa^2)^{-2}\rangle]$, increases with decreasing
temperature. Thus the spin relaxation time, $\tau_s\sim 1/(\langle
|{\bf \Omega}|^2-\Omega_z^2\rangle\tau_p^{ei})$, decreases with decreasing
temperature at such temperature interval. When the temperature is
further lowered down, both the screening and the inhomogeneous
broadening vary little with temperature and the spin relaxation time
saturates. These behaviors, together with the decrease of the spin
relaxation time with increasing temperature at high temperature
(mainly due to the increase of inhomogeneous broadening in
nondegenerate regime), lead to the {\em nonmonotonic} temperature
dependence. The peak is roughly around the electron Fermi temperature,
$T\sim T_F$.

It is noted from Fig.~\ref{fig:T_p2}(b) that the temperature
dependence of the total spin relaxation time is still monotonic: it decreases with
increasing temperature. This is due to the contribution of the Bir-Aronov-Pikus
mechanism. It is expected that in other materials where the Bir-Aronov-Pikus
mechanism is less important than that in GaAs, such as GaSb, the
temperature dependence of the total spin relaxation time is nonmonotonic.

The question arises that whether there is a nonmonotonic temperature
dependence of the spin relaxation time in $n$-type semiconductors due to screening. A simple
estimation may help to illustrate the problem: the electron-impurity
scattering\footnote{The electron-impurity scattering is the cause of 
  the monotonic temperature dependence of the spin lifetime in
  $n$-type semiconductors.} rate varies as 
$1/\tau_p^{ei}\propto \langle
k V_q^2\rangle \propto \langle k (q^2+\kappa^2)^{-2}\rangle\sim \langle
k\rangle (\kappa^2)^{-2}\sim T^{2.5}$,\footnote{The factor $\langle
k\rangle$ comes from the density of states. This factor varies little
in the $p$-type case in the above discussion at the relevant
temperature, as electron system is in the degenerate regime.}
whereas the inhomogeneous broadening varies as $\sim \langle |{\bf
  \Omega}|^2-\Omega_z^2\rangle\sim \langle
\varepsilon_{\bf k}^3\rangle \sim T^3$. Therefore, the spin relaxation time
$\tau_s\sim 1/[\langle |{\bf \Omega}|^2-\Omega_z^2\rangle\tau_p^{ei}]\sim
T^{-0.5}$, still decreases with increasing temperature. Such estimation
applies for nondegenerate regime. For other regimes, the
inhomogeneous broadening still varies with temperature faster
than the scattering rate does. Therefore the the spin relaxation time always decreases
with increasing temperature in $n$-type III-V semiconductors in
metallic regime.

However, the situation may change when the strain-induced spin-orbit
coupling dominates. In such case, the inhomogeneous broadening $\sim
\langle |{\bf \Omega}|^2-\Omega_z^2\rangle\sim \langle
\varepsilon_{\bf k}\rangle\sim T$ varies with temperature slower than the
electron-impurity scattering does. One readily obtains that
$\tau_s\sim T^{1.5}$ in nondegenerate regime. As shown in
Fig.~\ref{fig:T_peak_screen}(b), the temperature dependence is indeed
{\em nonmonotonic} in strained $n$-GaAs with
$n_e=2\times 10^{15}$~cm$^{-3}$. However, the screening effect is only
important for low electron density $n_e\lesssim 10^{16}$~cm$^{-3}$. At
high electron density, the screening plays a marginal role in the
electron-impurity scattering even at low temperature. This is because
$1/\tau_p^{ei}\propto \langle k (q^2+\kappa^2)^{-2}\rangle$. When
electron density is high, the $q^2$ factor becomes larger than
$\kappa^2$ on average.

\begin{figure}[htb]
\begin{minipage}[h]{0.45\linewidth}
    \centering
    \includegraphics[height=5cm]{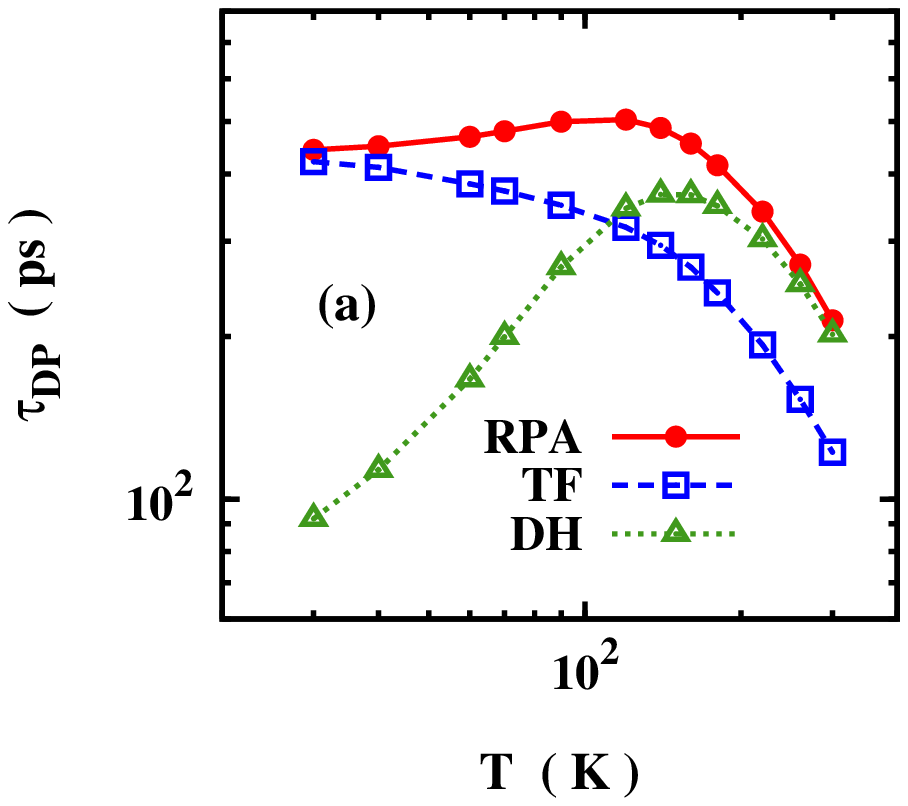}
  \end{minipage}\hfill
  \begin{minipage}[h]{0.45\linewidth}
    \centering
    \includegraphics[height=5cm]{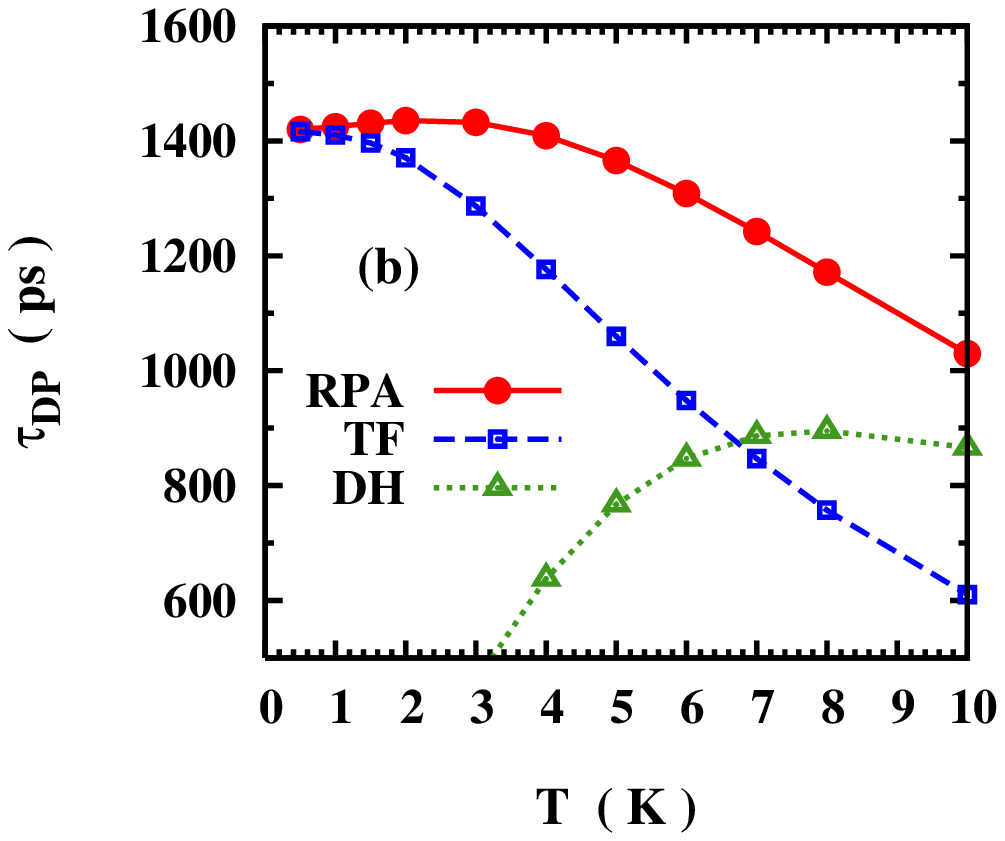}
  \end{minipage}\hfill
\caption{ (a) $p$-GaAs with hole density $n_h=3\times
  10^{18}$~cm$^{-3}$, $n_i=n_h$ and $N_{\rm ex}=0.1n_h$. Temperature
  dependence of the D'yakonov-Perel' spin relaxation times calculated
  with the Thomas-Fermi (TF) (curve
  with $\square$) and Debye-Huckle (DH) (curve with $\triangle$)
  screenings as well as the screening with the
  random-phase-approximation (RPA) (curve with $\bullet$).
  (b) $n$-GaAs with electron density $n_e=2\times 10^{15}$~cm$^{-3}$,
  $n_i=n_e$ and $N_{\rm ex}=4\times 10^{13}$~cm$^{-3}$. The spin
  relaxation times calculated with the Thomas-Fermi and Debye-Huckle
  screenings as well as the screening with the
  random-phase-approximation {\sl vs.} temperature. From Jiang and
  Wu \cite{jiang:125206}.} 
\label{fig:T_peak_screen}
\end{figure}

{\bf Photo-excitation density dependence.}
After analyzing the relative efficiency of the D'yakonov-Perel' and Bir-Aronov-Pikus mechanisms in
the two limiting cases of low and high photo-excitation, one would be
eager to see the crossover between the two limits. In
Fig.~\ref{fig:exci} the photo-excitation density dependence of the
Bir-Aronov-Pikus and D'yakonov-Perel' spin relaxation times as well as their ratio are plotted for two hole
densities at 50~K. In the low excitation limit, the D'yakonov-Perel' (Bir-Aronov-Pikus) mechanism
is more important for the case in Fig.~\ref{fig:exci}~(a) [(b)]. For both
cases, the Bir-Aronov-Pikus and D'yakonov-Perel' spin relaxation times
first decrease slowly then rapidly with
increasing photo-excitation density. Moreover, the D'yakonov-Perel' spin relaxation time
decreases faster than the Bir-Aronov-Pikus one. Hence the importance of the Bir-Aronov-Pikus
mechanism decreases with photo-excitation density. 

To understand such behavior, one notices that only the electron
density varies significantly with the photo-excitation density. The
photo-excitation density dependence of the Bir-Aronov-Pikus spin relaxation time mainly comes from
the fact that  $1/\tau_{\rm BAP}\propto \langle v_{\bf
  k}\rangle\propto \langle \varepsilon_{\bf   k}^{1/2}\rangle$ [see
Eq.~(\ref{bapapp2})], whereas that of the D'yakonov-Perel' spin relaxation time originates from the
inhomogeneous broadening $1/\tau_{\rm   DP}\propto \langle |{\bf
  \Omega}|^2-\Omega_z^2\rangle \propto \langle \varepsilon_{\bf
  k}^3\rangle$. At low photo-excitation density, the electron system
is nondegenerate, thus both the Bir-Aronov-Pikus and D'yakonov-Perel'
spin relaxation times vary slowly. At higher photo-excitation density,
both the Bir-Aronov-Pikus and D'yakonov-Perel' spin relaxation times
decrease rapidly with photo-excitation density but the
D'yakonov-Perel' spin relaxation time decreases faster.

\begin{figure}[htb]
  \begin{minipage}[h]{0.45\linewidth}
    \centering
    \includegraphics[height=5cm]{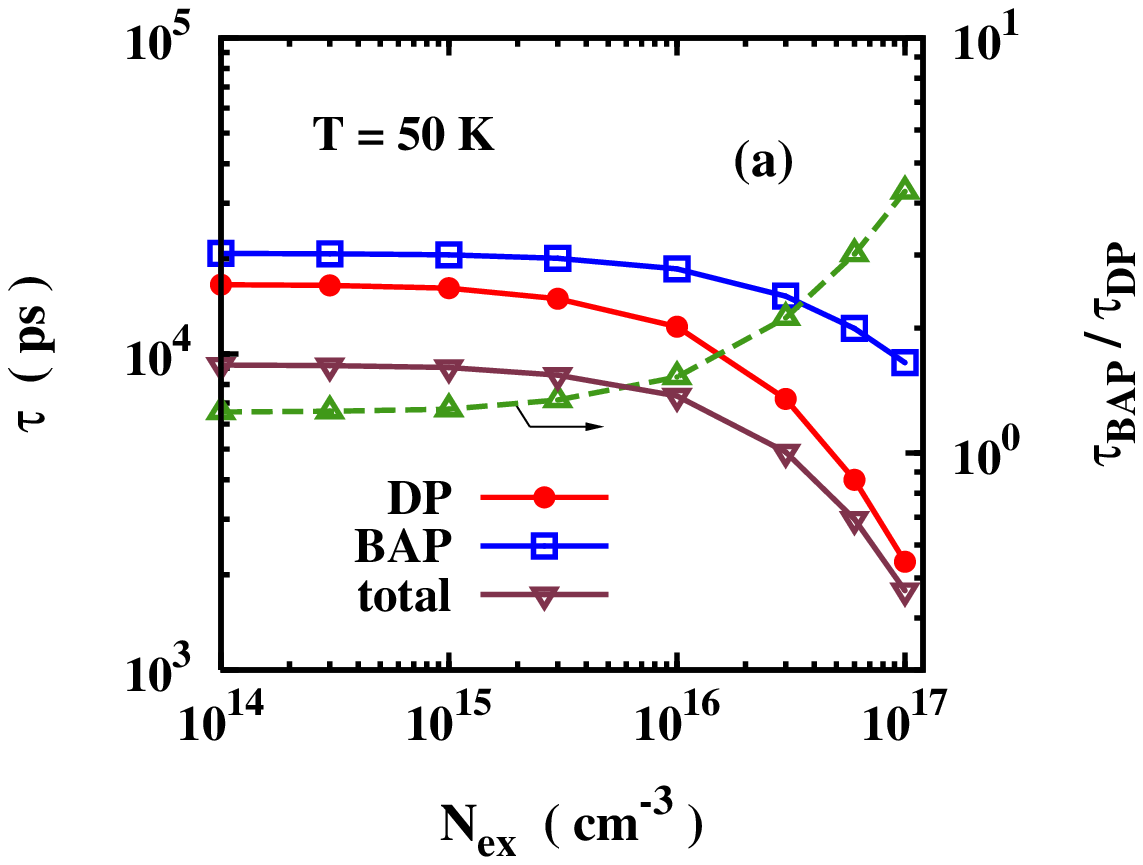}
  \end{minipage}\hfill
  \begin{minipage}[h]{0.45\linewidth}
    \centering
    \includegraphics[height=5cm]{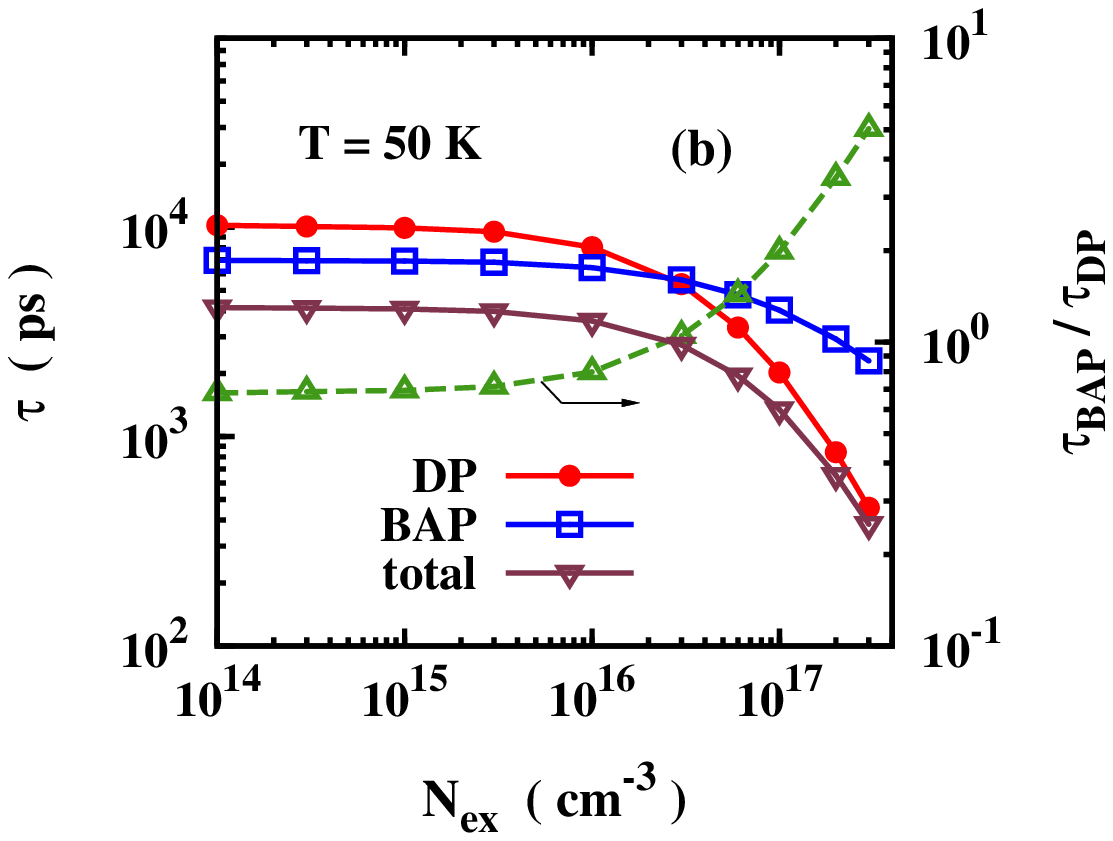}
  \end{minipage}\hfill
\caption{ $p$-GaAs. Spin relaxation times $\tau$ due to
  the Bir-Aronov-Pikus (BAP) and D'yakonov-Perel' (DP) mechanisms
  together with the total spin relaxation time {\sl vs.} the
  photo-excitation density $N_{\rm ex}$. The ratio of the two is
  plotted as dashed curve (note that the scale is on the right hand
  side of the frame). (a): $n_i=n_h=3\times 10^{17}$~cm$^{-3}$. (b):
  $n_i=n_h=3\times 10^{18}$~cm$^{-3}$. $T=50$~K. From Jiang
  and Wu \cite{jiang:125206}.}
\label{fig:exci}
\end{figure}

{\bf Hole density dependence.}
The hole density dependence of both the D'yakonov-Perel' and Bir-Aronov-Pikus spin relaxation times together with
their ratio are plotted in Fig.~\ref{fig:nh_p}. It is noted that the
Bir-Aronov-Pikus spin relaxation time decreases as $1/n_h$ at low hole density, which is consistent
with $\tau_{\rm BAP}\propto 1/n_h$ [see Eq.~(\ref{bapapp1})]. At high hole
density, $\tau_{\rm BAP}$ decreases slower than $1/n_h$ due to the
Pauli blocking of holes. The density dependence of the D'yakonov-Perel' spin relaxation time is not
so obvious: the spin relaxation time first increases, then decreases and again
increases with the hole density. As the electron distribution (hence
the inhomogeneous broadening) does not change with the hole 
density, the variation of the D'yakonov-Perel' spin relaxation time comes solely from the momentum
scattering (dominated by the electron-impurity scattering). Jiang and
Wu found that the nonmonotonic hole density dependence of the D'yakonov-Perel' spin relaxation time
is again related to the screening.

\begin{figure}[htb]
  \begin{minipage}[h]{0.45\linewidth}
    \centering
    \includegraphics[height=5cm]{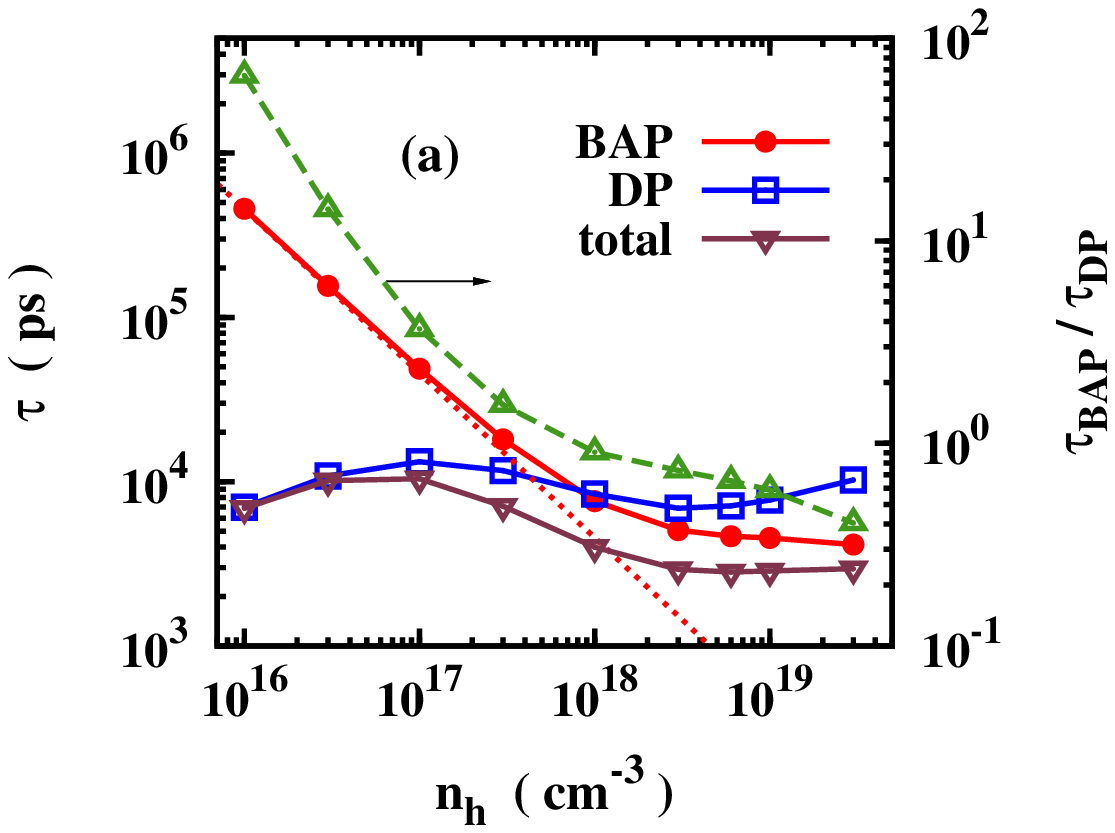}
  \end{minipage}\hfill
  \begin{minipage}[h]{0.45\linewidth}
    \centering
    \includegraphics[height=5cm]{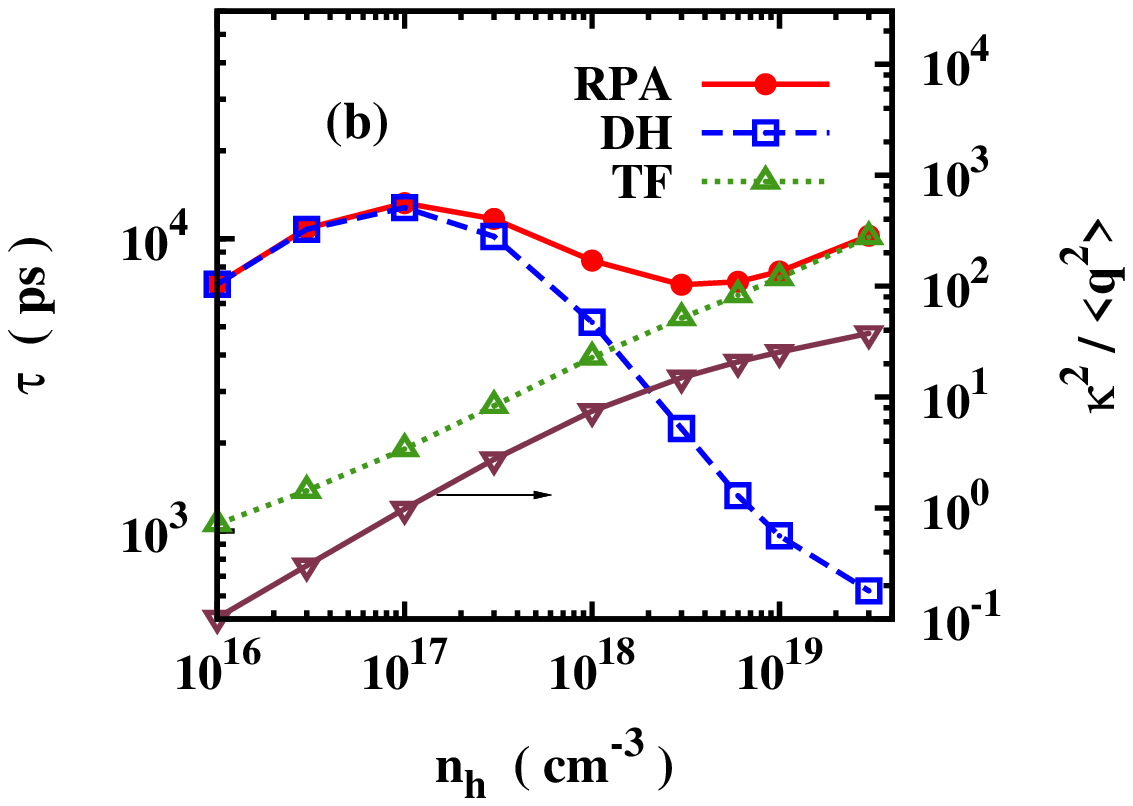}
  \end{minipage}\hfill
\caption{ $p$-GaAs. (a): spin relaxation times $\tau$
  due to the Bir-Aronov-Pikus (BAP) and D'yakonov-Perel' (DP)
  mechanisms together with the total spin relaxation time against hole
  density $n_h$. $N_{\rm ex}=10^{14}$~cm$^{-3}$, $n_i=n_h$, and $T=60$~K.
  The dotted curve denotes a fitting of the curve with $\bullet$
  using $1/n_h$ scale. The curve with $\triangle$ denotes the ratio
  $\tau_{\rm BAP}/\tau_{\rm DP}$ (note that the scale is on the right
  hand side of the frame). (b): spin relaxation times due to the
  D'yakonov-Perel' mechanism with the Debye-Huckle (DH) (curve with
  $\square$), Thomas-Fermi (TF) (curve with $\triangle$), and the
  random-phase-approximation (RPA) (curve with $\bullet$)
  screenings. The ratio $\kappa^2/\langle q^2\rangle$ is plotted as
  curve with $\triangledown$ (note that the scale is on the right hand
  side of the frame). From Jiang and Wu \cite{jiang:125206}.}
\label{fig:nh_p}
\end{figure}

To elucidate the underlying physics, the D'yakonov-Perel' spin
relaxation time calculated with the random-phase-approximation
screening together with those calculated with the Thomas-Fermi
screening \cite{haugkoch} and the Debye-Huckle screening
\cite{haugkoch} are plotted in Fig.~\ref{fig:nh_p}(b). From the figure
it is seen that the first increase and the decrease is related to the
Debye-Huckle screening, whereas the second increase is connected with
the Thomas-Fermi screening. The underlying physics is understood as
follows: In the low hole density regime, the
screening (mainly from holes) is small and the Coulomb potential, $V_q
\propto 1/(\kappa^2+q^2)$, changes slowly with the screening constant
$\kappa$ as well as hole density. Hence the electron-impurity
scattering increases with $n_h$ as $1/\tau_p^{ei}\propto n_i = n_h$. For
higher hole density ($n_h>10^{17}$~cm$^{-3}$), the screening constant
$\kappa$ becomes larger than the transfered momentum $q$. 
Hence the
electron-impurity scattering decreases with $n_h$ because
$1/\tau_p^{ei}\propto n_i \langle (\kappa^2+q^2)^{-2} \rangle \sim
n_h \kappa^{-4} \propto n_h^{-1}$ as $\kappa^2 \propto n_h$ for the Debye-Huckle
screening. As the hole density increases, the hole system gradually
enters into the degenerate regime, where the Thomas-Fermi screening applies and
$\kappa^2\propto n_h^{1/3}$. Hence, the electron-impurity scattering
increases with the hole density as $1/\tau_p^{ei}\propto n_h
\kappa^{-4} \propto n_h^{1/3}$. Consequently, the D'yakonov-Perel' spin relaxation time first
increases, then decreases and again increases with increasing hole
density.

\section{Spin diffusion and transport in semiconductors}
\label{sec:drift_diffusion}
\subsection{Introduction}

\begin{figure}[htbp]
  \centering
  \includegraphics[width=6cm]{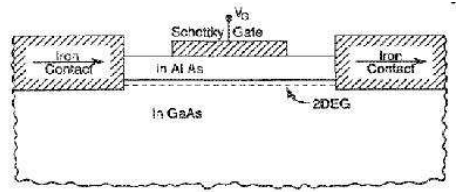}
  \caption{Schematic of Datta-Das transistor. 
    From Datta and Das \cite{datta:665}. 
  }
  \label{fig:datta-das}
\end{figure}

The implementation of spintronic devices also relies on
the understanding of spin resolved transport phenomena. 
The first prototype of the spin field-effect-transistor,
proposed by Datta and Das \cite{datta:665} and named after them,
utilizes the Rashba spin-orbit coupling for operation.
As shown in Fig.~\ref{fig:datta-das}, the so called Datta-Das
transistor 
consists of a three-layer structure with magnetic source and
collector connected by a semiconductor channel, and a voltage gate.  
The spin polarized carriers are injected into the conducting channel
from the source and driven towards and detected by the collector.
The carriers can flow freely through the drain (``on'' state) if their 
spins are parallel to that of drain or are blocked (``off'' state) if
anti-parallel. 
Similar to the traditional field-effect-transistor, the on/off {states}
of spin field-effect-transistor are switched by the gate voltage which
controls the spin direction of the passing carriers via the Rashba
spin-orbit coupling acting as an effective magnetic field, with its
strength  controlled by the gate voltage. 
It is believed that spin field-effect-transistor has the advantages of
low energy consumption and fast switching speed 
since it does not involve
creating or eliminating the electrical conducting channel during the
switching like the traditional field effect transistor
\cite{S.A.Wolf11162001,awschalomlossSamarth}.\footnote{Although 
operations of the spin field-effect-transistor do not 
create or eliminate conducting channel, the
electrons have to travel between the source and drain when the
transistor switches on/off states. Therefore, the
switching time of spin field-effect-transistor is limited by the
mobility of the channel. The energy dissipation and the switching time
of traditional and spin field-effect-transistors were compared
theoretically by Hall and Flatt\'{e} \cite{PhysRevB.68.115311}. They
pointed out that the energy dissipation of a typical spin
field-effect-transistor can be two orders of magnitude smaller than
that of a typical traditional transistor, but the switching time 
is usually longer \cite{PhysRevB.68.115311}. }

As one can see from the operation of the spin field-effect-transistor
prototype, the topics of spin transport include the generation of
nonequilibrium spin polarized carriers in conductors, the transfer of
these carriers from one place to another insider the conductor and/or
across the interfaces into other conductors, as well as the detection
of the spin signal of these carriers.  

Generation of the spin polarization can be achieved by optical
orientation of electron spin through transfer of angular momentum
of the circularly polarized photons to the carriers \cite{opt-or} and
electric injection of spin polarized carriers from ferromagnetic
conductor into non-magnetic conductor
\cite{aronov_76,johnson_85}. When electrons transfer from spin
polarized regime to unpolarized one, say from ferromagnetic metal or
semiconductors across the interface into non-magnetic conductors,
nonequilibrium spin polarization accumulates near the interface
\cite{johnson_08,dyakonovbook_08}. 
Due to spin relaxation, the spin polarization of the carriers in the
nonmagnetic conductor is not spacial uniform but decays as 
carriers move away from the interface, characterized by the spin injection
length. The spacial inhomogeneity of the spin polarization also
results in the spin diffusion, whose rate is described by the spin
diffusion coefficient. In the presence of the electric field, spins
are dragged by the electric field along with the carriers. The
response to the electric field is characterized by the conductivity or
mobility. An important task is to reveal the relations among these
characteristic parameters and how these parameters change under
different conditions. Spin detection is to extract the information of
spin polarization by sensing the change of the magnetic, optical
and/or electrical signals due to the nonequilibrium spin polarization
in the non-magnetic conductor. In experiments, optical methods are
usually more powerful as one can generate high spin polarization in
semiconductor and detect the spin signals with spacial resolution
optically. However, it is more desirable to generate and detect the
spin signal electrically in real spintronic devices. 

Even though the Datta-Das transistor is conceptually simple, there are
some crucial obstacles which make it hard to be realized: The low spin
injection/detection efficiency between the semiconductor and
ferromagnetic source and drain due to the conductance mismatch 
\cite{schmidt_00,rashba_02,PhysRevB.71.235327,crooker_05} (see
Sec.~\ref{sec:trans:efficiency} for details) and 
spin-flip scattering at the interfaces
\cite{galinon:182502,0953-8984-19-18-183201,
PhysRevLett.94.176601,PhysRevLett.96.136601}, 
spin relaxation in the semiconductor channel due to the joint effect
of the spin-orbit coupling and (spin conversing) scattering
\cite{dp,DP2,dyakonovperelbook} (see Sec.~5 for more details), 
and the precise control of the gate controlled  spin-orbit coupling. 
Due to these difficulties the Datta-Das transistor has not been
realized until Koo {et al.} claimed so recently
\cite{HyunCheolKoo09182009}, although there are still controversies on
their claim \cite{bandyopadhyay09,zainuddin10,agnihotri10}.

To overcome these obstacles one must first understand 
spin transport inside the semiconductor channel and 
across the interfaces. A lot of understandings on spin transport, 
to be reviewed in following sections, have been gained 
 in the past decade. Based on these understandings, 
some variants of the Datta-Das transistor
have been proposed. 
Schliemann {et al.} first proposed a nonballistic
spin field-effect-transistor based on the spin-orbit coupling of both the
Rashba and the Dresselhaus types \cite{PhysRevLett.90.146801}.
Due to the interplay of these two 
types of spin-orbit coupling, the spin diffusion length in the
semiconductor channel becomes larger or even infinity for spin
polarized along some special directions
\cite{PhysRevB.60.15582,PhysRevB.69.045317,0953-8984-14-12-202,PhysRevB.71.035319,
PhysRevB.70.241302} or for spin polarized current flowing along some
special directions \cite{PhysRevLett.90.146801,cheng:205328,
bernevig:236601,Koralek:nature458.610,Fabian2009,shen_09}
 (see Sec.~\ref{sec:trans:anisotropy} for details). 
This type of spin field-effect-transistor is robust against the spin
conversing scattering and is a more realistic alternative to the
Datta-Das spin field-effect-transistor.

\subsection{Theory of spin injection based on drift-diffusion model}

Although spin transport in semiconductor is a {relatively} new
problem, study of the spin transport in
ferromagnetic material 
has much longer history. It is thus natural to borrow the concepts and
methods of
spin transport in ferromagnetic metal to study the 
spin transport in semiconductor. 
In ferromagnetic conductors, the carriers and current are
spontaneously spin polarized below the Curie temperature. 
The basic transport properties of ferromagnetic metal 
can be understood by two-current model, where majority (say, spin-up) and
minority (spin-down) carriers contribute unequally and independently
to the total  
conductance \cite{mott_36a,mott_36b,campbell_67,fert_68,valet_93}. 
In two-current model, the spin-up and -down electrons transport 
independently except for weak spin-flip scattering that flips
carriers of one spin to the other
\cite{fert_68,hershfield_97,schmidt_00,rashba_02}. 
The carriers of different spins usually have different diffusion
coefficients $D_{\up/\down}$ and 
mobilities $\mu_{\up/\down}$. 
The electric currents of the spin-up and -down electrons are
determined by the drift caused by the electric field and diffusion due
to the spacial inhomogeneity of the carrier densities. 
Under linear approximation and in the diffusive limit, the currents are 
\cite{hershfield_97,schmidt_00,rashba_02,Fabianbook}
\begin{equation}
  \label{eq:trans:current}
  \mathbf{j}_{\eta}= -en_{\eta}\langle \mathbf{v}_{\eta}\rangle{=}
  e\mu_{\eta}n_{\eta}\mathbf{E}+e
  D_{\eta}\bnabla  n_{\eta}
  =\sigma_\eta\mathbf{E}+e
  D_{\eta}\bnabla  n_{\eta},
  \qquad (\eta=\up,\down),
\end{equation}
where $n_{\eta}$ and $\sigma_{\eta}=e\mu_{\eta}n_{\eta}$
are the electron density and conductance of spin $\eta$,
respectively. The electric field $\mathbf{E}$ is determined by the
Poisson equation  
\begin{equation}
  \label{eq:trans:poisson}
  \bnabla \cdot\mathbf{E}
  =-\nabla^2 \phi
  =e(n_{\up}+n_{\down}-n_0)/\varepsilon,
\end{equation}
with $\phi$, $-e$, 
$\varepsilon$ and $n_0$ being the electric potential, 
electron charge, 
dielectric constant and the
background positive charge density respectively. 
Since the total electron number is conservative, one can write down
the continuity equations for these two kinds of electrons:
\begin{eqnarray}
  \label{eq:trans:continuity}
  {\partial n_{\eta}\over\partial t}
&=&{1\over e}\bnabla \cdot\mathbf{j}_{\eta}
-{\delta n_{\eta}\over \tau_{\eta\bar{\eta}}}+
{\delta n_{\bar{\eta}}\over\tau_{\bar{\eta}\eta}}, 
\end{eqnarray}
with $\delta n_{\eta}= n_{\eta}-n_{\eta}^0$ and 
$\tau_{\eta\bar{\eta}}$ standing for deviation of the nonequilibrium charger
density $n_{\eta}$ from the equilibrium one $n_{\eta}^0$ and
the average carrier spin-flip time,
respectively. 

Combined with different initial and boundary conditions, the above
drift-diffusion equation or its equivalence
is widely used in the study of spin-related transport in
semiconductors, including in the understanding of the existing experimental spin
injection/extraction results and in the proposing of new schemes of spintronic
devices. 
However, it should be pointed out that although these equations give 
qualitatively correct results in many cases, the validity of the
drift-diffusion model should be carefully examined when applied
to spin transport in semiconductors. This is because the spin
transport in ferromagnetic metal and semiconductor can be quite different: 
First, the applied electric field and spacial gradient of 
electron density in metal are usually small due to large conductance and high
carrier density, hence electrons are usually not far away from
the equilibrium. Therefore Eq.~(\ref{eq:trans:current}) describes the
current in metals accurately in most cases. In semiconductors,
however, both the applied 
electric field and the gradient of carrier density can be very large,
and electrons can be easily driven to states far away from the
equilibrium. As a result,
Eq.~(\ref{eq:trans:current}) may no longer hold. 
More importantly, the spin-up and \mbox{-down}
branches in ferromagnetic metal are well separated and the 
coherence between these two
branches is very small. While in nonmagnetic semiconductor the spin-up
and \mbox{-down} electrons are usually degenerate, an applied
or effective (from the spin-orbit coupling) magnetic field can
cause spins to precess and result in large spin coherence. 

Therefore quantitative calculation 
based on the drift-diffusion model 
is questionable when the spin coherence is essential to the spin
kinetics or when the electrons involve in the
transport are far away from the equilibrium. It has been shown that in
the presence of the magnetic field, either the external applied one or 
the intrinsic effective one from the spin-orbit coupling, the drift-diffusion
model is inadequate in accounting for the spin transport in 
semiconductors \cite{PhysRevB.66.235109,weng:410,jiang:113702,%
cheng:073702,weng:063714}. 
Nevertheless, the drift-diffusion model is still useful in qualitative
study of spin transport since its simplicity and flexibility to add
new factors of physics.

\subsubsection{Spin transport in nonmagnetic semiconductor using
drift-diffusion model}
\label{sec:trans:nm}

One can easily obtain some basic
properties of the spin transport using drift-diffusion model to study
the transport property of the magnetic momentum 
inside nonmagnetic semiconductors in the simplest case. 
Assuming that the charge density in nonmagnetic
semiconductor is uniform and 
the electronic transport coefficients 
are not affected by the spin polarization, one can then
write down the transport equation for the magnetic momentum $\mathbf{S}$
in a uniform electric field $\mathbf{E}$ and magnetic field
$\mathbf{B}$ \cite{dyakonovperelbook},
\begin{equation}
\label{eq:trans:S}
  {\partial \mathbf{S}\over \partial t}=
  D\bnabla ^2 \mathbf{S}
  -e\mu \mathbf{E}\cdot\bnabla  \mathbf{S} + 
  g\mu_B\mathbf{B}\times\mathbf{S}
  -{\mathbf{S}\over\tau_s},
\end{equation}
with $\mu$, $D$, and $1/\tau_s$
being the charge mobility, diffusion coefficient and spin relaxation
time, respectively. Note that the drift-diffusion equation here has
been modified to include the Larmor precession of the spin around the
applied magnetic field $\mathbf{B}$. 

{\bf Spin accumulation in the steady state.}
When a semiconductor is in contact with a spin polarization source
at $x=0$, $\mathbf{B}=0$ 
and the electric field is along the $x$-direction, the
drift-diffusion model predicts 
spin accumulation with exponential decay in the semiconductor
$S(x)=S_0\exp[-x/L_s(E)]$
in which the electric-field-dependent spin injection length reads
\cite{aronov_76,PhysRevB.66.201202,PhysRevB.66.235302}
\begin{equation}
  \label{eq:trans:Ls}
  L^{-1}_s(E)={1\over L_s}\sqrt{1+
    {L_d^2\over 4L_s^2}}
  -{E\over |E|}{L_d\over 2L_s^2},
\end{equation}
where $L_s=\sqrt{D\tau_s}$ is the diffusion length, $L_d=|e\mu 
E|\tau_s=|v_d|\tau_s$ is the drifting length over spin relaxation time
with drift velocity $v_d$ determined by the electric field. 
Without the
electric field, the spin injection length is the
diffusion length $L_s$. The applied electric field can
significantly change the injection length by dragging or pulling the
electron \cite{PhysRevB.66.201202,PhysRevB.66.235302}: For large downstream 
electric field, the electron spin 
injection is the distance that the electrons 
move with drift velocity within the spin
lifetime time, $L_s(E)=L_d$; For
large upstream field, $L_s(E)=L_s^2/L_d$. 

{\bf Evolution of the spin polarized electron package.} 
The shape of a spin polarized
carrier packet changes with time due to the drift and diffusion. 
The temporal evolution of a $\delta$ spin polarized packet 
of height $S_0$ at $x=0$ is  
\begin{equation}
  \label{eq:trans:delta}
  S(x,t)={S_0\over \sqrt{2\pi Dt}}  
  \exp\left[-{t\over \tau_s}-{(x-v_dt)^2\over 4Dt}\right], 
\end{equation}
which has the form of Gaussian function whose center is determined by the
drifting, $v_dt$, and width determined by the diffusion,
$\sqrt{Dt}$. 

\subsubsection{Hanle effect in spin transport}
\label{sec:hanle}

In the spacial homogeneous system with a spin excitation source, such
as circularly polarized light 
\cite{vekua_76,PhysRevLett.23.1152,Kikkawa01212000}
or spin resonant microwave radiation 
\cite{PhysRevB.56.7422,hu:204},
the total spin is the accumulation of survival spin from the past. 
When a perpendicular magnetic field is presented, spin undergoes Larmor
precession. Since spins excited at different time have different
precession phases and tend to cancel each other, the perpendicular
magnetic field reduces the total accumulated spin. 
This is known as the Hanle effect
\cite{Hanle_24,dyakonovperelbook}. 
In spin transport, the Hanle effect also appears
\cite{Appelbaum:447.295}: 
Spin accumulation at position $x$
is the sum of the electron spins travel from different places.
Since different electrons have different transit time
when they reach $x$, their phases are different and tend to cancel
each other. 
As shown in Fig.~\ref{fig:trans:L_EB_schema}, for
a constant spin pumping/injection starting from $t=0$ at $x=0$, when a constant
magnetic field $B$ is applied along the $\hat{y}$ direction, 
the spin accumulation at time $t$ and position $x$ is 
\begin{eqnarray}
  \label{eq:trans:sz_th}
  S_z(x,t) = \int_0^t {S_0 \over \sqrt{2\pi t'}} e^{-t'/\tau_s-(x-e\mu
      Et')^2/(4Dt')}\cos\omega t' \, d\,t', \\
  S_x(x,t) = \int_0^t {S_0 \over \sqrt{2\pi t'}} e^{-t'/\tau_s-(x-e\mu
      Et')^2/(4Dt')}\sin\omega t' \, d\,t',
  \label{eq:trans:sx_th}
\end{eqnarray}
with $\omega=g\mu_BB$ being the Larmor frequency.
For the steady state spin injection under a magnetic field, the spin
accumulation 
can be calculated by letting time in
Eqs.~(\ref{eq:trans:sz_th}) and (\ref{eq:trans:sx_th}) to be infinity or
can be solved directly from Eq.~(\ref{eq:trans:S}). The results are
\begin{eqnarray}
  \label{eq:trans:sz_h}
  S_z(x) &=& S_0 e^{-x/L_s(E,B)}\cos [x/L_0(E,B)],\\
  S_x(x) &=& S_0 e^{-x/L_s(E,B)}\sin [x/L_0(E,B)],
\end{eqnarray}
where
\begin{eqnarray}
\label{eq:trans:LsEB}
  L^{-1}_s(E,B)
&=&{E\over |E|}{L_d\over 2L_s^2}
+{1\over L_s}\sqrt{\left(1+{L^2_d\over
      4L_s^2}\right)^2+(\omega\tau_s)^2}
\cos{\theta\over 2},\\
\label{eq:trans:L0EB}
L^{-1}_0(E,B)&=&
{1\over L_s}\sqrt{\left(1+{L^2_d\over
      4L_s^2}\right)^2+(\omega\tau_s)^2}
\sin{\theta\over 2}, \\
\label{eq:trans:LsL0t}
\tan\theta &=& {\omega\tau_s\over 1+L_d^2/(4L_s^2)}.
\end{eqnarray}
One can see that the spin polarizations change with the position as a
damped oscillation with the electric and magnetic
field dependent injection length $L_s(E,B)$ and the spacial oscillation 
``period'' $L_0(E,B)$ defined by
Eqs.~(\ref{eq:trans:LsEB}--\ref{eq:trans:LsL0t}). 
Even when there is no spin relaxation mechanism, {
  i.e.}, $\tau_s=\infty$, the spin injection length is still finite in
the presence of a perpendicular magnetic field,
$L^{-1}_s(0,B)=L^{-1}_0(0,B)=\sqrt{\omega/2D}$.\footnote{This effect
  was first predicted by Weng and Wu microscopically from the kinetic
  spin Bloch equation approach \cite{PhysRevB.66.235109}.}
Without the driving
of the electric field, the  oscillation ``period'' is larger or close
to the injection length, which means that the oscillation may not be
easy to detect.  
Under a strong downstream electric field, 
$L_s(E,B)\simeq L_d$, and $L_0(E,B)\simeq L_d/\omega \tau_s$, it is
then possible to observe many oscillations in spin accumulation. 

\begin{figure}[htbp]
  \centering
\includegraphics[width=4cm]{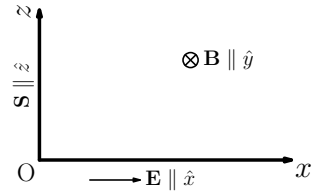}
  \caption{Orientations of spin transport and electric/magnetic fields.}
  \label{fig:trans:L_EB_schema}
\end{figure}

\subsubsection{Spin injection theory}

To study the spin injection from ferromagnetic electrode to 
semiconductor, it is usually more convenient to use the
drift-diffusion equation in the form of electrochemical
potential \cite{hershfield_97,schmidt_00,rashba_02,Fabianbook}. 
For the state near the equilibrium, 
$\delta n_{\eta}(\mathbf{r})=
g_{\eta}\delta\tilde{\zeta}_{\eta}(\mathbf{r})$,
with 
$\delta \tilde{\zeta}_{\eta}(\mathbf{r})=
\tilde{\zeta}_{\eta}(\mathbf{r})-\tilde{\zeta}_{\eta}^0(\mathbf{r})$
standing for the 
deviation in the chemical potential
$\tilde{\zeta}_{\eta}(\mathbf{r})$ away from the
equilibrium one $\tilde{\zeta}_{\eta}^0(\mathbf{r})$
of the electron with spin $\eta$ at position $\mathbf{r}$.
$g_{\eta}=(\partial n_{\eta}/\partial \tilde{\zeta}_{\eta})$
is the temperature dependent density of states. The current can be
rewritten as 
\cite{hershfield_97,schmidt_00,rashba_02,Fabianbook}
\begin{equation}
  \label{eq:trans:current_mu}
  \mathbf{j}_{\eta}=-\sigma_{\eta}
  \bnabla 
  \phi(\mathbf{r})+
  eD_{\eta} g_{\eta}\bnabla \delta\tilde{\zeta}
  =
  {\sigma_{\eta} \over e}\bnabla \zeta_{\eta}(\mathbf{r}),
\end{equation}
in which we have used the Einstein relation, 
$\sigma_{\eta}=e^2D_{\eta}g_{\eta}$,
and introduced the electrochemical potential, 
$\zeta(\mathbf{r})=\tilde{\zeta}(\mathbf{r})-e\phi(\mathbf{r})$. 

Introducing the notations, $g=g_{\up}+g_{\down}$,
$g_s=g_{\up}-g_{\down}$, 
$\sigma=\sigma_{\up}+\sigma_{\down}$, 
$\sigma_s=\sigma_{\up}-\sigma_{\down}$, 
$\zeta=(\zeta_{\up}+\zeta_{\down})/2$, 
$\zeta_s=(\zeta_{\up}-\zeta_{\down})/2$, 
$D=(D_{\up}+D_{\down})/2$ and $D_s=(D_{\up}-D_{\down})/2$,
and enforcing the charge neutrality, 
{i.e.} $n_{\up}+n_{\down}\equiv n^0_{\up}+n^0_{\down}$, 
one obtains 
\begin{equation}
  \label{eq:trans:charge_neutr}
  \zeta(x)+e\phi(x)=-g_s\zeta_s(x)/g.
\end{equation}
The non-equilibrium
spin accumulation is expressed as
\begin{equation}
  \label{eq:trans:spin_accum}
  \delta S = {4g_{\up}g_{\down}\over g}\zeta_s.
\end{equation}
From Eq.~(\ref{eq:trans:current_mu}) one can further write down the
charge and spin currents expressed in form of the electrochemical
potential,  
\begin{eqnarray}
  \label{eq:trans:charge_current}
  \mathbf{j}_{e}&=&\mathbf{j}_{\up}+\mathbf{j}_{\down} 
  ={\sigma\over e}\bnabla  \zeta + {\sigma_s\over e}
  \bnabla \zeta_s,\\
  \mathbf{j}_{s}&=&\mathbf{j}_{\up}-\mathbf{j}_{\down} =
  {\sigma_s\over e}\bnabla \zeta
  +{\sigma\over e}\bnabla \zeta_s,\\
  {\partial \zeta_s\over \partial t}&=&\overline{D}\bnabla ^2\zeta_s
  -{\zeta_s\over\tau_s},
\end{eqnarray}
with $\overline{D}=g(g_{\up}/D_{\down}+g_{\down}/D_{\up})^{-1}$ being the
effective spin diffusion coefficient. 

From these equations, one can obtain the spin transport properties
in conductors. 
For nonmagnetic conductor, the spin-up and -down bands are degenerate,
one recovers the results obtained in Sec. \ref{sec:trans:nm} for the
spin transport in nonmagnetic conductor under weak electric field. 
In ferromagnetic conductor, the dynamics of nonequilibrium spin accumulation is
similar to that in nonmagnetic conductor, 
{i.e.}, it exponentially decays with the 
position, with the spin diffusion length being
$L_s=\sqrt{\overline{D}\tau_s}$. 
For a uniform ferromagnetic conductor without nonequilibrium spin
accumulation, the current 
polarization is determined by the difference of the conductivities of the
two spin branches, 
\begin{equation}
  P_{\sigma}={j_s/j_e}={\sigma_s/\sigma}.
\end{equation}
In the presence of the spin
accumulation, the current polarization is modified as
\begin{equation}
  P_{j} = P_{\sigma} + {L_s \nabla \zeta_s/(j_e R)},
\end{equation}
where $R=\sigma L_s/(4\sigma_{\up}\sigma_{\down})$ is the effective
resistance of the conductor with unit surface area over a distance of
the diffusion length.

\subsubsection{Spin injection efficiency of ferromagnetic
  conductor/nonmagnetic conductor junction}
\label{sec:trans:efficiency}

To discuss the spin injection from ferromagnetic conductor
into nonmagnetic conductor, one needs to study the
transport through their interface. In drift-diffusion model, the
effect of interface is phenomenally described by spin selective
interface conductance. 
Assuming that there is no strong spin-flip scattering at the 
interface, the current of each spin branch is conservative. 
However, the electrochemical
potential can be discontinuous if the contact is not ohmic. 
Introducing the contact conductance $\Sigma_{\eta}$ for spin $\eta$, 
the boundary condition at the interface (x=0) is then written as
\cite{aronov_76,johnson_85,schmidt_00,rashba_02,crooker_05}
\begin{equation}
  \label{eq:trans:contact}
  j_{\eta}(0)=\Sigma_{\eta}[\zeta_{\eta F}(0)-\zeta_{\eta N}(0)]. 
\end{equation}
Similarly, one can introduce $\Sigma=\Sigma_{\up}+\Sigma_{\down}$,
$\Sigma_s=\Sigma_{\up}-\Sigma_{\down}$ and rewrite the boundary
conditions for charge current and spin current.

Using the above drift-diffusion equation, the matching condition of
interface at $x=0$, and the boundary conditions
$\zeta_{sF}(-\infty)=0$ and $\zeta_{sN}(\infty)=0$ at the far left end
of ferromagnetic electrode
and the far right end of nonmagnetic electrode, one is able to write down
the spin injection efficiency across the 
ferromagnet/nonmagnetic conductor
junction. The current
polarization at the interface is \cite{schmidt_00,rashba_02}
\begin{equation}
  \label{eq:trans:efficiency}
  P_j(0)={j_s(0)\over j_e(0)}={R_F P_{\sigma F}+R_c P_{c}
    \over R_F+R_c+R_N}, 
\end{equation}
where $R_F$, $R_N$ and $R_{c}={\Sigma/(4\Sigma_{\up}\Sigma_{\down})}$,  are
the effective resistances of ferromagnetic conductor, 
nonmagnetic conductor and contact, respectively.
$P_{\sigma F}$ and $P_{c}=(\Sigma_{\up}-\Sigma_{\down})/\Sigma$ are
the conductance polarizations of the ferromagnetic conductor and
contact respectively.  
Since the ferromagnetic conductor usually has much larger conductance
and shorter 
spin diffusion length, $R_F\ll R_N$. Therefore, for transparent
contact, the polarization of the injected current 
$P_j(0)=P_{\sigma\,F}R_{F}/(R_{F}+R_N)$ is much smaller than
$P_{\sigma\,F}$, the current polarization in ferromagnetic conductor, 
due to the conductance mismatch
\cite{schmidt_00,rashba_02,PhysRevB.71.235327,crooker_05}.  
For {large} contact resistance, the current polarization is determined
by the contact polarization $P_j(0)\simeq P_c$. 
Therefore the
conduction mismatch can be reduced by inserting an spin selective
layer with large resistance
\cite{hershfield_97,schmidt_00,PhysRevB.64.184420,rashba_02,
PhysRevB.71.235327,crooker_05,Fabianbook}. 
One can also avoid the conductance mismatch by using spin filter
\cite{PhysRevLett.80.4578,PhysRevB.64.195319,PhysRevLett.85.1962} to
replace the ferromagnetic metal as spin injector. 

\subsubsection{Silsbee-Johnson spin-charge coupling}

In electrical spin injection, the spin polarized current flows from a
ferromagnetic electrode into nonmagnetic conductor and produces a
non-equilibrium spin 
accumulation near the interface in the nonmagnetic conductor. 
The inverse of the above effect also holds: The presence of
non-equilibrium spin accumulation in nonmagnetic conductor near the
interface will produce an electromotive force in the circuit which
drives a charge current to flow in a close circuit or 
results in a voltage drop in an open circuit. 
This effect is called Silsbee-Johnson spin charge coupling
\cite{silsbee_80,johnson_85}. 
This spin-charge coupling can affect the spin transport
properties in return. In the spin injection, the electromotive force
caused by the spin accumulation impedes the charge/spin current and
{reduces} the overall conductivity, which is called the
spin bottleneck effect
\cite{silsbee_80,johnson_85,son_87,johnson_93,johnson_93b}. 

\begin{figure}[htbp]
    \centering
    \includegraphics[width=7cm]{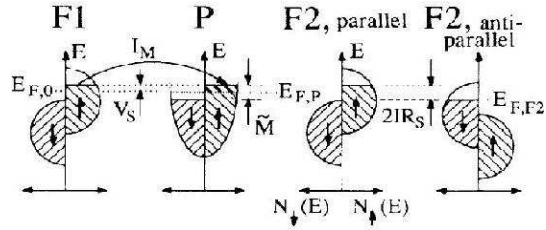}   
    \caption{Schema of Silsbee-Johnson effect and all electrical spin
      detection. From Johnson \cite{johnson_93}.
}
    \label{fig:trans:silbee}
\end{figure}
Physically speaking, this spin-charge coupling can be
understood by the simplified model shown in
Fig.~\ref{fig:trans:silbee}, proposed by Silsbee  
\cite{silsbee_80}. In this model, it is assumed that the ferromagnetic 
electrode is a
half-metal so that only one spin branch (say, spin-up) electrons carry
the current and Fermi levels in ferromagnetic
and nonmagnetic conductors are aligned before they are
connected. 
After they are connected, 
the spin accumulation near the interface raises the
(electro)chemical potential of spin-up electrons in nonmagnetic electrode,
and lowers
that of spin-down electrons to ensure the neutrality of charge, 
as shown in the left part of Fig.~\ref{fig:trans:silbee}.
As a result of this electrochemical potential shifting in 
nonmagnetic electrode, an
electromotive force must be produced to raise the electrochemical
potential in ferromagnetic electrode so that it aligns with that 
of spin-up electrons in nonmagnetic electrode. 
Quantitative calculation of Silsbee-Johnson coupling based on
drift-diffusion model can be obtained from 
Eqs.~(\ref{eq:trans:charge_neutr}--\ref{eq:trans:contact}). 
In close circuit the change in the resistance caused by the spin
bottleneck effect is
\begin{equation}
  \delta R={R_N(R_cP_c^2+R_FP_F^2)+R_FR_c(P_F-P_c)^2
\over R_F+R_c+R_N}.
\end{equation}
For more complicated structures like 
ferromagnetic/nonmagnetic/ferromagnetic conductors,
the resistance change due
to the spin accumulation depends on the length 
of nonmagnetic conductor
and the spin polarizations of 
ferromagnetic electrodes. When the length of 
nonmagnetic conductor is much larger than the spin
diffusion length, the resistance change is simply the sum of that of
two ferromagnetic/nonmagnetic 
contacts. When the length of nonmagnetic conductor 
is smaller than or of the order of
spin diffusion length, the change in resistance depends on the alignment
of the magnetic momentum of the two electrodes, as shown in the right
part of Fig.~\ref{fig:trans:silbee}. This is known as
spin valve effect. It can be quantitatively calculated by matching the
boundary conditions on these two contacts \cite{rashba_02}. 

In open circuit, for a spin accumulation at $x=0$, the voltage drop
caused by the electromotive force over a distance far longer than the
spin diffusion length is given by
\begin{equation}
  \label{eq:trans:sj_coupling}
  \Delta\phi = 
  \bigl({g_s/g}-{\sigma_s/\sigma}\bigr) {\zeta_s(0)/e}.
\end{equation}
If a second ferromagnetic electrode is attached to the 
nonmagnetic electrode, it can be used to
detect the spin accumulation at the interface of the nonmagnetic electrode
and the second ferromagnetic electrode. If there
is spin accumulation, the voltage drop between the nonmagnetic and 
the second ferromagnetic electrodes depends on
the relative alignments of the spin momentums in these two electrodes, 
as shown in
the right part of Fig.~\ref{fig:trans:silbee}. 
Silsbee-Johnson spin-orbital coupling is particular useful in the
non-local spin detection \cite{johnson_93,johnson_93b,jedema_01,%
jedema_03,erve:212109,lou_07,staa_08,ciorga_09}.

\subsection{Spin injection through Schottky contacts}

In the drift-diffusion model, the contact of ferromagnetic conductor 
and semiconductor is
usually described as a sheet layer with finite or zero resistance. 
Studies of the 
spin injection through Schottky barrier, which appears naturally at
the interface of ferromagnetic metal and semiconductor,
have also been carried out using drift-diffusion
model \cite{albrecht_02,albrecht_03} and Monte Carlo simulation
\cite{shen_04,saikin_05,saikin:1769,wang_prb_05,saikin_06,lopez_08}. 

In the framework of {the} drift-diffusion model, the tunneling through
the Schottky barrier is still described by (spin selective) interface
resistance. However, the conductivity inside the semiconductor is no
longer a constant, since it is proportional to carrier concentration
which varies dramatically with position near the Schottky barrier. 
Taking the position dependence of the conductivity into account, it is
shown that the existence of a depletion region of the Schottky barrier
greatly reduces the spin accumulation.
For a ferromagnetic metal/semiconductor 
contact with significant depletion region, the spin
accumulation is reduced almost to zero in a distance less than 100~nm
\cite{albrecht_02,albrecht_03}. Outside the depletion region, the spin
accumulation decays with the distance at a slower rate of spin
injection length (at the order of 1~$\mu$m). 
The existence of the Schottky barrier also greatly reduces the spin 
polarization of the injected current. Although the spin polarization
of injected current decays at the rate of spin injection length
throughout the semiconductor side, it drops rapidly inside the 
ferromagnetic conductor near the 
ferromagnetic metal/semiconductor surface when a high barrier is
presented. 

In the Monte Carlo simulations, studies focus on the transport
inside the semiconductor. The ferromagnetic conductor
is treated as source that
provides spin polarized carriers and is not affected by the
semiconductor. The current is injected into the semiconductor through
the barrier by both thermionic emission and direct tunneling
\cite{sze,sun_03}. In this model, the tunneling probability through
the barrier, proportional to the density of states for
spin-up and -down carriers in the ferromagnetic conductor, is
naturally spin dependent. Therefore, this model is beyond the
drift-diffusion model. In the simulation, the spacial profile of the
Schottky barrier is determined by solving the charge motion and the
Poisson equation self-consistently \cite{martin_96}. The transport in
the semiconductor is based on semiclassical approximation that
includes ``drifting'' and ``scattering'' processes. The electron
spin is subjected to the influent of the Rashba and/or Dresselhaus 
spin-obit interaction during the ``drifting'' process. The Monte Carlo
simulations of the spin injection from Fe contact into GaAs quantum well
{showed} that, although injected spin polarization is large right next to
the Schottky barrier, similar to the results of {the} drift-diffusion model,
the total spin polarization drops to nearly zero in a few tens of
nanometers, a distance of the order of depletion region, due to large
unpolarized carriers in the semiconductor. 
Moreover, the average magnetic momentum of the 
{\em injected} electrons also decays dramatically in the depletion
region but much more slowly beyond that. 
This is because the Rashba and Dresselhaus spin-orbit couplings depend
on the electron momentum, the 
larger the kinetic energy, 
the larger the effective magnetic field. In the depletion region, 
electrons have much larger kinetic energy, therefore suffer much faster
decay rate. Outside the depletion region, the decay rate is much
slower as electrons lose their kinetic energy to overcome the
electrical potential barrier as well as due to the inelastic scattering
\cite{shen_04,saikin_05,saikin:1769,wang_prb_05,saikin_06,lopez_08}.  
Usually, the Rashba effect considered in the spin kinetics in 
quantum wells is determined by the electric field perpendicular to the
quantum well plane (due to the built-in electric field or the gate
voltage). In  the presence of the Schottky barrier, there is a strong
in-plane electric field in the depletion region induced by the
barrier. This barrier induced electric field also leads to the Rashba
effect and further reduces the magnetic momentum of the injected
electrons
\cite{wang_prb_05}. 

As one can see from the above studies that, a high Schottky barrier
with significant depletion region is highly undesirable for spin
injection. In order to increase the spin injection efficiency, one
needs to {modify} the Schottky barrier. A thin heavily doped $n^{+}$
layer, which sharply reduces the height and thickness of the barrier,
was proposed to be inserted between the ferromagnetic electrode and
semiconductor to increase the injection efficiency
\cite{hanbicki_02,hanbicki_03,osipov_04b,bratkovsky_04,%
bratkovsky_05,osipov_05,adelmann_05,saha:142504}. 
Using this setup, 
highly efficient electrical spin injection can been achieved. 
It has been reported that in a wide range of temperature,
up to 30\% of the spin polarization
is injected from Fe into $n$-type AlGaAs and GaAs and survives over a
distance of 100~nm
\cite{hanbicki_02,hanbicki_03,adelmann_05}. 
Up to 50\% spin polarization has been reported to survive over a
distance of 300~nm in the experiments of spin injection from Fe into
$n$-type ZnSe when the temperature is around or
less than 100~K \cite{hanbicki_09}. 

Theoretical studies of spin injection through the interface of
ferromagnetic conductor  
and semiconductor with a $\delta$-doped layer near the interface were
carried out in
Refs.~\cite{osipov_04b,bratkovsky_04,bratkovsky_05,osipov_05}. In
these studies, it was assumed that the depletion region is only a few
nanometers and the tunneling electrons do not lose their spin polarization
over this region. By assuming that the depletion region is of triangular
shape, a direct tunnel current was calculated and the transport outside
the depletion region was performed using {the} drift-diffusion model. 
Unlike the traditional drift-diffusion model, the
spin polarization of the tunneling carriers depends on the band
structure of ferromagnetic conductor, but is independent on the
transport properties of the ferromagnetic metal. Hence, 
there is no conductance mismatch in this model, and it is possible to
achieve high spin polarization at the interface. 
Moreover, it was
shown that the spin injection efficient is not a constant but strongly
depends on the injection current, the larger the current amplitude,
the higher the injection efficiency. In order to achieve high
efficiency, a high electric field is required and therefore 
injection length also increases with the injection current as 
predicted by Eq.~(\ref{eq:trans:Ls})
\cite{osipov_04b,bratkovsky_04,bratkovsky_05,osipov_05}. 
However, the applied electric field also enhances the Rashba
spin-orbit coupling and reduces the injected spin polarization as
pointed out by Wang and Wu \cite{wang_prb_05}. 
Recent experiment by Kum {et al.} showed that the 
competing effects of the electric field cancel each other, leading to
an almost negligible decrease of magnetoresistance with bias current
\cite{kum:212503}. 

\subsection{Spin detection and 
experimental study of spin transport}

Experimental study of spin transport also requires an effective method
to extract the spin information from semiconductors. In some senses,
spin detection and extraction are reverse of the spin
generation and injection. Each spin generation/injection mechanism can
also be reversely used to extract the spin information. Generally
speaking, there are two categories of spin detection: optical and
electrical spin detections.  
Optical spin detection, such as the Faraday/Kerr rotation and 
circular polarization of electroluminescence, provides reliable 
information of spin polarization in semiconductor and has been proven
to be a powerful experimental tool in the study of the spin transport. 
While high efficiency electrical spin detection is essential to
the spin transistor.

\subsubsection{Spin detection using electroluminescence}


The electroluminescence is the reverse of the optical orientation. 
In the usual electroluminescence
experiment setup, the electrons (holes) to be detected are driven to a
structure like a light emitting diode where they recombine with
unpolarized holes (electrons) and emit photons. Due to the
selection rule, the emitted photons are circularly polarized if the
carriers to be detected are spin polarized. By measuring the
polarization of the electroluminescence, one obtains the spin
polarization. 
This technique was employed in the first experimental observation of
spin injection from ferromagnetic nickel tip of a scanning tunneling
microscope into nonmagnetic $p$-type GaAs, and  
the injected spin polarization of about $40\%$ at the surface was
reported \cite{alvarado_92}. 
However, the early attempts on high spin injection efficiency in the
real devices were not so successful due to spin-flip scattering at 
the interface and the conductance mismatch \cite{schmidt_00,rashba_02}. 
Spin polarization injected from traditional ferromagnetic metals,
such as Fe, Ni, Co and their alloys, 
into GaAs or InAs was usually only a few percents
\cite{filip_00,zhu_01,JJAP.42.L87}.
By various means, the spin injection efficiency has been improved over
the years. Here we list some important results in the following,
loosely cataloged by the structures used in the experiments. 
\begin{itemize}
\item {
    Spin injection from ferromagnetic or diluted magnetic
  semiconductors into GaAs: }
Ferromagnetic or diluted magnetic and nonmagnetic semiconductors
have similar conductance, therefore spin injection in these structures
surfers less conductance mismatch. However, since 
ferromagnetic
semiconductor has low Curie temperature, the experiments are usually
performed at low temperature.
Spin injection from $n$-type II-V diluted magnetic semiconductor 
into $n$-GaAs under applied magnetic field is very efficient. 
Injected spin polarization from Be$_x$Mn$_y$Zn$_{1-x-y}$Se into GaAs 
achieves 90\% at low temperature ($<5$~K), but drops to 20\% when the
temperature rises to 35~K
\cite{fiederling_99,Schmidt2001202}. 
Spin injection from $p$-type GaMnAs or MnAs into $n$-type GaAs is more
difficult. The efficiency of direct spin injection from GaMnAs or MnAs
into $n$-type GaAs usually is only a few percent even at low
temperature 
\cite{Ohno1999,ploog:7256,ramsteiner_02,PhysRevB.66.081304}. 
Injection from GaMnAs or MnAs via interband tunneling in Esaki diode
is much more efficient
\cite{kohda_01,PhysRevB.65.041306,andresen:3990,dorpe:3495, 
kita_06,ciorga_06,kohda:012103,einwanger:152101}. 80\% injected spin
polarization in Esaki diode has been realized when temperature is 10~K
\cite{kohda:012103}. 
Molenkamp group has proposed and demonstrated to use a magnetic double
barrier resonant tunneling diode as spin injector \cite{gruber:1101,PhysRevLett.90.246601,0268-1242-19-7-029,scherbakov:162104,supp07}. 
A spin polarization up to 80\% has been injected from 
BeTe/Zn$_{1-x}$Se/BeTe into GaAs under 1.6~K
temperature and 7.5~T magnetic field 
\cite{gruber:1101}.\footnote{
It is  noted that 
the direction of the spin polarization 
through magnetic double barrier resonant 
diode can be changed by an external voltage which selects the
spin dependent resonant tunneling level
\cite{gruber:1101,PhysRevLett.90.246601,0268-1242-19-7-029,scherbakov:162104,supp07}.
This is quite different from the spin injection from ferromagnetic material 
where the injected spin polarization is the same as the majority
spin of the injector. To change the spin direction injected from the
ferromagnetic injector, a magnetic field should be applied to change
its spin direction. While using magnetic double barrier resonant
diode, it can be achieved electrically. 
}

\item {
    Spin injection from traditional ferromagnetic metal
  through a Schottky barrier into GaAs or AlGaAs:} 
By constructing a Schottky barrier with narrow depletion regime at the
interface of the ferromagnetic metal and semiconductor, spin injection
efficiency is greatly improved
\cite{kawaharazuka:3492,ploog:7256,PhysRevB.71.121301,
hickey:193204,hanbicki:1240,hanbicki:082507}.
Injected spin polarization of 30\% from Fe into GaAs 
at room temperature was reported \cite{hanbicki:1240}. 
\item
{Spin injection from traditional ferromagnetic metal through an
insulator layer into GaAs or AlGaAs: }
By inserting an insulator tunneling layer,
the injected spin polarization can be further increased. 
The tunneling layer improves the surface condition by preventing the
magnetic atoms from diffusing into semiconductor and thus reduces the
spin-flip scattering at the interface. Moreover, this tunneling layer
can be regarded as a spin-selective contact with low conductance and
hence reduces the conductance mismatch 
\cite{hershfield_97,schmidt_00,PhysRevB.64.184420,rashba_02,Fabianbook}.
Typical tunneling layers are composed of 
AlO$_x$ \cite{manago:694,motsnyi:265,erve:4334,oh:043515}
and MgO  
\cite{PhysRevLett.94.056601,Parkin2004,wang:052901,salis:262503,
JJAP.46.L4,lu:152102,truong:141109,hovel:021117,APEX.2.023006}, 
although GaO$_x$ has also been used to improve the charge injection
efficiency \cite{APEX.2.083003}.
Using AlO$_x$ as tunneling layer
\cite{manago:694,motsnyi:265,erve:4334,oh:043515}, the injection
efficiency can achieve 40\% when the temperature is lower than 180~K
\cite{erve:4334,oh:043515}. 
MgO tunneling layer is even more efficient in improving the spin
injection. 
Due to the band structures and symmetry, tunneling from Fe, Co and
their alloy through MgO barrier is strongly spin selective. The
conductance of majority spin is orders of magnitude higher than that
of minority spin
\cite{PhysRevLett.85.1088,PhysRevB.63.054416,PhysRevB.63.220403,
PhysRevB.70.172407,Parkin2004}. In this case, the injected spin
polarization is determined by the tunneling spin polarization. 
Injected spin polarization of 70-80\% 
from CoFe electrode through MgO layer into GaAs 
at room temperature was reported 
\cite{PhysRevLett.94.056601,salis:262503}. 
%
\item {Spin injection from Heusler alloys into GaAs or AlGaAs:}
In additional to the spin injection from traditional ferromagnetic
metal, there are also experiments on using Heusler alloys (some of
them have been predicted to be half-metals
\cite{PhysRevLett.50.2024,PhysRevB.70.205114}) as 
spin injector in hoping that both high spin and charge injection
efficiencies can be realized
\cite{dong:102107,kawagishi:07A703,hickey:232101, 
ramsteiner:121303,damsgaard:124502}. Over 50\% injection efficiency
from Heusler alloy Co$_2$FeSi into GaAs up to 100~K 
has been demonstrated 
\cite{ramsteiner:121303}. 
\end{itemize}

The electroluminescence
technique has also been used to demonstrate that spin
polarized carriers can be injected and sustain the polarization 
over a microscopic distance by constructing the 
light emitting diode structure away from
the surface of ferromagnetic and nonmagnetic conductors
\cite{fiederling_99,ohno_99,kohda_01,zhu_01,
hanbicki_02,motsnyi_02,ramsteiner_02,kohda_06,ciorga_09}
or away from optical injection position \cite{hagele:1580}.
Considering the optical absorption of the semiconductor, the photon
emitted at position away from the surface should be weighted as
$\exp[-\alpha(\lambda) x]$. By further combined with the fact that
absorption coefficient is wave-length dependent, experimental
measurement of wave-length dependence of the electroluminescence
polarization was used to 
estimate the spin injection length \cite{dzhioev_97}.

\subsubsection{Spin imaging}

The spin detection using the Faraday/Kerr effect measures the
Faraday/Kerr rotation angle of transmitted/reflected linearly
polarized photons.  These angles 
are proportional to the spin
component (magnetization) along the direction defined by the
propagation of probe light beam \cite{PhysRevLett.72.717}, 
and thus provide direct information of the magnetization along this
direction. Usually the
probe light is nearly normal to the sample surface, therefore only the
magnetization normal to the surface can be measured by this technique
directly. The information of spin coherence can be obtained by
applying an in-plane magnetic field to rotate the in-plane spin
component to the normal direction. 
Recently developed tomographic Kerr rotation technique further enables
the measurement of the spin coherence at any direction
\cite{kosaka_08,nature.457.702}. 

\begin{figure}[htbp]
  \begin{minipage}{0.435\linewidth}
  \centering
  \includegraphics[width=6cm]{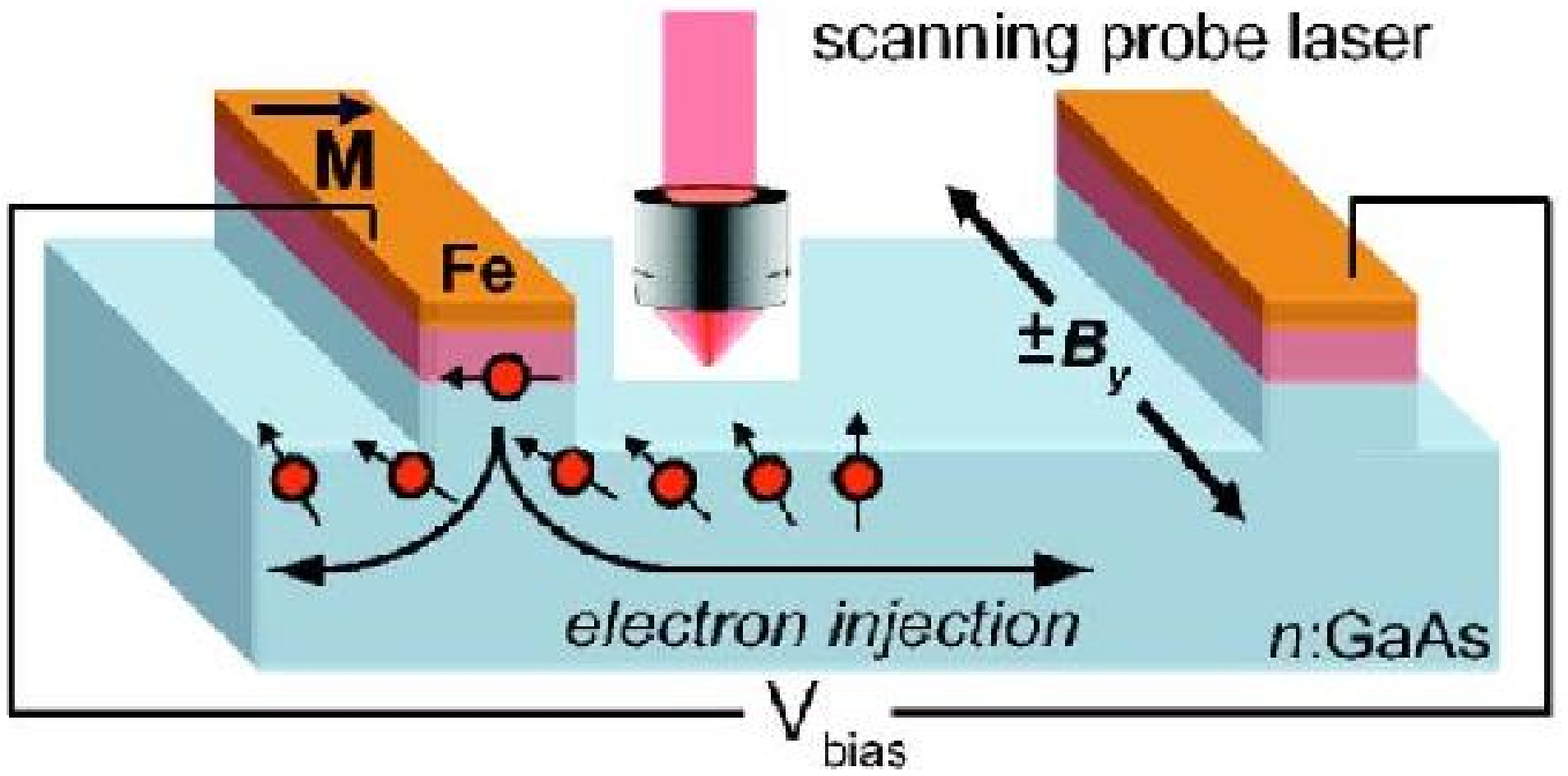}
  \includegraphics[width=6cm]{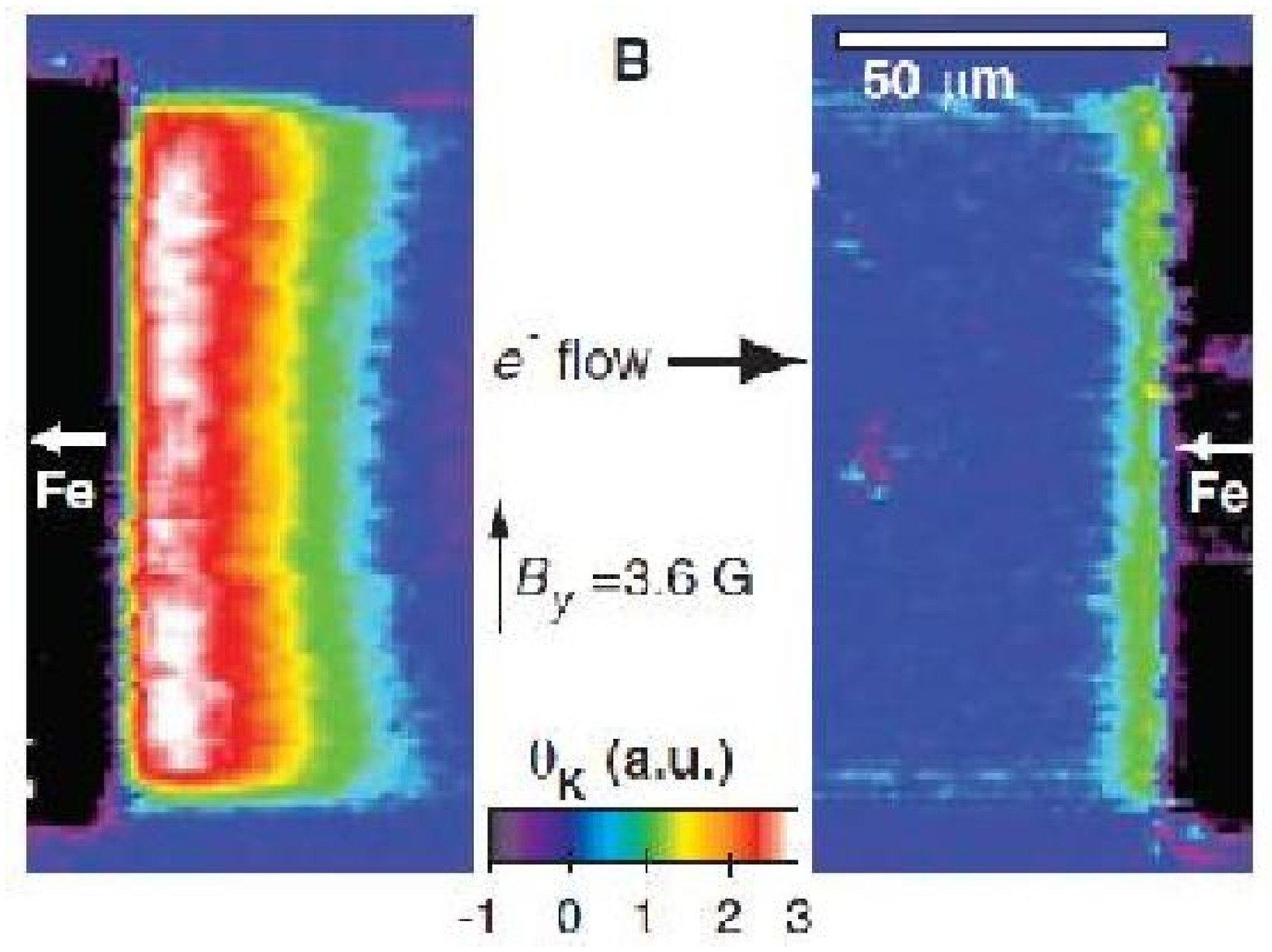}
  \end{minipage}
  \hspace{1cm}
  \begin{minipage}{0.435\linewidth}
  \centering
  \includegraphics[width=6cm]{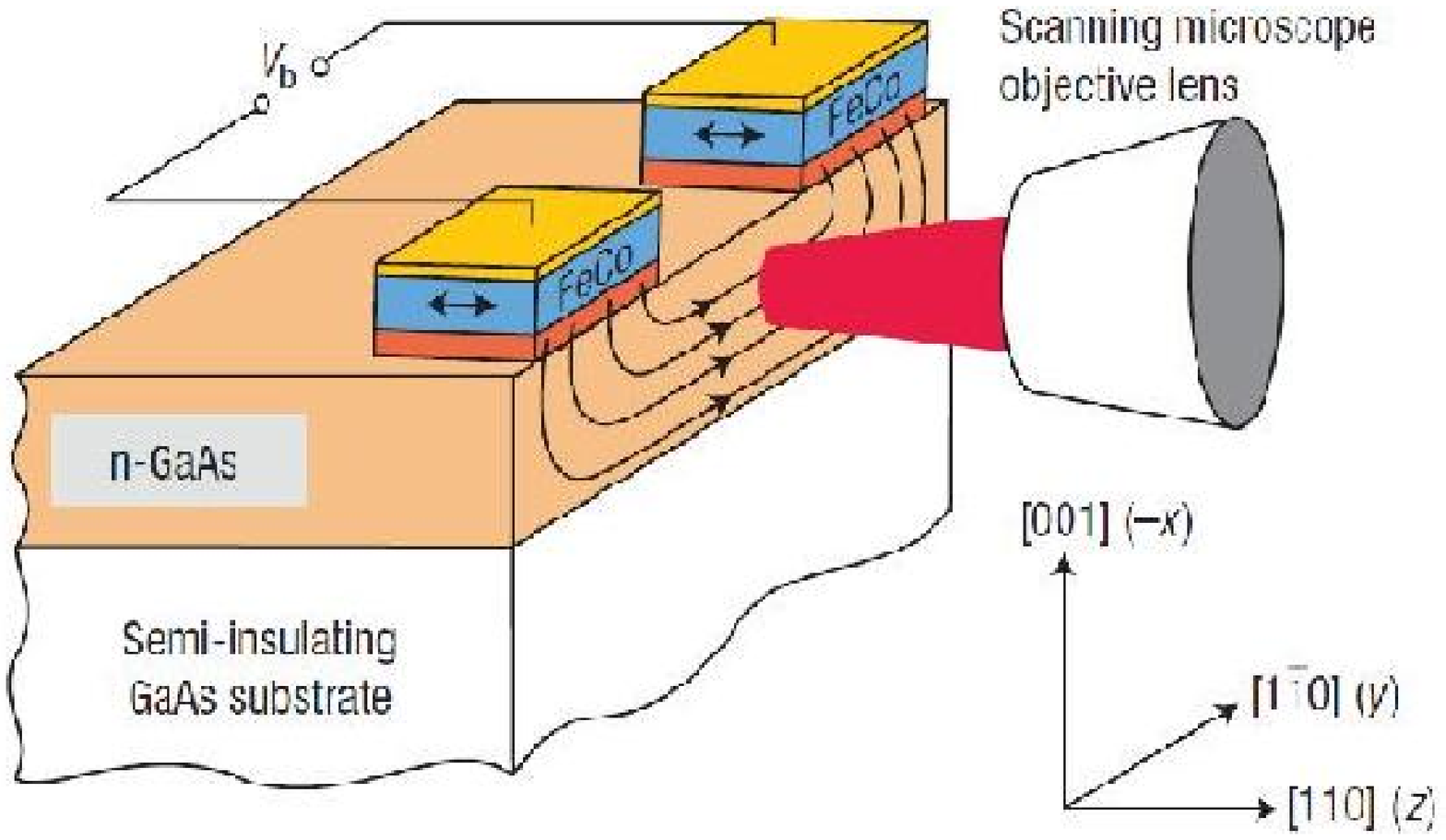}
  \includegraphics[width=6cm]{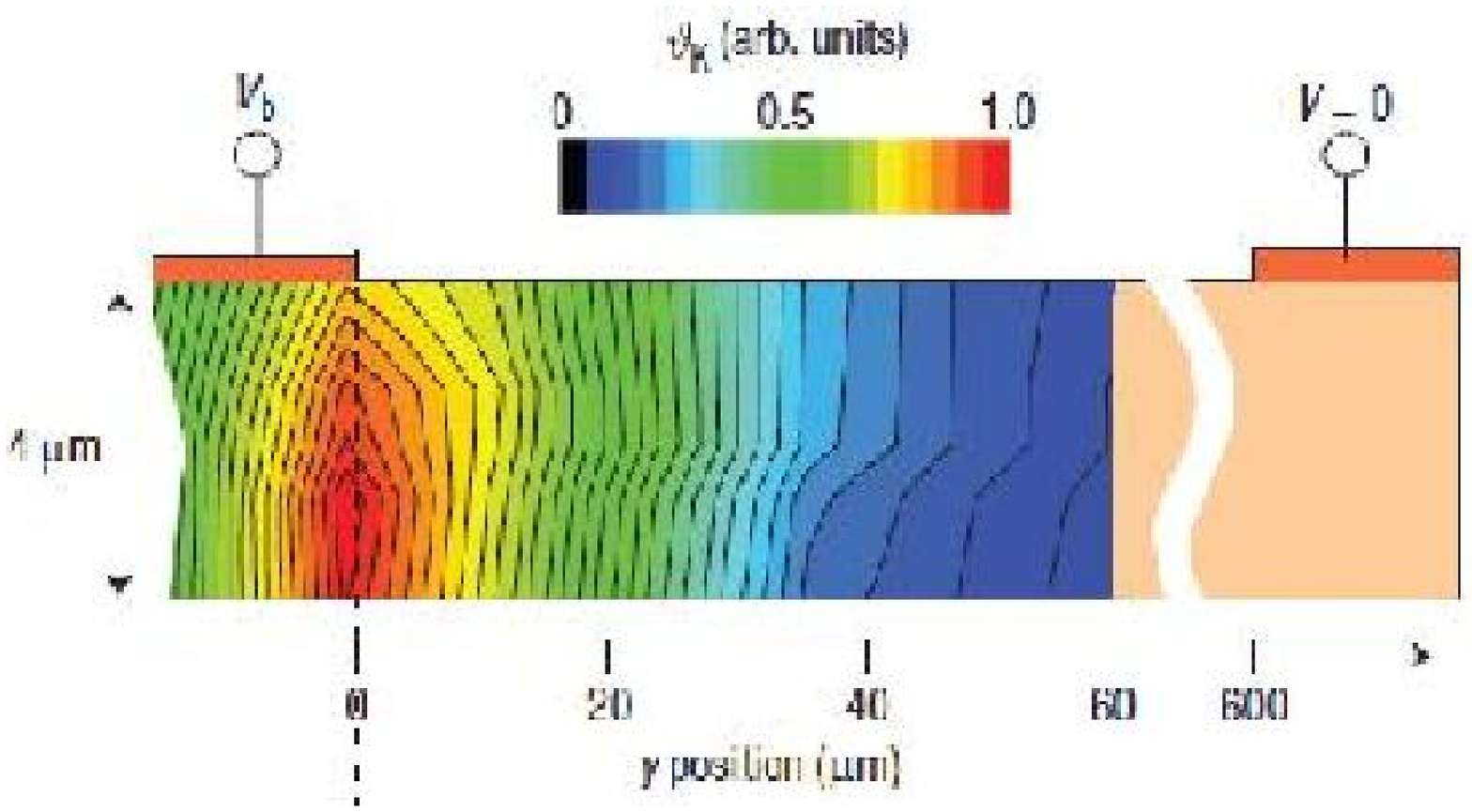}
  \end{minipage}
  \caption{ Schematic experiment setup for spin image of
    electrical spin injection. 
  Left: Surface scanning, an in-plane magnetic field is applied to
  rotate the injected spin to out of plane for Kerr rotation
  measurement. From Crooker et al. \cite{crooker_07} ;  
  Right: Side-surface scanning. From Kotissek et al. \cite{kotissek_07}.}
  \label{fig:trans:imaging}
\end{figure}

By moving the focus of the probe light and performing the measurement
at different time, one obtains the space
and temporal resolved information of the spin polarization, and thus images 
the transport of spin polarization. 
Typical schemes of experiment setups and results
for spin imaging are shown in
Fig.~\ref{fig:trans:imaging}. 
In the experiments, spin polarized carriers are usually injected
from ferromagnetic films into the
semiconductor. Depending on the experiment setup, 
the probe laser can be focused on the top surface of the semiconductor 
(left side of Fig.~\ref{fig:trans:imaging})
\cite{kikkawa_99,crooker_05,PhysRevLett.94.236601,hruska:075306,%
furis:102102,furis_07,crooker_07}
or on the side-surface
(right side of Fig.~\ref{fig:trans:imaging})
\cite{kotissek_07}.
Since the easy axis of the ferromagnetic film is usually parallel to
the film plane, an in-plane magnetic field is required to rotate the
injected spin to the normal of the surface
\cite{crooker_05,PhysRevLett.94.236601,hruska:075306,%
furis:102102,furis_07,crooker_07}.  
In the side surface scanning experiments, the easy axis of the
ferromagnetic film can be arranged to be parallel to the probe laser
and no magnetic field is required \cite{kotissek_07}. The spin
polarized electrons in ferromagnetic electrode are 
driven through interface into
the semiconductors at low temperature. The polarization of the
injected spin is estimated to be a few percents. Spin transport over
macroscopic distance ($>1\mu$m) has been clearly demonstrated by these
experiments 
\cite{crooker_05,PhysRevLett.94.236601,hruska:075306,
furis:102102,furis_07,crooker_07,kotissek_07}.
The spin image technique has also been adopted to study the spin
accumulations caused by the extrinsic spin Hall effect 
\cite{dyakonov_71,Dyakonov1971459,PhysRevLett.83.1834,
PhysRevLett.85.393,Kato12102004,sih_05,schliemann_06,
rashba_book_07,0953-8984-20-8-085209,dyakonov_08,
dyakonov:70360R,0953-8984-21-25-253202,vignale_09}.

\subsubsection{Electrical spin detection} 

There are also many works on all electrical spin
injection/transport/detection. The simplest electrical spin detection
method is to utilize the magnetoresistive effect by driving the spin
polarized carriers from one ferromagnetic electrode on one side 
across the
interface to semiconductor and extract the spin using another ferromagnetic 
electrode on the other side as  the Datta-Das transistor. 
Due to spin-valve effect, the resistances should be different when the
spin momentum of left electrode is parallel and anti-parallel to that
of the right electrode if the spin polarization injected from the left
electrode survives at the right electrode. 
This effect is quantitatively characterized by magnetoresistance,
the relative change of the resistance when the magnetic momentums change
from parallel to anti-parallel alignments. 
Early works using this simple 
ferromagnetic/nonmagnetic/ferromagnetic
sandwich structure indeed showed
that the resistance does have the hysteresis loop when an magnetic
field is applied to change the direction of the magnetic momentum on
the right electrode \cite{lee_99,hammar_99,filip_00}.
Due to conductance mismatch, magnetoresistance of the 
ferromagnetic/semiconductor/ferromagnetic structure 
is not very large, usually less than $1\%$ for ohmic 
ferromagnetic metal/semiconductor 
contact \cite{lee_99,hammar_99,filip_00}. For the system suffered less
from the conductance mismatch,  magnetoresistance of $8.2\%$ in MnAs/GaAs/Ga:MnAs
structure has been archived at low temperature \cite{hai_08}. 
However, since magnetoresistance in this structure is usually low and the spin
polarized current flows through ferromagnetic electrode, the spin
accumulation signal  
can be masked by other magnetoresistance effects 
such as the anisotropic magnetoresistance of the magnetic electrode 
and local Hall effect caused
by the fringe magnetic field near the contact 
\cite{monzon_97,monzon_99,monzon_99b,monzon_00,tang_00}.
These direct measurements of spin valve effect in the sandwich
structure are not decisive.  

A more sophisticated all electrical spin detection is to use non-local
spin-valve effect \cite{johnson_93,johnson_93b,jedema_01,%
jedema_03,erve:212109,lou_07,staa_08,ciorga_09}, first proposed by Johnson
based on the Silsbee-Johnson spin charge coupling
\cite{silsbee_80,johnson_85,johnson_93,johnson_93b}. 
The schematic of non-local spin-valve 
setup used in the experiment by Lou {et al.} is represented in
Fig.~\ref{fig:trans:nonlocal_spin_valve}. 
In the experiment different contacts
are grown on top of a GaAs layer. 
The spin polarized electrons are driven to form the spin polarized
current flow between contacts 1 and 3, while non-local voltage between
contacts 4 and 5 is measured. 
Due to Silsbee-Johnson spin charge coupling, if there is a spin
accumulation at the interface of GaAs layer and contact 4,  
a change in the non-local voltage can be detected when the direction of
magnetization in contact 4 changes. 
This non-local geometry separates the injection
and detection paths and thus avoids or reduces the spurious effects
caused by the other magnetoresistance effects. Using this technique,
clearer signals of spin accumulation were demonstrated through 
hysteresis loop behavior in the resistance change with the magnetic
field \cite{jedema_01,jedema_03,erve:212109,lou_07,staa_08,yang2008,ciorga_09}. 
More recently, Koo {et al.} employed this method to
  demonstrate the oscillatory channel conductance due to 
  gate voltage controlled spin precession 
  \cite{HyunCheolKoo09182009}. 
  This experiment is of
  particular interest since it is claimed to be the first experimental
  realization of electric spin injection, spin detection and
  coherent spin manipulation in one device, although there are some
  controversies on the claim
  \cite{bandyopadhyay09,zainuddin10,agnihotri10}. 

\begin{figure}
  \centering
  \includegraphics[height=7cm]{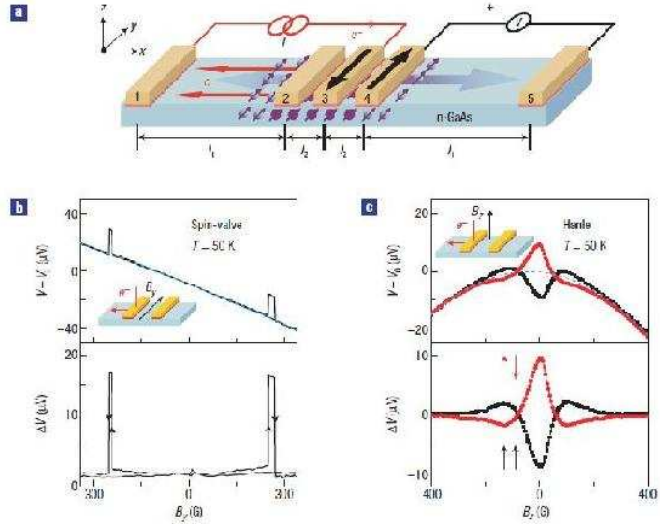}
  \caption{
    (a) A schematic diagram of the non-local experiment (not to
    scale). The five 
    contacts have magnetic easy axes along $y$-axis.
    The large arrows indicate the magnetizations of the source and
    detector.  Electrons are injected along the path shown in red. The
    injected spins (purple) diffuse in either direction from contact
    3. The non-local voltage is detected at contact 4.
    (b) Non-local voltage, $V_{4,5}$, versus in-plane magnetic field,
    $B_y$ , (swept in both directions) for sample A at a current
    $I_{1,3}$=1.0~mA at T=50~K.  
    The raw data are shown in the upper panel, the lower panel shows the data
    with this background subtracted. 
    (c) Non-local voltage, $V_{4,5}$, versus perpendicular magnetic field, $B_z$,
    for the same contacts and bias conditions as
    in (b). The data in the lower panel have the background
    subtracted. The data shown in black are 
    obtained with the magnetizations of contacts 3 and 4 parallel, and the
    data shown in red are obtained in the antiparallel configuration. 
    From Lou et al. \cite{lou_07}.}
  \label{fig:trans:nonlocal_spin_valve}
\end{figure}

\subsubsection{Spin transport in silicon}

\begin{figure}[htbp]
  \begin{minipage}{0.435\linewidth}
  \centering
  \includegraphics[width=6cm]{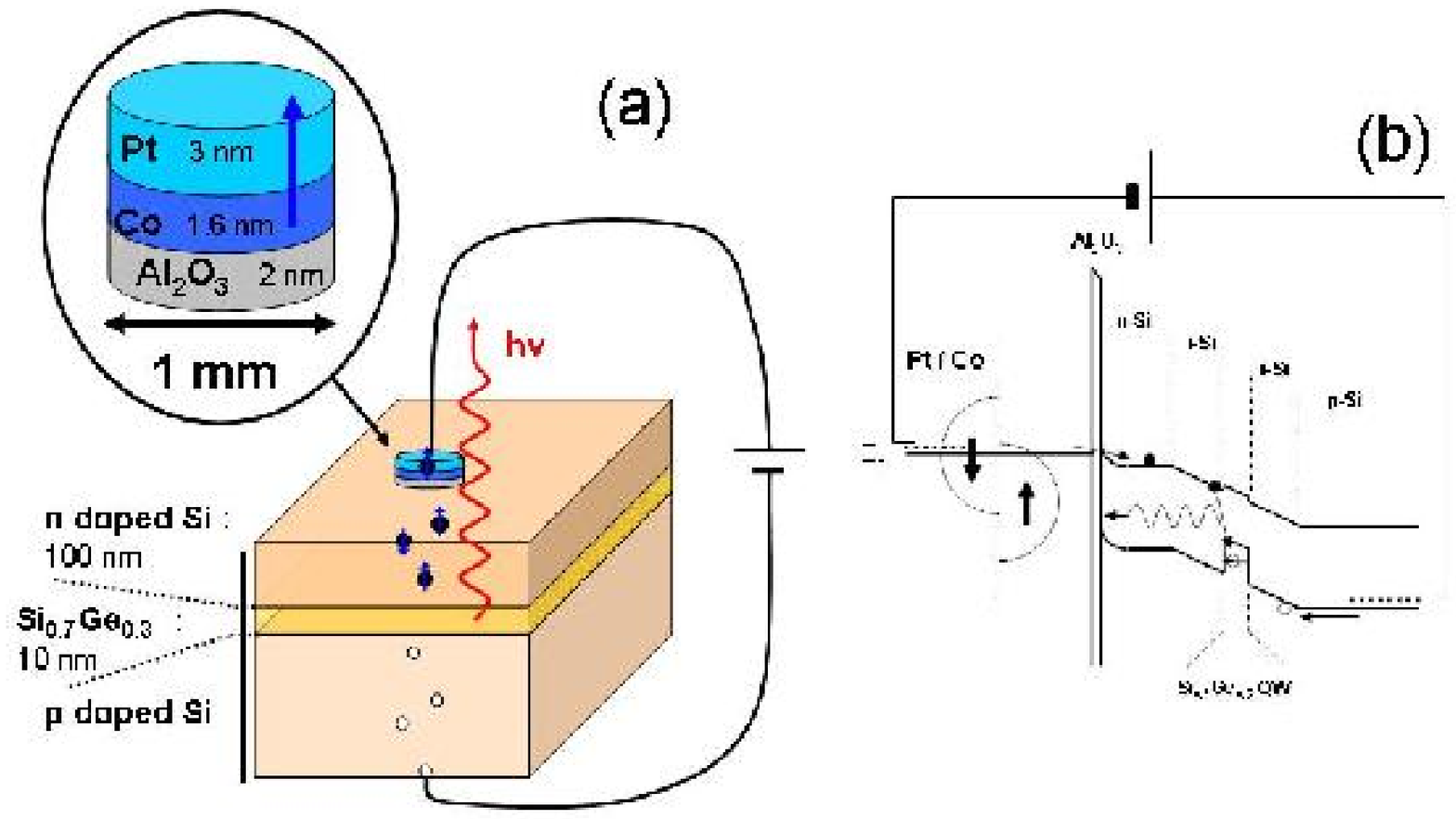}
  \caption{ 
    (a) Schematic device structure of the SiGe spin light-emitting
    diode with the Co/Pt
    ferromagnetic top electrode and (b) associated simplified band diagram
    under bias in the injection regime. 
    From Grenet et al. \cite{grenet:032502}.}
  \label{fig:trans:si_el1}    
  \end{minipage}\hspace{1cm}
  \begin{minipage}{0.435\linewidth}
    \includegraphics[width=6cm]{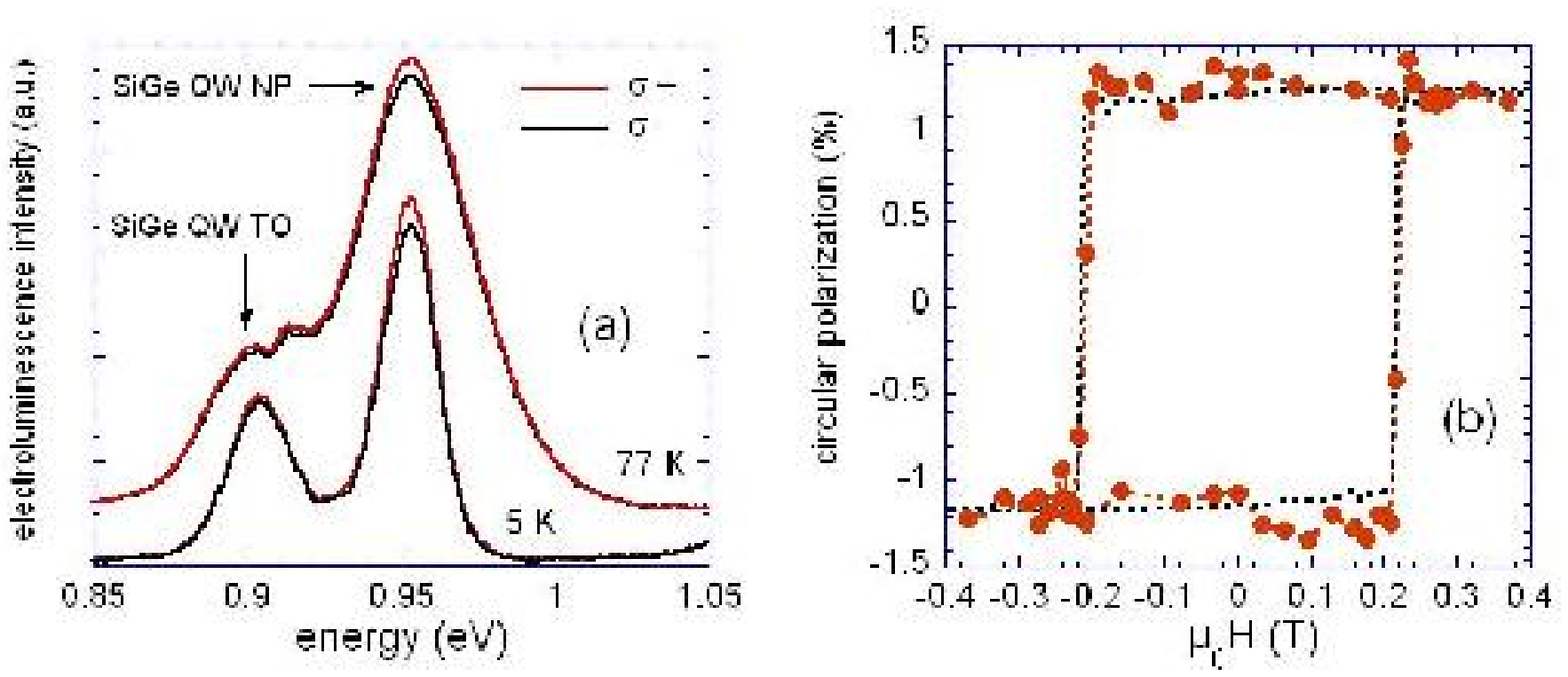}
  \caption{ 
    (a) Electroluminescence spectra recorded at 5~K and 77~K for a
    saturated remanent state of the ferromagnetic Co/Pt contact. The
    applied voltage and current are respectively around 10~V and
    10~mA. The no-phonon (NP) peak is clearly circularly polarized
    indicating an efficient spin injection into the silicon top
    layer. The transverse optical (TO) phonon replica is 
    also slightly circularly polarized. (b) Electroluminescence
    circular polarization
    recorded at the maximum of the non-phonon line for different
    remanent states of the Co/Pt layer (red dots). 
    From Grenet et al.~\cite{grenet:032502}.
  }
  \label{fig:trans:si_el2}
  \end{minipage}
\end{figure}

Spin related phenomenons in silicon has also attracted much attention 
recently because of the long spin life time in silicon and the
compatibility to the current industrial semiconductor technologies 
\cite{PhysRevB.59.13242,Jantsch2002504,PhysRevB.66.195315,
PhysRevB.69.115333,PhysRevB.71.075315,zutic_06,zutic_06b,
Appelbaum:447.295,appelbaum:262501,huang_07,huang_07a,
huang:072501,huang_07c,jansen_07,Jonker:nphys673,huang_08,
li_08,jang_08,tashpour_09,sousa_09,jang_09,grenet:032502,ando_09}. A major
difficulty for studying spin related properties in silicon is the spin
generation and detection. 
Optical spin generation/detection is not 
effective since silicon is an indirect-band semiconductor
\cite{zuticrmp,zutic_06}.  
Early studies on the Faraday/Kerr effects in silicon were limited in very
narrow frequency ranges in microwave or infrared region
\cite{furdyna_60,furdyna_61,palik_70}. Although the limitation was
lifted by using terahertz time domain spectroscopy
\cite{morikawa_06,ikebe_08}, which covers a 
wider frequency range, spin detection using  the Faraday/Kerr rotation is
yet to be conducted. 
There are a few experiments using electroluminescence to obtain the
information of the spin polarization in silicon 
\cite{Jonker:nphys673,grenet:032502}. 
By constructing surface-emitting light emitting diodes with n-i-p
silicon layers, Jonker {et al.} demonstrated the electrical spin
injection from Fe through an Al$_2$O$_3$ layer into $n$-type
silicon. The injected spin polarization under 3~T magnetic field was
estimated to be 30\% at 5~K, with significant spin polarization
extending to at least 125~K.  
\cite{Jonker:nphys673}. 
Using a fully strained Si$_{0.7}$Ge$_{0.3}$ quantum well to replace
the intrinsic silicon layer, Grenet {et al.} demonstrated the
electrical spin 
injection from CoFe into $n$-type silicon at zero magnetic
field \cite{grenet:032502}. Their experiment setup and results  
are shown in Figs.~\ref{fig:trans:si_el1} and
~\ref{fig:trans:si_el2}. 
The advantage of the strained SiGe quantum well over bulk silicon is
the appearance of the direct optical transition through the so called
non-phonon transition \cite{PhysRevLett.58.729,eberl_87,PhysRevB.40.5683}.
Due to the non-phonon transition and its phonon replica 
\cite{PhysRevB.40.5683}, radiative recombination in SiGe quantum well
is much faster than that in silicon. The radiative relaxation time in
SiGe quantum well is about tens of nanoseconds \cite{grenet:032502} 
compared to hundreds of microseconds in silicon 
\cite{Jonker:nphys673,Thewalt1982573,yubook}. 
Moreover the presence of germanium and stress also enhances the
spin-orbit coupling and raises the circular polarization of the
emitted light from the recombination of the spin polarized electrons
with holes \cite{PhysRevB.3.2623,Aleshkin1997,Olajos1996283}. 
These factors improve the efficiency of the electroluminescence. Using
this structure to detect the electrical spin injection from CoFe 
into silicon, as shown in Fig.~\ref{fig:trans:si_el2},
circular polarization of 3\% was achieved at 5~K
and remained almost constant up to 200~K at zero magnetic field
\cite{grenet:032502}.

Most spin related experiments in silicon use electrical 
injection/detection 
\cite{Appelbaum:447.295,appelbaum:262501,huang_07,huang_07a,
huang:072501,huang_07c,jansen_07,huang_08,li_08,
jang_08,tashpour_09,sousa_09,jang_09,ando_09,APEX.2.053003}.
Robust electric spin injection through a hot-electron spin valve 
\cite{Appelbaum:447.295,appelbaum:262501,huang_07,huang_07a,
huang:072501,huang_07c,jansen_07,Jonker:nphys673,huang_08,
li_08,jang_08,grenet:032502}
or Schottky tunneling barrier contact
\cite{erve:212109,nahid_09,ando_09,uhrmann_09} 
into silicon has been proposed and realized. The
schematic electronic band diagram and experimental setup 
of spin injection and detection using the
hot-electron spin valve are shown in Fig.~\ref{fig:trans:si_band}: 
Driven by the voltage $V_E$, 
unpolarized electrons are injected from nonmagnetic emitter (Al
electrode in the original experiment) into ferromagnetic base (FM1)
through a tunneling barrier.  
FM1 base acts as a spin filter in which
electrons with majority spin pass through the Schottky barrier
into silicon ballistically, but electrons with minority spin can not
pass as they quickly lose energy in the FM1 base due to the spin
selective scattering. Under the voltage $V_{c1}$, 
electrons transport inside the silicon and arrive at the second
ferromagnetic layer (FM2). 
FM2 serves as spin
detector and makes the current $I_{c2}$ depend on the current spin
polarization by allowing only electrons with majority spin of FM2 pass
through into the second collector. 
The alignment of magnetic momentum of FM1 and FM2 can be changed by an
in-plane magnetic field. 
The success of spin injection is demonstrated by spin valve effect and
Hanle effect, shown in Fig.~\ref{fig:trans:si_exp2}. 
Spin valve effect, that is the change in the spin momentum alignments
of FM1 and FM2 results in the change in $I_{c2}$, is quantitatively
described by magnetocurrent, $(I_{c2}^{P}-I_{c2}^{AP})/I_{c2}^{AP}$,
the relative change in the current when the alignment of
magnetizations in FM1 and FM2 changes. 
2\% magnetocurrent was observed in the original
experiment \cite{Appelbaum:447.295}. By relocating the ferromagnetic
bases away from 
silicon interfaces to eliminate the ``magnetically-dead'' silicide layer
\cite{veuillen_87,tsay_99,tsay_99b}, the magnitude of magnetocurrent 
was substantially increased by more than one order of magnitude
\cite{huang:072501,huang_07c}. 
To further confirm that the change in $I_{c2}$ is indeed caused by
spin valve effect instead of other spurious effects, non-local
spin detection was also used \cite{erve:212109}. More conclusive results
showing the spin precession and spin dephasing were obtained: 
A perpendicular magnetic field, which rotates the injected spin,
changes the relative alignment of the injected spin and
spin in FM2. As a result $I_{c2}$ oscillates with the strength of
the perpendicular magnetic field 
\cite{Appelbaum:447.295,erve:212109,appelbaum:262501,huang_07,
huang_07a,huang:072501,huang_07c,jansen_07,Jonker:nphys673,
huang_08,li_08,jang_08}. 

\begin{figure}[htbp]
  \begin{minipage}{0.435\linewidth}
  \centering
  \includegraphics[width=5cm]{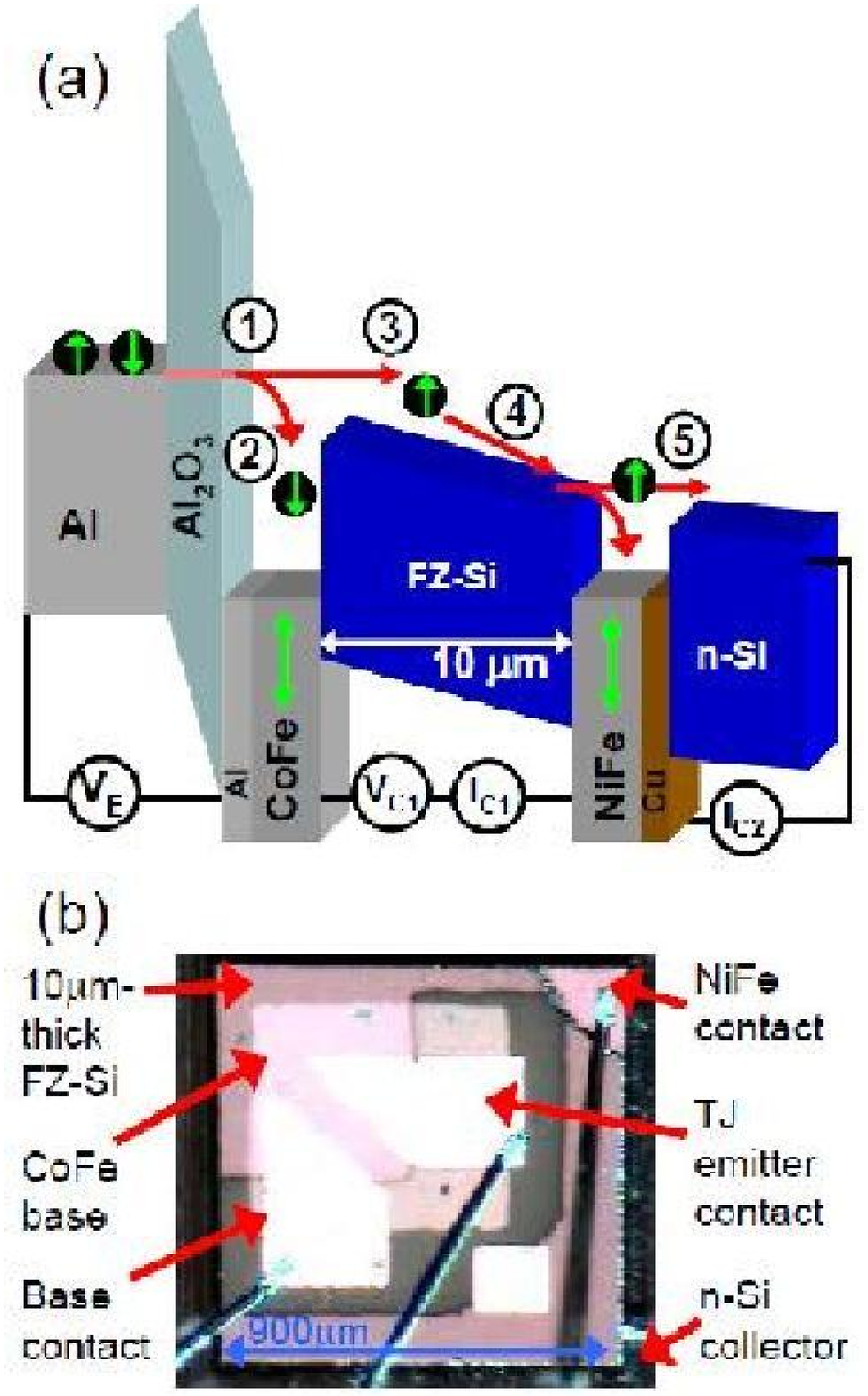}
  \caption{ Schematic electronic band diagram (a) and 
    experimental setup (b) of the hot-electron spin valve. 
    From Appelbaum et al. \cite{Appelbaum:447.295}.}
  \label{fig:trans:si_band}    
  \end{minipage}\hspace{1cm}
  \begin{minipage}{0.435\linewidth}
  \includegraphics[width=5cm]{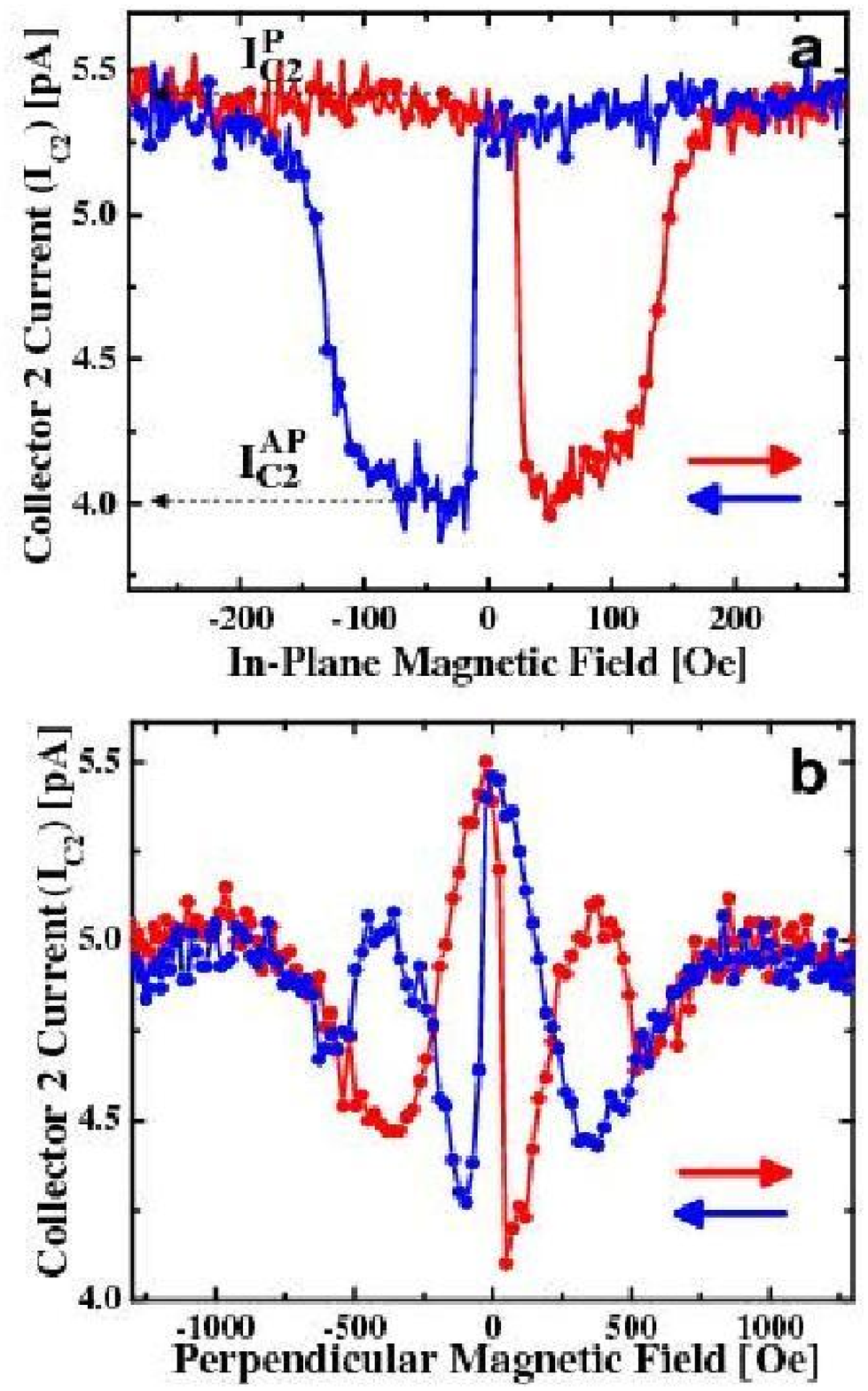}
  \caption{ (a)
    In-plane spin-valve effect for the silicon spin transport device with
    emitter tunnel junction bias $V_E=-$1.6V and $V_{C1}=$0V at 85K.
    (b) Spin precession and dephasing (Hanle effect), measured by
    applying a perpendicular magnetic field. 
    From Huang et al.~\cite{huang_07c}.
  }
  \label{fig:trans:si_exp2}
  \end{minipage}
\end{figure}

Due to the spin relaxation and the interference of the spin precession
around the magnetic field
\cite{PhysRevB.66.235109,weng:410,huang_08,zhang:075303}, 
the injected spin polarization decays with
the distance. The spin injection length decreases with the increase of
the magnetic field, and was shown to become very small when
the magnetic field becomes slightly strong \cite{Appelbaum:447.295}. 
Since $I_{c2}$ is proportional to the spin polarization at FM2,
I$_{c2}$ also decreases with the magnetic field. 
This decay, first predicted by Weng and Wu \cite{PhysRevB.66.235109}
fully microscopically, 
was explained by the Hanle effect in Ref.~\cite{Appelbaum:447.295}.

\section{Microscopic theory of spin transport}

In Section \ref{sec:drift_diffusion}, the phenomenal theory of spin
transport based on the drift-diffusion model is reviewed. In that model,
all characteristic parameters such as the mobility $\mu$,
diffusion coefficient $D$ in spin transport, and spin relaxation time
$\tau_s$ can not be determined within the framework of model, but rather need
to be experimentally measured or calculated from other theoretical
model. It is usually assumed that $\mu$ and $D$ are simply the charge
mobility and diffusion coefficient. However, there is no reason that
these assumptions should be universally true aside from the na\"ive
intuition, and the relations between the characteristic parameters of
charge transport and spin transport remain yet to be revealed. In the
presence of the spin-orbit coupling, the charge transport and spin
transport are coupled. 
A fully microscopic theory is essential to understand this coupled
transport. 
In this section, we will review spin transport inside the
semiconductor from the fully microscopic point of view. 
The main results reviewed in this section are also based on
the kinetic spin Bloch equation approach. The equations have been laid
out in Sec.~5, but in Sec.~5 we focuses on reviewing the results 
of the spacial uniform
system. The results reviewed in this section are focused on the spacial
nonuniform system. 
There are several related
researches based on the linear response theory
\cite{PhysRevB.70.155308,culcer_04} and kinetic 
theory\footnote{Equations of this theory are the same as the kinetic
  spin Bloch equations but without electron-electron Coulomb
  interaction.} 
\cite{halperin_04,bleibaum_06,bleibaum_06b,bleibaum_06c,
bryksin:165313,bryksin:205317,bryksin:075340,kleinert:073314}, whose
results will also be briefly addressed. 

\subsection{Kinetic spin Bloch equations with spacial gradient}

The kinetic spin Bloch equations for the spin kinetics in the presence
of spacial {inhomogeneity} are given by 
Eqs.~(\ref{eq5.3-10}-\ref{eq5.3-17}). 
The electric field $\mathbf{E}$ in these equations is 
determined from the Poisson equation [Eq.~(\ref{eq:trans:poisson})].
As pointed out in Sec.~\ref{sec5.3}, the kinetic spin Bloch equations
include all the factors important to the spin dynamics, including the
drifting driven by an electric field, diffusion due to the spacial
inhomogeneity, spin precession around the total magnetic field as
well as all the relevant scattering. 
By solving these equations, all the measurable quantities, such as
mobility, charge diffusion length, spin diffusion length, can be
obtained self-consistently without any fitting parameter. 

More importantly, from this fully microscopic approach, 
some important issues overlooked by the phenomenological 
drift-diffusion model are recovered. 
Weng and Wu
\cite{PhysRevB.66.235109} performed the first fully microscopic
investigation on spin diffusion and transport by setting up and
solving the kinetic spin Bloch equations. Unlike the previous spin
transport theories \cite{hershfield_97,schmidt_00,rashba_02,
PhysRevB.66.165301,PhysRevB.66.201202,PhysRevB.66.235302,
PhysRevB.67.014421,PhysRevLett.84.4220} where only the diagonal
elements of the density matrix
$\rho_{\sigma\sigma}(\mathbf{r},\mathbf{k},t)$ are included in the
theory, they showed that it is of crucial importance to include the
off-diagonal terms $\rho_{\sigma-\sigma}(\mathbf{r},\mathbf{k},t)$ in
studying the spin diffusion and transport \cite{weng:410,PhysRevB.66.235109}. 
They predicted spin
oscillations in GaAs quantum well along the spin diffusion in the
absence of any applied magnetic field \cite{weng:410,PhysRevB.69.125310}.
These oscillations were later observed in experiments
\cite{beck_06,PhysRevLett.94.236601}. 
The spin oscillations without any applied magnetic field are beyond the
two-current drift-diffusion model widely used in the spin transport
study \cite{hershfield_97,schmidt_00,rashba_02,
PhysRevB.66.165301,PhysRevB.66.201202,PhysRevB.66.235302,
PhysRevB.67.014421,PhysRevLett.84.4220,Fabianbook}. 
Moreover, by introducing the off-diagonal terms, Weng and Wu showed
that there is an additional inhomogeneous broadening associated with
the spacial gradient of the spin polarization
\cite{weng:410,cheng:073702,PhysRevB.66.235109}. 
As shown in Eq.~(\ref{eq5.3-12}), the coefficient of the spacial
gradient is proportional to $\nabla_{\mathbf{
  k}}\bar{\varepsilon}_{\mathbf{k}}(\mathbf{r},t)$, which is momentum
dependent.  
It does not lead to any significant 
results for charge transport aside from the diffusion in
space. However, when there is spin precession this
momentum-dependent coefficient introduces an additional
inhomogeneous broadening since the spacial spin precession is
momentum dependent
\cite{weng:410,cheng:073702,PhysRevB.66.235109}. 
This inhomogeneous broadening, combined with any spin-conserving
scattering, leads to an irreversible spin relaxation and dephasing in
the spin transport. 

In the following subsections, we will first
present some simplified cases where analytical results can be
obtained. Then we review the numerical results from the kinetic spin
Bloch equations. 

\subsection{Longitudinal spin decoherence in spin diffusion}
\label{subsec:longitudinal}

The importance of the off-diagonal terms of the density matrix in the
study of spin
transport can be qualitatively understood 
by studying the spin diffusion in GaAs quantum well
in a much simplified case where there
is no electric field, no scattering and no Hartree-Fock self-energy.
Assuming that the transport is along the $x$-direction, 
in the steady state, the kinetic spin Bloch equations
[Eqs.~(\ref{eq5.3-10}-\ref{eq5.3-17})] can be 
simplified as
\begin{equation}
{k_x\over m^\ast}{\partial \rho(x,\mathbf{k},t)\over \partial x}
+i\bigl[\mathbf{h}(\mathbf{k})\cdot{\bsigma \over 2},
\rho(x,\mathbf{k},t)\bigr]=0. 
\label{eq:trans:simplified}
\end{equation}
In order to avoid the complexity of the spin injection, it is assumed 
that the electron at $x=0$ 
is polarized along the $z$-axis, and the boundary
conditions are written as  
\begin{equation}
  \rho(0,\mathbf{k}) = \left ( 
    \begin{array}{cc}
      f^0_{\up}(\mathbf{k}) & 0\\
      0 & f^0_{\down}(\mathbf{k})
    \end{array}
    \right ),
\end{equation}
where
$f_{\sigma}(0,\mathbf{k}) =
f^0_{\sigma}(\mathbf{k})=\{\exp[(\varepsilon_k-\mu_{\sigma})/T]+1\}^{-1}$
is the Fermi distribution function with chemical potential $\mu_{\sigma}$. 
The solution to this equation in the steady state reads
\cite{cheng:073702}
\begin{equation}
  \rho(x,\mathbf{k}) = 
    e^{-i(m^{\ast}/2k_x)\mathbf{h}(\mathbf{k})\cdot\bsigma x}
    \rho(0,\mathbf{k}) 
    e^{i(m^{\ast}/2k_x)\mathbf{h}(\mathbf{k})\cdot\bsigma x}.
    \label{eq:trans:rhob}
\end{equation}
This equation describes the spacial spin precession around the
total magnetic field $\mathbf{h}(\mathbf{k})$. For electron with
momentum $\mathbf{k}$, the 
precession period is characterized by
\begin{equation}
  \label{eq:trans:Omega}
  \bomega(\mathbf{k}) = \mathbf{h}(\mathbf{k})m^{\ast}/{k_x}.
\end{equation}
$\mathbf{h}(\mathbf{k})=g\mu_B\mathbf{B}+\bOmega(\mathbf{k})$ with
$\bOmega(\mathbf{k})$ being the D'yakonov-Perel' term. 
Eq.~(\ref{eq:trans:rhob}) clearly shows the effect of the inhomogeneous
broadening in the spacial spin precession on the spin transport.  
For each electron moving along the $x$-direction with fixed velocity
$v_{k_x}=k_x/m^{\ast}$, its spin precesses with fixed spacial period
without any decay. However, electrons with different momentums
have different spacial precession periods and can have different
precession axes if the Dresselhaus and/or Rashba terms are present. 
As a result, spins
of different electrons eventually become out of phase
and cancel each other as they moving along the transport direction.
Due to this interference among electrons of different $\mathbf{k}$,
the total spin momentum decays along the diffusion 
distance. 
Comparing with the spin kinetics in the time domain, where the
inhomogeneous broadening is determined by {inhomogeneity} in the
precession frequencies and direction determined by 
$\mathbf{h}(\mathbf{k})$ 
\cite{wu:epjb.18.373,JPSJ.70.2195,PhysRevB.68.075312,
zhou:045305,lue:125314}, 
in spin transport 
it is determined by the inhomogeneous in the spacial oscillation
``frequency'' and direction given by $\bomega(\mathbf{k})$
in Eq.~(\ref{eq:trans:Omega}) 
\cite{PhysRevB.66.235109,weng:410,cheng:073702}. 
Even for a uniform magnetic field in the Voigt configuration, 
which does not induce any inhomogeneous broadening in the time domain, the
spacial precession ``frequencies'' $\bomega(\mathbf{k})$ 
are still different for different
electrons. Therefore, a spacial uniform magnetic field {\em alone} can
provide an inhomogeneous broadening in the spin transport as first
pointed out by Weng and Wu back to 2002
\cite{PhysRevB.66.235109,weng:410}.
This effect is illustrated in Fig.~\ref{fig:trans:B} where
the electron densities $N_{\sigma}$ in a GaAs quantum well subject to
a uniform in-plane magnetic field are plotted as functions of
position $x$. In order to demonstrate the effect of this new
inhomogeneous broadening, the spin-orbit
coupling is assumed to be zero. 
One can clearly see the decay in spin polarization due
to the inhomogeneous broadening caused by the difference in spacial
precession periods. 
The stronger the magnetic field is, the quicker the spin polarization
decays \cite{PhysRevB.66.235109,weng:410}.
It is noted that although the inhomogeneous broadening alone can
reduce the total spin momentum, it is just an interference effect.
There is no irreversible loss of spin
coherence since there is no scattering (hence no true dissipation) in
system. This can be seen from the fact that the incoherently summed spin
coherence $\rho$, also plotted in Fig.~\ref{fig:trans:B},
does not decay with the position. 
It should be pointed out that this
additional spin decoherence is beyond the Hanle effect since 
for the latter the spin signal does not decay when there is no
scattering according to Eq.~(\ref{eq:trans:LsEB}).  

\begin{figure}[htbp]
  \centering
  \includegraphics[width=6cm]{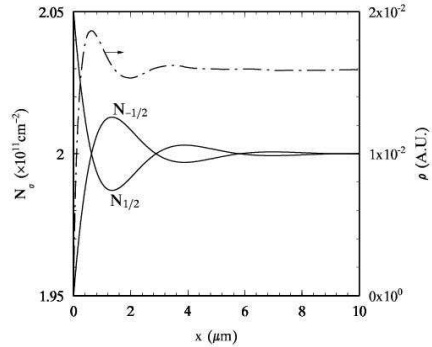}
  \caption{Electron densities of up spin and down spin (solid curves)
    and incoherently summed spin coherence $\rho$ (dashed curve) versus
    the diffusion length $x$ 
in $n$-type GaAs quantum well without spin-orbit coupling but with
 $B=1$~T.  Note the scale of the spin
    coherence is on the right side of the figure.
    From Weng and Wu \cite{PhysRevB.66.235109}.
  }
  \label{fig:trans:B}
\end{figure}

When the scattering is turned on, it provides a channel which 
combined with the inhomogeneous broadening leads to 
irreversible spin
dephasing
\cite{PhysRevB.61.2945,wu:epjb.18.373,JPSJ.70.2195,
PhysRevB.68.075312,PhysRevB.70.195318,PhysRevB.69.245320}. 
As a result both spin polarization and spin coherence decay
with the distance 
\cite{PhysRevB.66.235109,weng:410,PhysRevB.69.125310,jiang:113702,
cheng:073702,zhang:075303}. 
Similar to the spin evolution in the time domain, the scattering also
has counter effect that suppresses the inhomogeneous broadening
in $\bomega(\mathbf{k})$ given by Eq.~(\ref{eq:trans:Omega})
\cite{wu:epjb.18.373,PhysRevB.47.6807,PhysRev.154.737,
PhysRevB.4.1285,PhysRevLett.51.130,PhysRevB.32.6965,PhysRevB.44.9048}.
The spin dephasing/relaxation mechanism in spin transport has been
realized experimentally by Appelbaum {et al.} 
\cite{Appelbaum:447.295,appelbaum:262501,huang_07,huang_07a,
huang:072501,huang_07c,jansen_07,huang_08,li_08} in bulk silicon, 
where there is no D'yakonov-Perel' spin-orbit coupling
due to the center inversion symmetry. 
For spin transport in bulk silicon in the presence of a magnetic
field, the decay of the spin 
polarization should be mainly caused by the above mentioned
inhomogeneous broadening 
if other spin-relaxation mechanisms are ignored. Moreover, the
inhomogeneous broadening in this situation is particularly simple
since all electrons have same precession axis, the only difference
is the magnitude of the spacial precession frequencies
$\bomega_{\mathbf{k}}$ given by Eq.~(\ref{eq:trans:Omega}). 
Therefore bulk silicon provides an ideal platform to study this spin
relaxation/decoherence due to the additional inhomogeneous
broadening. 
As shown in Fig.~\ref{fig:trans:si_exp2}, indeed the spin injection
length decreases with the magnetic field, consistent with the
theoretical prediction 
\cite{weng:410,cheng:073702,weng:063714,zhang:075303}.

Although the full kinetic spin Bloch equations with scattering are
complicated, one can still obtain some analytical results under some
approximations 
\cite{dyakonovperelbook,weng:410,cheng:073702,weng:063714,zhang:075303}. 
Assuming that there are no applied electric field, inelastic
scattering and the Hartree-Fock term, 
the kinetic spin Bloch equations for spin
transport along the $x$-axis can be written as  
\cite{weng:410,cheng:073702,weng:063714,zhang:075303}
\be
 {\partial \rho^l(x,k,t)\over\partial t}+
{v_k\over 2}{\partial\over \partial
  x}[\rho^{l+1}(x,k,t)+\rho_k^{l-1}(x,k,t)]
=
-i\sum_{m}[\mathbf{h}^{l-m}(k)\cdot{\bsigma/2},
\rho^m(x,k,t)]
-{\rho^l(x,k,t)/\tau_k^l},
\label{eq:trans:diffusive}
\ee
where $v_k=k/m^{\ast}$ is the velocity of the electron with momentum
$\mathbf{k}$. 
$\rho^l(x,k,t)$ and $\mathbf{h}^l(k)$ are the $l$-th order of the Fourier
components of $\rho(x,\mathbf{k},t)$ and $\mathbf{h}(\mathbf{k})$ with
respect to the angle of $\mathbf{k}$ and the $x$-axis,
respectively. 
$\tau_k^l$ is the $l$-th order momentum relaxation time due to
the electron-impurity scattering. Noted that 
$1/\tau^0_k=0$ and $\tau^l_k=\tau^{-l}_k$.

In the diffusive limit where the momentum relaxation time is small,
the higher order components $\rho^{l}(x,k,t)$ with $|l|>1$ are small and
can be neglected. 
If there is no spin-orbit coupling and the applied magnetic field is
along the $y$-axis, by neglecting higher order Fourier components 
with $|l|>1$, in the steady state the kinetic spin Bloch equations can
be simplified as \cite{zhang:075303}
\begin{equation}
  \label{eq:trans:si_diffusive}
  {\partial^2 \rho^0(x,k)\over \partial x^2}=
  -2\left({\omega \over 2v_k}\right)^2
[\sigma_y,[\sigma_y,\rho^0(x,k)]]
+i{\omega\over v_k^2\tau^1_k}[\sigma_y,\rho^0(x,k)],
\end{equation}
where $\omega=g\mu_BB$ is spin precession frequency in time domain
under the uniform magnetic field $\mathbf{B}$. 
In term of the spin momentum, 
the solution to this equation for the spin injection at $x=0$ is a
damped oscillation 
\begin{equation}
\label{eq:trans:si-simplified}
S_z(x,k)=S_z(0,k)
\cos\left({x\over v_k\tau^1_k\Delta}\right)
e^{-{x\omega\Delta/v_k}}
\end{equation}
with
\begin{equation}
\label{eq:trans:si-simplified1}
\Delta=\left(\sqrt{1+{1\over (\omega\tau^1_k)^2}}-1\right)^{-1/2}.
\end{equation}

At low temperature, when the injected spin only polarized at the Fermi
level, the spin injection length $L_d^{-1}=\omega\Delta/v_f$, with
$v_f$ standing for the Fermi velocity. The injection length obtained
here decreases monotonely with the increase of magnetic field and
scattering strength. Therefore, the scattering enhances the spin
dephasing/relaxation in spin diffusion. The counter effect of the
electron-impurity scattering on the inhomogeneous broadening 
here is not significant \cite{zhang:075303}. 

\subsubsection{Spin diffusion in Si/SiGe quantum wells}

In order to gain
a deeper insight into the spin relaxation along the spin diffusion
caused by the additional inhomogeneous broadening addressed in the
previous subsection, Zhang and Wu
studied the spin transport in a symmetric silicon quantum well with even
number of monoatomic silicon layers through the kinetic spin Bloch equation
approach \cite{zhang:075303}. There is no D'yakonov-Perel'
spin-orbital coupling in this kind of quantum well
\cite{PhysRevB.69.115333}. 
Since the Rashba term is very small in the asymmetric Si/SiGe quantum well
\cite{0022-3719-17-33-015,PhysRevB.59.13242,Jantsch2002504,
PhysRevB.66.195315,PhysRevB.69.115333,PhysRevB.71.075315}, 
it was argued that their results also hold for the asymmetric Si/SiGe
quantum wells as long as the D'yakonov-Perel' term is weak enough
\cite{zhang:075303}.  


The kinetic spin Bloch equations in (001) Si/SiGe quantum well are similar
to those in GaAs quantum well \cite{zhang:075303}. 
In the equations the drifting under the
electric field, diffusion due to the spacial inhomogeneity and precession
around the magnetic field (including the contribution of the Hartree-Fock
term), as well as the spin-conserving scattering are all addressed. 
Numerical solutions of full kinetic spin Bloch equations for  spin
transport in silicon quantum well at $T=80$~K 
with different scattering mechanisms and under different magnetic
fields are shown in
Figs.~\ref{fig:trans:si_mag} and \ref{fig:trans:si_scatt}.

Figure~\ref{fig:trans:si_mag} clearly shows that the spin
injection length decreases with increasing magnetic
field, in accordance with the prediction by Weng and Wu
\cite{PhysRevB.66.235109,weng:410}. This 
result is easy to understand since 
$\bomega_{\mathbf{k}}=m^{\ast}g\mu_B\mathbf{B}/k_x$ 
is proportional to the magnetic field. Increase of the applied
magnetic field results in
the increase of
inhomogeneous broadening.

Figure~\ref{fig:trans:si_scatt} shows the effect of scattering on the
spin diffusion. 
It is seen from the figure that
adding any new scattering leads to a shorter spin diffusion length. 
This is quite different to the results in the system with the
D'yakonov-Perel' spin-orbit coupling.  
When the D'yakonov-Perel' spin-orbit coupling is present,
the scattering 
not only opens a spin dephasing/relaxation channel, 
but also has a counter effect on the 
inhomogeneous broadening
\cite{wu-review,lue:125314,weng:410,cheng:205328,cheng:073702}. 
In the strong scattering limit, adding a new scattering 
suppresses the inhomogeneous broadening and prolongs the spin diffusion
length \cite{weng:410,cheng:073702}. However, for the system without the
D'yakonov-Perel' spin-orbit coupling, the counter effect of the
scattering on the inhomogeneous broadening is marginal.
The scattering affects the spin diffusion only through the momentum
relaxation time and hence only reduces the spin injection length as
shown by
Eqs.~(\ref{eq:trans:si-simplified}) and (\ref{eq:trans:si-simplified1})
\cite{zhang:075303}. 

The electron density dependence of the spin diffusion was also
investigated 
at different temperatures. It was found that at relatively high
temperature or low density, when the electrons are nondegenerate, the
spin diffusion length and spacial precession period are insensitive to
the electron density. At low 
temperate or high density when the electrons are degenerate, the spin
diffusion length and spacial precession period increase with 
increasing density. This is understood from the Fermi wavevector
dependence of the
damping rate and precession period 
as shown in
Eqs.~(\ref{eq:trans:si-simplified}) and (\ref{eq:trans:si-simplified1}).   


\begin{figure}[htbp]
  \begin{minipage}{0.435\linewidth}
  \centering
  \includegraphics[width=6cm]{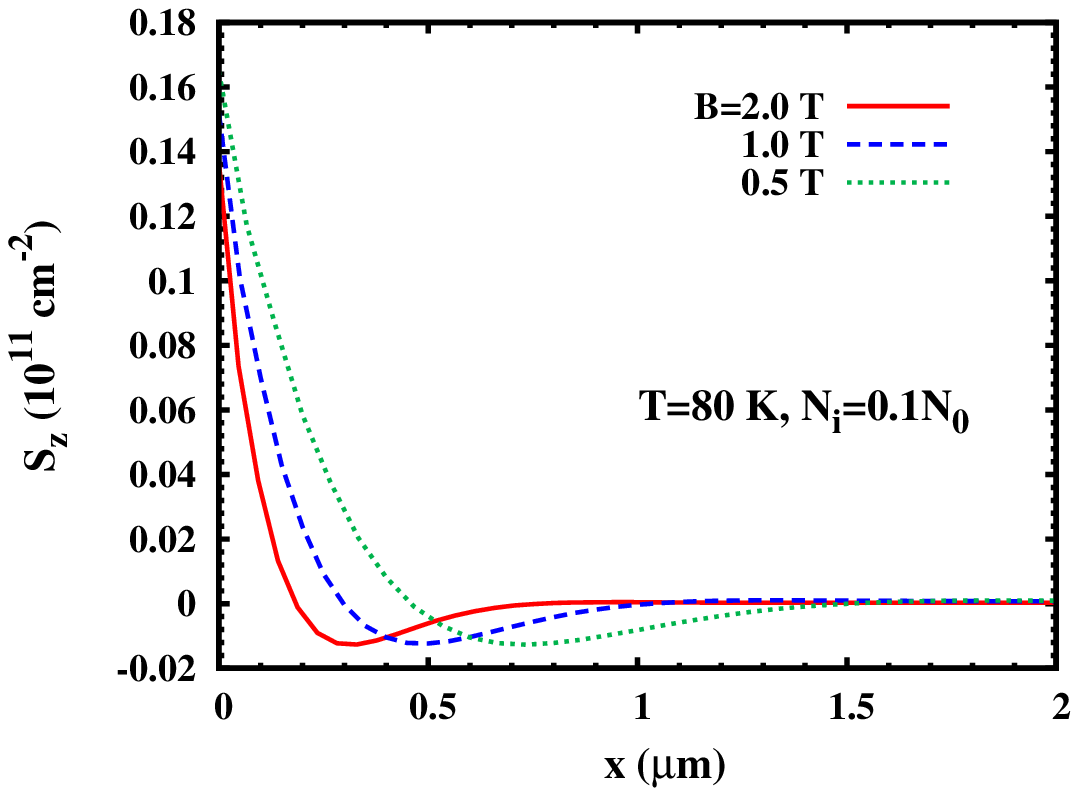}
  \caption{
     Spin polarization 
    $S_z$ {\it vs}. position $x$ in the steady state
    for intrinsic silicon quantum well under
    different magnetic field strengths. 
    Solid curve: $B=2$~T; Dashed curve: $B=1$~T;
    Dotted curve: $B=0.5$~T. $T=80$~K and $N_{i}=0.1N_0$.    
    From Zhang and Wu \cite{zhang:075303}.
  }
  \label{fig:trans:si_mag}
  \end{minipage}\hspace{1cm}
  \begin{minipage}[htbp]{0.435\linewidth}
  \centering
  \includegraphics[width=6cm]{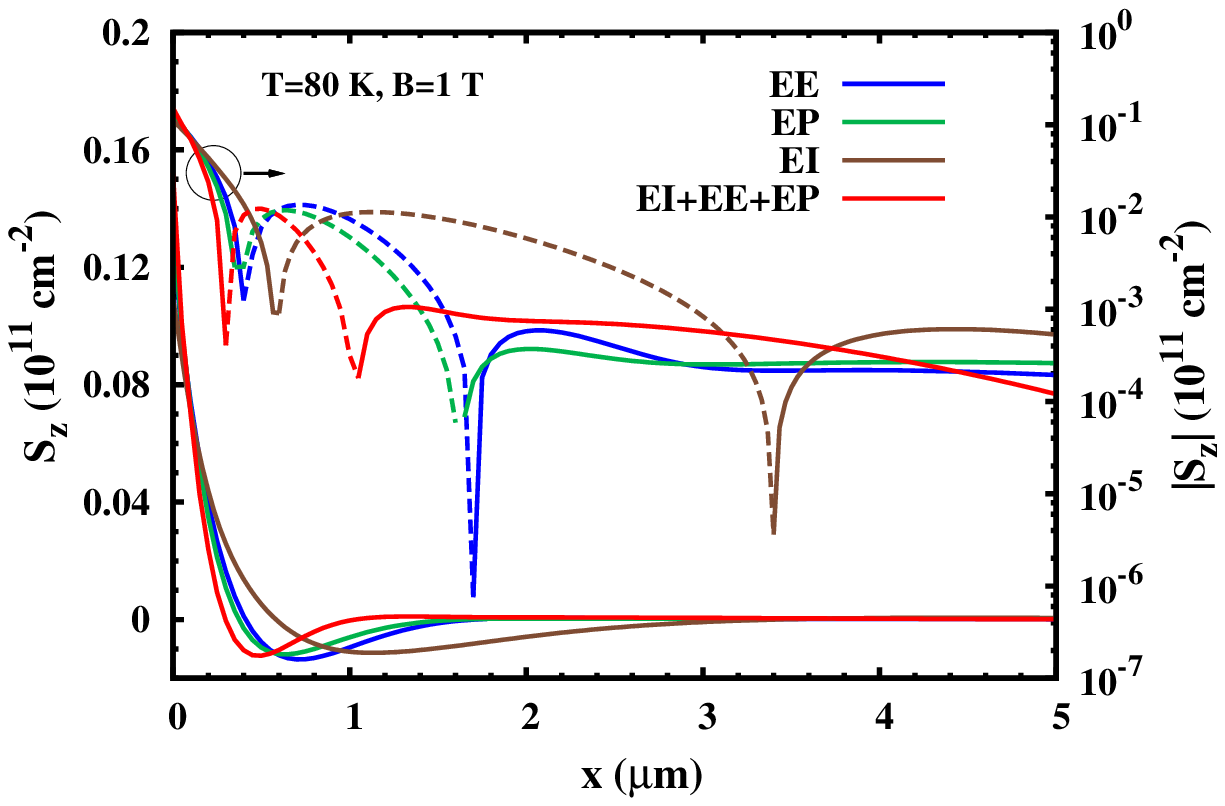}
  \caption{ The steady-state spacial distributions of spin
    signal $S_z$ in silicon quantum well
    with different scatterings included. The
    curves labeled with ``EE'', ``EP'' or ``EI'' stand
    for the calculations with  the electron-electron,
    electron-phonon or electron-impurity scattering, respectively,
    while the curve labeled with
    ``EI+EE+EP'' stands for the calculation with all the
    scatterings. 
    In order to get a clear view of the decay
    and precession of $S_z$, the
    corresponding absolute value of $S_z$ against $x$ is also plotted 
    on a log-scale (Note the scale is on the right hand of the frame).
    The dashed curves correspond to the part with  
    $S_z<0$.
    From Zhang and Wu \cite{zhang:075303}.
  }
  \label{fig:trans:si_scatt}
  \end{minipage}
\end{figure}

\subsection{Spin oscillations in the absence of magnetic field
  along the diffusion}

When the D'yakonov-Perel' term is present, the problem
becomes more complicated since in this case both the spacial precession
frequencies and axis are momentum dependent. Studies of spin
transport in the system with the spin-orbit coupling from the kinetic spin
Bloch equation approach have predicted many novel results 
\cite{PhysRevB.66.235109,weng:410,jiang:113702,%
cheng:073702,weng:063714,zhang:075303}.
Some of them have been verified experimentally 
\cite{beck_06,PhysRevLett.94.236601}.

\begin{figure}[htbp]
  \centering
  \includegraphics[width=5cm]{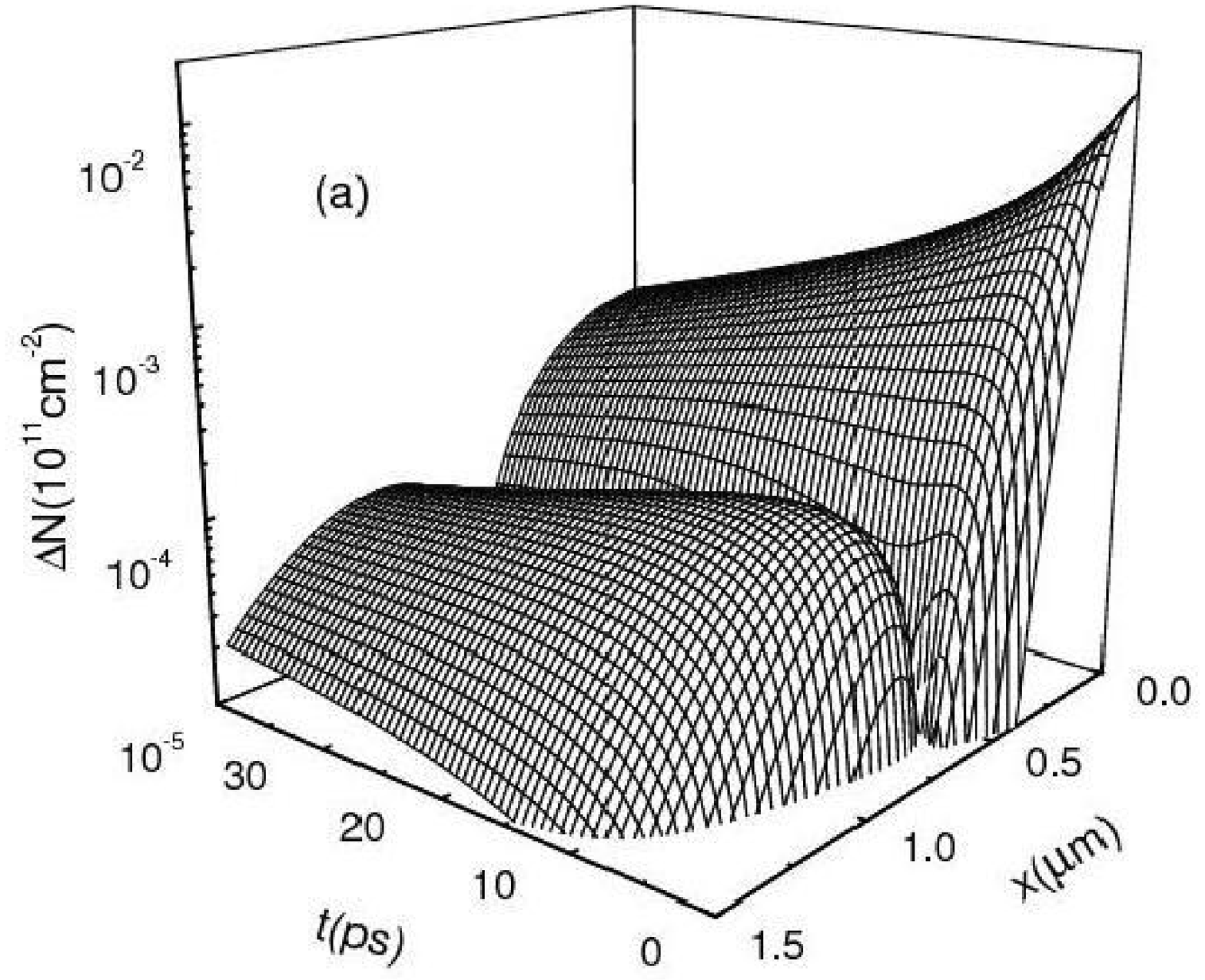}
  \includegraphics[width=6cm]{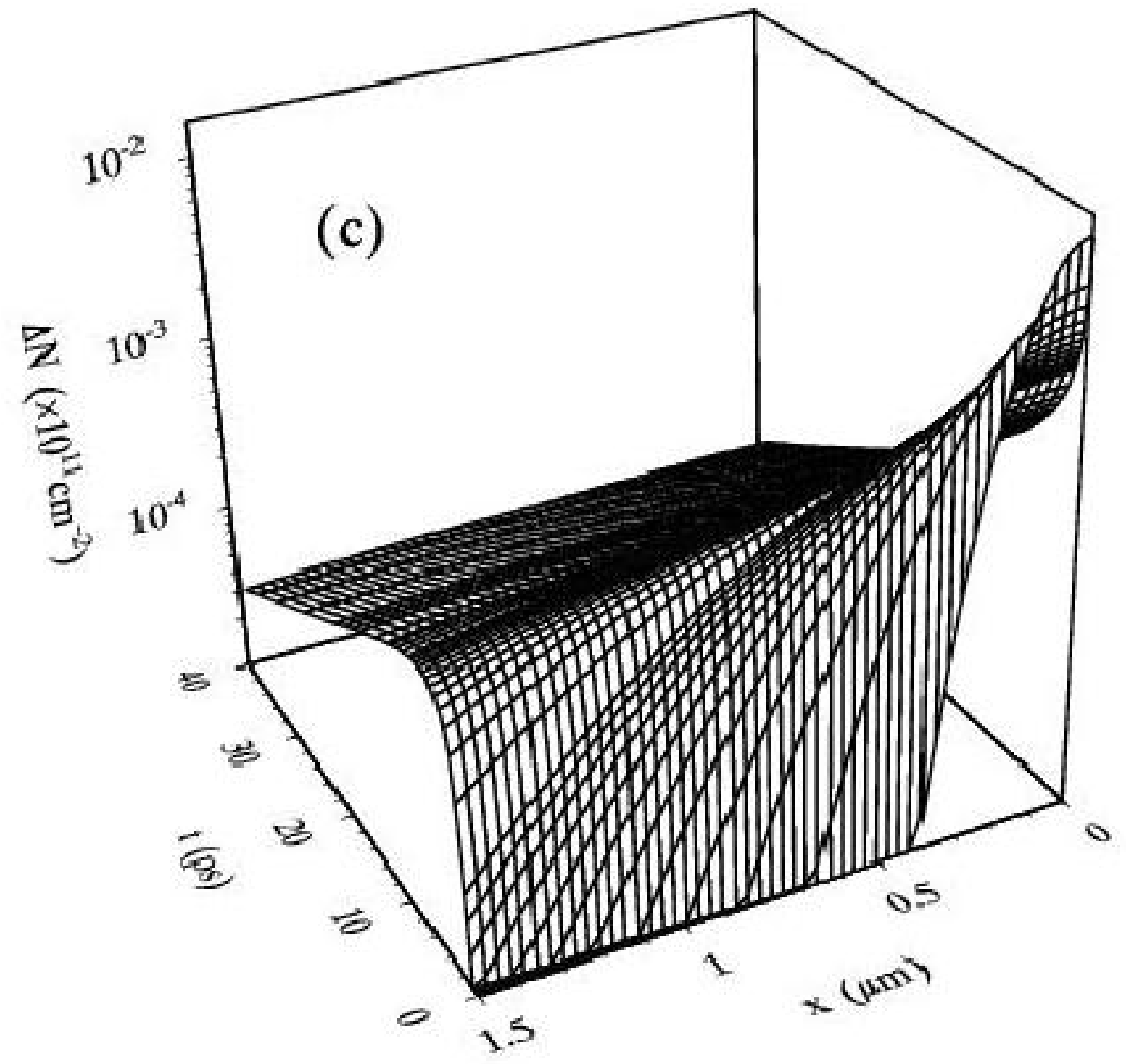}
  \caption{Theoretical predictions 
    of the absolute value of the spin
    imbalance $|\Delta N|$ 
    {\sl vs.} the position $x$ and the time $t$ for a spin pulse 
 in $n$-type GaAs (001) quantum well based on
    the kinetic spin Bloch equation approach (left) and
    drift-diffusion model (right).  The initial pulse width
    $\delta x=0.15$~$\mu$m. 
    From Weng and Wu \cite{weng:410}. 
  } 
  \label{fig:trans:pulse1}
\end{figure}

One interesting result is the spin oscillation in the absence of any 
applied magnetic
field along the spin diffusion when the Dresselhaus and/or Rashba
terms are present. In the spacial homogeneous system, spin oscillation
without any applied magnetic field can only be observed in time domain 
at very low
temperature ($T<2$~K), {i.e.}, in the weak scattering limit
\cite{PhysRevLett.89.236601}. 
In high temperature
regime where scattering is strong enough, 
the Dresselhaus and/or Rashba terms result in
a monotonically decay of spin polarization versus time in the
spacial uniform system. If one adopts simple relaxation approximation
to describe the effect of the spin-orbit coupling, one should not
expect any spin oscillation in the diffusive limit where the scattering
is strong enough. Indeed, from the drift-diffusion
equation Eq.~(\ref{eq:trans:S}), there is no spin oscillation either in
spin injection in the steady state or in transient spin transport if the
applied magnetic field is absent.  

However according to the kinetic spin Bloch equation approach, in the
spin transport the spacial spin precession is 
characterized by $\bomega_{\mathbf{k}}$ [Eq.~(\ref{eq:trans:Omega})]. 
If one only considers the Dresselhaus effective magnetic
field, the average of $\bomega_{\mathbf{k}}$ is 
$\langle \bomega_{\mathbf{k}}\rangle
=m^{\ast}\gamma(\langle k_y^2\rangle -\langle
k_z^2\rangle,0,0)$. For electrons in quantum well, this value is not
zero. Therefore, the spacial spin oscillation due to the Dresselhaus
effective magnetic field survives even at high temperature when the
scattering is strong enough.

By solving the kinetic spin Bloch equations, the spin oscillation
without any applied magnetic field was first predicted in the transient
diffusion of a spin pulse in GaAs quantum wells  
\cite{weng:410,PhysRevB.69.125310,jiang:113702}. 
A typical time evolution of spin pulse is 
shown in Fig.~\ref{fig:trans:pulse1} where the absolute value of
the spin imbalance $|\Delta N_{\sigma}(x, t)|$ 
is plotted as a function of the position $x$ along the diffusion
direction and time $t$. 
The initial state is a spin pulse of Gaussian spacial
profile, $\Delta N(x,0) = \Delta N_0\exp(-x^2/\delta x^2)$ with
$\delta x=0.15$~$\mu$m. 
For comparison, the result based on the drift-diffusion model is also
shown. 
One can see that the kinetic spin Bloch equation approach and the
drift-diffusion model give qualitatively similar results for the
evolution at the center of the pulse, where it decays monotonically 
with time as a result of spin diffusion and spin dephasing. 
However, the results from these two theories are quite different for
the evolution outside the original spin pulse. The result from the
drift-diffusion model shows that spin polarization at large $x$ first
increases with time and then decays to zero. The sign of the spin
polarization does not change throughout the space and time. On the
contrary, the kinetic spin Bloch equation theory predicts that spin
polarization oscillates with time. 
The oscillation is the combined result of the spin
precession caused by the Dresselhaus effective magnetic field and the
spin diffusion.

\begin{figure}[htbp]
  \centering 
  \includegraphics[width=5.6cm]{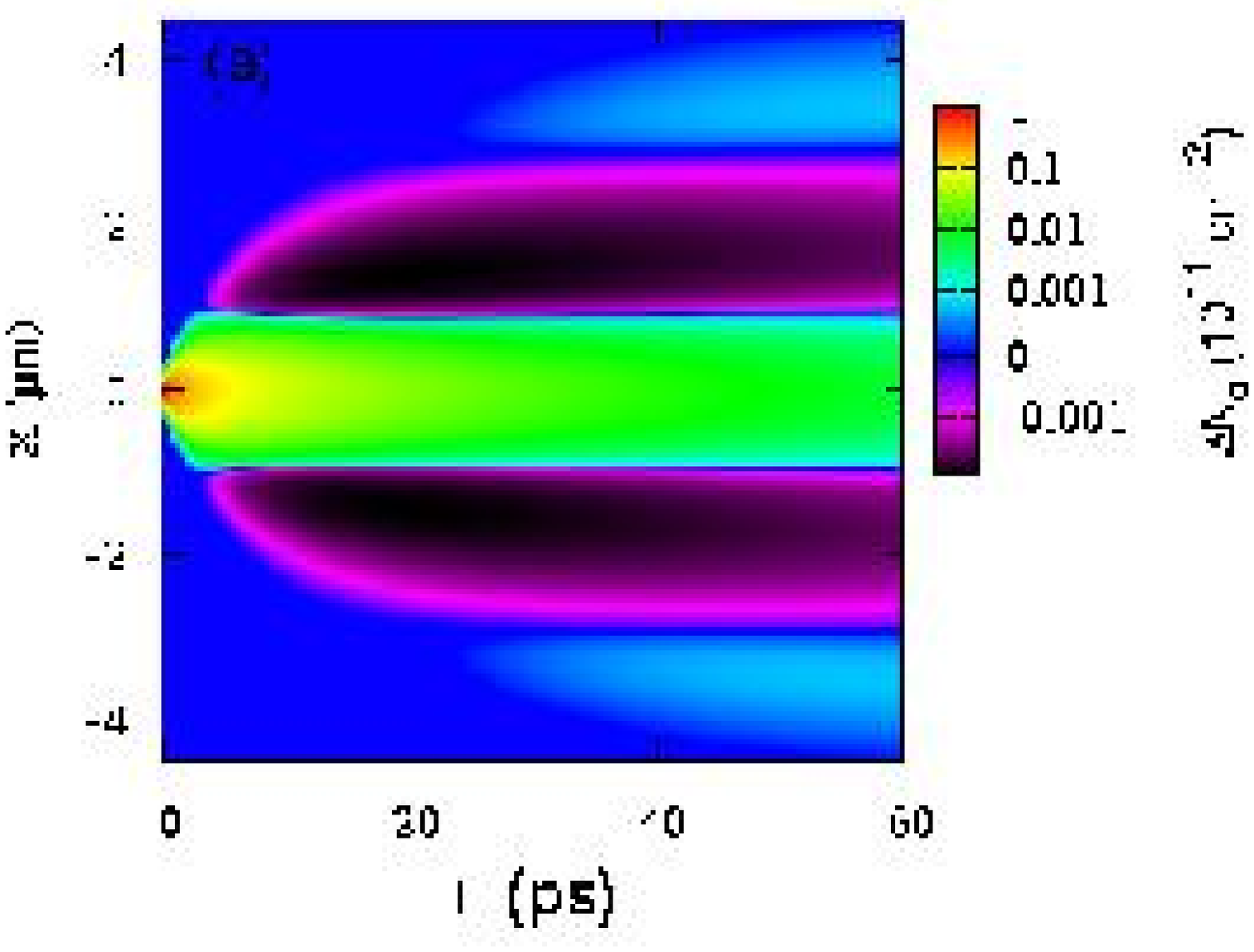}
  \includegraphics[width=5.5cm]{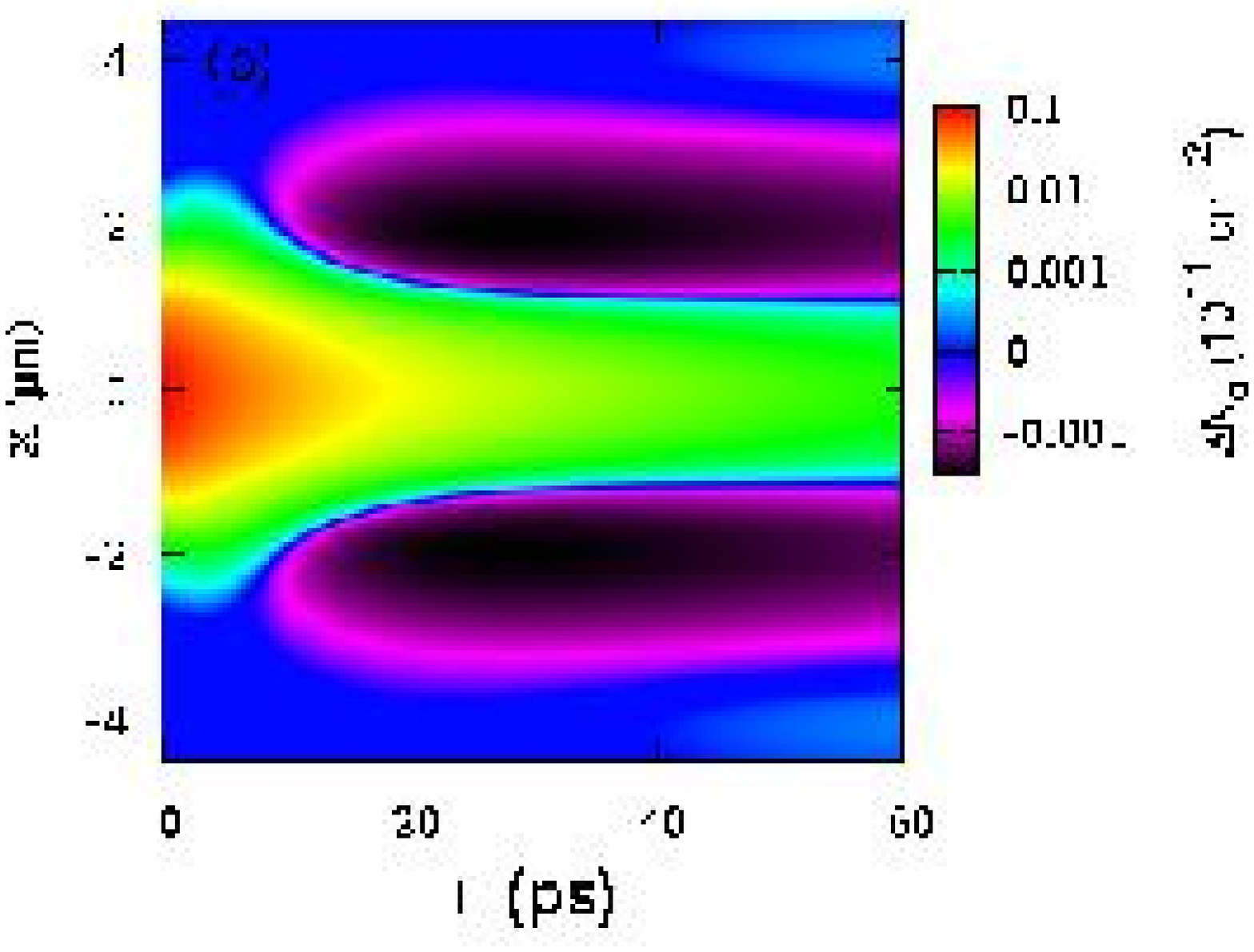}
  \caption{
     The contour plots of spin imbalance $\Delta
    N$ {\sl vs.}
    the  position $x$ and the time $t$ for different pulse widths
in $n$-type GaAs (001) quantum well. ($a$):
    $\delta x = 0.1~\mu$m; ($b$): $\delta x = 1.0~\mu$m. $T=200$\ K.    
    From Jiang et al.~\cite{jiang:113702}.    
  }  
  \label{fig:trans:pulse2}
  \end{figure}

The evolutions of spin pulse with different pulse widths are shown in
Fig.~\ref{fig:trans:pulse2}. 
Since the oscillation is a result of diffusion
and spin precession around the effective magnetic field, the system
with narrower spin pulse shows stronger spin oscillation in the
transient spin transport.
For the narrower pulse, the peak of the reversed spin polarization
appears in shorter time. Moreover, more oscillation can be observed
with narrower pulse \cite{jiang:113702}. 

The effects of the scattering and the electric field on the
oscillation without any magnetic field were also studied.
It was shown that the
oscillation is robust against the scattering.
Adding the electron-impurity
scattering slows down the diffusion but does not eliminate the
oscillation. In some regime, the electron-impurity scattering even
boosts the oscillation as it helps to sustain the spin coherence 
\cite{PhysRevB.69.125310,jiang:113702}. The
Coulomb scattering has a similar effect on the spin diffusion and
oscillation \cite{jiang:113702}.
Without any electric field, the spin signals diffuse symmetrically 
in the space around the original pulse center. When an electric field
is applied along the diffusion direction, the diffusion is no longer
symmetrical. The pulse is dragged against the electric field and
stronger opposite spin polarization appears at the side against the
electric field \cite{jiang:113702}.
}

\begin{figure}[htbp]
  \centering
  \includegraphics[width=6cm]{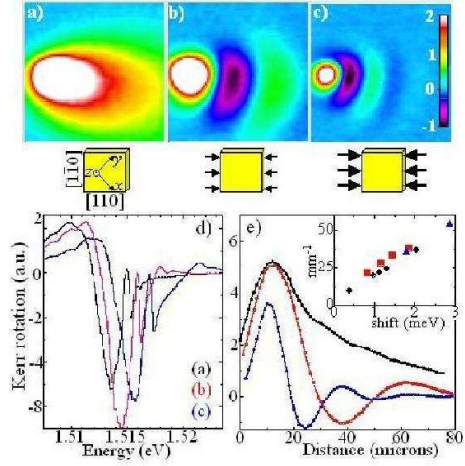}
  \caption{Experimental result of spin precession in bulk GaAs without magnetic
    field. (a)-(c) images of two-dimensional spin flow (E=10 V/cm) at 4 K, showing
    induced spin precession with increasing [110] uniaxial stress. (d)
    Kerr rotation {\sl vs.} probe photon energy for the images,
    showing blueshift of 
    GaAs band edge with stress. (e) Line cuts through the
    images. Inset: Spacial frequency of spin precession {\sl vs.} band edge
    shift. From Crooker and Smith \cite{PhysRevLett.94.236601}.}
  \label{fig:trans:crooker}
\end{figure}

The spin oscillation without any applied magnetic field in the
transient spin transport was later 
observed experimentally by Crooker and Smith in 
strained bulk system \cite{PhysRevLett.94.236601}, which is shown in
Fig.~\ref{fig:trans:crooker}. 
Unlike two-dimensional case, in bulk the average of
$\bomega_{\mathbf{k}}$ from {the} Dresselhaus term is zero, since 
$\langle\bomega_{\mathbf{k}}\rangle=m^\ast\gamma(\langle
k_y^2\rangle-\langle k_z^2\rangle,0,0)=0$ due to the symmetry in 
the $y$- and $z$-directions. This is consistent 
with the experimental
result that there is no spin oscillation for the system without
stress. However, when the stress is applied, an additional spin-orbit
coupling, 
namely the coupling of electron spins to the strain tensor, appears
\cite{opt-or,PhysRevB.16.2822,Cardona1984701,PhysRevB.38.1806,
PhysRevB.72.033311,strainbook,PhysRevLett.94.236601}. 
This additional spin-orbit coupling also acts as an effective magnetic
field. In the experiment setup, the stress applied is along the $[110]$
axis, the spacial oscillation caused by this additional
effective magnetic field is characterized by 
\begin{equation}
  \label{eq:trans:strain}
  \langle\bomega_{\mathbf{k}}\rangle=-c_3\epsilon_{yx}(0,1,0),   
\end{equation}
where $c_3$ is a
constant depends on the interband deformation potential and
$\epsilon_{yx}$ is in-plane shear
\cite{opt-or,PhysRevB.16.2822,Cardona1984701,PhysRevB.38.1806,%
strainbook,PhysRevLett.94.236601}. Therefore, once the stress is
applied, one can observe spacial spin oscillation as shown in
Fig.~\ref{fig:trans:crooker}, even when there is no applied magnetic field. 

The spin oscillation without any applied magnetic field was later
shown theoretically to survive even in the steady state spin
injection 
\cite{PhysRevB.70.155308,cheng:073702,shen_04,
saikin_05,saikin:1769,wang_prb_05,saikin_06,lopez_08}.

The oscillation without an applied magnetic field can also be
understood from the simplified kinetic spin Bloch equations
Eq.~(\ref{eq:trans:diffusive}).  
In the diffusive limit, 
by neglecting high order of the momentum relaxation time, 
Eq.~(\ref{eq:trans:diffusive}) can be further simplified as 
\begin{equation}
  {\partial{\mathbf{S}}(x,t)\over \partial t}=
  D {\partial^2{\mathbf{S}}(x,t)\over \partial x^2}
  - \bar{\mathbf{h}}\times
  {\partial{\mathbf{S}}(x,t)\over \partial x} 
  -\overleftrightarrow{\Gamma}{\mathbf{S}}(x,t) 
\label{eq:diffusive_limit}
\end{equation}
in (001) GaAs quantum well where the dominant spin-orbit coupling
is the Dresselhaus term.
Here
${\mathbf{S}}(x,k,t)=\sum_{\mathbf{k}}\mathtt{Tr}\{\bsigma\rho^0(x,k,t)\}$
is the spin momentum. $D=\langle k^2\tau_k^1/2m^{\ast 2}\rangle$ is the
diffusion coefficient. 
\begin{equation}
\bar{\mathbf{h}}= \langle k^2\tau^1_k/m^{\ast}\rangle(\alpha,
0, 0),
\label{eq:hq}
\end{equation}
is the net Dresselhaus effective magnetic field in the spin diffusion. 
The last term in Eq.~(\ref{eq:diffusive_limit})
is the spin relaxation caused by the Dresselhaus term
together with the spin conserving scattering. The relaxation matrix 
reads
\begin{equation}
\overleftrightarrow{\Gamma} = 
\left( 
    \begin{array}{{ccc}}
      {1\over \tau_{\parallel}} & 0 & 0\\
      0 & {1\over \tau_{\parallel}} & 0\\
      0 & 0 & {1\over \tau_{\perp}}
    \end{array}
\right),
\end{equation}
where $1/\tau_{\parallel}= \alpha^2\langle k^2\tau_k^1\rangle/2
+{\langle(\gamma k^3)^2\tau_k^3\rangle}/{32}$ and
$1/\tau_{\perp}=2/\tau_{\parallel}$ are the spin relaxation times of
in-plane and out-of-plane components respectively. 
Similar equations have also been obtained from linear response
theory \cite{PhysRevB.70.155308,culcer_04}. 
One can clearly see the spin precession around the net effective
magnetic field $\bar{\mathbf{h}}$ from 
Eq.~(\ref{eq:diffusive_limit}). 
In the transient spin transport, the precession around the effective
magnetic field results in spin oscillations in both time and space domains
even when there is no applied magnetic field 
\cite{PhysRevB.66.235109,weng:410,jiang:113702,
cheng:073702,weng:063714}.
Moreover, due to the precession, the steady-state spin injection is no
longer a simple exponential decay but a damped oscillation 
\cite{PhysRevB.70.155308,culcer_04,cheng:073702,weng:063714}.

\subsection{Steady-state spin transport in GaAs quantum well}

Due to spin oscillation, in the steady state the polarization of 
injected spin in GaAs quantum well is a damped oscillation $\Delta N(x)\propto
\exp(-x/L_d)\cos(x/L_0+\phi)$. Here $L_d$ and $L_0$ stand for the
diffusion length and the spacial oscillation period, respectively. 
How these two parameters change with the scattering mechanism,
temperature, applied magnetic or electric field are important to the
realization of spintronic devices. 

\subsubsection{Effect of the scattering on the spin diffusion}

Similar to the spin kinetics in the spacial uniform system, 
the scattering also affects the spin transport in
complicated ways.
On one hand, each scattering mechanism provides additional channel
of spin relaxation/dephasing in the presence of inhomogeneous broadening. 
Adding new scattering mechanism tends to 
increase spin relaxation/dephasing. 
However, scattering also affects the charge and spin transport
parameters such as diffusion coefficient and mobility. 
Electron-impurity and electron-phonon scattering can directly change the
electron momentum and reduce the spin and charge diffusion
coefficients and mobilities. For electron-electron Coulomb scattering,
although it does 
not directly change the charge diffusion coefficient and mobility, it
reduces the 
spin diffusion coefficient through 
the inhomogeneous broadening $\bomega_{\mathbf{k}}$ and the 
Coulomb drag effect 
\cite{amico_00,amico_01,flensberg_01,PhysRevB.65.085109,amico_03,amico_04,
amico_04b,jiang:113702,vignale_05,nature.437.1330,tse_07,takahashi_08,
badalyan_08}. 
All of these effects 
result in shorter spin injection length $L_d$. 
On the other hand, the scattering tends to suppress the inhomogeneous
broadening in time domain as well as in space domain.
This helps to prolong the diffusion length $L_d$. 
These competing effects on spin diffusion and spin transport were
investigated using 
the kinetic spin Bloch equation approach 
\cite{PhysRevB.66.235109,weng:410,PhysRevB.69.125310,jiang:113702,
cheng:073702,weng:063714}. The main
results are reviewed in the following. 

\begin{figure}[htbp]
  \centering
  \includegraphics[width=6cm]{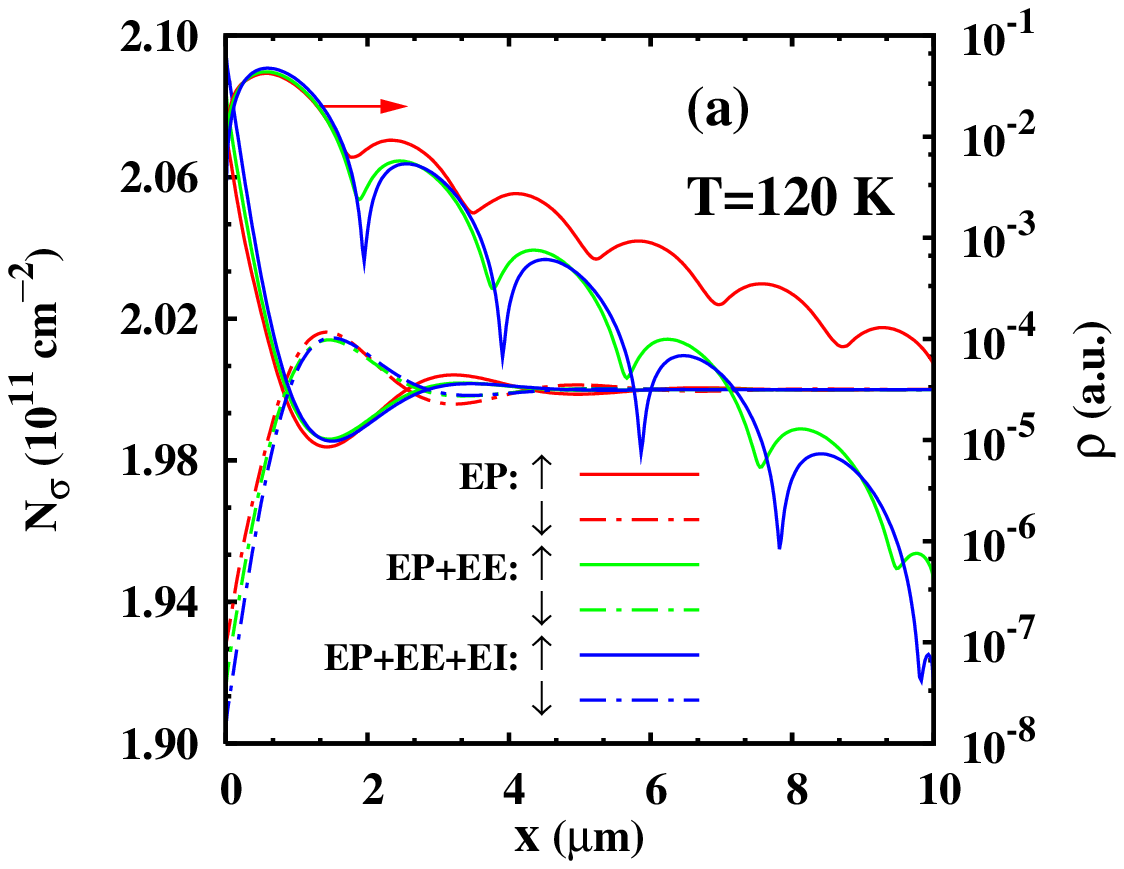}
  \includegraphics[width=6cm]{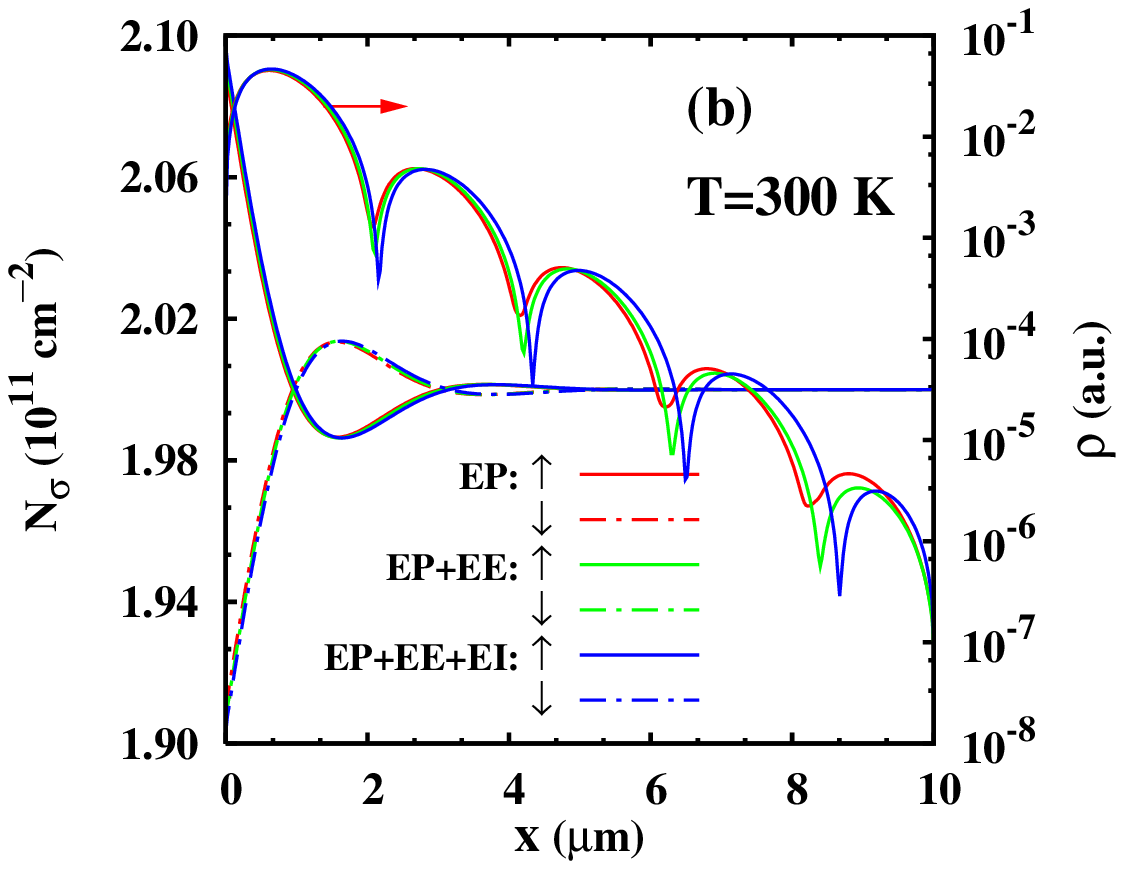}
  \caption{ Effect of the scattering on spin diffusion in the
steady state at (a) $T=120$~K and (b) $T=300$~K in GaAs quantum well with
$a=7.5$~nm. Red curves: with 
only the electron--longitudinal-optical-phonon (EP) scattering; Green curves: with both
the electron-electron (EE) and electron--longitudinal-optical-phonon scattering; Blue
curves: with all 
the scattering, {i.e.}, the 
electron-electron, electron--longitudinal-optical-phonon and electron-impurity (EI)
scattering. The impurity density $N_i=N_e$. Note the scale of the
incoherently summed spin coherence is on the right hand side of the
figure.
From Cheng and Wu \cite{cheng:073702}.
}
  \label{fig:trans:scatt}
\end{figure}

In Fig.~\ref{fig:trans:scatt}, 
the spin-resolved electron density $N_\sigma$
and the incoherently summed spin
coherence $\rho$ 
in the steady state are plotted against the position $x$
by first including only the electron--longitudinal-optical-phonon scattering (red curves),
then adding the electron-electron scattering (green curves) and
finally adding the electron-impurity scattering (blue curves) with
$N_i=N_e$ at $T=120$~K (a) and 300~K (b). In the calculation there
is no applied magnetic field. One can see that the scattering affects
the transport in a complex way: At $T=120$~K, the spin injection
length $L_d$ always decreases when a new scattering mechanism is
added: $L_d$ is significantly reduced when the Coulomb scattering is
added; adding  electron-impurity scattering further reduces
$L_d$, but only slightly. However, when $T=300$~K, $L_d$ becomes
slightly shorter when the Coulomb scattering is first added, but becomes
slightly longer when electron-impurity scattering is further
added. These results show that for GaAs quantum well with width
$a=7.5$~nm, all scattering mechanisms are not strong
enough at $T=120$~K 
and the system falls into weak scattering limit. The counter effect of
the scattering to the inhomogeneous broadening is weak. 
Therefore
adding a new scattering reduces the injection length. At
$T=300$~K, the scattering becomes stronger.
As a result the competing effects of the
scatterings cancel each other and result in marginal change in
diffusion length when a new scattering is added.  

\begin{figure}
  \begin{minipage}{0.435\linewidth}
\centerline{\includegraphics[width=5cm]{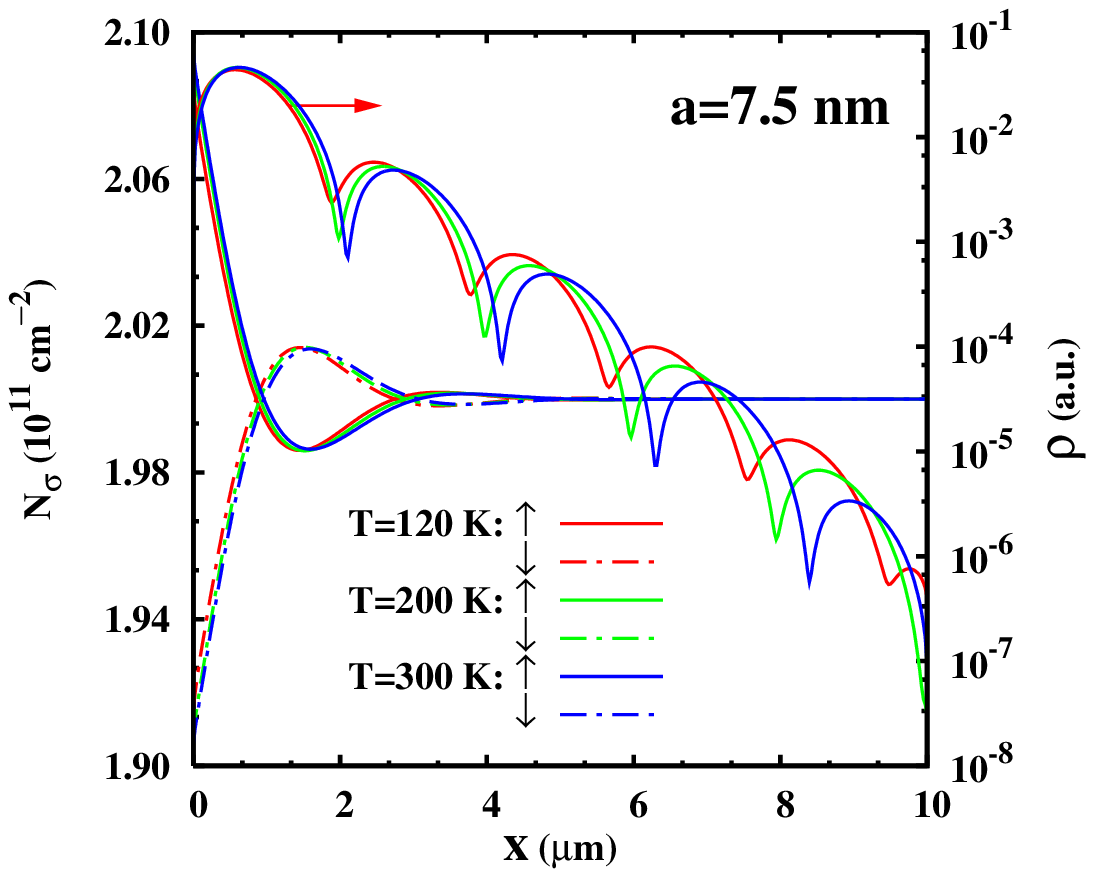}}
\caption{  $N_{\sigma}$ and  $\rho$
  {\sl vs.} the position $x$ at
  $T=120$, 200 and 300~K in $n$-type GaAs (001) quantum well
  with width $a=7.5$~nm and $L=10$~$\mu$m.
  $N_i=0$.
  Note the scale of the incoherently summed spin coherence is on the
  right hand side of the figure. 
  From Cheng and Wu \cite{cheng:073702}.
}
\label{fig:trans:temp}
\end{minipage}\hspace{1cm}
  \begin{minipage}{0.435\linewidth}
  \centerline{\includegraphics[width=5cm]{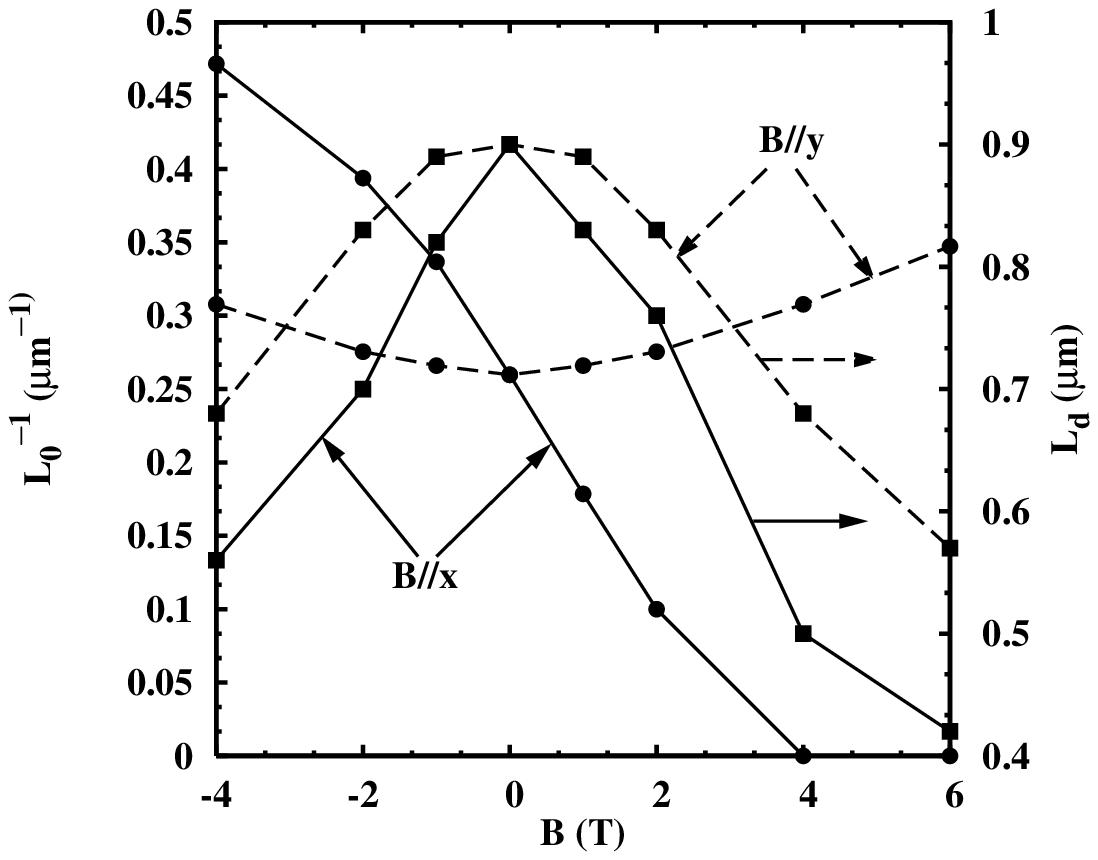} }
  \caption{Inverse of the period of the spin oscillation $L_0^{-1}$
    (curves with $\bullet$)
    and the spin diffusion length $L_d$
    (curves with $\blacksquare$) for $n$-type GaAs (001) quantum well
    {\it vs.} the external magnetic
    field $B$  which is applied either vertical ($\| y$, dashed curves)
    or parallel ($\| x$, solid curves) to the diffusion
    direction. $T=120$\ K and $N_i=0$.
    The dashed curves are for the vertical magnetic field and the
    solid curves are for the parallel one.
    Note the scale of the diffusion length $L_d$ is
    on the right hand side of the figure. 
    From Cheng and Wu \cite{cheng:073702}.
  }
  \label{fig:trans:LB}    
  \end{minipage}

\end{figure}

In Fig.~\ref{fig:trans:temp}, the results of steady-state spin
injection at different temperatures are shown. It is seen that the
spin injection length $L_d$ slightly decreases as the temperature
rises. The change in $L_d$ is mild due to the cancellation of the
opposite effects in scattering. The spacial period $L_0$ 
on the other hand systematically increases with the
temperature. This is because when there is only the Dresselhaus term, 
$L_0^{-1}\simeq |\langle\bomega_{\mathbf{k}}\rangle|=|m^\ast\gamma(\langle k_y^2\rangle
-\langle k_z^2\rangle)|$. In small quantum well, $\langle k_z^2\rangle$
is
larger than $\langle k_y^2\rangle$. 
As the temperature increases, 
$\langle k_y^2\rangle$ becomes larger and consequently $L_0$ becomes larger.

\subsubsection{Spin transport in the presence of magnetic and electric fields}

The magnetic field effect on the steady-state spin diffusion is shown
in Fig.~\ref{fig:trans:LB} where the spin diffusion length
$L_d$ and the spin oscillation period $L_0$ as functions of the applied
magnetic field  at $T=120$~K are plotted. The direction of the
magnetic field is either vertical (along the $y$-axis)  or parallel
(along the $x$-axis) to the diffusion direction. 
As the magnetic field increases the inhomogeneous broadening in spin
diffusion,  it leads to additional spin relaxation/decoherence and therefore the
spin diffusion length 
decreases with the magnetic field, regardless of the direction of the
magnetic field. However, it is interesting to see that the
spin diffusion length has different symmetries when the direction of the
magnetic field changes: When
$\mathbf{B}$ is parallel to the $y$-axis, $L_d(B)=L_d(-B)$; However,
when $\mathbf{B}$ is parallel to the $x$-axis, $L_d(B)\not=L_d(-B)$.
This can be understood from the fact  that the density matrices
from the kinetic spin Bloch equations 
have the symmetry $\rho_{k_x,k_y,kz}(B)=\rho_{k_x,-k_y,k_z}(-B)$
when $\mathbf{B}$ is along the $y$-axis. This symmetry is
broken if $\mathbf{B}$ is along the $x$-axis.

\begin{figure}[htbp]
  \begin{minipage}{0.425\linewidth}
    \centerline{\includegraphics[width=5cm]{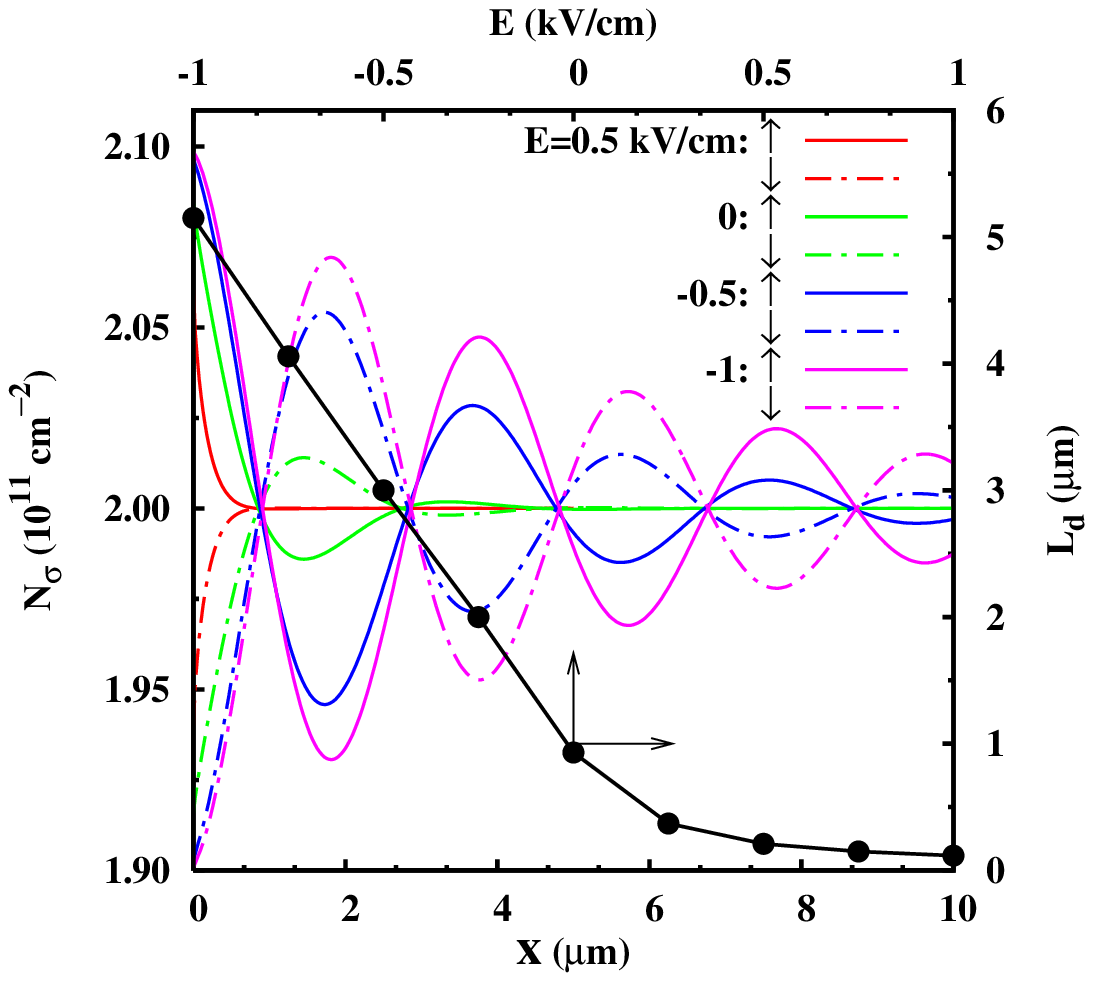}}
    \caption{ Spin-resolved electron density
      $N_{\sigma}$  {\sl vs.} position $x$ at different electric field
      $E=0.5$, 0, $-0.5$ and $-1$\ kV/cm and the spin diffusion length
      $L_d$ for $n$-type GaAs (001) quantum well
      against the electric field $E$ at $T=120$\ K. $N_i=0$. It
      is noted that although $x$ is plotted up to 10\ $\mu$m, $L=25$\
      $\mu$m when $E=-1.0$ and $-0.75$\ kV/cm; 20\ $\mu$m when
      $E=-0.5$ and $-0.25$\ kV/cm; 10\ $\mu$m when $E=0$; and 5\
      $\mu$m when $E=0.5$, 0.75 and 1.0\ kV/cm. Note the scales of the
      spin diffusion length are on the top and the right hand side of
      the figure.
      From Cheng and Wu \cite{cheng:073702}.
    } 
    \label{fig:trans:LE}
  \end{minipage}\hspace{1cm}
  \begin{minipage}{0.435\linewidth}
    \centerline{\includegraphics[width=6.cm]{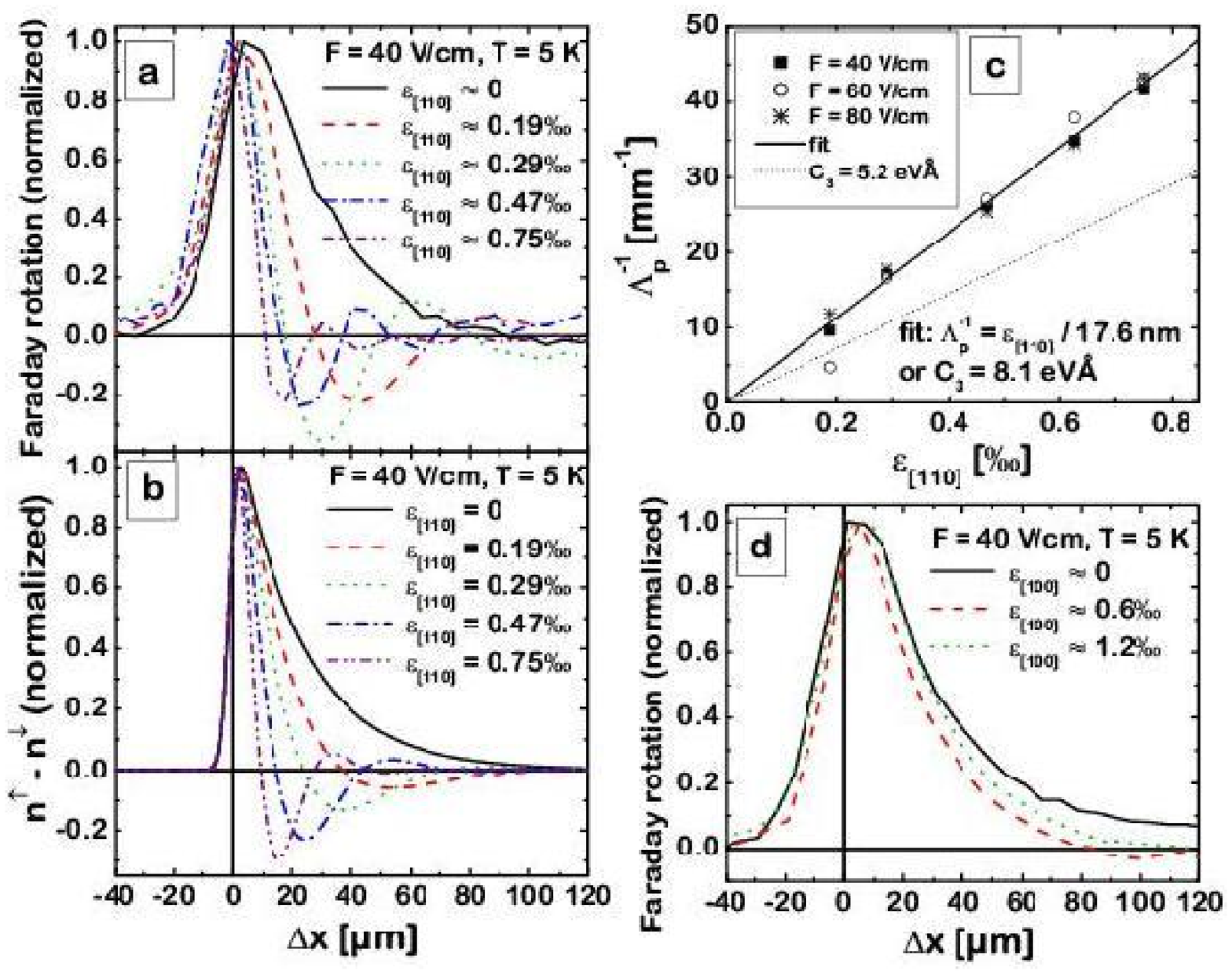}}
    \caption{
      Spin transport at $F=40$~V/cm and $N_D=1.4\times10^{16}\,$cm$^{-3}$ in
      strained bulk GaAs. (a) measured Faraday rotation for drift along
      $[1\overline10]$ and strain along $[110]$; (b) calculated spin
      polarization with the parameters of (a) (using $C_3=5.2$~eV\AA); 
      (c) Spacial oscillation ``frequency'' $\Lambda_p^{-1}$ 
      versus strain $\epsilon_{[110]}$, linear
      fit corresponding to $C_3=8.1$~eV\AA, and expected result for
      $C_3=5.2$~eV\AA\ (dotted line); (d) measured Faraday rotation
      for drift and strain along $[100]$.  
      From Beck et al.~\cite{beck_06}.
    }
    \label{fig:trans:beck}
  \end{minipage}
  \end{figure}

Differing from the magnetic field dependence of the spin diffusion length,
the spin oscillation period $L_0$  decreases monotonically with the
vertical magnetic field, regardless the
sign of the field. However $L_0$ increases with
the parallel magnetic field when the field is along the diffusion direction
and less than 4~T (the spin
precession disappears when the field is larger than 4~T)
but $L_0$ increases with the magnetic field when it is
anti-parallel to the diffusion direction. 
In the presence of the Dresselhaus term and the
applied magnetic field $\mathbf{B}$, the spacial
period is determined by  
$m^\ast\gamma(\langle k_y^2\rangle-\langle k_z^2\rangle,0,0)+
m^\ast g\mu_B\mathbf{B}\langle 1/k_x\rangle$. 
$\langle 1/k_x\rangle$ represents
the average of $1/k_x$. Due to the spin transport it does
not vanish and can be roughly estimated by 
$\langle 1/k_x\rangle^{-1}\approx 
\langle k_x\rangle=\sum_{\mathbf{k}}k_x\Delta 
f_{\mathbf{k}}/\sum_{\mathbf{k}}\Delta f_{\mathbf{k}}$,
which is a positive value for spin transport along $x$-direction.
For the vertical
magnetic field,  the average of  the 
Dresselhaus effective magnetic field
and the  applied one
are perpendicular to each other, so the magnitude
of the spin oscillation period
is determined by ${\bigl[(m^\ast g\mu_B{B})^2/\langle k_x\rangle^2
+m^{\ast 2}\gamma^2(\langle k_y^2\rangle-\langle k_z^2\rangle)^2\bigr]}^{-1/2}$ 
which always decreases with the
magnetic field. However, for the
parallel field, these two vectors are in the same direction and
the period is determined by $|m^\ast\gamma(\langle k_y^2\rangle-\langle
k_z^2\rangle)+m^\ast g\mu_{\mathtt{B}}B/\langle k_x\rangle|^{-1}$. 
These two vectors can enhance or cancel each other depending on their 
relative strength and sign. For the particular case in
Fig.~\ref{fig:trans:LB}, the  
the inverse of period $L_0^{-1}$ decreases with the magnetic field,
until $B\sim$~4~T when 
 $|\gamma(\langle k_y^2\rangle-\langle
k_z^2\rangle)+g\mu_{\mathtt{B}}B/\langle k_x\rangle|\sim0$. 
Further increase of the magnetic field does not lead to clear spin
oscillation since the inhomogeneous broadening caused by the magnetic
field becomes too large. 

The effect of electric field on the spin transport is shown in
Fig.~\ref{fig:trans:LE} where the spin densities in the
steady state are plotted against the position $x$ at different electric
fields $E=0.5$, 0, $-0.5$ and $-1$\ kV/cm. In the calculations, the
temperature is $120$~K and only the intrinsic scattering mechanisms,
i.e. the electron-phonon and electron-electron Coulomb
scatterings, are included. 
It is seen from the figure that the spin oscillation period almost
does not change when the electric field varies from $-1$~kV/cm to
0.5~kV/cm. The null effect of the electric field on spacial precession
is quite different from the effect of electric field on the precession
in time domain \cite{PhysRevB.69.245320,PhysRevB.72.033311}. 
In the spacial uniform system, due to the drifting of
the electrons driven by the electric field, the Dresselhaus or Rashba
term gives an effective magnetic 
field proportional to the electric field. This effective magnetic
field in turn gives a precession frequency proportional to the
electric field \cite{PhysRevB.69.245320,PhysRevB.72.033311}.  
However, in the spin transport, the average of 
the inhomogeneous broadening $\langle\bomega_{\mathbf{k}}\rangle
=m^\ast\gamma(\langle k_y^2\rangle-\langle k_z^2\rangle,0,0)$
does not depend on the drifting velocity,
therefore is not affected by the electric field. This is consistent
with the previous experimental observation in the strained bulk system
\cite{beck_06}, which is shown in Fig.~\ref{fig:trans:beck}. 
In the experiment, the oscillation is induced by the additional
spin-orbit coupling caused by the strain. As shown in
Eq.~(\ref{eq:trans:strain}), the spacial oscillation
frequency [denoted as $\Gamma_{p}^{-1}$ in
Fig.~\ref{fig:trans:beck}(c)] is $C_3\epsilon_{yx}$, which is
proportional to the strain but does not  depend on the drifting
velocity or electric field.

It is further noted from the figure that the spin diffusion length
is markedly affected by the electric field. 
To reveal this effect, the spin
diffusion length is plotted as function of electric field in the same
figure. One finds when the electric field varies from $-1$~kV/cm to
$+1$~kV/cm, the spin diffusion length decreases monotonically,
which is consistent with the prediction of the drift-diffusion model
\cite{PhysRevB.66.201202,PhysRevB.66.235302}.  
From inhomogeneous broadening point of view, 
this can be easily understood from the fact that when an electric field is
applied along $-x$-direction ($+x$-direction), the drift velocity
caused by the electric field is along the $x$-direction
($-x$-direction), which enhances (cancels) the velocity driven by the
spin gradient. Therefore, $\langle k_x\rangle$ is increased (reduced)
and the inhomogeneous broadening is decreased (enhanced). A longer
(shorter) spin diffusion length is then observed \cite{cheng:073702}.

\subsection{Anisotropy of spin transport in the presence of
  competing Rashba and Dresselhaus fields}
\label{sec:trans:anisotropy}

When the Dresselhaus and Rashba terms are both important in
semiconductor quantum well, the total effective magnetic field can be
highly anisotropic and spin kinetics is also highly anisotropic in
regards to the direction of the spin 
\cite{PhysRevB.60.15582}. 
For some special polarization direction, 
spin relaxation time is extremely large
\cite{PhysRevB.69.045317,0953-8984-14-12-202,PhysRevB.60.15582,
PhysRevB.71.035319,PhysRevLett.90.146801,PhysRevB.70.241302}.
For example, if the coefficients of the
linear Dresselhaus and Rashba terms are equal 
to each other in (001) quantum well of small well width
and the cubic Dresselhaus term is not important, 
the effective magnetic field is along the [110] direction for all
electrons.
For the spin components perpendicular to the [110] direction, this
effective magnetic field flips the spin and leads to a finite spin
dephasing time. 
For spin along the [110] direction, this effective
magnetic field can not flip it. Therefore, when the spin
polarization is along the [110] direction, 
the Dresselhaus and Rashba terms can not cause any spin dephasing.
When the cubic Dresselhaus term is taken into
account, the spin dephasing time for spin polarization along the [110]
direction is finite but still much larger than other directions
\cite{cheng:083704}. 

The anisotropy in the spin direction is also expected in spin transport. 
When the Dresselhaus and Rashba terms are comparable, the spin
injection length $L_d$ for the spin polarization 
perpendicular to [110] direction is usually much shorter than
that for the spin polarization along [110] direction. 
In the ideal case when there are only the linear Dresselhaus
and Rashba terms with identical strengths, 
spin injection length for spin polarization parallel to the
[110] direction becomes infinity
\cite{0953-8984-14-12-202,PhysRevLett.90.146801,PhysRevB.70.241302}. 
This effect has promoted  
Schliemann {et al.} to propose the nonballistic spin-field-effect
transistor \cite{PhysRevLett.90.146801}. In such a transistor, a gate
voltage is used to tune the strength of the Rashba term and therefore
control the spin injection length. 


However, spin transport actually involves both the spin polarization
and spin transport 
directions. The latter has long been overlooked in the literature of
spin transport. In the
kinetic spin Bloch equation approach, this direction corresponds to the
spacial gradient in the diffusion term [Eq.~(\ref{eq5.3-12})]
and the electric field in the drifting term [Eq.~(\ref{eq5.3-11})]. 
The importance of the spin transport direction has not been realized until
Cheng {et al.} pointed out that 
the spin transport is highly anisotropic not only in the sense of
the spin polarization direction but also in the spin transport direction 
when the Dresselhaus and Rashba effective magnetic fields
are comparable \cite{cheng:205328}. 
They even predicted that in (001) GaAs quantum well with identical 
linear Dresselhaus and Rashba coupling strengths, 
the spin injection along $[\bar{1}10]$ or
$[110]$\footnote{$[\bar{1}10]$ or $[110]$ depends on the relative
  signs of the Dresselhaus and Rashba coupling strengths.} 
can be infinite {\em regardless of} the
direction of the spin polarization
\cite{cheng:205328}.
In the following we review the results
on the anisotropic spin transport from the kinetic spin Bloch equation
approach. 

\begin{figure}
  \begin{center}
    \includegraphics[width=6cm]{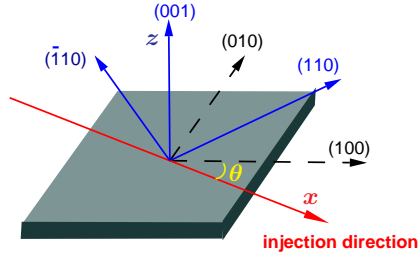}
  \end{center}
  \caption{ Schematic of the different directions considered
    for the spin polarizations ([110], [$\bar{1}10$] and [001]-axes
    and spin diffusion/injection ($x$-axis).
    From Cheng et al.~\cite{cheng:205328}.
  }
\label{fig:trans:schem}
\end{figure}

The schematic of spin transport in (001) GaAs quantum well is shown in
Fig.~\ref{fig:trans:schem}. The transport direction is chosen to be
along the 
$x$-axis, the angle between $x$-axis and [100] crystal direction is
$\theta$. In this coordinate system, the 
Dresselhaus effective magnetic field is 
\be
{\mathbf{h}}_D(\mathbf k)=
\beta
\bigl(  -k_x\cos
2\theta +k_y\sin 2\theta,\;
k_x\sin 2\theta + k_y\cos 2\theta,\;
0\bigr) 
 +\gamma(\frac{k_x^2-k_y^2}{2}\sin
2\theta+k_xk_y\cos 2\theta)
\bigl( k_y,\;
-k_x,\;
0\bigr), 
\ee
while the Rashba field reads
\begin{equation}
{\mathbf{h}}_R(\mathbf k)=\alpha\bigl(
  k_y,\; -k_x, \;
0\bigr),
\end{equation}
with $\beta=\gamma\langle k_z^2\rangle$. For the special case with
$\alpha=\beta$, the total magnetic field can be written as
\begin{equation}
  {\mathbf{h}}(\mathbf k) = 2 \beta
  \left[\sin(\theta-\frac{\pi}{4})k_x
    +\cos(\theta-\frac{\pi}{4})k_y\right] \hat{\mathbf n}_0 
  + \gamma\left[\frac{k_x^2-k_y^2}{2}\sin
  2\theta+k_xk_y\cos 2\theta\right]
\left(
k_y,\;
-k_x,\;
0\right),
\label{eq:total}
\end{equation}
with the special direction $\hat{\mathbf{n}}_0
=\bigl(\cos(\pi/4-\theta),\; \sin(\pi/4-\theta),\;
  0\bigr)$
representing the crystal direction $[110]$.
As given by Eq.~(\ref{eq:trans:Omega}), the spin precession around the
effective field magnetic field is characterized by $\bomega_{\mathbf{k}}
=m^{\ast}\mathbf{h}(\mathbf{k})/k_x$ for spin transport along
the $x$-axis.

\subsubsection{Spin injection under the identical Dresselhaus and Rashba
  strengths $\alpha=\beta$} 
\label{sec:trans:infinite}

When $\alpha=\beta$, 
\be
\bomega_{\mathbf{k}} = m^{\ast}
\biggl\{2\beta 
\left(
  \sin(\theta-\frac{\pi}{4})+\cos(\theta-\frac{\pi}{4})\frac{k_y}{k_x}\right)
\hat{\mathbf n}_0  +
\gamma(\frac{k_x^2-k_y^2}{2}\sin
2\theta+k_xk_y\cos 2\theta)
\bigl(k_y/k_x,\;
-1, \;
0\bigr)\biggr\}.
\label{eq:trans:omega_ab}
\ee
It can be splitted into 
two parts: the zeroth-order 
term (on $k$) which is always along the same direction of 
$\hat{\mathbf  n}_0$ and the second-order term which comes from the
cubic Dresselhaus term.  
If the cubic Dresselhaus term is omitted, the effective magnetic fields
for all $\mathbf{k}$ states align along $\hat{\mathbf{n}}_0$ (crystal
[110]) direction. Therefore, if the spin polarization is along
$\hat{\mathbf{n}}_0$, there is no spin relaxation even in
the presence of scattering since there is no spin precession. 
%
Nevertheless, it is interesting to see from
Eq.~(\ref{eq:trans:omega_ab}) that when 
$\theta=3\pi/4$, 
{i.e.}, the spin transport is along the [$\bar{1}10$] direction,
$\bomega_{\mathbf{k}}=2m^{\ast}\beta\hat{\mathbf{n}}_0$ is independent
on $\mathbf{k}$ if the cubic Dresselhaus term is neglected. 
Therefore, in this special spin transport direction, there is no
inhomogeneous broadening in the spin transport for {\em any} spin
polarization. 
The spin injection length is therefore infinite 
regardless of the direction of spin polarization.
This result is highly counterintuitive, considering that the
spin relaxation times for the spin components perpendicular to the
effective magnetic field are
finite in the spacial uniform system. 
The surprisingly contradictory results, 
{i.e.}, the finite spin relaxation/dephasing time versus the infinite spin
injection length, are due to the difference in the inhomogeneous
broadening in spacial uniform and non-uniform systems.  
In spacial uniform system, the inhomogeneous broadening is determined
by 
the D'yakonov-Perel' term, {i.e.}, 
$\mathbf{h}(\mathbf{k})$ [Eq.~(\ref{eq:total})],
which is momentum dependent. However in
the spin transport problem the inhomogeneous broadening is determined by
$\bomega_{\mathbf{k}}$, which is momentum-independent when the
transport direction is along $[\bar{1}10]$ here.
This prediction has not yet been realized experimentally. However,
very recent experimental findings on spin helix
\cite{bernevig:236601,Koralek:nature458.610,Fabian2009,shen_09} have
provide strong evidence to support this prediction.

\subsubsection{Spin-diffusion/injection--direction and
spin-polarization--direction dependence at $\alpha=\beta$}

\begin{figure}[htb]
  \centerline{\includegraphics[width=6cm]{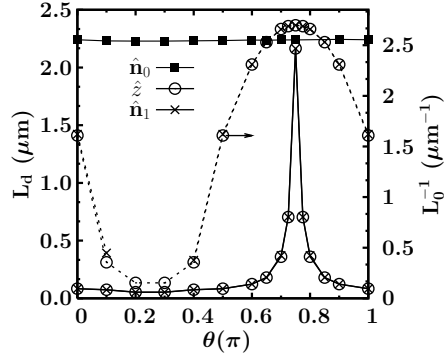}}
\caption{Spin diffusion length $L_d$ (solid curves) and the inverse of
the spin oscillation period $L_0^{-1}$ (dashed curves)
for $n$-type GaAs (001)  quantum well with
$\alpha=\beta$ as functions of the injection
direction for different spin polarization directions
$\hat{\mathbf n}_0$, $\hat{\mathbf z}$ and $\hat{\mathbf n}_1$ at
$T=200$\ K. It is noted that the scale of the spin oscillation
period is on the right hand side of the frame.
From Cheng et al.~\cite{cheng:205328}.
}
\label{fig:trans:angle}
\end{figure}

When the cubic Dresselhaus term 
is taken into account, 
$\bomega_{\mathbf{k}}$ 
becomes momentum 
dependent again for all spin transport directions. 
This inhomogeneous broadening results in a finite but still highly
anisotropic diffusion length for both spin polarization and
spin transport directions. The anisotropy is shown in 
Fig.~\ref{fig:trans:angle}, where 
the spin diffusion length $L_d$ and the inverse of the
spin oscillation period $L_0^{-1}$ are plotted against the
spin diffusion/injection angle $\theta$ for spin polarizations
along $\hat{\mathbf{n}}_0$,  $\hat{\mathbf{z}}$ and
$\hat{\mathbf n}_1=\hat{\mathbf z}\times\hat{\mathbf n}_0$ (crystal
direction $[\bar{1}10]$), respectively. 

It is seen that the spin
injection length becomes finite but independent on the direction of
the spin diffusion/injection if the  spin polarization is along [110]
($=\hat{\mathbf n}_0$). Meanwhile, the spin polarization along this
direction has an oscillation period approaching infinity.
This can be understood
because the spin polarization is in the same direction as the effective
magnetic field given by the zeroth-order term in
$\bomega_{\mathbf{k}}$
and, thus, results in no oscillation. It is noted that
the effective magnetic field from the second-order term
of $\bomega_{\mathbf{k}}$ is ${\mathbf
  k}$-dependent and the average of the second-order term over 
$\mathbf{k}$ states is nearly zero.
Therefore the cubic term does not lead to any oscillation at high
temperature due to the scattering 
\cite{cheng:083704,PhysRevLett.89.236601,lue:125314}. 
However, the inhomogeneous broadening due to the second-order term 
leads to a finite spin diffusion length.
By rewriting the second-order term of
$\bomega_{\mathbf{k}}$ into
$\frac{m^{\ast}\gamma}{2\hbar^2}{k^2}
\sin(2\theta_{\mathbf k}+2\theta)\bigl(k_y/k_x,\;
-1,\; 0)$ with $k$ and $\theta_{\mathbf k}$ denoting the
magnitude and the direction of the wavevector $\mathbf k$ 
respectively, one
finds that the magnitude of the inhomogeneous broadening does not change with
the spin diffusion/injection direction $\theta$. Consequently
the spin diffusion length does not change with the spin
diffusion/injection direction.

When the injected spin momentum is perpendicular to
$\hat{\mathbf{n}}_0$, $L_0$ and $L_d$ become anisotropic in
regard to the spin transport direction. The spin injection length is
small except around $\theta=3\pi/4$. $L_0$ and
$L_d$ are almost identical for two spin polarization 
directions $\hat{z}$ and $\hat{\mathbf{n}}_1$.
Because in this case, as $\gamma \langle k^2\rangle\ll \beta$, 
the kinetics of systems are determined by the
zeroth order terms of $\bomega_{\mathbf{k}}$
which are the same
for these two directions.
The oscillation period in the diffusion $L_0$ is determined by the
$\mathbf k$-independent component of
$\bomega_{\mathbf{k}}$ \cite{cheng:205328}, {i.e.},
${2m^{\ast}\beta\sin(\theta-\pi/4)}/{\hbar^2}$ which is in good agreement
with the results shown in Fig.~\ref{fig:trans:angle}.
Moreover, when the injection direction is along $[\bar{1}10]$
($\theta=3\pi/4$), the spin diffusion lengths for all of the
three spin polarization directions are almost identical. The
anisotropy in the direction of the spin polarization disappears. 
This is because when the spin diffusion/injection direction
is $\theta=3\pi/4$, the  $\mathbf k$-dependent component
in the zeroth order term of
$\bomega_{\mathbf{k}}$ disappears. The spin
injection length is determined by the inhomogeneous broadening in the
quadratic terms in $\bomega_{\mathbf{k}}$, which are identical for any
spin polarization directions.  

\begin{figure}[htb]
\begin{minipage}{0.435\linewidth}
  \begin{center}
    \includegraphics[width=6cm]{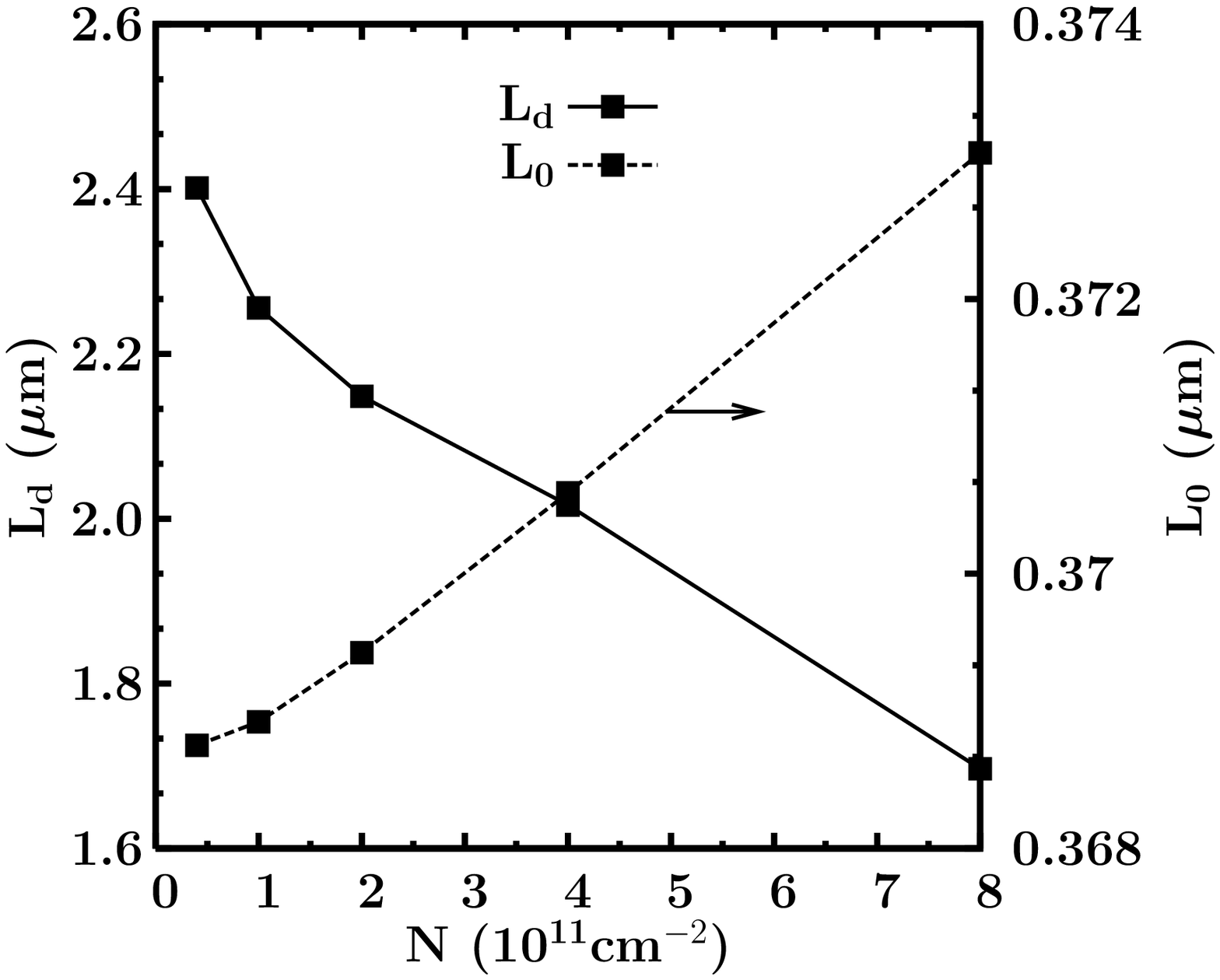}
  \end{center}
\caption{Spin diffusion length $L_d$  and spin oscillation period $L_0$
at $T=200$~K for spin transport in $n$-type GaAs (001) quantum well along
    $\theta=3\pi/4$ direction as functions of the
  electron density. Note the scale of the oscillation period is on the
  right hand side of the frame.
  From Cheng et al.~\cite{cheng:205328}.
}
  \label{fig:trans:angleN}
\end{minipage}
\hspace{1cm}
\begin{minipage}{0.435\linewidth}
  \begin{center}
    \includegraphics[width=6cm]{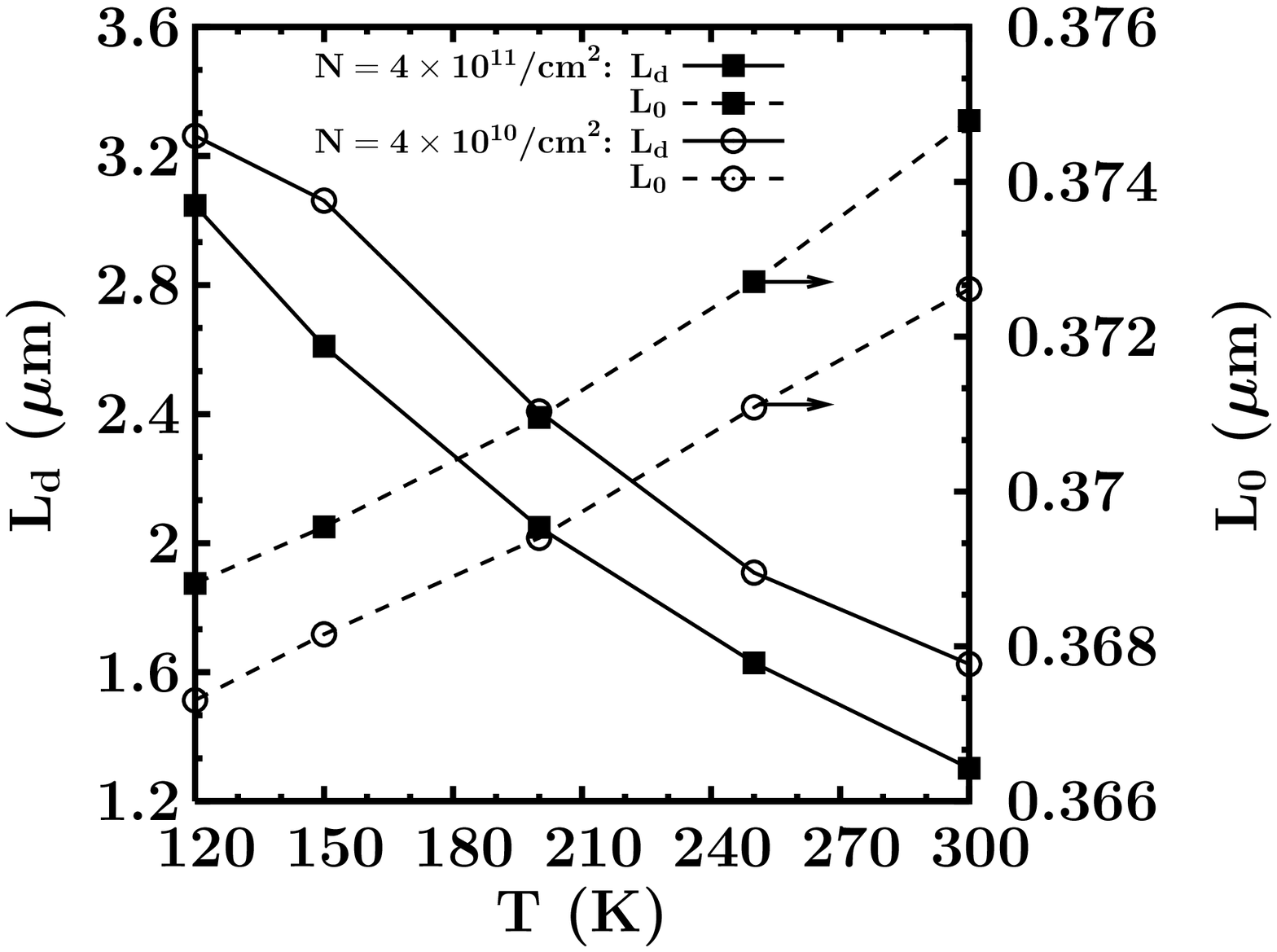}
  \end{center}
  \caption{Spin diffusive length $L_d$  and spin oscillation period
    $L_0$ in $n$-type GaAs (001) quantum well
    at two different electron densities,
    $4\times 10^{11}$\ cm$^{-2}$ and
 $4\times 10^{10}$\ cm$^{-2}$ {\sl vs.} the temperature. The spin
diffusion/injection direction $\theta=3\pi/4$. Note the scale of the
oscillation period is on the right hand side of the frame.
From Cheng et al.~\cite{cheng:205328}.
}
\label{fig:trans:angleT}
\end{minipage}
\end{figure}

\subsubsection{Temperature and electron-density dependence with
  injection along  $[\bar{1}10]$ and $\alpha=\beta$}

The  electron-density and temperature dependences of spin injection
along $[\bar{1}10]$ in the case $\alpha=\beta$ are shown in 
Figs.~\ref{fig:trans:angleN} and \ref{fig:trans:angleT} respectively. 
It is seen that the diffusion length $L_d$ decreases with electron
density and temperature, whereas the oscillation period $L_0$
increases with them. However, the change observed in the oscillation
period is almost negligible (1\%) in contrast to the pronounced
changes observed in the diffusion length (more than 40\%).  
This is quite different from the spin injection in GaAs quantum well when
there is only the Dresselhaus term, where $L_d$ and $L_0$ only
change slightly with the electron density and temperature
\cite{cheng:073702,weng:410}.  

This is
understood by the difference in the behavior of the inhomogeneous
broadening
$\bomega_{\mathbf{k}}$.
Unlike case of $\alpha=0$ where the zeroth order term dominates the
inhomogeneous broadening,
when $\alpha=\beta$ and $\theta=3\pi/4$, the second-order term alone
determines the inhomogeneous broadening. 
This term increases 
effectively with the
temperature and electron density due to the increase of the
average value of $k^2$. This is the reason for the obtained marked decrease
of the spin diffusion length. The oscillation period $L_0$
is determined by the $\mathbf k$-independent zeroth-order term of
$\bomega_{\mathbf k}$,
which does not change with the 
electron density and the temperature. Therefore one
observes only a  slight change in the
oscillation period, originating from
the second-order term and the scattering.

\subsubsection{Gate voltage dependence with injection direction
along  $[\bar{1}10]$}

Spin transport along $[\bar{1}10]$ under different gate voltages,
{i.e.}, different Rashba coupling strengths $\alpha$ since it can be
controlled by the gate voltage, was also studied
\cite{cheng:205328}. It was 
pointed out that when the 
cubic Dresselhaus term is present, the longest spin injection length
no longer takes place at 
$\alpha=\beta$ \cite{cheng:205328}.\footnote{Similar conclusion has
  also been given by Stanescu and Galitski in Ref.~\cite{stanescu:125307}.}
This can also be understood from the inhomogeneous broadening. 
For spin transport along $[\bar{1}10]$ with Dresselhaus and
Rashba terms, 
$\bomega_{\mathbf{k}}$ can be written as
\begin{equation}
  \bomega_{\mathbf{k}}
  = 2[(-\beta+\alpha +\frac{\gamma k^2}{2}
  )\frac{k_y}{k_x} - \gamma k_xk_y]\hat{\mathbf n}_1 
  - 2[\beta+\alpha -
  \frac{\gamma}{2}(k_x^2-k_y^2)]\hat{\mathbf n}_0.
  \label{3piov4}
\end{equation}
It is seen that the effect of $\alpha$ on the inhomogeneous
broadening is determined by the first term. 
In order to get the longest spin injection length, $\alpha$ should
be tuned to make the first term become minimal. Due to the cubic
Dresselhaus term, the optimized condition is $\alpha
=\hat{\beta}=\beta-\gamma \langle k^2\rangle/2$ instead of
$\alpha=\beta$ 
\cite{stanescu:125307,cheng:205328}. 
This prediction was later confirmed by the experiment
\cite{Koralek:nature458.610}. 

\subsection{Transient spin grating}

The drift-diffusion model tells that the spin signal exponentially
decays with the distance in spin injection and the injection length $L_d$
is related with the spin diffusion coefficient $D_s$ and spin relaxation
time $\tau_s$ by the formula $L_d=\sqrt{D_s\tau_s}$
[Eq.~(\ref{eq:trans:Ls})]. However, 
as pointed out by Weng and Wu \cite{PhysRevB.66.235109}, 
the drift-diffusion model
oversimplifies spin transport in semiconductors by neglecting the
off-diagonal terms of the density matrix. When these terms are included, the
spin signal in the spin injection was shown to be a damped oscillation,
characterized by spin injection length $L_d$ and spacial spin
oscillation period $L_0$, instead of simple exponential decay
\cite{weng:410,cheng:073702,weng:063714}. 
In the extreme case of spin transport along the $[\bar{1}10]$-axis in
(001) GaAs quantum well under the identical linear Dresselhaus and
Rashba spin orbit coupling strengths, the spin injection length
becomes infinite even when 
the spin diffusion coefficient and the spin relaxation time are
finite \cite{cheng:205328}. This result is totally beyond the
drift-diffusion model as both $D_s$ and $\tau_s$ are finite. 
Therefore, the relations among the characteristic parameters 
given by the drift-diffusion model need modification
\cite{cheng:205328,stanescu:125307,PhysRevB.70.155308,weng:063714}. 
Weng {et al.} presented the correct relations among the characteristic
parameters, which hold even under the extreme case
like the case in Sec.~\ref{sec:trans:anisotropy}, by studying the
transient spin grating \cite{weng:063714}.  

Transient spin grating, whose spin polarization varies periodically in
real space, is excited optically by two non-collinear coherent light
beams with orthogonal linear polarization 
\cite{PhysRevLett.76.4793,nature.437.1330,carter:136602,
weber:076604,Koralek:nature458.610}. 
Transient spin grating technique 
is well suited to study the spin transport since it can directly probe
the decay rate of nonuniform spin distributions. 
Spin diffusion coefficient $D_s$ can be obtained from the transient
spin grating experiments 
\cite{PhysRevLett.76.4793,nature.437.1330,carter:136602,weber:076604}. 
In the literature, the drift-diffusion model was employed to extract
$D_s$ from the experimental data. 
With the drift-diffusion model, 
the transient spin grating was predicted to decay exponentially
with time with a decay rate of 
$\Gamma_q=D_sq^2+1/\tau_s$, where $q$ is the wavevector of the spin
grating \cite{PhysRevLett.76.4793,nature.437.1330}. 
However, this result is not accurate since it neglects the spin
precession which plays an important role in spin
transport. 
Indeed, experimental results show that the decay of transient
spin grating takes a double-exponential form instead of single
exponential one
\cite{nature.437.1330,weber:076604,Koralek:nature458.610}. 
Therefore, it is worthy to study transient spin grating using the
kinetic spin Bloch equation approach.  

\subsubsection{Simplified solution for transient spin grating}
\label{sec:trans:simplified_grating}

The transient spin grating can be qualitatively understood in a 
simplified case where analytical solution can be obtained. 
By neglecting the Hartree-Fock term, the inelastic scattering such as
the electron-phonon and the electron-electron Coulomb scatterings and
the coupling to the Poisson equation, in the diffusive limit, 
one is able to write the kinetic spin Bloch equations as
Eq.~(\ref{eq:diffusive_limit}), with 
$\bar{\mathbf{h}}$ and  
$\overleftrightarrow{\Gamma}$ being 
\begin{equation}
\bar{\mathbf{h}}=\langle k^2\tau^1_k/m\rangle(-\hat{\beta}\cos 2\theta,
\hat{\beta}\sin 2\theta-\alpha,0),
\end{equation}
and 
\begin{equation}
\label{eq:trans:Gamma_grating}
\overleftrightarrow{\Gamma} = 
{1\over 2}\left(
    \begin{array}{ccc}
      (\alpha^2 +\hat{\beta}^2- 2\alpha\hat{\beta}\sin
      2\theta){\langle k^2\tau_k^1\rangle}\atop
      +{\langle(\gamma k^3)^2\tau_k^3\rangle}/{16}
      &
      2\alpha\hat{\beta}{\langle k^2\tau_k^1\rangle}\cos 2\theta
      & 0\\
      2\alpha\hat{\beta}{\langle k^2\tau_k^1\rangle}\cos 2\theta &
      (\alpha^2 +\hat{\beta}^2+2\alpha\hat{\beta}\sin
      2\theta){\langle k^2\tau_k^1\rangle}\atop
      +{\langle(\gamma k^3)^2\tau_k^3\rangle}/{16}
      & 0\\
      0&0 &
      {2(\alpha^2+\hat{\beta}^2) \langle k^2\tau_k^1\rangle
        \atop
      +{\langle(\gamma k^3)^2\tau_k^3\rangle/8}
    }
    \end{array}
  \right)
\end{equation}
for the system with both the Dresselhaus and Rashba terms. 
One can obtain the analytical solution of the above kinetic spin Bloch
equations for transient spin grating as following \cite{weng:063714}
\begin{equation}
\label{eq:trans:szt}
S_z(q,t)=S_z(q,0)(\lambda_+e^{-t/\tau_+}+\lambda_-e^{-t/\tau_-}),
\end{equation}
with relaxation rates
\begin{equation}
  \label{eq:trans:decay}
  \Gamma_{\pm}=\frac{1}{\tau_{\pm}}=Dq^2+{1\over 2}
  ({1\over\tau_{s 1}}+{1\over\tau_s})
  \pm{1\over 2\tau_{s 2}}
  \sqrt{1+\frac{16Dq^2 \tau_{s 2}^2}{\tau^{\prime}_{s 1}}},
\end{equation}
in which
\begin{equation}
  \label{eq:trans:lambda}
  \lambda_{\pm}={1\over 2}\left(1\pm
  {1\over\sqrt{1+16Dq^2\tau^2_{{s}2}/\tau^{\prime}_{{s}1}}}\right).
\end{equation}
Here $\tau_{s1}$ ($\tau_{s2}$) is the spin relaxation time of the in-plane spin
which mixes (does not mix) with the out-of-plane spin due to the net
effective magnetic field.  For  spin injection/diffusion along the $[110]$
axis, $\tau_{s1}= \langle(\alpha-\hat{\beta})^2
k^2\tau_k^1\rangle/2+\gamma ^2\langle k^6\tau_k^3\rangle/32$
and $\tau_{{s}1}^{\prime}=\langle(\alpha-\hat{\beta})^2
k^2\tau_k^1\rangle/2$.
For spin injection/diffusion along the $[\bar{1}10]$
axis, $\tau_{s1}= \langle(\alpha+\hat{\beta})^2
k^2\tau_k^1\rangle/2+\gamma ^2\langle k^6\tau_k^3\rangle/32$
and
$\tau_{{s}1}^{\prime}=\langle(\alpha+\hat{\beta})^2
k^2\tau_k^1\rangle/2$.
In the
long wavelength limit ($q\ll 1$),
$\Gamma_+\simeq 1/\tau_s+(1+4\tau_{{s}2}/\tau_{{s}1}^{\prime})Dq^2$
and $\Gamma_-\simeq
1/\tau_{{s}1}+(1-4\tau_{{s}2}/\tau^{\prime}_{{s}1})Dq^2$
become quadratic functions of $q$,
roughly correspond to the out-of-plane and the in-plane relaxation rates
respectively.
In general cases both of these two decay rates
[Eq.~(\ref{eq:trans:decay})] are not simple
quadratic functions of $q$. If one uses the quadratic fitting
to yield the spin diffusion coefficient, one gets a value that is
either larger (for $\Gamma_+$) or smaller (for $\Gamma_-$) than the
true one. The accurate way to get the 
information of spin diffusion coefficient should be from
the average of these two rates
\begin{equation}
\Gamma=(\Gamma_++\Gamma_-)/2=Dq^2+1/\tau'_s
\label{eq:trans:Gamma}
\end{equation}
with 
$1/\tau'_s=(1/\tau_s+1/\tau_{{s1}})/2$,
which differs from
the current widely used formula by replacing the spin decay rate by the
average of the in- and out-of-plane relaxation rates. 
The difference of these two decay rates is a linear function of the
wavevector $q$ when $q$ is relatively large:
\begin{equation}
  \label{eq:trans:DeltaGamma}
  \Delta\Gamma=cq+d. 
\end{equation}
For the simplified solution $c=2\sqrt{D_s/\tau^{\prime}_{{s}1}}$ and
$d$ is a value close to zero. 

The steady-state spin injection can be extracted from the  transient
spin grating signal 
by integrating the  transient spin grating signal
Eq.~(\ref{eq:trans:szt}) over the time 
from 0 to $\infty$ and the wave-vector from $-\infty$ to
$\infty$. From the simplified solution 
[Eqs.~(\ref{eq:trans:szt}), (\ref{eq:trans:Gamma}) and
(\ref{eq:trans:DeltaGamma})], 
the integrated transient spin grating reads
$S_z(x)=S_z(0)e^{-x/L_s}\cos (x/L_0+\psi)$  with
\begin{eqnarray}
\label{eq:trans:ls}
L_s&=&2D_s/\sqrt{|c^2-4D_s(1/\tau^{\prime}_s-d)|},\\
L_0&=&2D_s/c.
\label{eq:trans:l0}
\end{eqnarray}
It is noted that if one only considers the Rashba term or the linear
Dresselhaus term
one can 
recover the result $L_s=2\sqrt{D_s\tau_s}$ from linear response
theory \cite{PhysRevB.70.155308}. 
It is seen that the spin precession actually prolongs the out-of-plane
spin injection length by mixing the fast decaying out-of-plane
spin with  the slow decaying in-plane spin. 
Equations~(\ref{eq:trans:ls}) and (\ref{eq:trans:l0})
give the right spin injection length and
the spin oscillation period in the presence of the spin-orbit coupling.
From the experiment point of view, one can monitor the
time evolutions of transient spin grating with different wavevectors
$q$ and obtain the corresponding decay rates $\Gamma_\pm$. 
By fitting $\Gamma_\pm$ with Eqs.~(\ref{eq:trans:Gamma}) and 
(\ref{eq:trans:DeltaGamma}), 
one can then
calculate the spin injection length and spin oscillation period
accurately from Eqs.~(\ref{eq:trans:ls}) and (\ref{eq:trans:l0}), 
instead of using the inaccurate formula from the drift-diffusion model. 
Weng {et al.} demonstrated that the relations defined by 
Eqs.~(\ref{eq:trans:ls}) and (\ref{eq:trans:l0}) are true even in the
extreme case when the drift-diffusion model totally fails 
\cite{weng:063714}.

\subsubsection{Kinetic spin Bloch equation solution for transient spin grating}

\begin{figure}[htpb]
  \begin{minipage}{0.435\linewidth}
  \centering
  \includegraphics[width=5cm]{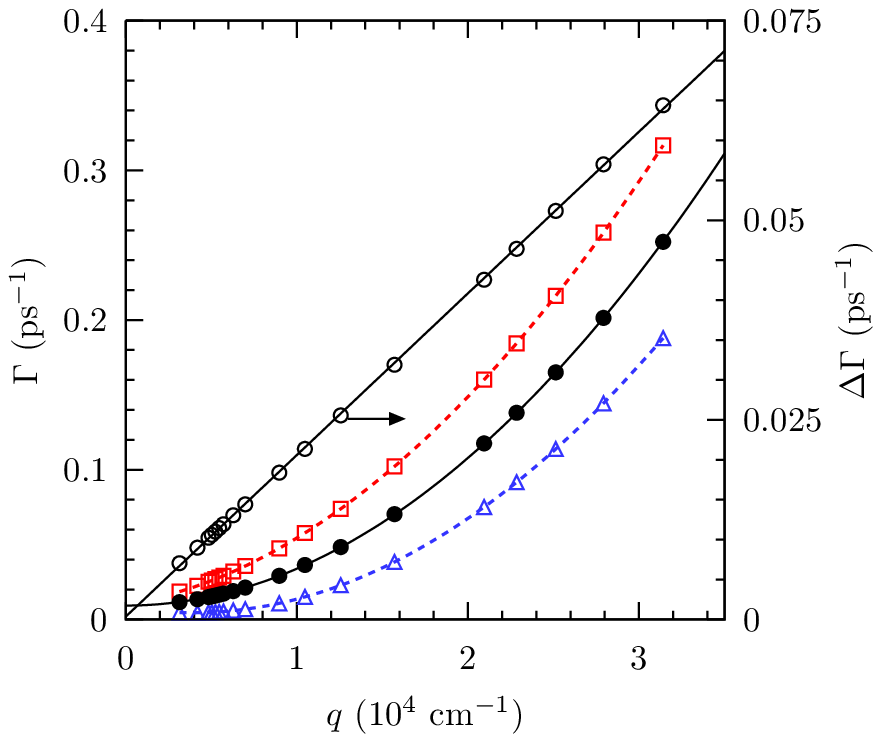}
   \caption{
      $\Gamma=(\Gamma_++\Gamma_-)/2$ and
     $\Delta\Gamma=(\Gamma_+-\Gamma_-)/2$
     {\em vs}. $q$ at  $T=295$\ K in $n$-type GaAs (001) quantum well.
     Open boxes/triangles are the relaxation rates $\Gamma_{+/-}$
     calculated from the full kinetic spin Bloch equations.
     Filled/open circles represent $\Gamma$ and $\Delta\Gamma$ respectively.
     Noted that the scale for $\Delta\Gamma$ is on the
     right hand side of the frame. The solid curves are the fitting to
     $\Gamma$ and $\Delta\Gamma$ respectively. The dashed curves are
     guide to eyes.
     From Weng et al.~\cite{weng:063714}.
   }
   \label{fig:trans:gamma}
  \end{minipage}
  \hspace{1cm}
  \begin{minipage}{0.435\linewidth}
  \centering
  \includegraphics[width=8cm]{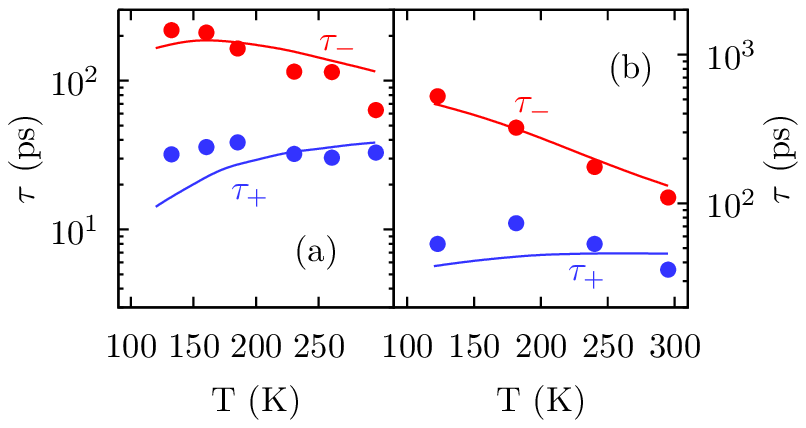}
   \caption{ Spin relaxation times $\tau_\pm$
     {\sl vs.} temperature in $n$-type GaAs (001) quantum well
     for (a) high-mobility sample with $q=0.58\times 10^4$\ cm$^{-1}$
     and (b) low-mobility sample with $q=0.69\times
     10^4$\ cm$^{-1}$. The dots are the experiment data from
     Ref.~\cite{weber:076604}.
     From Weng et al.~\cite{weng:063714}.
 }
   \label{fig:trans:fitting}
 \end{minipage}
 \end{figure}

The spin diffusion coefficient together with the relations 
Eqs.~(\ref{eq:trans:ls}) and (\ref{eq:trans:l0}) obtained by the above
simplified
kinetic spin Bloch equations are derived with only the elastic
scattering approximation. It does not take into account of the inelastic
electron-phonon and electron-electron Coulomb scattering. Therefore
the diffusion coefficient obtained from Eq.~(\ref{eq:trans:Gamma})
does not include the contribution of the spin Coulomb drag. 
Moreover, this approach oversimplifies the complex effect of the
scattering on the spin transport.  
Whether these relations remain valid in the genuine situation remains yet
to be checked. To answer this, 
Weng {et al.} 
numerically solved the full kinetic spin Bloch equations with all the
scattering explicitly included \cite{weng:063714}.  
Their numerical results showed that all of the important qualitative
results presented in Sec.~\ref{sec:trans:simplified_grating} are still
valid. Especially, the temporal evolution of the transient spin 
grating with all the scattering included 
still shows a double-exponential decay,
which agrees with the experimental result
\cite{nature.437.1330,weber:076604,Koralek:nature458.610} but 
is contrary
to the result from the drift-diffusion model.
Moreover the two decay rates can still
be fitted with Eqs.~(\ref{eq:trans:Gamma}) and
(\ref{eq:trans:DeltaGamma}) pretty well \cite{weng:063714}.
The fitting is 
shown in Fig.~\ref{fig:trans:gamma}, where
the relaxation rates 
$\Gamma_{\pm}=1/\tau_{\pm}$, their average
$\Gamma$
and difference
$\Delta\Gamma$ 
of a typical transient spin grating are plotted as functions of
$q$. The numerical results suggested that the two decay rates
$\Gamma_{\pm}=1/\tau_{\pm}$ are not quadratic functions of $q$, but
their average decay rate $\Gamma$ is. Using quadratic function to fit
the decay rates $\Gamma_{\pm}(q)$, one gets poor result. In contrast,
the average decay rate $\Gamma$ fits pretty well by a quadratic
function $\Gamma=D_sq^2+1/\tau_s^{\prime}$. The resident error of the
quadratic fitting for $\Gamma$ is two orders of magnitude smaller than
those of $\Gamma_{\pm}$. Moreover, for $\Gamma$ the constant term
$\tau_s^{\prime}$ is very close to $4\tau_s/3$, inverse of the average
of the in-plane and out-of-plane spin relaxation rates. 
Inspired by Eq.~(\ref{eq:trans:Gamma}), the coefficient of the
quadratic term $D_s$ can be reasonably assumed to be the spin
diffusion coefficient. In this way one can calculate the spin
diffusion coefficient with the effect of spin Coulomb drag
included. 
For $\Delta\Gamma$, it can be accurately fitted by a linear function
of $q$. It is therefore concluded that even with all the scatterings
included, 
one can still extract the spin diffusion coefficient $D_s$ 
and obtain the injection length and oscillation period 
using Eqs.~(\ref{eq:trans:ls}) and (\ref{eq:trans:l0}) 
from the transient spin grating experiments.


From Eqs.~(\ref{eq:trans:Gamma}) and (\ref{eq:trans:DeltaGamma}) one
can see that $\tau_{+}$ decreases monotonically as $q$
increases while $\tau_{-}$ has a peak at some wavevector $q_0$. 
Earlier theoretical works, which only considered linear Dresselhaus
term, predicted that the ratio of these two
relaxation times $\tau_{-}/\tau_{+}$ 
and the position of the peak $q_0$ do not vary with
temperature \cite{PhysRevB.70.155308,bernevig:236601}. However, 
Weng {et al.} showed that once the cubic Dresselhaus term is
included, both $\tau_{-}/\tau_{+}$ and $q_0$ depend on
temperature. 
With the cubic Dresselhaus term,
the peak position $q_0$ moves from $q_0=\sqrt{15}m^{\ast}\beta/2$ (which
is independent of temperature) to
$q_0=\sqrt{15}m^{\ast}\hat{\beta}/2=\sqrt{15}m^{\ast}(\beta-\gamma
k^2/2){/2}$. Since $\langle k^2\rangle$ increases with increasing
temperature, thus $q_0$ decreases with it. 
The temperature dependence of $\tau_+/\tau_-$
also originates from the contribution of the cubic Dresselhaus term. 
These theoretical results qualitatively agree with 
experiment ones \cite{weber:076604}.
%
%
%
In additional to the
correct qualitative agreement between the 
theoretical and experimental results, the quantitative accuracy of
results from the kinetic spin Bloch equation approach is also good.
The results are shown in Fig.~\ref{fig:trans:fitting}, in which spin
relaxation times are plotted as function of temperature together with
the experimental data from Ref.~\cite{weber:076604}. 
In the calculation the finite
square well assumption \cite{zhou:045305} was used, and all the
parameters were chosen to be the experimental values if available.  
The only
adjustable parameters were the spin-orbit coupling coefficients
$\gamma$ and $\alpha$. In the calculation, $\gamma$ was chosen to be
$11.4$~eV$\AA^3$ and $13.8$~eV$\AA^3$ for the high
[Fig.~\ref{fig:trans:fitting}(a)] and low mobility
[Fig.~\ref{fig:trans:fitting}(b)] samples respectively and $\alpha$ was
set to be $0.3\beta$, close to the choice in the experimental work
\cite{weber:076604}. 


\subsubsection{
Spin Coulomb drag}

\begin{figure}[htpb]
  \begin{minipage}{0.435\linewidth}
    \includegraphics[width=6cm]{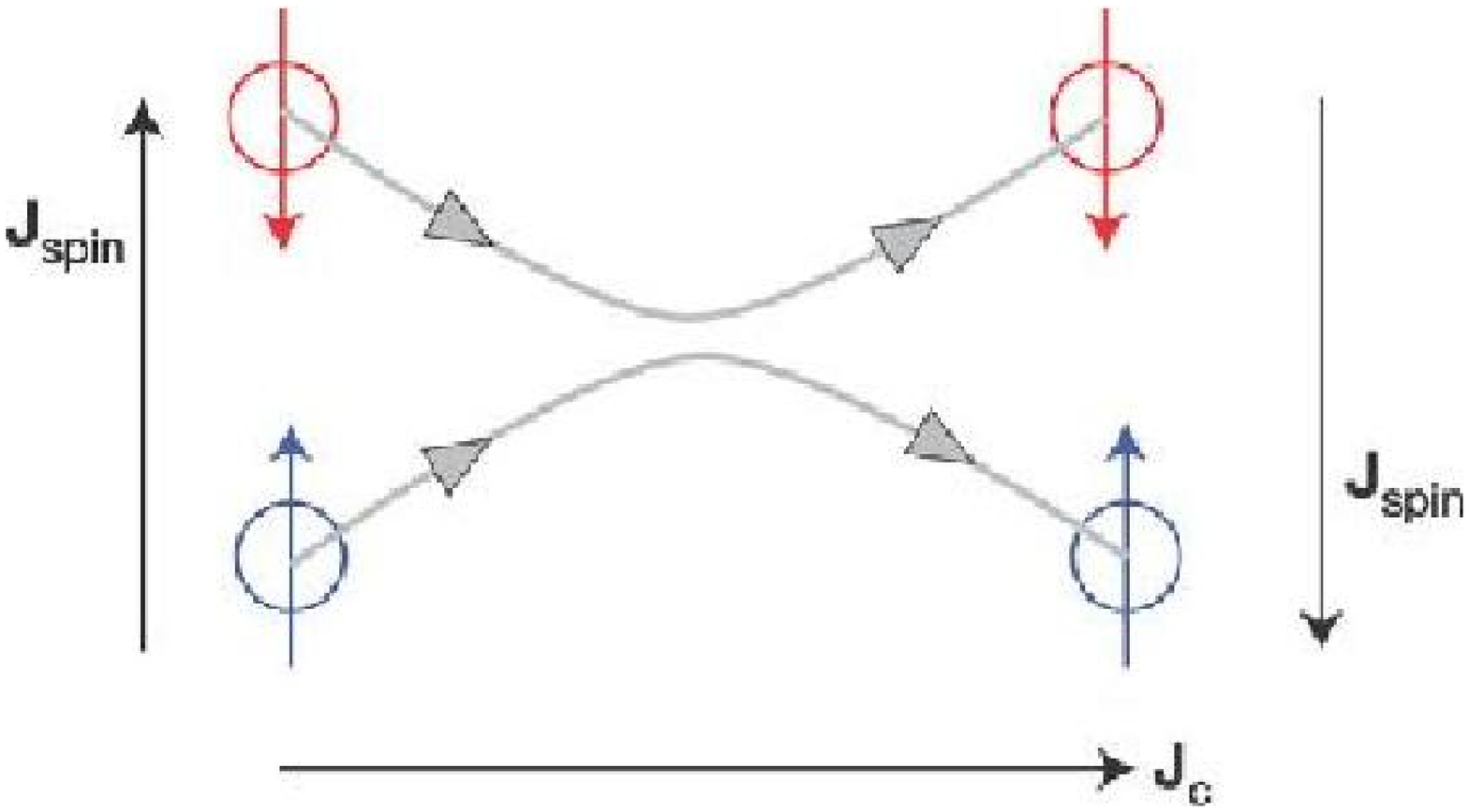}
    \caption{
      A representation of electron-electron Coulomb
      scattering that does not conserve spin-current. Before the
      collision the spin-current, $\mathbf{J}_{\mathtt{spin}}$, is positive; after,
      it is negative. The charge current, $\mathbf{J}_c$, does not
      change. Colors correspond to spin states.
      From Weber et al.~\cite{nature.437.1330}.
    } 
    \label{fig:trans:drag}
  \end{minipage}
  \hspace{1cm}
  \begin{minipage}{0.435\linewidth}
   \centering
   \includegraphics[width=6cm]{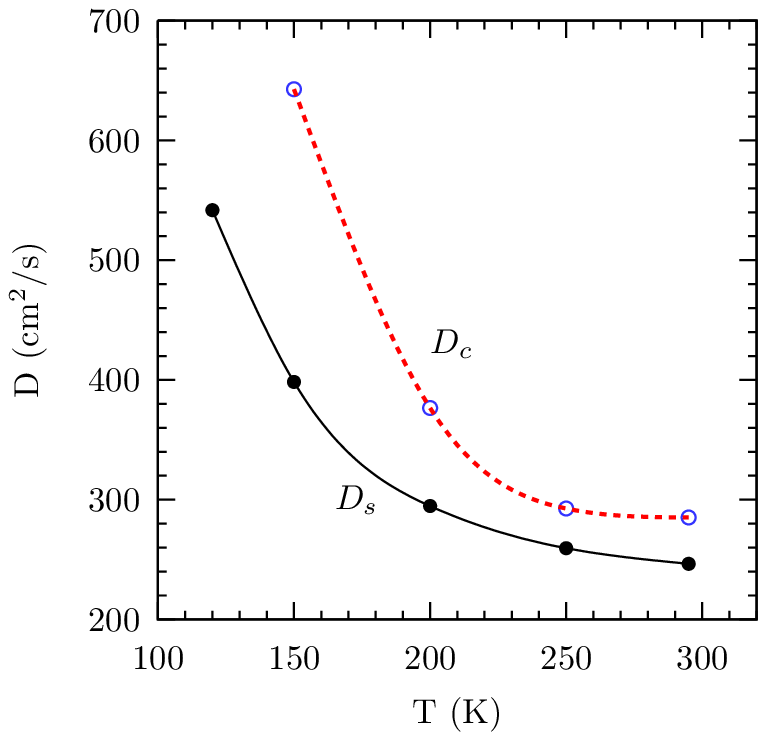}
   \caption{ Diffusion coefficient as a function of temperature
 in $n$-type GaAs (001) quantum well.
     Solid circles: Spin diffusion constant $D_s$ with Coulomb drag;
     Open circles: Charge diffusion constant $D_c$.
     From Weng et al.~\cite{weng:063714}.
   }
   \label{fig:trans:DS}
  \end{minipage}
 \end{figure}

From the evolution of the transient spin grating one can observe 
the Coulomb drag effect on the spin diffusion directly. 
The effect of electron-electron Coulomb scattering on the spin
diffusion coefficient is illustrated in Fig.~\ref{fig:trans:drag}. 
For spin-unpolarized charge diffusion, the electrons move along
the same direction and the Coulomb scattering does not change the
center-of-mass motion, therefore it does not change the charge
diffusion coefficient directly. However, for the spin 
transport, spin-up and -down electrons move against each other.
Therefore the Coulomb scattering between them 
slows down the relative motion of these
two spin species and thus reduces spin diffusion coefficient.  This is the
so-called spin drag or spin Coulomb drag effect
\cite{amico_00,amico_01,flensberg_01,PhysRevB.65.085109,amico_03,amico_04,
amico_04b,jiang:113702,vignale_05,nature.437.1330,tse_07,takahashi_08,
badalyan_08},
which was first
proposed by D'Amico and Vignale 
\cite{amico_00,amico_01,PhysRevB.65.085109,amico_03,amico_04,
amico_04b,vignale_05}.

In Fig.~\ref{fig:trans:DS} the spin diffusion coefficient calculated
in the above mentioned method as a function of temperature was
presented. For comparison, the charge diffusion coefficient, which is
calculated by solving the kinetic spin Bloch equations with the
initial condition being the charge gradient instead of the spin
gradient, is also included.  It is clearly seen from the figure that
$D_s<D_c$.  

From the figure one can
tell that in the relative high temperature regime, as the temperature
increases, the spin and charge diffusion coefficients 
and their difference all decrease.
Therefore the Coulomb drag is
stronger in the low temperature regime. However, even at room
temperature the Coulomb drag is still strong enough to reduce the
diffusion coefficient by 30\%.  These results quantitatively agree
with those of Refs.~\cite{PhysRevB.65.085109,amico_03}.
It should be pointed out that the reduction of spin diffusion
coefficient mostly comes from the Coulomb drag. The spin-orbit
coupling only has small effect on the diffusion coefficient since the
spin-orbit coupling is very 
small compared to the Fermi energy. The numerical result shows that
the removal of the spin-orbit coupling only changes spin diffusion
coefficient by one
tenth percent for the system studied, which is consistent with the
result from the linear response theory \cite{tse_07}.

\begin{figure}
  \begin{minipage}{0.435\linewidth}
  \centering
  \includegraphics[width=4cm,angle=-90]{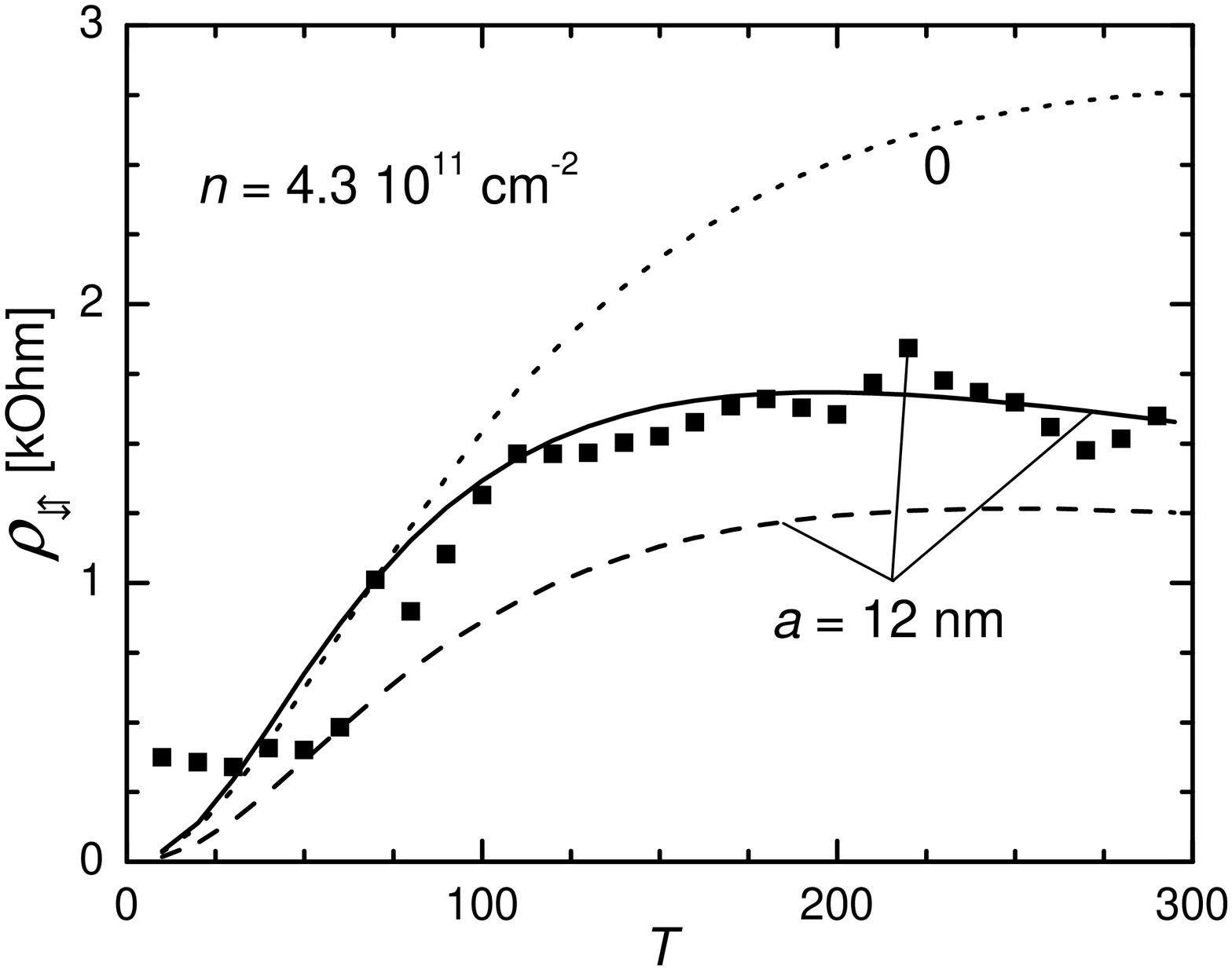}
  \caption{
    Spin drag resistivity $\rho_{\uparrow\downarrow}$ as a 
    function of temperature for
in $n$-type GaAs (001)  quantum well with
    the electron density $n=4.3\times 10^{11}$~cm$^{-2}$. The dashed
    and dotted curves correspond to $\rho_{\uparrow\downarrow}$,
    calculated within the random phase approximation for quantum well
    width $a=0$ and $12$ nm, respectively.  
    The solid curve correspond to $\rho_{\uparrow\downarrow}$, 
    calculated beyond the random phase approximation 
    for $a=12$ nm. 
    The symbols represent $\rho_{\uparrow\downarrow}$, deduced from 
    the experimental data for $a=12$ nm of
    Ref.~\cite{nature.437.1330}. 
    From Badalyan et al.~\cite{badalyan_08}.
  }
  \label{fig:trans:drag_rho}    
  \end{minipage}\hspace{1cm}
  \begin{minipage}{0.435\linewidth}
  \centering
  \includegraphics[width=6cm]{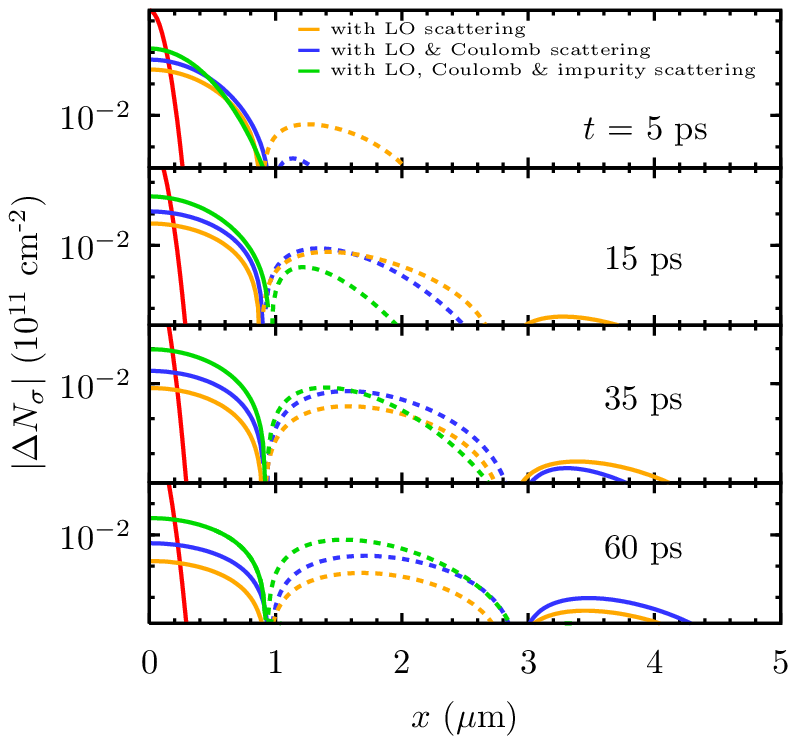}
  \caption{ The absolute values of the spin imbalance
    $|\Delta N_{\sigma}|$ {\sl vs.}  the position $x$ at certain times $t$
in $n$-type GaAs  (001) quantum well
    for the cases with only the electron-phonon scattering (orange);
    the electron-phonon and electron-electron scattering (blue); and
    the electron-phonon, electron-electron and electron-impurity
    scattering (green). The red curve shows the initial spin
    pulse. $\Delta N_{\sigma}(0)=10^{11}$\ cm$^{-2}$, $\delta x=0.1$\ $\mu$m
    and $N_i=0.5N_e$. Solid curves: $\Delta N > 0$; Dashed curves:
    $\Delta N < 0$. 
    From Jiang {et al.} \cite{jiang:113702}.
}
  \label{fig:trans:drag-dynamics}    
  \end{minipage}
\end{figure}

Phenomenologically, the spin Coulomb drag modifies 
Ohm's law as
$\mathbf{E}_{\eta}=\sum_{\eta'}\rho_{\eta\eta'}\mathbf{j}_{\mathbf{\eta'}}$, 
in which $\mathbf{E}_{\eta}$ is the ``electric'' field on 
electrons with spin $\eta$. The field also includes the contribution
from the gradient of the local chemical potential, which can be spin
dependent. The off-diagonal terms of the resistivity $\rho_{\up\down}$
correspond to the spin drag resistivity, defined by 
$\rho_{\up\down}=E_{\up}/{j}_{\down}$ when $j_{\up}=0$. 
Using the generalized Einstein relations with the spin Coulomb drag
effect taken into account, the spin diffusion coefficient
reads \cite{amico_00,amico_01,PhysRevB.65.085109,nature.437.1330} 
\begin{equation}
  {D_s\over D_c} = {\chi_0 \over \chi_s}{1\over
    1+|\rho_{\up\down}|/\rho}, 
\label{eq:trans:DS_drag}
\end{equation}
where $D_c$ and $\rho$ are the charge diffusion coefficient and
mobility, respectively and
$\chi_0/\chi_s$ is the many-body enhancement of spin
susceptibility of the electron gas. In normal electron density and
temperature, $\chi_0/\chi_s$ is close to 1
\cite{yarlagadda_89,kwon_94,giuliani_05,nature.437.1330}. 
Spin drag resistivity can be calculated from the linear response
theory. In two dimension, it takes the form
of \cite{amico_00,amico_01,PhysRevB.65.085109,badalyan_08}:
\begin{equation}
\rho_{\up\down}={-1\over 2 e^2n_{\up}n_{\down}k_BT}
\int {d^2q\over (2\pi)^2} q^2\int_0^{\infty}{d\omega\over 2\pi}
|V_{\up\down}(\mathbf{q},\omega)|^2
{\mathtt{Im}\Pi_{0\up}(\mathbf{q},\omega)
  \mathtt{Im}\Pi_{0\down}(\mathbf{q},\omega) \over
  \sinh^2(\omega/2T)}, 
\label{eq:trans:transresistivity}
\end{equation} 
in which 
$V_{\up\down}(\mathbf{q},\omega)$ is the dynamically screened
effective Coulomb interaction between spin $\up$ and
$\down$ electrons. $\Pi_{0\eta}$ is the non-interacting spin-resolved
polarization function. Spin-orbit coupling is shown to have small
enhancement to spin drag resistivity \cite{tse_07}. 
$\rho_{\up\down}$ increases with temperature
as $(T/T_F)^2$ in low temperature regime ($T\ll T_F$) but decreases with 
temperature as $1/T$ in high temperature regime ($T\gg T_F$)
\cite{flensberg_01,amico_03}. The quantitative comparison of
$\rho_{\up\down}$ between theoretical calculation and experiment
results is shown in Fig.~\ref{fig:trans:drag_rho}. 
With the modification of finite quantum well width and local field
correlations, $\rho_{\up\down}$ calculated from
Eq.~(\ref{eq:trans:transresistivity}) is in good agreement with 
the experiment data extracted from Ref.~\cite{nature.437.1330} using
Eq.~(\ref{eq:trans:DS_drag}) \cite{badalyan_08}.

Jiang {et al.} showed that the spin 
Coulomb drag also plays an important role in the transient spin
transport \cite{jiang:113702}. 
In Fig.~\ref{fig:trans:drag-dynamics}, the spacial profiles of the spin
imbalance at different times are plotted for a Gaussian spin pulse with
different scattering. It is clearly seen that when the Coulomb
scattering is included, both the diffusion of the spin pulse and the
decay of the spin polarization at the center of the pulse become much
slower. When there is only electron-phonon scattering, a third peak in
$\Delta N_{\sigma}$  appears after about 15 picosecond
diffusion. The Coulomb scattering delays the appearance of the
third peak to 35 picoseconds. Moreover, since the Coulomb scattering
reduces the spin dephasing, the magnitudes of the spin
polarization and the peaks with the Coulomb scattering are higher than
those without. 

\begin{figure}[htbp]
\end{figure}

\subsubsection{Infinite spin injection length in Sec.~\ref{sec:trans:anisotropy}
revisited from the point of view of transient spin grating}

For most of the cases, $L_s$ determined by Eq.~(\ref{eq:trans:ls}) 
is in the same order of $\sqrt{D_s\tau_s}$,
although the former is usually larger. 
However, there are some special cases, say 
the denominator in Eq.~(\ref{eq:trans:ls}) approaches zero, 
where $L_s$ can be  
orders of magnitude larger than $\sqrt{D_s\tau_s}$.
Specifically, according to Cheng {et al.} \cite{cheng:205328},
reviewed in Sec.~\ref{sec:trans:anisotropy}, 
when the spin  injection/transport direction
is  along $[\bar{1}10]$ in (001) quantum well, $L_s$ becomes larger and larger
as $\alpha$ approaches $\beta$, regardless of the direction
of spin polarization.
In the limit of $\alpha=\beta$, 
the spin injection length  tends to infinity when the cubic
Dresselhaus term is ignored and the spin polarization oscillates
with a spacial period of $2\pi/(2m^{\ast}\beta)$ without any decay
\cite{cheng:205328}.  

The non-decay spin oscillation model can also be understood from the
point of view of transient spin grating. From the simplified solution
in Sec.~\ref{sec:trans:simplified_grating},  
one obtains that the fitting
parameters for $\Delta\Gamma$ are
$d\rightarrow 0$ and $c\rightarrow2\sqrt{D_s/\tau'_s}$
when $\alpha\rightarrow\beta$ in the system without the cubic
Dresselhaus term. 
Consequently $\tau_{\pm}=(\sqrt{D}q\pm
1/\sqrt{\tau_s})^{-2}$ when $\alpha=\beta$. It is then straightforward
to see that $\tau_-$ becomes infinite  provided
$q=q_0=1/\sqrt{D\tau_s}=2m^{\ast}\beta$. Therefore the steady-state spin
injection along $[\bar{1}10]$ axis is dominated by this non-decay
transient spin grating 
mode which is responsible for the infinite spin injection length and
the spacial oscillation period $2\pi/q_0=2\pi/(2m^{\ast}\beta)$
\cite{weng:063714}.

\subsubsection{Persistent spin helix state}

\begin{figure}[htpb]
  \begin{minipage}{0.435\linewidth}
    \includegraphics[width=5.5cm]{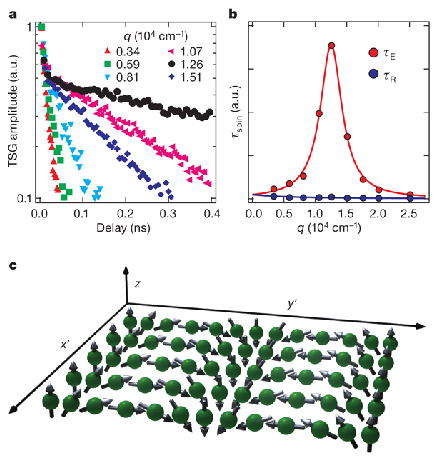}
    \caption{ a, Transient spin grating decay curves at various
      wavevectors, q, for an asymmetrically doped 
 GaAs quantum well with a mixture
      of Rashba and Dresselhaus spin-orbit couplings. b, Lifetimes for
      the spin-orbit-enhanced 
      [$\tau_{\mathtt{E}}$, corresponding to
      $\tau_{-}$ in Eq.~(\ref{eq:trans:decay})]
      and -reduced
      [$\tau_{\mathtt{R}}$, corresponding to
      $\tau_{+}$ in  Eq~(\ref{eq:trans:decay})]
      helix modes extracted from 
      double-exponential fits to the data in a. The solid lines are a
      theoretical fit (see text) using a single set of spin-orbit
      parameters for both helix modes. Error bars (s.d.) are the size
      of the data points. c, Illustration of a helical spin wave,
      which is one of the normal modes. In this picture, $z$ is the
      growth direction [001], and the axes $x'$ and $y'$ respectively
      refer to the $[110]$ and $[1\bar{1}0]$ directions in the plane
      of the quantum well. The green spheres represent electrons whose
      spin directions are given by the arrows.         
      From Koralek et al.~\cite{Koralek:nature458.610}.
    } 
    \label{fig:trans:psh1}
  \end{minipage}
  \hspace{1cm}
  \begin{minipage}{0.435\linewidth}
   \centering
   \includegraphics[width=5.2cm]{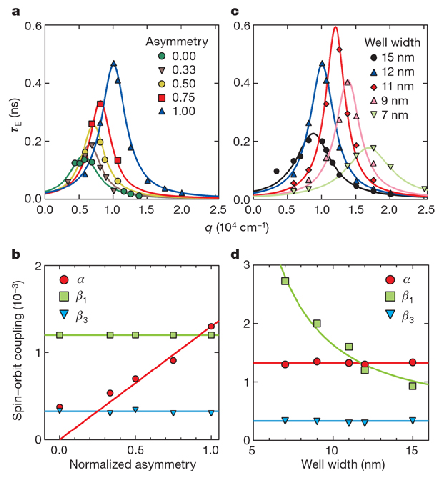}
   \caption{
     a, c, Lifetimes of the enhanced helix mode are shown for 
 GaAs quantum wells
     with varying degrees of doping asymmetry (a) and well width
     (c). The normalized asymmetry is the difference between the
     concentrations of dopants on either side of the well, divided by
     the total dopant concentration. b, d, Plots summarizing the
     spin-orbit parameters from these fits in a and c.  
     In this figure, the spin-orbit coupling strengths are expressed
     as dimensionless quantities by normalizing to the Fermi
     velocity. 
     $\beta_1$ and $\beta_3$ are the coefficients of the linear and
     cubic Dresselhaus terms, corresponding to $\beta$ and $\gamma$ in
     the text, respectively. 
     From Koralek et al.~\cite{Koralek:nature458.610}.
   }
   \label{fig:trans:psh2}
  \end{minipage}
 \end{figure}

The spin
transport model with infinite spin injection length in the
Sec.~\ref{sec:trans:infinite} \cite{cheng:205328}, 
or the non-decay transient spin grating with wavevector
$\mathbf{q_0}=(q_0,0,0)$ in the previous subsection \cite{weng:063714}, 
is related to the persistent spin helix, 
which was first proposed by Bernevig {et al.}
\cite{bernevig:236601,Koralek:nature458.610,Fabian2009,shen_09}. 
Persistent spin helix has an SU(2) symmetry which is robust against the
spin-conserving scattering. 
The SU(2) symmetry of the persistent spin helix 
is generated by the following operators \cite{bernevig:236601}:
\begin{equation}
  \label{eq:trans:su2}
  S_{\mathbf{q}_0}^{-}=
  \sum_{\mathbf{k}}c^{\dag}_{\mathbf{k};-}
  c_{\mathbf{k}+\mathbf{q}_0;+}\,
  ,\;\;
  S_{\mathbf{q}_0}^{+}=
  \sum_{\mathbf{k}}
  c^{\dag}_{\mathbf{k}+\mathbf{q}_0;+}
  c_{\mathbf{k};-}\,
  ,\;\;
  S_0^z=
  \sum_{\mathbf{k}}
  (c^{\dag}_{\mathbf{k};+}
  c_{\mathbf{k};+}-
  c^{\dag}_{\mathbf{k};-}
  c_{\mathbf{k};-}).
\end{equation}
Here
$c^{\dag}_{\mathbf{k};\lambda}$ ($c_{\mathbf{k};\lambda}$) is the creation
(annihilation) operator of electron with momentum $\mathbf{k}$ and
spin $\lambda$. $\lambda=\pm$ are the index of spin eigenstates with
spin-orbit coupling. It is noted that nonzero $\langle
S_0^z\rangle$ corresponds to spin polarization along $[110]$
direction. While nonzero $\langle S_{\mathbf{q}_0}^{\pm}\rangle$
correspond to spin grating with wave-vector $\mathbf{q}_0$. 
The spin profile of nonzero $\langle S_{\mathbf{q}_0}^{\pm}\rangle$ 
in the real space is helical wave. The out-of-plane mode of the
spin helix is illustrated in Fig.~\ref{fig:trans:psh1}c. 

The generating operators of persistent spin helix obey the
communication relations for angular momentum, 
\begin{equation}
  [S_0^z,S_{\mathbf{q}_0}^{\pm}]=\pm 2 S_{\mathbf{q}_0}^{\pm},\;\;
  [S_{\mathbf{q}_0}^+,S_{\mathbf{q}_0}^-]=S_0^z. 
\end{equation}
Moreover, since $|\mathbf{k};+\rangle$ and
$|\mathbf{k}+\mathbf{q}_0;-\rangle$ are degenerate, these operators also
commute with the electron Hamiltonian. 
This symmetry is robust against
spin conserving scattering, such as electron-impurity, electron-phonon and
electron-electron Coulomb scatterings, since these three operators also
commute with the finite wave-vector particle densities
$\rho_{\mathbf{q}}=\sum_{\sigma}c^{\dag}_{\mathbf{k}+\mathbf{q}\sigma}c_{\mathbf{k}\sigma}$,
which are the components that appear in the scattering Hamiltonian. 
This SU(2) symmetry means that the life time of expectation values of
$S_0^z, S_{\mathbf{q}_0}^{\pm}$ are infinite, and hence the transient
spin grating with $\mathbf{q}_0$ does not decay with time even when
there is spin conserving scattering. In spin transport, this
corresponds to the infinite spin injection length proposed by Cheng
{et al.}
\cite{cheng:205328,weng:063714,shen_09}.  
When the cubic Dresselhaus term is present, the SU(2) symmetry is
broken, the lifetimes of the spin helix become finite, which corresponds
to the finite spin injection length in the presence of the cubic
Dresselhaus term \cite{cheng:205328,weng:063714,shen_09}. 

Recently, Koralek {et al.} have observed 
the emergence of the persistent spin helix experimentally 
in the asymmetrically modulation-doped GaAs quantum wells, where both the
Dresselhaus and Rashba spin-orbit couplings are important, by using
the transient spin-grating spectroscopy 
\cite{Koralek:nature458.610}. Their results are shown in
Figs.~\ref{fig:trans:psh1} and \ref{fig:trans:psh2}. 
The experiment results clearly show that the
transient spin grating signal is in a double exponential decay form
and can be fitted by two lifetimes, $\tau_{\mathtt{E}}$ and $\tau_{\mathtt{R}}$ 
(corresponding to $\tau_{-}$ and $\tau_{+}$ in the previous
subsections). For some special wavevectors, $\tau_{\mathtt{E}}$ is
much larger than $\tau_{\mathtt{R}}$ when the Dresselhaus and the
Rashba terms are comparable. All of these experimental results consist
with the theoretical ones \cite{weber:076604,weng:063714}. By
constructing a series of quantum wells with different well widths (to
tune the linear Dresselhaus spin-orbit coupling) and doping
asymmetries (to change the Rashba spin-orbit coupling), Koralek {et
  al.} demonstrated how the lifetimes of spin helix depend on the
Dresselhaus and Rashba terms as well as temperature. They found that
the optimized condition to observe the peak spin helix lifetime is
$\alpha=\hat{\beta}=\beta-\gamma k^2/2$ instead of $\alpha=\beta$ due
to the existence of 
the cubic Dresselhaus term. This finding is consistent with the
previous theoretical predictions
\cite{cheng:205328,stanescu:125307,weng:063714}.

\section{Summary}
The main concern of this review was to give an overview of the latest
developments of the spin dynamics of semiconductors. We focused on two
main issues: (i) spin relaxation and dephasing and (ii) spin diffusion
and transport. We reviewed both experimental and theoretical
progresses in these two directions. In theoretical part, besides the results
based on the single-particle approach, we reviewed our systematic
investigation based on fully microscopic many-body kinetic spin Bloch
equation approach. Many novel effects predicted by this approach have
been realized experimentally. Also many widely adopted common beliefs
based on the single-particle theory in the literature were shown to be
incorrect. We provided a comprehensive understanding of the spin dynamics of
semiconductors and their confined structures from a fully microscopic
many-body point of view. 

\section*{Acknowledgements}

This work was supported by the National Natural Science Foundation of
China under Grant  No.\ 10725417, the National Basic
Research Program of China under Grant No.\ 2006CB922005 and the
Knowledge Innovation Project of Chinese Academy of Sciences. One of the 
authors (MWW) would like to thank L. J. Sham and J. M. Kikkawa who first brought
this exciting field into his attention back to 1999. He would
also like to thank H. Haug, H. Metiu, J. Fabian, C. Sch\"uller, T. Korn,
X. Marie and E. L. Ivchenko for many discussions and/or collaborations.
Many interesting discussions with H. Z. Zheng, T. S. Lai, Y.
Ji and X. D. Cui on the experimental findings are also
acknowledged. Last but not the least, he would like to thank
his former and current students M. Q. Weng, J. L. Cheng, J. H. Jiang, K.
Shen, P. Zhang and Y. Zhou for their excellent collaborations and P. Zhang
for the typesetting of part of this manuscript and the proofreading reading 
of the whole manuscript.


\end{document}